\documentclass[prd,showpacs,letterpaper,10pt,aps,notitlepage,nofootinbib,groupedaddress]{revtex4-2}
\usepackage{changepage}
\usepackage[utf8]{inputenc}
\usepackage{amsfonts}
\usepackage{amsmath}
\usepackage{graphicx}
\usepackage{placeins}
\usepackage{mathrsfs}
\usepackage{bm}
\usepackage[utf8]{inputenc}
\usepackage{multirow}
\usepackage{slashed}
\usepackage[export]{adjustbox}
\usepackage[caption=false]{subfig}
\usepackage{hyperref}
\usepackage[table]{xcolor}

\hypersetup{colorlinks, linkcolor = [rgb]{0,0.0,0.75}, citecolor = [rgb]{0,0.0,0.75}, urlcolor = [rgb]{0,0.0,0.75}}

\usepackage{adjustbox}
\usepackage{physics}
\usepackage{float}
\usepackage{comment}
\usepackage{orcidlink}

\begin{document}

\title{Properties of the positive and negative parity charm-strange and bottom-strange mesons \texorpdfstring{$\bm{D_s}$}{Ds}, \texorpdfstring{$\bm{D_s^*}$}{Ds*}, \texorpdfstring{$\bm{D_{s0}^*}$}{Ds0*}, \texorpdfstring{$\bm{D_{s1}}$}{Ds1}, \texorpdfstring{$\bm{B_s}$}{Bs}, \texorpdfstring{$\bm{B_s^*}$}{Bs*}, \texorpdfstring{$\bm{B_{s0}^*}$}{Bs0*}, \texorpdfstring{$\bm{B_{s1}}$}{Bs1} from lattice QCD: \\ masses, decay constants, and compositeness}

\newcommand{\wm}{\phantom{-}}

\author{Forrest Guyton \orcidlink{0009-0005-8953-2517}}
\author{Stefan Meinel \orcidlink{0000-0003-1034-1004}}
\affiliation{Department of Physics, University of Arizona, Tucson, AZ 85721, USA}

\begin{abstract}
We present a lattice-QCD determination of properties of the lightest scalar, pseudoscalar, vector, and axial-vector heavy-strange mesons $D_s$, $D_s^*$, $D^*_{s0}$, $D_{s1}$, $B_s$, $B_s^*$, $B^*_{s0}$, and $B_{s1}$. This includes the decay constants of all mesons, and the binding energies and Weinberg compositeness parameters of the positive-parity states. The calculations are performed with domain-wall fermions for the light and strange quarks and anisotropic clover actions for the charm and bottom quarks. We use seven different ensembles of gauge configurations generated by RBC/UKQCD with pion masses ranging from 431 MeV to 139 MeV and lattice spacings ranging from 0.114 fm to 0.073 fm, which allows us to perform combined chiral and continuum extrapolations. For the negative-parity mesons, we obtain $f_{D_s}=251.4(2.1)(0.4)(2.5)\:{\rm MeV}$, $f_{D_s^*}=272.5(4.3)(1.0)(2.7)\:{\rm MeV}$, $f_{B_s}=228.4(5.8)(0.5)(2.3)\:{\rm MeV}$, $f_{B_s^*}=229.2(4.4)(0.8)(2.3)\:{\rm MeV}$, $f_{D_s^*}/f_{D_s}=1.086(14)(11)$, and $f_{B_s^*}/f_{B_s}=1.003(22)(10)$. In the positive-parity sector, the finite-volume energies and decay constants are extracted using the generalized eigenvalue problem from correlation matrices with three different types of hadron interpolating operators, including operators with covariant derivatives and meson-meson-scattering operators at both source and sink. After extrapolation to the physical point, we obtain $f_{D^*_{s0}}=136.6 (8.0)(4.0)(1.4)$ MeV, $f_{D_{s1}}=200 (33)(24)(2)$ MeV,  $f_{B^*_{s0}}=207 (12)(8)(2)$ MeV, and $f_{B_{s1}}= 196 (16)(11)(2)$ MeV. Our results for $f_{B^*_{s0}}$ and $f_{B_{s1}}$ are the first from lattice QCD. L\"uscher's method is used to find the infinite-volume bound-state masses. At the physical point, we obtain $m_{D^*_{s0}}-m_D-m_K=-48 (14)(4)$ MeV, $m_{D_{s1}}-m_{D^*}-m_K=-61 (15)(2)$ MeV, $m_{B^*_{s0}}-m_B-m_K= -69 (13)(4)$ MeV, and $m_{B_{s1}}-m_{B^*}-m_K=-77 (10)(5)$ MeV.  Our analysis shows consistency with an interpretation of the positive-parity states as predominantly molecular. A limitation of our present analysis is that left-hand cuts in the scattering amplitudes that are relevant at low pion mass are neglected, and corrections to the weak-binding limit used in Weinberg's criterion are expected to be large in some cases.
\end{abstract}

\maketitle

\tableofcontents

\FloatBarrier
\section{Introduction}
\FloatBarrier

There has been considerable interest in the positive-parity heavy-strange mesons since the 2003 discovery of the $D^*_{s0}(2317)^\pm$ by the BaBar collaboration \cite{implications,babar}. The $D^*_{s0}(2317)^\pm$ state lies $\sim 45 $ MeV below the $DK$ kinematic threshold, which is far from the $n^{2s+1} \ell_J=1^3 P_0$  $q\bar{q}$ state expected at the time from potential models (see, for example, Ref.~\cite{quarkmodelexpectations}). This discrepancy prompts an interpretation of the $D^*_{s0}(2317)^\pm$ as an exotic state. Further possibly non-$q\bar{q}$ states are known from experiment, including the $J^P=1^+$ $D_{s1}(2460)^\pm$ sitting below and near the $D^* K$ threshold. Table \ref{tab:Ds} summarizes the lowest known $J=0$ and $J=1$ $D_s$ and $B_s$ mesons. The physical states with a given $J^P$ are a mixture of different $^{2s+1} \ell$ states, and the $s$ and $\ell$ values shown in the table are expected from quark models to be the dominant components. However, it has been suggested the $J^P=1^+$ mesons, in particular, are subject to large mixing angles between $1^3 P_1$ and $1^1 P_1$ \cite{Li:2018eqc}.

\begin{table}
\centering
\begin{tabular}{ |c | c | c| c | } 
 \hline
 $n^{2s+1} \ell_J$  & $J^P$ & State & Mass\\
 \hline
 $1^1 S_0$ & $0^-$ & $D_s^\pm$ & $1968.35 \pm 0.07$ MeV\\
 $1^3 S_1$ & $1^-$ &  $D_s^{*\pm}$ & $2106.6 \pm 3.4$ MeV\\
 $1^3 P_0$ & $0^+$ & $D_{s0}^*(2317)^\pm$ & $2317.8 \pm 0.5$ MeV\\
    & $0^+$ & isospin-averaged $D K$ & $2362.8 \pm 0.2$ MeV\\
 $1^3 P_1$ & $1^+$ &  $D_{s1}(2460)^\pm$  & $2459.5 \pm 0.6$ MeV\\
   & $1^+$ & isospin-averaged $D^* K$ & $ 2504.20\pm 0.04$ MeV\\
 $1^1 P_1$ & $1^+$ &  $D_{s1}(2536)^\pm$  & $2535.11 \pm 0.06$ MeV\\
 \hline
\end{tabular}
\quad\quad
 \begin{tabular}{ |c | c | c| c | } 
 \hline
 $n^{2s+1} \ell_J$  & $J^P$ & State & Mass\\
 \hline
 $1^1 S_0$ & $0^-$ &  $B_s^0$ & $5366.91 \pm 0.11$ MeV\\
 $1^3 S_1$ & $1^-$ &  $B_s^{*}$ & $5415.4 \pm 1.4$ MeV\\
$1^3 P_0$ & $0^+$ & $B_{s0}^*$ & \\
  & $0^+$ & isospin-averaged $B K$ & $5775.16 \pm 0.11$ MeV \\
 $1^3 P_1$ & $1^+$ &  $B_{s1}$  & \\
  & $1^+$ & isospin-averaged $B^* K$ & $5820.39 \pm 0.10$ MeV\\
 $1^1 P_1$ & $1^+$ &  $B_{s1}(5830)^0$  & $5828.65 \pm 0.24$ MeV\\
 \hline
\end{tabular}

\caption{Lowest $J=0$ and $J=1$ $D_s$ and $B_s$ states predicted by the quark model, as well as associated kinematic thresholds,  and experimentally observed meson masses where available~\cite{ParticleDataGroup:2026mpi}. Rows labeled ``isospin averaged $H^{(*)}K$'' are the isospin-averaged kinematic thresholds for $H^{(*)}$-$K$ scattering. The associated pseudoscalar(vector)-pseudoscalar scattering state has total positive parity.}
\label{tab:Ds}
\end{table}

More specifically, the $D^*_{s0}(2317)$ and $D_{s1}(2460)$ states have been identified as candidates for a molecular structure \cite{implications,guo}, that is, a four-quark state consisting of two shallowly bound $q\bar{q}$ color-singlet pairs. This is based on key features such as nearness to the $D^{(*)}K$ threshold and a relatively strong coupling to the meson-meson scattering-state channel \cite{implications}, and is also supported by the application of Weinberg's compositeness criterion \cite{guo,weinberg}. 
A $D^{(*)} K$ molecular structure also naturally explains the relation 
$m_{D_{s1}}-m_{D^*_{s0}} \approx m_{D^*}- m_D$, which in a simple $\bar{q}q$ model would be unexpected, given the different $\ell$ quantum numbers. In the molecular interpretation, the $D_{s1}-D^*_{s0}$ splitting can be thought of as caused primarily by the splitting between the $D^*$ and $D$ components. 

Belle II has recently measured the branching fractions of the radiative $D^*_{s0}(2317)^\pm \rightarrow D_s^{*\pm} \gamma$ decay and the isospin-violating strong decay $D^*_{s0}(2317)^\pm \rightarrow D_s^\pm \pi^0$ \cite{Belle-II:2025dzk}. The ratio of these branching fractions may be used to characterize the structure of the $D^*_{s0}(2317)^\pm$ \cite{Faessler:2007gv}. The Belle II measurement $\frac{\mathcal{B}(D^*_{s0}(2317)^\pm \rightarrow D_s^{*\pm} \gamma)}{\mathcal{B}(D^*_{s0}(2317)^\pm \rightarrow D_s^\pm \pi^0)}\approx 7\%$ suggests neither a purely molecular interpretation nor a pure $q\bar{q}$-state, but indicates an admixture \cite{Belle-II:2025dzk}. Additionally, a recent amplitude analysis of 
$D_{s1}(2460)^+\to D_s^+ \pi^+\pi^-$ and $D_{s1}(2536)^+\to D_s^+ \pi^+\pi^-$ decays by LHCb points to a sizable molecular component of the $D_{s1}(2460)$ that is absent for the $D_{s1}(2536)$ \cite{LHCb:2026wks}.

Lattice studies of the positive-parity $D_s$ spectrum date back to 1997, preceding the $D^*_{s0}(2317)^\pm$ discovery \cite{earlywork, Dougall:2003hv, Bali:2003jv, Hein:2000qu, Lewis:2000sv}. However, these and other studies \cite{Kalinowski:2015bwa,Cichy:2016bci} using only quark-antiquark interpolating fields  failed to resolve the below-threshold state. After introducing interpolating fields of the form of two-meson pairs, a number of works have found a spectrum consistent with the BaBar measurement \cite{latticeDs, bali, Alexandrou:2019tmk,Cheung:2020mql}. Non-lattice approaches to these states can be found, for example, in Refs.~\cite{Mehen:2005hc, Dai:2006uz,Guo:2006fu,Altenbuchinger:2013vwa,Wang:2015mxa,Fu:2021wde}.

In the bottom-strange sector, less is known of the positive-parity spectrum from experiment. The $B_{s1}(5830)$, as well as more massive states which have not yet been assigned a spin, have been identified above the $B^* K$ threshold \cite{ParticleDataGroup:2026mpi}, but the expected ground states have not yet been observed. The ground state $B^*_{s0}$ and $B_{s1}$ have been studied on the lattice \cite{Burch:2008qx,Gregory:2010gm,lang,mohler,Hudspith:2026vxt, Wurtz:2015mqa} and with other methods \cite{Cleven:2010aw, Wang:2015mxa, Yang:2022vdb, DiPierro:2001dwf, Bardeen:2003kt,Guo:2006fu,Altenbuchinger:2013vwa,Fu:2021wde,Colangelo:2012xi,Badalian:2007yr,Guo:2021rjv,Cheng:2017oqh,Alhakami:2020vil}. Their proximity to the $B^{(*)}K$ thresholds also raises the possibility of molecular structures.

The internal structure of mesons can also be probed through their decay constants. For the positive-parity heavy-strange mesons, the decay constants that we consider here are defined through\footnote{Throughout this work we adopt the notation $H=D$ or $B$ and $Q=c$ or $b$. We label the positive-parity heavy-strange  $J^P=0^+$ particle $H^*_{s0}$, and the $J^P=1^+$ particle $H_{s1}$.} $\bra{0} J_V^\mu \ket{H^*_{s0}(p)}=i f_{H^*_{s0}} p^\mu$ and $\bra{0} J_A^\mu \ket{H_{s1}(p,\epsilon)}=f_{H_{s1}} m_{H_{s1}} \epsilon^\mu$, where $J_V^\mu=\bar{Q}\gamma^\mu s$ and $J_A^\mu=\bar{Q}\gamma^\mu\gamma_5 s$.
One may expect $f_{D^*_{s0}}/f_{D_s}<1$ due to a molecular $D^*_{s0}$ being less compact than a $\bar{q} q$ state \cite{Herdoiza:2006qv}. Apart from direct calculations, discussed further below, the decay constants $f_{D^*_{s0}}$ and $f_{D_{s1}}$ have also been extracted from $B \to D D_{sJ}$ decays using factorization, yielding $f_{D^*_{s0}}/f_{D_{s1}}\sim 0.5$ \cite{Hwang:2004kga}, which deviates significantly from the heavy-quark symmetry prediction $f_{H^*_{s0}} = f_{H_{s1}}$. Such large deviations might be expected, given the orbitally excited nature of the positive-parity system and the relatively small size of the charm quark mass. 

The masses and decay constants of heavy-strange mesons also appear in dispersive bounds on $Q\to s$ semileptonic form factors. 
Currents for $b\rightarrow s$ (or $c\rightarrow s$) transitions have overlap with heavy-strange two-meson scattering pairs, and the form factors inherit the pole structure of the $B^{(*)} K$ ($D^{(*)} K$) amplitudes. Precise knowledge of the pole locations is helpful in parameterizing these form factors, and dispersive bounds on the shapes of the form factors can be strengthened by subtracting the single-particle contributions, which are proportional to the squares of the decay constants (see, e.g., Refs.~\cite{dispersivebounds,Amhis:2022vcd,Gubernari:2023puw,Gubernari:2026sqc,Farrell:2026swf} for recent applications to $b\to s$ transitions). Improving the precision of $b\to s$ form factors is particularly important in light of observed discrepancies between SM predictions and LHC measurements of $b\to s \ell^+\ell^-$ branching fractions and angular observables \cite{banomolies}.
The decay constants are known from lattice QCD with high precision for ground-state pseudoscalar heavy-strange mesons \cite{FlavourLatticeAveragingGroupFLAG:2024oxs}, but are less well known for the other quantum numbers, in particular for the positive-parity case. 

The positive-parity decay constants $f_{D^*_{s0}}$ and $f_{D_{s1}}$ have been studied using several continuum approaches \cite{Li:2018eqc, Wang:2015mxa, DiPierro:2001dwf, Cheng:2003kg, Bardeen:2003kt, Narison:2003td, Cheng:2003sm, Hwang:2004kga, Mehen:2005hc, Colangelo:2005hv, Cheng:2006dm, Veseli:1996kn}, and there is one lattice-QCD calculation \cite{bali}. The bottom-strange positive-parity decay constants $f_{B^*_{s0}}$ and $f_{B_{s1}}$ have been estimated using sum rules, the Bethe-Salpeter equation, and the ``mock-meson approach'' \cite{sr, Li:2018eqc, Wang:2007tu, Wang:2015mxa, Pullin:2021ebn}, but there appears to be no published lattice calculation.

Here, we present a comprehensive new lattice-QCD analysis of the positive and negative parity heavy-strange mesons $D_s$, $D_s^*$, $D^*_{s0}$, $D_{s1}$, $B_s$, $B_s^*$, $B^*_{s0}$, and $B_{s1}$. This includes the decay constants of all of these mesons, and the binding energies and Weinberg compositeness parameters of the positive-parity states.
The calculations are performed in the exact isospin limit with 2+1 flavors of domain-wall fermions and using nonperturbatively tuned anisotropic clover actions for the charm and bottom quarks. We use seven different ensembles of gauge configurations generated by RBC/UKQCD with pion masses ranging from 431 MeV to 139 MeV and lattice spacings ranging from 0.114 fm to 0.073 fm, which allows us to perform combined chiral and continuum extrapolations. 
In the positive-parity sector, the finite-volume energies and decay constants are extracted using the generalized eigenvalue problem from correlation matrices with three different types of hadron interpolating operators, including operators with covariant derivatives and meson-meson-scattering operators at both source and sink. L\"uscher's method is used to determine the $H^{(*)}K$ scattering amplitudes, which we parametrize using effective-range expansions (ERE). From the effective-range parameters, we obtain the infinite-volume bound-state masses and evaluate the couplings and compositeness criteria.

The remainder of this paper is organized as follows. Section \ref{sec:lattice} contains details on the lattice actions and parameters. Section \ref{sec:negparity} presents our analysis of the decay constants of the negative-parity mesons. This is followed by several sections on the positive-parity mesons: Sec.~\ref{sec:opbasis} on the operator basis and correlators, Sec.~\ref{sec:data} on the extractions of the finite-volume spectra and decay constants, Sec.~\ref{sec:luscher} on the application of L\"uscher's method, Sec.~\ref{sec:chiralcontinuum} on the chiral-continuum extrapolations of the binding energies, decay constants, and ERE parameters, Sec.~\ref{sec:heavyquark} on heavy-quark symmetry, and Sec.~\ref{sec:molecular} discussing the compositeness. We give conclusions in Sec.~\ref{sec:conclusions}. Appendix \ref{sec:symcorr} presents an alternative analysis without the meson-meson operators. Appendix \ref{sec:C44} gives details on the implementation of the correlation matrix element with meson-meson operators at both source and sink. Additional plots of correlators and chiral-continuum extrapolations are included in Appendix \ref{sec:additional}.

\section{Lattice Action and Ensemble Details}
\label{sec:lattice}
Our computations use gauge configurations produced by the RBC/UKQCD collaboration with the Iwasaki gauge action and 2+1 flavors of dynamical domain-wall fermions \cite{RBC-UKQCD:2008mhs,RBC:2010qam,ukqcd, boyle}.
Table \ref{tab:ensembles} contains the lattice spacings, sizes, light and (partially quenched) strange-quark masses for each of the seven ensembles. The C00078 and F1M ensembles use the M{\"o}bius domain-wall fermion formulation \cite{mobius}, and the others use the formulation of \textit{Shamir} \cite{shamir}. Ensemble C00078 has a near-physical pion mass, and all ensembles have large volumes, corresponding to $3.86 \lesssim m_\pi L \lesssim 6.09$. Note that C005LV is simply a larger volume version of C005, which provides a convenient test of volume dependence.
Correlation functions are constructed via covariant approximation averaging \cite{AMA} over different source locations, using a total of 
$N_{\rm ex}$ and $N_{\rm sl}$ ``exact'' and ``sloppy'' samples on each ensemble (a lower number of samples is used for the correlation-matrix element with two-meson operators at both source and sink, as discussed in Sec.~\ref{sec:opbasis}).
The sloppy samples were obtained using propagators with reduced conjugate-gradient iteration count, combined with low-mode deflation for the light quarks \cite{Meinel:2016dqj,Meinel:2020owd,Meinel:2021rbm,Meinel:2023wyg}.

\begin{table}[h]
    \centering
    {\footnotesize
\begin{tabular}{|l|c|c|c|c|c|c|c|c|c|}
    \hline
    Label & $N_s^3\times N_t $  & $a^{-1}$ [\text{GeV}]  & DW Type  &  $m_{u,d}$ &  $m_\pi$ (\text{GeV}) & $m_{s}^{(\mathrm{sea})}$ 
    & $m_{s}^{(\mathrm{val})}$  & $N_{\rm ex}$ & $N_{\rm sl}$ \\
    \hline
    C00078 & $48^3\times96$  & $1.7295(38)$ & M{\"o}bius   & $0.00078$  & $0.13917(35)$  & $0.0362$  & $0.0362$  & 158  & 5056 \\
    C005LV & $32^3\times64$  & $1.7848(50)$ & Shamir  & $0.005$    & $0.3398(12)$   & $0.04$    & $0.0323$  & 186 & 5022 \\
    C005   & $24^3\times64$  & $1.7848(50)$ & Shamir  & $0.005$    & $0.3398(12)$   & $0.04$    & $0.0323$  & 311 & 9952 \\
    C01    & $24^3\times 64$ & $1.7848(50)$ & Shamir  & $0.01$     &  $0.4312(13)$       & $0.04$    & $0.0323$  & 283 & 9056 \\
    F004   & $32^3\times64$  & $2.3833(86)$ & Shamir  & $0.004$    & $0.3036(14)$   & $0.03$    & $0.0248$  & 251 & 8032 \\
    F006   & $32^3\times64$  & $2.3833(86)$ & Shamir  & $0.006$    & $0.3607(16)$   & $0.03$    & $0.0248$  & 445 & 14240 \\
    F1M    & $48^3\times96$  & $2.708(10)$ & M{\"o}bius  & $0.002144$ & $0.2320(10)$   & $0.02144$ & $0.02217$ & 226 & 7232 \\
    \hline
    \end{tabular}}
    \caption{Parameters of the ensembles \cite{ukqcd, boyle} and domain-wall propagators used in this work. Quark masses are given in lattice units.}
    \label{tab:ensembles}
    \end{table}

\begin{table}
 \begin{tabular}{|lcccccc|}
\hline
 & $\wm a m_Q^{(c)}$ & $\nu^{(c)}$ & $c_{E,B}^{(c)}$ & $\wm a m_Q^{(b)}$ & $\nu^{(b)}$ & $c_{E,B}^{(b)}$  \\
\hline
C00078   & $\wm0.27514$        & $1.1883$ & $2.0712$ & $8.1476$ & $3.3743$ & $5.3944$  \\
C005(LV), C01 & $\wm0.15410$ & $1.2004$ & $1.8407$    & $7.3258$ & $3.1918$ & $4.9625$ \\
F004, F006    & $-0.05167$   & $1.1021$ & $1.4483$   & $3.2823$   & $2.0600$ & $2.7960$   \\
F1M           & $-0.05874$  & $1.0941$ & $1.5345$  & $2.3867$  & $1.8323$ & $2.4262$ \\
\hline
\end{tabular}
\caption{\label{tab:HQ_params}Parameters of the heavy-quark actions for charm and bottom \cite{Meinel:2023wyg}. }
\end{table}

The heavy quarks are handled with an anisotropic clover action tuned to remove discretization errors \cite{El-Khadra:1996wdx,Chen:2000ej,Aoki:2001ra,Aoki:2003dg,Christ:2006us,Lin:2006ur,RBC:2012pds}, which allows us to work directly at the physical charm and bottom masses for all lattice spacings. Our action is of the form 

\begin{equation}
    S_Q=a^4\!\sum_x \bar{Q}\! \left[ m_Q + \gamma_0 \nabla_0 -\frac{a}{2} \nabla_0^{(2)} + \nu \!\sum_{i=1}^3  \left(\gamma_i \nabla_i - \frac{a}{2} \nabla_i^{(2)}\right) - c_E \frac{a}{2} \!\sum_{i=1}^3 \sigma_{0i} F_{0i} - c_B \frac{a}{4} \!\sum_{i,j=1}^3 \sigma_{ij} F_{ij} \right]\! Q.
\end{equation}
The bare mass $am_Q$, anisotropy coefficient $\nu$, and clover coefficients $c_B=c_E$ are tuned by matching the $D_s^{(*)}$ and $B_s^{(*)}$ dispersion relations and hyperfine splittings \cite{RBC:2012pds,Meinel:2023wyg}. The values of these parameters were determined in Ref.~\cite{Meinel:2023wyg} (which labels the ``C00078'' ensemble ``CP'') and are shown again here in Table \ref{tab:HQ_params}.

The currents used to calculate the decay constants are renormalized according to the mostly-nonperturbative method introduced in Refs.~\cite{Hashimoto:1999yp, El-Khadra:2001wco}, and have the form
\begin{subequations}\label{eq:currents}
\begin{align}
    J_{V_0}&=\sqrt{Z_V^{ss}Z_V^{QQ}}\rho_{V_0}  \left[\bar{s} \gamma_0 Q + 2a \left(c^R_{V_0} \Bar{s} \gamma_0 \gamma_j \nabla_j Q + c^L_{V_0} \Bar{s} \overleftarrow{\nabla}_j \gamma_0 \gamma_j  Q    \right)\right],\\
    J_{A_0}&=\sqrt{Z_V^{ss}Z_V^{QQ}}\rho_{A_0}  \left[\bar{s} \gamma_0\gamma_5 Q + 2a \left(c^R_{A_0} \Bar{s} \gamma_0\gamma_5 \gamma_j \nabla_j Q + c^L_{A_0} \Bar{s} \overleftarrow{\nabla}_j \gamma_0\gamma_5 \gamma_j  Q    \right)\right],\\
    J_{V_i}&=\sqrt{Z_V^{ss}Z_V^{QQ}}\rho_{V_i}  \Big[\bar{s} \gamma_i  Q  + 2a \left(c^R_{V_i} \Bar{s} \gamma_i  \gamma_j \nabla_j Q + c^L_{V_i} \Bar{s} \overleftarrow{\nabla}_j \gamma_i  \gamma_j  Q  + d^R_{V_i} \bar{s}  \overrightarrow{\nabla_i} Q +  d^L_{V_i} \bar{s} \overleftarrow{\nabla}_i  Q \right)\Big], \\
    J_{A_i}&=\sqrt{Z_V^{ss}Z_V^{QQ}}\rho_{A_i}  \Big[\bar{s} \gamma_i \gamma_5 Q  + 2a \left(c^R_{A_i} \Bar{s} \gamma_i \gamma_5 \gamma_j \nabla_j Q + c^L_{A_i} \Bar{s} \overleftarrow{\nabla}_j \gamma_i \gamma_5 \gamma_j  Q  + d^R_{A_i} \bar{s} \gamma_5 \overrightarrow{\nabla_i} Q +  d^L_{A_i} \bar{s} \overleftarrow{\nabla}_i \gamma_5 Q \right)\Big].
\end{align}
\end{subequations}
Here, the factors $Z_V^{qq}$ are computed nonperturbatively \cite{ukqcd,Boyle:2017jwu,Meinel:2016dqj,Meinel:2021rbm,Meinel:2023wyg, Marshall:2024pfg}, while the residual matching factors $\rho_J$ and $\mathcal{O}(a)$-improvement terms are calculated to one loop in lattice perturbation theory \cite{detmold, lehner}. The values are given in Table \ref{tab:matchingfactors}.

\begin{table}
\begin{center}
\small
\begin{tabular}{|cllllllll|}
\hline
 Parameter         & \hspace{2ex}            & \hspace{1ex} C00078   & \hspace{2ex} & C005(LV), C01     & \hspace{2ex} & \hspace{1ex} F004, F006   & \hspace{2ex} &  \hspace{1ex} F1M \\
 \hline 
 \hline
 \multicolumn{9}{|c|}{charm}\\
 \hline
 $\rho_{V_0}=\rho_{A_0}$  & &  $\wm1.0027(11)$       & & $\wm1.0027(11)$      & &  $\wm1.00195(59)$    & &  $\wm1.00152(73)$      \\[0.5ex] 
 $\rho_{V_i}=\rho_{A_i}$  & &  $\wm0.9948(20)$      & & $\wm0.9948(20)$      & &  $\wm 0.99675(99)$    & &  $\wm0.9978(15)$       \\[0.5ex] 
 $c_{V_0}^R=c_{A_0}^R$    & &  $\wm0.0402(72)$      & & $\wm0.0402(72)$       & &  $\wm 0.0353(53)$     & &  $\wm0.0326(60)$       \\[0.5ex] 
 $c_{V_0}^L=c_{A_0}^L$    & &  $-0.0048(19)$      & & $-0.0048(19)$          & &  $-0.00270(82)$        & &  $-0.0016(14)$         \\[0.5ex] 
 $c_{V_i}^R=c_{A_i}^R$    & &   $\wm0.0346(50)$    & & $\wm0.0346(50)$       & &  $\wm0.0283(32)$      & &  $\wm0.0249(47)$       \\[0.5ex] 
 $c_{V_i}^L=c_{A_i}^L$    & & $\wm0.00012(23)$       & & $\wm0.00012(23)$      & &  $\wm0.00040(12)$     & &  $\wm0.00055(19)$      \\[0.5ex] 
 $d_{V_i}^R=-d_{A_i}^R$   & &  $-0.0041(16)$       & & $-0.0041(16)$         & &  $-0.0039(12)$        & &  $-0.0038(12)$         \\[0.5ex] 
 $d_{V_i}^L=-d_{A_i}^L$   & &   $\wm0.00210(82)$      & & $\wm0.00210(82)$          & &  $\wm0.00260(79)$      & &  $\wm0.00287(84)$      \\[0.5ex] 
 $Z_V^{cc}$ & & $\wm1.40756(17)$  & & $\wm1.35761(16)$  & & $\wm1.160978(74)$  & & $\wm1.112316(61)$  \\[0.5ex]
 $Z_V^{ss}$ & & $\wm0.71076(25)$  & & $\wm0.71273(26)$  & & $\wm0.7440(18)$   & &  $\wm0.7639(42)$ \\[0.5ex]
\hline
 \hline 
 \multicolumn{9}{|c|}{bottom}\\
 \hline
 $\rho_{V_0}=\rho_{A_0}$  & &  $\wm1.027(10)$     & & $\wm1.027(10)$       & &  $\wm 1.0166(51)$      & &  $\wm 1.0112(74)$     \\[0.5ex] 
 $\rho_{V_i}=\rho_{A_i}$  & &  $\wm0.9972(11)$   & & $\wm0.9972(11)$      & &  $\wm 0.9940(18)$      & &  $\wm 0.9922(25)$     \\[0.5ex] 
 $c_{V_0}^R=c_{A_0}^R$    & &  $\wm0.0558(73)$   & & $\wm0.0558(73)$      & &  $\wm 0.0547(57)$      & &  $\wm 0.0541(58)$     \\[0.5ex] 
 $c_{V_0}^L=c_{A_0}^L$    & & $-0.0099(39)$    & & $-0.0099(39)$        & &  $-0.0095(29)$         & &  $-0.0093(29)$        \\[0.5ex] 
 $c_{V_i}^R=c_{A_i}^R$    & &  $\wm0.0485(45)$      & & $\wm0.0485(45)$      & &  $\wm 0.0480(37)$      & &  $\wm 0.0477(37)$     \\[0.5ex] 
 $c_{V_i}^L=c_{A_i}^L$    & &  $-0.0033(13)$    & & $-0.0033(13)$        & &  $-0.00200(61)$        & &  $-0.00129(93)$       \\[0.5ex] 
 $d_{V_i}^R=-d_{A_i}^R$   & &  $-0.00079(31)$    & & $-0.00079(31)$       & &  $-0.00120(37)$        & &  $-0.00142(43)$       \\[0.5ex] 
 $d_{V_i}^L=-d_{A_i}^L$   & &  $\wm0.00180(70)$    & & $\wm0.00180(70)$     & &  $\wm 0.00047(14)$     & &  $-0.00026(74)$       \\[0.5ex] 
$Z_V^{bb}$ & & $\wm9.9128(81)$  & & $\wm9.0631(84)$  & & $\wm4.7449(21)$  & & $\wm3.7777(23)$  \\[0.5ex]
 $Z_V^{ss}$ & & $\wm0.71076(25)$  & & $\wm0.71273(26)$  & & $\wm0.7440(18)$  & & $\wm0.7639(42)$  \\[0.5ex]
 \hline
\end{tabular}\vspace{-2ex}
\end{center}
\caption{\label{tab:matchingfactors}The residual matching factors, $\mathcal{O}(a)$-improvement coefficients, and nonperturbative renormalization factors of the flavor-conserving temporal vector currents \cite{lehner,ukqcd,Boyle:2017jwu,Meinel:2016dqj,Meinel:2021rbm,Meinel:2023wyg,Marshall:2024pfg,Farrell:2025gis,Farrell:2026swf}. See Refs.~\cite{Farrell:2025gis,Farrell:2026swf} for details on the perturbative uncertainty estimates. For C00078, we use the same perturbative coefficients as for C005(LV) and C01, given that the lattice spacing is almost the same.} 
\end{table}

In the hadron interpolating fields, the light and strange quarks are Gaussian-smeared using APE-smeared links, while the heavy quarks are Gaussian-smeared using stout-smeared links \cite{Morningstar:2003gk} with $\rho_\text{stout}=0.8, \  N_\text{stout}=10$. The other smearing parameters are listed in Table \ref{tab:smearingparams}.

\begin{table}[H]
\centering
	\begin{tabular}{|l|ccccc|ccc|ccc|cc|} \hline 
		Ensemble   & \multicolumn{4}{c}{Up and down quarks} & \hspace{1ex} & \multicolumn{2}{c}{Bottom quarks} & \hspace{1ex} & \multicolumn{2}{c}{Charm quarks} & \hspace{1ex} & \multicolumn{2}{c|}{Strange quarks} \\
        \hline
               & $N_\textrm{Gauss}$ & $\sigma_\textrm{Gauss}$ & $N_\textrm{APE}$ & $\alpha_\textrm{APE}$ && $N_\textrm{Gauss}$ & $\sigma_\textrm{Gauss}$ && $N_\textrm{Gauss}$ & $\sigma_\textrm{Gauss}$ && $N_\textrm{Gauss}$ & $\sigma_\textrm{Gauss}$ \\ 
    C00078     & $100$           & $7.171$           & $25$ & $2.5$ && $10$ & $2.0$  && $20$ & $3.0$ && $30$ & $4.350$ \\
    C005LV, C005, C01  & $\phantom{0}30$ & $4.350$ & $25$ & $2.5$ && $10$ & $2.0$  && $20$ & $3.0$  && $30$ & $4.350$ \\
    F004, F006 &         $\phantom{0}60$ & $5.728$ & $25$ & $2.5$ && $16$ & $2.667$ && $20$ & $3.0$ && $60$ & $5.728$\\
    F1M & $130$ & $8.9$           & $25$ & $2.5$ && $20$ & $3.0$ && $35$ & $4.5$ && $70$ & $6.600$ \\
    \hline
	\end{tabular}
	\caption{\label{tab:smearingparams} Parameters for the smearing of the quark fields in the hadron interpolating operators. For the heavy quarks, the gauge links used in the Gaussian smearing are stout-smeared with $\rho_\text{stout}=0.8, \  N_\text{stout}=10$. See Eq.~(8) of Ref.~\cite{Leskovec:2019ioa} for the Gaussian definitions, Eq.~(8) of Ref.~\cite{Bonnet:2000dc} for the APE-smearing definitions, and Ref.~\cite{Morningstar:2003gk} for the stout-smearing definitions.}
\end{table}

\FloatBarrier
\section{Decay Constants of the Negative-Parity Mesons}
\FloatBarrier
\label{sec:negparity}

To determine the decay constants of the negative-parity $D^{(*)}_s$ and $B^{(*)}_s$ mesons, we use the interpolating fields
\begin{gather}
    \Phi_{H_{s}}=\Bar{Q}\gamma_5 s, \\
    \Phi^i_{H^*_{s}}=\Bar{Q}\gamma^i s.
\end{gather}
We compute the correlation functions
\begin{align}
C_{\Phi\Phi}(t) &=\sum_{\vec{z}}\langle \Phi_{H_s^{(*)}}(\vec{z},t+t_s)\Phi_{H_s^{(*)}}^\dagger(\vec{x}_s,t_s)\rangle, \\
C_{J\Phi}(t) &=\sum_{\vec{z}}\langle J^\dag(\vec{z},t+t_s) \Phi_{H_s^{(*)}}^\dagger(\vec{x}_s,t_s)\rangle,
\end{align}
where $J=J_{A_0}$ for the pseudoscalar case and $J=J_{V_i}$ for the vector case, and  $(\vec{x}_s,t_s)$ is the source position.. We average over the spatial directions $i$ and over forward and backward propagation. We then fit these correlation functions in the ground-state-dominated time region using
\begin{align}
C_{\Phi\Phi}(t) &= A_{\Phi\Phi} \left(e^{-Et}+e^{-E(T-t)}\right), \\
C_{J^\Phi}(t) &= A_{J\Phi}  \left(e^{-Et}+e^{-E(T-t)}\right),
\end{align}
where $T$ is the time extent of the lattice; effective-energy plots showing the fit ranges and fitted energies on all ensembles are given in Appendix \ref{sec:negpardecayconstfitplots}. We then obtain the decay constants as
\begin{equation}
    f=A_{J\Phi} \sqrt{\frac{2 }{E\:A_{\Phi\Phi}}}, \label{eq:fnegparity}
\end{equation} 
where we propagate the uncertainties in a correlated way using statistical bootstrap. We also compute the ratios $f_{H_s^*}/{f_{H_s}}$ using bootstrap.

To estimate the systematic uncertainties due to the choices of fit ranges, we perform a second set of fits with all $t_{\rm min}$ values increased by 1 on the coarse lattices and 2 on the fine lattices, and repeat the bootstrap calculations of the decay constants and their ratios using the shifted fits. We then estimate the systematic uncertainty in each decay-constant or ratio result as the maximum of (i) the shift in the central value or (ii) the increase in the statistical uncertainty, computed as $\sqrt{\sigma_{\rm shifted}^2-\sigma^2}$. We add this systematic uncertainty to the statistical uncertainty in quadrature.

Our results for the decay constants and the ratios $f_{H_s^*}/{f_{H_s}}$ on each ensemble are given in Table \ref{tab:negpardecayconst}.

\begin{table}[h]
\begin{tabular}{|l|c|c|c|c|c|c|}
\hline
Ensemble &  $f_{D_s}$ (MeV) & $f_{D_s^*}$ (MeV) & $f_{B_s}$ (MeV) & $f_{B_s^*}$ (MeV) & $f_{D_s^*}/f_{D_s}$ & $f_{B_s^*}/f_{B_s}$ \\
\hline
C00078 & 255.38(98) & 276.0(1.7) & 252.7(3.9) & 255.3(2.6) & 1.0808(71) & 1.011(16) \\
F1M & 254.08(76) & 275.8(2.3) & 244.7(3.8) & 243.2(3.1) & 1.086(11) & 0.994(16) \\
F004 & 258.7(1.0) & 284.3(2.5) & 246.9(4.6) & 250.3(2.6) & 1.0989(95) & 1.014(20) \\
C005 & 260.0(1.4) & 285.7(2.5) & 261.8(4.4) & 260.6(2.3) & 1.099(14) & 0.996(16) \\
C005LV & 260.9(1.2) & 286.4(1.9) & 262.9(4.4) & 262.3(2.7) & 1.0980(73) & 0.998(18) \\
F006 & 260.27(66) & 286.1(4.5) & 248.6(3.8) & 248.0(2.4) & 1.099(18) & 0.997(14) \\
C01 & 263.6(1.8) & 290.8(1.9) & 266.2(4.5) & 263.9(2.7) & 1.1034(99) & 0.992(17) \\
\hline
\end{tabular}
\caption{\label{tab:negpardecayconst}Decay constants of the negative-parity heavy-strange mesons. Only the statistical uncertainties are shown.}
\end{table}

\begin{figure}

\includegraphics[width=0.48\linewidth]{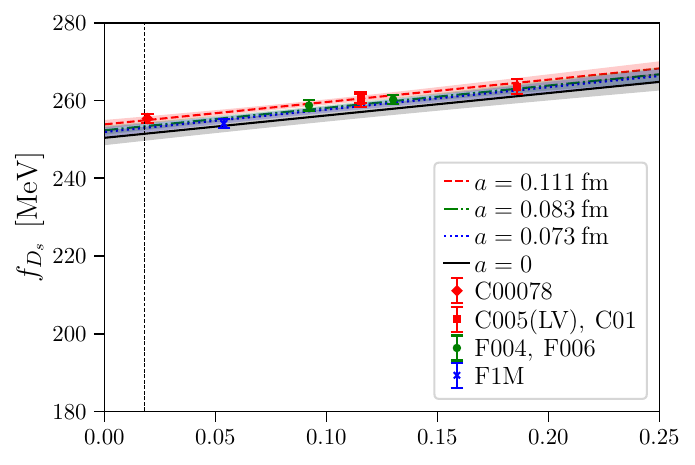} \hfill \includegraphics[width=0.48\linewidth]{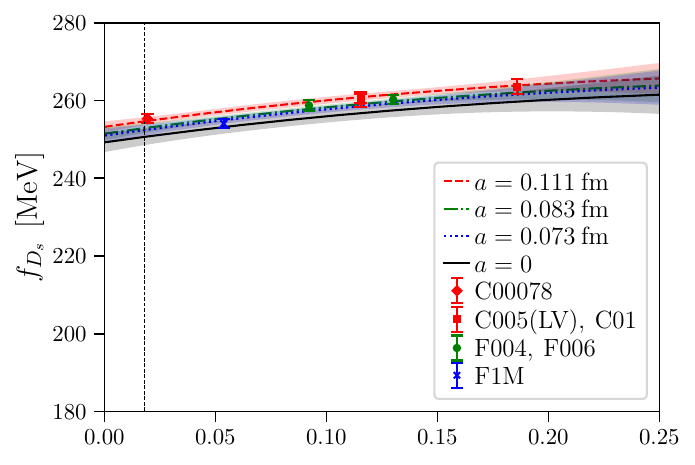} 

\vspace{-2ex}

\includegraphics[width=0.48\linewidth]{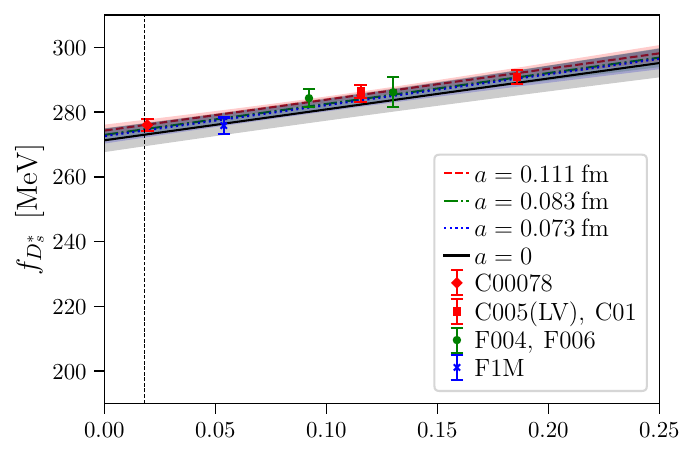} \hfill \includegraphics[width=0.48\linewidth]{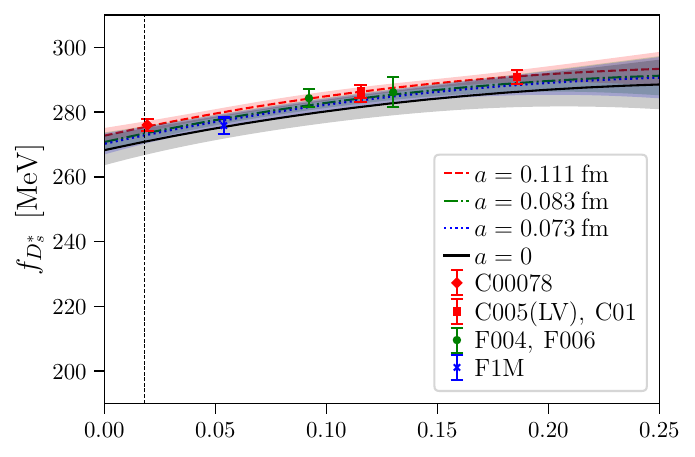} 

\vspace{-2ex}

\includegraphics[width=0.48\linewidth]{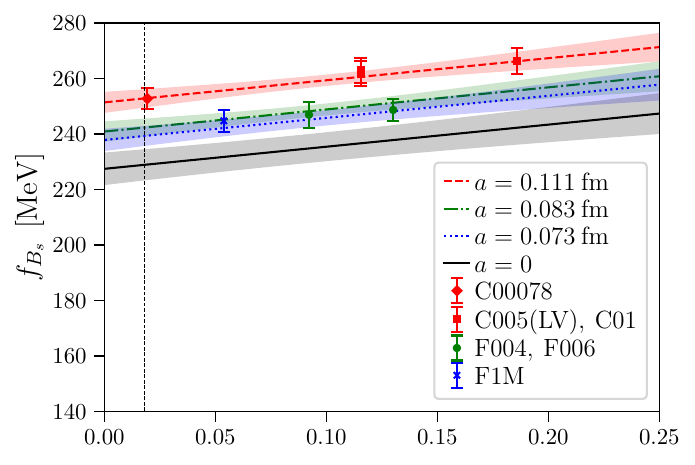} \hfill \includegraphics[width=0.48\linewidth]{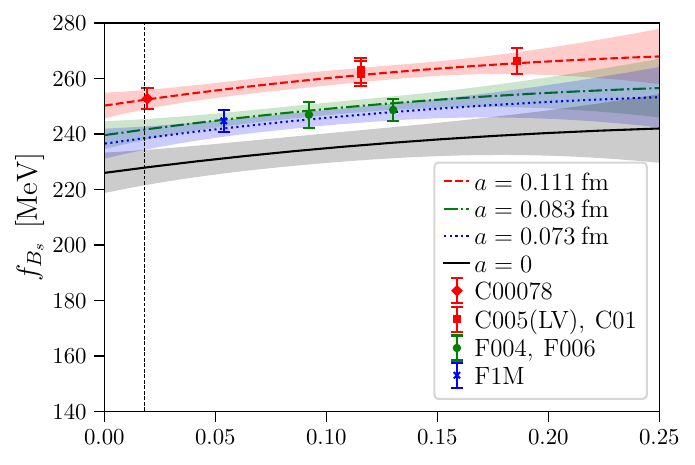} 

\vspace{-2ex}

\includegraphics[width=0.48\linewidth]{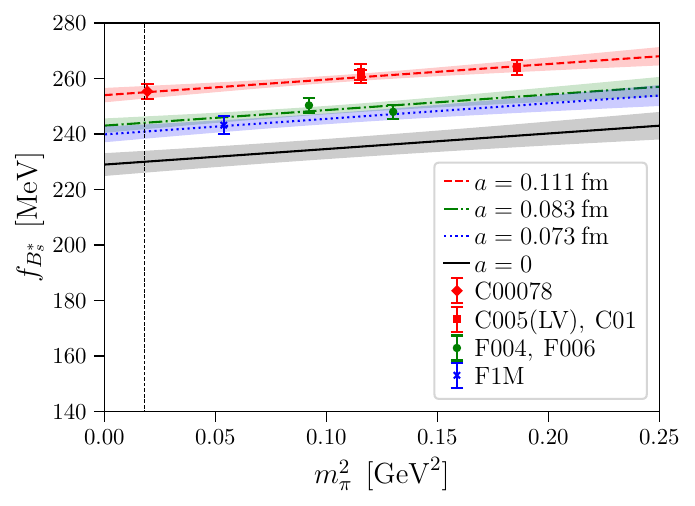} \hfill \includegraphics[width=0.48\linewidth]{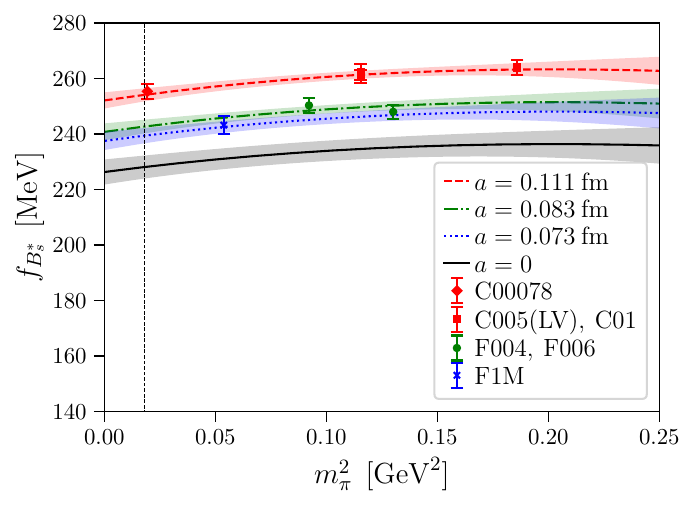} 
\caption{\label{fig:negparchiralcont}Chiral-continuum extrapolations of the negative-parity decay constants using the linear model (left) and the ``$\chi$PT B'' model (right). The curves show the model evaluated at the three different lattice spacings corresponding to the C005/C01 (and $\approx$ C00078), F004/F006, and F1M ensembles, and in the continuum limit. The vertical dashed line indicates the physical pion mass.}
\end{figure}

\begin{figure}

\includegraphics[width=0.48\linewidth]{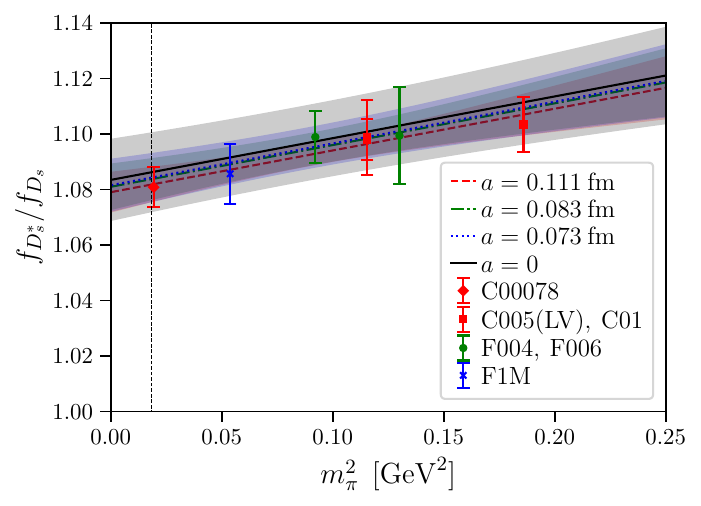} \hfill \includegraphics[width=0.48\linewidth]{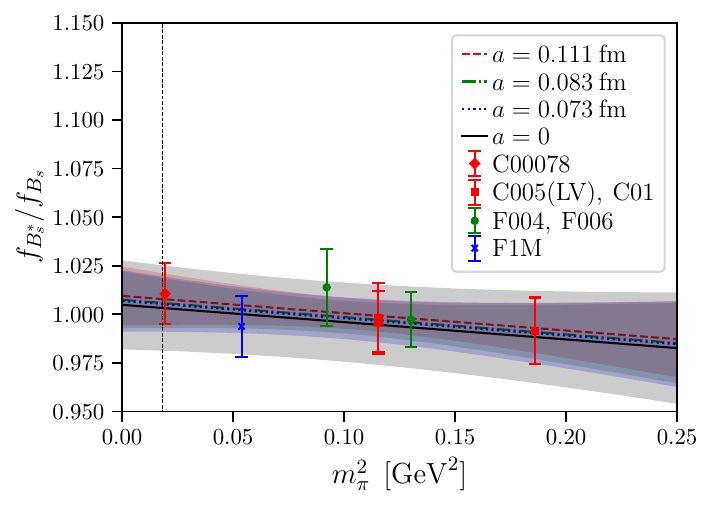} 
\caption{\label{fig:negparratiochiralcont}Chiral-continuum extrapolations of the vector-to-pseudoscalar decay-constant ratios. The curves show the model evaluated at the three different lattice spacings corresponding to the C005/C01 (and $\approx$ C00078), F004/F006, and F1M ensembles, and in the continuum limit. The vertical dashed line indicates the physical pion mass.}
\end{figure}

To extrapolate the data to the physical point, we consider the following two functional forms:
\begin{align}
    \text{linear in }m_\pi^2\text{ :  }f&=f_{\text{phys}}+\frac{C}{4\pi f_\pi} (m_{\pi,{\rm lat}}^2 - m_{\pi, \text{phys}}^2)+d\Lambda^3 a^2 , \\
    \chi\text{PT:  }f&=f_0 \left[ 1 - \frac{1+3\,g^2}{ (4\pi f_\pi)^2 }\left( m_K^2 \ln\frac{m_K^2}{\mu^2} + \frac{1}{3 }m_\eta^2 \ln\frac{m_\eta^2}{\mu^2}  \right) + c_K \frac{m_K^2}{(4\pi f_\pi)^2} + c_\eta \frac{m_\eta^2}{(4\pi f_\pi)^2}  \right]\left( 1 + d\,\Lambda^2 a^2 \right). \label{eq:chPT}
\end{align}
The latter is based on NLO $SU(3)$ heavy-meson chiral perturbation theory \cite{Goity:1992tp}. We set $f_\pi=130.5$ MeV and $\Lambda=500$ MeV. For the axial coupling, we use a Gaussian prior $g=0.5\pm 0.1$, based on Refs.~\cite{Detmold:2011bp,Bernardoni:2014kla,Flynn:2015xna,Gerardin:2021jch}, while the other parameters are unconstrained. To evaluate the meson masses in Eq.~(\ref{eq:chPT}), we consider two schemes, labeled A and B in the following. For ``$\chi$PT A,'' we set
\begin{align}
m_K^2 &= m_{K, {\rm lat}}^2, \\
m_\eta^2 &= \frac{4}{3} m_{K, {\rm lat}}^2 - \frac{1}{3} m_{\pi, {\rm lat}}^2,
\end{align}
where $m_{K, {\rm lat}}$ and $m_{\pi, {\rm lat}}$ are the lattice kaon and pion masses on each ensemble. For ``$\chi$PT B,'' we express everything in terms of the lattice pion masses by setting
\begin{align}
m_K^2 &= \frac12\left(m_{\eta_s}^2 + m_{\pi,{\rm lat}}^2\right), \\
m_\eta^2 &=\frac13\left(2\,m_{\eta_s}^2 + m_{\pi,{\rm lat}}^2\right)
\end{align}
with the mass of the fictitious $s\bar{s}$ pseudoscalar meson fixed to the physical value $m_{\eta_s}=686$ MeV \cite{Davies:2009tsa}; this is justified because the valence strange-quark masses are tuned to the physical value for each ensemble. The results for the decay constants at the physical point $m_\pi=135$ MeV and $a=0$ from the three different types of fits are given in Table \ref{tab:negparcontinuum}, and plots of the ``linear in $m_\pi^2$'' and ``$\chi$PT B'' fits are shown in Fig.~\ref{fig:negparchiralcont}. We also considered fits with an additional $d^\prime\,\Lambda^3 a^3$ term included, with a Gaussian prior $0\pm1$ for the additional parameter $d^\prime$; we found that adding this term has negligible effect.

\begin{table}
\begin{tabular}{|ll|c|c|c|}
\hline
Quantity & Model & $\chi^2/{\rm d.o.f.}$ & Weight (\%) & Result (MeV) \\
\hline
$f_{D_s}$   & Linear     &  0.45 & 53 & 251.5(1.8)  \\
            & $\chi$PT A &  0.43 & 25 & 251.8(2.5)  \\
            & $\chi$PT B &  0.51 & 23 & 250.7(2.1)  \\
\hline
$f_{D_s^*}$ & Linear     &  0.54 & 49 & 273.0(3.6)  \\
            & $\chi$PT A &  0.55 & 23 & 273.1(5.5)  \\
            & $\chi$PT B &  0.38 & 28 & 270.9(4.1)  \\
\hline
$f_{B_s}$   & Linear     &  0.38 & 54 & 228.9(5.6)  \\
            & $\chi$PT A &  0.43 & 23 & 227.8(5.9)  \\
            & $\chi$PT B &  0.43 & 23 & 227.9(6.4)  \\
\hline
$f_{B_s^*}$ & Linear     &  0.55 & 51 & 230.0(3.9) \\
            & $\chi$PT A &  0.62 & 23 & 228.6(5.5)  \\
            & $\chi$PT B &  0.52 & 25 & 228.1(4.1)  \\
\hline
\end{tabular}
\caption{\label{tab:negparcontinuum}Results of chiral-continuum extrapolations of the negative-parity decay constants, for the three different fit models discussed in the main text. The parameter $g$, which is constrained by a prior, is not counted in the calculation of the number of degrees of freedom. The uncertainties shown here include the statistical and lattice-spacing uncertainties. Also shown are the weights with which each result enters in the model average.}
\end{table}

To obtain our final estimates, we perform model averages \cite{bays} over the three different types of fits, which yields
\begin{eqnarray}
f_{D_s}&=&251.4(2.1)(0.4)(2.5)\:{\rm MeV}, \\
f_{D_s^*}&=&272.5(4.3)(1.0)(2.7)\:{\rm MeV}, \\
f_{B_s}&=&228.4(5.8)(0.5)(2.3)\:{\rm MeV}, \\
f_{B_s^*}&=&229.2(4.4)(0.8)(2.3)\:{\rm MeV}.
\end{eqnarray}
Here, the first uncertainty is the combination of statistical and lattice-spacing uncertainties, the second uncertainty is from the variation across fit models, and the third uncertainty is an estimate of the uncertainty due to missing higher-order corrections to the residual matching factors. We take the latter to be 1\% for both charm and bottom, which is more conservative than the scale-variation-based uncertainties shown in Table \ref{tab:matchingfactors}.

Given the large values of $m_\eta L$ and $m_K L$ for our ensembles, we expect finite-volume effects to be negligible compared to the above uncertainties. Systematic errors can also arise from heavy-quark discretization effects that are not fully removed through the continuum extrapolation, due to their nontrivial mass dependence \cite{Oktay:2008ex,Christ:2014uea,El-Khadra:1996wdx}. These effects were estimated for matrix elements of $b\to s$ currents with our choice of bottom-quark parameters in Appendix E of Ref.~\cite{Farrell:2026swf} and were found to be below 1\%. For the charm case, the additive mass renormalization is no longer negligible compared to the heavy-quark mass and must be taken into account when evaluating the mismatch functions, but we have not determined the critical bare masses. However, given how close the anisotropy parameters are to 1 in this case \cite{Meinel:2023wyg}, we expect the discretization errors to behave similarly to the standard (light-quark) clover case.

For the vector-to-pseudoscalar decay-constant ratios, we perform chiral-continuum extrapolations linear in $m_\pi^2$ and $a^2$, which are found to have $\chi^2/{\rm d.o.f.}$ values of 0.14 and 0.23 for charm and bottom, respectively, and are shown in Fig.~\ref{fig:negparratiochiralcont}. The fits yield the physical-point results
\begin{eqnarray}
f_{D_s^*}/f_{D_s}&=&1.086(14)(11), \\
f_{B_s^*}/f_{B_s}&=&1.003(22)(10), 
\end{eqnarray}
where the first uncertainty is statistical and the second uncertainty is the estimate of the uncertainty due to missing higher-order corrections to the residual matching factors (this uncertainty does not cancel in the ratio, as the matching coefficients for the temporal axial and spatial vector currents are different).

Our results for $f_{D_s}$ and $f_{B_s}$ are consistent with the FLAG world averages \cite{FlavourLatticeAveragingGroupFLAG:2024oxs} (both for $N_f=2+1$ and $N_f=2+1+1$), which are based on Refs.~\cite{Davies:2010ip,FermilabLattice:2011njy,Boyle:2017jwu,Yang:2014sea,Na:2012iu,Carrasco:2014poa,Kuberski:2024pms,FermilabLattice:2011njy,McNeile:2011ng,Na:2012kp,Aoki:2014nga,Christ:2014uea,Dowdall:2013tga,ETM:2016nbo,Hughes:2017spc,Bazavov:2017lyh,Frezzotti:2024kqk} (see also Ref.~\cite{Black:2022eph} for more recent preliminary work by the RBC/UKQCD collaboration). A comparison of our results for the vector-pseudoscalar ratios with previous lattice-QCD results \cite{Donald:2013sra,Colquhoun:2015oha,Lubicz:2017asp,Blossier:2018jol,Balasubramamian:2019wgx,Cai:2026xja} is shown in Fig.~\ref{fig:negparratiocomp} in Sec.~\ref{sec:conclusions}. For the charmed mesons, our result agrees with those of Refs.~\cite{Donald:2013sra,Lubicz:2017asp}, but is lower than the value obtained in Ref.~\cite{Blossier:2018jol}. For the bottom mesons, we find agreement with Ref.~\cite{Balasubramamian:2019wgx}, but our result is higher than those of Refs.~\cite{Colquhoun:2015oha,Cai:2026xja}.

\section{Positive-Parity Mesons}

\subsection{Operator Basis and Calculation of Correlators}
\label{sec:opbasis}

To resolve the low-lying $J^P=0^+$ heavy-strange states of interest, we use the operator basis
\begin{subequations}    
   \begin{gather}
    \Phi^{(1)}=\Bar{Q}s, \label{eq:spin0simple}\\
    \Phi^{(2)}=\Bar{Q}\gamma^i \nabla_i s, \\
    \Phi^{(4)}(\vec{x},t)=\sum_{\vec{y}} \left[\Phi_{\Bar{K}^0}(\vec{x},t)\Phi_{\Bar{H}^0}(\vec{y},t) + \Phi_{K^-}(\vec{x},t)\Phi_{H^+}(\vec{y},t) \right], 
    \end{gather}
\end{subequations}
where $\Phi_{K^-}=\Bar{u}\gamma_5 s$, $\Phi_{\Bar{K}^0}=\Bar{d}\gamma_5 s$, $\Phi_{H^+}=\Bar{Q}\gamma_5 u$, and $\Phi_{\Bar{H}^0}=\Bar{Q}\gamma_5 d$ (the $^+$ and $^0$ superscripts correspond to the charges for $H=B$). For $J^P=1^+$, we use
\begin{subequations}
    \begin{gather}
    \Phi^{(1)i}=\Bar{Q}\gamma^i\gamma_5 s,\\
    \Phi^{(2)i}=\Bar{Q}\gamma_5 \nabla^i s,\\
    \Phi^{(4)i}(\vec{x},t)=\sum_{\vec{y}}  \left[\Phi_{\Bar{K}^0}(\vec{x},t)\Phi_{\Bar{H}^{*0i}}(\vec{y},t) + \Phi_{K^-}(\vec{x},t)\Phi_{H^{*+i}}(\vec{y},t) \right],\label{eq:spin1mesonmeson}
    \end{gather}
\end{subequations}
where  $\Phi_{H^{*+i}}=\Bar{Q}\gamma^i u$, and $\Phi_{\Bar{H}^{*0i}}=\Bar{Q}\gamma^i d$. For the covariant derivatives in $\Phi^{(2)}$ and $\Phi^{(2)i}$, we used stout-smeared gauge links (matching the links used in the heavy-quark smearing). Inspecting Table \ref{tab:Ds} for the $J^P=1^+$ particle, we expect the quark-model spin triplet $D_{s1}(2460)$ and its bottom partner to couple strongly to $\Phi^{(1)i}$, and the quark-model spin singlet $D_{s1}(2536)$ and its bottom partner $B_{s1}(5830)$ to couple strongly to $\Phi^{(2)i}$. The meson-meson operators $\Phi^{(4)(i)}$ are included to help resolve the $H^{(*)}K$ scattering states. 

We additionally introduce the convention that
\begin{align}
\Phi^{(3)}&=J^{\dag}_{V_0},    \\ 
\Phi^{(3)i}&=J^{\dag}_{A_i}.
\end{align}
\begin{figure}[b]
    \centering
    \includegraphics[width=0.99\linewidth]{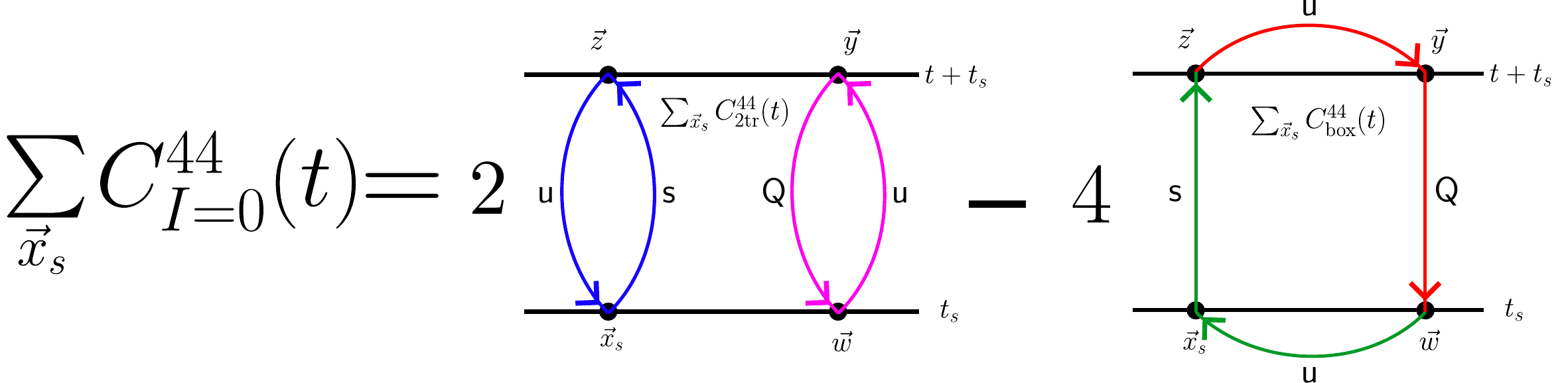}
    \caption{Decomposition of the isospin-projected $C^{44}(t)$ in terms of quark propagators. The red and green propagator combinations in the box diagram on the right-hand side are sequential-stochastic-timeslice propagators, where the propagators that serve as the sources are themselves first sourced by the same $\xi_{\text{box}}$ at time $t_s$. The blue and pink loops in the left diagram are computed by the one-end-trick from distinct sources $\xi_{\text{blue}}$ and $\xi_{\text{pink}}$ \cite{stochasitc}.}
    \label{fig:quarkline}
\end{figure}
Using the above operators, we compute zero-momentum projected $(4\times 3)$ correlation matrices 
\begin{equation}
    C^{lm}(t)=\sum_{\vec{z}} \langle \Phi^{(l)}(\vec{z}, t+t_s) \Phi^{(m)\dagger}(\vec{x}_s,t_s)\rangle. \label{eq:corr}
\end{equation}

\begin{figure}
    \includegraphics[width=0.49\linewidth]{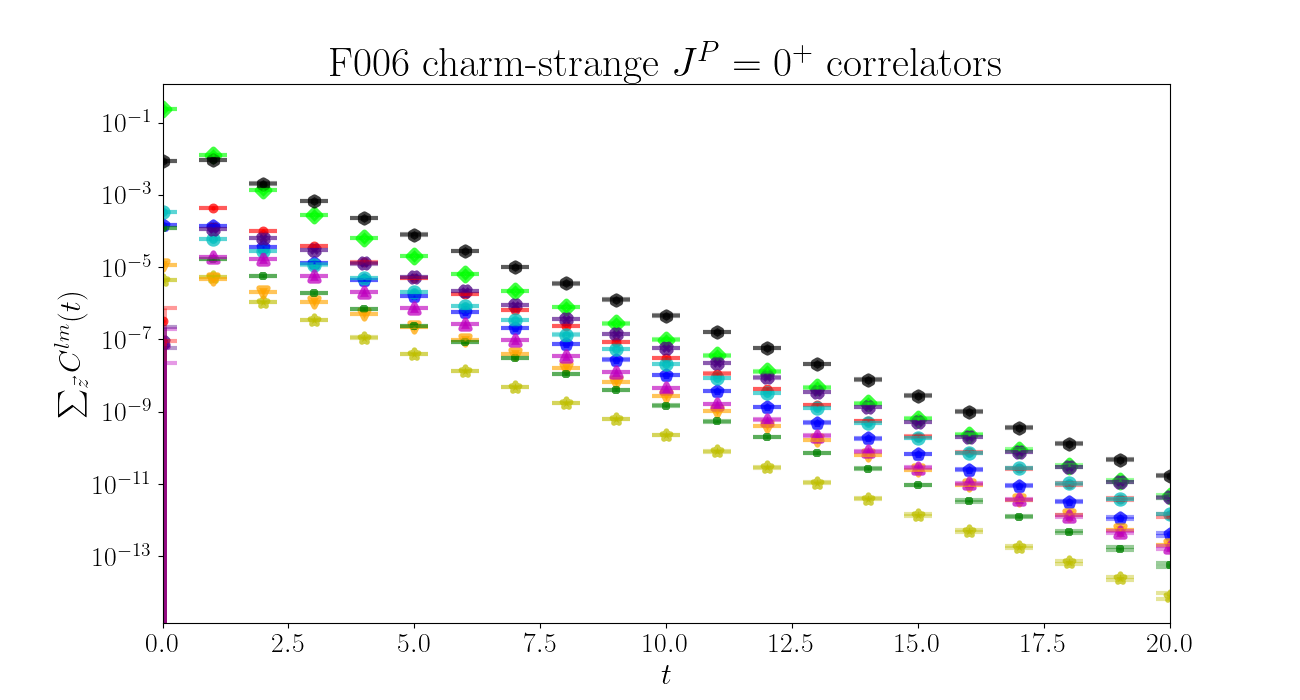}
    \hfill
    \includegraphics[width=0.49\linewidth]{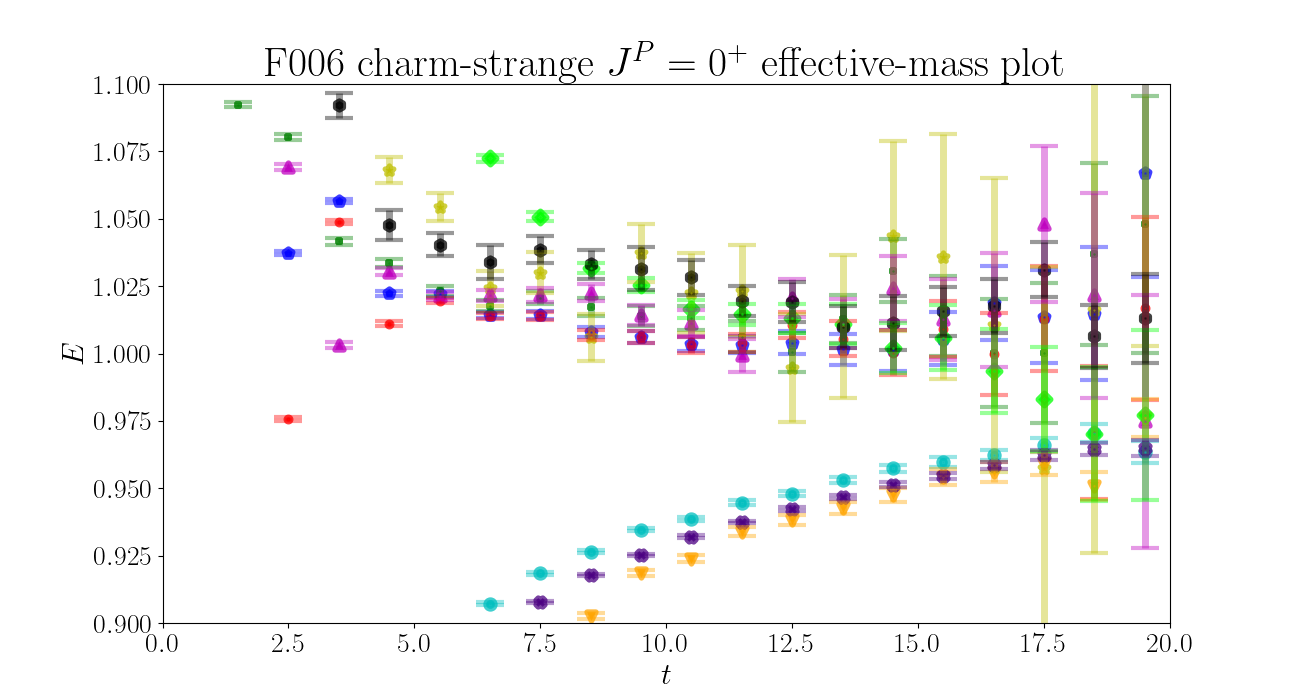}
    \hfill
        
    \includegraphics[width=0.49\linewidth]{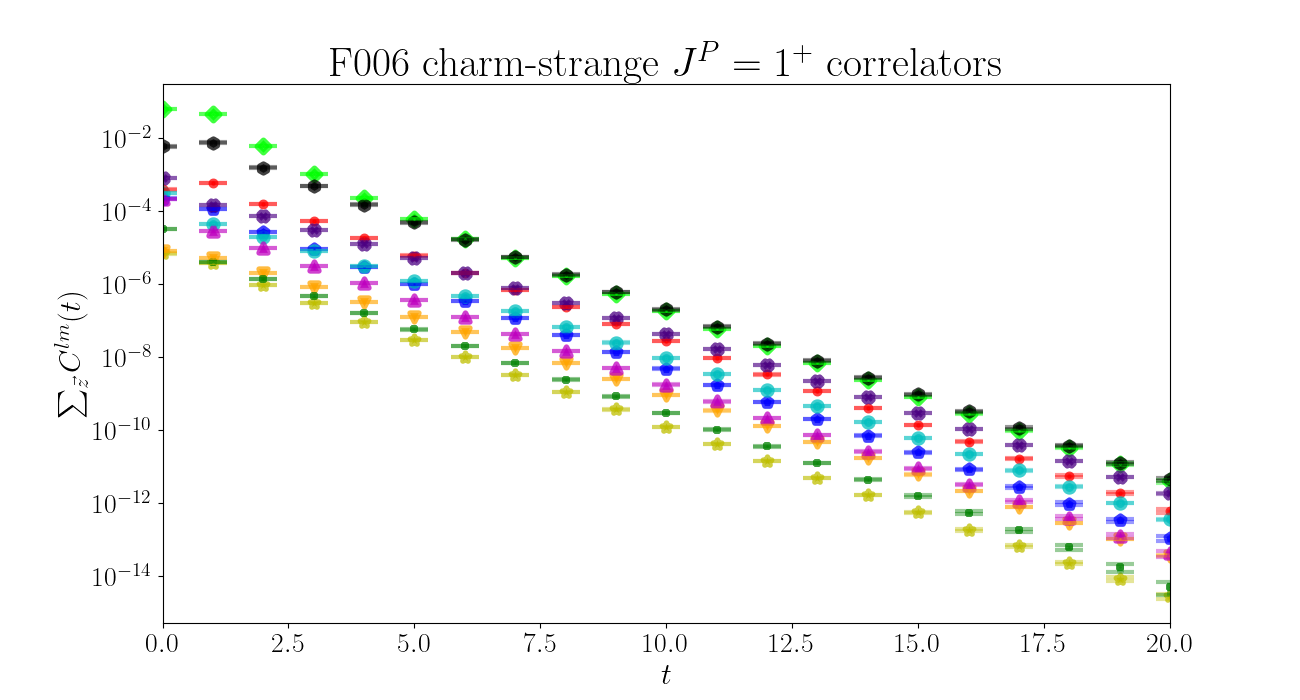}
    \hfill
    \includegraphics[width=0.49\linewidth]{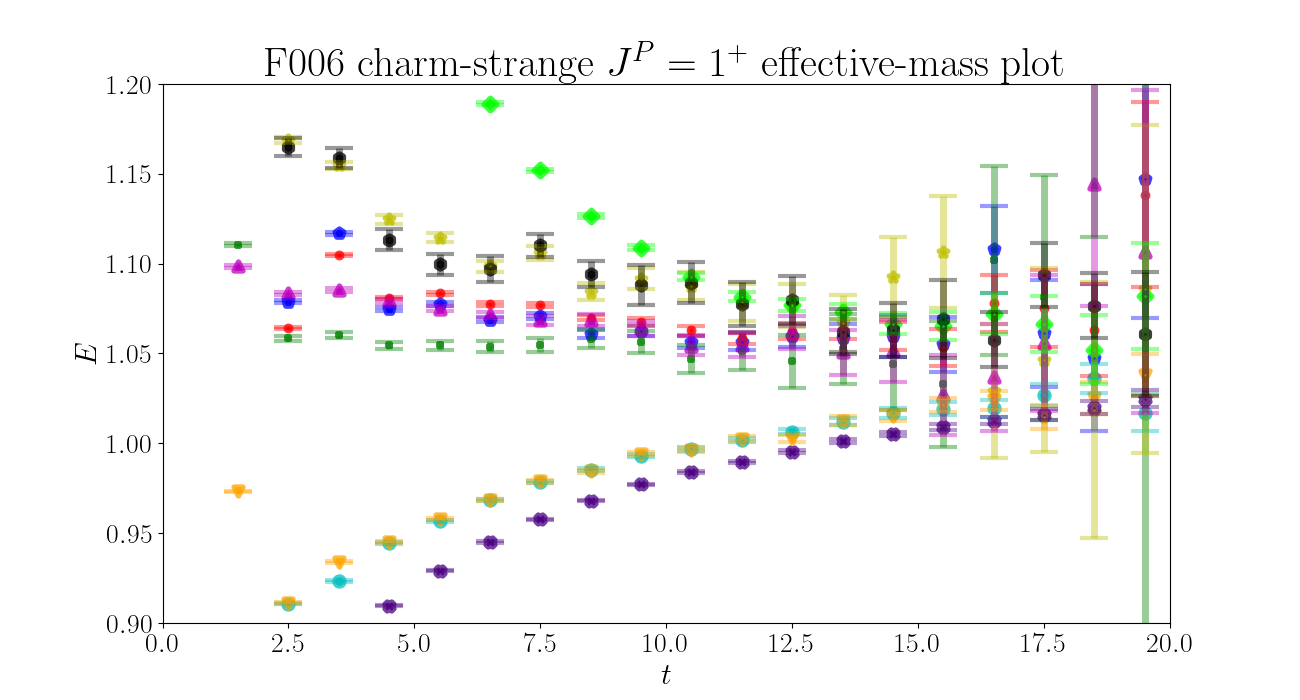}

    \includegraphics[width=0.49\linewidth]{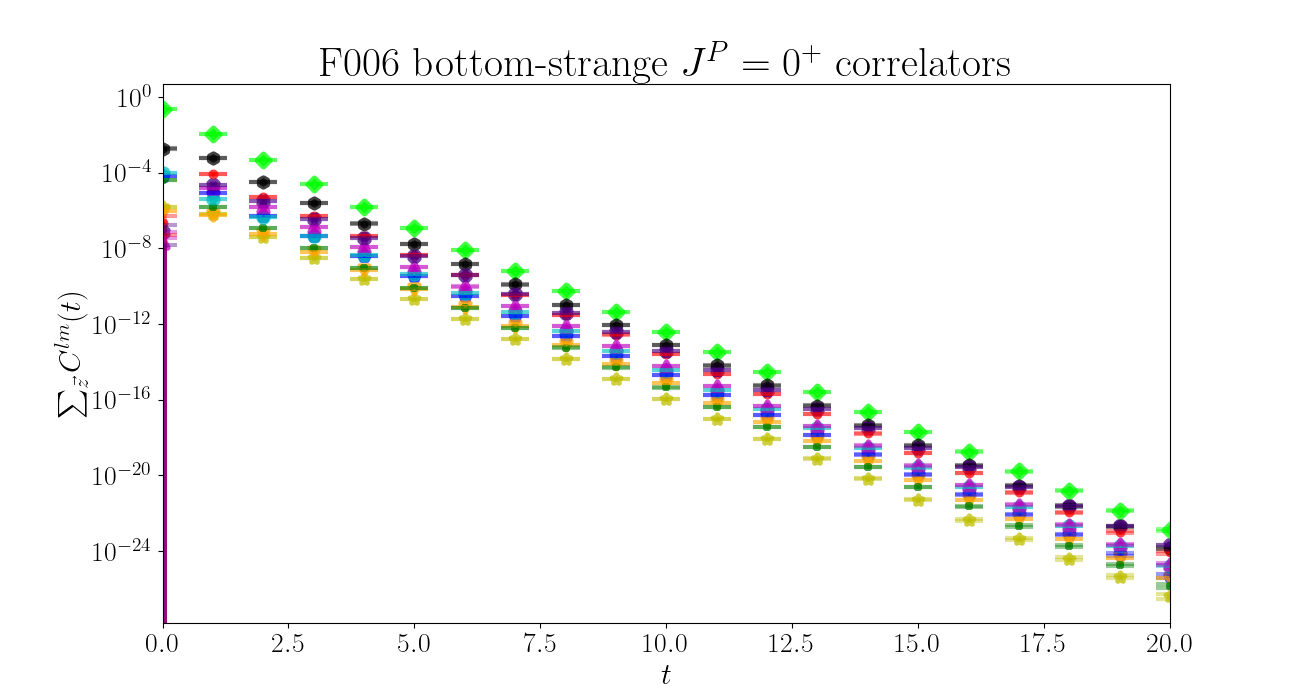}
    \hfill
    \includegraphics[width=0.49\linewidth]{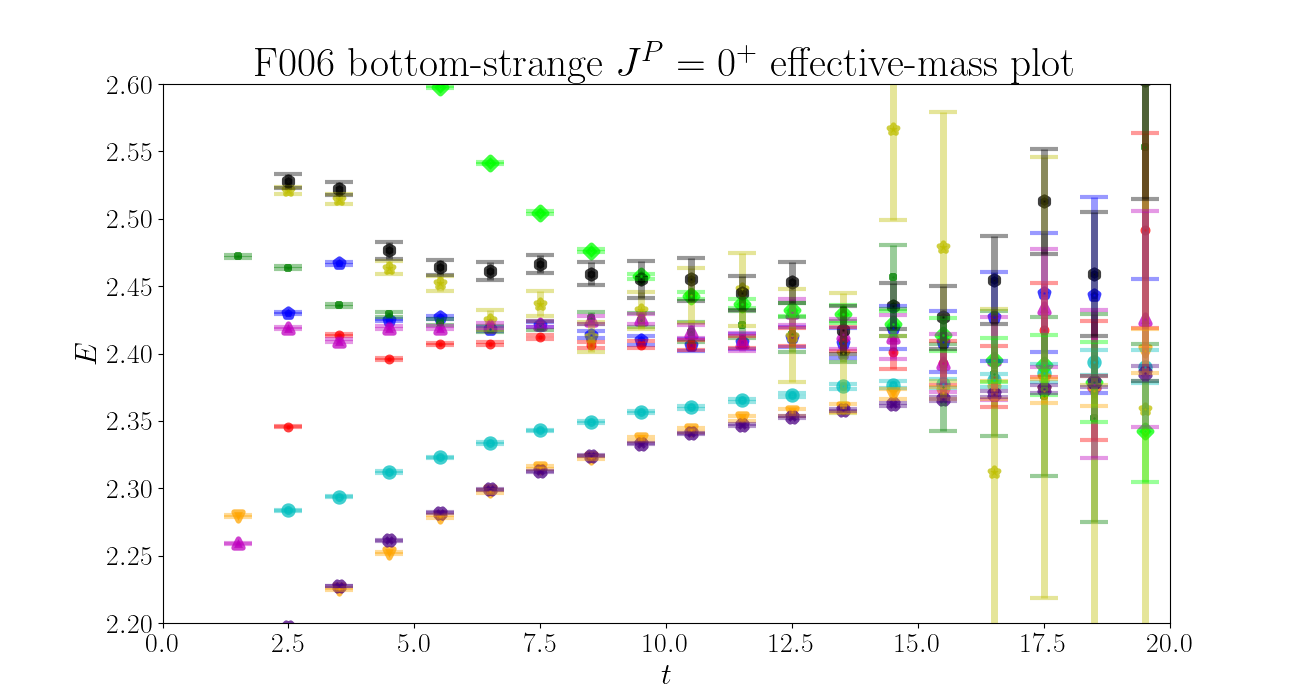}
    \hfill
        
    \includegraphics[width=0.49\linewidth]{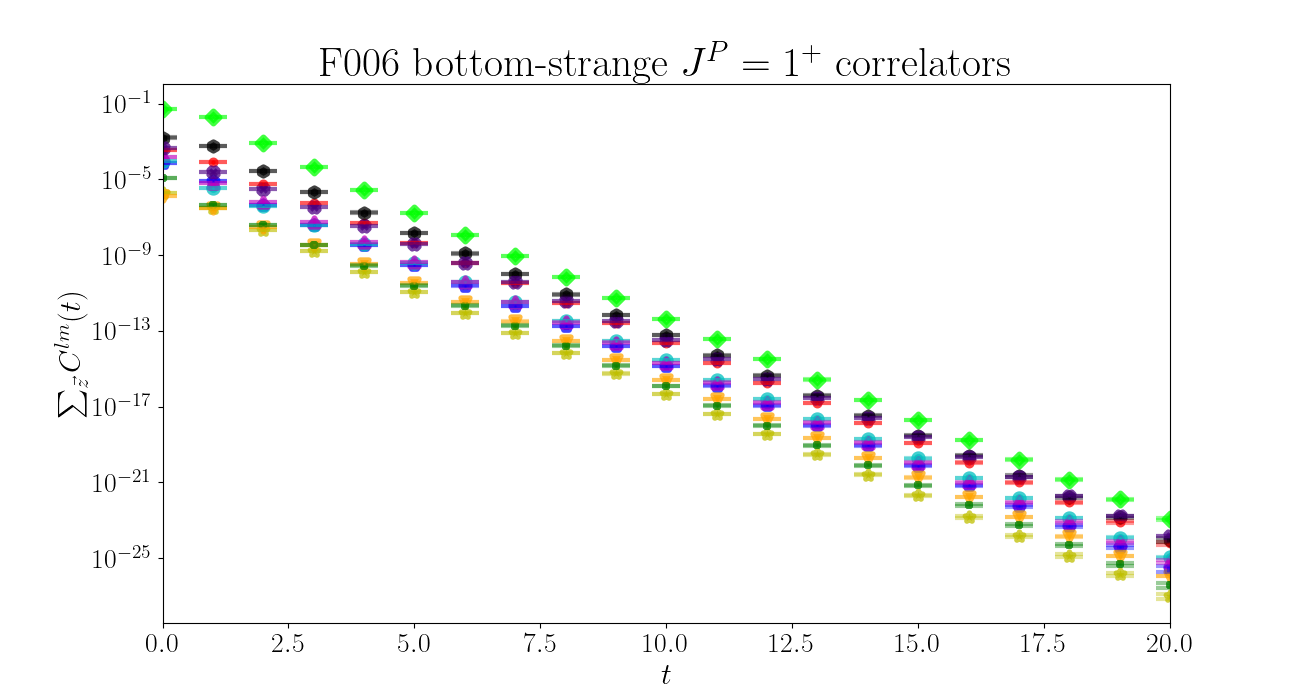}
    \hfill
    \includegraphics[width=0.49\linewidth]{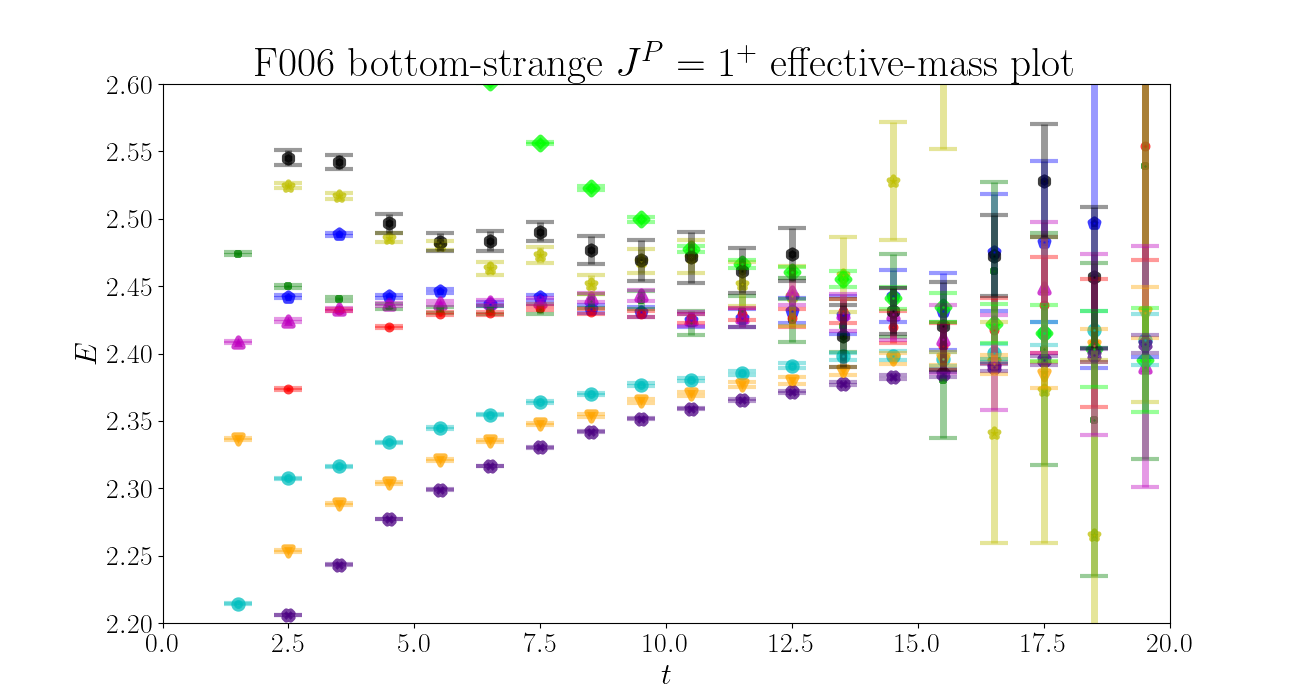}
    \centering
    \includegraphics[width=0.6\linewidth]{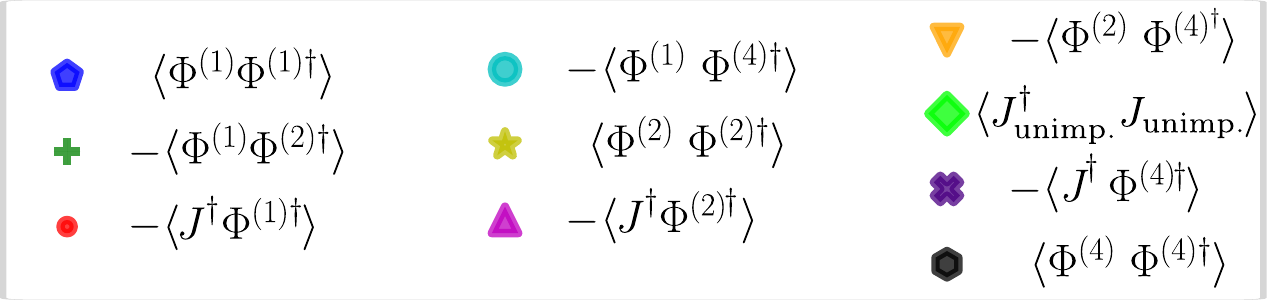}
    \caption{Elements of $C^{lm}(t)$ on the F006 ensemble. Correlators are arranged above with their associated effective-energy plots, with $E$ in lattice units. For the correlator plots, all of the negative off-diagonal elements have their sign flipped. A correlation function involving two unimproved currents $J_\text{unimp.}$ is additionally shown.$J_\text{unimp.}$ is given by dropping $\mathcal{O}(a)$-terms from Eq.~(\ref{eq:currents}). For the other ensembles see Appendix \ref{sec:corrplots}.}
    \label{fig:Clmplots}
\end{figure}

We additionally calculate the $D^{(*)}$, $B^{(*)}$, and $K$ correlators $C_M(t)=\sum_{\vec{z}} \langle \Phi_M(\vec{z}, t+t_s) \Phi^\dagger_M(\vec{x}_s,t_s) \rangle$, where $M=D^{(*)},B^{(*)}, K$ in order to extract the $H^{(*)} K$ kinematic thresholds $E_{th}=E_{H^{(*)}}+E_K$. As we are working in the limit of exact isospin symmetry, we do not consider the $H_s^{(*)}\pi$ thresholds.\\

From previous projects \cite{Meinel:2016dqj,Meinel:2020owd,Meinel:2021rbm,Meinel:2023wyg} we have at our disposal precomputed light and strange quark propagators with Gaussian-smeared sources at $(\vec{x}_s,t_s)$. These are used to compute all $C^{lm}$ where $l \ne 4$. The elements $\sum_{\vec{z}}\langle \Phi^{(l)}(\vec{z}, t+t_s) \Phi^{(4)^\dagger}(\vec{x}_s,t) \rangle$ and $\sum_{\vec{z}}\langle \Phi^{(l)i}(\vec{z}, t+t_s) \Phi^{(4)i^\dagger}(\vec{x}_s,t) \rangle$ are computed using the light-quark propagator as a source for a sequential heavy-quark propagator, $T(z,x_s)\equiv \sum_{\vec{y}} G_Q(z;\vec{y},t_s) \gamma_5 G_u(\vec{y},t_s;\vec{x}_s,t_s) $ and $T(z,x_s)^{i}\equiv \sum_{\vec{y}} G_Q(z;\vec{y},t_s) \gamma^i G_u(\vec{y},t_s;\vec{x}_s,t_s) $, where the smearing is not explicitly shown and $u$ here denotes the light quarks, both up and down. The correlation functions $\sum_{\vec{z}}\langle \Phi^{(l)}(\vec{z}, t+t_s) \Phi^{(2)^\dagger}(\vec{x}_s,t_s) \rangle$ and $\sum_{\vec{z}}\langle \Phi^{(l)i}(\vec{z}, t+t_s) \Phi^{(2)i^\dagger}(\vec{x}_s,t_s) \rangle$ with the derivative operators at the source are computed using heavy-quark propagators with derivative sources.

Of the elements with the meson-meson operator at the sink, we need only compute $C^{44}$, because we can use $C^{4l}=C^{l4}$. We make the effort to include $C^{44}$ to enable the construction of a generalized eigenvalue problem (Sec.~\ref{sec:gevp}) and avoid a negative bias in the fitted energy. The latter can occur when using correlation functions with two-hadron operators at one end only and local operators at the other end, due to the possible emergence of false plateaus that arise when the amplitudes $\bra{0}\Phi^l(t+t_s)\ket{n}\bra{n}\Phi^{4\dagger}(t_s)\ket{0}$ vary in sign for different $n$ \cite{Horz:2020zvv, Iritani:2016jie}.

The computation of $C^{44}$ requires new light and strange quark propagators. The contractions, which are given in Appendix \ref{sec:C44}, split $C^{44}$ into a piece containing the product of two traces, and a piece represented by a box quark-line diagram (see Fig.~\ref{fig:quarkline}).  
 
 While the correlation matrix elements other than $C^{44}$ use a fixed $\vec{x}_s$, for $C^{44}$ we instead estimate the sum over all $\vec{x}_s$ by introducing  stochastic propagators \cite{stochasitc} resulting from the inversion of the Dirac equation from stochastic sources, as explained in detail in Appendix \ref{sec:C44}. These sources have the form $\xi^{t_s, n}(\vec{y},t)=\delta_{t_s, t}\xi^n(\vec{y})$ (multiplied by identity matrices in color and Dirac indices that are not explicitly shown) with independent random values $\xi^n(\vec{y})\in \mathbb{Z}_2$ at each $\vec{y}$ and for each noise sample $n$, which leads to
 \begin{equation}
	 \frac{1}{N} \sum_{n=1}^N \xi^n(\vec{x})^* \xi^n(\vec{y})=\delta_{\vec{x},\vec{y}}+\mathcal{O}(\frac{1}{\sqrt{N}}). \label{eq:stochsource}
\end{equation}
Therefore, two of these stochastic propagators, when combined, approximate the combination of the associated all-to-all propagators. For the box diagram, we also compute new sequential light-quark propagators, sourced by the stochastic heavy-quark propagators at time $t_s+t$, for each value of $t$. The correlation functions constructed using stochastic propagators are unbiased for any value of $N$, and we use $N=1$. Furthermore, to save resources, we computed $C^{44}$ only on a reduced number of gauge configurations: 21 configurations for C00078 and C01, 18 on F004, and 20 on the other ensembles. We combine $C^{44}$ with the other elements by generating 1000 bootstrap samples for all elements.

Since we have computed $\sum_{\vec{x}_s} C^{44}$, we divide out the spatial volume to obtain $C^{44}(t)=\frac{1}{L^3} \sum_{\vec{x}_s}C^{44}(t)$. Where applicable, we average over the Lorentz indices. We also average the off-diagonal $C^{lm}$'s with $C^{ml}$'s where both are available, as they have the same expectation values. Finally, we average over forward and backward propagation in time. 

As an example, the correlation functions and their associated effective-energy plots from the F006 ensemble are shown in Fig. ~\ref{fig:Clmplots}. The correlators and effective-energy plots for the other ensembles can be found in Appendix \ref{sec:corrplots}.

\subsection{Finite Volume Spectra And Decay Constants}\label{sec:data}
\subsubsection{The Generalized Eigenvalue Problem}
\FloatBarrier

\label{sec:gevp}
Upon generating 1000 bootstrap samples, we extract the spectrum from the generalized eigenvalue problem (GEVP) \cite{Michael:1985ne,Luscher:1990ck,Blossier:2009kd}
\begin{equation}
C^{lm}(t) v^m_n(t, t_0)=\lambda_n(t, t_0) C^{lm}(t_0) v^m_n(t, t_0)
\end{equation}
where we solve for generalized eigenvalues $\lambda_n(t, t_0)$ and eigenvectors $v^m_n(t, t_0)$. Here and in the following, we use the convention that the repeated indices $l$ and $m$ are summed over 1,2,4 (corresponding to the three different hadron interpolating operators for each system).
The functions $\lambda_n(t, t_0)$ are also referred to as principal correlators, and are expected to behave as
\begin{equation}
\lambda_n(t, t_0) \propto e^{-E_n(t-t_0)}
\end{equation}
up to contamination from higher excited states. In our implementation, reference eigenvectors and eigenvalues are first computed at a reference time $t=t_\text{ref}>t_0$, sorted by the eigenvalues, and then rescaled,
\begin{equation}
v^k_n(t_\text{ref},t_0) \rightarrow v'^k_n(t_\text{ref},t_0)=\frac{v_n^k(t_\text{ref},t_0)}{\sqrt{v_n^l(t_\text{ref},t_0) C^{lm}(t_0) v_n^m(t_\text{ref},t_0)}}.
\end{equation}
\begin{table}[H]
    \centering
    \begin{tabular}{|lc|cc|}
    \hline
        Ensemble &  Particle & $t_0$ & $t_\textrm{ref}$ \\ \hline
        C00078   & $D_{s0}^*$   & 6 & 7 \\
        C00078   & $B_{s0}^*$, $D_{s1}$, $B_{s1}$ & 5 & 7 \\\hline
        C005LV, C005, C01 & all & 3 & 6\\ \hline
        F004, F006 & all & 4 & 6\\ \hline
        F1M &  $D_{s0}^*$, $B_{s0}^*$ & 5 & 7\\
        F1M & $D_{s1}$  & 7  & 9\\
        F1M & $B_{s1}$ & 6  & 9\\
        \hline
    \end{tabular}
    \caption{The values of $t_0$ and $t_\textrm{ref}$ (in lattice units) used in the GEVP and sorting of states.}
    \label{tab:t0tref}
\end{table}
The GEVP is solved for all $t$, and the eigenvectors are rescaled according to
\begin{equation}
v^k_n(t,t_0) \rightarrow v'^k_n(t,t_0)=\frac{v^k_n(t,t_0)}{\sqrt{v^l_n(t,t_0) C^{lm}(t_0)v^m_n(t,t_0)}}.
\end{equation}
Normally, all states are sorted by finding the maximum overlap $v'^l_n(t,t_0) C^{lm}(t_0) v'^m_{n'}(t_\text{ref},t_0)$, which identifies state $n$ with reference state $n'$ \cite{Dudek:2010wm}. However, in the spin-1 case, where the spectrum includes two nearby excited states, we sort the two excited states by overlap with the two-meson operator $\Phi^{(4)(i)}$, as we expect one $H^{*}K$-like state near the threshold in addition to the excited $^1 P_1$ state. In the following, we label the state with the maximum $\Phi^{(4)(i)}$ overlap ``1'' and the other excited state ``2'' (the ground state is labeled ``0'').

\begin{figure}
    \centering
    \includegraphics[width=0.3\linewidth]{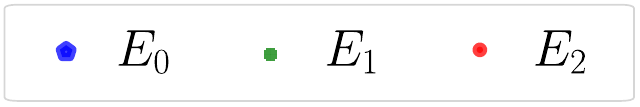}
    
    \includegraphics[width=0.49\linewidth]{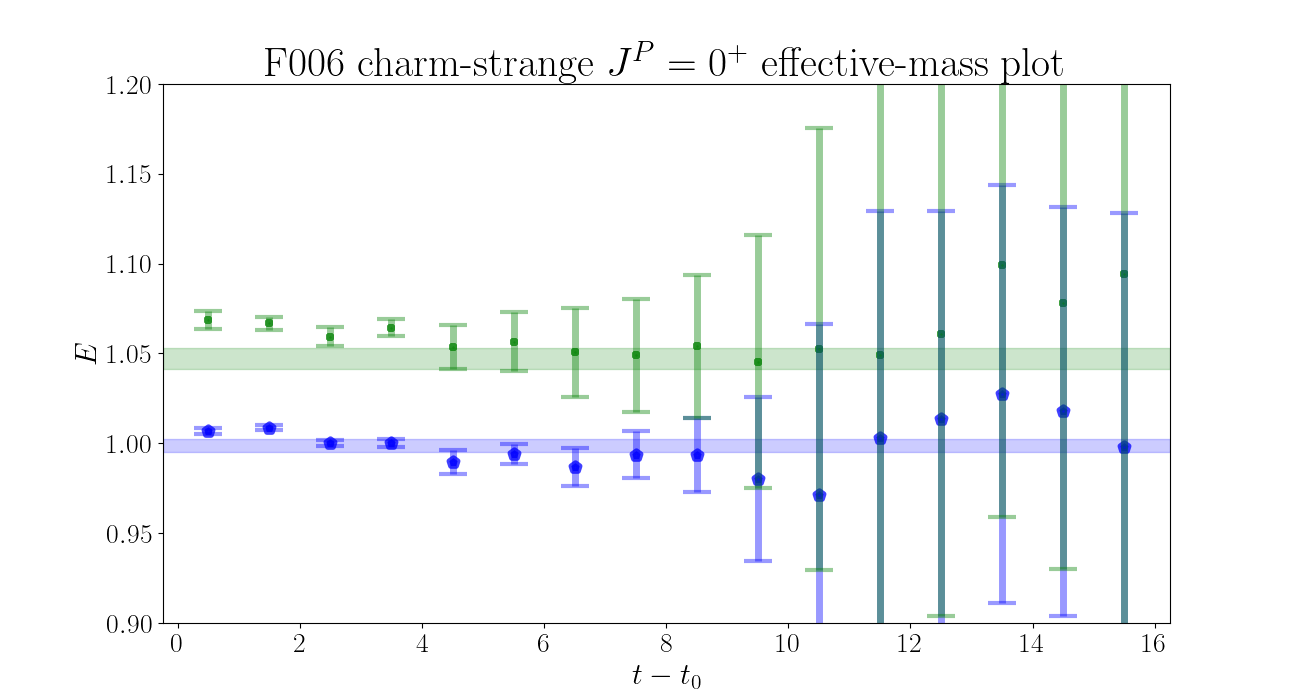}
    \hfill
    \includegraphics[width=0.49\linewidth]{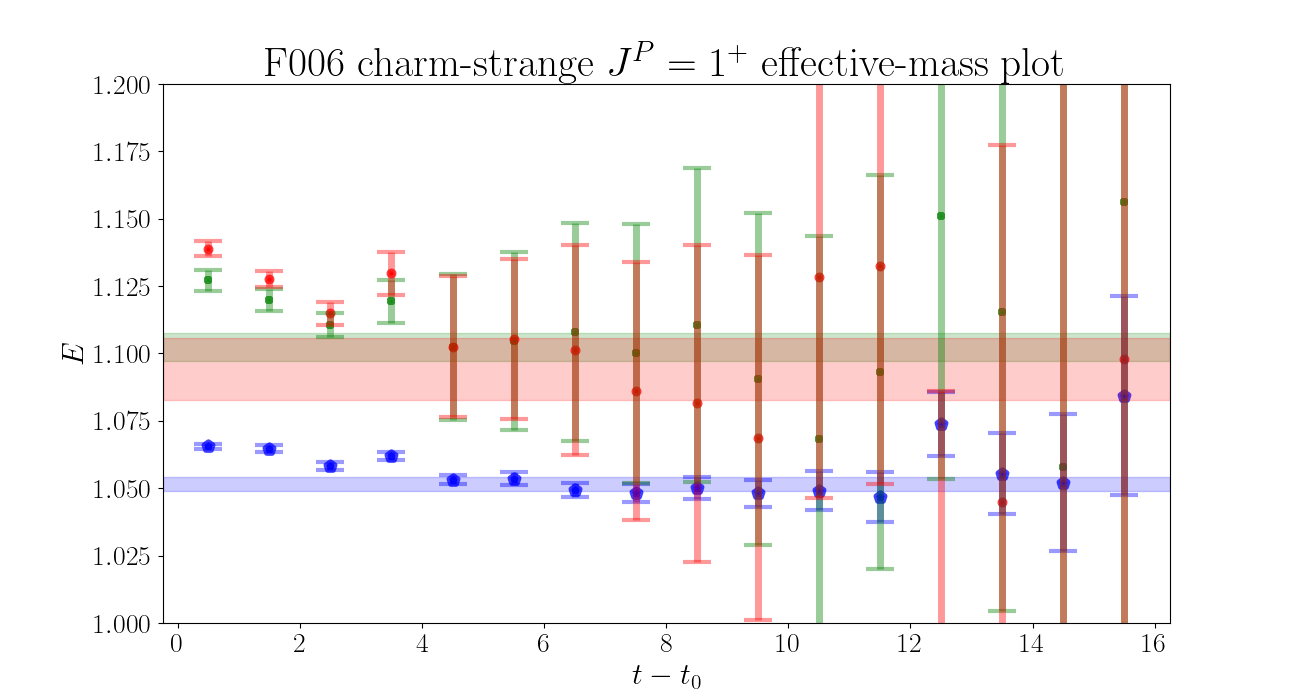}
    
    \includegraphics[width=0.49\linewidth]{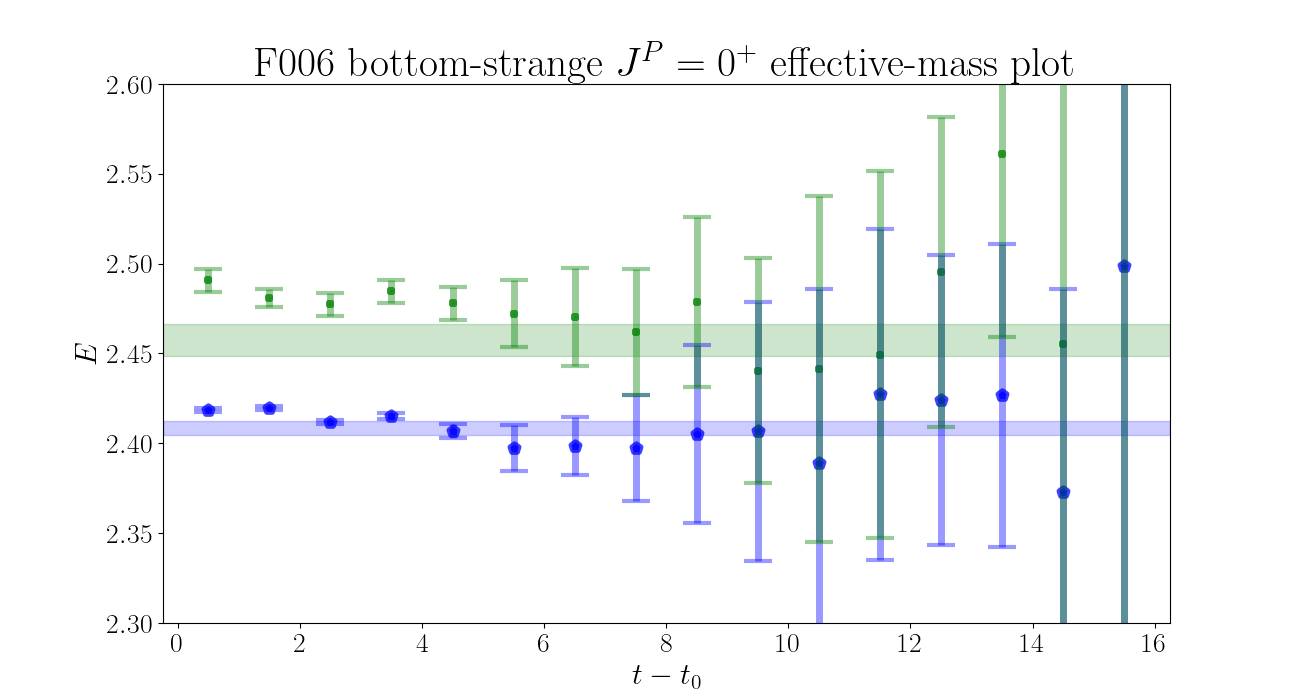}
    \hfill
    \includegraphics[width=0.49\linewidth]{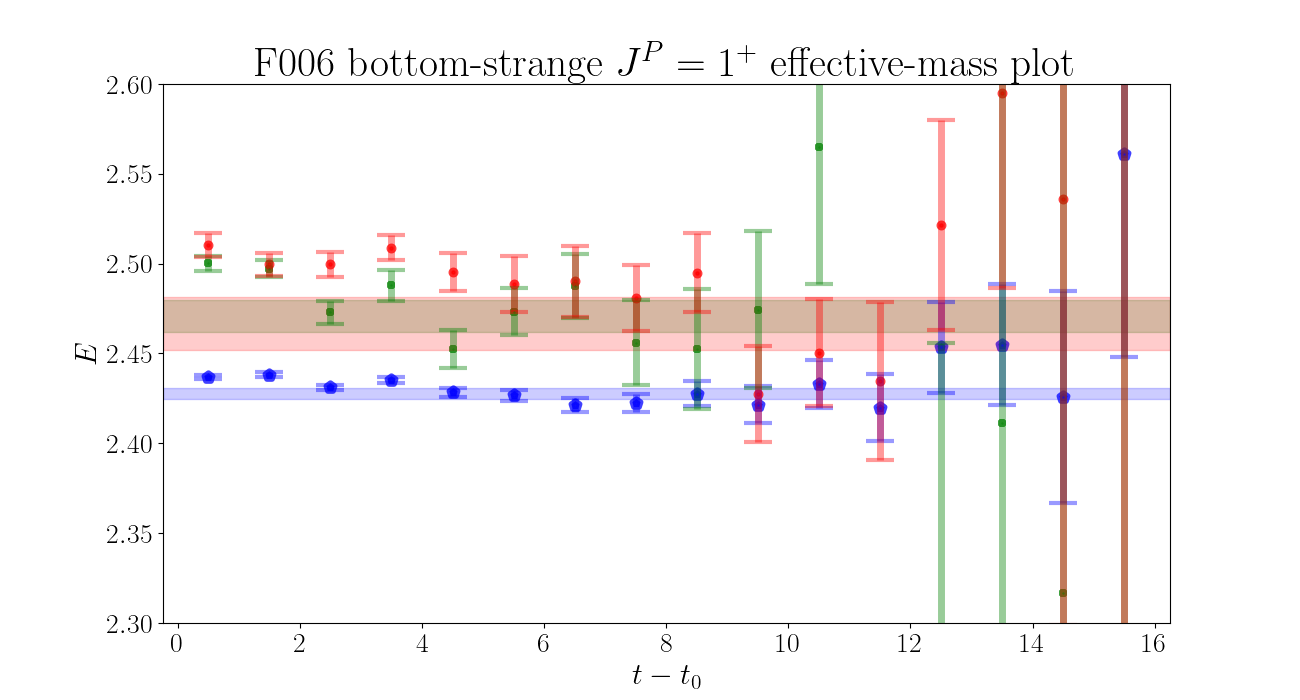}
    \caption{Effective-energy plots of the positive-parity principal correlators for the lowest two (three) states of the $J=0$ ($J=1$) system on the F006 ensemble. The bands correspond to the final model-averaged energies obtained from these principal correlators. The energy and time are given in lattice units. For the other ensembles, see Appendix \ref{sec:principalcorrplots}.}
    \label{fig:principalcorr}
\end{figure}

The choices of $t_0$ and $t_\text{ref}$ adopted in this work is given in Table \ref{tab:t0tref}. Examples of the principal correlators resulting from the GEVP on F006 are provided in Fig.~\ref{fig:principalcorr}, where we show the effective energies
\begin{equation}
E_{n,\rm eff}(t) = \ln \frac{\lambda_n(t,t_0)}{\lambda_n(t-1,t_0)}
\end{equation}
along with with the model-averaged fit results. The corresponding plots for the other ensembles are provided in Appendix \ref{sec:principalcorrplots}.

\subsubsection{Extracting $E_n$ and $f_n$ in Finite Volume from the GEVP}
\label{sec:extracting}

The positive-parity finite-volume energy levels $E_n$ are obtained from single-exponential fits to the principal correlators. We perform model averaging over the results from multiple  $t_\text{min}$'s in some range around the onset of the plateau (at fixed $t_\text{max}$), weighted according to the Akaike information criterion with a penalty term included for less data \cite{bays}. We typically use 3-4 different $t_\text{min}$'s on the coarse lattices and 5-6 on the fine lattices. Following Ref.~\cite{bays}, we compute the model weight for each $t_\text{min}$ as
\begin{equation}
    w = A\exp(-\frac{1}{2}(\chi^2+2k+2N_\text{cut})), \label{eq:aic}
\end{equation}
where $k$ is the number of fit parameters, $N_\text{cut}$ is the number of data points not included in the fit, and $A$ is a normalization constant chosen such that the weights sum to 1.
A weighted average is then computed for each bootstrap sample.

\begin{figure}
\centering

\includegraphics[height=0.299\textwidth]{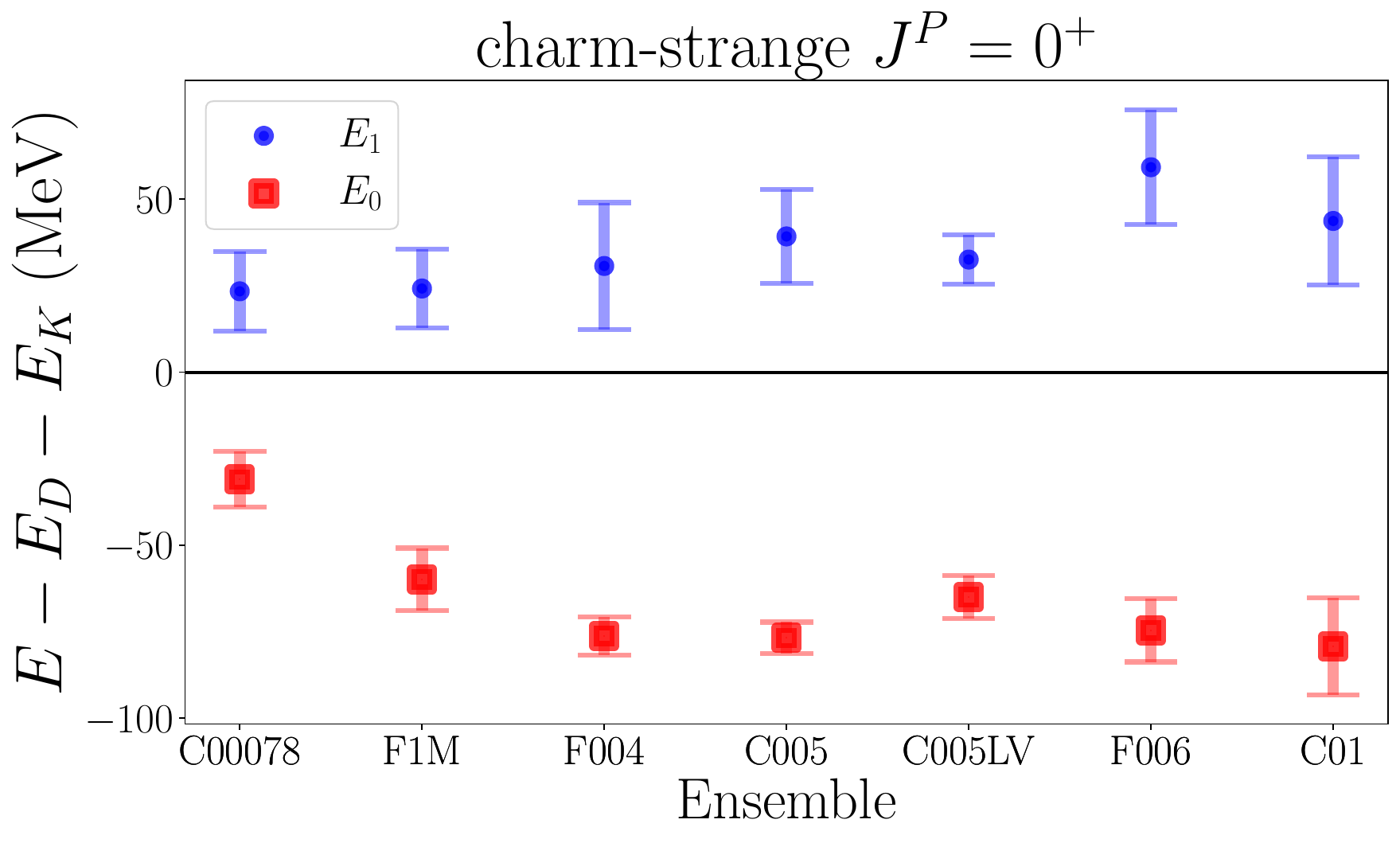}
\hfill
\includegraphics[height=0.299\textwidth]{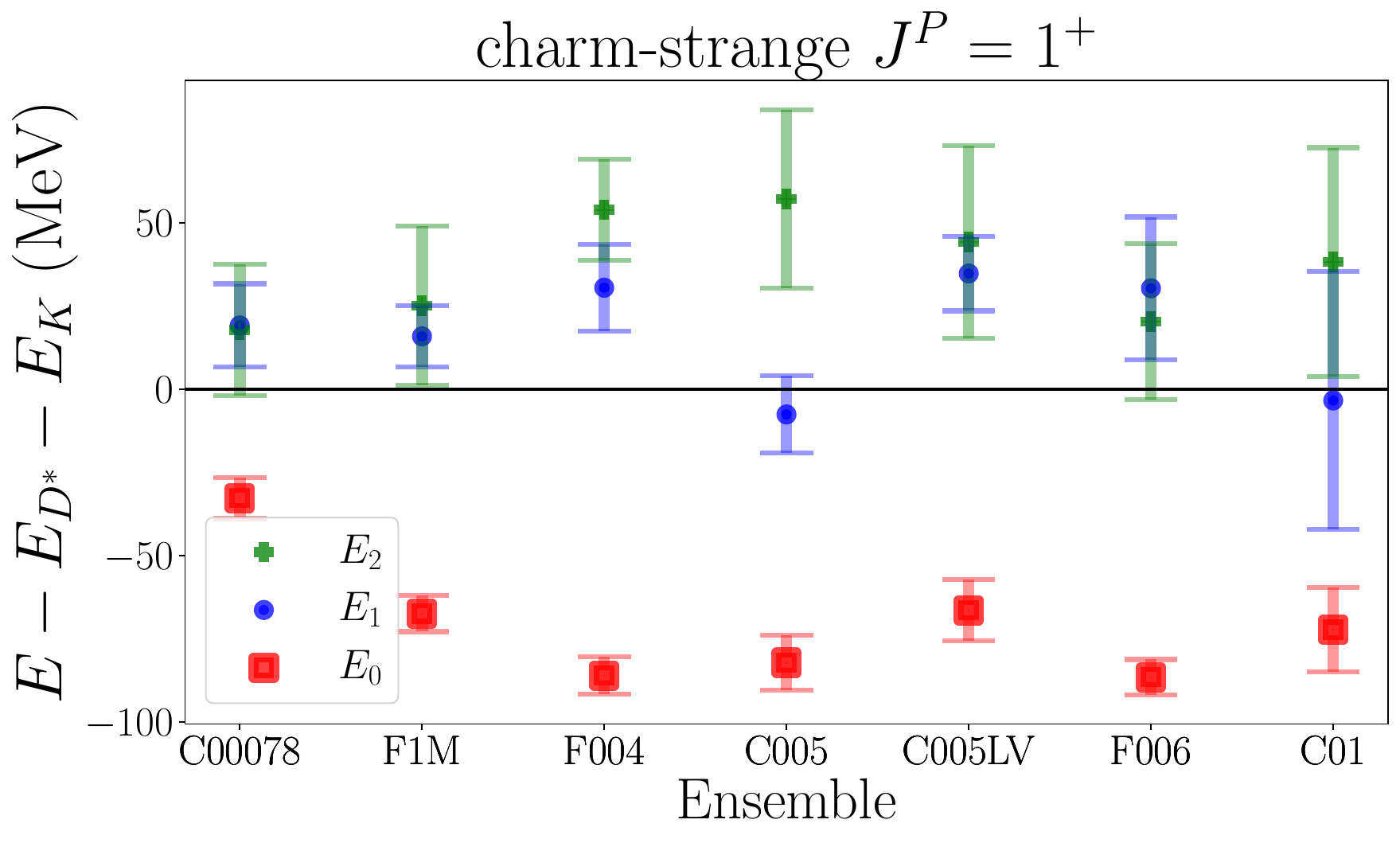}
\caption{Finite-volume spectra of the positive-parity charm-strange states relative to the $DK$ and $D^* K$ thresholds.}
\label{fig:Dsspec}
\end{figure}

\begin{figure}
\includegraphics[height=0.299\textwidth]{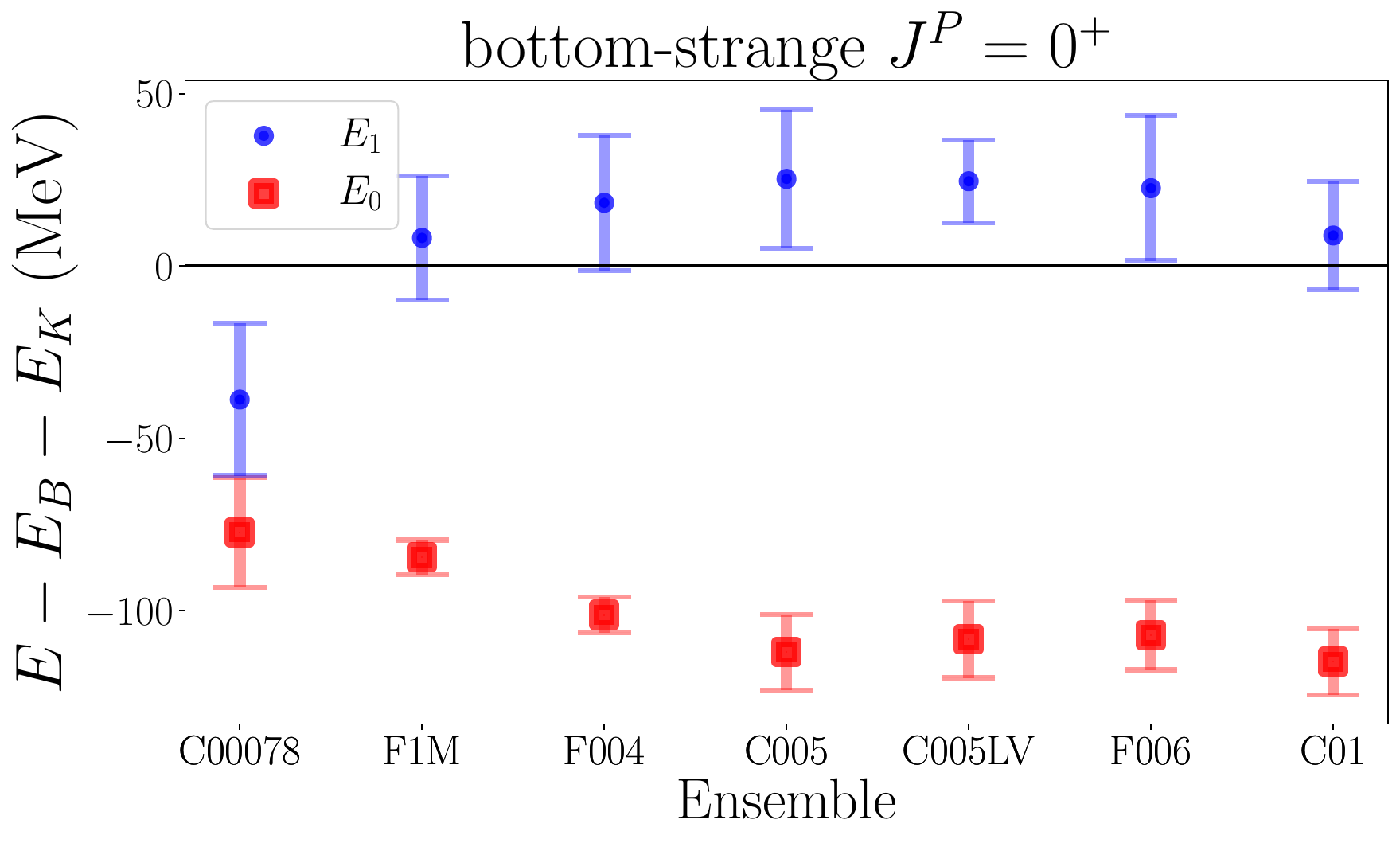}
\hfill
\includegraphics[height=0.299\textwidth]{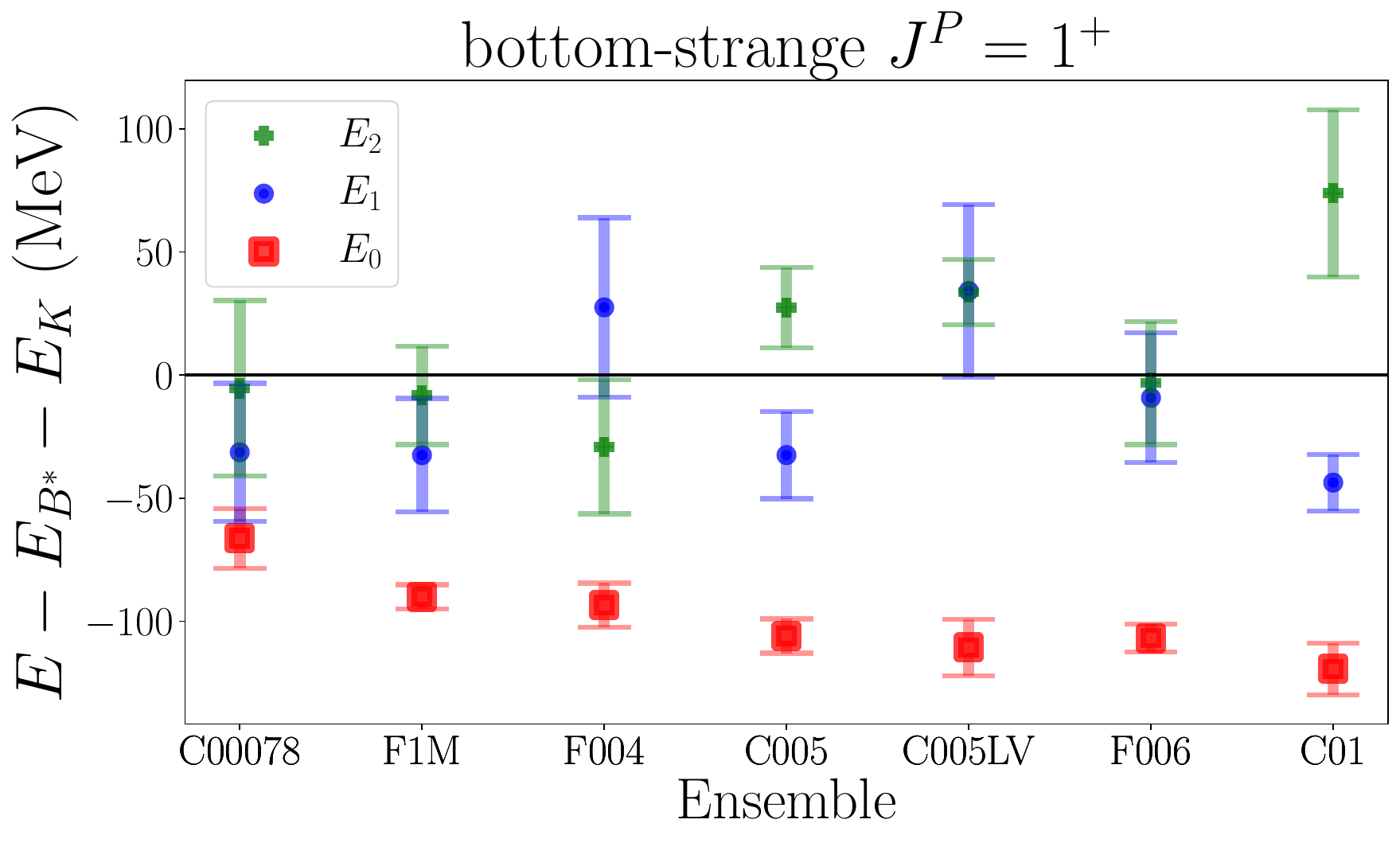}

\caption{Finite-volume spectra of the positive-parity bottom-strange states relative to the $BK$ and $B^* K$ thresholds.}
\label{fig:Bsspec}
\end{figure}

This same procedure (with somewhat wider ranges for $t_{\rm min}$) is performed for the $K$, $D^{(*)}$ and $B^{(*)}$ masses. We then calculate the difference to the $H^{(*)}K$ kinematic threshold $\Delta E_n=E_n-E_{H^{(*)}}-E_K$ for each bootstrap sample. The masses of the $K$, $D^{(*)}$ and $B^{(*)}$ on each ensemble are given in Table \ref{tab:DKmasses}.
The finite-volume spectra are shown in Figs.~\ref{fig:Dsspec}-\ref{fig:Bsspec} and listed in Table \ref{tab:spec}. In evaluating the bootstrap means and standard deviations, some outlier samples were dropped because those samples produced numerical exceptions or unphysical results in the later analyses using L\"uscher's method and the effective-range expansions (Sec.~\ref{sec:luscher}), as well as Weinberg's compositeness criterion (Sec.~\ref{sec:molecular}).

\begin{table}[H]
    \centering
    \begin{tabular}{|l|c|c|c|c|c|c|}
    \hline
        Ensemble & \multicolumn{2}{|c|}{$K$} & \multicolumn{2}{|c|}{$D$} & \multicolumn{2}{|c|}{$D^*$} \\
        \hline
         & mass & mass (MeV) & mass & mass (MeV) & mass & mass (MeV) \\
		C00078 & $0.28838 \pm 0.00026$ & $498.7 \pm 1.5$ & $1.0829 \pm 0.0011$ & $1872.8 \pm 5.7$ & $1.1635 \pm 0.0023$ & $2012.1 \pm 7.0$ \\
		F1M & $0.19080 \pm 0.00019$ & $516.7 \pm 2.0$ & $0.69144 \pm 0.00041$ & $1872.4 \pm 7.0$ & $0.74729 \pm 0.00070$ & $2023.7 \pm 7.7$ \\
		F004 & $0.22485 \pm 0.00058$ & $535.8 \pm 2.4$ & $0.79496 \pm 0.00049$ & $1894.4 \pm 7.2$ & $0.85676 \pm 0.00075$ & $2041.7 \pm 7.9$ \\
		C005LV & $0.30573 \pm 0.00049$ & $545.7 \pm 1.8$ & $1.06081 \pm 0.00076$ & $1893.3 \pm 5.5$ & $1.1461 \pm 0.0012$ & $2045.6 \pm 6.1$ \\
		C005 & $0.30704 \pm 0.00047$ & $548.0 \pm 1.8$ & $1.06093 \pm 0.00062$ & $1893.5 \pm 5.4$ & $1.1454 \pm 0.0012$ & $2044.3 \pm 6.1$ \\
		F006 & $0.23180 \pm 0.00043$ & $552.4 \pm 2.3$ & $0.79681 \pm 0.00036$ & $1898.8 \pm 7.2$ & $0.85791 \pm 0.00076$ & $2044.4 \pm 7.9$ \\
		C01 & $0.32451 \pm 0.00045$ & $579.2 \pm 1.8$ & $1.07120 \pm 0.00060$ & $1911.9 \pm 5.5$ & $1.1568 \pm 0.0011$ & $2064.7 \pm 6.1$ \\
        \hline
    \end{tabular}
    \begin{tabular}{|l|c|c|c|c|}
    \hline
        Ensemble & \multicolumn{2}{|c|}{$B$} & \multicolumn{2}{|c|}{$B^*$} \\
        \hline
         & mass & mass (MeV) & mass & mass (MeV) \\
		C00078 & $3.0636 \pm 0.0062$ & $5298 \pm 19$ & $3.0832 \pm 0.0060$ & $5332 \pm 19$ \\
		F1M & $1.9543 \pm 0.0011$ & $5292 \pm 20$ & $1.9717 \pm 0.0012$ & $5339 \pm 20$ \\
		F004 & $2.2251 \pm 0.0010$ & $5302 \pm 20$ & $2.2445 \pm 0.0011$ & $5349 \pm 20$ \\
		C005LV & $2.9723 \pm 0.0016$ & $5305 \pm 15$ & $3.0001 \pm 0.0021$ & $5355 \pm 15$ \\
		C005 & $2.9710 \pm 0.0021$ & $5303 \pm 15$ & $2.9980 \pm 0.0027$ & $5351 \pm 16$ \\
		F006 & $2.22667 \pm 0.00092$ & $5306 \pm 20$ & $2.2458 \pm 0.0012$ & $5352 \pm 20$ \\
		C01 & $2.9791 \pm 0.0014$ & $5317 \pm 15$ & $3.0064 \pm 0.0019$ & $5366 \pm 15$ \\
            \hline
    \end{tabular}
    \caption{Masses of the kaon and negative-parity heavy-light mesons on each ensemble. The first sub-column for each mass is in lattice units, while the second sub-column is in MeV and accounts for the uncertainty in the determination of the lattice spacing.}
    \label{tab:DKmasses}
\end{table}

Using the generalized eigenvector averaged over all bootstrap samples at the reference time $t_\text{ref}$, we form the optimized operators
\begin{equation}
W^n= v^n_l \Phi^{(l)},
\end{equation}
which  overlap only with the state $n$ and high-lying states beyond the subset that is diagonalized by the GEVP. From the original correlation matrices, we then construct the optimized correlaters  
\begin{align}
    C_{JW^n}(t)&=\sum_{\vec{z}} \langle J^\dagger(\vec{z}, t+t_s) W^{n\dagger}(\vec{x}_s,t_s)\rangle= v^n_m C^{3m}(t),\\
    C_{W^nW^n}(t)&=\sum_{\vec{z}} \langle W^n(\vec{z}, t+t_s) W^{n\dagger}(\vec{x}_s,t_s)\rangle= v^{n}_l C^{lm}(t)v^n_m.
\end{align}
Effective-energy plots of the $\langle J^\dagger W^\dagger \rangle$ and $\langle WW^\dagger \rangle$ effective-energy correlators are shown in Fig.~\ref{fig:F006wplots} for the F006 ensemble and in Appendix \ref{sec:wplots} for the other ensembles. We fit the above correlators in the region of single-state dominance using
\begin{align}
C_{JW^n}(t) &= A_{JW^n} e^{-E_n t}, \\
C_{W^n W^n}(t) &= A_{W^n W^n} e^{-E_n t},
\end{align}
and calculate the decay constants as 
\begin{equation}
    f_n=A_{JW^n} \sqrt{\frac{2 }{E_n\:A_{W^n W^n}}}. \label{eq:fposparity}
\end{equation} 
In Eq.~(\ref{eq:fposparity}), we set $E_n$ equal to the values obtained from the fits to the principal correlators.

\begin{figure}
    \centering
    \centering
    \includegraphics[width=0.49\linewidth]{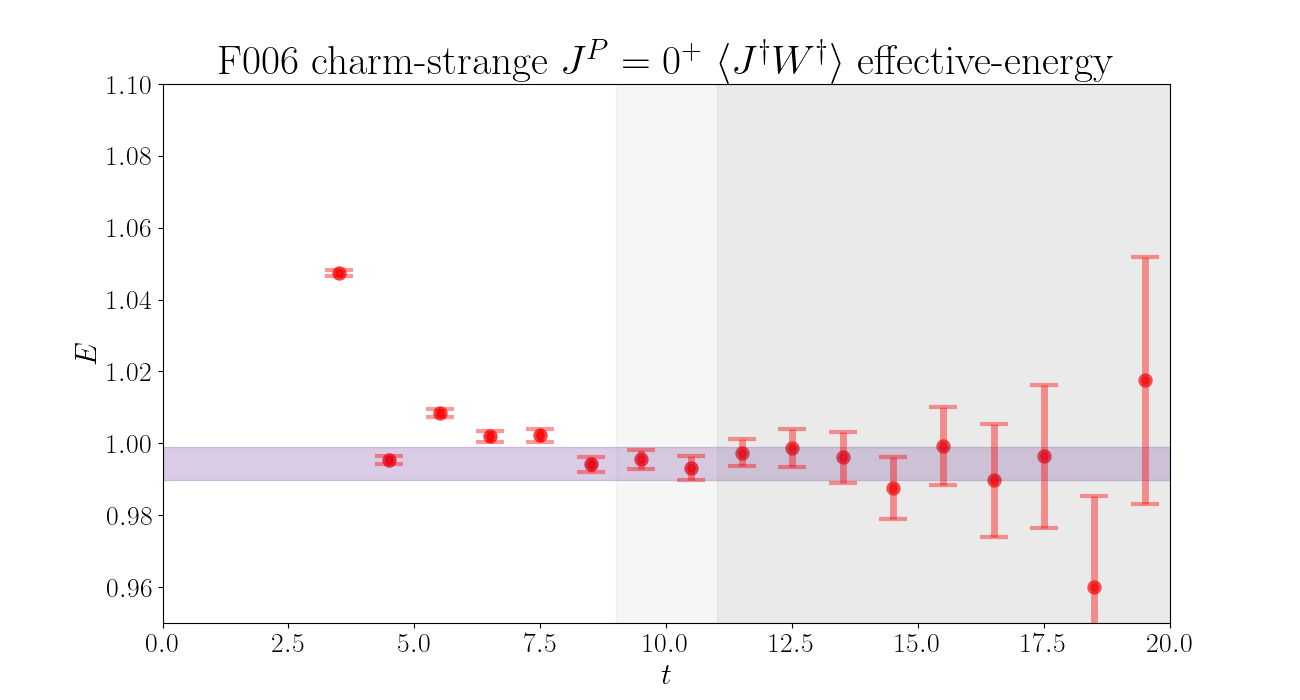}
    \hfill
    \includegraphics[width=0.49\linewidth]{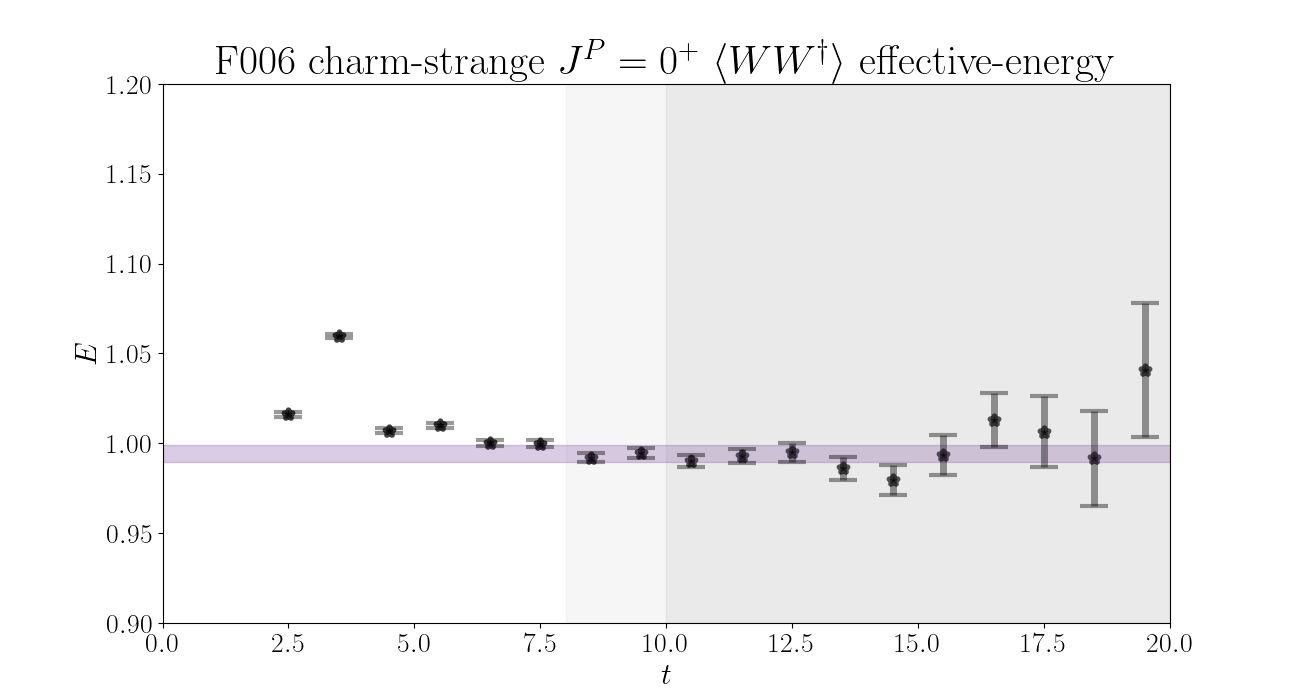}
    
    \includegraphics[width=0.49\linewidth]{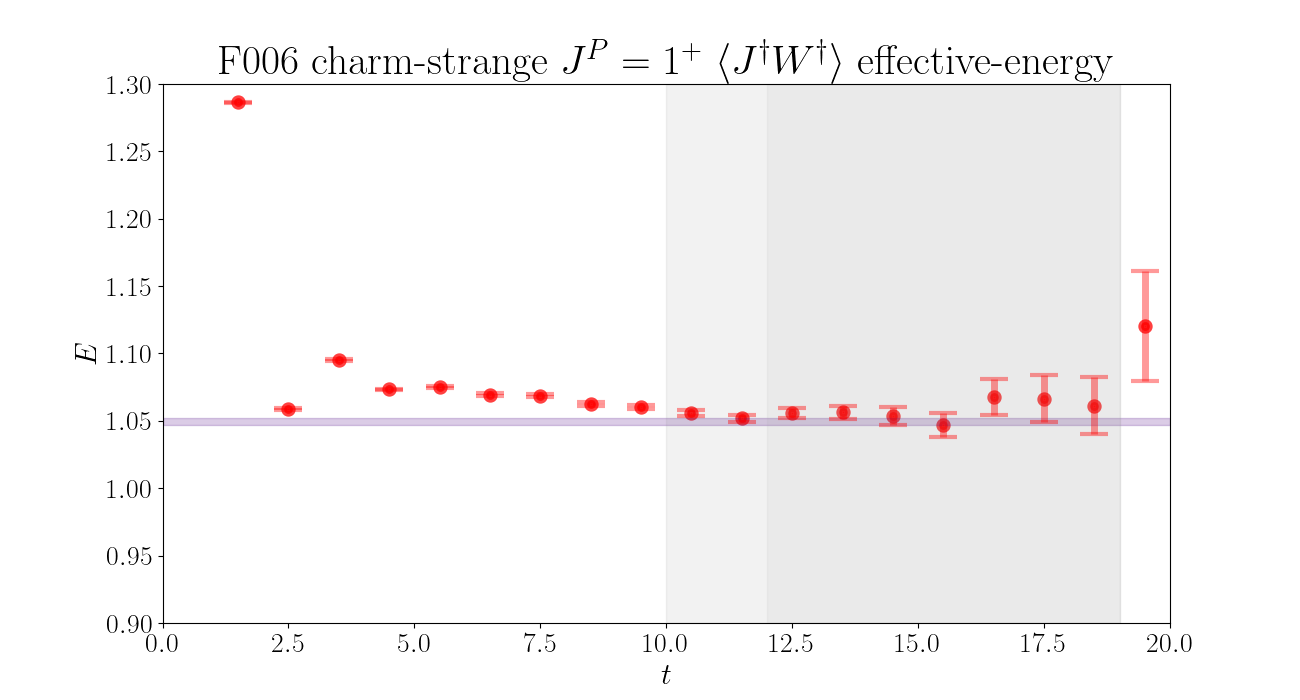}
    \hfill
    \includegraphics[width=0.49\linewidth]{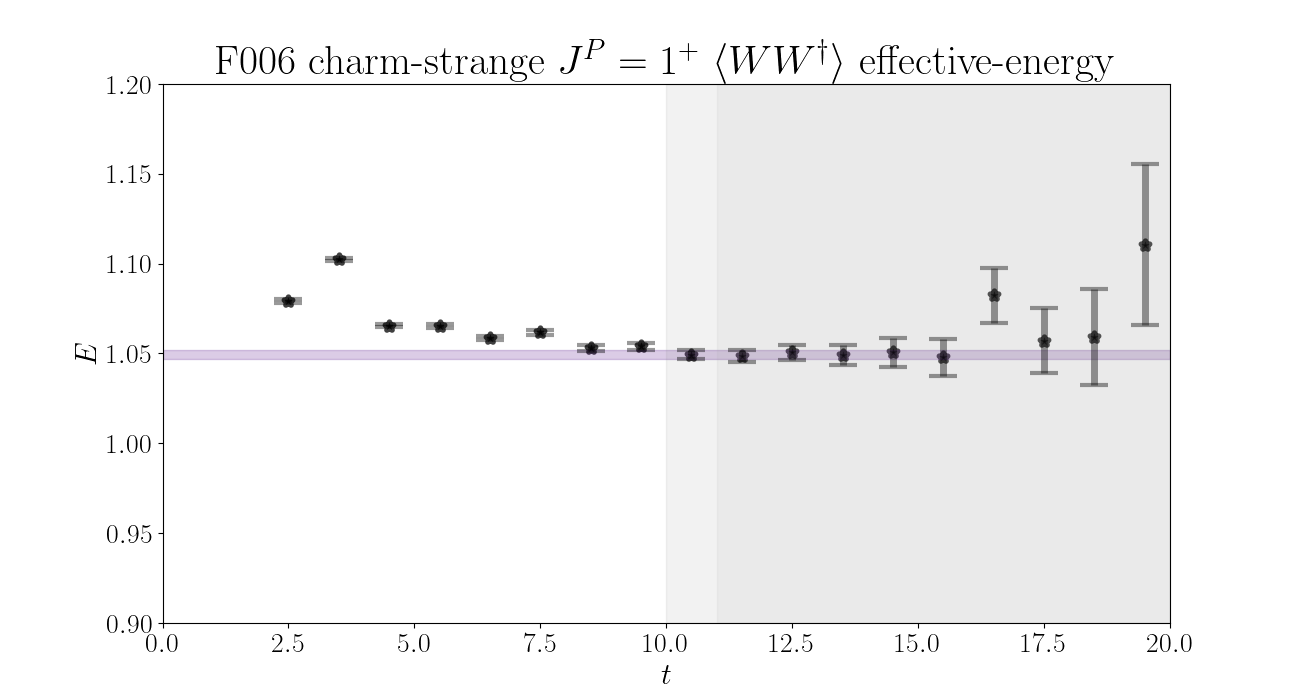}

    \includegraphics[width=0.49\linewidth]{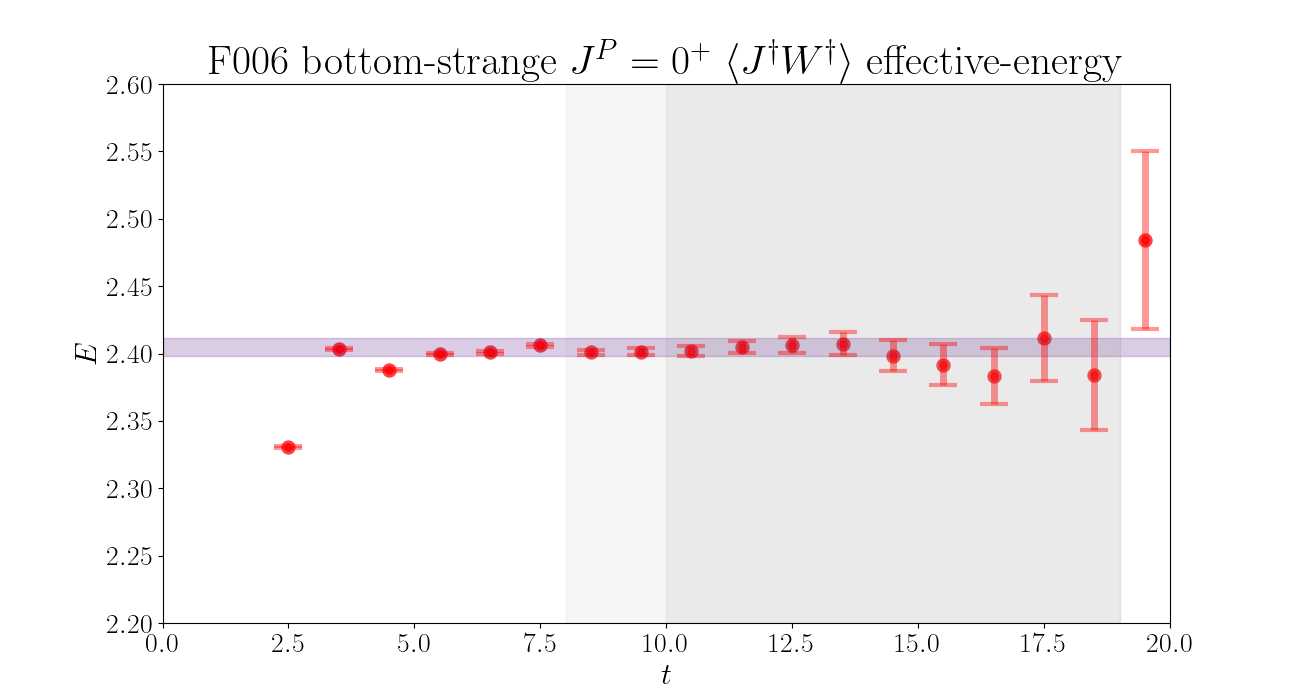}
    \hfill
    \includegraphics[width=0.49\linewidth]{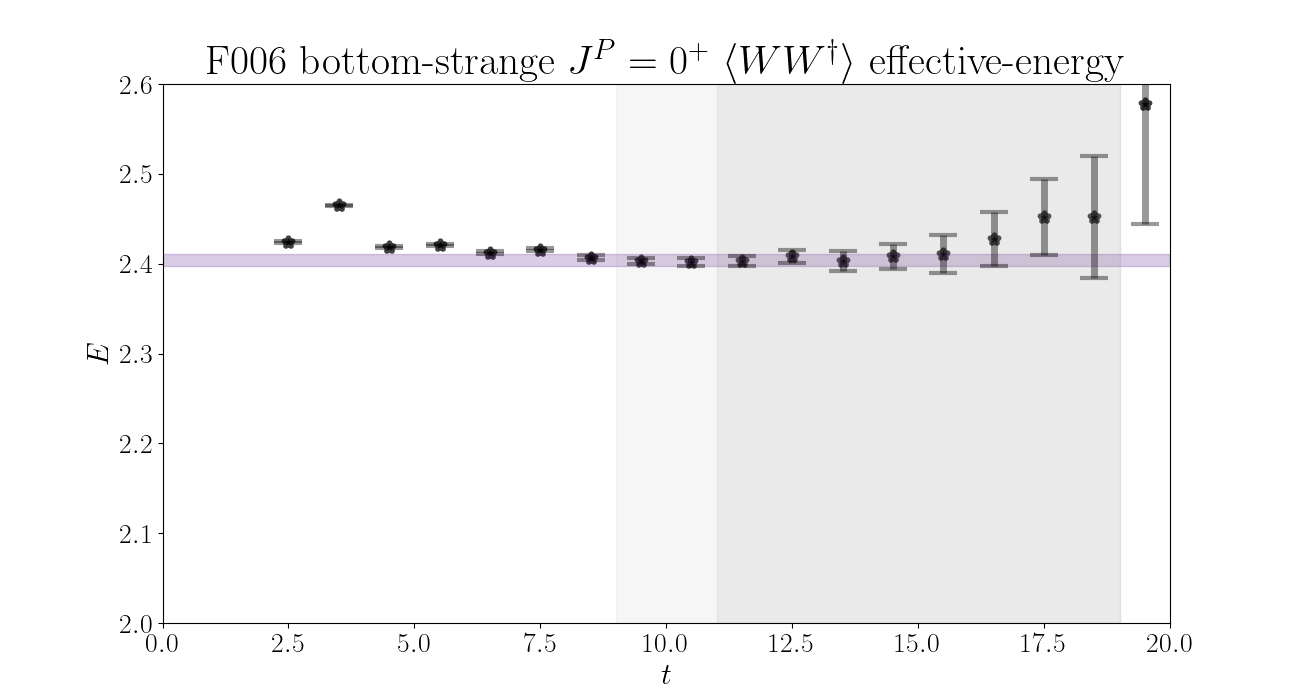}
    
    \includegraphics[width=0.49\linewidth]{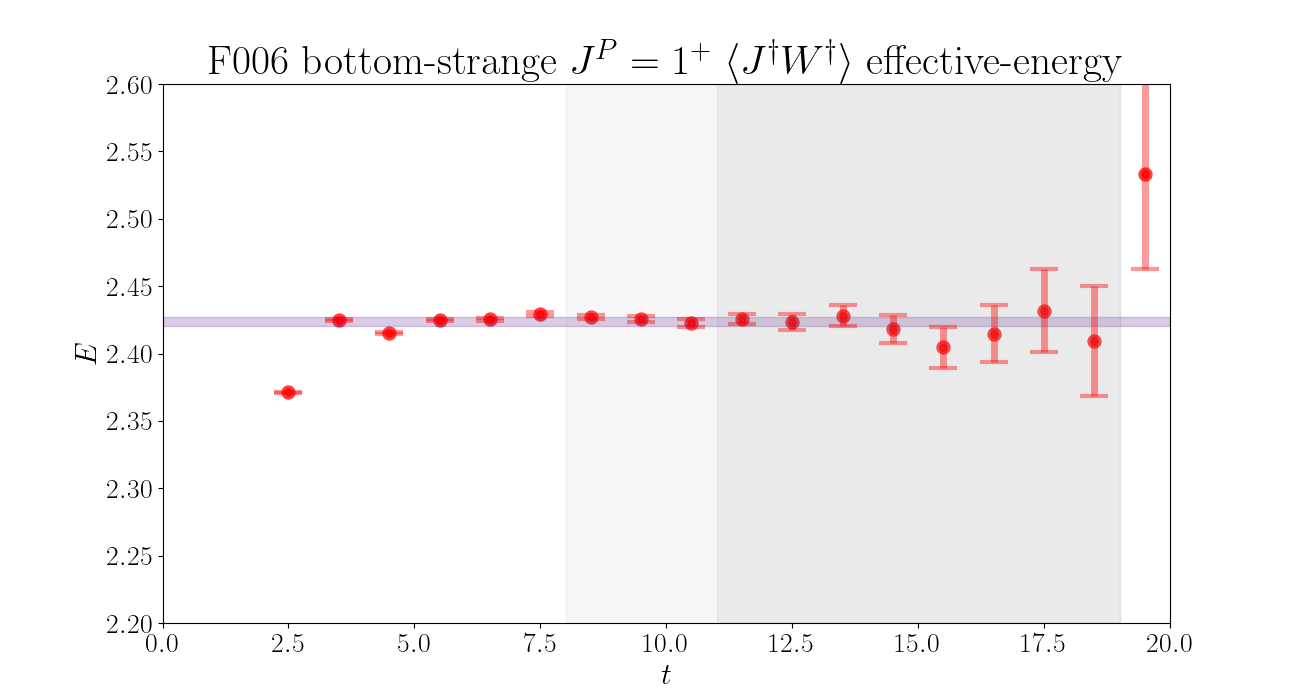}
    \hfill
    \includegraphics[width=0.49\linewidth]{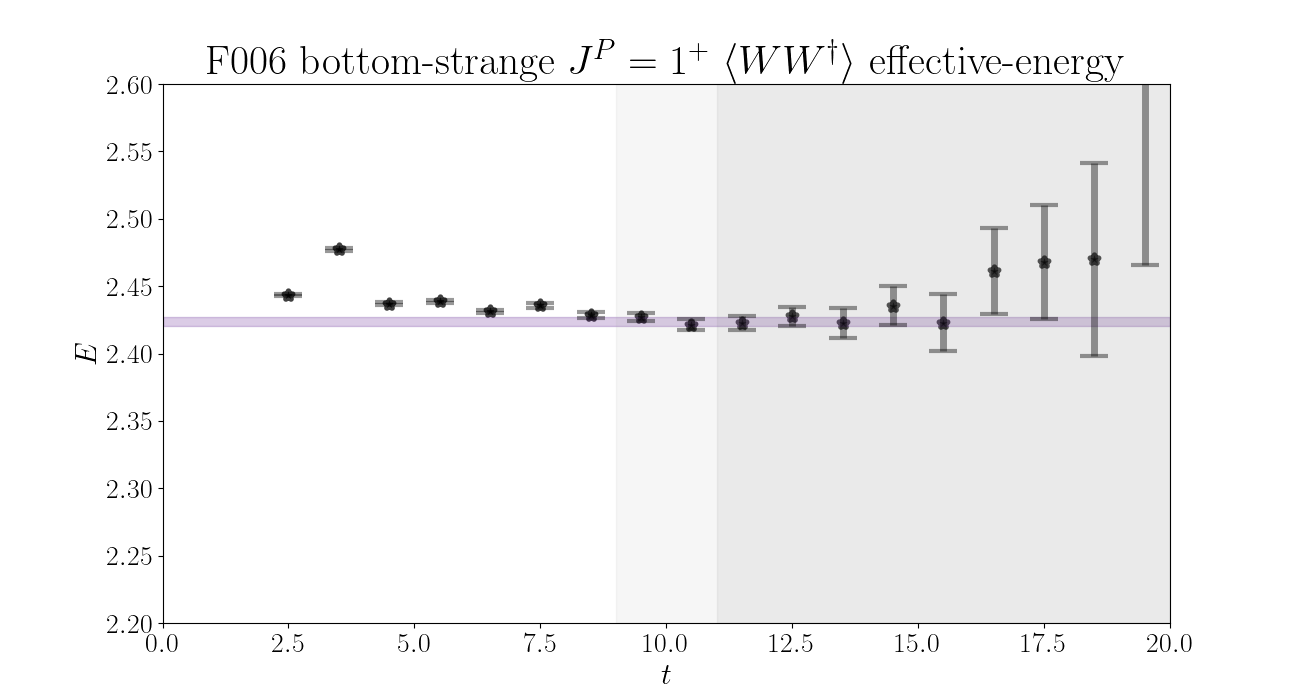}
    \caption{Effective-energy plots (in lattice units) for the $\langle J^\dagger W^\dagger \rangle$ and $\langle WW^\dagger \rangle$ correlators for each system on F006. For plots of the other ensembles, see Appendix \ref{sec:wplots}. The horizontal purple bands correspond to the model-averaged ground-state energies extracted from the principal correlators. The combinations of the light and dark shaded regions indicate the fit ranges used to obtain the central values and statistical uncertainties of the amplitudes, while the dark shaded regions indicates the fit ranges used in the calculation of the fit-range systematic uncertainties.}
    \label{fig:F006wplots}

\end{figure}

\begin{table}
    \centering
    \begin{tabular}{|l|c|c|}
    \hline
         Ensemble & $\Delta E$ (MeV) &  $\Delta E_1$ (MeV)\\ \hline
         \hline
    \multicolumn{3}{|c|}{Charm-strange $J^P=0^+$}\\
         \hline
		C00078 & $-31.0 \pm 8.1$ & $23 \pm 12$ \\
		F1M & $-59.9 \pm 9.0$ & $24 \pm 13$ \\
		F004 & $-76.2 \pm 5.5$ & $31 \pm 18$ \\
		C005LV & $-65.0 \pm 6.2$ & $32.6 \pm 7.1$ \\
		C005 & $-76.8 \pm 4.6$ & $39 \pm 14$ \\
		F006 & $-74.6 \pm 9.1$ & $59 \pm 17$ \\
		C01 & $-79 \pm 14$ & $44 \pm 19$ \\
         \hline
    \hline
    \multicolumn{3}{|c|}{Bottom-strange $J^P=0^+$}\\
    \hline
		C00078 & $-77 \pm 16$ & $-39 \pm 22$ \\
		F1M & $-84.5 \pm 5.0$ & $8 \pm 18$ \\
		F004 & $-101.3 \pm 5.1$ & $18 \pm 20$ \\
		C005LV & $-108 \pm 11$ & $25 \pm 12$ \\
		C005 & $-112 \pm 11$ & $25 \pm 20$ \\
		F006 & $-107 \pm 10$ & $23 \pm 21$ \\
		C01 & $-114.9 \pm 9.5$ & $9 \pm 16$ \\
    \hline
    \end{tabular}
         \qquad\qquad
    \begin{tabular}{|l|c|c|c|}
    \hline
         Ensemble & $\Delta E$ (MeV) &  $\Delta E_1$ (MeV) & $\Delta E_2$ (MeV)\\
         \hline
                 \hline
    \multicolumn{4}{|c|}{Charm-strange $J^P=1^+$}\\
    \hline
		C00078 & $-32.7 \pm 6.2$ & $19 \pm 13$ & $18 \pm 20$ \\
		F1M & $-67.4 \pm 5.4$ & $15.8 \pm 9.2$ & $25 \pm 24$ \\
		F004 & $-86.0 \pm 5.6$ & $31 \pm 13$ & $54 \pm 15$ \\
		C005LV & $-66.5 \pm 9.2$ & $35 \pm 11$ & $44 \pm 29$ \\
		C005 & $-82.2 \pm 8.2$ & $57 \pm 27$ & $-8 \pm 12$ \\
		F006 & $-86.5 \pm 5.3$ & $30 \pm 21$ & $20 \pm 24$ \\
		C01 & $-72 \pm 13$ & $38 \pm 34$ & $-3 \pm 39$ \\
         \hline
    \hline
    \multicolumn{4}{|c|}{Bottom-strange $J^P=1^+$}\\
    \hline
		C00078 & $-66 \pm 12$ & $-31 \pm 28$ & $-5 \pm 36$ \\
		F1M & $-90.0 \pm 4.9$ & $-33 \pm 23$ & $-8 \pm 20$ \\
		F004 & $-93.4 \pm 9.0$ & $28 \pm 36$ & $-29 \pm 27$ \\
		C005LV & $-111 \pm 11$ & $34 \pm 13$ & $34 \pm 35$ \\
		C005 & $-105.9 \pm 7.0$ & $27 \pm 16$ & $-33 \pm 18$ \\
		F006 & $-106.8 \pm 5.6$ & $-3 \pm 25$ & $-9 \pm 26$ \\
		C01 & $-119 \pm 11$ & $74 \pm 34$ & $-44 \pm 12$ \\
         \hline
    \end{tabular}

\caption{Finite-volume energy differences from the $H^{(*)}K$ thresholds in MeV for each of the states and each ensemble.}
    \label{tab:spec}
\end{table}

\begin{figure}[H]
\includegraphics[height=0.28\textwidth]{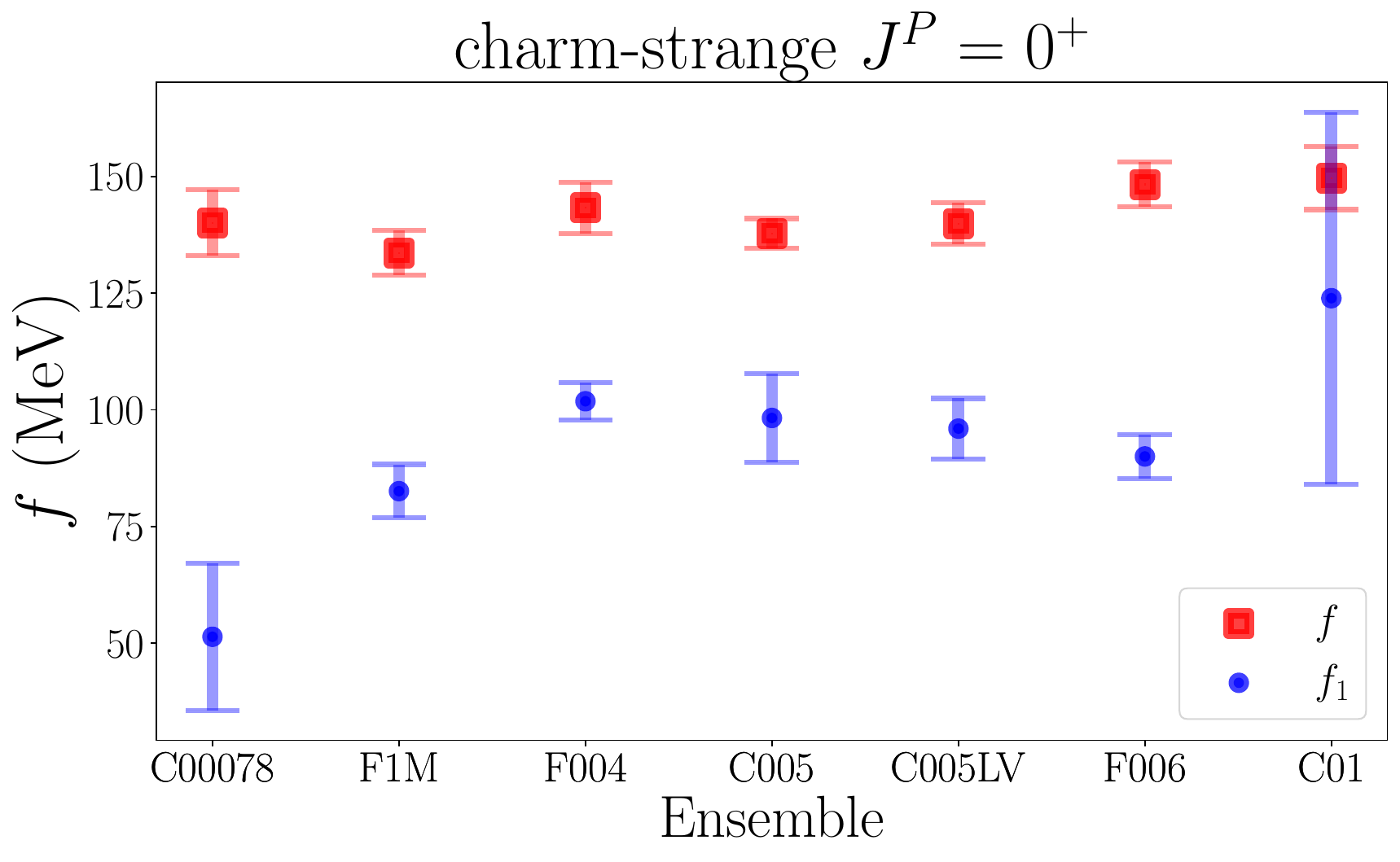}
\hfill
\includegraphics[height=0.28\textwidth]{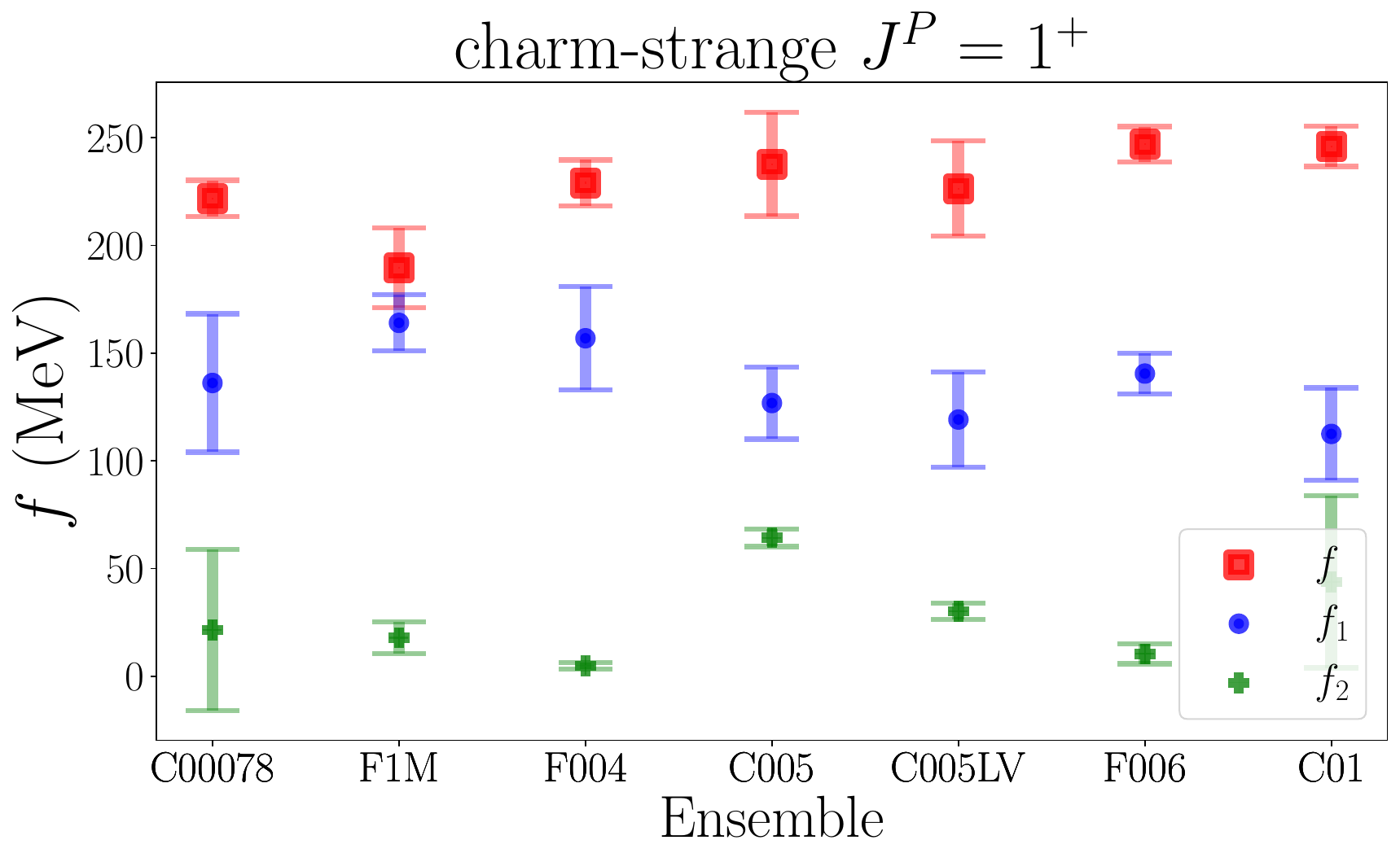}

\caption{Finite-volume decay constants of the positive-parity charm-strange states.}
\label{fig:Dsdecay}
\end{figure}

\begin{figure}[H]
\includegraphics[height=0.28\textwidth]{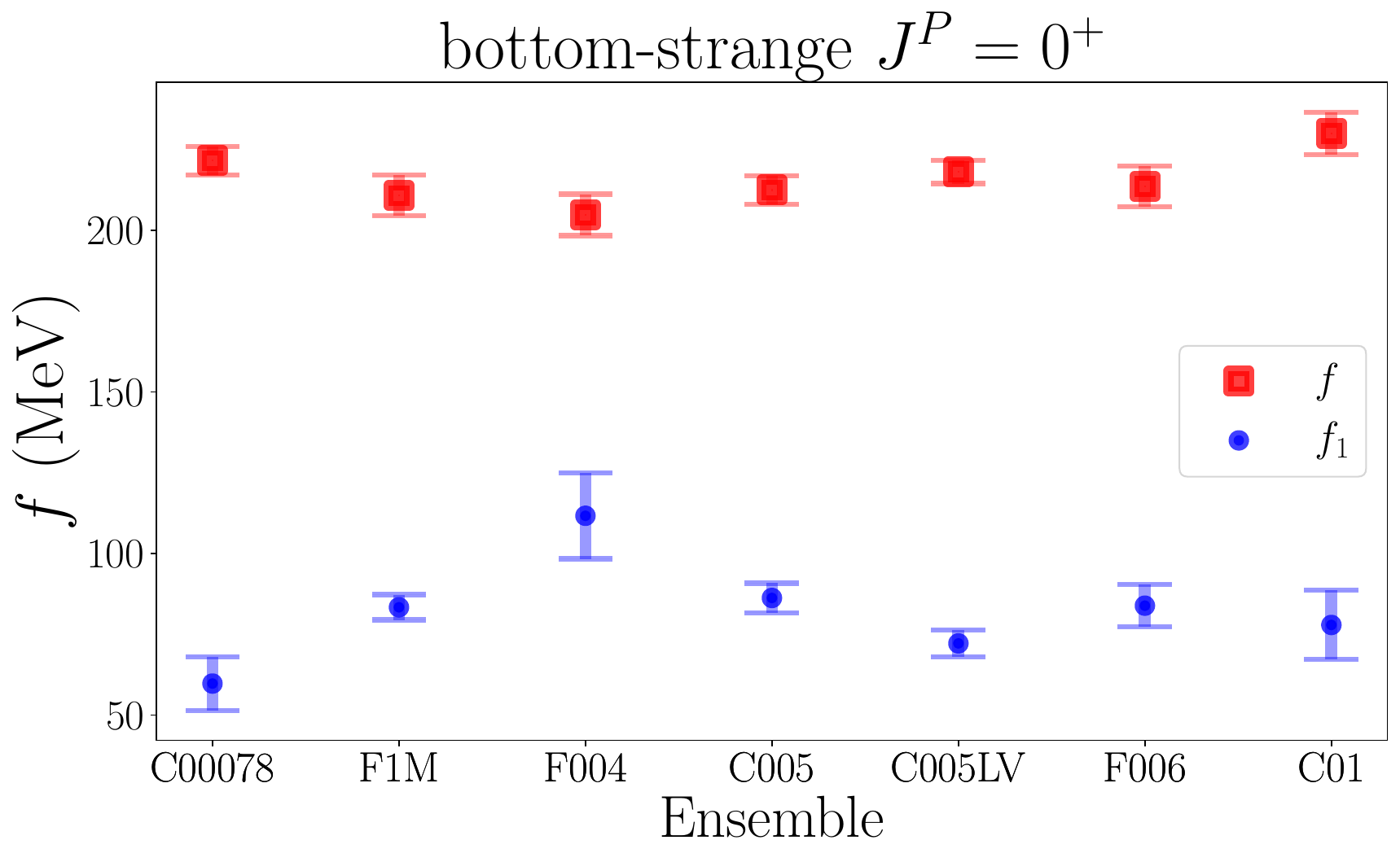}
\hfill
\includegraphics[height=0.28\textwidth]{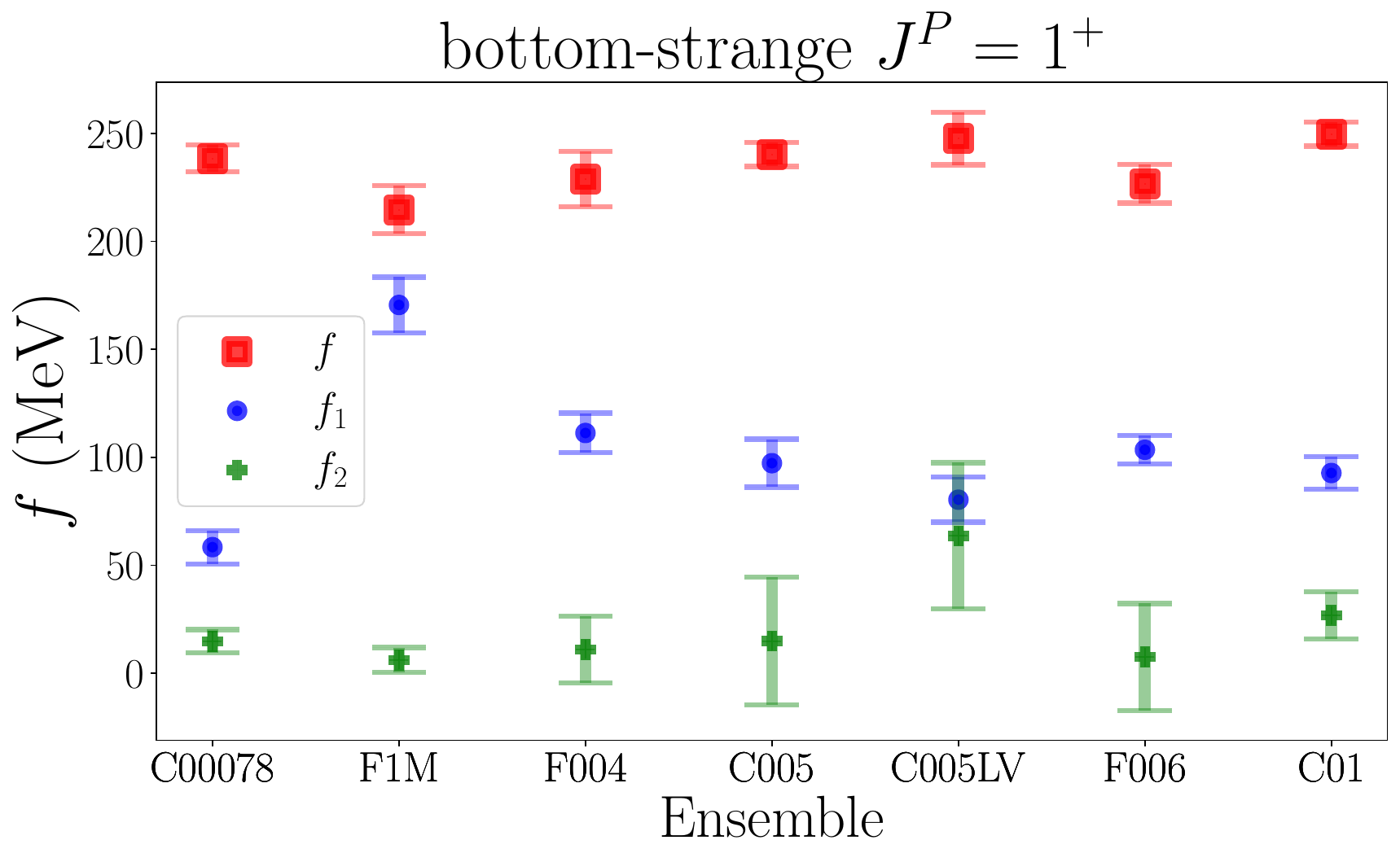}

\caption{Finite-volume decay constants of the positive-parity bottom-strange states.}
\label{fig:Bsdecay}
\end{figure}

While the model-averaging procedure described above provides a comprehensive approach to varying the choice of fit range, we found that it resulted in unreasonably small uncertainty estimates for the the decay constants. This was caused by the larger $t_{\rm min}$ values getting very small weights as a result of local statistical fluctuations in the $\langle J^\dagger W^\dagger\rangle$ correlators.
For this reason, we choose only a single fit to contribute to the central value and statistical uncertainty of the decay constant, and included an additional systematic error in the decay constant equal to the statistical uncertainty of a fit where $t_\text{min}\rightarrow t_\text{min}+t_\text{shift}$. The chosen single fit for the central value actually matched the fit given the highest weight by the AIC, except for the $D_{s1}$ on F1M and F006 and the $B_{s1}$ on C005LV. The $t_\text{shift}$ values are chosen individually for each correlation function, and typically range from $t_\text{shift}\sim 0.1 - 0.3$ fm.
The results for the decay constants are shown in Figs.~\ref{fig:Dsdecay}-\ref{fig:Bsdecay} and are given in Table  \ref{tab:decay}.

Note that the finite-volume ``decay constants'' of the excited states do not have a simple interpretation. We expect that $f_1$ corresponds to a matrix element with an interacting $H^{(*)}K$ finite-volume state, and that $f_2$ also cannot be directly identified as corresponding to the predominantly $^1 P_1$ resonance, which, even in infinite-volume QFT, would not be an asymptotic state.

\begin{table}[H]
    \centering
    \begin{tabular}{|l|c|c|}
    \hline
    Ensemble & $f$ (MeV) &  $f_1$ (MeV)\\ \hline
    \hline
    \multicolumn{3}{|c|}{Charm-strange $J^P=0^+$}\\
         \hline
		C00078 & $140.1 \pm 7.1$ & $51 \pm 16$ \\
		F1M & $133.6 \pm 4.8$ & $82.6 \pm 5.7$ \\
		F004 & $143.3 \pm 5.5$ & $101.8 \pm 4.0$ \\
		C005LV & $139.9 \pm 4.5$ & $96.0 \pm 6.5$ \\
		C005 & $137.8 \pm 3.2$ & $98.2 \pm 9.6$ \\
		F006 & $148.3 \pm 4.7$ & $90.0 \pm 4.7$ \\
		C01 & $149.7 \pm 6.8$ & $124 \pm 40$ \\
         \hline
    \hline
    \multicolumn{3}{|c|}{Bottom-strange $J^P=0^+$}\\
    \hline
		C00078 & $221.5 \pm 4.3$ & $59.8 \pm 8.3$ \\
		F1M & $210.7 \pm 6.3$ & $83.4 \pm 3.9$ \\
		F004 & $204.7 \pm 6.4$ & $112 \pm 13$ \\
		C005LV & $218.0 \pm 3.6$ & $72.2 \pm 4.2$ \\
		C005 & $212.4 \pm 4.4$ & $86.2 \pm 4.6$ \\
		F006 & $213.5 \pm 6.2$ & $83.9 \pm 6.6$ \\
		C01 & $230.0 \pm 6.6$ & $78 \pm 11$ \\
         \hline
    \end{tabular}
         \qquad\qquad
    \begin{tabular}{|l|c|c|c|}
    \hline
             Ensemble & $f$ (MeV) &  $f_1$ (MeV) & $f_2$ (MeV)\\
    \hline
    \hline
        \multicolumn{4}{|c|}{Charm-strange $J^P=1^+$}\\
         \hline
		C00078 & $221.8 \pm 8.3$ & $136 \pm 32$ & $22 \pm 37$ \\
        F1M & $190 \pm 18$ & $164 13$ & $17.9 \pm 7.4$ \\
		F004 & $229 \pm 11$ & $157 \pm 24$ & $4.9 \pm 1.5$ \\
        C005LV & $226 \pm 22$ & $119 \pm 22$ & $30.2 \pm 3.8$ \\
		C005 & $238 \pm 24$ & $127 \pm 17$ & $64.3 \pm 4.1$ \\
		F006 & $246.8 \pm 8.2$ & $140.5 \pm 9.4$ & $10.4 \pm 4.7$ \\
		C01 & $245.9 \pm 9.3$ & $113 \pm 21$ & $44 \pm 40$ \\
         \hline
    \hline
    \multicolumn{4}{|c|}{Bottom-strange $J^P=1^+$}\\
    \hline
		C00078 & $238.5 \pm 6.2$ & $58.4 \pm 7.7$ & $14.8 \pm 5.4$ \\
		F1M & $215 \pm 11$ & $171 \pm 13$ & $6.1 \pm 5.8$ \\
		F004 & $229 \pm 13$ & $111.3 \pm 9.2$ & $11 \pm 15$ \\
		C005LV & $248 \pm 12$ & $80 \pm 10$ & $64 \pm 34$ \\
		C005 & $240.3 \pm 5.6$ & $97 \pm 11$ & $15 \pm 30$ \\
		F006 & $226.7 \pm 9.0$ & $103.5 \pm 6.6$ & $8 \pm 25$ \\
		C01 & $249.7 \pm 5.5$ & $92.8 \pm 7.6$ & $27 \pm 11$ \\
         \hline
    \end{tabular}

    \caption{Finite-volume decay constants in MeV for each of the states, calculated on each ensemble as shown in Fig.~\ref{fig:Dsdecay}-\ref{fig:Bsdecay}.}
    \label{tab:decay}
\end{table}

\subsection{Application of Lüscher's Method to the Energy Levels}\label{sec:luscher}
\FloatBarrier

\begin{table}
\begin{tabular}{|l|cccc|}
\hline
        &  $DK$   &    $D^*K$ &      $BK$ &   $B^*K$ \\
\hline
Phys.  &  $-24$  &  $-24$  & $-21$  & $-21$ \\
C00078 &  $-25$  &  $-25$  & $-22$  & $-22$  \\
F1M    &  $-69$  &  $-68$  & $-60$  & $-60$ \\
F004   &  $-119$  &  $-117$  & $-103$  & $-103$  \\
C005s  &  $-149$  &  $-147$  & $-129$  & $-129$  \\
F006   &  $-169$  &  $-166$  & $-146$  & $-146$  \\
C01    &  $-242$  &  $-238$  & $-210$  & $-210$  \\
\hline
\end{tabular}
\caption{\label{tab:LHC}Center-of-mass energies of the onsets of the two-pion-exchange left-hand cuts relative to the thresholds, in MeV.}
\end{table}

In infinite volume, the positive-parity mesons of interest appear as subthreshold poles in the $S$-wave $H^{(*)}K$ scattering amplitudes on the physical Riemann sheet.
The center-of-momentum energy $\sqrt{s}$ and scattering momentum $p$ are related as
\begin{gather}
    \sqrt{s}=\sqrt{m_K^2+p^2}+\sqrt{m_{H^{(*)}}^2+p^2}, \label{eq:sqrts}\\
    {p}^2 = \frac{1}{4}\left[s-2(m_K^2+m_{H^{(*)}}^2)+\frac{(m_{H^{(*)}}^2-m_K^2)^2}{s}\right], \label{eq:mom}
\end{gather}
and the elastic $S$-wave scattering amplitude is related to the scattering phase shift $\delta(p)$ through
\begin{equation}
 \mathcal{M}_0=\frac{-8\pi\sqrt{s}}{p\cot\delta(p)-ip}. \label{eq:M0}
\end{equation}
Bound-state poles occur at $p=p_B$ with $p_B\cot\delta(p_B)-ip_B=0$ at purely imaginary $p_B=i|p_B|$.

We apply Lüscher's Method \cite{Luscher:1990ux,Briceno} to extract these scattering phase shifts from the finite-volume lattice spectra, assuming that mixing with higher partial waves is negligible. We then fit these scattering phase shifts using effective-range expansions (ERE) and find the bound-state poles.
These steps are exact only if there are no left-hand cuts in the energy region considered. The closest left-hand cut affecting the $H^{(*)}K$ scattering amplitudes below threshold is expected to be due to two-pion exchange. The energies at which these cuts end are listed in Table \ref{tab:LHC}. For the lighter pion masses, these cuts extend above the ground-state finite-volume energies, which means that in these cases both the L\"uscher quantization condition and the effective-range expansion (ERE) are not strictly valid. Modifications to the L\"uscher quantization condition and ERE (some of which are limited to single-particle exchange and hence not applicable to our case) have recently been proposed \cite{Raposo:2023oru,Dawid:2023jrj,Meng:2023bmz,Hansen:2024ffk,Bubna:2024izx,Du:2024gzw,Dawid:2024oey,Yu:2025gzg,Raposo:2025dkb,Bubna:2025gsd,Pang:2026ktm}. Here, we neglect the left-hand cuts and proceed with the standard methods, leaving the implementation of more advanced formalisms to future work.

\begin{table}
    \centering
    \begin{tabular}{|l|c|c|c|c|}
    \hline
    Ensemble & $\frac{1}{a_0}$ for $D^*_{s0}$ (MeV) & $\frac{1}{a_0}$ for $D_{s1}$ (MeV) & $\frac{1}{a_0}$ for $B^*_{s0}$ (MeV) & $\frac{1}{a_0}$ for $B_{s1}$ (MeV) \\
    \hline
C00078 & $ -152 \pm 25 $ & $ -159 \pm 18 $ & $ -272 \pm 32 $ & $ -251 \pm 25 $\\
F1M & $ -215 \pm 19 $ & $ -234 \pm 11 $ & $ -289 \pm 10 $ & $ -299.3 \pm 9.1 $\\
F004 & $ -237 \pm 13 $ & $ -261 \pm 12 $ & $ -318.4 \pm 9.9 $ & $ -303 \pm 18 $\\
C005s& $ -236.8 \pm 8.1 $  & $ -247 \pm 13 $ & $ -340 \pm 14 $ & $ -337 \pm 12 $\\
F006 & $ -236 \pm 22 $ & $ -266 \pm 11 $ & $ -334 \pm 19 $ & $ -334 \pm 10 $\\
C01 & $ -252 \pm 35 $ & $ -239 \pm 32 $ & $ -355 \pm 17 $ & $ -365 \pm 14 $\\
\hline
    \end{tabular}
   \caption{Inverse scattering length $\frac{1}{a_0}$ extracted from zeroth-order fits to Eq.~(\ref{eq:LOERE}). ``C005s'' denotes the combination of C005 and C005LV.}
    \label{tab:ERE}
\end{table}

\begin{table}
\centering

\begin{tabular}{|c|c|c|}
        \hline
         Ensemble & $E$ (MeV) & $\Delta E$ (MeV)\\
         \hline
          \multicolumn{3}{|c|}{$D^*_{s0}$} \\
         \hline
         \hline
C00078 & $2401.0 \pm 9.2$ & $-29.5 \pm 8.8$ \\
F1M & $2445 \pm 10$ & $-55.4 \pm 9.9$ \\
F004 & $2495.2 \pm 7.6$ & $-64.9 \pm 6.8$ \\
C005s & $2504.2 \pm 4.5$ & $-63.9 \pm 4.2$ \\
F006 & $2515 \pm 12$ & $-63 \pm 11$ \\
C01 & $2561 \pm 17$ & $-70 \pm 17$ \\
         \hline

        \hline
        \multicolumn{3}{|c|}{$D_{s1}$} \\
        \hline
        \hline
C00078 & $2542.3 \pm 9.4$ & $-31.5 \pm 6.7$ \\
F1M & $2604.5 \pm 6.5$ & $-64.1 \pm 5.9$ \\
F004 & $2654.7 \pm 7.6$ & $-77.2 \pm 6.6$ \\
C005s & $2659.8 \pm 7.3$ & $-68.0 \pm 7.0$ \\
F006 & $2675.1 \pm 7.1$ & $-78.3 \pm 6.2$ \\
C01 & $2705 \pm 17$ & $-62 \pm 16$ \\
         \hline
\end{tabular}
\qquad\qquad
\begin{tabular}{|c||c|c|}
        \hline
         Ensemble & $E$ (MeV) & $\Delta E$ (MeV)\\
         \hline
          \multicolumn{3}{|c|}{$B^*_{s0}$} \\
         \hline
         \hline
C00078 & $5874 \pm 24$ & $-77 \pm 16$ \\
F1M & $5892.2 \pm 6.9$ & $-83.3 \pm 5.2$ \\
F004 & $5935.4 \pm 7.0$ & $-97.1 \pm 5.6$ \\
C005s & $5958.9 \pm 8.9$ & $-108.2 \pm 8.2$ \\
F006 & $5962 \pm 13$ & $-104 \pm 11$ \\
C01 & $6008 \pm 11$ & $-112 \pm 10$ \\
         \hline

        \hline
        \multicolumn{3}{|c|}{$B_{s1}$} \\
        \hline
        \hline
C00078 & $5897 \pm 20$ & $-66 \pm 12$ \\
F1M & $5945.2 \pm 7.0$ & $-88.9 \pm 5.0$ \\
F004 & $5973 \pm 11$ & $-88 \pm 10$ \\
C005s & $6005.7 \pm 7.9$ & $-106.1 \pm 6.7$ \\
F006 & $6007.4 \pm 7.4$ & $-103.3 \pm 6.0$ \\
C01 & $6062.8 \pm 9.7$ & $-117.7 \pm 8.6$ \\
         \hline
\end{tabular}
   \caption{Infinite-volume masses and binding energies from the zeroth order ERE. ``C005s'' denotes the average of C005 and C005LV.}
    \label{tab:InfVmass}
\end{table}

Because our lattice calculations are done at zero spatial momentum, we have
\begin{equation}
\sqrt{s_n} = E_n.
\end{equation}
For each $E_n$, we compute $p_n$ using Eq.~(\ref{eq:mom}) and obtain the phase shift as
\begin{equation}
     p_n \cot \delta(p_n) = \frac{2}{\sqrt{\pi} L} \mathcal{Z}_{00} \left(1; \left(\frac{L}{2\pi}p_n\right)^2 \right), \label{eq:luscher}
\end{equation}
where $\mathcal{Z}_{00}$ is the generalized zeta function \cite{Luscher:1990ux}. A general integral expression for $\mathcal{Z}_{lm}$ is given in Ref.~\cite{Leskovec:2012gb}. For $S$-wave scattering it reduces to
\begin{gather}
    \mathcal{Z}_{00}(1 ; q^2)=  \frac{\pi}{2} \sum_{\substack{\vec{n}\in \mathbb{Z}^3 \\ \vec{n}\ne 0}}  \int_0^1 dt  \ e^{tq^2}\frac{\exp(-\pi^2 \vec{n}^2)}{t^{3/2}}+ 
    \frac{1}{\sqrt{4\pi}}\int_0^1 dt \  (e^{tq^2}-1)\left(\frac{\pi}{t}\right)^{3/2} - \pi+ \frac{1}{2\pi} \sum_{\vec{n}\in \mathbb{Z}^3} \frac{e^{-(\vec{n}^2-q^2)}}{\vec{n}^2-q^2}.\label{eq:zeta}
\end{gather}
To evaluate Eq.~(\ref{eq:zeta}) we truncate the sums over $\vec{n}=(i,j,k)$ where $-50<i,j,k<50$. 

We then consider two parametrizations of the momentum dependence of $p\cot\delta(p)$: (i), the zeroth-order ERE
\begin{equation}
    p \cot \delta(p) = \frac{1}{a_0}, \label{eq:LOERE}
\end{equation}
which we fit to only the $n=0$ ground-state data point, and (ii), the first-order ERE
\begin{equation}
    p \cot \delta(p) = \frac{1}{a_0} + \frac{r}{2}p^2,\label{eq:ERE}
\end{equation}
which we fit to the $n=0$ and $n=1$ data points. Using these fits, we then determine the infinite-volume binding momenta $p_B$ and the corresponding bound-state masses and binding energies through Eq.~(\ref{eq:sqrts}). The uncertainties are propagated using bootstrap.

\begin{table}
\centering
\begin{tabular}{|c|c|c|}
        \hline
         Ensemble & $1/a_0$ (MeV) & $r$ (MeV$ ^{-1}$) \\
         \hline
         \hline
          \multicolumn{3}{|c|}{$D^*_{s0}$} \\
         \hline
C00078  & $ -100 \pm 47 $ &  $ 0.0046 \pm 0.0031 $  \\
F1M  & $ -245 \pm 115 $ &  $ -0.0011 \pm 0.0037 $  \\
F004  & $ -403 \pm 256 $ &  $ -0.0048 \pm 0.0069 $  \\
C005s  & $ -243 \pm 41 $ &  $ 0.0000 \pm 0.0010 $ \\
F006  & $ -224 \pm 47 $ &  $ 0.0004 \pm 0.0011 $  \\
C01  & $ -274 \pm 77 $ &  $ -0.0007 \pm 0.0019 $ \\
             \hline
        \hline
        \multicolumn{3}{|c|}{$D_{s1}$} \\
        \hline
C00078  & $ -132 \pm 105 $ &  $ 0.0025 \pm 0.0066 $  \\
F1M  & $ -339 \pm 208 $ &  $ -0.0035 \pm 0.0066 $  \\
F004  & $ -385 \pm 175 $ &  $ -0.0031 \pm 0.0040 $ \\
C005s & $ -220 \pm 53 $ &  $ 0.0010 \pm 0.0012 $ \\
F006  & $ -463 \pm 490 $ &  $ -0.005 \pm 0.012 $ \\
C01  & $ -351 \pm 346 $ &  $ -0.0032 \pm 0.0092 $  \\
         \hline
\end{tabular}
\qquad\qquad
\begin{tabular}{|c|c|c|}
        \hline
         Ensemble & $1/a_0$ (MeV) & $r$ (MeV$ ^{-1}$) \\
         \hline
         \hline
          \multicolumn{3}{|c|}{$B^*_{s0}$} \\
         \hline
C00078  & $ -110 \pm 40 $ &  $ 0.0044 \pm 0.0013 $  \\
F1M  & $ -459 \pm 544 $ &  $ -0.004 \pm 0.013 $  \\
F004  & $ -540 \pm 541 $ &  $ -0.004 \pm 0.010 $ \\
C005s  & $ -348 \pm 206 $ &  $ -0.0001 \pm 0.0034 $ \\
F006  & $ -540 \pm 694 $ &  $ -0.004 \pm 0.011 $ \\
C01  & $ -828 \pm 1045 $ &  $ -0.007 \pm 0.016 $\\

        \hline
        \hline

        \multicolumn{3}{|c|}{$B_{s1}$} \\
        \hline
C00078  & $ -108 \pm 51 $ &  $ 0.0045 \pm 0.0019 $  \\
F1M  & $ -602 \pm 466 $ &  $ -0.007 \pm 0.011 $ \\
F004  & $ -436 \pm 515 $ &  $ -0.003 \pm 0.010 $ \\
C005s  & $ -320 \pm 216 $ &  $ 0.0003 \pm 0.0037 $ \\
F006  & $ -818 \pm 1039 $ &  $ -0.009 \pm 0.018 $ \\
C01  & $ -249 \pm 91 $ &  $ 0.0017 \pm 0.0013 $ \\
\hline
\end{tabular}
   \caption{Effective-range parameters extracted from the first-order fits using Eq.~(\ref{eq:ERE}). ``C005s'' denotes the combination of C005 and C005LV. }
    \label{tab:firstERE}
\end{table}

\begin{table}
\centering
\begin{tabular}{|c|c|}
        \hline
         Ensemble &  $\Delta E$ (MeV)\\
         \hline
         \hline
          \multicolumn{2}{|c|}{$D^*_{s0}$} \\
         \hline
C00078 & $-24 \pm 11$ \\
F1M & $-54.4 \pm 9.6$ \\
F004 & $-68.3 \pm 8.4$ \\
C005s & $-66.1 \pm 5.7$ \\
F006 & $-61 \pm 11$ \\
C01 & $-70 \pm 16$ \\
        \hline
        \hline
        \multicolumn{2}{|c|}{$D_{s1}$} \\
        \hline
C00078 & $-27.6 \pm 8.1$ \\
F1M & $-64.3 \pm 6.3$ \\
F004 & $-79.9 \pm 7.7$ \\
C005s & $-68 \pm 10$ \\
F006 & $-79.9 \pm 8.1$ \\
C01 & $-61 \pm 16$ \\
         \hline
\end{tabular}
\qquad\qquad
\begin{tabular}{|c|c|}
        \hline
         Ensemble & $\Delta E$ (MeV)\\
         \hline
         \hline
          \multicolumn{2}{|c|}{$B^*_{s0}$} \\
         \hline
C00078 & $-36 \pm 18$ \\
F1M & $-81.8 \pm 5.8$ \\
F004 & $-96.8 \pm 6.2$ \\
C005s & $-108 \pm 10$ \\
F006 & $-102 \pm 11$ \\
C01 & $-111 \pm 10$ \\
        \hline
        \hline
        \multicolumn{2}{|c|}{$B_{s1}$} \\
        \hline
C00078 & $-35 \pm 19$ \\
F1M & $-85.6 \pm 6.0$ \\
F004 & $-85 \pm 11$ \\
C005s & $-102.7 \pm 8.0$ \\
F006 & $-102.7 \pm 6.3$ \\
C01 & $-108 \pm 15$ \\
\hline
\end{tabular}
   \caption{Infinite-volume binding energies from the first-order ERE. ``C005s'' denotes the average of C005 and C005LV. These values are used later to estimate systematic errors in our final reported $\Delta E$, whose central values are from the zeroth-order ERE.}
    \label{tab:InfVmassfirstorder}
\end{table}

\begin{figure}
\subfloat[The inverse scattering lengths extracted from the phase-shift data for the ground-state $D^*_{s0}$ on all ensembles.]{
\includegraphics[width=0.48\linewidth]{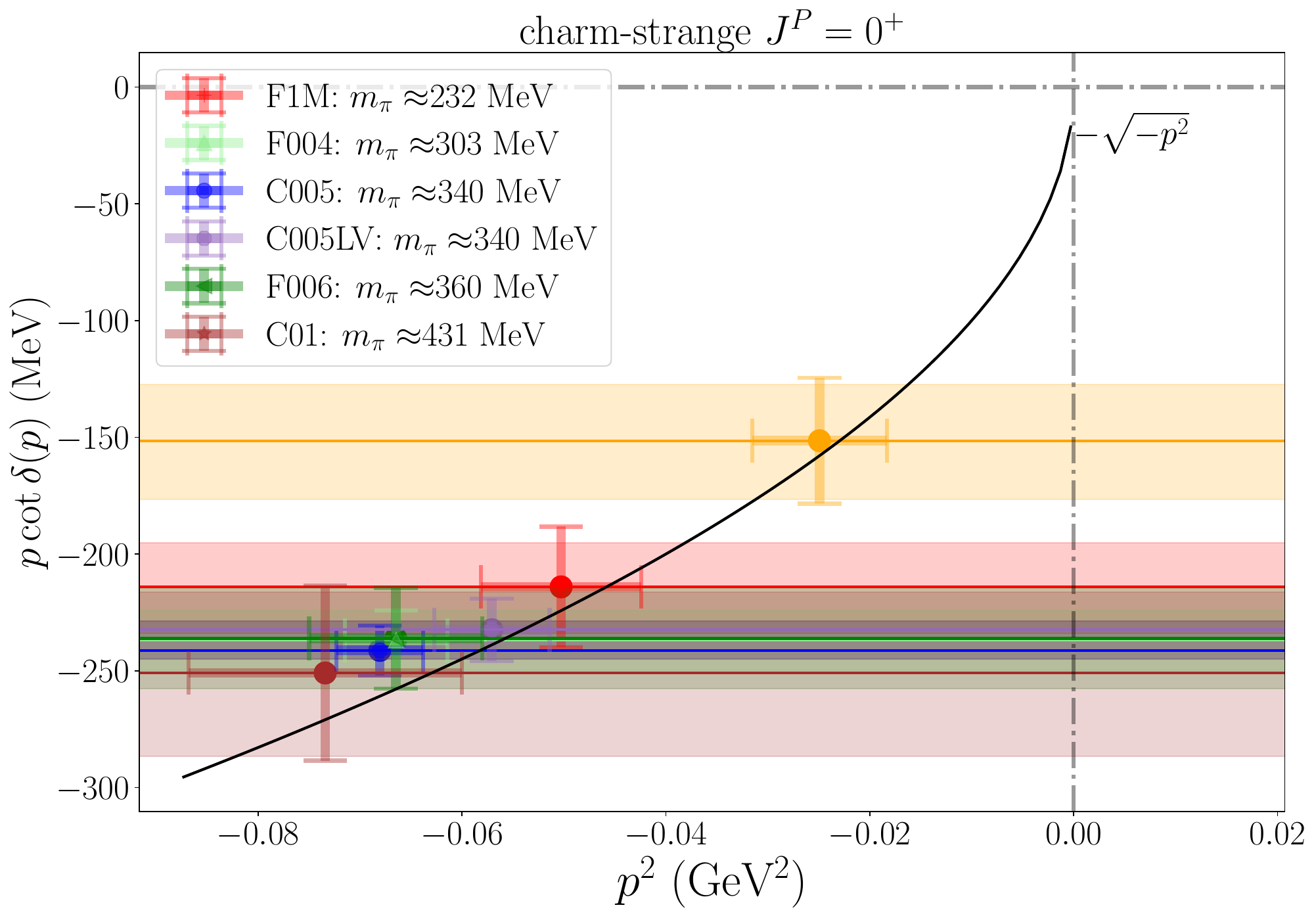}
}
\hfill
\subfloat[The inverse scattering lengths extracted from the phase-shift data for the ground-state $D_{s1}$ on all ensembles.]{\includegraphics[width=0.48\linewidth]{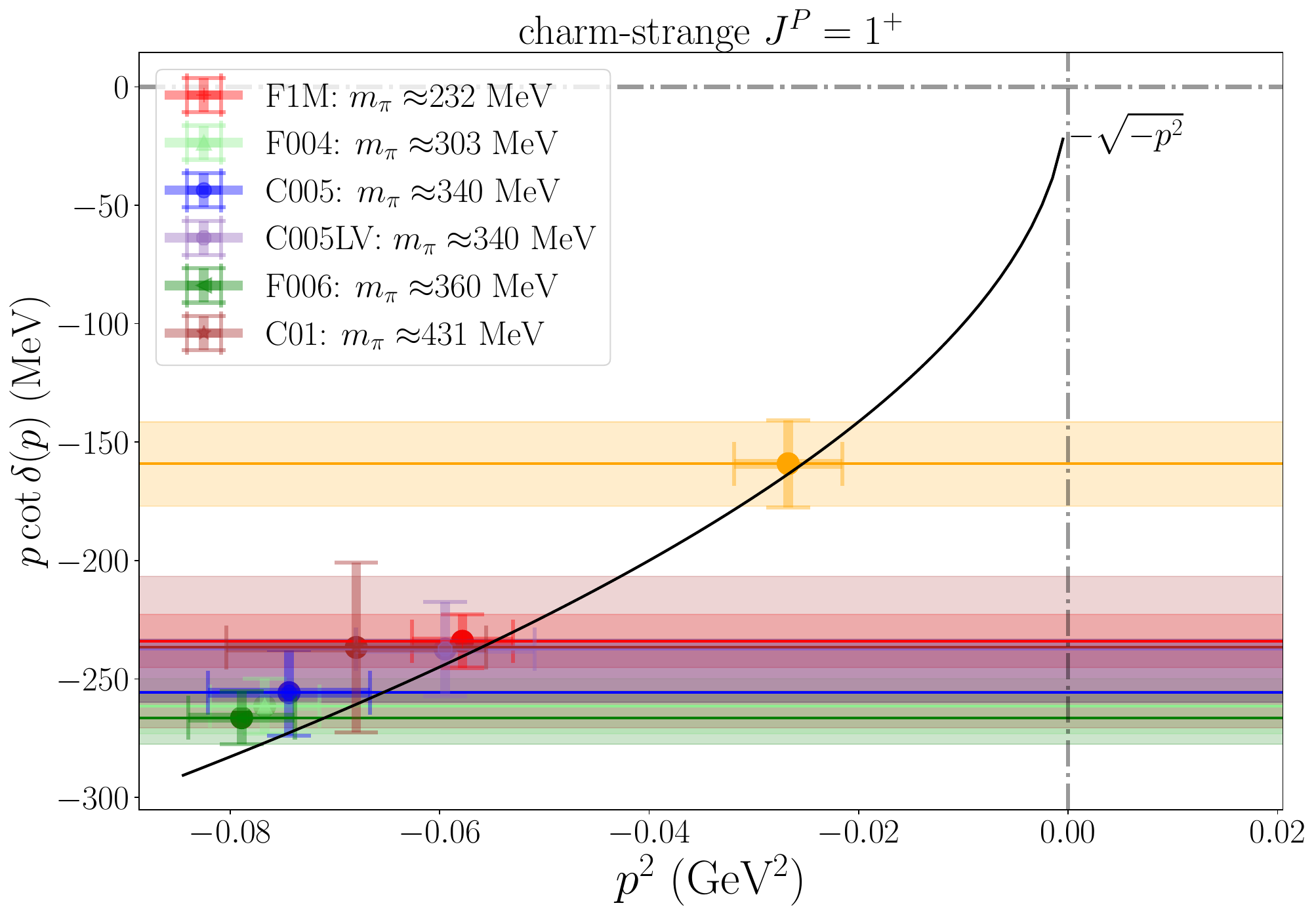}
}

\subfloat[The inverse scattering lengths extracted from the phase-shift data for the ground-state  $B^*_{s0}$ on all ensembles.]{
\includegraphics[width=0.48\linewidth]{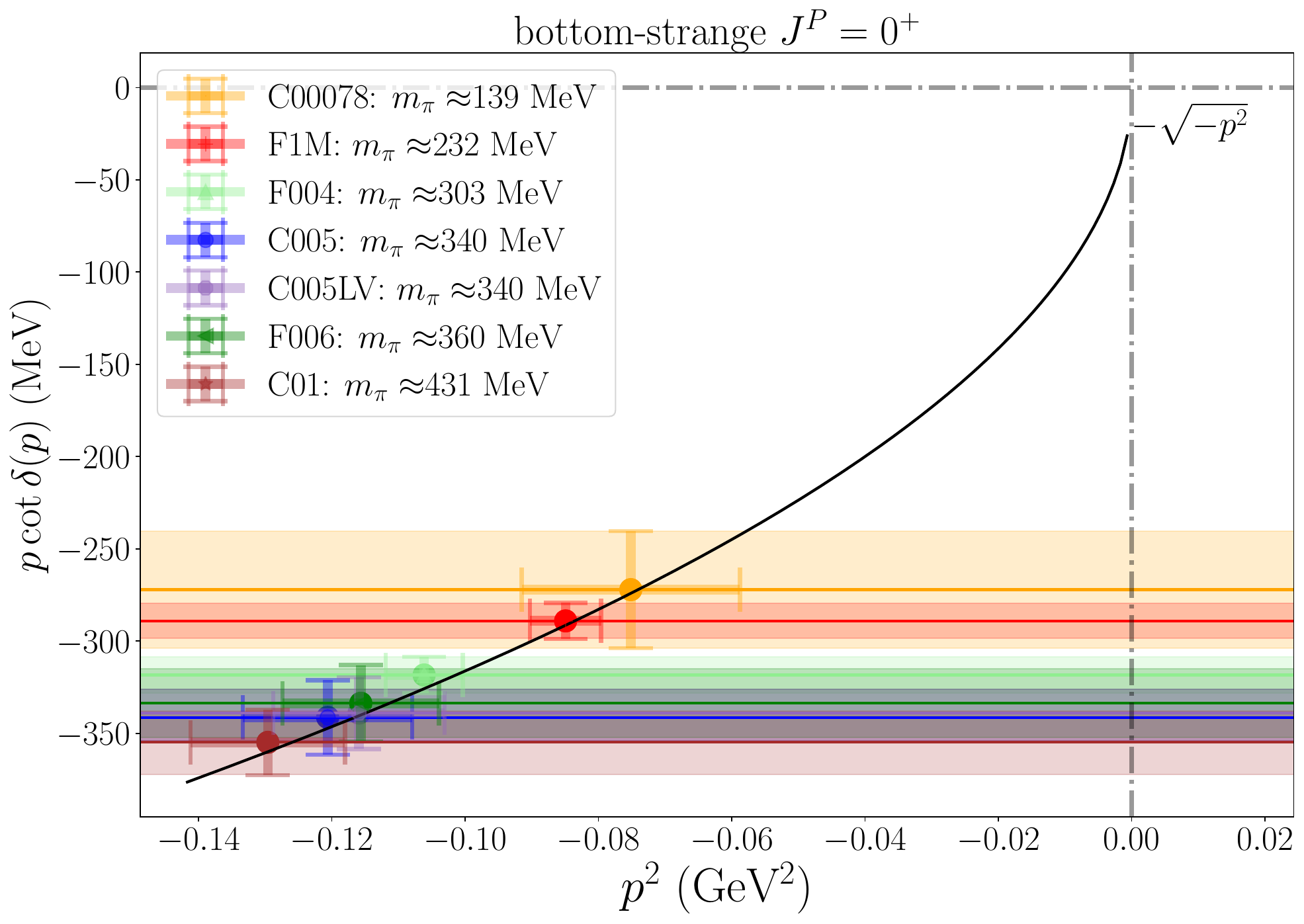}
}
\hfill
\subfloat[The inverse scattering lengths extracted from the phase-shift data for the ground-state $B_{s1}$ on all ensembles.]{
\includegraphics[width=0.48\linewidth]{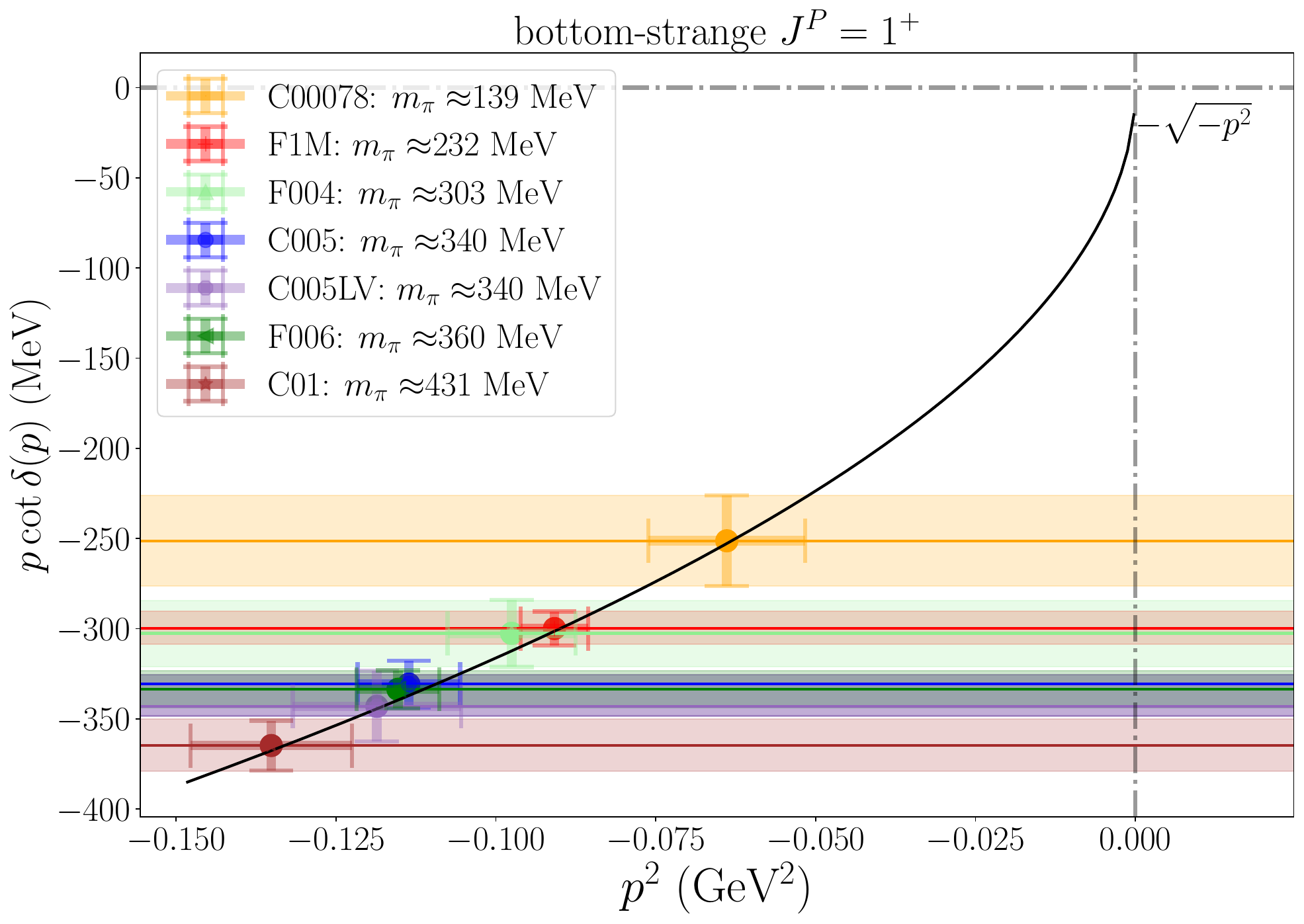}
}
\caption{Examples of the extraction of bound state pole mass from $p\cot\delta(p)$. The intersection with the black curve corresponds to the solution for the bound state pole in infinite volume (though still at finite lattice spacing and at the pion mass given in the legend). The purple band is for both C005 and C005LV combined.}
\label{fig:ERE}
\end{figure}

To obtain the central values of the binding energies, we use the zeroth-order ERE, because the $n=1$ energy level have large uncertainties. The results for $1/a_0$ and for the infinite-volume masses and binding energies from these zeroth-order fits are given in Tables \ref{tab:ERE} and \ref{tab:InfVmass}, and the fits are shown in  Fig.~\ref{fig:ERE}. Results for $\Delta E$ from the first-order ERE fits are subsequently used to estimate a systematic uncertainty, as explained in Sec.~\ref{sec:chiralcontinuum}. 
Besides providing this estimate of the systematic uncertainty in $\Delta E$, the first-order ERE fits are also useful in assessing Weinberg's compositeness criterion (Sec.~\ref{sec:molecular}). 

\begin{figure}
\subfloat[ The effective-range expansion fitted to the the phase-shift data for the lowest two charm-strange $J^P=0^+$ states on all ensembles. ]{
\includegraphics[width=0.48\linewidth]{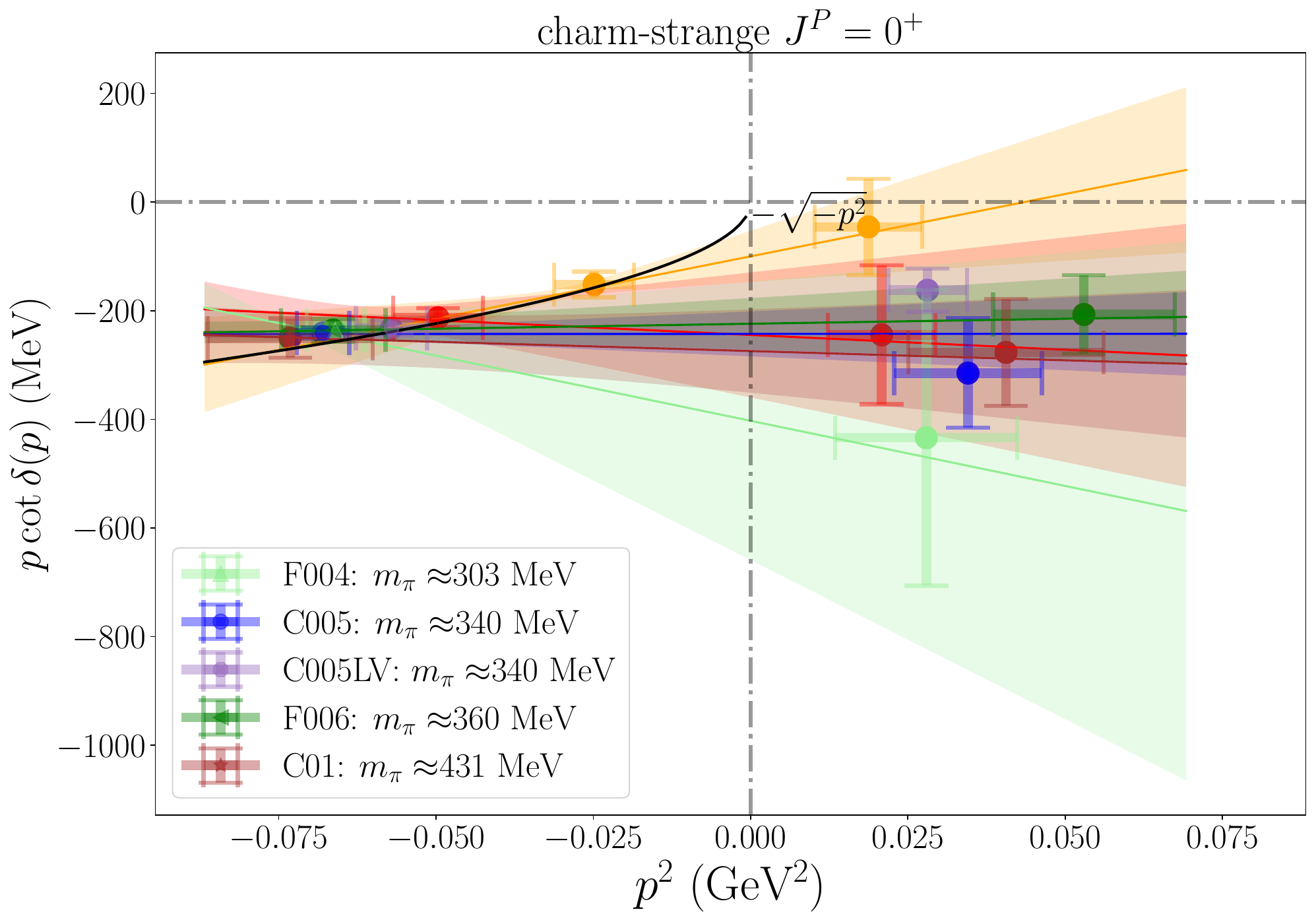}
}
\hfill
\subfloat[ The effective-range expansion fitted to the the phase-shift data for the lowest two charm-strange $J^P=1^+$ states on all ensembles. ]{
\includegraphics[width=0.48\linewidth]{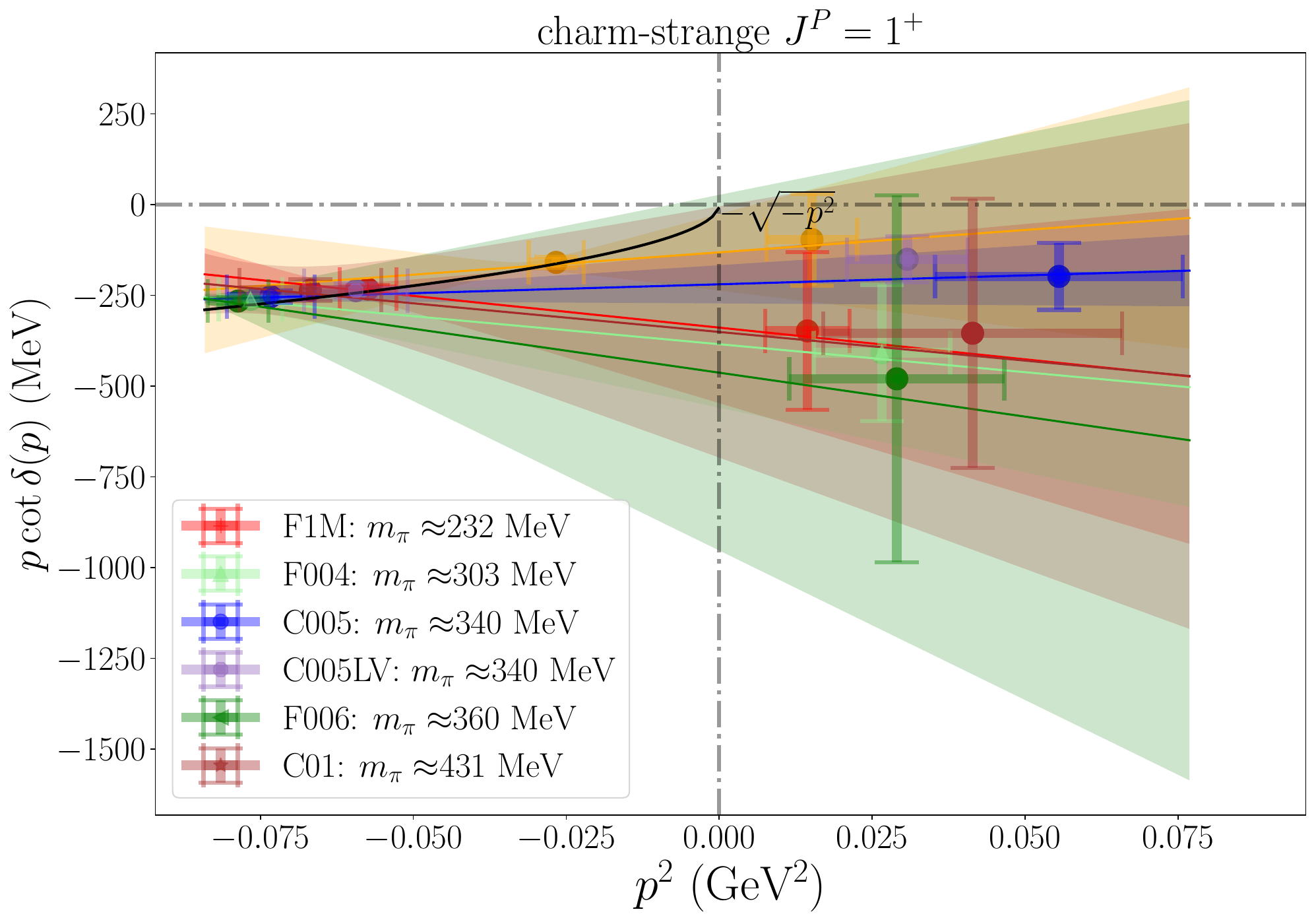}
}

\subfloat[ The effective-range expansion fitted to the the phase-shift data for the lowest two bottom-strange $J^P=0^+$ states on all ensembles. ]{
\includegraphics[width=0.48\linewidth]{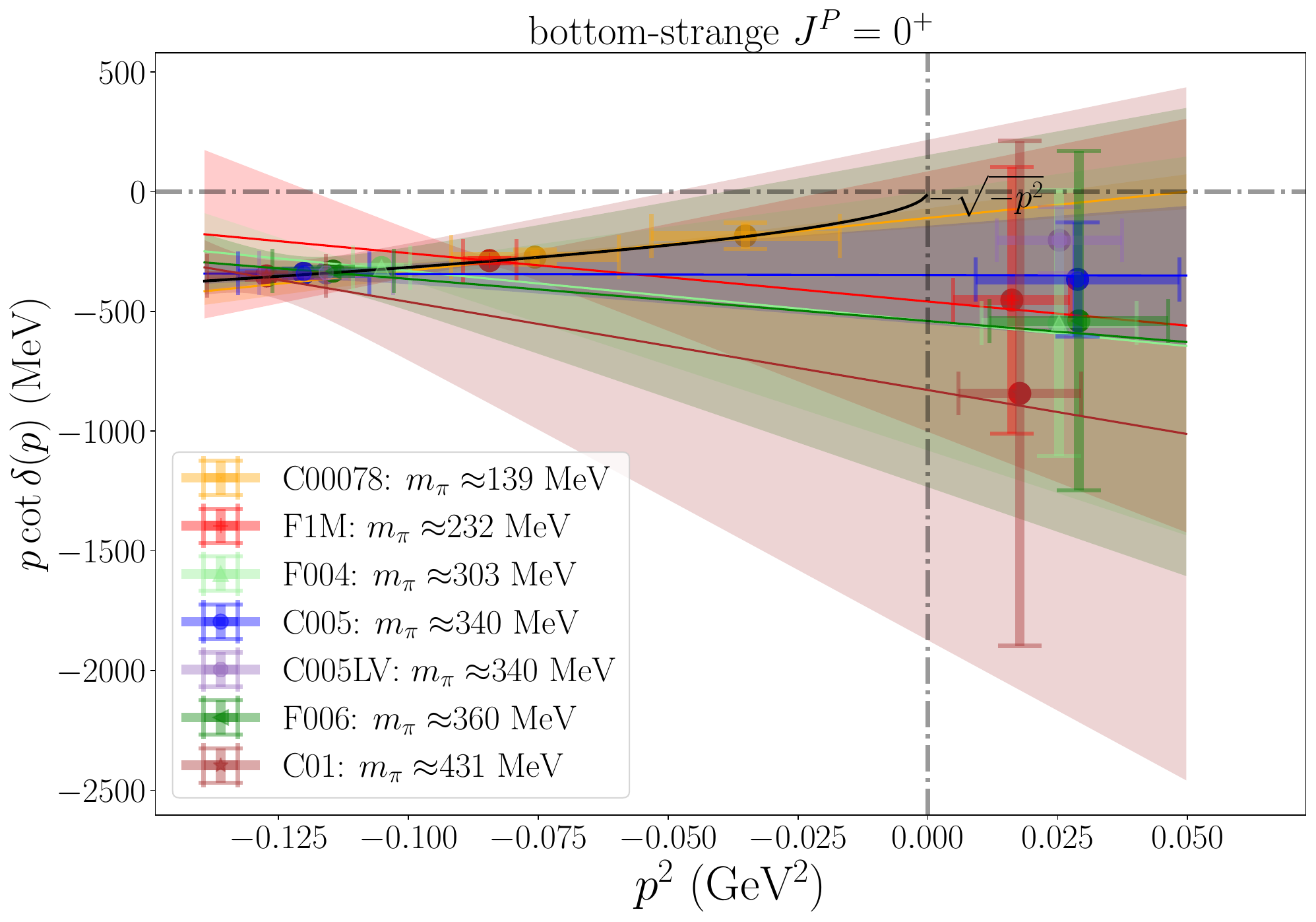}
}
\hfill
\subfloat[ The effective-range expansion fitted to the the phase-shift data for the lowest two bottom-strange $J^P=1^+$ states on all ensembles. ]{
\includegraphics[width=0.48\linewidth]{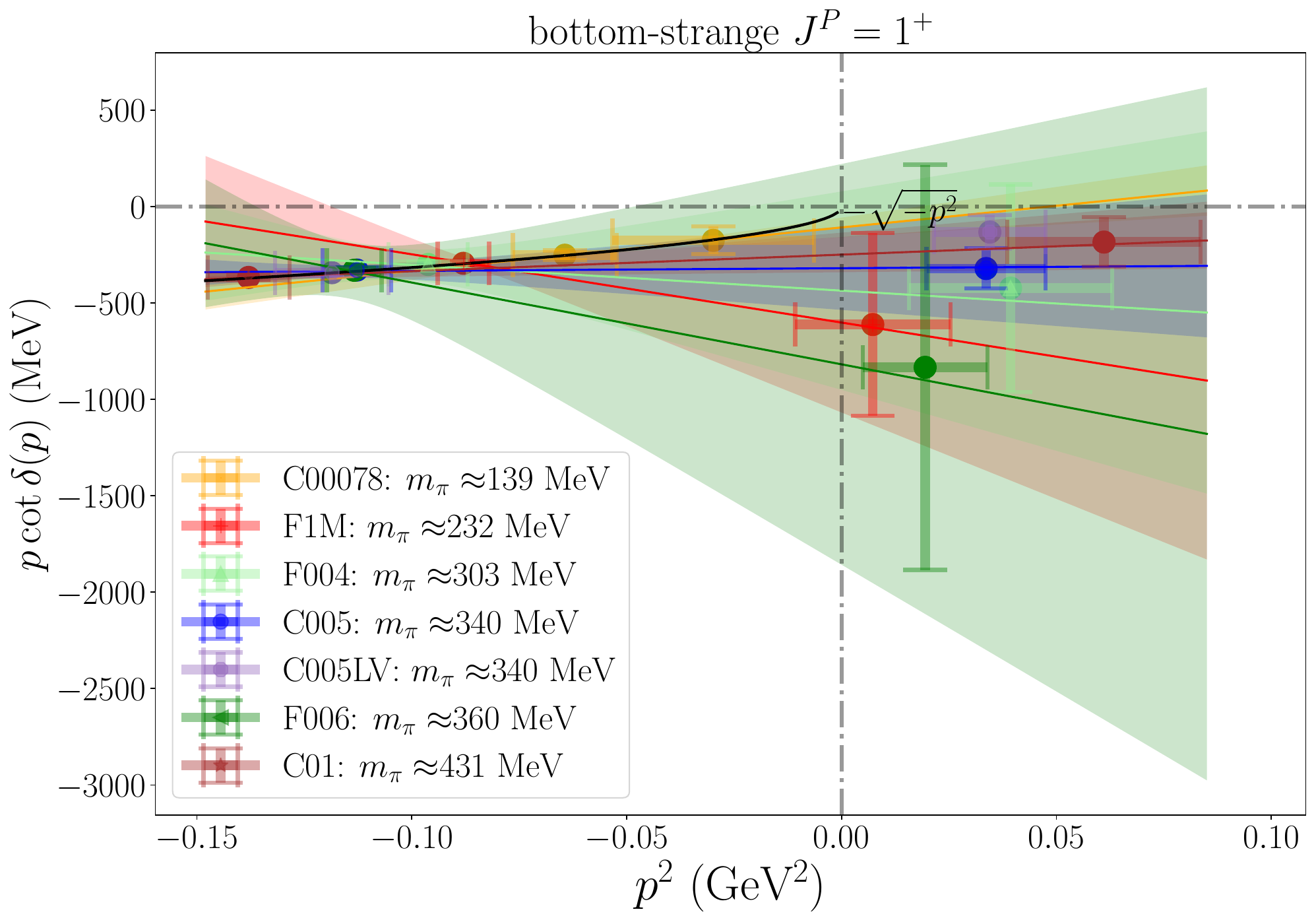}
}
\caption{Extraction of the bound state pole mass from $p\cot\delta(p)$ via a first-order fit to the effective-range expansion. The intersection with the black curve corresponds to the solution for the bound state pole in infinite volume (though still at finite lattice spacing and at the pion mass given in the legend). The purple band is for both C005 and C005LV combined.}
\label{fig:EREfirstorder}
\end{figure}

To obtain results at first order in $p^2$, it becomes necessary to address two subtleties which are not relevant in the zeroth-order fits. Firstly, there may be a two-fold ambiguity in the solution for $p_B$, corresponding to the choice of intersection point of the ERE fit to $p\cot\delta$ and $-\sqrt{-p^2}$  (shown in Fig.~\ref{fig:EREfirstorder}) in the negative $p^2$ region. Reference \cite{Iritani:2017rlk} describes the conditions necessary to ensure a positive residue in the scattering amplitude, allowing the identification of the physical pole. In our case, it so happens that the physical bound-state pole is always determined by the intersection nearest to the ground-state data point.

The second issue arises due to the large statistical fluctuations in the energy levels for the first excited state, which result in poor convergence of Eq.~(\ref{eq:zeta}) on some bootstrap samples. This manifests as a non-Gaussian tail in the distribution of the $\Delta E$ bootstrap samples.  Both the fact that the finite-volume excited-state energies are subject to aggressive statistical fluctuations (as shown above in Figs.~\ref{fig:Dsspec}-\ref{fig:Bsspec}), and the extreme sensitivity of the intersection point to $r$ with a slight upward slope of the ERE, make filtering out samples in the non-Gaussian tail necessary. This is accomplished manually by identifying the physical Gaussian distribution in the $\Delta E$ samples and imposing appropriate bounds on $\Delta E$, where the physical part of the distribution always contains the $\Delta E$ extracted from the zeroth-order ERE fit.

The fits to the first-order EREs are plotted in Fig.~\ref{fig:EREfirstorder}, with the corresponding effective-range parameters given in Table \ref{tab:firstERE}, and binding energies given in Table \ref{tab:InfVmassfirstorder}. The $\frac{1}{a_0}$ values from the first-order fits are consistent with the values from the zeroth-order fits (Table \ref{tab:ERE}).

It is less straightforward to determine finite-volume corrections for the decay constants. Lellouch and Lüscher constructed a factor relating finite-volume matrix elements of a weak Hamiltonian to infinite-volume matrix elements with two-particle final states in Ref.~\cite{Lellouch:2000pv}, with application to $K\rightarrow \pi\pi$ decays.
A general procedure for relating infinite-volume and finite-volume matrix elements of an arbitrary local current for arbitrary-spin $0\rightarrow 2$ transitions has been developed along these lines in Ref.~\cite{Briceno:2015csa}. In principle, this formalism may be used to account for finite-volume corrections in the decay constants presented here via the generalized Lellouch-Lüscher factor $\mathcal{R}$ [Eqs.~(8-9) in Ref.~\cite{Briceno:2015csa}]. In practice, however, this is difficult, as the calculation of $\mathcal{R}$ and the determination of the infinite-volume decay constant require a parametrization of the $0\rightarrow 2$ amplitudes, and their analytic continuation below the $H^{(*)} K$ threshold. Parameterizing the relevant form factors is beyond the scope of this work, and fitting the form factors from lattice data would require more precise results for the above-threshold excited states. In Sec.~\ref{sec:chiralcontinuum}, we use another, less rigorous method to estimate finite-volume systematics for the decay constants.

\subsection{Chiral-Continuum Extrapolations and Characterization of Additional Energy and Decay Constant Systematics}
\label{sec:chiralcontinuum}

In this section, we present our extrapolations of the positive-parity decay constants and infinite-volume binding energies to zero lattice spacing and to the physical pion mass. For the binding energies, heavy-meson chiral-perturbation-theory predictions are available in principle \cite{Mehen:2005hc}, but we expect chiral perturbation theory to be less reliable than for the negative-parity case due to the larger energy scales involved and the closeness of the two-meson thresholds. Therefore, we use simple analytic extrapolations for both the binding energies and the decay constants, which describe the data well.

We consider the following functional forms:
\begin{align}
    \text{linear in }m_\pi^2: \hspace{4ex}&\Delta E = \Delta E_{\text{phys}}+\frac{C}{4\pi f_\pi} (m_\pi^2 - m_{\pi \text{phys}}^2)+d\,\Lambda^3 a^2, \label{eq:chiralcontE}\\
    &\hspace{2.5ex}f=f_{\text{phys}}+\frac{C}{4\pi f_\pi} (m_\pi^2 - m_{\pi \text{phys}}^2)+d\,\Lambda^3 a^2, \label{eq:chiralcontf}\\
    \text{quadratic in }m_\pi^2: \hspace{4ex}&\Delta E=\Delta E_{\text{phys}}+\frac{C}{4\pi f_\pi} (m_\pi^2 - m_{\pi \text{phys}}^2)+ \frac{C_2}{(4\pi f_\pi)^3} (m_\pi^4 - m_{\pi \text{phys}}^4)+d\,\Lambda^3 a^2, \\
   & \hspace{2.5ex}f=f_{\text{phys}}+\frac{C}{4\pi f_\pi} (m_\pi^2 - m_{\pi \text{phys}}^2)+ \frac{C_2}{(4\pi f_\pi)^3} (m_\pi^4 - m_{\pi \text{phys}}^4)+d\,\Lambda^3 a^2.\label{eq:chiralcontf-nextorder}
\end{align}
Here, $\Lambda=500 \ \text{MeV}$ is a typical hadronic scale, $f_\pi=130.5$ MeV and $m_{\pi \text{phys}}=135.0$ MeV are the physical iso-QCD pion decay constant and pion mass \cite{FlavourLatticeAveragingGroupFLAG:2024oxs}, and 
 $C$, $C_2$, and $d$ are the dimensionless fit parameters.

In order to account for the systematics in model choice, we proceed with model averaging the two fits, weighted by Eq.~(\ref{eq:aic}), with the systematic from choice of model included in the variance of the model-averaged value $a=f, \Delta E$ by 
\begin{equation}
 \sigma_a^2=\sum_{i=1}^{N_m} \sigma_{a,i}^2 w_i+\sum_{i=1}^{N_m} \langle a\rangle_i^2 w_i-\left(\sum_{i=1}^{N_m} \langle a\rangle_i w_i \right)^2,
\end{equation}
with $N_m=2$ the number of models and $w_i$ the $i$th-model weight \cite{bays}.

\begin{figure}[h]
    \centering
    \includegraphics[width=0.9\linewidth]{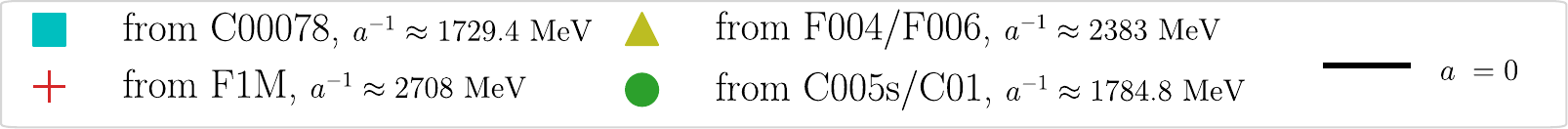}
    \includegraphics[width=0.49\linewidth]{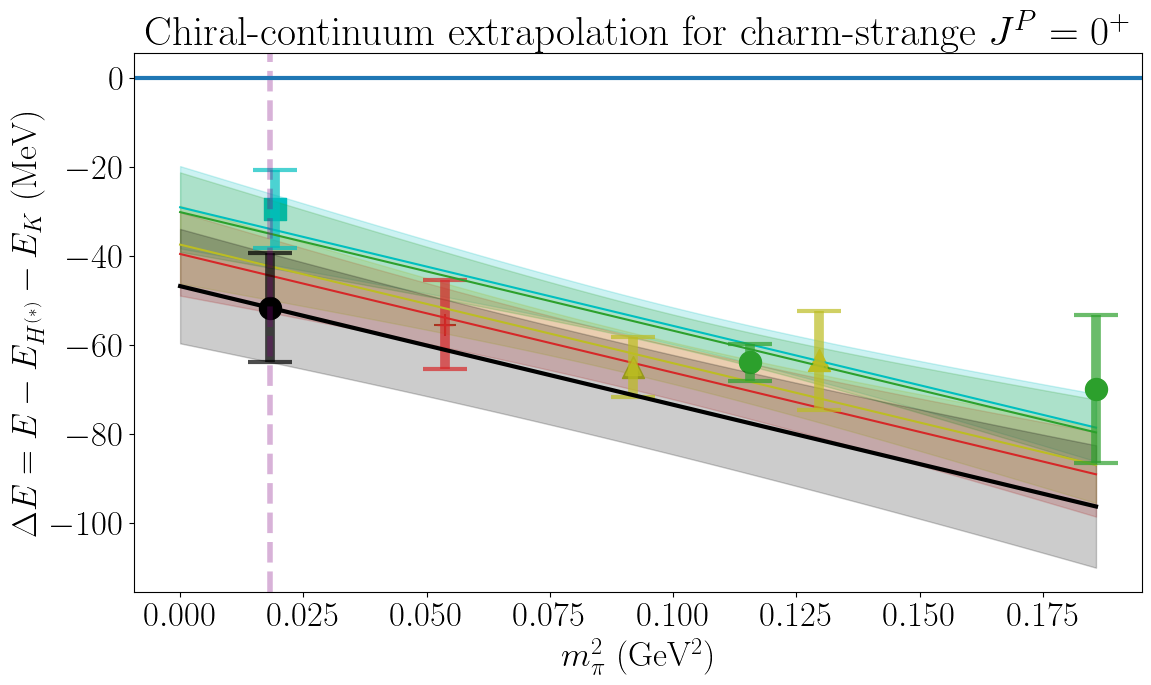}
    \hfill
    \includegraphics[width=0.49\linewidth]{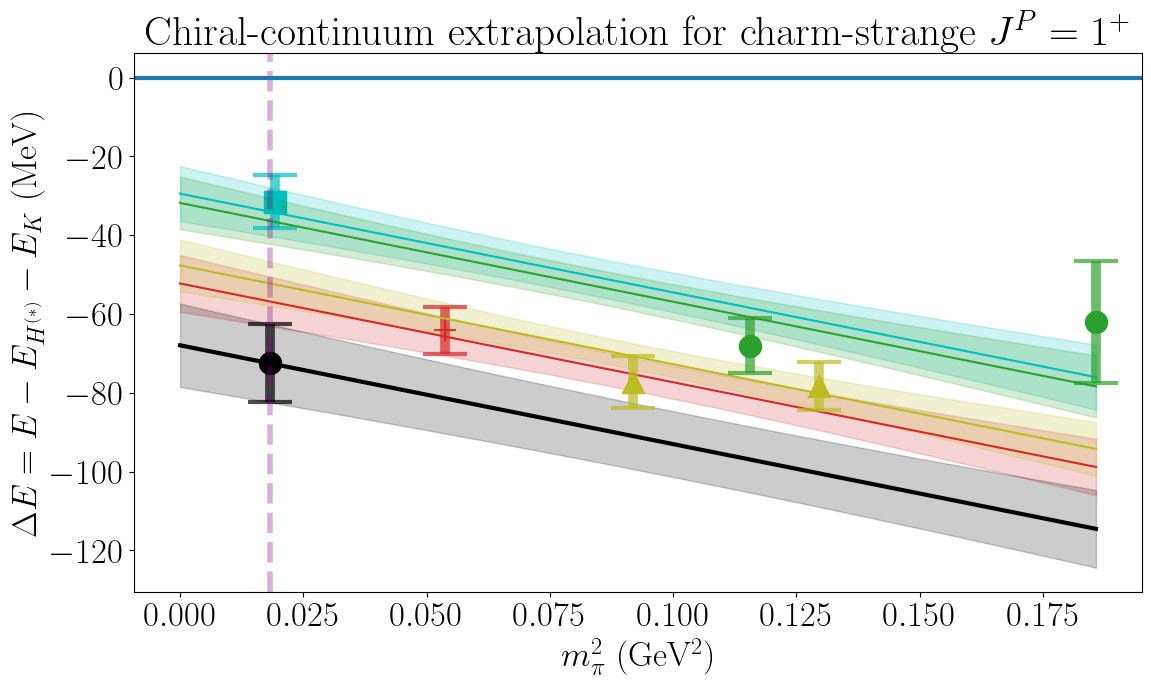}
    
    \includegraphics[width=0.49\linewidth]{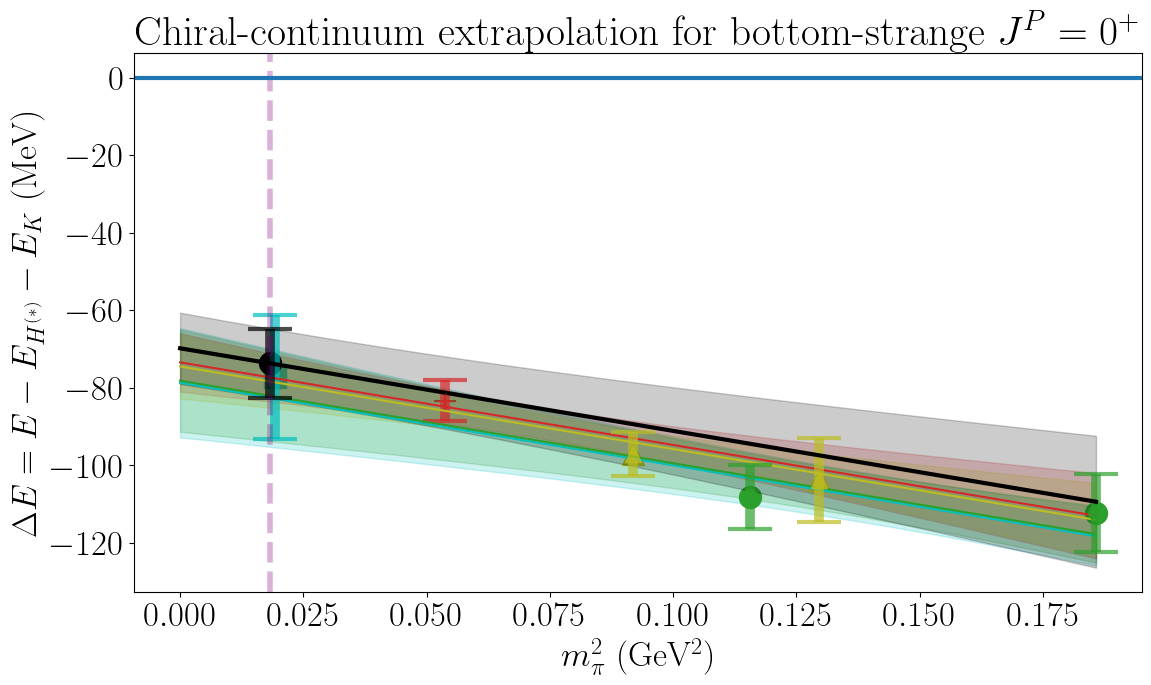}
    \hfill
    \includegraphics[width=0.49\linewidth]{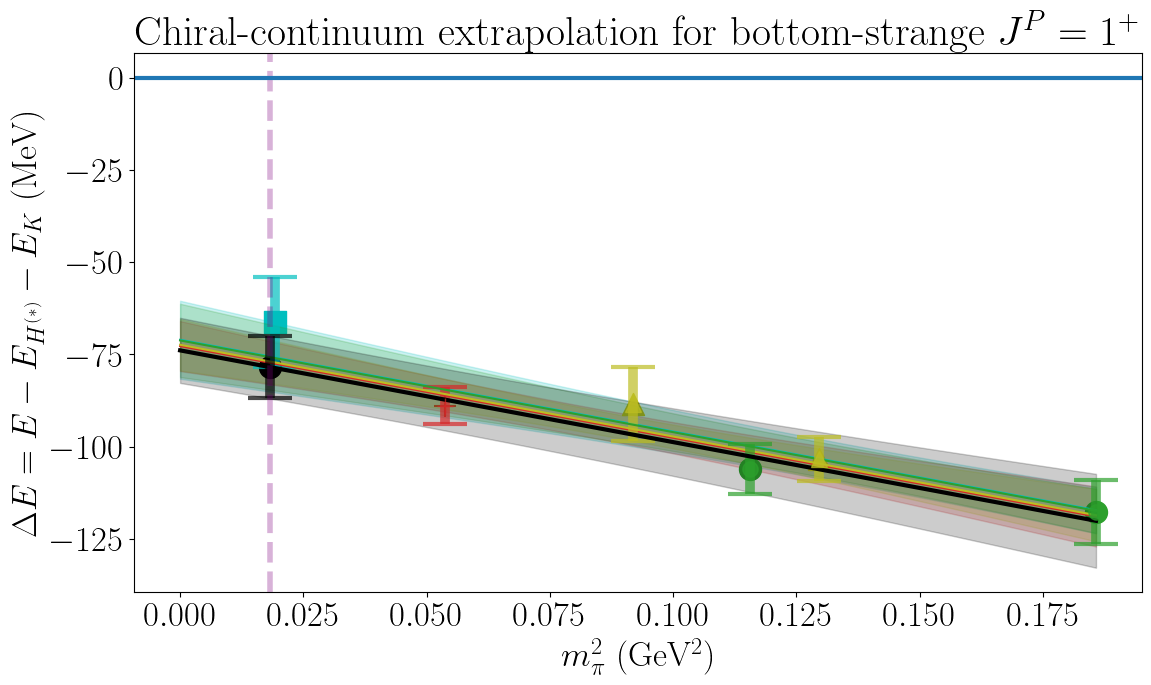}
    \caption{Linear-in-$m_\pi^2$ chiral-continuum extrapolations of the ground state binding energies from the zeroth-order ERE. The bands represent chiral extrapolations at fixed lattice spacing. The The physical pion mass is indicated by the dashed line. The black band is the chiral extrapolation at $a=0$ and the black data point is $\Delta E_\text{phys}$. }
    \label{fig:chiralcontE}
\end{figure}

\begin{figure}[h]
    \centering
    \includegraphics[width=0.9\linewidth]{positive_parity/chiralcontiumlegend.pdf}
    \includegraphics[width=0.49\linewidth]{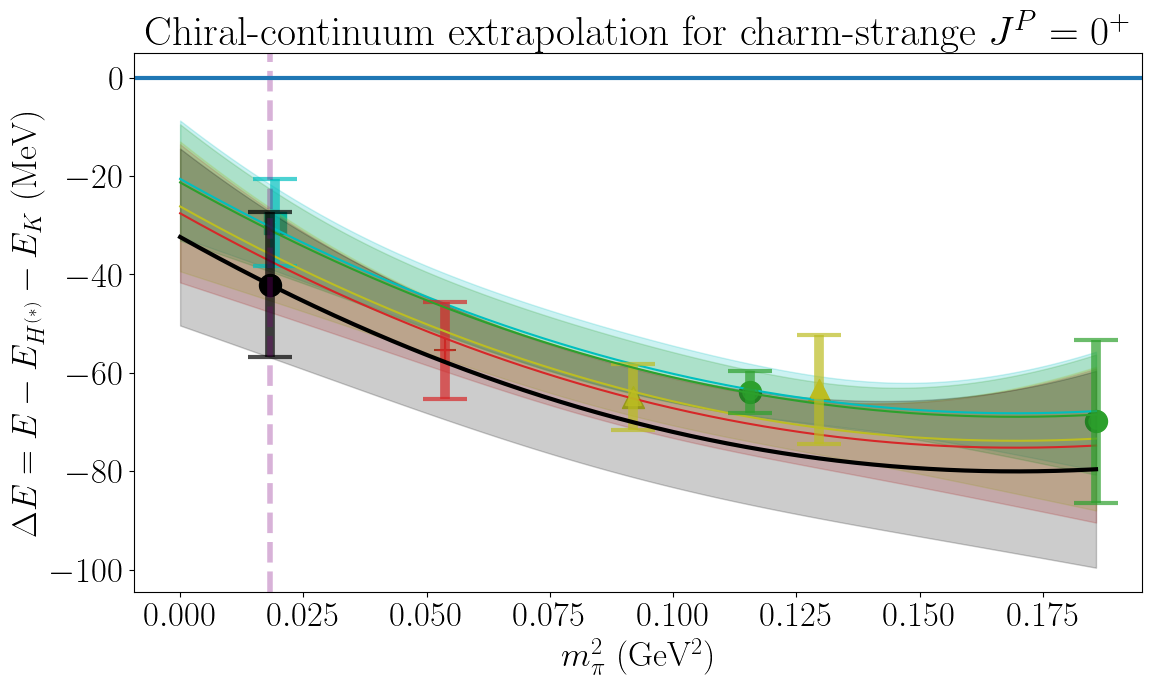}
    \hfill
    \includegraphics[width=0.49\linewidth]{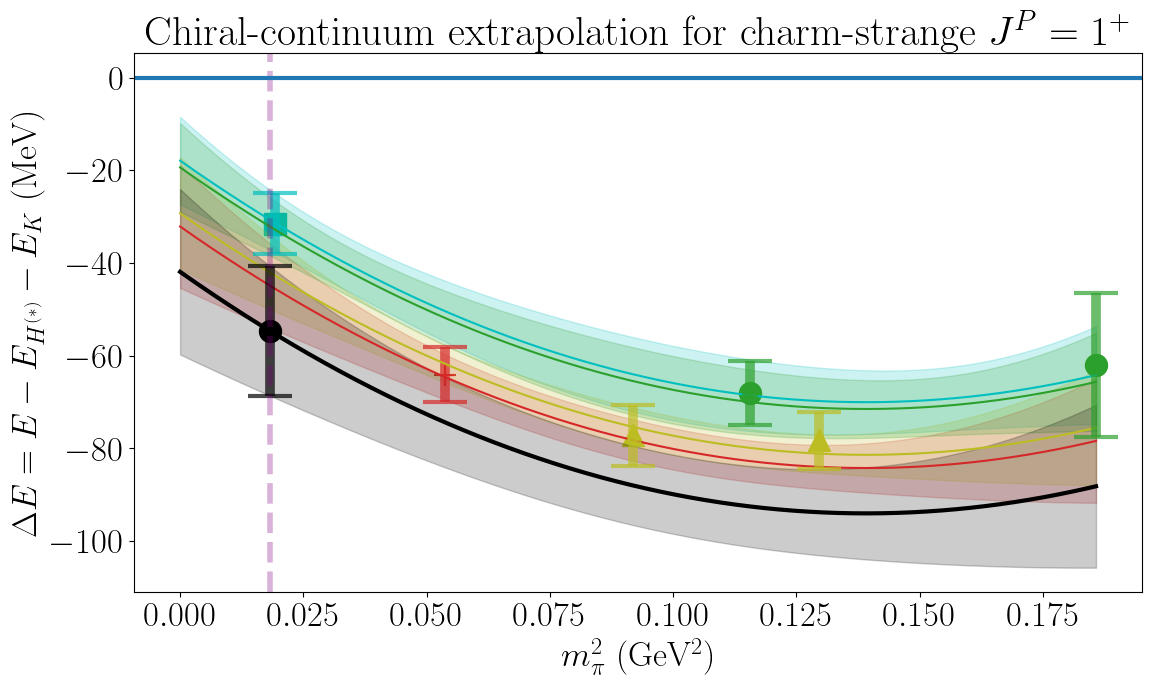}
    
    \includegraphics[width=0.49\linewidth]{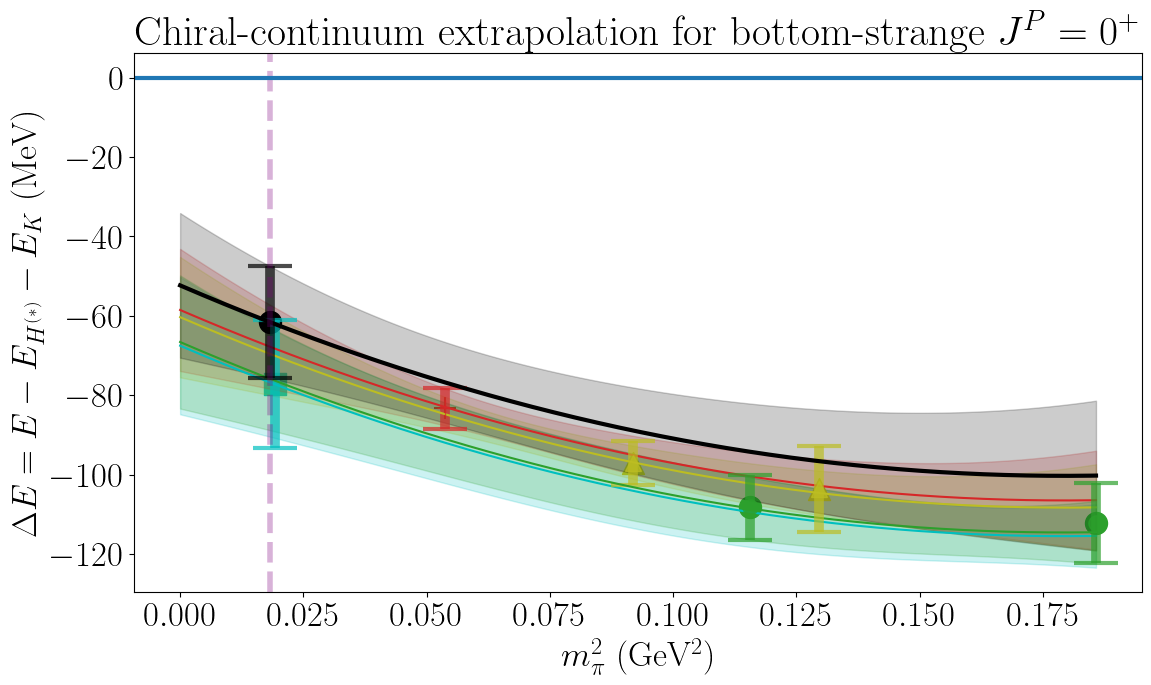}
    \hfill
    \includegraphics[width=0.49\linewidth]{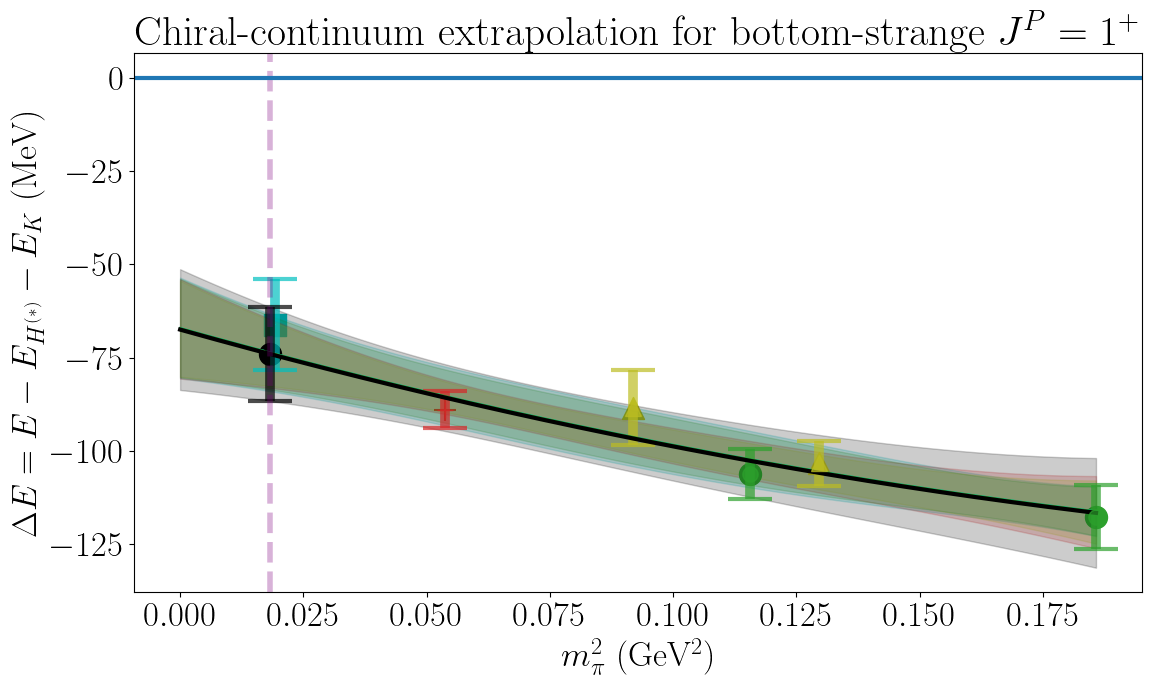}
    \caption{Quadratic-in-$m_\pi^2$ chiral-continuum extrapolation fits of the ground-state binding energies. The bands represent chiral extrapolations at fixed lattice spacing. The physical pion mass is indicated by the dashed line. The black band is the chiral extrapolation at $a=0$ and the black data point is $\Delta E_\text{phys}$.}
    \label{fig:chiralcontEnextorder}
\end{figure}

The extrapolations of infinite-volume binding energies extracted from the zeroth-order fit to the ERE are shown in Fig.~\ref{fig:chiralcontE}-\ref{fig:chiralcontEnextorder}, with corresponding fit parameters and model-averaged $ \Delta E_{\text{phys}}$ given in Table \ref{tab:chiralfitE}. Figures \ref{fig:chiralcontEnextorderERE}-\ref{fig:chiralcontEnextorderEREnextorder} in Appendix \ref{sec:chiralplots} show the chiral-continuum extrapolations of the infinite-volume binding energies extracted from the first-order ERE fits, with the fit parameters and model-averaged $ \Delta E_{\text{phys}}$ listed in Table \ref{tab:chiralfitEnextorder}.

\begin{table}[h]
    \centering
    \resizebox{\textwidth}{!}{
    \begin{tabular}{|l|c|c|c|c||c|c|c|c|c||c|}
    \hline
     & \multicolumn{4}{|c|}{Linear-in-$m_\pi^2$ fit (Fig. \ref{fig:chiralcontE})} & \multicolumn{5}{|c|}{Quadratic-in-$m_\pi^2$ fit (Fig. \ref{fig:chiralcontEnextorder})} & Model averaged\\
    \hline
    $H_{sJ}$ & $C$ & $d$ & $\Delta E_\text{phys}$(MeV) & weight & $C$ & $d$ & $C_2$ & $\Delta E_\text{phys}$(MeV) & weight & $\Delta E_\text{phys}$(MeV)\\ \hline
$D^*_{s0}$  & $ -0.45 \pm 0.13$ & $ 0.42 \pm 0.33$ & $ -52 \pm 12$ & 0.59 & $ -0.95 \pm 0.46$ & $ 0.28 \pm 0.35$ & $ 8.1 \pm 7.1$ & $ -42 \pm 15$ & 0.41 & $ -48 \pm 14 $ \\\hline
$D_{s1}$  & $ -0.43 \pm 0.11$ & $ 0.92 \pm 0.28$ & $ -72.5 \pm 9.9$ & 0.34 & $ -1.27 \pm 0.48$ & $ 0.57 \pm 0.34$ & $ 13.1 \pm 7.3$ & $ -55 \pm 14$ & 0.66 & $ -61 \pm 15 $ \\\hline
$B^*_{s0}$  & $ -0.36 \pm 0.15$ & $ -0.21 \pm 0.42$ & $ -73.7 \pm 9.0$ & 0.59 & $ -0.91 \pm 0.52$ & $ -0.36 \pm 0.44$ & $ 7.3 \pm 6.6$ & $ -62 \pm 14$ & 0.41 & $ -69 \pm 13 $ \\\hline
$B_{s1}$   & $ -0.42 \pm 0.11$ & $ 0.07 \pm 0.33$ & $ -78.4 \pm 8.5$ & 0.71 & $ -0.63 \pm 0.45$ & $ 0.01 \pm 0.35$ & $ 2.8 \pm 5.9$ & $ -74 \pm 13$ & 0.29 & $ -77 \pm 10 $ \\\hline
         \end{tabular} }
    \caption{Fit parameters of the chiral-continuum extrapolations of the infinite-volume binding energies $\Delta E$ obtained from the zeroth-order ERE.}
    \label{tab:chiralfitE}
\end{table}

\begin{table}[h]
    \centering
        \resizebox{\textwidth}{!}{
    \begin{tabular}{|l|c|c|c|c||c|c|c|c|c||c|}
    \hline
     & \multicolumn{4}{|c|}{Linear-in-$m_\pi^2$ fit (Fig. \ref{fig:chiralcontE})} & \multicolumn{5}{|c|}{Quadratic-in-$m_\pi^2$ fit (Fig. \ref{fig:chiralcontEnextorder})} & Model averaged\\
    \hline
    $H_{sJ}$ & $C$ & $d$ & $\Delta E_\text{phys}$(MeV) & weight & $C$ & $d$ & $C_2$ & $\Delta E_\text{phys}$(MeV) & weight & $\Delta E_\text{phys}$(MeV)\\ \hline
$D^*_{s0}$ & $ -0.47 \pm 0.16$ & $ 0.45 \pm 0.38$ & $ -51 \pm 13$ & 0.49 & $ -1.21 \pm 0.53$ & $ 0.25 \pm 0.41$ & $ 11.2 \pm 7.7$ & $ -37 \pm 17$ & 0.51 & $ -44 \pm 17 $ \\\hline
$D_{s1}$ & $ -0.45 \pm 0.13$ & $ 1.13 \pm 0.34$ & $ -77 \pm 11$ & 0.38 & $ -1.44 \pm 0.59$ & $ 0.66 \pm 0.44$ & $ 15.0 \pm 8.7$ & $ -54 \pm 17$ & 0.62 & $ -63 \pm 19 $ \\\hline
$B^*_{s0}$  & $ -0.60 \pm 0.16$ & $ 0.54 \pm 0.48$ & $ -79 \pm 10$ & 0.38 & $ -1.55 \pm 0.58$ & $ 0.22 \pm 0.52$ & $ 12.5 \pm 7.3$ & $ -58 \pm 16$ & 0.62 & $ -65 \pm 18 $ \\\hline
$B_{s1}$ & $ -0.53 \pm 0.16$ & $ 0.40 \pm 0.42$ & $ -78 \pm 11$ & 0.59 & $ -1.20 \pm 0.62$ & $ 0.28 \pm 0.44$ & $ 9.4 \pm 8.4$ & $ -65 \pm 15$ & 0.41 & $ -73 \pm 14 $ \\\hline
         \end{tabular}}
    \caption{Fit parameters of the chiral-continuum extrapolations of the infinite-volume binding energies $\Delta E$ obtained from the first-order ERE.}
    \label{tab:chiralfitEnextorder}
\end{table}

In order to account for the systematic errors in the binding energies associated with the choice of zeroth-order as opposed to first-order fits to the ERE, we give a systematic error on the binding energy obtained via the shift in central value $\lvert \Delta E^{(1)}_{\rm phys}-  \Delta E^{(0)}_{\rm phys} \rvert$. This results in the final binding energies
\begin{align}
m_{D^*_{s0}}-m_D-m_K&=-48 (14)(4)\text{ MeV},\\
m_{D_{s1}}-m_{D^*}-m_K&=-61 (15)(2)\text{ MeV},\\
m_{B^*_{s0}}-m_B-m_K&= -69 (13)(4)\text{ MeV},\\
m_{B_{s1}}-m_{B^*}-m_K&=-77 (10)(5)\text{ MeV.}
\end{align}
Here, the first uncertainty is the model-averaged statistical/extrapolation uncertainty, and the second is the systematic uncertainty associated with choice of ERE order. The experimental values are $m_{D^*_{s0}}-m_D-m_K=-45.0(0.6)$ MeV and $m_{D_{s1}}-m_{D^*}-m_K=-44.7(0.6)$ MeV \cite{ParticleDataGroup:2026mpi}; our results are consistent with them. A comparison with other calculations is made in Sec.~\ref{sec:conclusions}.

\begin{figure}[h]
    \centering
    \includegraphics[width=0.9\linewidth]{positive_parity/chiralcontiumlegend.pdf}
    \includegraphics[width=0.49\linewidth]{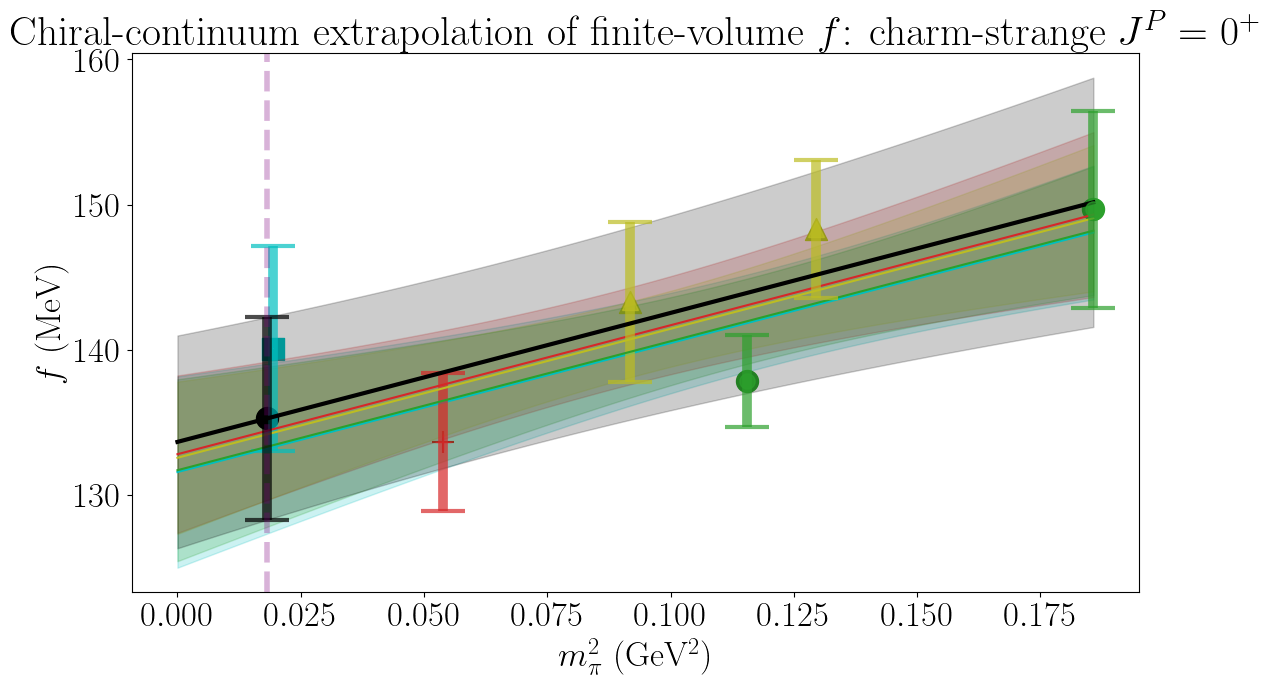}
    \hfill
    \includegraphics[width=0.49\linewidth]{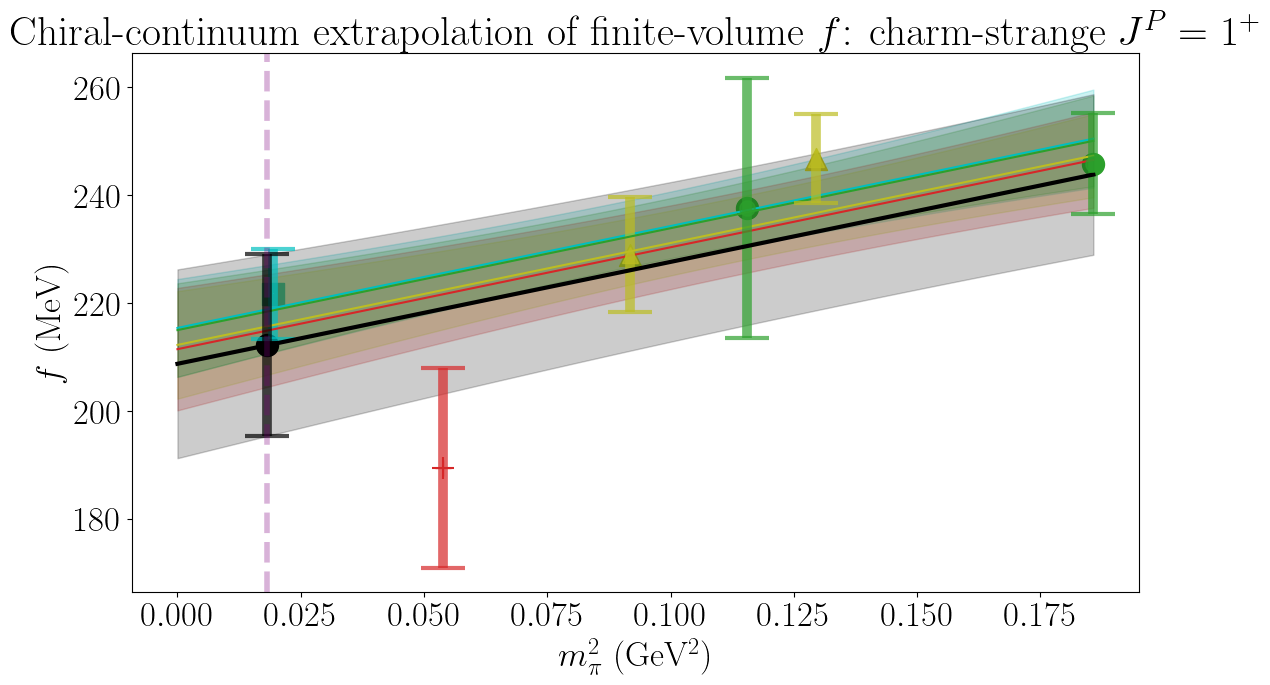}
    
    \includegraphics[width=0.49\linewidth]{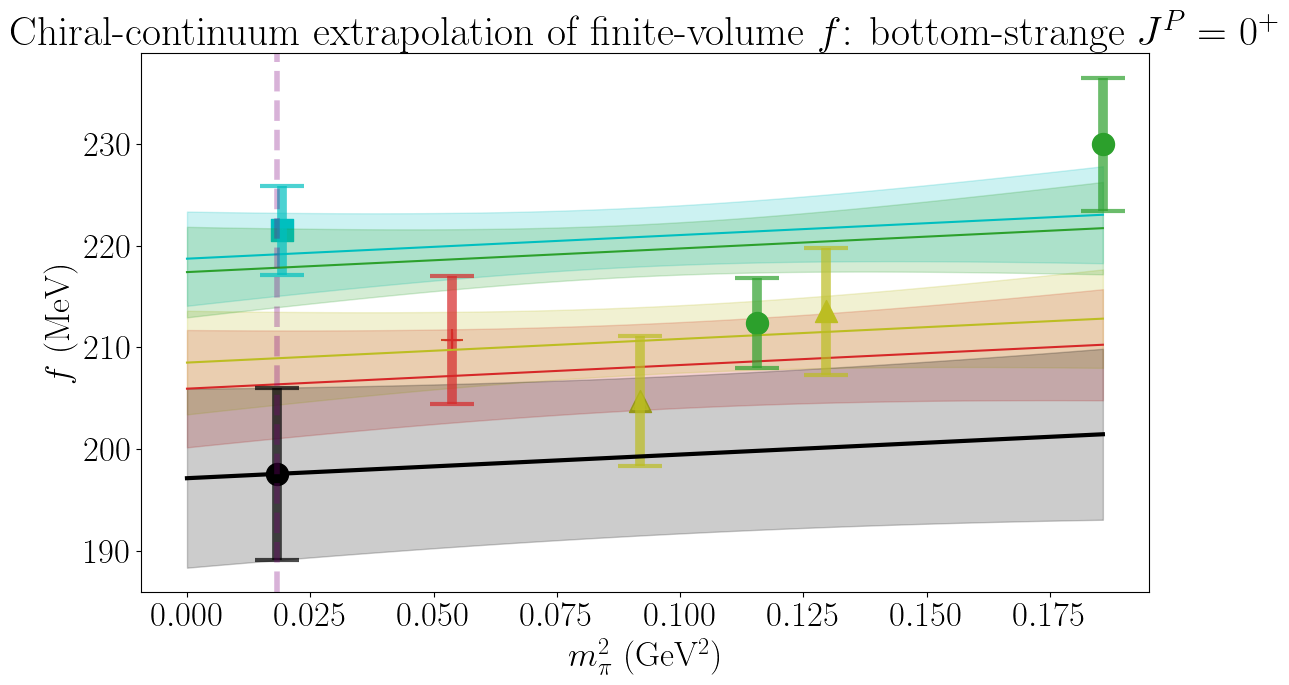}
    \hfill
    \includegraphics[width=0.49\linewidth]{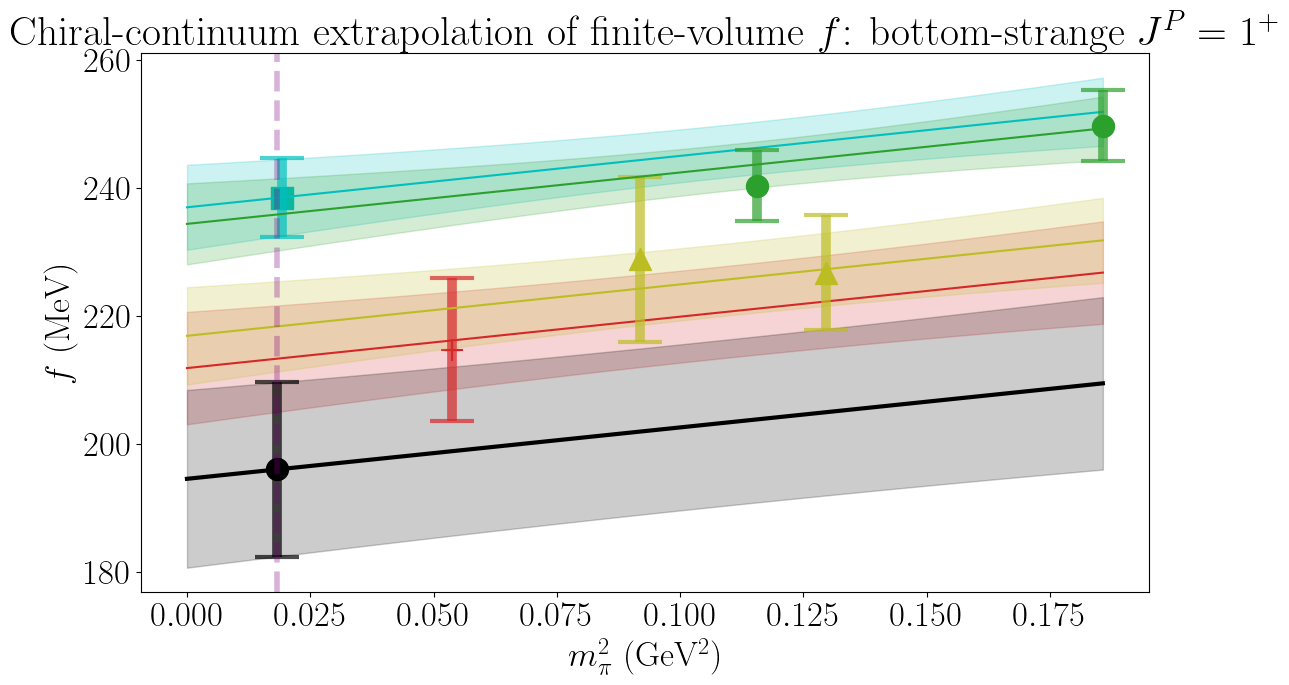}
    \caption{Linear-in-$m_\pi^2$ chiral-continuum extrapolation fits of the ground-state decay constants (without treating finite volume effects). The bands represent chiral extrapolations at fixed lattice spacing. The physical pion mass is indicated by the dashed line. The black band is the chiral extrapolation at $a=0$ and the black data point is  $f_\text{phys}$.}
    \label{fig:chiralcontf}
\end{figure}

\begin{figure}[h]
    \centering
    \includegraphics[width=0.9\linewidth]{positive_parity/chiralcontiumlegend.pdf}
    \includegraphics[width=0.49\linewidth]{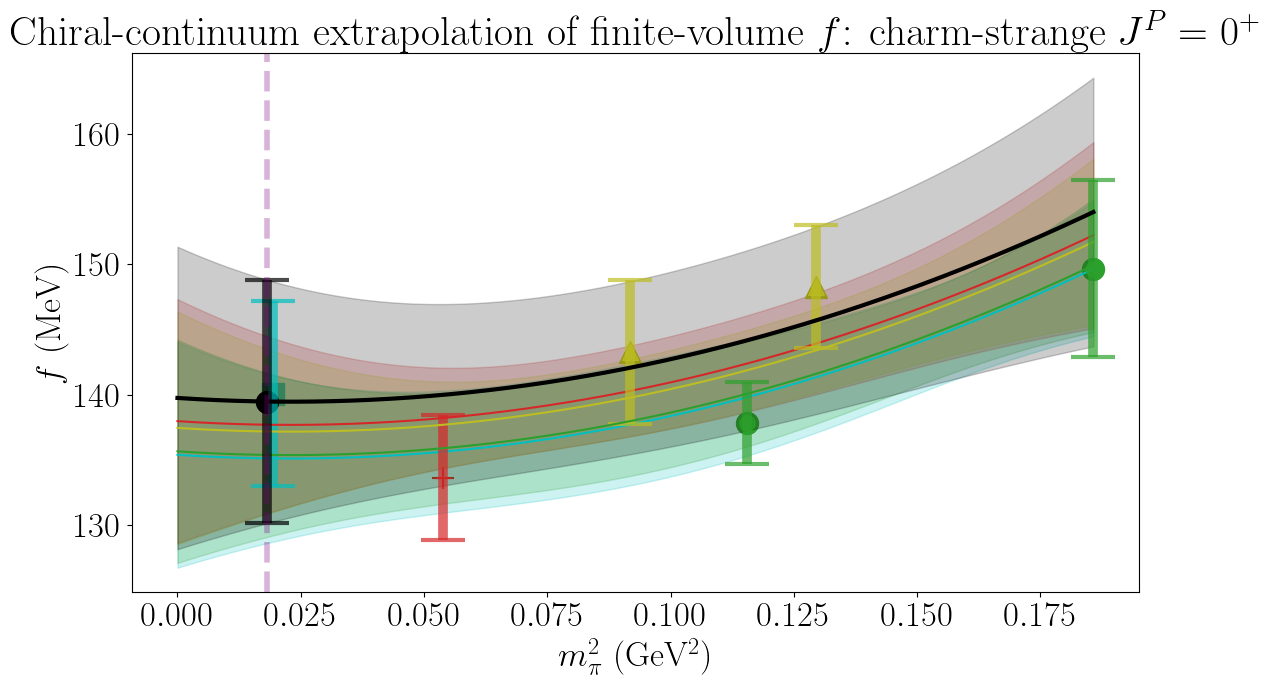}
    \hfill
    \includegraphics[width=0.49\linewidth]{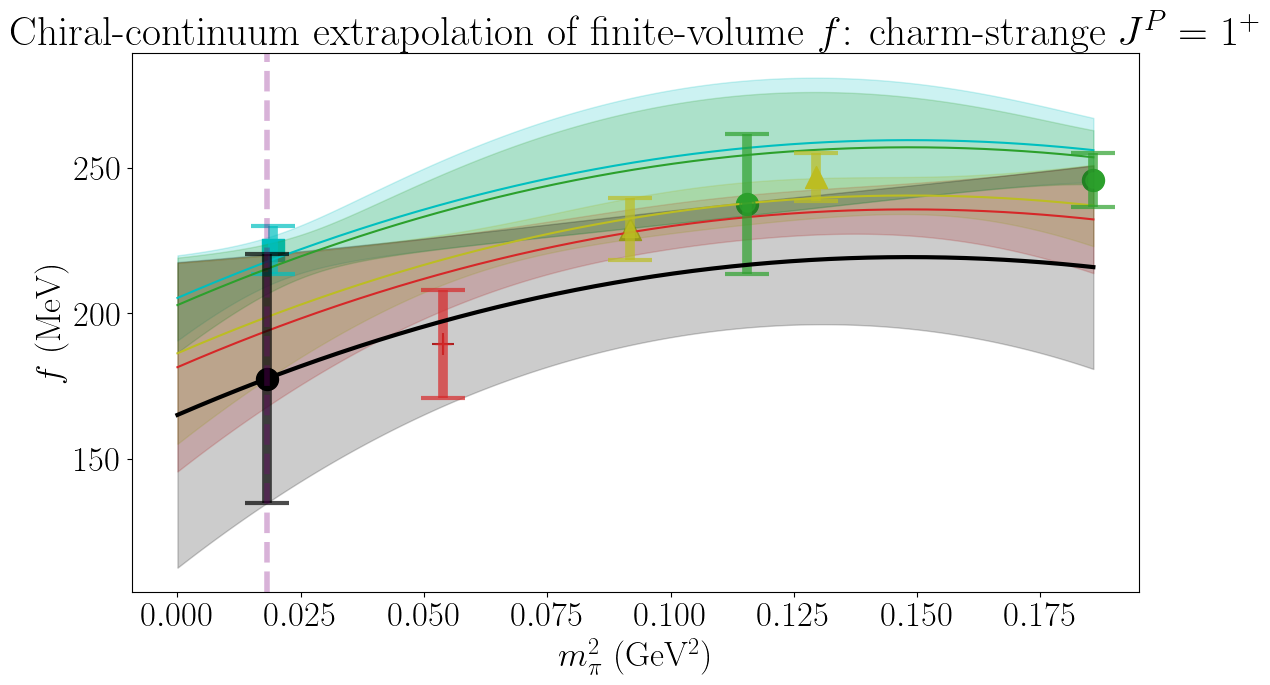}
    
    \includegraphics[width=0.49\linewidth]{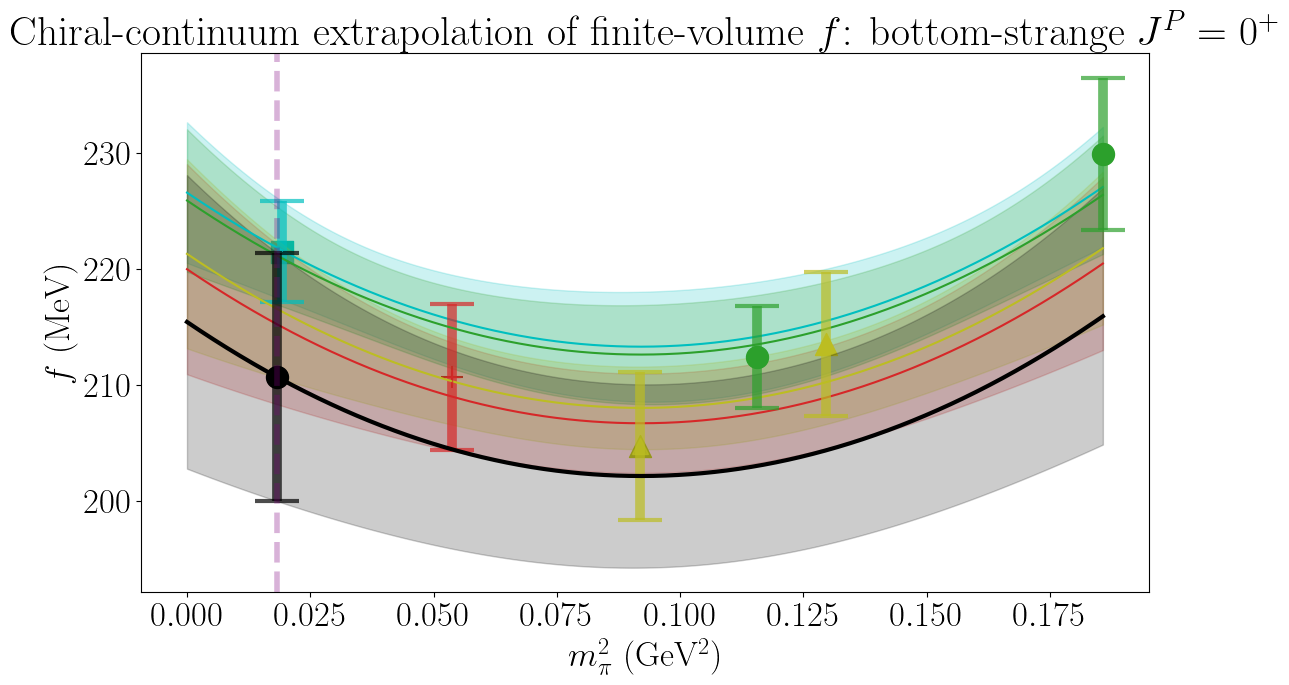}
    \hfill
    \includegraphics[width=0.49\linewidth]{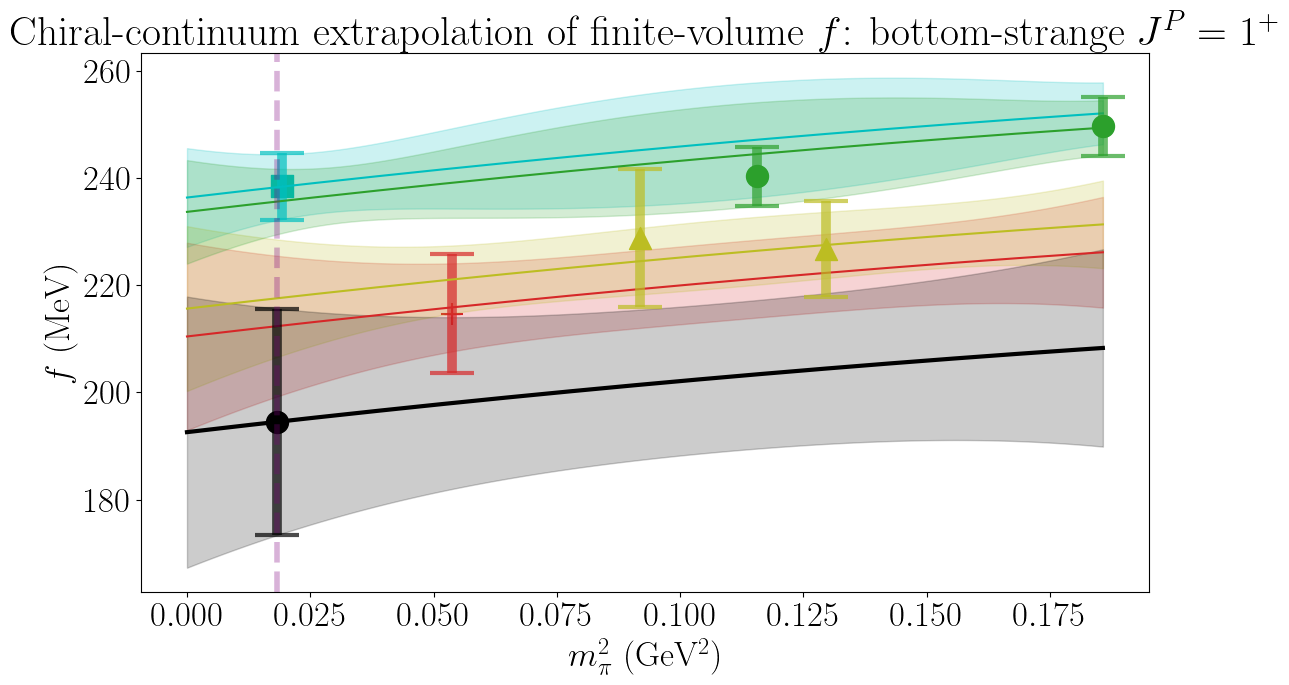}
    \caption{Quadratic-in-$m_\pi^2$ chiral-continuum extrapolation fits of the ground-state decay constants (without treating finite volume effects). The bands represent chiral extrapolations at fixed lattice spacing. The physical pion mass is indicated by the dashed line. The black band is the chiral extrapolation at $a=0$ and the black data point is  $f_\text{phys}$.}
    \label{fig:chiralcontfnextorder}
\end{figure}

Extrapolations of the decay constants are shown in Fig.~\ref{fig:chiralcontf}-\ref{fig:chiralcontfnextorder}, and the corresponding fit parameters and model-averaged $ f_{\text{phys}}$ are given in Table \ref{tab:chiralfitf}.

\begin{table}[h]
    \centering
        \resizebox{\textwidth}{!}{
    \begin{tabular}{|l|c|c|c|c||c|c|c|c|c||c|}
    \hline
     & \multicolumn{4}{|c|}{Linear-in-$m_\pi^2$ fit (Fig. \ref{fig:chiralcontf})} & \multicolumn{5}{|c|}{Quadratic-in-$m_\pi^2$ fit (Fig. \ref{fig:chiralcontfnextorder})} & Model averaged\\
    \hline
    $H_{sJ}$ & $C$ & $d$ & $f_\text{phys}$(MeV) & weight & $C$ & $d$ & $C_2$ & $f_\text{phys}$(MeV) & weight & $f_\text{phys}$(MeV)\\ \hline
$D^*_{s0}$ & $ 0.151 \pm 0.081$ & $ -0.05 \pm 0.22$ & $ 135.3 \pm 7.0$ & 0.68 & $ -0.04 \pm 0.30$ & $ -0.10 \pm 0.23$ & $ 2.7 \pm 4.0$ & $ 139.5 \pm 9.3$ & 0.32 & $ 136.6 \pm 8.0 $ \\\hline
$D_{s1}$ & $ 0.32 \pm 0.12$ & $ 0.16 \pm 0.45$ & $ 212 \pm 17$ & 0.65 & $ 1.2 \pm 1.1$ & $ 1.0 \pm 1.0$ & $ -12 \pm 14$ & $ 178 \pm 43$ & 0.35 & $ 200 \pm 33 $ \\\hline
$B^*_{s0}$ & $ 0.039 \pm 0.068$ & $ 0.52 \pm 0.23$ & $ 197.5 \pm 8.5$ & 0.26 & $ -0.49 \pm 0.27$ & $ 0.27 \pm 0.26$ & $ 7.7 \pm 3.8$ & $ 211 \pm 11$ & 0.74 & $ 207 \pm 12 $ \\\hline
$B_{s1}$ & $ 0.136 \pm 0.079$ & $ 1.02 \pm 0.36$ & $ 196 \pm 14$ & 0.73 & $0.18 \pm 0.51$ & $ 1.05 \pm 0.51$ & $ -0.6 \pm 6.6$ & $ 195 \pm 21$ & 0.27 & $ 196 \pm 16 $ \\
\hline
         \end{tabular}}
    \caption{Fit parameters of the chiral-continuum extrapolations of the decay constants.}
    \label{tab:chiralfitf}
\end{table}

To estimate the finite-volume effects in the decay constants, we apply the chiral-perturbation-theory-inspired corrections \cite{bali}
\begin{gather}
    f_\text{new}=f_\text{latt}-g(m_\pi^2) \frac{e^{-Lm_\pi}}{(L m_\pi)^{3/2}}\label{eq:fFVcorrections}
\end{gather} 
to the lattice results from each ensemble. To this end, we first obtain the values of the coefficients $g(m_\pi^2)$, independently for each system, at $m_\pi=m_{\pi,\text{C005}}$ by fitting Eq.~(\ref{eq:fFVcorrections}) to the data from the C005 and C005LV ensembles, which differ only in $L$. We then scale these coefficients to the pion masses of the other ensembles using \cite{bali}
\begin{gather}
    g(m_\pi^2)=\frac{m_\pi^2}{m^2_{\pi,\text{C005}}}g_\text{C005}.
\end{gather} 

We then repeat the chiral-continuum extrapolations using Eqs.~(\ref{eq:chiralcontf}) and (\ref{eq:chiralcontf-nextorder}) for the $f_\text{new}$, resulting in the model-averaged values 
$f_{D^*_{s0}\text{new,phys}}= 140(12)$ MeV, $f_{D_{s1}\text{new,phys}}=176(53)$ MeV,  $f_{B^*_{s0}\text{new,phys}}= 215(14)$ MeV, $f_{B_{s1}\text{new,phys}}=207(26) $ MeV.

To obtain the final results for the decay constants, we take the central values and statistical/extrapolation uncertainties from the fits without finite-volume corrections, and use the shifts $\lvert f_\text{new,phys}-f_\text{phys} \rvert$ as our estimates of the finite-volume errors. In addition, we include the 1\% estimate of the systematic uncertainty due to missing higher-order corrections in the residual matching factors. This gives
\begin{align}
f_{D^*_{s0}}&=136.6 (8.0)(4.0)(1.4)\text{ MeV},\\
f_{D_{s1}}&=200 (33)(24)(2)\text{ MeV},\\
f_{B^*_{s0}}&=207 (12)(8)(2)\text{ MeV},\\
f_{B_{s1}}&= 196 (16)(11)(2)\text{ MeV}.
\end{align}
 A comparison with other calculations is made in Sec.~\ref{sec:conclusions}.

We also perform chiral-continuum extrapolation of the effective-range parameters $a_0$ and $r$ using
\begin{gather}
        a_0^{-1}=a^{-1}_{0\text{phys}}+\frac{C}{4\pi f_\pi} (m_\pi^2 - m_{\pi \text{phys}}^2)+d\Lambda^3 a^2 \label{eq:achiral}\\
    r^{-1}=r^{-1}_{\text{phys}}+\frac{C}{4\pi f_\pi} (m_\pi^2 - m_{\pi \text{phys}}^2)+d\Lambda^3 a^2 .\label{eq:rchiral}
\end{gather} 
For $a_0$, we perform separate extrapolations of the results from the zeroth-order and first-order effective-range expansions. The values of $a_0$ and $r$ at the physical point are provided in Table \ref{tab:ar}.
As discussed in Sec.~\ref{sec:molecular}, the ERE parameters are broadly consistent with an $H^{(*)}K$-molecule interpretation for all the systems investigated.

\begin{table}[h]
    \centering
    \begin{tabular}{|l|c|c|c|c|c|}
    \hline
$H_{sJ}^{(*)}$ & ERE Order & $a_{0{\rm phys}}$ (MeV$^{-1}$) & $a_{0{\rm phys}}$ (fm) & $r_{\rm phys}$ (MeV$^{-1}$) & $r_{\rm phys}$ (fm) \\
\hline
$D_{s0}^*$ & 0th & $-0.00490 \pm 0.00057$ & $-0.97 \pm 0.11$ & --- & --- \\
$D_{s0}^*$ & 1st & $-0.0066 \pm 0.0051$ & $-1.3 \pm 1.0$ & $0.01 \pm 0.14$ & $1 \pm 27$ \\ \hline
$D_{s1}$ & 0th & $-0.00406 \pm 0.00031$ & $-0.800 \pm 0.061$ & --- & --- \\
$D_{s1}$ & 1st & $-0.0020 \pm 0.0011$ & $-0.40 \pm 0.22$ & $-0.0014 \pm 0.0018$ & $-0.28 \pm 0.36$ \\ \hline
$B_{s0}^*$ & 0th & $-0.00372 \pm 0.00022$ & $-0.734 \pm 0.044$ & --- & --- \\
$B_{s0}^*$ & 1st & $-0.0018 \pm 0.0023$ & $-0.35 \pm 0.46$ & $-0.0022 \pm 0.0044$ & $-0.44 \pm 0.86$ \\ \hline
$B_{s1}$ & 0th & $-0.00362 \pm 0.00020$ & $-0.714 \pm 0.039$ & --- & --- \\
$B_{s1}$ & 1st & $-0.00110 \pm 0.00078$ & $-0.22 \pm 0.15$ & $-0.0017 \pm 0.0012$ & $-0.34 \pm 0.23$ \\
\hline
    \end{tabular}
    \caption{Effective-range parameters extrapolated to the physical point.}
    \label{tab:ar}
\end{table}

\FloatBarrier
\subsection{Discussion of Heavy-Quark Symmetries}
\label{sec:heavyquark}

In this section, we compare our results with expectations from heavy-quark effective theory (HQET), which corresponds to an expansion in $\Lambda_{\rm QCD}/m_Q$. At leading order in this expansion, there exists a flavor symmetry relating the charm and bottom states, and there are no spin-dependent interactions with the heavy quark, such that the spin of the heavy quark $\mathbf{S}_Q$, and the spin of the light degrees of freedom, $\mathbf{S}_\ell=\mathbf{J}-\mathbf{S}_Q$, become conserved \cite{Isgur:1991wq,Manohar:2000dt}.
 Mesons with one heavy quark are thus labeled by the $j, s_\ell, s_Q$ eigenvalues, with states of a given $s_\ell$ appearing in mass-degenerate doublets.
In the charm sector, the lowest $s_\ell=\frac{1}{2}$ and $s_\ell=\frac{3}{2}$ doublets correspond to the $\big(D_{s0}^*(2317)^\pm , D_{s1}(2460)^\pm \big)$ and $\big( D_{s1}(2536)^\pm , D_{s2}^\pm \big)$, respectively. In the bottom sector, we identify the $s_\ell=\frac{1}{2}$ doublet as $(B^*_{s0}, B_{s1})$ and the $s_\ell=\frac{3}{2}$ doublet as $(B_{s1}(5830)^0, B_{s2}^0)$, using the same notation as Table \ref{tab:Ds}.

Heavy-quark spin dependence of the meson masses appears starting at next-to-leading order in $\Lambda_{\rm QCD}/m_Q$, in the form of a chromomagnetic term $H_{\rm CM}=g\bar{Q}_v\frac{\sigma_{\mu\nu}F^{\mu\nu}}{4m_Q} Q_v$, where, at tree level, $Q_v(x)= e^{im_Qv\cdot x} \frac{1+\slashed{v}}{2} Q(x)$ with the heavy-quark four-velocity $v$. The contribution of $H_{cm}$ to the meson mass is found to be 
\begin{gather}
    \bra{H_{sJ}}H_{\rm CM}\ket{H_{sJ}}=\frac{\lambda_2}{m_Q}2 \vec{S}_Q\cdot \vec{S}_\ell=\frac{\lambda_2}{m_Q}\left[j(j+1)-s_Q(s_Q+1)-s_\ell (s_\ell+1)\right],
\end{gather}    
where $\lambda_2 \sim \Lambda_{\rm QCD}^2\sim (340 \ \text{MeV})^2$.
This leads to the estimates
\begin{gather}
    m_{H_{s1}}-m_{H^*_{s0}}\sim \frac{5\lambda_2}{4m_Q} \sim \frac{5 \Lambda_{\rm QCD}^2}{4m_Q}\sim \begin{cases}
        113 \ \text{MeV}  &  \text{if }\  H=D,\\
        34 \ \text{MeV}  & \text{if }\ H=B.\\
    \end{cases}
\end{gather}
The experimental value for positive-parity charm-strange hyperfine splitting is $m_{D_{s1}}-m_{D^*_{s0}}\approx 142$ MeV, and our lattice result is $m_{D_{s1}}-m_{D^*_{s0}}\approx 128$ MeV. Using the experimentally measured $D_{s0}^*(2317)^\pm-D_{s1}(2460)^\pm$ hyperfine splitting as input, the HQET prediction for the $B^*_{s0}-B_{s1}$ splitting is $m_{B_{s1}}-m_{B^*_{s0}} \sim \frac{m_c}{m_b} (m_{D_{s1}}-m_{D^*_{s0}})\sim 43$ MeV (the prediction becomes $\sim 39$ MeV when using the splitting we have calculated on the lattice as input); this is close to our lattice result of $m_{B_{s1}}-m_{B^*_{s0}}\approx 37$ MeV.

The decay constants may also be studied in the heavy-quark limit.  In the HQET convention where the normalization of hadronic states is independent of $m_Q$, matrix elements must not depend on $m_Q$ in the heavy-quark limit. In the standard normalization, this requires $\bra{0}J_V^\mu\ket{H^*_{s0}}= i \sqrt{2m_Q}  C v^\mu$ and $\bra{0}J_A^\mu\ket{H_{s1}}=\sqrt{2m_Q}  C \epsilon^\mu$, with $C$ heavy-quark-mass-independent. Up to $\mathcal{O}\left(\frac{\Lambda_{\rm QCD}}{m_Q}\right)$ and radiative corrections, we therefore expect
\begin{gather}
    \frac{f_{B_{sJ}}}{f_{D_{sJ}}}=\sqrt{\frac{m_{D_{sJ}}}{m_{B_{sJ}}}}, \label{eq:fhqet}\\
    \frac{f_{H^*_{s0}}}{f_{H_{s1}}}=1. \label{eq:fspinhqet}
\end{gather}
Our lattice results for the bottom-to-charm decay-constant ratios are $f_{B^*_{s0}}/f_{D^*_{s0}}\approx 1.5$ and $f_{B_{s1}}/f_{D_{s1}}\approx 0.98$, which deviate significantly from the leading-order HQET predictions $\sqrt{m_{D^*_{s0}}/m_{B^*_{s0}}}\approx 0.64$,  $\sqrt{m_{D_{s1}}/m_{B_{s1}}}\approx 0.65$. For the $j=0/j=1$ ratios, our lattice results are $f_{D^*_{s0}}/f_{D_{s1}}\approx 0.68$ and $f_{B^*_{s0}}/f_{B_{s1}}\approx 1.0$, showing that heavy-quark symmetry is heavily broken for the charmed mesons, but provides a better description of the bottom mesons, as expected.

The decay constants $f_{D^*_{s0}}$ and $f_{D_{s1}}$ have been previously calculated on the lattice at a single lattice spacing in Ref.~\cite{bali}, with the result $f_{D^*_{s0}}/f_{D_{s1}}\approx 0.59$. 
Extractions of the positive-parity $D_{s0,1}^{(*)}$ decay constants from branching ratios of nonleptonic $B$ decays using the factorization hypothesis have also found deviations from the heavy-quark limit, with $f_{D^*_{s0}}/f_{D_{s1}}\sim 0.5$  \cite{Hwang:2004kga}.
The quark-model calculations of Refs.~\cite{Hwang:2004kga, Veseli:1996yg, Hwang:1996ha, LeYaouanc:1974da} also obtained $f_{D^*_{s0}}/f_{D_{s1}}\sim 0.5$.
In the quark-model picture of the positive-parity heavy-strange mesons, large deviations from $f_{H^*_{s0}}/f_{H_{s1}}=1$ can additionally be understood in in terms of the $^1P_1$ and $^3P_1$ mixing of the physical $J^P=1^+$ states, since the $^1 P_1$ component is not part of the HQET doublet. A fit of the experimental spectra and widths to predictions from potential models suggested that the $D_{s1}$ mixing angles are large compared to the heavy-light $D_1$ mixing angles \cite{Cahn:2003cw}. Reference \cite{Li:2018eqc} directly computed the mixing angles from solutions to the  Bethe-Salpeter-equations, giving large mixing also in the bottom sector, but not as large as in the charm sector.

\subsection{Assessing the Exotic Nature of the Positive-Parity Heavy-Strange Mesons}
\label{sec:molecular}

To determine if a $H_{sJ}$ meson is a candidate for molecular structure, one may analyze a few important features. The first basic requirement is that the  state must be proximal to, and below, the $H^{(*)}K$ kinematic threshold, such that the distance below threshold may be interpreted as a binding energy associated with the interaction between two bound color-singlets. The $H_{sJ}$ binding energies $\Delta E$ in Table \ref{tab:chiralfitE} indicate that the positive-parity heavy-strange ground states are such near-threshold states.

Near the $H_{sJ}$ bound-state pole at $s_0$, the $T$-matrix for $S$-wave $H^{(*)} K$ scattering behaves like
\begin{equation}
    \mathcal{M}_0\approx \frac{g^2}{s-s_0}, \label{eq:poleresidue}
\end{equation}
where the residue is the square of the coupling to the bound state, $g$.
To interpret the state as a molecule, the coupling should be large \cite{barnes}. 

To more precisely characterize these states we make use of the Weinberg compositeness criterion \cite{weinberg}. The wavefunction of a shallow bound state at large distances behaves as $\psi_b \sim \frac{e^{-\gamma r}}{r}$ where $\gamma=\sqrt{2\mu |\Delta E|}$ (with $\mu$ the $H^{(*)} K$ reduced mass) is the inverse characteristic length scale of such a state. Weinberg's criterion is applicable to shallow bound states, in which $1/\gamma$ is large compared to the range of the interaction, $1/\beta$. 
In the case of heavy-strange positive-parity states, this range can be be estimated to be of order $1/m_{\rho}$ \cite{guo,Gil-Dominguez:2023puj}.

Weinberg's compositeness criterion is derived by writing the bound state as \cite{Weinberg:1963zza,Baru:2003qq,guo}
\begin{equation}
    \ket{H_{sJ}}=\lambda \ket{n}+\int d^3k\: \chi(\mathbf{k}) \ket{h_1 h_2},
\end{equation} 
where $\ket{n}$ and $\ket{h_1 h_2}$ are one- and two-particle eigenstates of a free Hamiltonian, and $\ket{H_{sJ}}$ is the eigenstate of the full Hamiltonian, which includes an interaction term $V$. The above states are normalized nonrelativistically. From $\langle H_{sJ}|H_{sJ}\rangle=1$ it follows that  $\lambda^2=Z= \lvert  \bra{n}\ket{H_{sJ}} \rvert^2$, the ``field strength renormalization.''

By considering the completeness relation in the limit of weak binding (where $\lvert \bra{h_1 h_2}V\ket{H_{sJ}}\rvert$ is taken to be a constant) it follows that  \cite{weinberg,guo}, after converting to the relativistic normalization to match our Eqs.~(\ref{eq:M0}) and (\ref{eq:poleresidue}),
\begin{equation}
    g^2 = 16 \pi s_0 \frac{\gamma}{\mu} \left( 1 - Z\right). \label{eq:grel}
\end{equation}
Thus, Eq.~(\ref{eq:grel}) makes precise the relationship between the coupling and compositeness, $1-Z$. Observe that $g^2$ is maximized when $Z=0$, i.e., when the $H_{sJ}$ state has no compact part.
The solution to the Lippman-Schwinger equation in the weak-binding limit furnishes an amplitude that takes the form of the effective-range expansion through order $p^2$ \cite{weinberg}. Using Eq.~(\ref{eq:grel}), the coefficients are found to be
\begin{align}
    a_0 &=-2 \frac{1-Z}{2-Z} \frac{1}{\gamma} + \mathcal{O}\left(\frac{1}{\beta}\right),\\
    r &=-\frac{Z}{1-Z} \frac{1}{\gamma} + \mathcal{O}\left(\frac{1}{\beta}\right).
\end{align}
The coupling then depends on the scattering length as
\begin{gather}
    g^2\approx-4\pi \frac{ 4 s_0 }{\mu} \frac{\gamma}{1+\frac{2}{a_0 \gamma}}. \label{eq:gapprox}
\end{gather}
For a pure hadronic molecule\footnote{The interpretation of $1-Z$ as compositeness in terms of $H^{(*)} K$ constituents is complicated by the existence of the $H_s^{(*)} \eta$-thresholds, and $1-Z$ includes the probability to have an $H_s^{(*)} \eta$ structure, as well as the $H^{(*)} K$ structure. However, given that the $H_s^{(*)} \eta$ thresholds are about 140-150 MeV above the $H^{(*)} K$ thresholds, we expect these contributions to be small, and we do not consider them further.} with $Z=0$, one expects $r=0+ \mathcal{O}\left(\frac{1}{\beta}\right)\lesssim \frac{1}{m_\rho}$ to be small and positive, as well as
\begin{equation}
    \frac{1}{a_0} \sim - \gamma=-\sqrt{2\mu |\Delta E|}. \label{eq:weinberg}
\end{equation}
The values of $-\gamma$ are given in Table \ref{tab:expectedparams} and are indeed seen to be close to $1/a_0$ (cf.~Tables \ref{tab:ERE} and \ref{tab:firstERE}). Table \ref{tab:firstERE} also exhibits consistency with the compositeness expectation $r \sim \mathcal{O}\left(\frac{1}{m_\rho}\right)$, but $r$ is also compatible with being zero or even negative (though not large and negative, as would be the case for $Z$ close to 1).

\begin{table}
    \centering
    \begin{tabular}{|l|c|c|c|c|}
    \hline
Ensemble & $-\gamma^{(0)}$ for $D^*_{s0}$ (MeV) & $-\gamma^{(0)}$ for $D_{s1}$ (MeV) & $-\gamma^{(0)}$ for $B^*_{s0}$ (MeV) & $-\gamma^{(0)}$ for $B_{s1}$ (MeV) \\
\hline
C00078 & $-152 \pm 23$ & $-159 \pm 17$ & $-265 \pm 28$ & $-246 \pm 23$\\
F1M & $-212 \pm 19$ & $-230 \pm 11$ & $-280.0 \pm 8.7$ & $-289.4 \pm 8.2$\\
F004 & $-233 \pm 12$ & $-256 \pm 11$ & $-307.4 \pm 8.9$ & $-293 \pm 17$\\
C005s & $-232.9 \pm 7.7$ & $-242 \pm 12$ & $-328 \pm 12$ & $-324 \pm 10$\\
F006 & $-233 \pm 20$ & $-261 \pm 10$ & $-322 \pm 17$ & $-321.7 \pm 9.3$\\
C01 & $-249 \pm 30$ & $-237 \pm 30$ & $-342 \pm 15$ & $-351 \pm 13$\\
\hline
\hline
Ensemble & $-\gamma^{(1)}$ for $D^*_{s0}$ (MeV) & $-\gamma^{(1)}$ for $D_{s1}$ (MeV) & $-\gamma^{(1)}$ for $B^*_{s0}$ (MeV) & $-\gamma^{(1)}$ for $B_{s1}$ (MeV)\\
\hline
C00078 & $-138 \pm 32$ & $-148 \pm 22$ & $-182 \pm 45$ & $-178 \pm 49$\\
F1M & $-210 \pm 19$ & $-230 \pm 11$ & $-278 \pm 10$ & $-284 \pm 10$\\
F004 & $-239 \pm 15$ & $-260 \pm 13$ & $-307 \pm 10$ & $-288 \pm 19$\\
C005s & $-237 \pm 10$ & $-242 \pm 18$ & $-327 \pm 15$ & $-319 \pm 12$\\
F006 & $-229 \pm 21$ & $-264 \pm 13$ & $-320 \pm 18$ & $-321 \pm 10$\\
C01 & $-249 \pm 29$ & $-235 \pm 32$ & $-341 \pm 16$ & $-336 \pm 22$\\
    \hline
    \end{tabular}
    \caption{Values of $-\gamma^{(0)}=-\sqrt{2\mu |\Delta E^{(0)}|}$ ($-\gamma^{(1)}=-\sqrt{2\mu |\Delta E^{(1)}}|$), computed using the zeroth(first)-order ERE, on each ensemble. ``C005s'' denotes the combination of C005 and C005LV.}
    \label{tab:expectedparams}
\end{table}

We obtained the values of $g$ and $Z$ on each bootstrap sample for the effective-range parameters from both the zeroth-order and first-order ERE, where samples resulting in numerical exceptions are dropped (which most commonly occurs for the $B_{s1}$ on the C01 ensemble). We also consider the systematic uncertainty in $Z$ arising from missing corrections of order $\xi\equiv\gamma/\beta$ by evaluating the lower and upper bounds proposed in Ref.~\cite{Kinugawa:2022fzn},
\begin{align}
    Z_l=&1-\frac{-a_0\gamma+\xi}{2+a_0 \gamma -\xi},\\
    Z_u=&1-\frac{-a_0\gamma-\xi}{2+a_0 \gamma +\xi}.
\end{align}
Including such an estimation is particularly important in the bottom sector, and for ensembles at large pion mass, where states are more deeply bound. We compute $Z_l$ and $Z_u$ for every bootstrap sample, and consider the values of $Z_l-\sigma_{Z_l}$ and $Z_u+\sigma_{Z_u}$, where $Z_{l,u}$ are the means and $\sigma_{Z_{l,u}}$ are the standard deviations.  
The results for $Z$, $Z_u+\sigma_{Z_u}$, $\xi=\gamma/\beta$, and $g$ from each ensemble are given in Table \ref{tab:gZ}. The results for $Z_l-\sigma_{Z_l}$ were always negative, and are therefore not included in the table.

\begin{table}
    \centering
    \begin{tabular}{|l|c|c|c|c|c|c|c|c|}

\hline
Ensemble &  $Z$  &  $Z_u+\sigma_{Z_u}$ & $\xi$ & $g$ (GeV) &  $Z$  & $Z_u+\sigma_{Z_u}$ & $\xi$ & $g$ (GeV)\\
\hline
\hline
0th order   & \multicolumn{4}{|c|}{$D_{s0}^*$} & \multicolumn{4}{|c|}{$D_{s1}$}\\ \hline
C00078 & $0.0182 \pm 0.0053$ & 0.39 & 0.20 &  $10.07 \pm 0.78$& $0.0199 \pm 0.0041$ & 0.39 & 0.20 &  $10.87 \pm 0.56$\\
F1M & $0.0318 \pm 0.0053$ & 0.48 & 0.27 &  $11.59 \pm 0.42$& $0.0375 \pm 0.0033$ & 0.50 & 0.30 &  $12.65 \pm 0.24$\\
F004 & $0.0354 \pm 0.0034$ & 0.50 & 0.30 &  $11.91 \pm 0.25$& $0.0427 \pm 0.0033$ & 0.54 & 0.33 &  $13.05 \pm 0.22$\\
C005s & $0.0339 \pm 0.0021$ & 0.50 & 0.30 &  $11.84 \pm 0.16$& $0.0370 \pm 0.0036$ & 0.52 & 0.31 &  $12.67 \pm 0.27$\\
F006 & $0.0332 \pm 0.0055$ & 0.52 & 0.30 &  $11.77 \pm 0.43$& $0.0416 \pm 0.0031$ & 0.55 & 0.34 &  $13.01 \pm 0.21$\\
C01 & $0.0343 \pm 0.0078$ & 0.56 & 0.32 &  $11.84 \pm 0.84$& $0.0314 \pm 0.0075$ & 0.54 & 0.30 &  $12.22 \pm 0.70$\\
\hline
\hline
0th order   & \multicolumn{4}{|c|}{$B_{s0}^*$} & \multicolumn{4}{|c|}{$B_{s1}$}\\ \hline
C00078 & $0.059 \pm 0.011$ & 0.59 & 0.34 &  $29.7 \pm 1.5$& $0.0513 \pm 0.0088$ & 0.55 & 0.32 &  $29.0 \pm 1.2$\\
F1M & $0.0610 \pm 0.0035$ & 0.58 & 0.36 &  $30.14 \pm 0.39$& $0.0649 \pm 0.0033$ & 0.59 & 0.38 &  $30.78 \pm 0.36$\\
F004 & $0.0675 \pm 0.0035$ & 0.62 & 0.40 &  $30.88 \pm 0.36$& $0.0621 \pm 0.0064$ & 0.61 & 0.38 &  $30.50 \pm 0.72$\\
C005s & $0.0728 \pm 0.0050$ & 0.65 & 0.43 &  $31.42 \pm 0.48$& $0.0717 \pm 0.0041$ & 0.64 & 0.42 &  $31.54 \pm 0.40$\\
F006 & $0.0692 \pm 0.0065$ & 0.65 & 0.42 &  $31.08 \pm 0.64$& $0.0692 \pm 0.0036$ & 0.64 & 0.42 &  $31.32 \pm 0.36$\\
C01 & $0.0707 \pm 0.0057$ & 0.67 & 0.44 &  $31.30 \pm 0.56$& $0.0740 \pm 0.0047$ & 0.68 & 0.46 &  $31.85 \pm 0.44$\\
\hline
\hline
1st order   & \multicolumn{4}{|c|}{$D_{s0}^*$} & \multicolumn{4}{|c|}{$D_{s1}$}\\ \hline
C00078 & $-1.7 \pm 1.2$ &  $-0.08$ & 0.17 &  $15.1 \pm 4.3$& $-1.6 \pm 1.5$ &  0.21 & 0.19 &  $16.2 \pm 4.8$\\
F1M & $0.04 \pm 0.45$ &  0.72 & 0.27 &  $11.2 \pm 2.4$& $0.27 \pm 0.34$ &  0.85 & 0.30 &  $10.7 \pm 2.5$\\
F004 & $0.40 \pm 0.28$ &  0.91 & 0.31 &  $9.2 \pm 2.1$& $0.36 \pm 0.24$ &  0.88 & 0.34 &  $10.5 \pm 2.0$\\
C005s & $-0.01 \pm 0.20$ &  0.60 & 0.31 &  $12.1 \pm 1.1$& $-0.36 \pm 0.37$ &  0.50 & 0.31 &  $14.8 \pm 1.9$\\
F006 & $-0.13 \pm 0.32$ &  0.57 & 0.30 &  $12.5 \pm 1.7$& $0.29 \pm 0.40$ &  0.94 & 0.34 &  $10.7 \pm 3.1$\\
C01 & $0.06 \pm 0.39$ &  0.73 & 0.32 &  $11.5 \pm 2.3$& $0.13 \pm 0.46$ &  0.85 & 0.30 &  $11.0 \pm 2.9$\\
\hline
\hline
1st order   & \multicolumn{4}{|c|}{$B_{s0}^*$} & \multicolumn{4}{|c|}{$B_{s1}$}\\ \hline
C00078 & $-4.0 \pm 2.8$ &  $-0.69$ & 0.23 &  $55 \pm 22$& $-4.7 \pm 3.7$ &  $-0.63$ & 0.22 &  $58 \pm 27$\\
F1M & $-0.4 \pm 1.3$ &  0.98 & 0.36 &  $32 \pm 16$& $0.1 \pm 1.4$ &  1.19 & 0.37 &  $26 \pm 15$\\
F004 & $0.24 \pm 0.47$ &  0.97 & 0.40 &  $26.4 \pm 8.6$& $-0.16 \pm 0.77$ &  0.88 & 0.37 &  $31 \pm 11$\\
C005s & $-0.5 \pm 1.1$ &  0.80 & 0.42 &  $38 \pm 13$& $-0.60 \pm 0.86$ &  0.70 & 0.41 &  $40 \pm 11$\\
F006 & $0.03 \pm 0.68$ &  0.95 & 0.41 &  $30 \pm 11$& $0.37 \pm 0.51$ &  1.07 & 0.42 &  $23.6 \pm 9.8$\\
C01 & $0.39 \pm 0.43$ &  1.05 & 0.44 &  $23.6 \pm 8.9$& $-2.3 \pm 1.7$ &  0.37 & 0.44 &  $57 \pm 15$\\
    \hline
    \end{tabular}
    
    \caption{The results for the field-strength renormalization $Z$, the upper limit on the field-strength renormalization $Z_u+\sigma_{Z_u}$, the ratio $\xi=\gamma/\beta$ that governs higher-order corrections, and the coupling of the bound states to the two-meson systems, $g$, using both the 0th and 1st-order ERE fits.}
    \label{tab:gZ}
\end{table}

The results based on the zeroth-order ERE results suggest compatibility with a molecular interpretation (and incompatibility with a pure compact state) for all lattice spacings and quark masses. This is also the case for the results from the first-order ERE, although in some cases the uncertainties there are so large that they cover the entire physical region $0\le Z\le 1$. For C00078, the results for $Z$ from the first-order ERE are entirely unphysical.

\begin{table}
    \centering
    \begin{tabular}{|l|c|c|c|c|}
    \hline
         & $D_{s0}^*$ & $D_{s1}$ & $B_{s0}^*$ & $B_{s1}$\\ \hline
        $Z_\text{phys}^{(0)}$ &  $0.0311\pm 0.0063$ & $0.0436 \pm 0.0052$ & $0.0578 \pm 0.0057$ & $0.0606\pm 0.0054$\\
        $Z_\text{phys - chiral}^{(0)}$ &  $0.0248 \pm 0.0040$ & $0.0294 \pm 0.0030$ &  $0.0599 \pm 0.0040$ & $0.0600 \pm 0.0038$\\
        $Z_\text{phys}^{(1)}$ & $0.28\pm 0.49$ & $0.84\pm 0.48$ & $0.17 \pm 0.96$ & $1.4 \pm 1.6$\\ 
        $Z_\text{phys - chiral}^{(1)}$ & $0.064 \pm 0.39$ & $0.24\pm0.36$ & $-0.46 \pm 0.74$ & $0.1 \pm 1.3$\\ \hline
        $g_\text{phys}^{(0)}$ & $11.49 \pm 0.51$ GeV & $13.04 \pm 0.41$ GeV & $29.80 \pm 0.60$ GeV & $30.28 \pm 0.56$ GeV\\
        $g_\text{phys - chiral}^{(0)}$ & $11.23 \pm 0.42$ GeV & $12.29 \pm 0.28$ GeV & $30.02 \pm 0.44$ GeV & $30.29 \pm 0.42$ GeV\\
        $g_\text{phys}^{(1)}$ & $10.3 \pm 2.9$ GeV & $7.8 \pm2.9$ GeV & $28.6\pm 6.8$ GeV & $4.7\pm 7.0$ GeV\\
        $g_\text{phys-chiral}^{(1)}$ & $11.6 \pm 2.1$ GeV & $12.3 \pm 2.4$ GeV & $40 \pm 11$ GeV & $26.0 \pm 7.0$ GeV \\
        \hline
        
    \end{tabular}
    \caption{The results of chiral-continuum or chiral-only extrapolations of the couplings $g$ and field-strength renormalizations $Z$, where the superscripts indicate the order of the ERE.}
    \label{tab:gZfinal}
\end{table}

\begin{figure}
    \centering
    \includegraphics[width=0.9\linewidth]{positive_parity/chiralcontiumlegend.pdf}
    \includegraphics[width=0.49\linewidth]{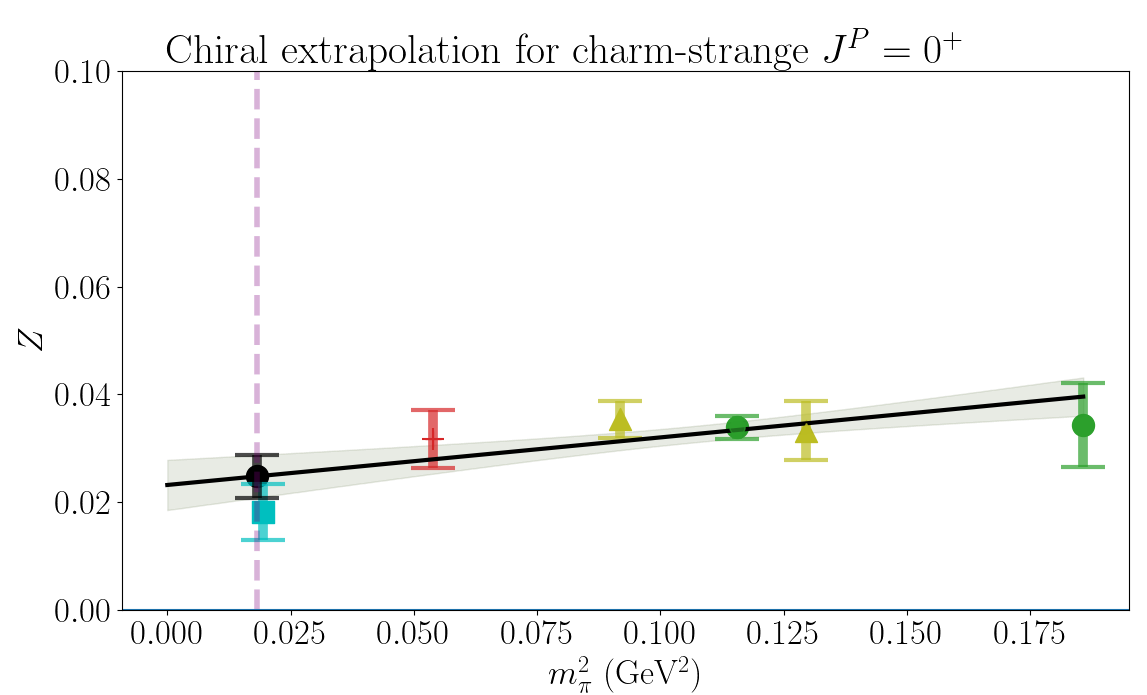}
    \hfill
    \includegraphics[width=0.49\linewidth]{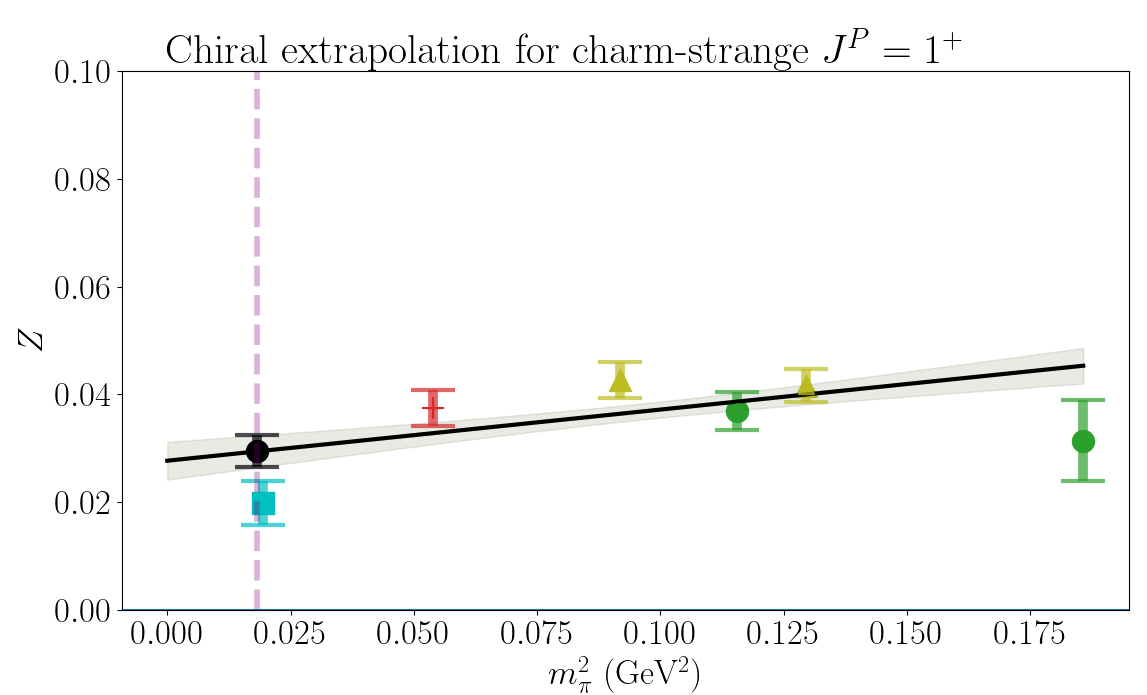}
    
    \includegraphics[width=0.49\linewidth]{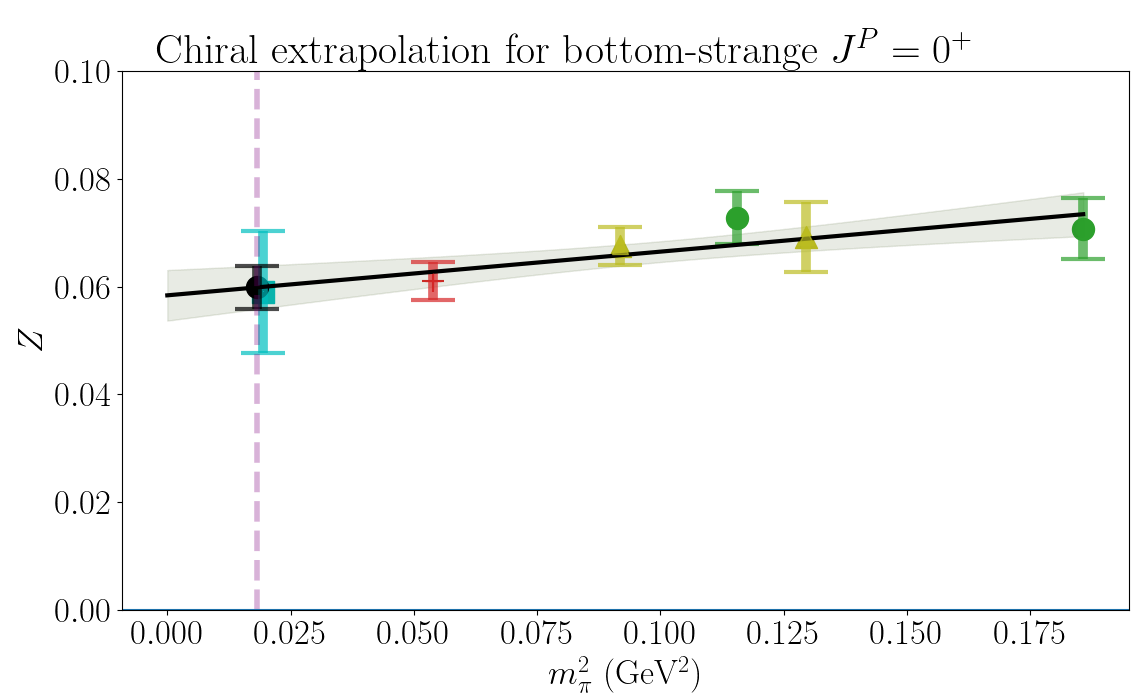}
    \hfill
    \includegraphics[width=0.49\linewidth]{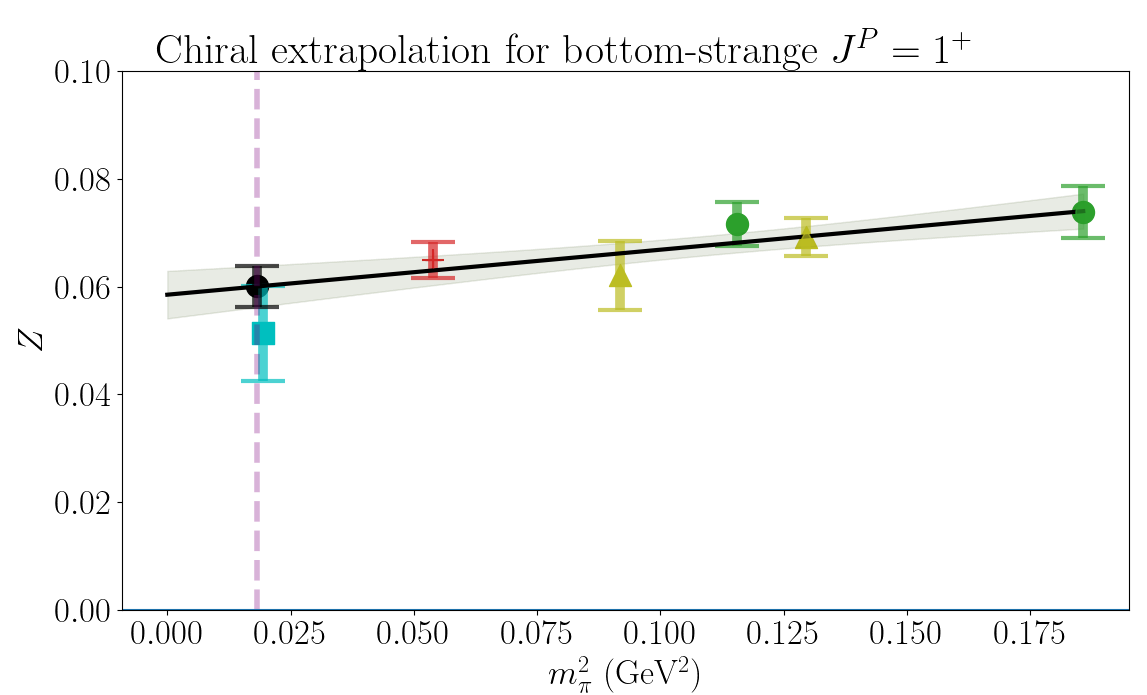}
    \caption{Chiral-only extrapolations of the field-strength renormalizations $Z$ obtained from zeroth-order ERE data. The physical pion mass is indicated by the dashed line, and the black data point is $Z^{(0)}_\text{phys - chiral}$ in Table \protect\ref{tab:gZfinal}.}
    \label{fig:Zchiralonly}
\end{figure}

\begin{figure}
    \includegraphics[width=0.9\linewidth]{positive_parity/chiralcontiumlegend.pdf}
    \centering
    \includegraphics[width=0.49\linewidth]{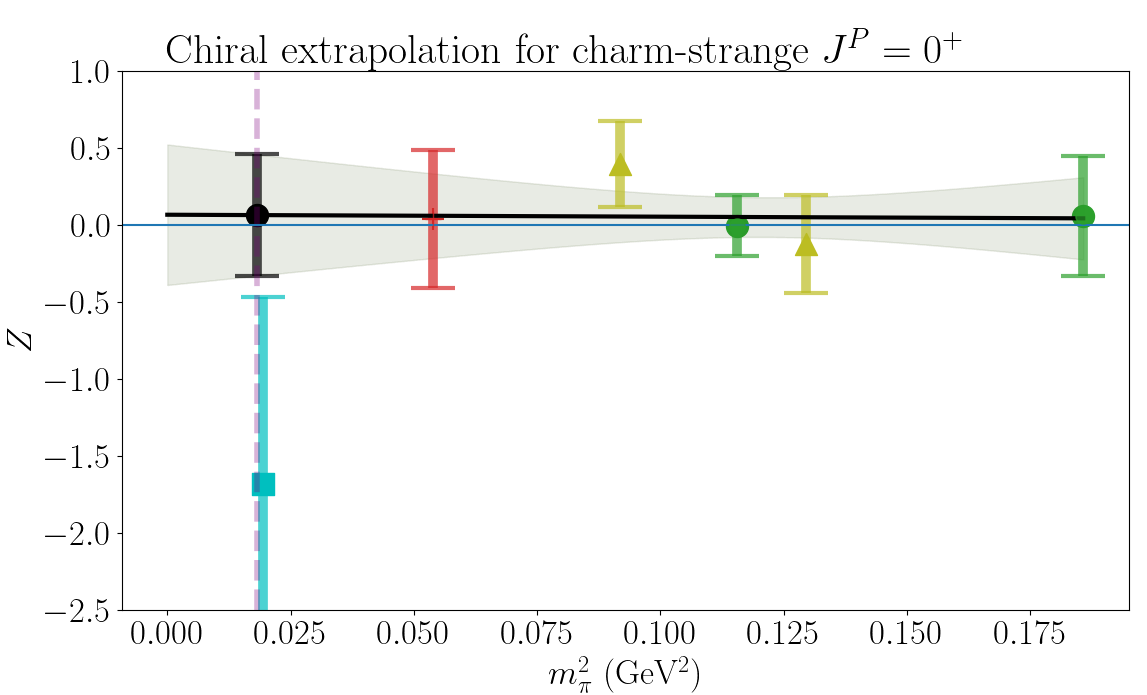}
    \hfill
    \includegraphics[width=0.49\linewidth]{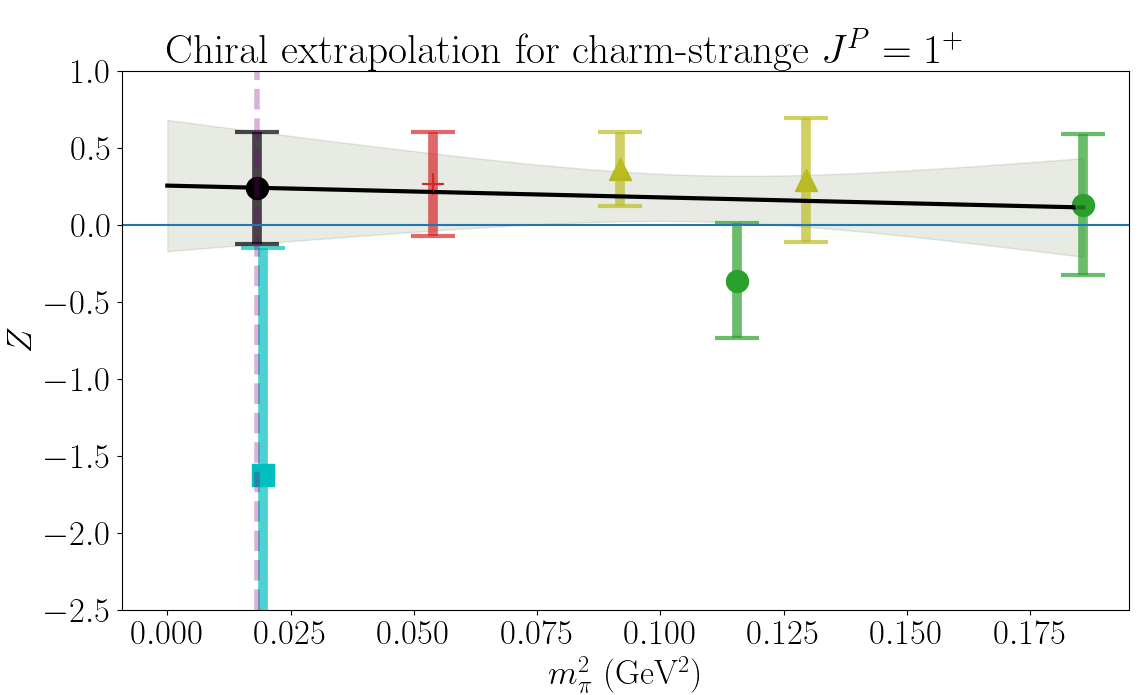}
    
    \includegraphics[width=0.49\linewidth]{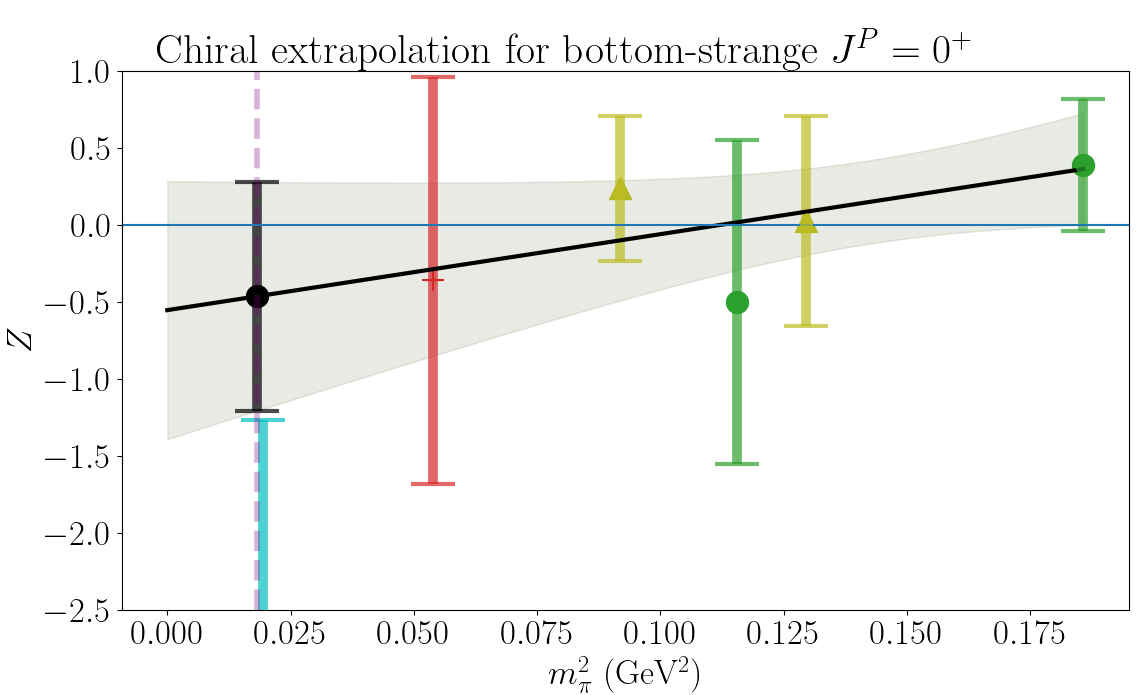}
    \hfill
    \includegraphics[width=0.49\linewidth]{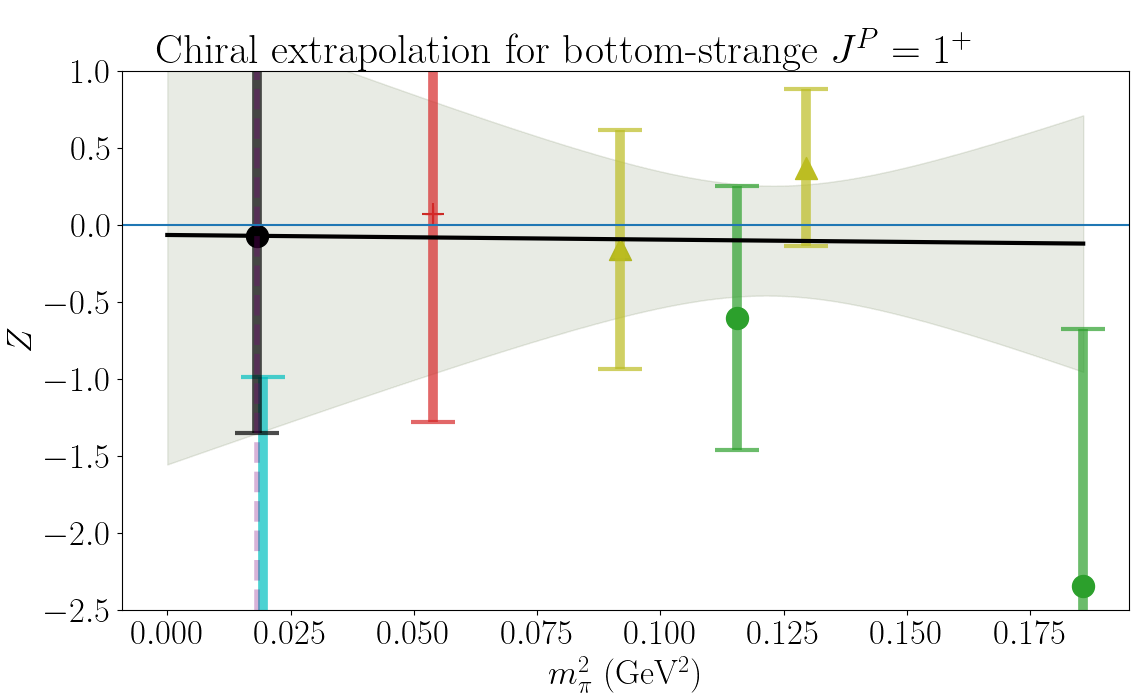}
    \caption{Chiral-only extrapolations of $Z$ obtained from first-order ERE data. The physical pion mass is indicated by the dashed line, and the black data point is $Z^{(1)}_\text{phys - chiral}$ in Table \protect\ref{tab:gZfinal}.}
    \label{fig:Zchiralonlyfirstorder}
\end{figure}

We perform chiral-continuum extrapolations, as well as chiral-only extrapolations, of $g$ and $Z$ using the simple forms
\begin{align}
    Z&=Z_\text{phys}+\frac{C}{(4\pi f_\pi)^2} (m_\pi^2 - m_{\pi \text{phys}}^2)+d\Lambda^2 a^2,\\
    Z&=Z_\text{phys-chiral}+\frac{C}{(4\pi f_\pi)^2} (m_\pi^2 - m_{\pi \text{phys}}^2),\\
    g&=g_\text{phys}+\frac{C}{4\pi f_\pi} (m_\pi^2 - m_{\pi \text{phys}}^2)+d\Lambda^3 a^2,\\
    g&=g_\text{phys-chiral}+\frac{C}{4\pi f_\pi} (m_\pi^2 - m_{\pi \text{phys}}^2).
\end{align}
The results for both the zeroth and first-order ERE cases are given in Table \ref{tab:gZfinal}, and plots of the chiral-only extrapolations of $Z$ are shown in Figs.~\ref{fig:Zchiralonly} and \ref{fig:Zchiralonlyfirstorder}.
The fit parameters $d$ always came out consistent with zero, so we expect that the chiral-only extrapolations also give reasonable estimates of the physical-limit parameters.
Dropping the unphysical C00078 data points for $Z^{(1)}$ and $g^{(1)}$ from the extrapolations changes the results less than 1$\sigma$.

Another potential probe of molecular structure of mesons is through their decay constants. Since molecular states are less spatially localized than $\bar{q} q$-states, we expect $f_{H_{sJ}}/f_{H_s} < 1$ for a molecular meson $H_{sJ}$. We find these ratios to be $\frac{f_{D^*_{s0}}}{f_{D_s}}\approx 0.55$, $\frac{f_{D_{s1}}}{f_{D^*_s}}\approx 0.75$, $\frac{f_{B^*_{s0}}}{f_{B_s}}\approx 0.88$, and $\frac{f_{B_{s1}}}{f_{B^*_s}}\approx 0.84$. It is notable that the hierarchy of these ratios broadly tracks how deeply bound each of the $H_{sJ}$ particles are, as one would expect for molecular states. However, the general observation that $\frac{H_{sJ}}{H_s^{(*)}}<1$ may also be related to the fact that we are comparing states that in the quark model are $P$-wave and $S$-wave. As such, these decay-constant ratios are certainly not a definitive indication of molecular structure.

Overall, the results in this section hint at a considerable molecular content for the positive-parity heavy-strange mesons studied in this work, most significantly for the $D^*_{s0}$.

\FloatBarrier
\section{Conclusions}
\label{sec:conclusions}

We have presented a comprehensive lattice-QCD study of the lowest-lying charm-strange and bottom-strange scalar, pseudoscalar, vector, and axial-vector mesons, giving results for decay constants, masses, and compositeness parameters. Our calculation uses seven different ensembles, including one with a near-physical pion mass, and all results are extrapolated to the continuum limit and physical pion mass.

\begin{figure}

\null\hspace{0.13\linewidth}\includegraphics[height=0.2\textheight]{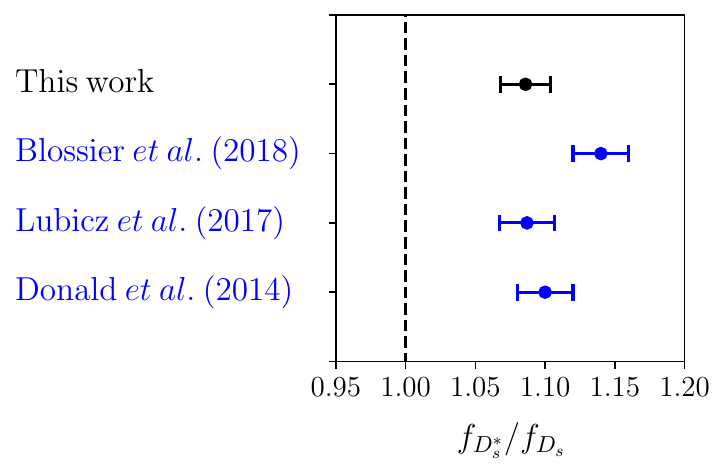} 

\includegraphics[height=0.2\textheight]{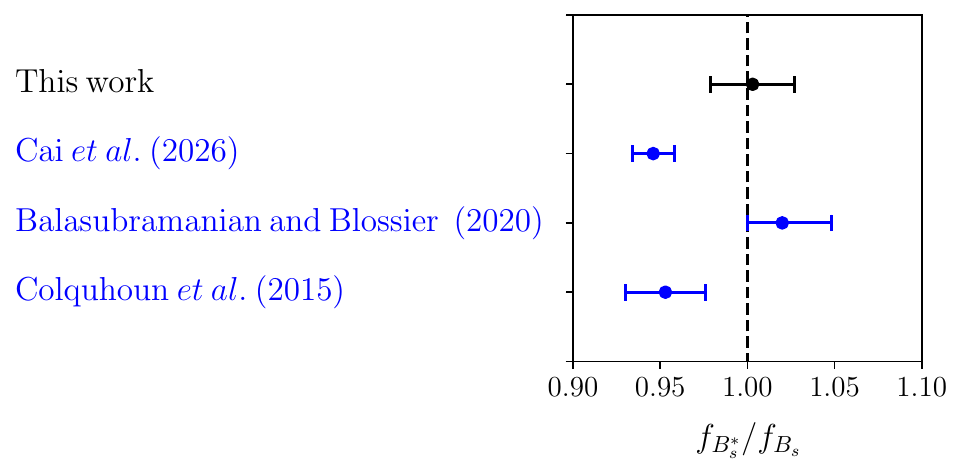} 
\caption{\label{fig:negparratiocomp}Comparison of our lattice-QCD results for $f_{D_s^*}/f_{D_s}$ and $f_{B_s^*}/f_{B_s}$ with those from Refs.~\cite{Donald:2013sra,Colquhoun:2015oha,Lubicz:2017asp,Blossier:2018jol,Balasubramamian:2019wgx,Cai:2026xja}. }
\end{figure}

The negative-parity heavy-strange meson masses were used to tune the heavy-quark actions, and we predict the decay constants. Our results for $f_{D_s}$ and $f_{B_s}$ are compatible with the FLAG world averages \cite{FlavourLatticeAveragingGroupFLAG:2024oxs}. A comparison of our results for the ratios $f_{D_s^*}/f_{D_s}$ and $f_{B_s^*}/f_{B_s}$ with previous lattice calculations is shown in Fig.~\ref{fig:negparratiocomp}. While all calculations clearly give $f_{D_s^*}/f_{D_s}>1$, we find $f_{B_s^*}/f_{B_s}$ to be consistent with 1, and in some tension with the lower values predicted by Refs.~\cite{Colquhoun:2015oha,Cai:2026xja}.

\begin{figure}
    \centering
    \vspace{-0.5cm}
    \includegraphics[width=\textwidth]{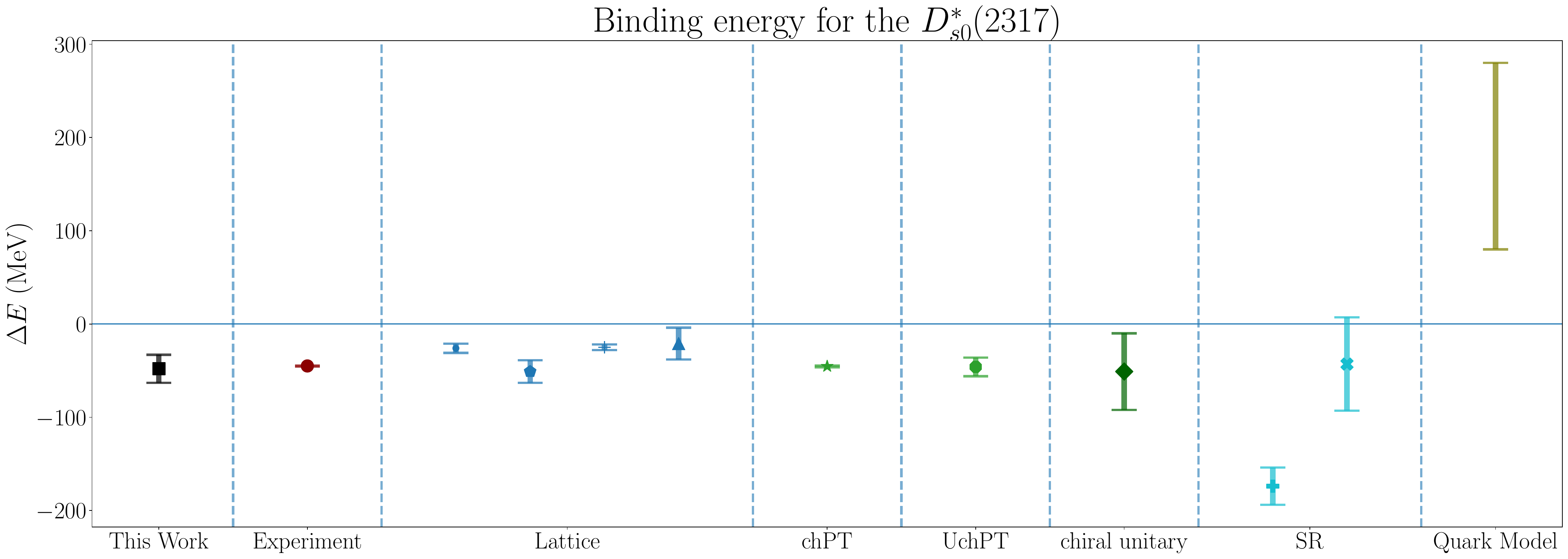}
        \includegraphics[width=\textwidth]{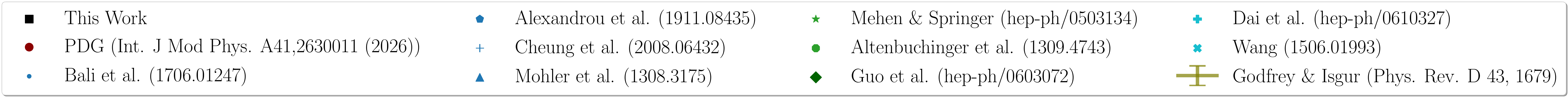}

    \vspace{4ex}
        
    \includegraphics[width=\textwidth]{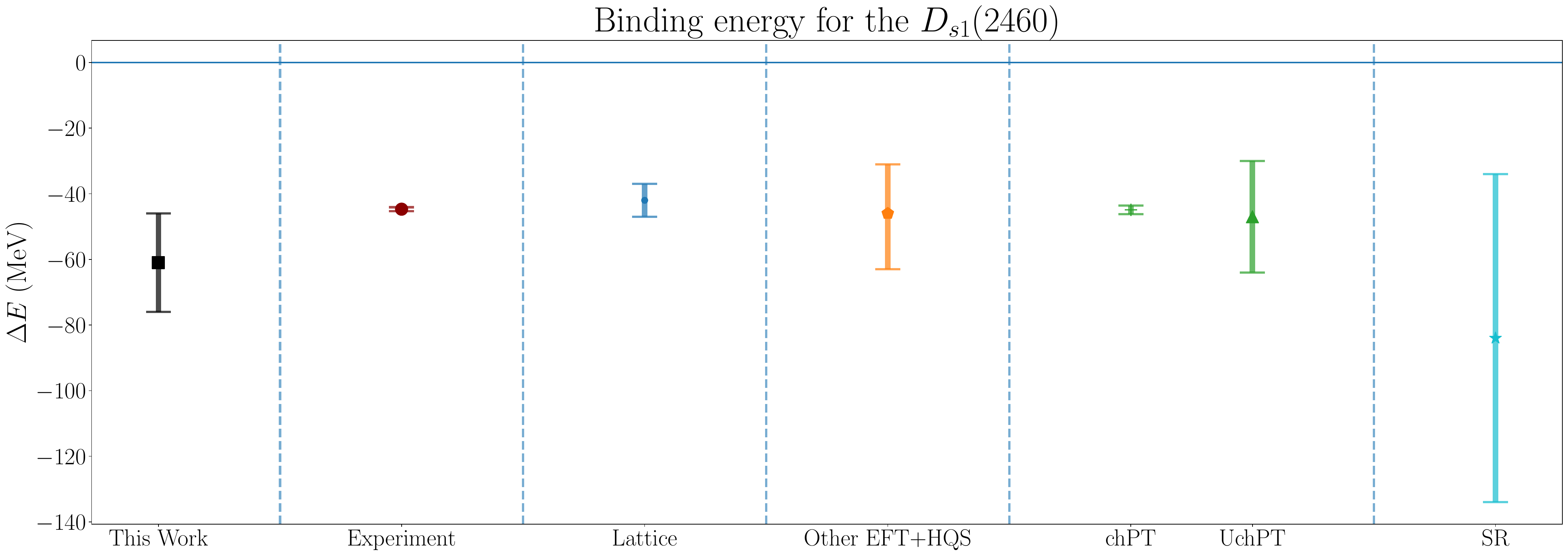}
            \includegraphics[width=\textwidth]{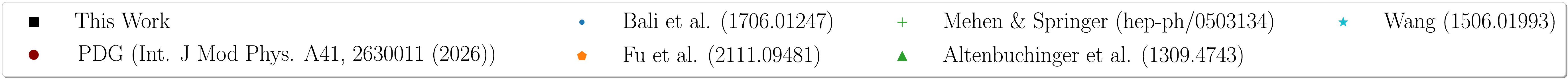}
    \caption{Comparison of our results for the $D_{s0}^*$ and $D_{s1}$ binding energies to the literature \cite{Alexandrou:2019tmk, Bardeen:2003kt, Burch:2008qx, Cheung:2020mql, Cleven:2010aw, Dai:2006uz, DiPierro:2001dwf, Gregory:2010gm, Mehen:2005hc, Wang:2015mxa, Yang:2022vdb, bali, lang, latticeDs, mohler,Hudspith:2026vxt, Alhakami:2020vil,Altenbuchinger:2013vwa,Badalian:2007yr,Cheng:2017oqh,Colangelo:2012xi,Fu:2021wde,Guo:2006fu,Guo:2021rjv,Wurtz:2015mqa}. When a work reports multiple sources of uncertainty, the uncertainties are combined in quadrature, and when only masses are given, we subtract the experimental isospin-averaged thresholds.}
    \label{fig:litcompED}
\end{figure}

\begin{figure}
    \vspace{-0.5cm}
    \centering
    \includegraphics[width=\textwidth]{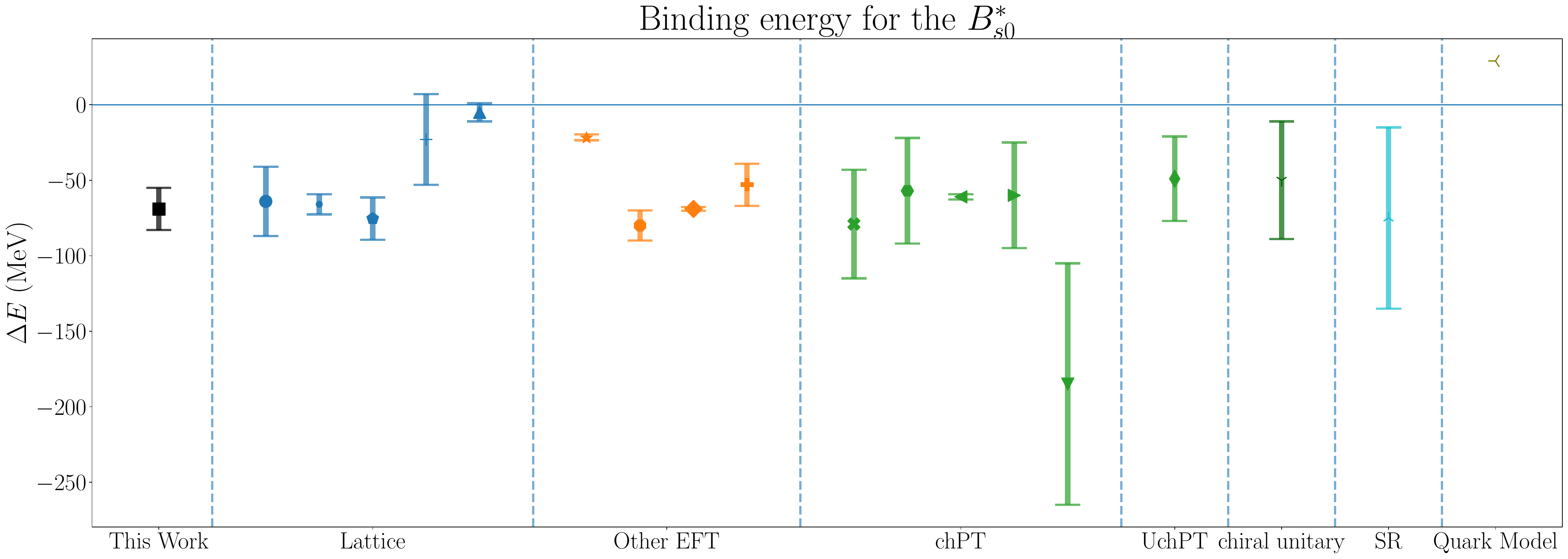}
    \includegraphics[width=\textwidth]{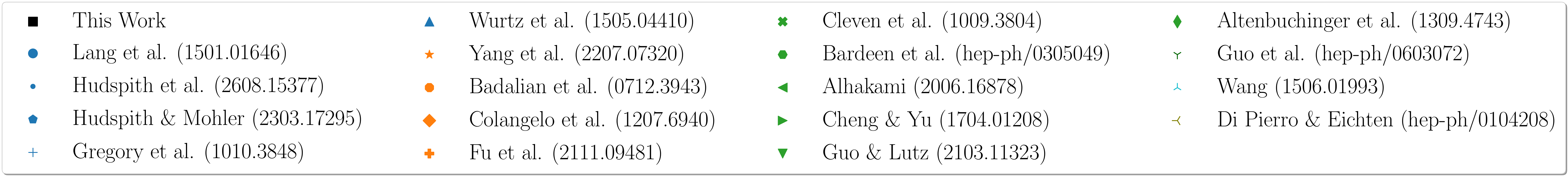}

\vspace{4ex}
    
    \includegraphics[width=\textwidth]{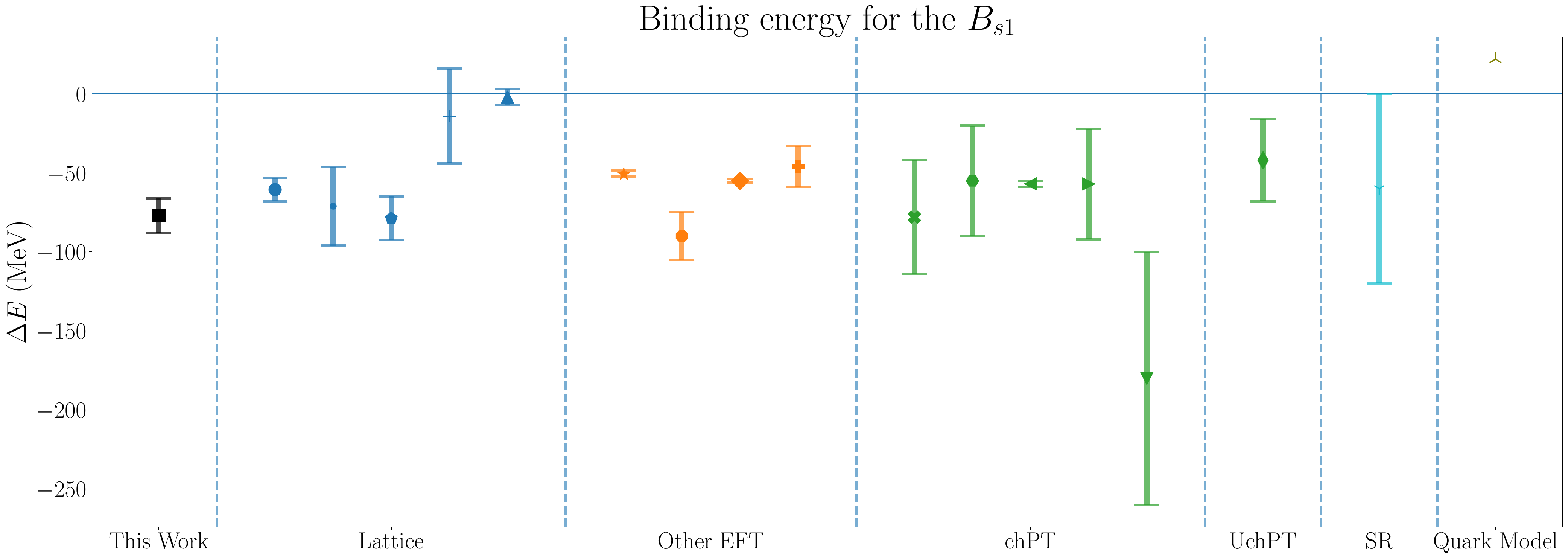}
    \includegraphics[width=\textwidth]{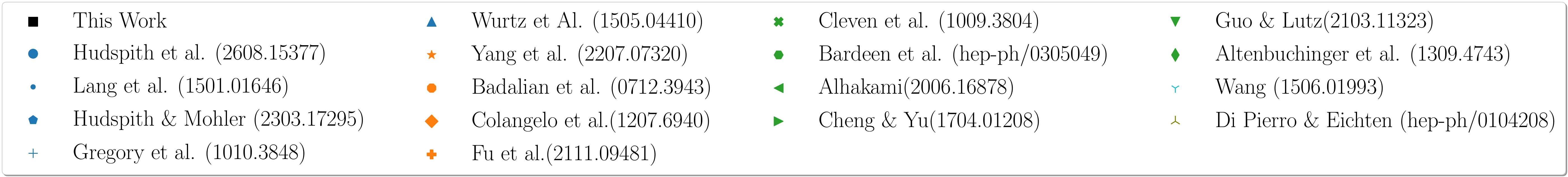}
    \caption{Comparison of our results for the $B_{s0}^*$ and $B_{s1}$ binding energies to the literature \cite{Alexandrou:2019tmk, Bardeen:2003kt, Burch:2008qx, Cheung:2020mql, Cleven:2010aw, Dai:2006uz, DiPierro:2001dwf, Gregory:2010gm, Mehen:2005hc, Wang:2015mxa, Yang:2022vdb, bali, lang, latticeDs, mohler,Hudspith:2026vxt, Alhakami:2020vil,Altenbuchinger:2013vwa,Badalian:2007yr,Cheng:2017oqh,Colangelo:2012xi,Fu:2021wde,Guo:2006fu,Guo:2021rjv,Wurtz:2015mqa}. When a work reports multiple sources of uncertainty, the uncertainties are combined in quadrature, and when only masses are given, we subtract the experimental isospin-averaged thresholds.}
    \label{fig:litcompEB}
\end{figure}

In the positive-parity sector, we used bases of interpolating fields containing $q\bar{q}$-operators, derivative operators, and meson-meson scattering-type operators.
This allowed the precise resolution of the near-threshold ground states $D^*_{s0}$,  $D_{s1}$, $B^*_{s0}$, and $B_{s1}$. 
The ground-state binding energies were corrected for finite-volume effects via the L\"uscher method prior to performing the chiral-continuum extrapolations, although we neglected left-hand cuts due to two-pion exchange. Our final results for the binding energies at the physical point are compared to results from the literature in 
Figs.~\ref{fig:litcompED}-\ref{fig:litcompEB}. In the charm sector, our results are consistent with the experimental values \cite{ParticleDataGroup:2026mpi}. In the bottom sector, our results for the binding energies are consistent with previous lattice results, except Ref.~\cite{Wurtz:2015mqa}. Additionally, our predicted hyperfine splittings are broadly consistent with expectations from HQET.

\begin{figure}
    \centering
    \vspace{-0.5cm}
    \includegraphics[width=\textwidth]{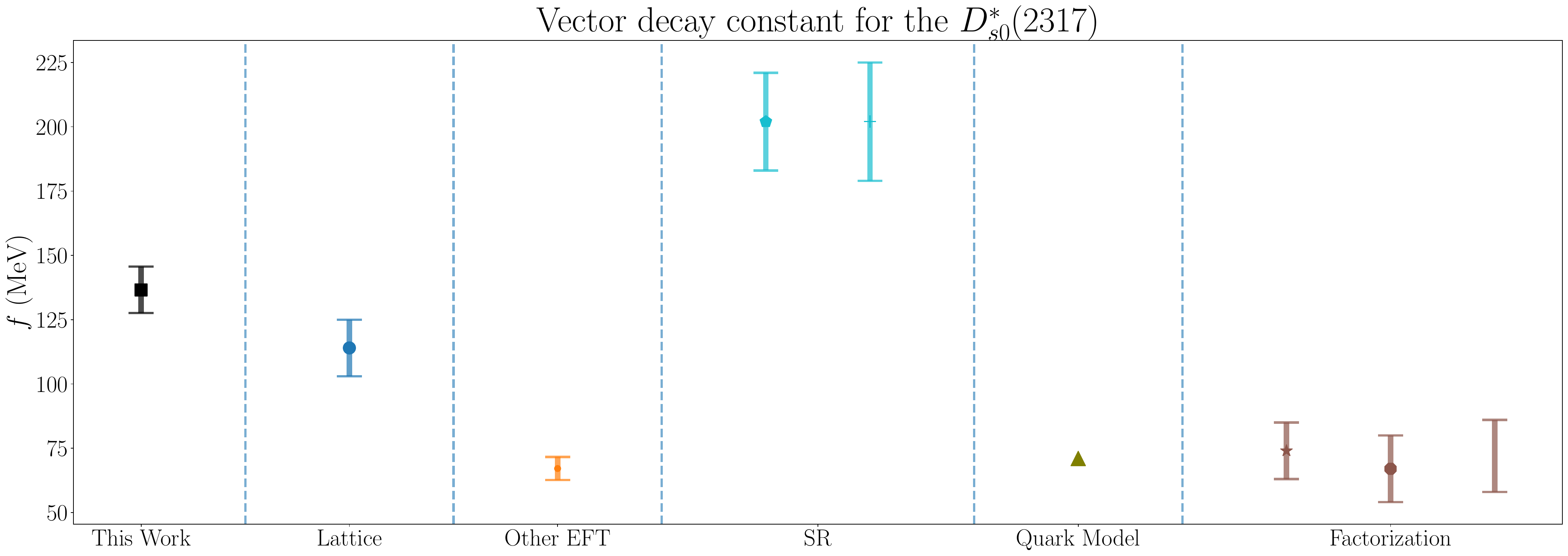}
        \includegraphics[width=\textwidth]{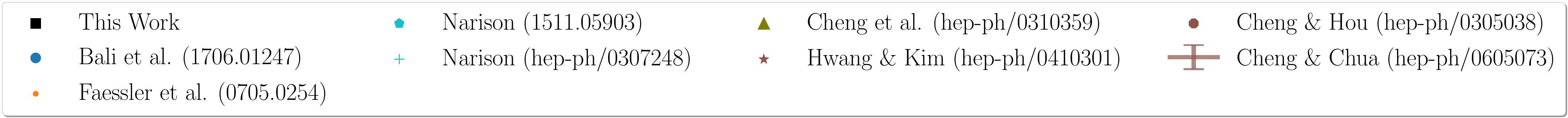}
    
    \vspace{4ex}
    \includegraphics[width=\textwidth]{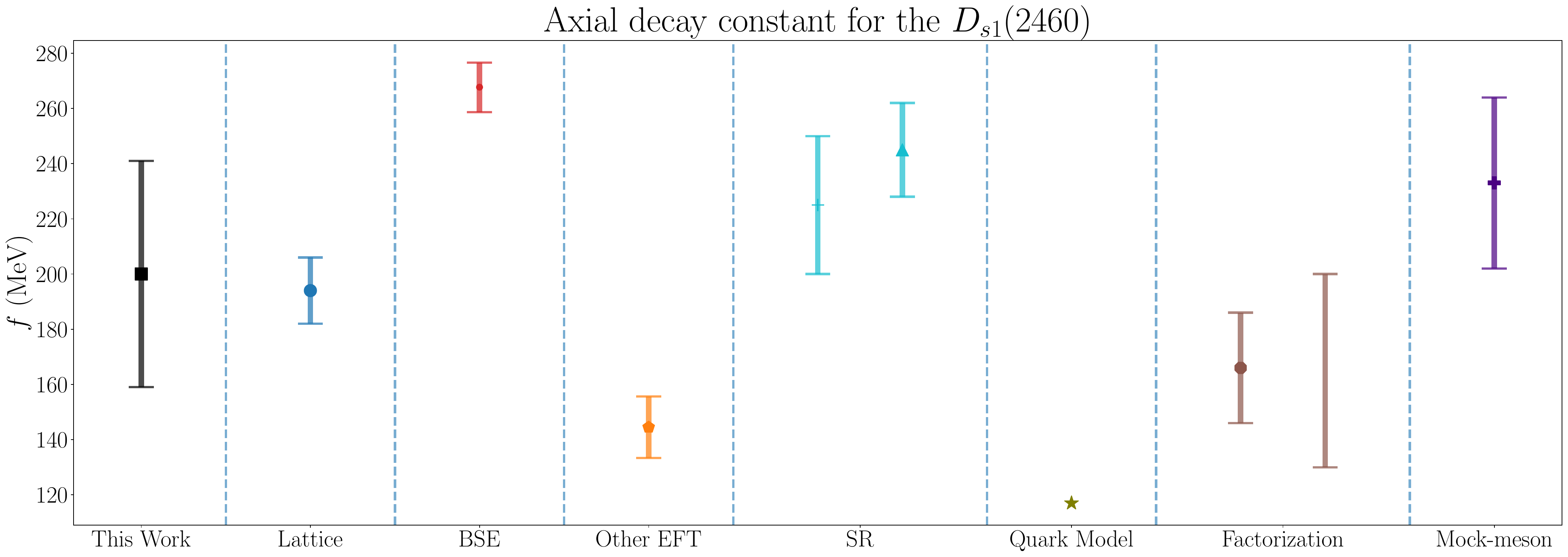}
        \includegraphics[width=\textwidth]{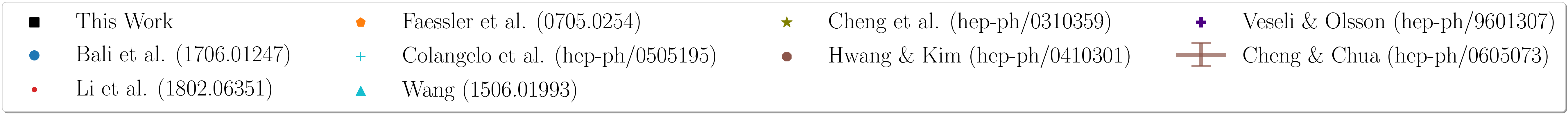}
    \caption{Comparison of our results for the $D_{s0}^*$ and $D_{s1}$ decay constants to the literature \cite{Bardeen:2003kt, Cheng:2003kg, Cheng:2003sm, Cheng:2006dm, Colangelo:2005hv, DiPierro:2001dwf, Hwang:2004kga, Li:2018eqc, Mehen:2005hc, Narison:2003td, Pullin:2021ebn, Veseli:1996kn, Wang:2007tu, Wang:2015mxa, sr}. When a work reports multiple sources of uncertainty, the uncertainties are combined in quadrature.}
    \label{fig:litcompfD}
\end{figure}

\begin{figure}
    \centering
    \vspace{-0.5cm}
    \includegraphics[width=\textwidth]{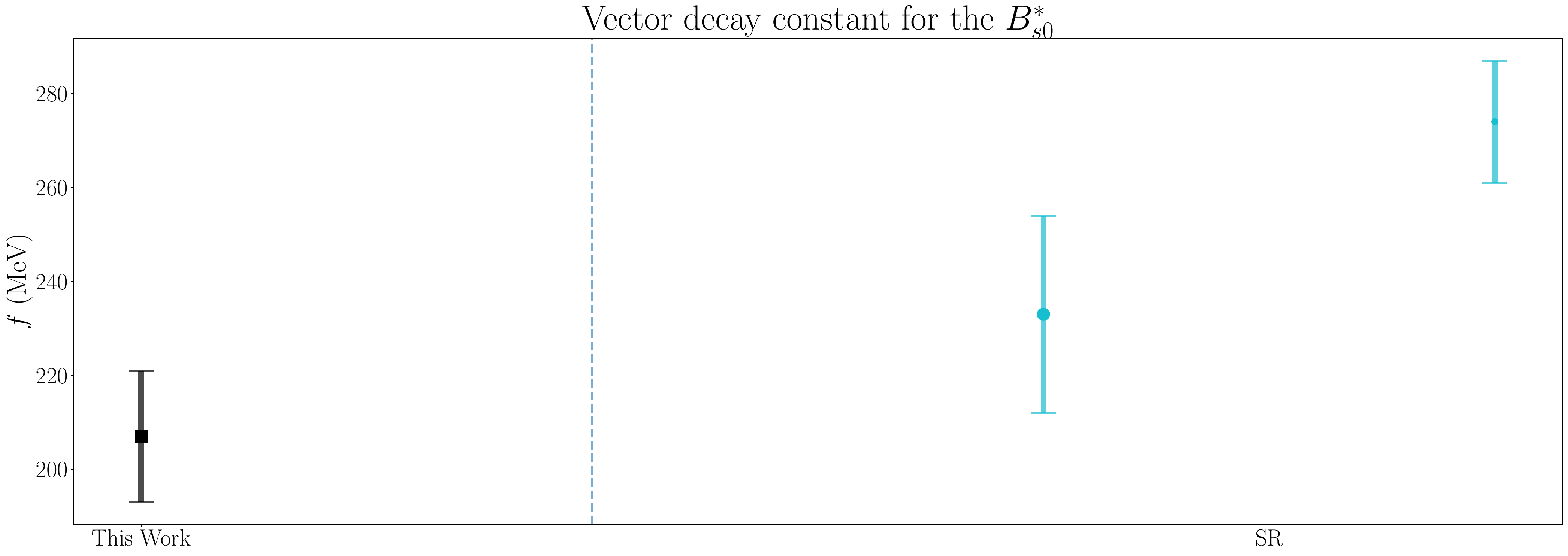}
    \includegraphics[width=0.5\textwidth]{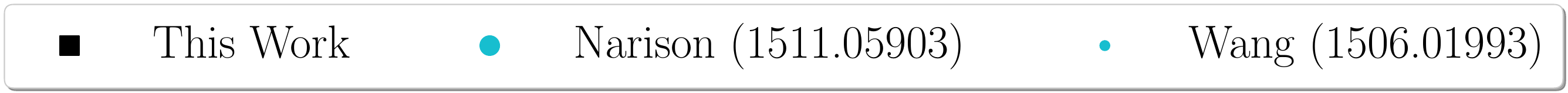}
    
    \vspace{4ex}
    \includegraphics[width=\textwidth]{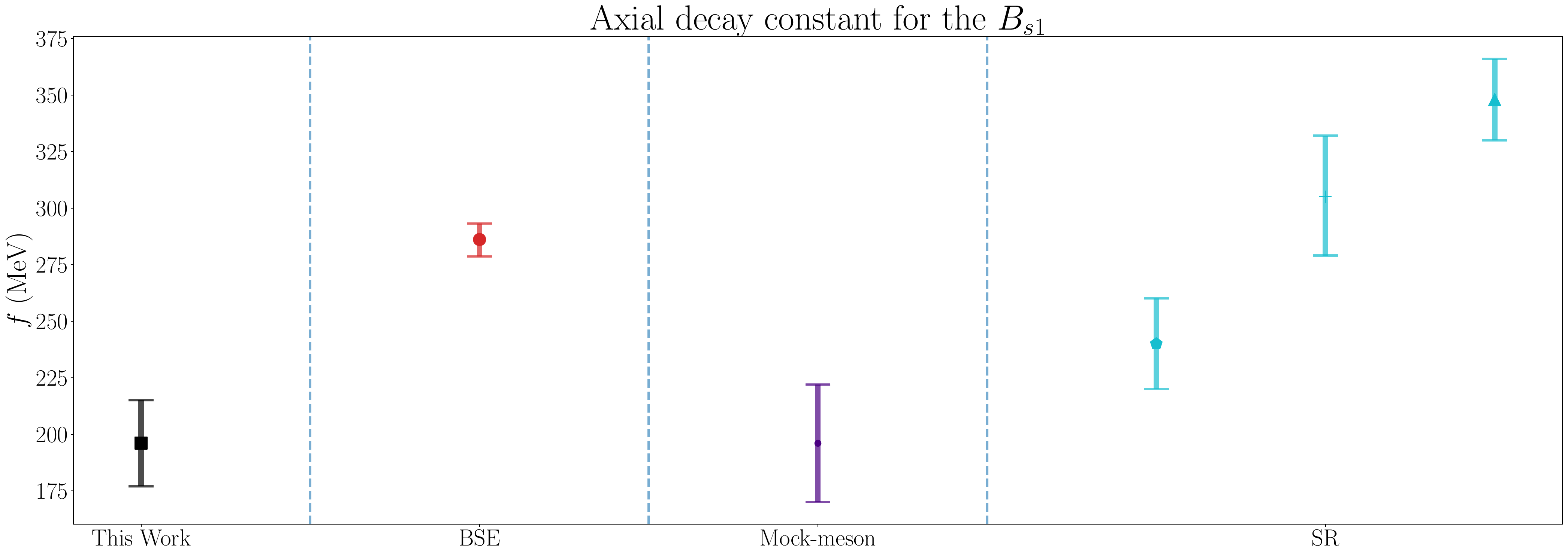}
    \includegraphics[width=\textwidth]{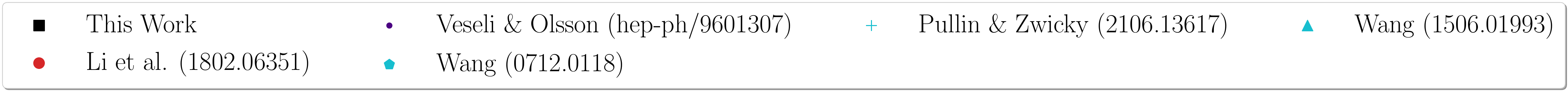}
    \caption{Comparison of our results for the $B_{s0}^*$ and $B_{s1}$ decay constants to the literature \cite{Bardeen:2003kt, Cheng:2003kg, Cheng:2003sm, Cheng:2006dm, Colangelo:2005hv, DiPierro:2001dwf, Hwang:2004kga, Li:2018eqc, Mehen:2005hc, Narison:2003td, Pullin:2021ebn, Veseli:1996kn, Wang:2007tu, Wang:2015mxa, sr}. When a work reports multiple sources of uncertainty, the uncertainties are combined in quadrature.}
    \label{fig:litcompfB}
\end{figure}

Using the results from L\"uscher's method, we also calculated the couplings $g$ of the $H_{sJ}$ states to the $H^{(*)}K$ channels, and used them to estimate the field-strength renormalization $Z$, with $1-Z$ being a measure of compositeness. We found large couplings $g$, effective ranges $r$ consistent with being small and positive, scattering lengths satisfying $\frac{1}{a_0}\sim -\gamma$, and small $Z$ values,  pointing in the direction of molecular components in the $H_{sJ}$ states. However, the results using the first-order ERE have large uncertainties, leaving the situation less clear, in particular for the $B_{s1}$. In addition, corrections to the weak-binding limit are expected to be non-negligible in many cases.

Comparisons of the decay constants of the positive-parity heavy-strange mesons are shown in Figs.~\ref{fig:litcompfD}-\ref{fig:litcompfB}. Our results for $f_{B^*_{s0}}$ and $f_{B_{s1}}$ are the first from lattice QCD. For $f_{D^*_{s0}}$, our result is slightly higher than that of the previous lattice calculation in Ref.~\cite{bali} (which did not include a continuum extrapolation), while our $f_{D_{s1}}$ value has large uncertainty and is consistent with that of Ref.~\cite{bali}. For the charmed mesons, we find a large deviation from the leading-order HQET prediction $f_{D_{s1}}=f_{D^*_{s0}}$, and the hierarchy $f_{D_{s1}}>f_{D^*_{s0}}$ agrees with the previous lattice-QCD calculation of Ref.~\cite{bali}. On the other hand, our results in the bottom sector are consistent with $f_{B_{s1}}=f_{B^*_{s0}}$. When comparing the charm and bottom decay constants, we again find large deviations from the leading-order scaling with the heavy-quark mass predicted by HQET.

\FloatBarrier
\section*{Acknowledgments}

We thank the RBC and UKQCD Collaborations for making their gauge-field ensembles available. This work was supported by the U.S. Department of Energy, Office of Science, Office of High Energy Physics under Award Number DE-SC0009913.
This research used resources of the National Energy Research Scientific Computing Center (NERSC), a U.S.~Department of Energy Office of Science User Facility supported by Contract Number DE-AC02-05CH1123. This research also used resources at Purdue University RCAC and at the University of Texas TACC through the Extreme Science and Engineering Discovery Environment (XSEDE) \cite{XSEDE} and the Advanced Cyberinfrastructure Coordination Ecosystem: Services \& Support (ACCESS) program \cite{10.1145/3569951.3597559}, which are supported by U.S. National Science Foundation grants ACI-154856, 2138259, 2138286, 2138307, 2137603, and 2138296.  We acknowledge the use of USQCD software \cite{USQCD}.

\section*{Data Availability}

Machine-readable files containing the finite-volume decay constants for both the negative and positive-parity heavy-strange mesons, the finite-volume binding energies of the positive-parity heavy-strange mesons, the $H^{(*)}$ and $K$ masses, the effective-range parameters, the couplings $g$, and field-strength renormalization values $Z$ are provided in the Supplemental Material \cite{Supplemental}. The other data are available from the authors upon reasonable request.

\section*{Note Added}

At the 43rd International Conference on High Energy Physics, a preliminary analysis by the LHCb collaboration finding a narrow bottom-strange state with a mass of $5698.9 \pm 1.6$ MeV was presented \cite{LHCbICHEP2026}. The mass difference from the isospin-averaged $BK$ threshold, $-76.3\pm1.6$ MeV, is consistent with our result for the $B_{s0}^*$.

\appendix
\FloatBarrier

\section{Considering the Sub-matrix of $C^{lm}$ without Meson-Meson Operators}
\label{sec:symcorr}

It is interesting to compare our results for the spectra and decay constants to those obtained by multi-exponential matrix fits of the symmetrized $C^{lm}(t)$ when dropping the meson-meson operators ($l,m=4$),
that is, simultaneous fits of the elements
\begin{equation}
    \begin{pmatrix}
        C^{11}(t) & \\
        C^{21}(t) & C^{22}(t)\\
        C^{31}(t) & C^{32}(t)
    \end{pmatrix}.\label{eq:matrixfit}
\end{equation}
Recall that we use the convention where $l=3$ corresponds to the fully $\mathcal{O}(a)$-improved current, which is included only at the sink. For $J=0$, we perform single-exponential fits at large $t$ using
\begin{equation}
C^{lm}(t)=A_l A_m \:e^{-Et},
\end{equation}
with fit parameters $A_1$, $A_2$, $A_3$, and $E$. For $J=1$, we instead perform two-exponential fits of the form 
\begin{equation}
    C^{lm}(t)=A_l A_m \left(e^{-Et} + B_{l} B_{m} e^{-(E+\Delta E)t}\right) \label{eq:symcorr}
\end{equation}
with additional parameters $B_1$, $B_2$, $B_3$, $\Delta E$ to account for the low-lying spin-singlet excited state expected from the quark model. Without the meson-meson operators, it is not feasible to fit a third (second) state for $J=1$ ($J=0$).

The decay constants are then obtained as
\begin{equation}
f=A_3 \sqrt{\frac{2}{E}}.
\end{equation}
The above procedure is what we termed the ``second type'' of analysis in our Lattice 2024 conference proceedings Ref.~\cite{proceedings}. Unfortunately, the results previously presented in Ref.~\cite{proceedings} were erroneous due to an error in forward-backward averaging the correlators in time. The corrected results are provided here, following the same methods as in Ref.~\cite{proceedings}, and are also included as an erratum in the latest arXiv version of Ref.~\cite{proceedings}.

\begin{figure}
    \centering
    \includegraphics[width=0.44\linewidth]{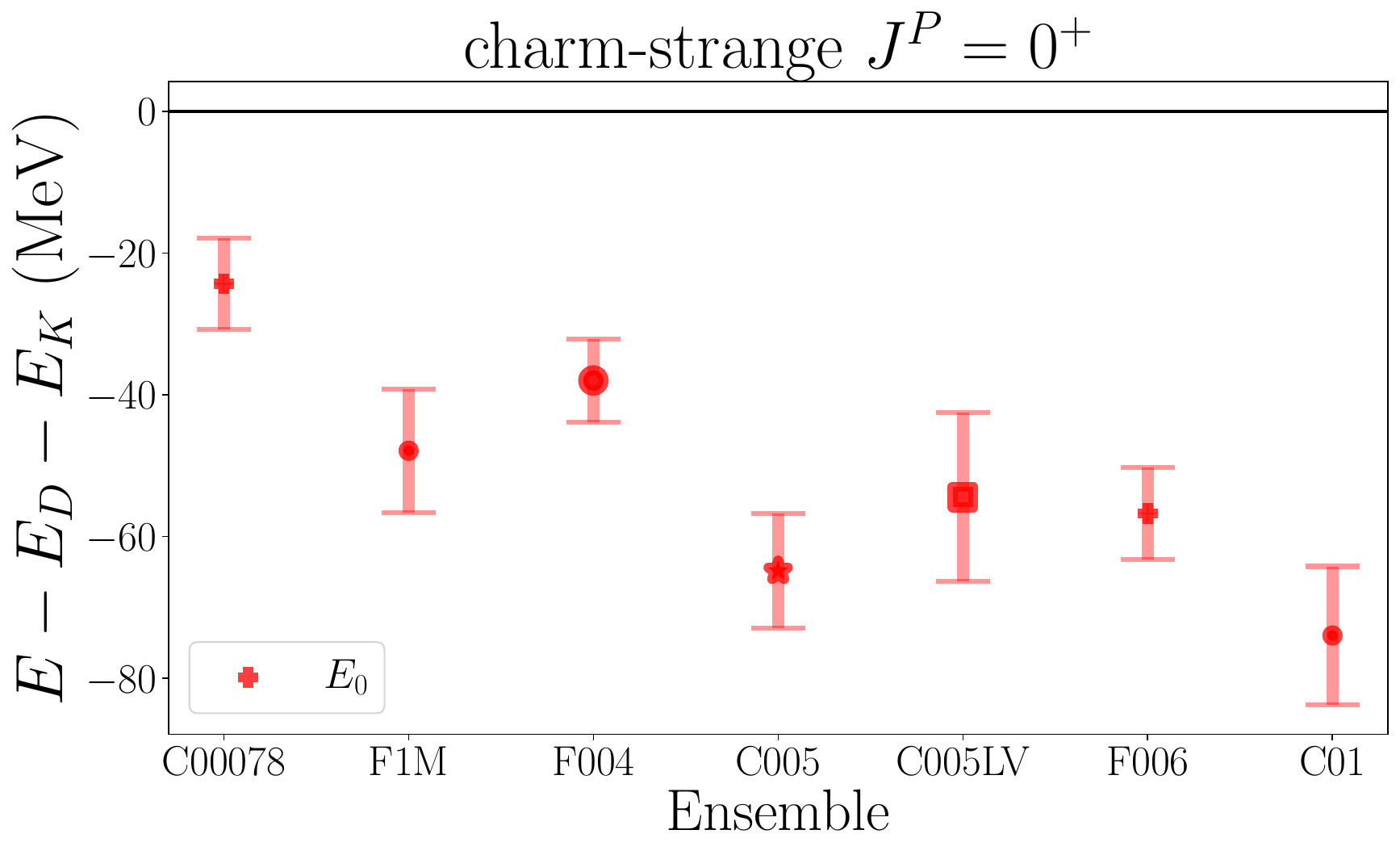}
    \hfill
    \includegraphics[width=0.44\linewidth]{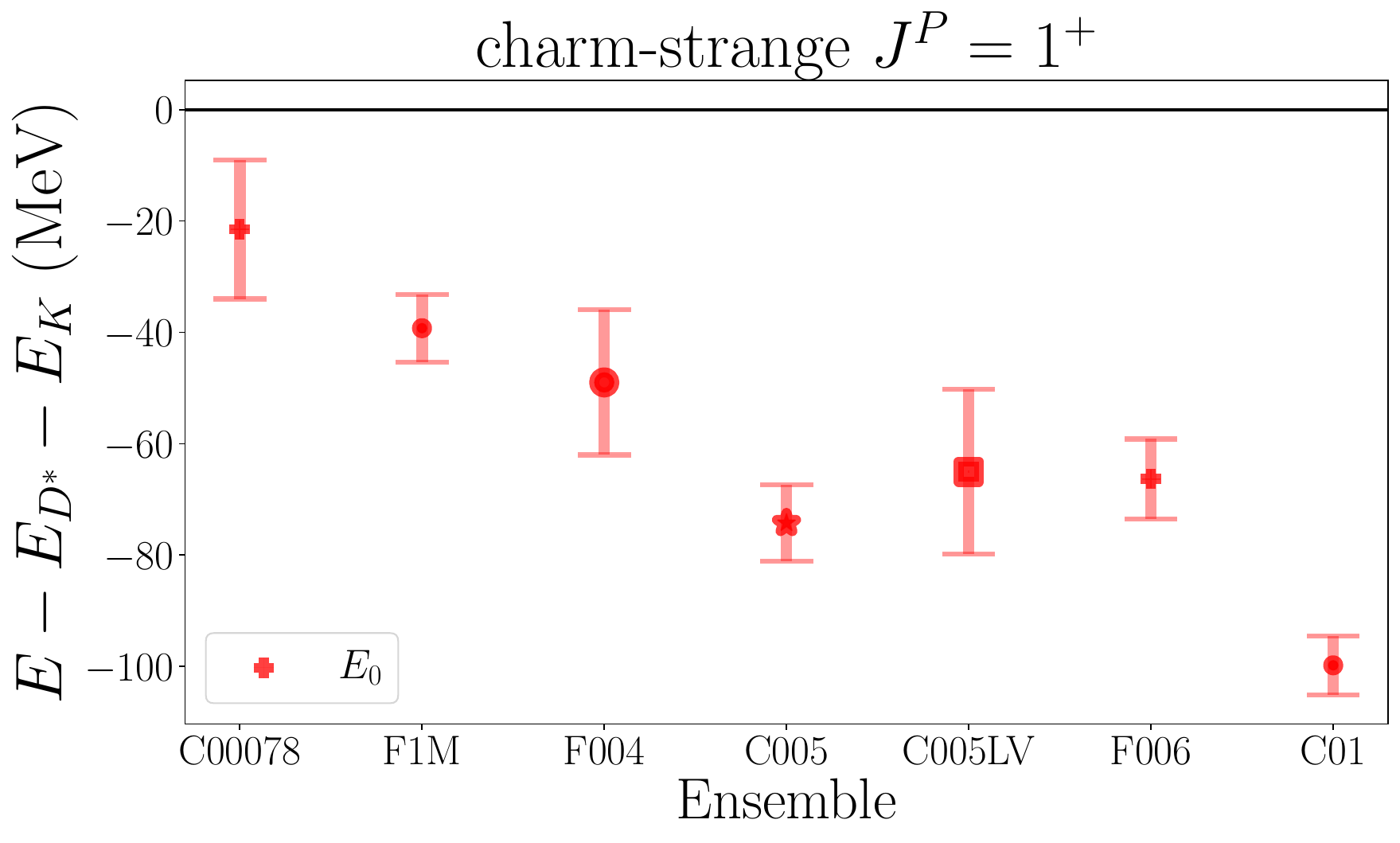}
    
    \includegraphics[width=0.44\linewidth]{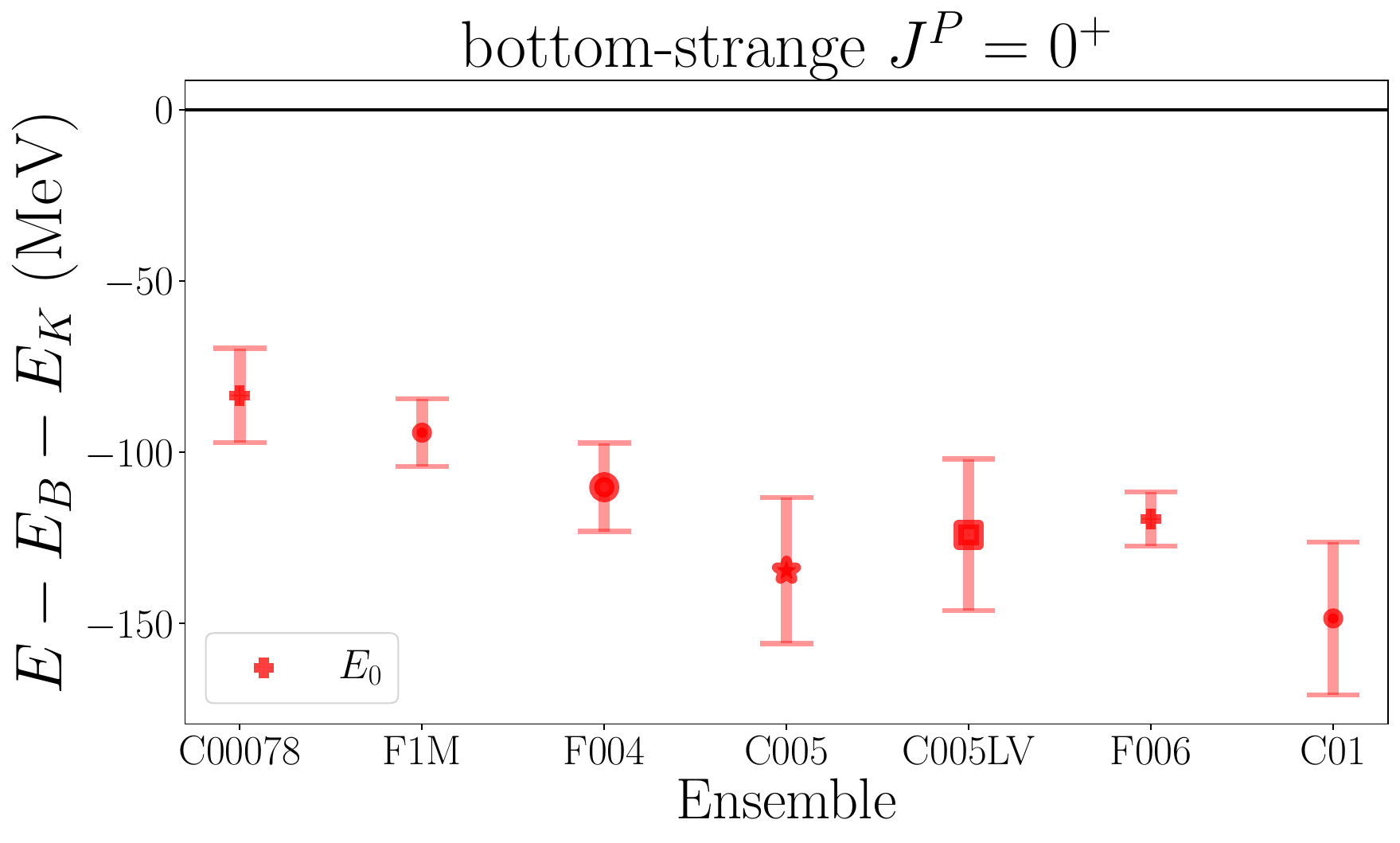}
    \hfill
    \includegraphics[width=0.44\linewidth]{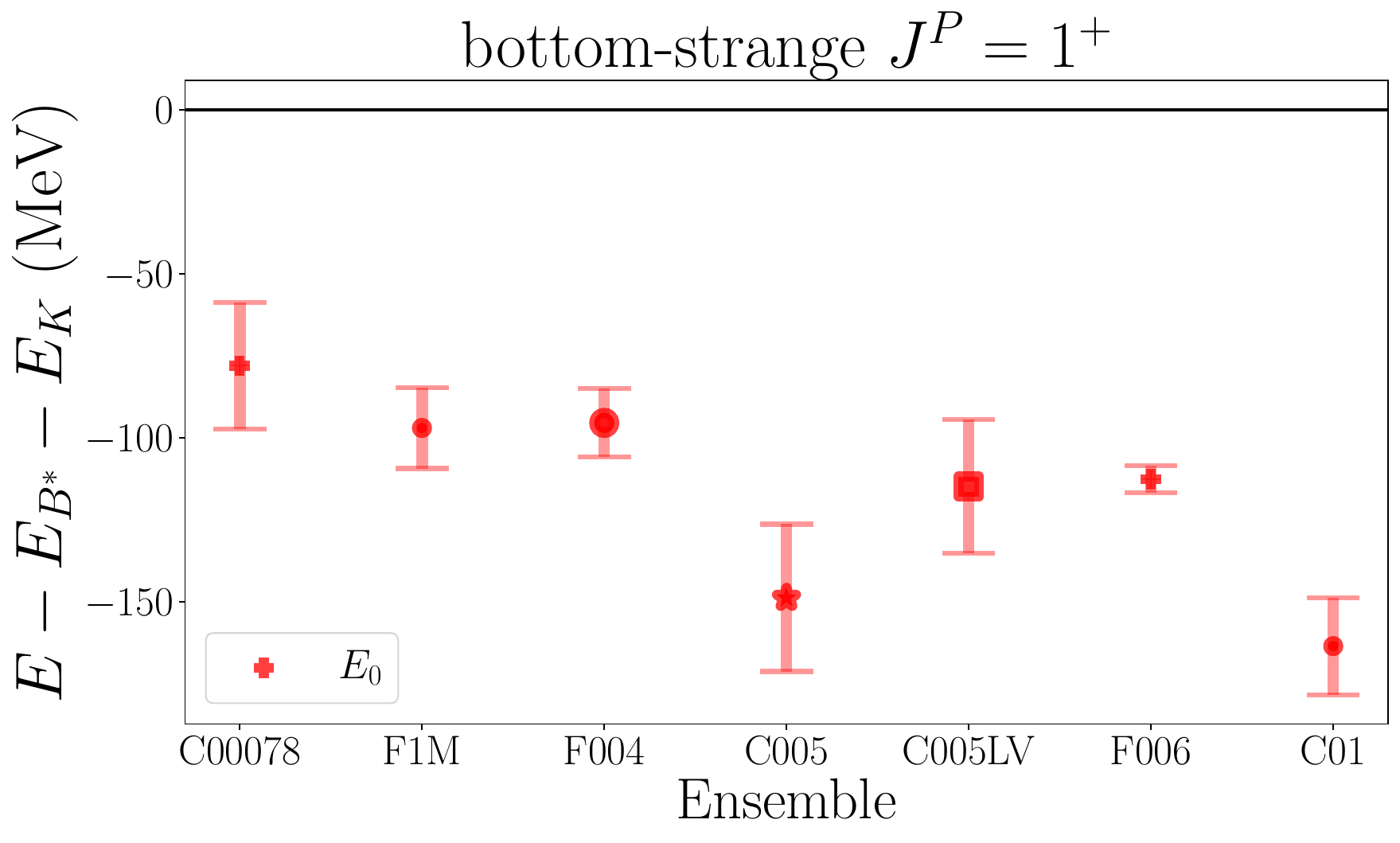}
    \caption{Ground-state finite-volume binding energies of the $H_{sJ}$ positive-parity heavy-strange mesons relative to the $H^{(*)} K$ thresholds from the symmetrized matrix fits of Eqs.~(\ref{eq:matrixfit}-\ref{eq:symcorr}).}
    \label{fig:symcorrE}
\end{figure}

\begin{figure}
    \centering
    \includegraphics[width=0.44\linewidth]{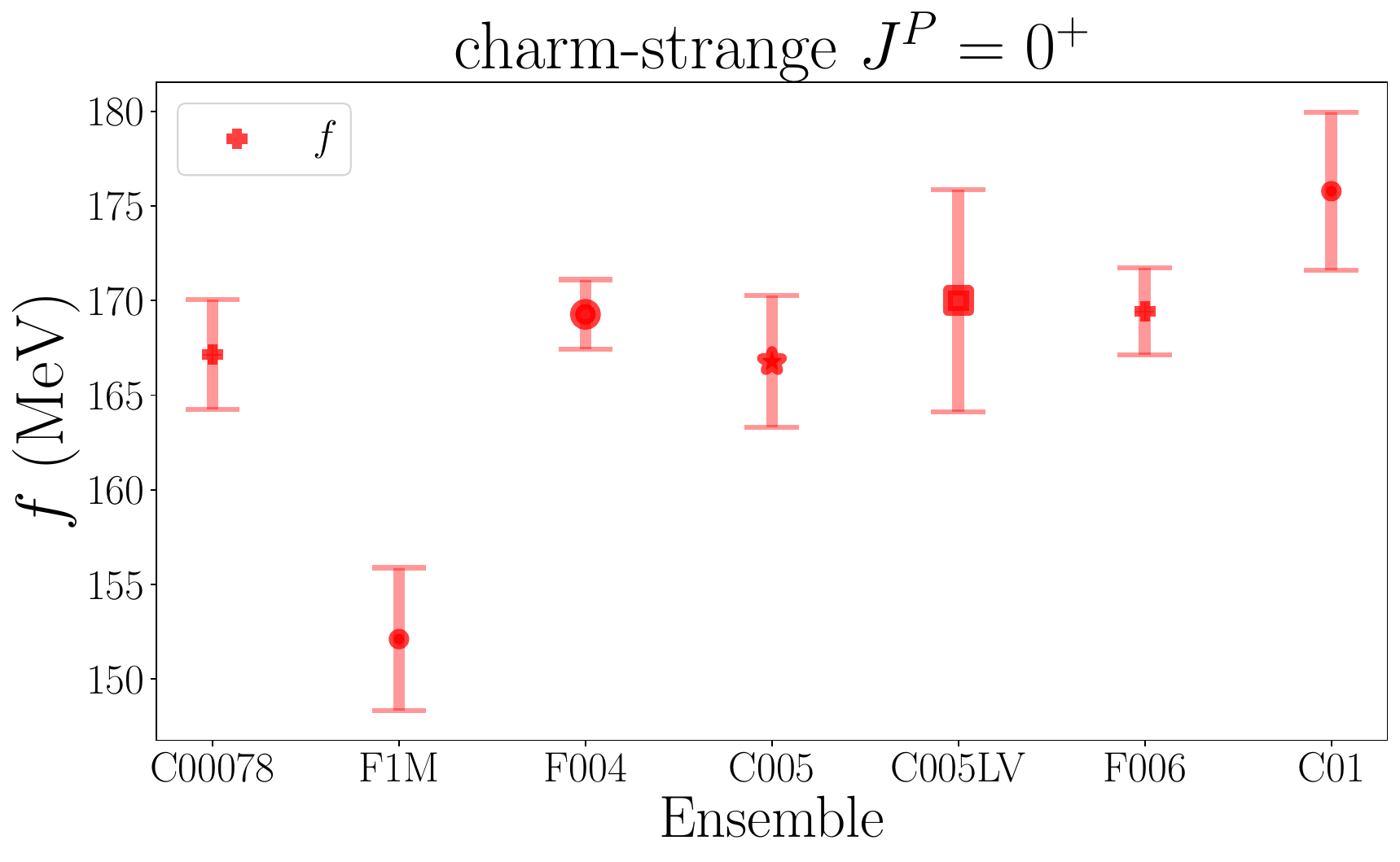}
    \hfill
    \includegraphics[width=0.44\linewidth]{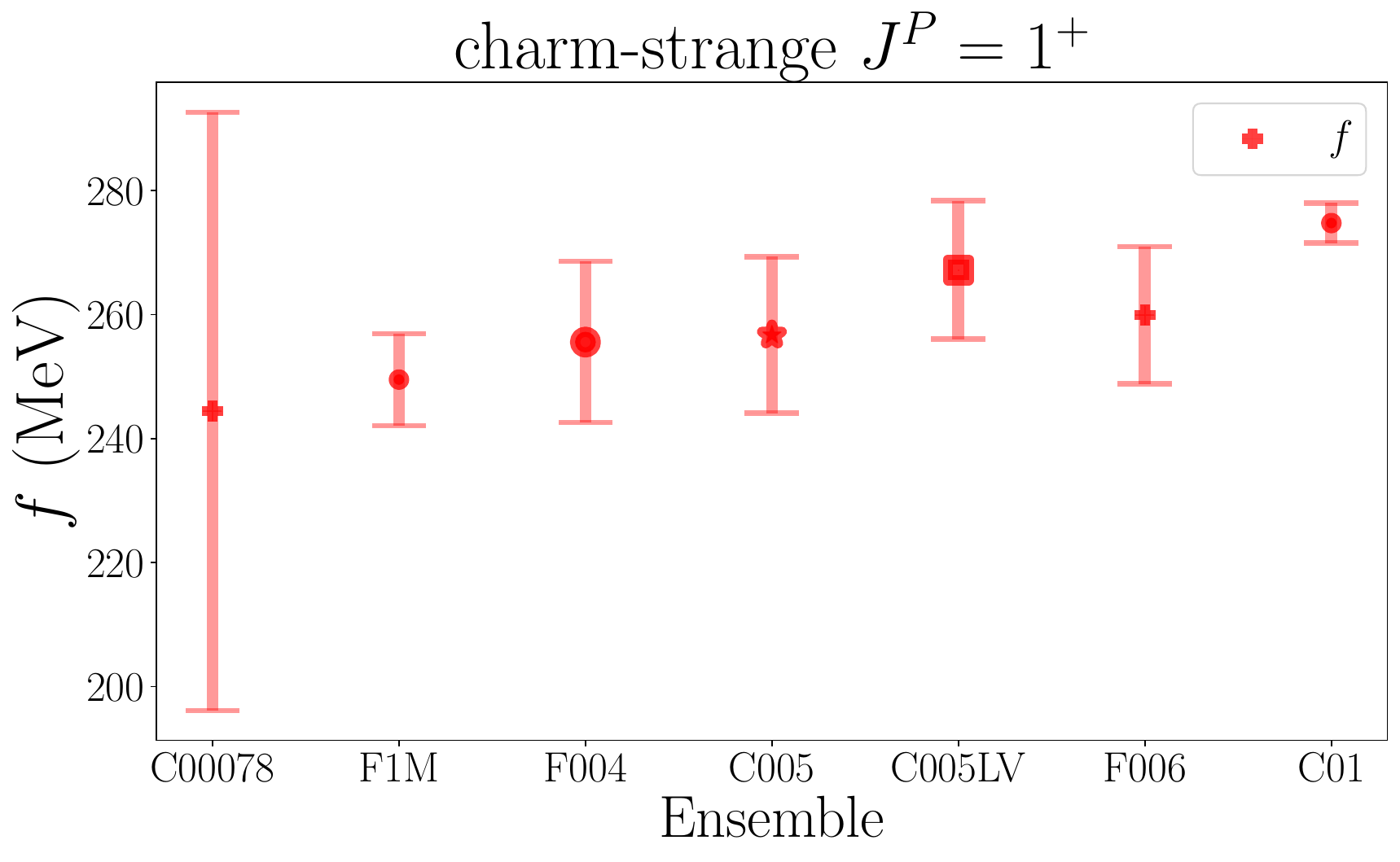}
    
    \includegraphics[width=0.44\linewidth]{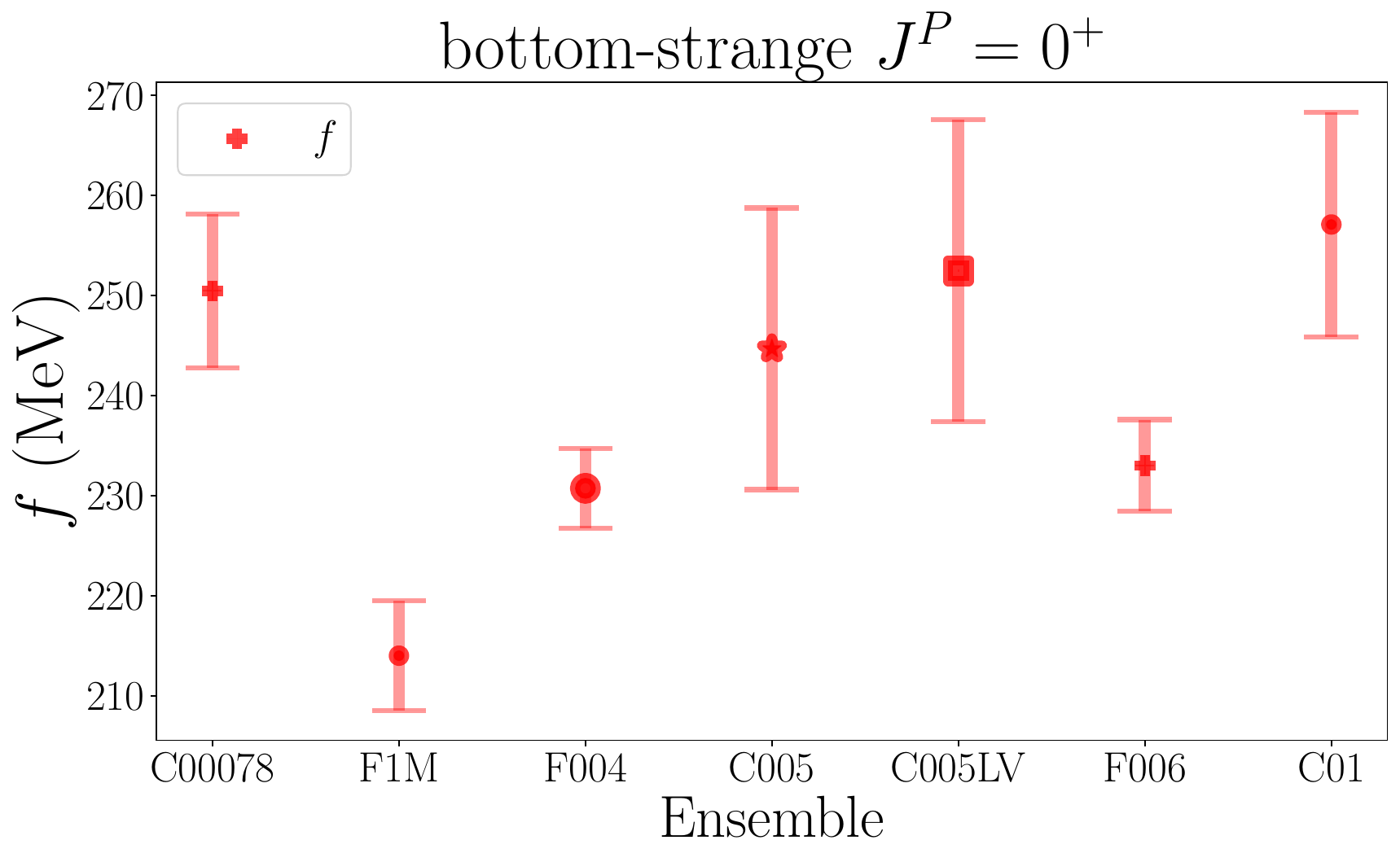}
    \hfill
    \includegraphics[width=0.44\linewidth]{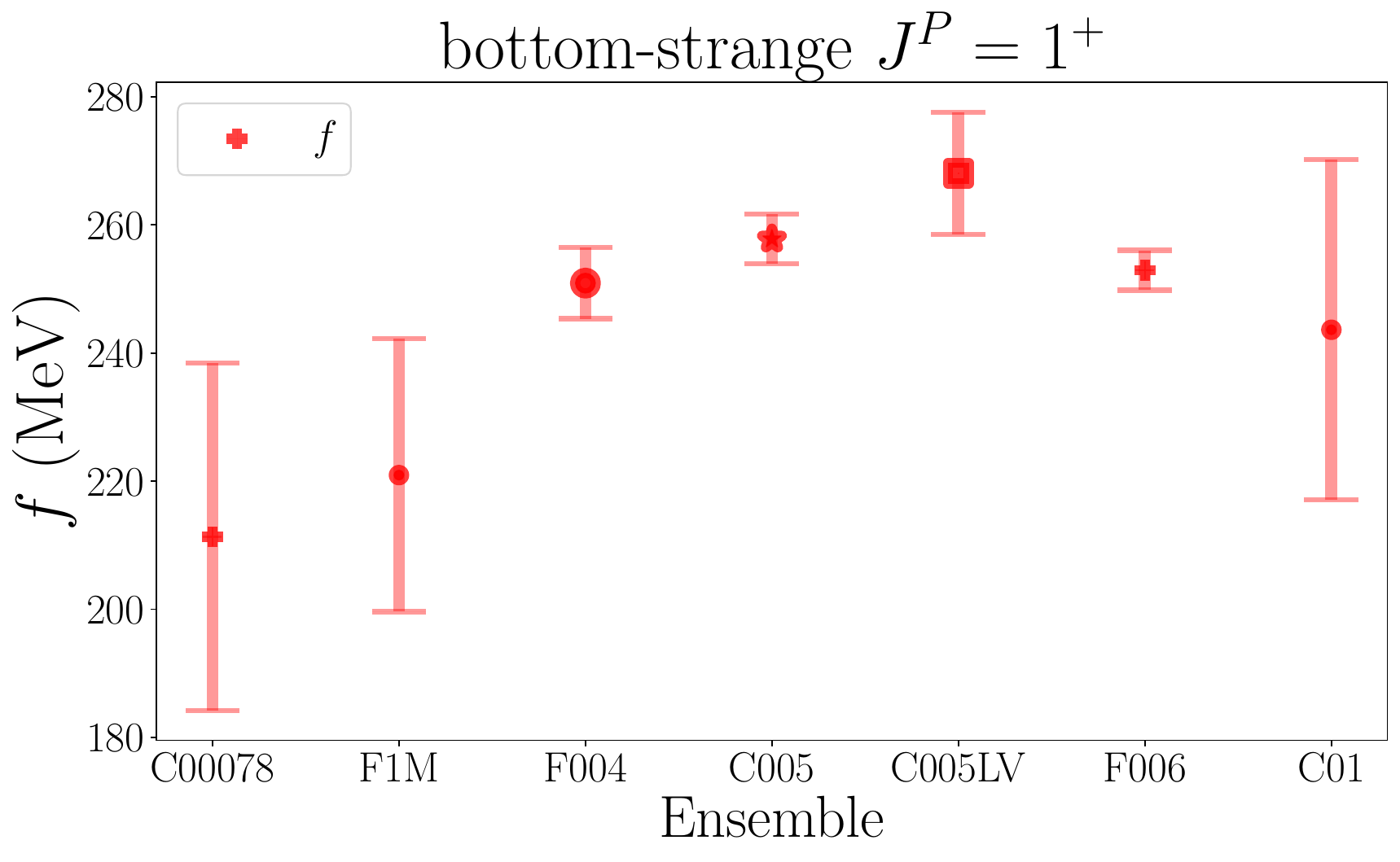}
    \caption{Ground-state finite-volume decay constants of the $H_{sJ}$ positive-parity heavy-strange mesons from the symmetrized matrix fit of Eqs.~(\ref{eq:matrixfit}-\ref{eq:symcorr}).}
    \label{fig:symcorrf}
\end{figure}

The finite-volume results for the ground-state binding energies and decay constants are plotted in Figs.~\ref{fig:symcorrE} and \ref{fig:symcorrf} respectively (the excited-state results from the $J=1$ matrix fit have very large uncertainties and are therefore not shown). Compared with our main results in Figs.~\ref{fig:Dsspec}, \ref{fig:Bsspec}, \ref{fig:Dsdecay}, \ref{fig:Bsdecay}, we observe that the results are qualitatively similar, but there are several significant discrepancies. Most notably, the $D_{s0}^*$ decay constants from the sub-matrix analysis are significantly higher on all ensembles compared to our main results, indicating that the fit uncertainties from the sub-matrix analysis are underestimated.

For comparison purposes, we nevertheless present chiral-continuum extrapolations of the sub-matrix results for the decay constants, and for the infinite-volume binding energies obtained using L\"uscher's method and the zeroth-order ERE. Again, we perform linear-in-$m_\pi^2$ and quadratic-in-$m_\pi^2$ fits and model-average the results from these two types of fits. The results of these extrapolations are provided in Table \ref{tab:symcorrE}.

After extrapolation to the physical point, the binding energies are roughly consistent between this approach and the main results of this work, but the uncertainties without the meson-meson operators are around twice as large (even though here we did not include the systematic uncertainty from a first-order ERE analysis), and the $D_{s1}$ binding energy becomes consistent with zero.

The chiral-continuum extrapolations of the decay constants are reasonably consistent with our main results, except for $f_{D^*_{s0}}$, where a significantly higher value is obtained. The chiral-continuum extrapolation of $f_{D^*_{s0}}$ exhibits a poor $\chi^2/\text{d.o.f}$, with $\chi^2/\text{d.o.f}>4$ for the dominant fit that enters the AIC model average. Even so, proceeding to further estimate finite-volume systematics on the decay constants of Table \ref{tab:symcorrf} using the methods of Sec.~\ref{sec:chiralcontinuum}, we arrive at $f_{D^*_{s0}}=155.7(5.4)(11)$ MeV, $f_{D_{s1}}=243(16)(24)$ MeV, $f_{B^*_{s0}}=190(13)(30)$ MeV, $f_{B_{s1}}=208(25)(27)$ MeV, with the first error statistical and the second systematic due to finite-volume effects.

What is interesting about the sub-matrix fit is that they provide corroboration that with contemporary lattice-QCD techniques, it is possible to resolve the below-threshold states of at least some positive-parity heavy-strange mesons, without employing meson-meson operators. This was first demonstrated for the $B_{sJ}$-mesons in Ref.~\cite{mohler} (and again in Ref.~\cite{Hudspith:2026vxt}), and is here also demonstrated for the more shallowly bound $D_{s0}^*$. This stands in contrast to the lattice-QCD calculations of the 1990s and 2000s \cite{earlywork, Dougall:2003hv, Bali:2003jv, Hein:2000qu, Lewis:2000sv}, as well as some more recent works without meson-meson type operators \cite{Kalinowski:2015bwa,Cichy:2016bci}. However, our analysis shows that fits without meson-meson operators may suffer from significant biases \emph{and} larger statistical uncertainties.

\begin{table}
    \centering
    \resizebox{\textwidth}{!}{
    \begin{tabular}{|l|c|c|c|c||c|c|c|c|c||c|}
    \hline
     & \multicolumn{4}{|c|}{Linear-in-$m_\pi^2$ fit} & \multicolumn{5}{|c|}{Quadratic-in-$m_\pi^2$ fit } & Model averaged\\
    \hline
    $H_{sJ}$ & $C$ & $d$ & $\Delta E_\text{phys}$(MeV) & weight & $C$ & $d$ & $C_2$ & $\Delta E_\text{phys}$(MeV) & weight & $\Delta E_\text{phys}$(MeV)\\ \hline
$D^*_{s0}$ & $-0.22 \pm 0.23$ & $0.15 \pm 0.72$ & $-21.9 \pm 26.6$ & 0.31 & $1.65 \pm 1.02$ & $1.18 \pm 0.90$ & $-26.96 \pm 14.24$ & $-72.4 \pm 37.7$ & 0.69 & $-57 \pm 42$\\\hline
$D_{s1}$ & $-0.72 \pm 0.27$ & $-0.30 \pm 0.96$ & $-3.4 \pm 24.4$ & 0.60 & $1.01 \pm 1.61$ & $0.82 \pm 1.41$ & $-22.75 \pm 20.95$ & $-52.5 \pm 51.3$ & 0.40 & $-23 \pm 45$\\\hline
$B^*_{s0}$ & $-0.52 \pm 0.32$ & $0.00 \pm 0.97$ & $-71.6 \pm 30.5$ & 0.72 & $0.06 \pm 1.68$ & $0.26 \pm 1.22$ & $-8.57 \pm 24.44$ & $-84.6 \pm 47.9$ & 0.28 & $-75 \pm 37$\\\hline
$B_{s1}$ & $-0.59 \pm 0.31$ & $-0.13 \pm 0.94$ & $-59.6 \pm 28.7$ & 0.56 & $1.14 \pm 1.44$ & $0.77 \pm 1.20$ & $-23.87 \pm 19.48$ & $-106.6 \pm 47.9$ & 0.44 & $-80 \pm 45$\\\hline
         \end{tabular}}
    \caption{Fit parameters, model weights, and model averages for the extrapolation of the infinite-volume binding energies from the symmetrized matrix fits of Eqs.~\ref{eq:matrixfit}-\ref{eq:symcorr}.}
    \label{tab:symcorrE}
\end{table}

\begin{table}
    \centering
    \resizebox{\textwidth}{!}{
    \begin{tabular}{|l|c|c|c|c||c|c|c|c|c||c|}
    \hline
     & \multicolumn{4}{|c|}{Linear-in-$m_\pi^2$ fit} & \multicolumn{5}{|c|}{Quadratic-in-$m_\pi^2$ fit } & Model averaged\\
    \hline
    $H_{sJ}$ & $C$ & $d$ & $f_\text{phys}$(MeV) & weight & $C$ & $d$ & $C_2$ & $f_\text{phys}$(MeV) & weight & $f_\text{phys}$(MeV)\\ \hline
    $D^*_{s0}$ & $ 0.118 \pm 0.042$ & $ 0.22 \pm 0.13$ & $ 156.2 \pm 4.4$ & 0.72 & $ 0.18 \pm 0.20$ & $ 0.25 \pm 0.18$ & $ -0.8 \pm 2.7$ & $ 154.4 \pm 7.4$ & 0.28 & $ 155.7 \pm 5.4 $ \\\hline
$D_{s1}$ & $ 0.33 \pm 0.22$ & $ -0.01 \pm 0.75$ & $ 241 \pm 13$ & 0.71 & $ -0.01 \pm 0.79$ & $ -0.06 \pm 0.76$ & $ 3.8 \pm 8.4$ & $ 249 \pm 21$ & 0.29 & $ 243 \pm 16 $ \\\hline
$B^*_{s0}$ & $ 0.16 \pm 0.10$ & $ 1.36 \pm 0.31$ & $ 191.4 \pm 9.3$ & 0.71 & $ 0.38 \pm 0.53$ & $ 1.51 \pm 0.47$ & $ -3.2 \pm 7.4$ & $ 185 \pm 18$ & 0.29 & $ 190 \pm 13$ \\\hline
$B_{s1}$ & $ 0.28 \pm 0.20$ & $ 0.32 \pm 0.27$ & $ 228 \pm 15$ & 0.37 & $ 1.88 \pm 0.94$ & $ 0.27 \pm 0.27$ & $ -21 \pm 12$ & $ 197 \pm 23$ & 0.63 & $ 208 \pm 25$ \\\hline
         \end{tabular}}
    \caption{Fit parameters, model weights, and model averages for the extrapolation of the finite-volume positive-parity decay constants from the symmetrized matrix fits of Eqs.~\ref{eq:matrixfit}-\ref{eq:symcorr}.}
    \label{tab:symcorrf}
\end{table}

\FloatBarrier
\section{$\sum_{\vec{x}_s} C^{44}$ Via Stochastic Propagators}
\label{sec:C44}
\subsection{Spin-0}
Instead of appealing to translational symmetry as we did with the other elements calculated with point-to-all propagators at fixed source, we estimate $C^{44}$ summed over all source positions, $\sum_{\vec{x}_s} C^{44}(t)$. 

The isospin-projected $\sum_{\vec{x}_s} C^{44}(t)$ is
\begin{align}
    \sum_{\vec{x}_s} C^{44}_{I=0}(t)&=\sum_{\vec{x}_s}(C^{44}_{\text{box}}(t)_{uu}+C^{44}_\text{box}(t)_{ud}+C^{44}_{\text{2tr}}(t)_{uu}) + (u \leftrightarrow d) \label{eq:C44a}
\end{align}
with
\begin{align}
	 \sum_{\vec{x}_s} C^{44}_{\text{2tr}}(t)_{qq'} &\equiv \sum_{\vec{z},\vec{w},\vec{y}, \vec{x}_s} 
	 \Tr{G_q^\dagger(\vec{z},t+t_s;\vec{x}_s,t_s)G_s(\vec{z},t+t_s;\vec{x}_s,t_s)}\Tr{G_Q^\dagger(\vec{y},t+t_s;\vec{w},t_s)G_{q'}(\vec{y},t+t_s;\vec{w},t_s)}, \\ 
	   \sum_{\vec{x}_s} C^{44}_{\text{box}}(t)_{qq'}&\equiv -  \sum_{\vec{z},\vec{w},\vec{y}, \vec{x}_s}  \Tr{G_s(\vec{z},t+t_s;\vec{x}_s,t_s)\gamma_5 G_q(\vec{x}_s,t_s;\vec{w},t_s) G_Q^\dagger(\vec{y},t+t_s;\vec{w},t_s)\gamma_5 G_{q'}^\dagger(\vec{z},t+t_s;\vec{y},t+t_s)}.
\end{align}
Using $G_u=G_d$, Eq.~(\ref{eq:C44a}) becomes
\begin{equation}
    \sum_{\vec{x}_s} C^{44}_{I=0}(t)=4 \sum_{\vec{x}_s}C^{44}_{\text{box}}(t)+ 2 \sum_{\vec{x}_s} C^{44}_{\text{2tr}}(t),
\end{equation}
corresponding to the two diagrams shown in Fig.~\ref{fig:quarkline}. We estimate these diagrams stochastically as explained in the following.

\subsubsection{Box Diagram}

We introduce the notation
\begin{gather}
	\sum_{\vec{x}_s} C^{44}_{\text{box}}(t)
	= - \sum_{\vec{z},\vec{w}} \Tr{T^{(s)}_{LP}(\vec{z},t+t_s;\vec{w},t_s) T^{(u)\dagger}_{RP}(\vec{z},t+t_s;\vec{w},t_s)} \label{eq:C44b}
\end{gather}
with the left and right ``paths''
\begin{align}
	T^{(s)}_{LP}(\vec{z},t+t_s;\vec{w},t_s)&=\sum_{\vec{x}_s} G_s(\vec{z},t+t_s;\vec{x}_s,t_s)\gamma_5 G_u(\vec{x}_s,t_s;\vec{w},t_s),\\
	T^{(u)}_{RP}(\vec{z},t+t_s;\vec{w},t_s)&=\sum_{\vec{y}}G_u(\vec{z},t+t_s;\vec{y},t+t_s)\gamma_5 G_Q(\vec{y},t+t_s,\vec{w},t_s).
\end{align}
Directly evaluating the sum over $\vec{w}$ using point sources at each $\vec{w}$ would be prohibitively expensive. Instead, we approximate Eq.~(\ref{eq:C44b}) by computing
\begin{align}
\frac{1}{N} \sum_{n=1}^N  \sum_{\vec{z}} \Tr{\widetilde{T}^{(s)n}_{LP}(\vec{z},t+t_s;t_s) \widetilde{T}^{(u)n\dagger}_{RP}(\vec{z},t+t_s;t_s)}
\end{align}
with the sequential-stochastic propagators 
\begin{align}
	\widetilde{T}^{(s)n}_{LP}(\vec{z},t+t_s;t_s)&=\sum_{\vec{x}_s} G_s(\vec{z},t+t_s;\vec{x}_s,t_s)\gamma_5 \widetilde{G}_u^n(\vec{x}_s,t_s;t_s),\\
	\widetilde{T}^{(u)n}_{RP}(\vec{z},t+t_s;t_s)&=\sum_{\vec{y}}G_u(\vec{z},t+t_s;\vec{y},t+t_s)\gamma_5 \widetilde{G}_Q^n(\vec{y},t+t_s;t_s),
\end{align}
which are sourced by $\gamma_5$ times the stochastic propagators
\begin{align}
\widetilde{G}_q^n(\vec{u},t_2;t_1)=\sum_{\vec{u},t'} G_q(\vec{u},t_2;\vec{v},t') \xi^{t_1, n}(\vec{v},t')
\end{align}
where
\begin{equation}
	\xi^{t_1, n}(\vec{v},t')=\delta_{t_1, t'}\xi^n(\vec{v}) \ \text{ with random } \xi^n(\vec{v})\in \mathbb{Z}_2 \ \text{ such that } \lim\limits_{N\rightarrow \infty} \frac{1}{N} \sum_{n=1}^N \xi^n(\vec{v})^* \xi^n(\vec{v}^{\,\prime})=\delta_{\vec{v},\vec{v}^{\,\prime}}.
\end{equation}
For finite $N$, this sum resolves to a Kronecker delta up to $\mathcal{O}(\frac{1}{\sqrt{N}})$. In practice, since we also average over gauge configurations, $N=1$ is sufficient and this is what we use. The stochastic and sequential-stochastic propagators are obtained by solving the sourced Dirac equations
\begin{align}
	D^{(q)} \widetilde{G}_q^n(\vec{u},t_2;t_1)&=\xi^{t_1, n}(\vec{u},t_2),\\
	D^{(s)} \widetilde{T}^{(s)n}_{LP}(\vec{z},t+t_s;t_s)&=\gamma_5 \widetilde{G}^n_u(\vec{z},t_s;t_s), \\
	D^{(u)}\widetilde{T}^{(u)n}_{RP}(\vec{z},t+t_s;t_s)&=\gamma_5 \widetilde{G}_Q^n(\vec{z},t+t_s;t_s).
\end{align}
The $\mathbb{Z}_2$ noise sources $\xi$ are Gaussian smeared with the smearing parameters associated with the relevant quark. Upon calculation, the propagators are again smeared at the sink.
The heavy-quark propagator serving as the source for $\widetilde{T}^{(u)n}_{RP}$ is first smeared with the same heavy-quark smearing as the $\mathbb{Z}_2$-source from which it was calculated, and then again smeared with the light-quark smearing parameters in preparation for calculating the propagator with the light-quark Dirac equation. Finally, this heavy-quark source is restricted to the sink timeslice before proceeding to solve for $\widetilde{T}^{(u)n}_{RP}$. The light-quark propagator serving as the source for $\widetilde{T}^{(s)n}_{LP}$ is first smeared with the same light-quark smearing parameters as its $\mathbb{Z}_2$-source, then smeared again with the strange-quark smearing parameters in preparation for calculating the propagator with the strange-quark Dirac equation. Finally, this light-quark source is restricted to the source timeslice, before proceeding to solve for $\widetilde{T}^{(s)n}_{LP}$.
After producing the propagators, $\widetilde{T}^{(s)n}_{LP}$ ($\widetilde{T}^{(u)n}_{RP}$) is then smeared at the sink with strange(light)-quark smearing parameters.

\subsubsection{2-Trace Part}

To estimate the 2-trace part
\begin{gather}
	\sum_{\vec{x}_s} C^{44}_{\text{2tr}}(t) =\sum_{\vec{z},\vec{w},\vec{y},\vec{x}_s}  \Tr{G_u^\dagger(\vec{z},t+t_s;\vec{x}_s,t_s)G_s(\vec{z},t+t_s;\vec{x}_s,t_s)}\Tr{G_Q^\dagger(\vec{y},t+t_s;\vec{w},t_s)G_u(\vec{y},t+t_s;\vec{w},t_s)},
\end{gather}
we apply the one-end trick twice (once for each trace) \cite{stochasitc}. We introduce independent $\mathbb{Z}_2$ noise sources $\xi$ and $\eta$ and solve the Dirac equations
\begin{align}
	D^{(u)} \widetilde{G}_u^n(\vec{u},t_2;t_s)&=\xi^{t_s, n}(\vec{u},t_2),\\
	D^{(s)} \widetilde{G}_s^n(\vec{u},t_2;t_s)&=\xi^{t_s, n}(\vec{u},t_2),\\
	D^{(Q)} \widehat{G}_Q^m(\vec{u},t_2;t_s)&=\eta^{t_s, m}(\vec{u},t_2),\\
	D^{(u)} \widehat{G}_s^m(\vec{u},t_2;t_s)&=\eta^{t_s, m}(\vec{u},t_2),
\end{align}
with the appropriate smearing applied to $\xi$ and $\eta$ and after computing the propagators. Then, each of the above traces is obtained as
\begin{align}
\sum_{\vec{z},\vec{x}_s}\Tr{G_u^\dagger(\vec{z},t+t_s;\vec{x}_s,t_s)G_s(\vec{z},t+t_s;\vec{x}_s,t_s)}&=\lim\limits_{N\rightarrow \infty} \frac{1}{N}  \sum_{n=1}^N \sum_{\vec{z}} \Tr\left\{ \widetilde{G}_u^{n\dag}(\vec{z},t+t_s;t_s)  \widetilde{G}_s^n(\vec{z},t+t_s;t_s) \right\},\\
\sum_{\vec{y},\vec{w}} \Tr{G_Q^\dagger(\vec{y},t+t_s;\vec{w},t_s)G_u(\vec{y},t+t_s;\vec{w},t_s)} &=\lim\limits_{M\rightarrow \infty} \frac{1}{M}  \sum_{m=1}^M \sum_{\vec{y}}  \Tr\left\{ \widehat{G}_Q^{m\dag}(\vec{y},t+t_s;t_s) \widehat{G}_s^m(\vec{y},t+t_s;t_s) \right\}.
\end{align}
Again, $N=1$ and $M=1$ are sufficient and are what we use.

\subsection{Spin-1}
Similarly to the spin-0 case, we have
\begin{align}
    \sum_{\vec{x}_s} C^{44ij}_{I=0}(t)&=4 \sum_{\vec{x}_s}C^{44ij}_{\text{box}}(t)+ 2 \sum_{\vec{x}_s} C^{44ij}_{\text{2tr}}(t),\\
    \sum_{\vec{x}_s} C^{44 ij}_\text{2tr}(t)&\equiv\sum_{\vec{x}_s,\vec{y},\vec{z},\vec{w}} \Tr{G_u^\dagger(\vec{z},t+t_s;\vec{x}_s,t_s) G_s(\vec{z},t+t_s;\vec{x}_s,t_s)}\Tr{\gamma_5 G_Q^\dagger(\vec{y},t+t_s;\vec{w},t_s)\gamma_5 \gamma^i G_u(\vec{y},t+t_s;\vec{w},t_s)\gamma^j},\\
    \sum_{\vec{x}_s} C^{44 ij}_\text{box}(t)&\equiv \sum_{\vec{x}_s,\vec{y},\vec{z},\vec{w}} \Tr{G_s(\vec{z},t+t_s;\vec{x}_s,t_s)\gamma_5 G_u(\vec{x}_s,t_s;\vec{w},t_s)\gamma^j \gamma_5 G_Q^\dagger(\vec{y},t+t_s;\vec{w},t_s)\gamma^i G_u^\dagger(\vec{z},t+t_s;\vec{y},t+t_s)}.
\end{align}

\vspace{-10ex}

\subsubsection{Box Diagram}

The box diagram has the form
\begin{align}
	\sum_{\vec{x}_s} C^{44 ij}_\text{box}(t)&=
	\sum_{\vec{z},\vec{w}} \Tr{T^{(s)}_{LP}(\vec{z},t+t_s;\vec{w},t_s)\gamma^j \gamma_5 T_{RP1}^{(u)i \dagger}(\vec{z},t+t_s;\vec{w},t_s)}, \label{eq:boxspin1}
\end{align}
where $T^{(s)}_{LP}$ is the same as for the spin-0 case, and
\begin{align}
T_{RP1}^{i(u) \dagger}(\vec{z},t+t_s;\vec{w},t_s)=\sum_{\vec{y}} G_u(\vec{z},t+t_s;\vec{y},t+t_s)\gamma^{i\dagger} G_Q(\vec{y},t+t_s;\vec{w},t_s).
\end{align}
We estimate Eq.~(\ref{eq:boxspin1}) as
\begin{align}
\frac{1}{N} \sum_{n=1}^N  \sum_{\vec{z}} \Tr{\widetilde{T}^{(s)n}_{LP}(\vec{z},t+t_s;t_s)\gamma^j\gamma_5 \widetilde{T}^{i(u)n\dagger}_{RP1}(\vec{z},t+t_s;t_s)},
\end{align}
where $\widetilde{T}^{i(u)n}_{RP1}$ is the solution to
\begin{align}
	D^{(u)}\widetilde{T}^{i(u)n}_{RP1}(\vec{z},t+t_s;t_s)&=\gamma^{i\dag} \widetilde{G}_Q^n(\vec{z},t+t_s;t_s).
\end{align}
The smearing steps are analogous to the spin-0 case, and we again use $N=1$.

\subsubsection{2-Trace Part}

Note that the factor $\sum_{\vec{z},\vec{x}_s}\Tr{G_u^\dagger(\vec{z},t+t_s;\vec{x}_s,t_s) G_s(\vec{z},t+t_s;\vec{x}_s,t_s)}$ has already been calculated for the spin-0 case. We obtain the other factor as
\begin{align}
\sum_{\vec{y},\vec{w}}\Tr{ G_Q^\dagger(\vec{y},t+t_s;\vec{w},t_s)\gamma_5 \gamma^i G_u(\vec{y},t+t_s;\vec{w},t_s)\gamma^j \gamma_5} &=\lim\limits_{M\rightarrow \infty} \frac{1}{M}  \sum_{m=1}^M \sum_{\vec{y}} \Tr{ \widehat{G}_Q^{m\dagger}(\vec{y},t+t_s;t_s)\gamma_5 \gamma^i \widehat{P}_u^{mj}(\vec{y},t+t_s;t_s) },
\end{align}
where $\widehat{P}_u^{mj}$ is the solution to
\begin{align}
D^{(u)} \widehat{P}_u^{mj}(\vec{u},t_2;t_s) = \gamma^j\gamma_5 \eta^{t_s,m}(\vec{u},t_2).
\end{align}
Again, we use $M=1$ and perform smearing steps analogous to the spin-0 case.

\FloatBarrier
\section{Additional Plots}
\label{sec:additional}
\subsection{Negative-Parity Correlators}
\label{sec:negpardecayconstfitplots}
\FloatBarrier

\begin{figure}[H]
    \centering
    \centering
    \includegraphics[width=0.49\linewidth]{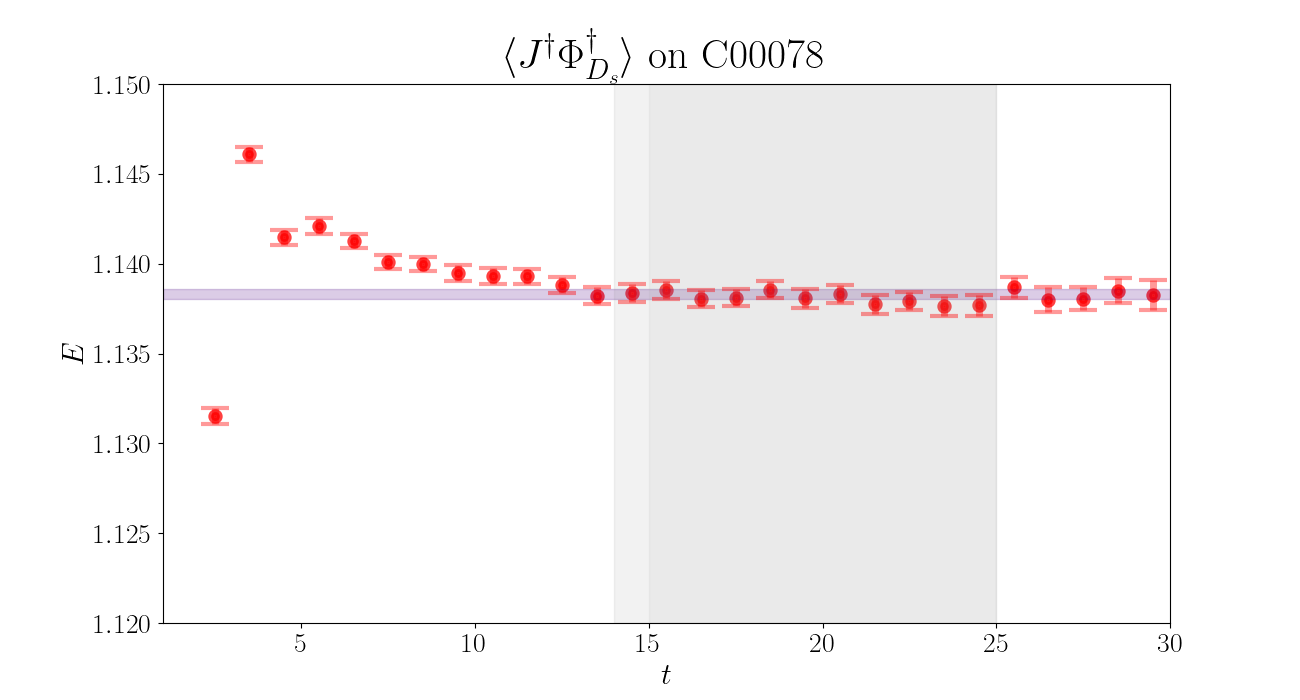}
    \hfill
    \includegraphics[width=0.49\linewidth]{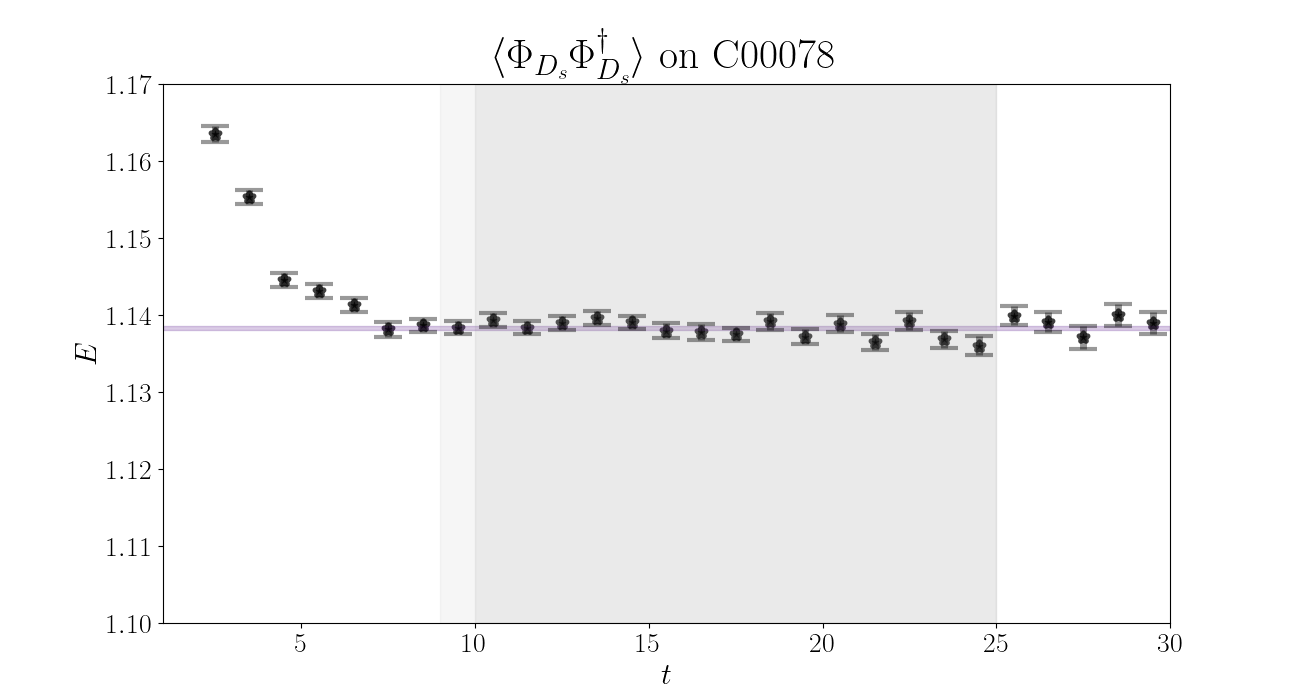}
    
    \includegraphics[width=0.49\linewidth]{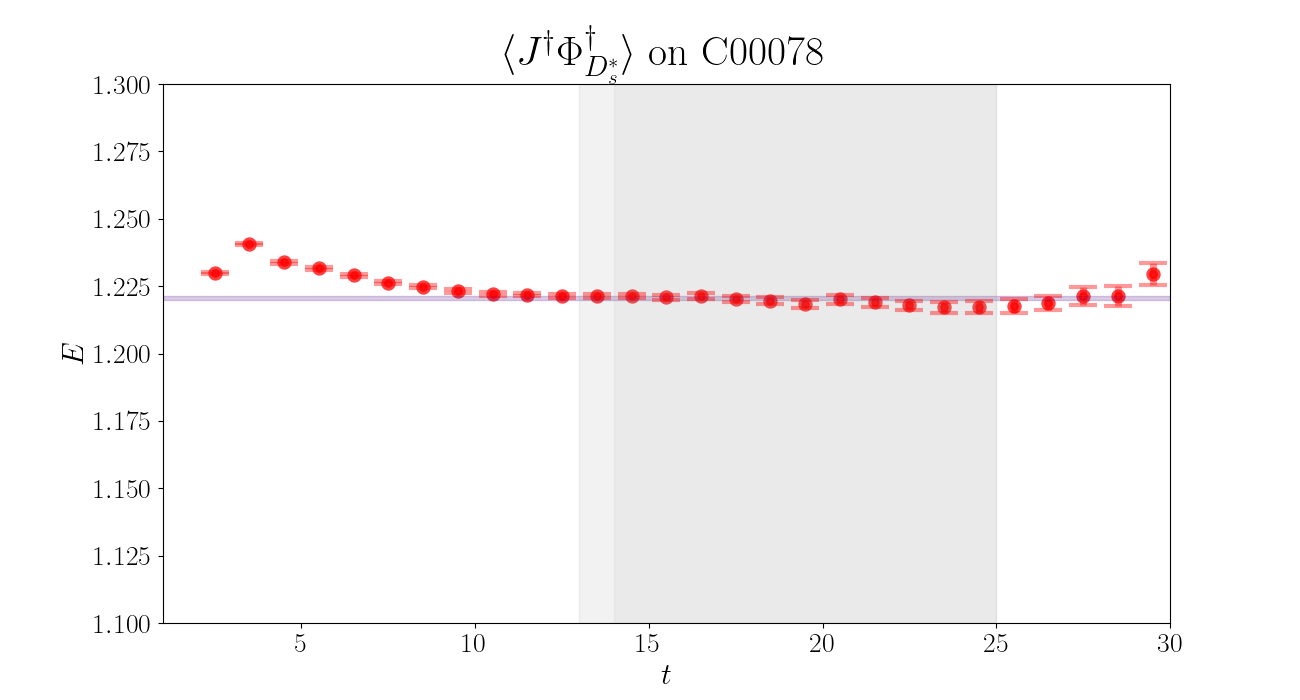}
    \hfill
    \includegraphics[width=0.49\linewidth]{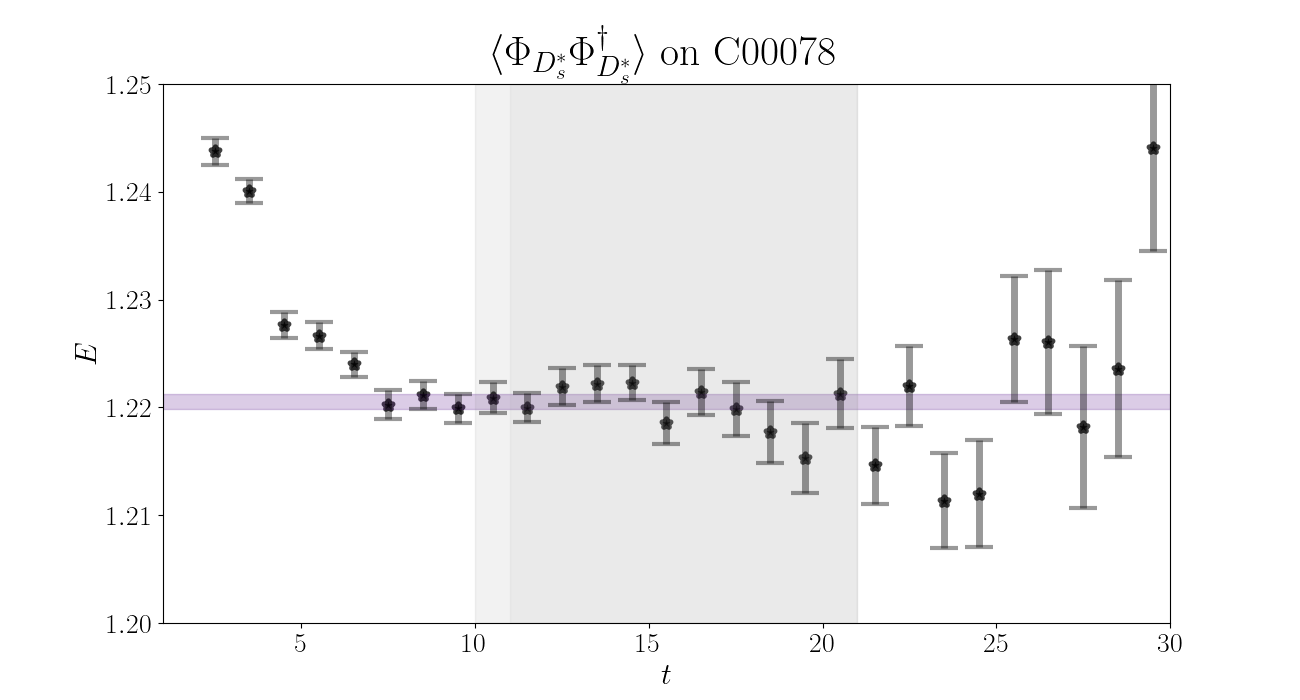}

    \includegraphics[width=0.49\linewidth]{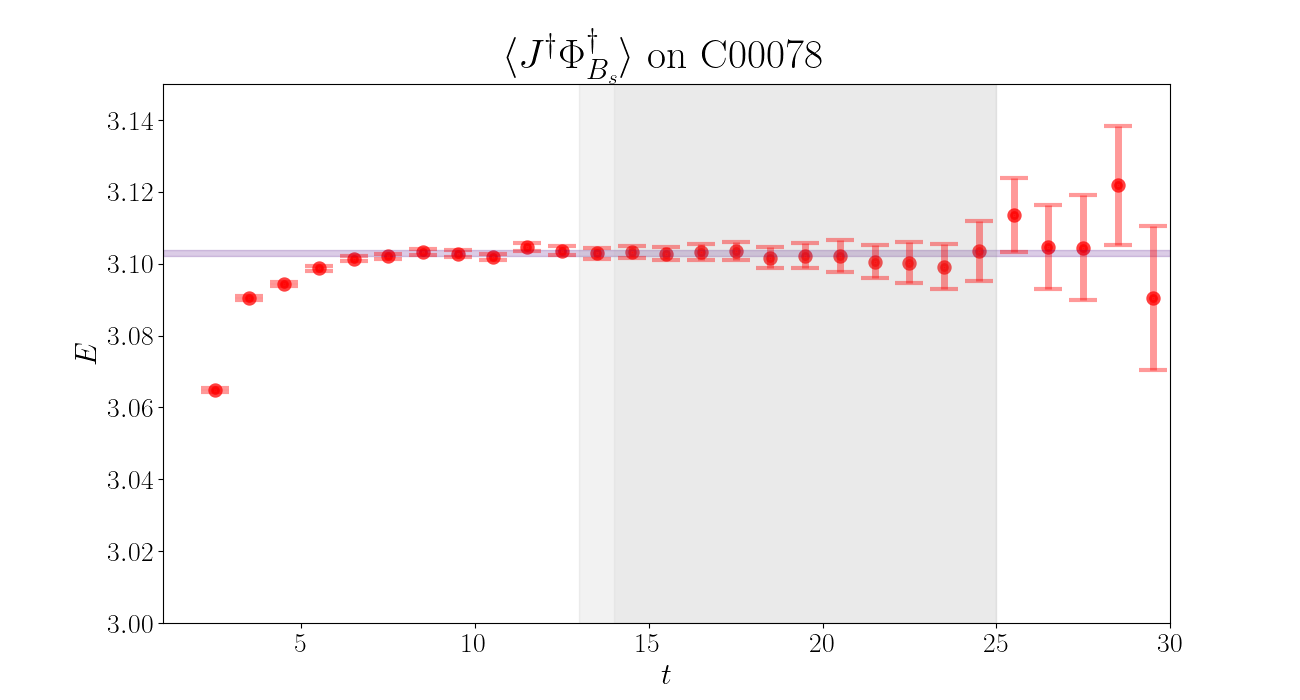}
    \hfill
    \includegraphics[width=0.49\linewidth]{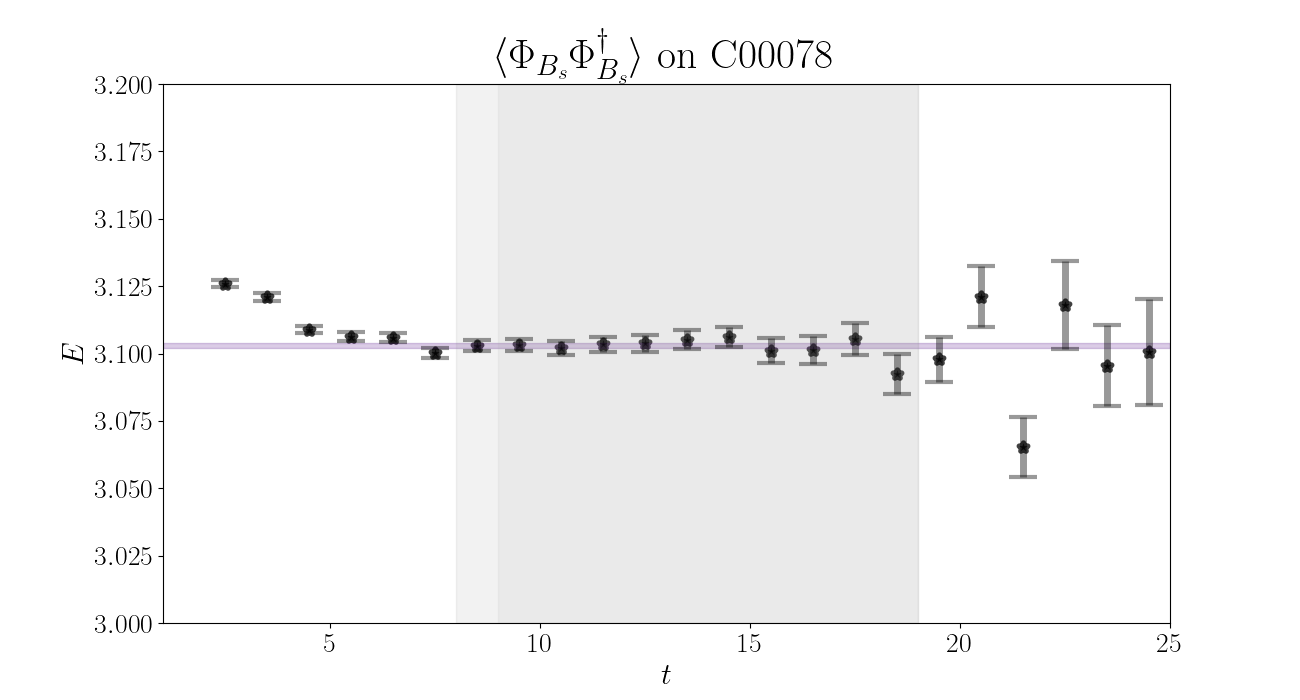}
    
    \includegraphics[width=0.49\linewidth]{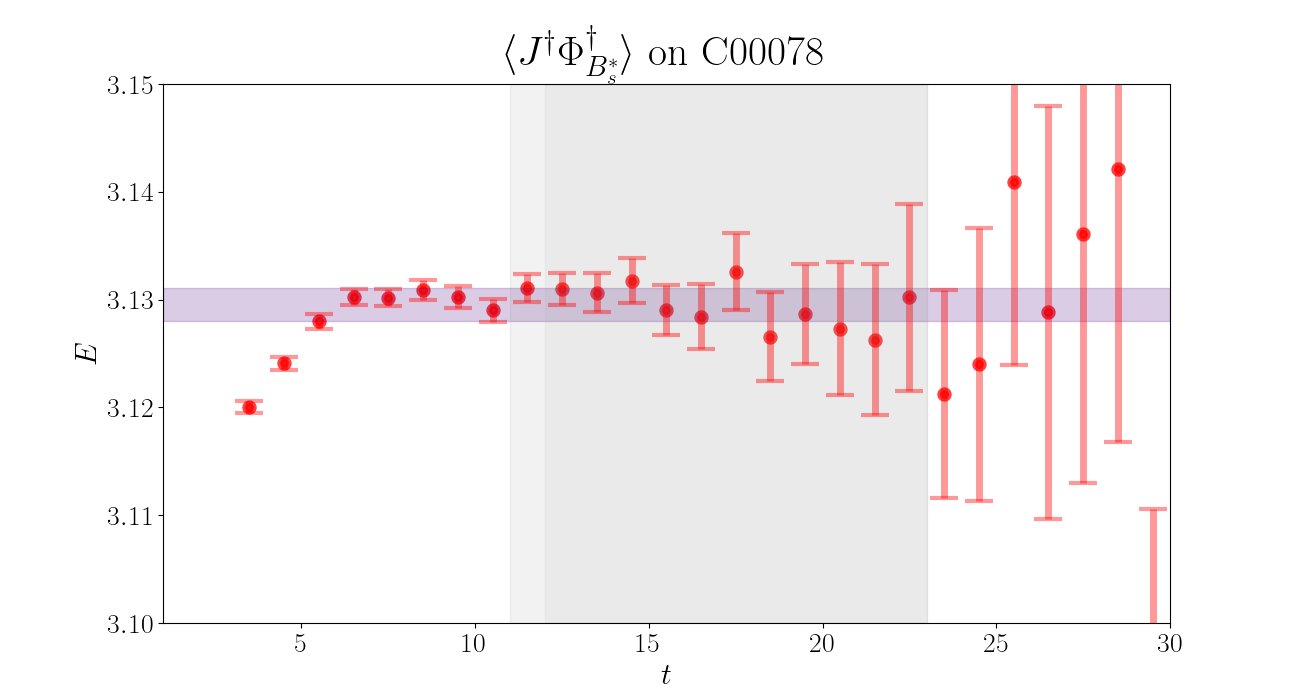}
    \hfill
    \includegraphics[width=0.49\linewidth]{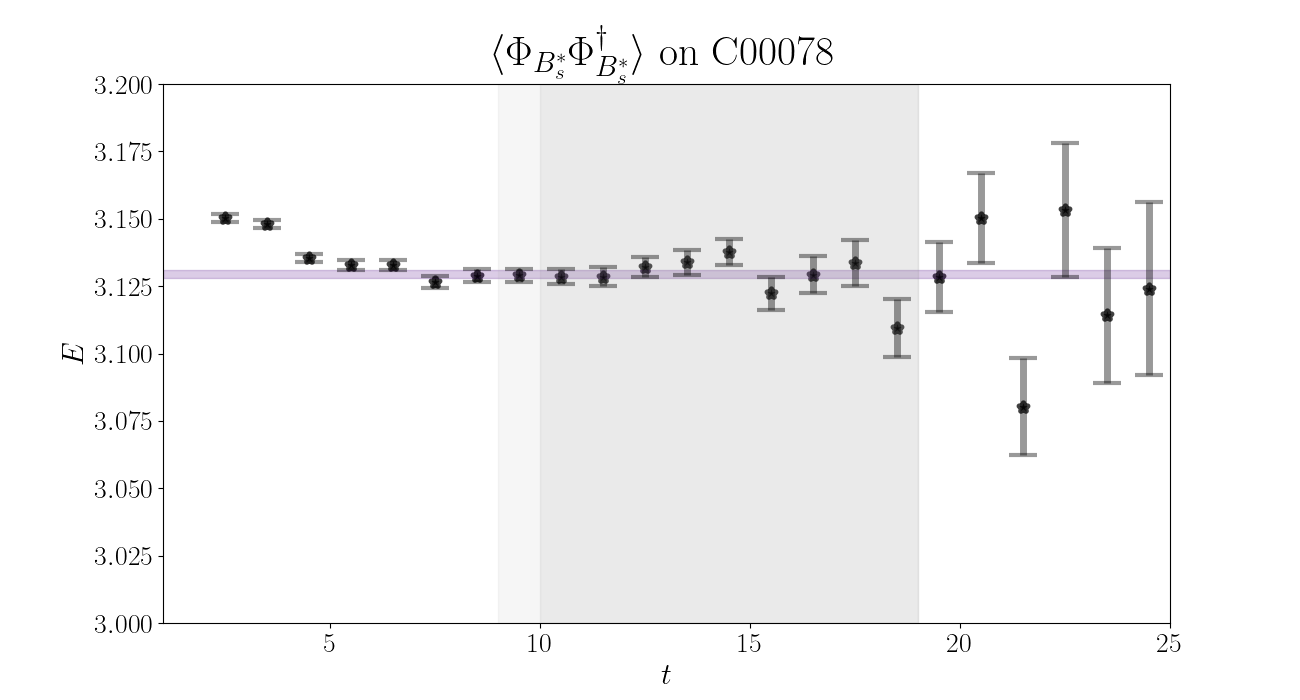}
    \caption{Effective-energy plots (in lattice units) for the negative-parity hadron-current and hadron-hadron correlators for each system on C00078. The horizontal purple band corresponds to the ground-state energy extracted from the hadron-hadron correlator. The combinations of the light and dark shaded regions indicate the fit ranges used to obtain the central values and statistical uncertainties of the energies and amplitudes, while the dark shaded regions indicates the fit ranges used in the calculation of the fit-range systematic uncertainties. \label{fig:negpardecayconstfitplots-C00078}}
\end{figure}

\begin{figure}[H]
    \centering
    \centering
    \includegraphics[width=0.49\linewidth]{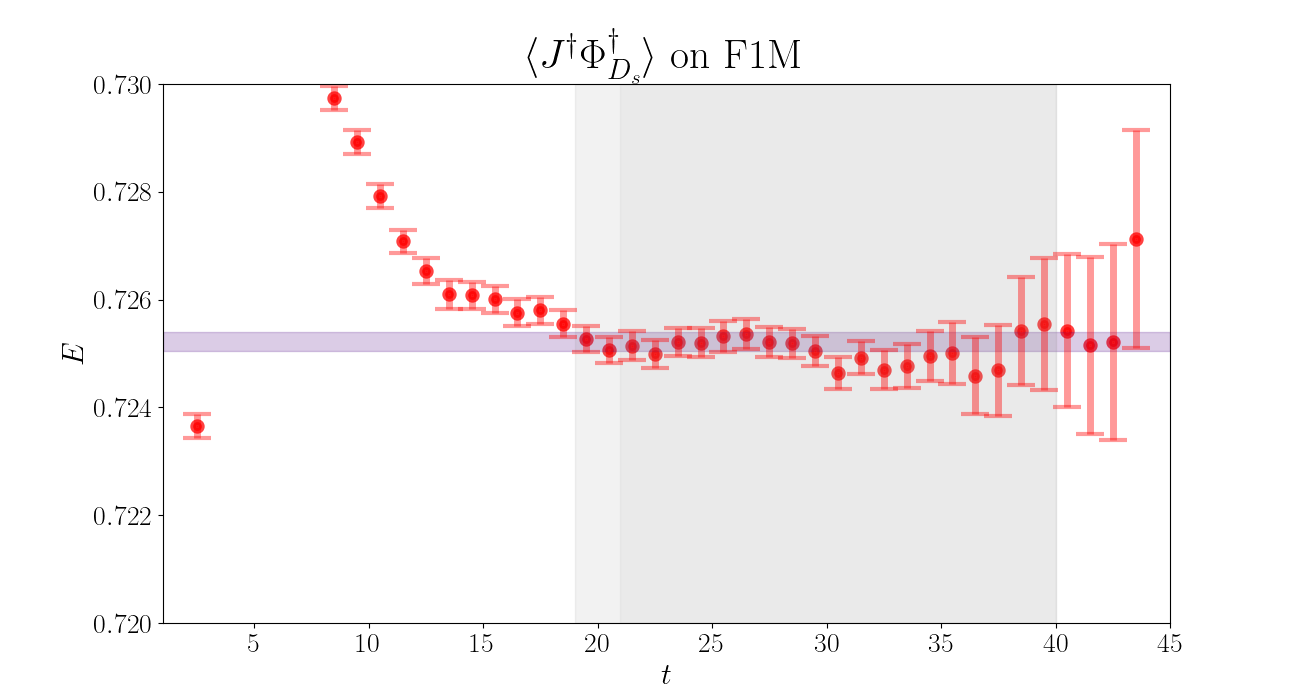}
    \hfill
    \includegraphics[width=0.49\linewidth]{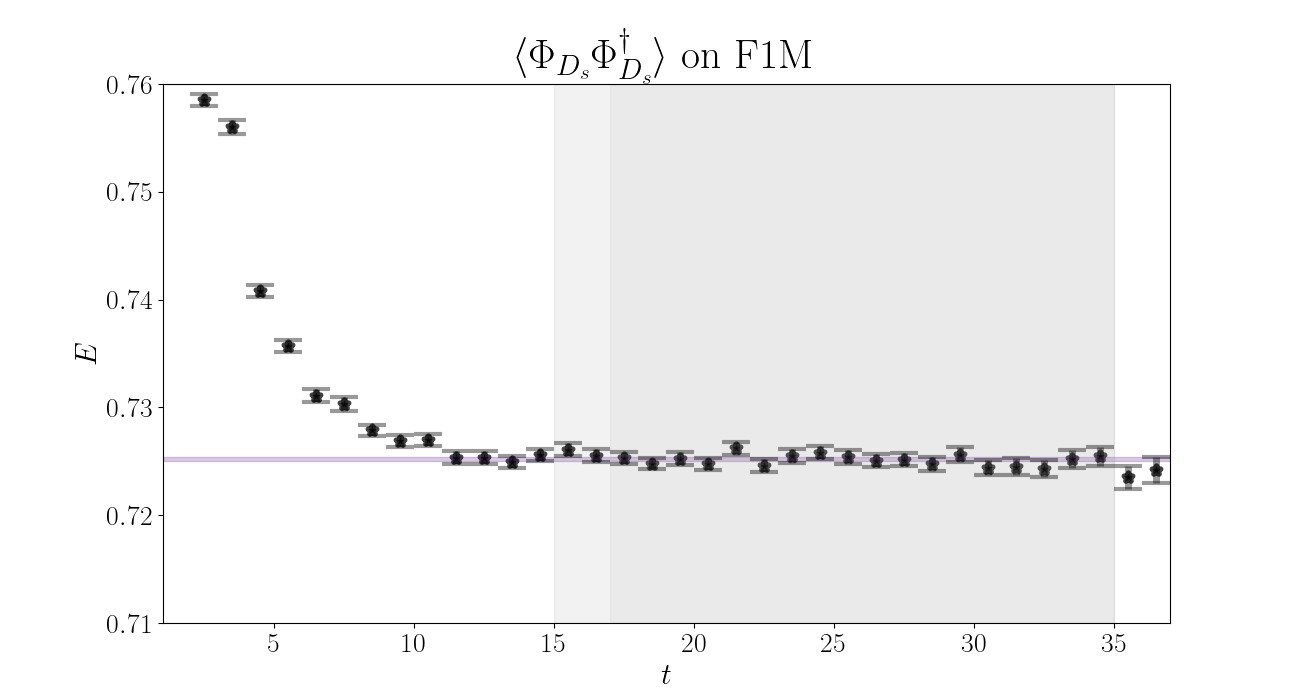}
    
    \includegraphics[width=0.49\linewidth]{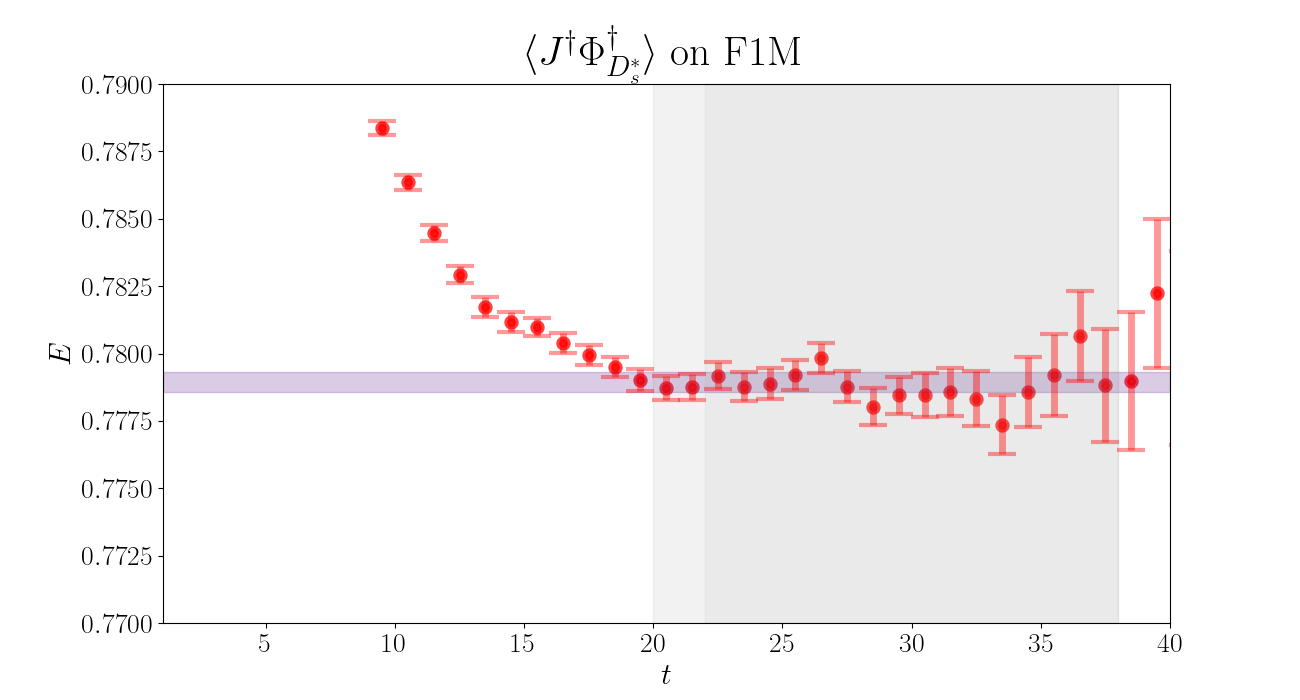}
    \hfill
    \includegraphics[width=0.49\linewidth]{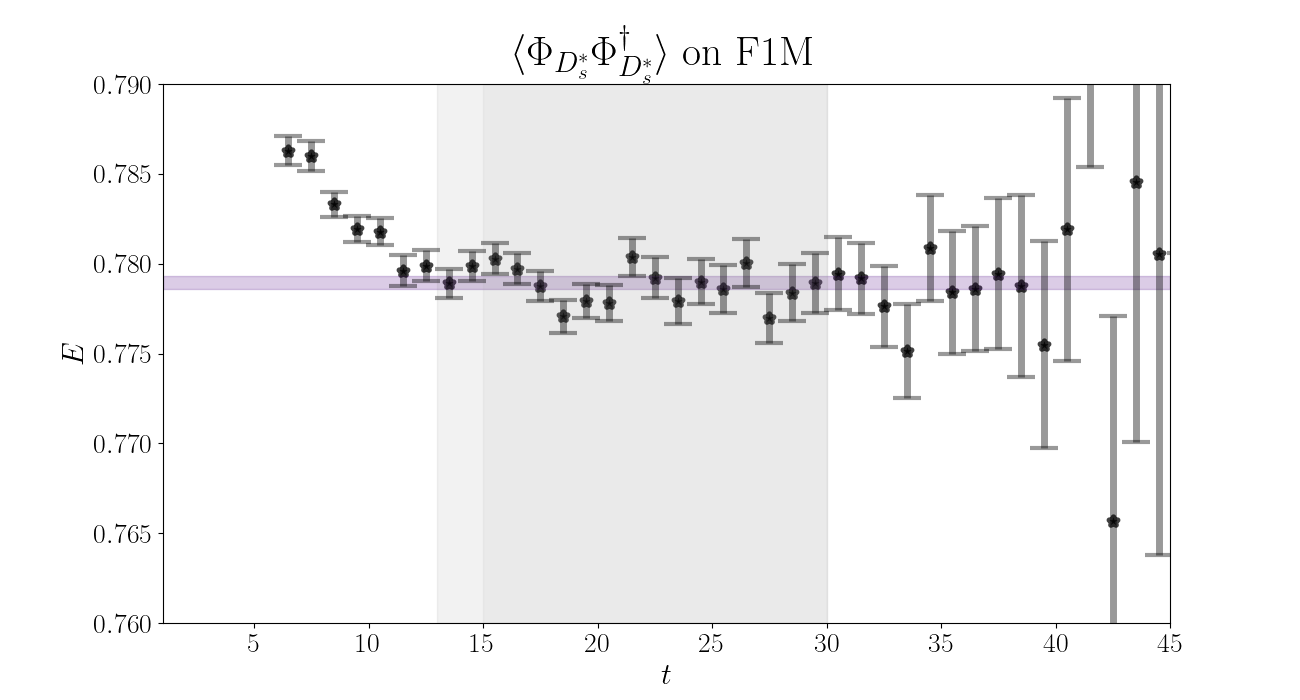}

    \includegraphics[width=0.49\linewidth]{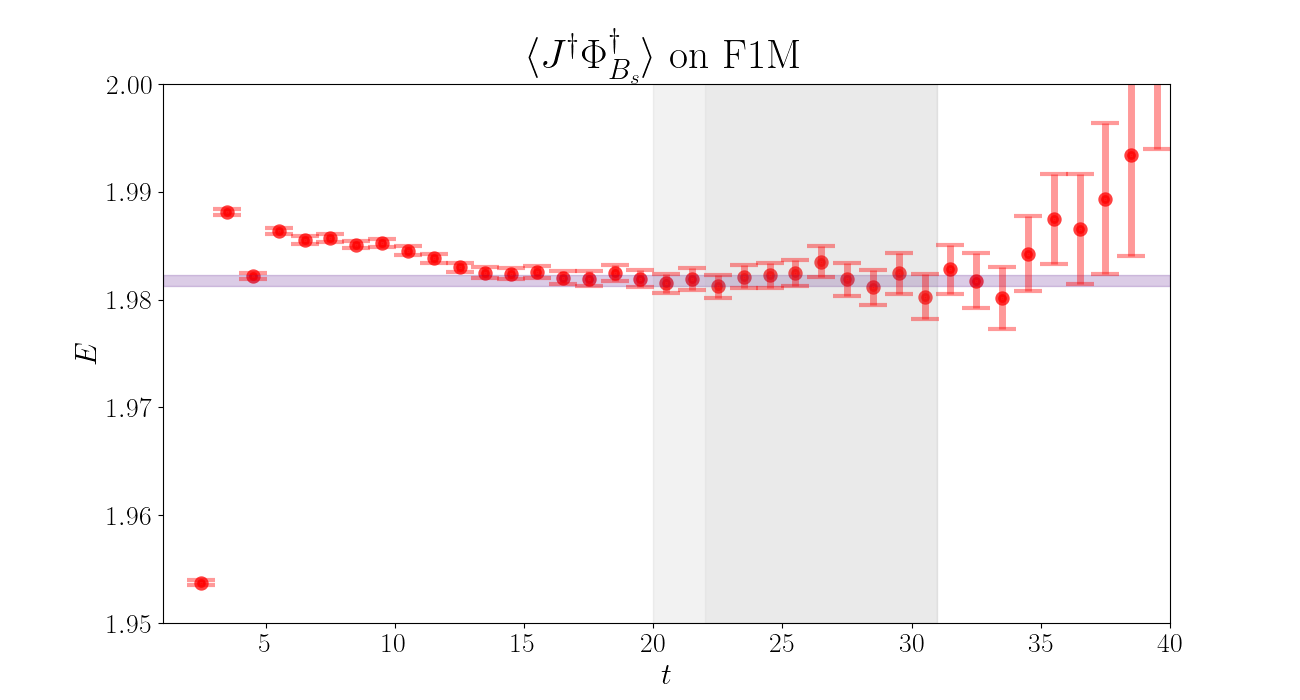}
    \hfill
    \includegraphics[width=0.49\linewidth]{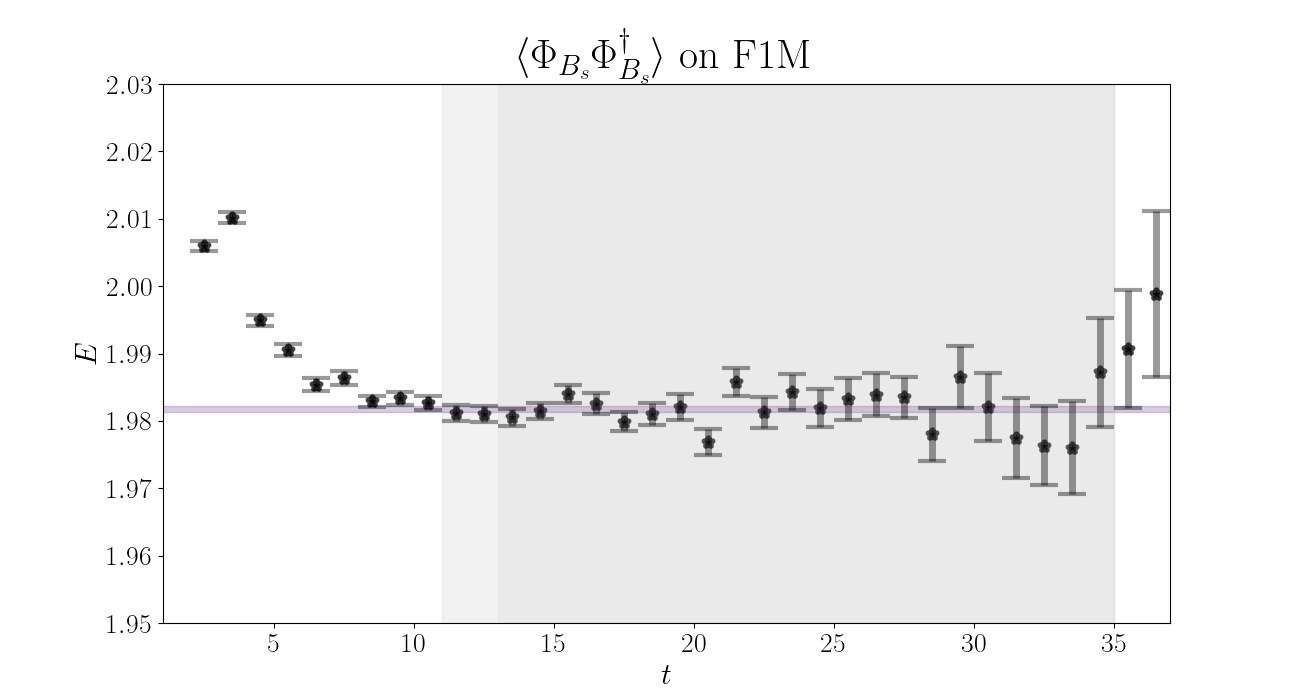}
    
    \includegraphics[width=0.49\linewidth]{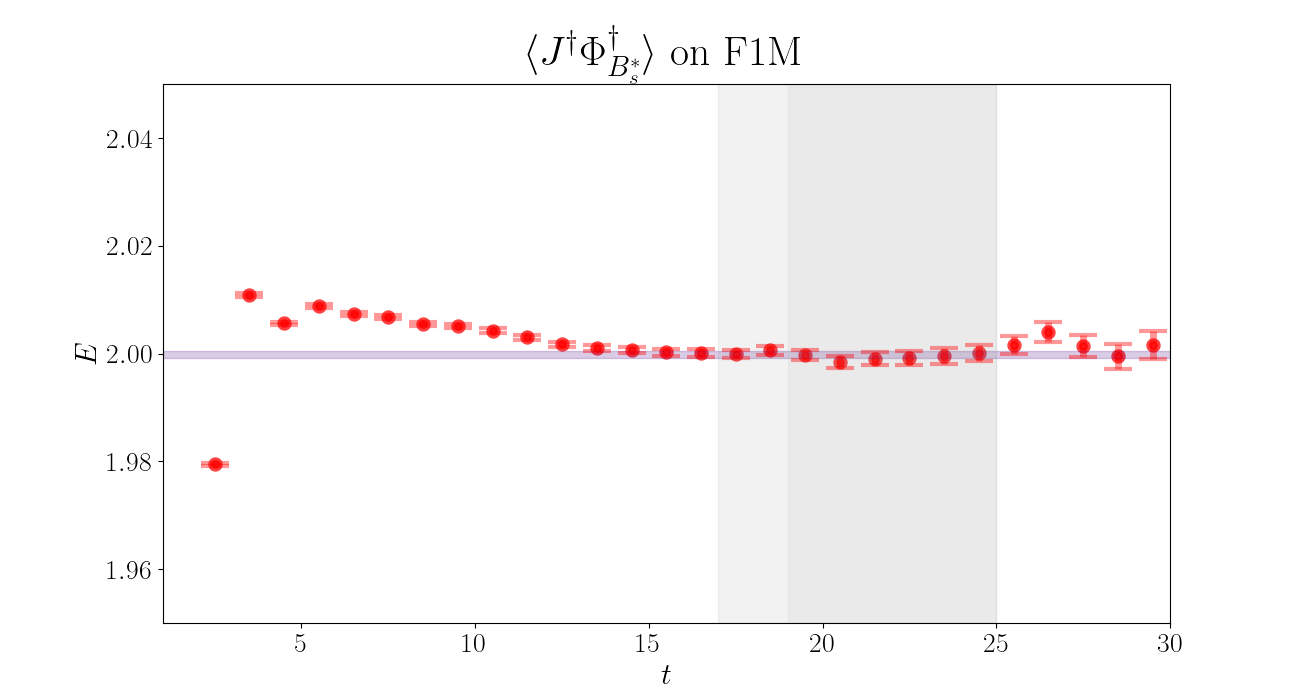}
    \hfill
    \includegraphics[width=0.49\linewidth]{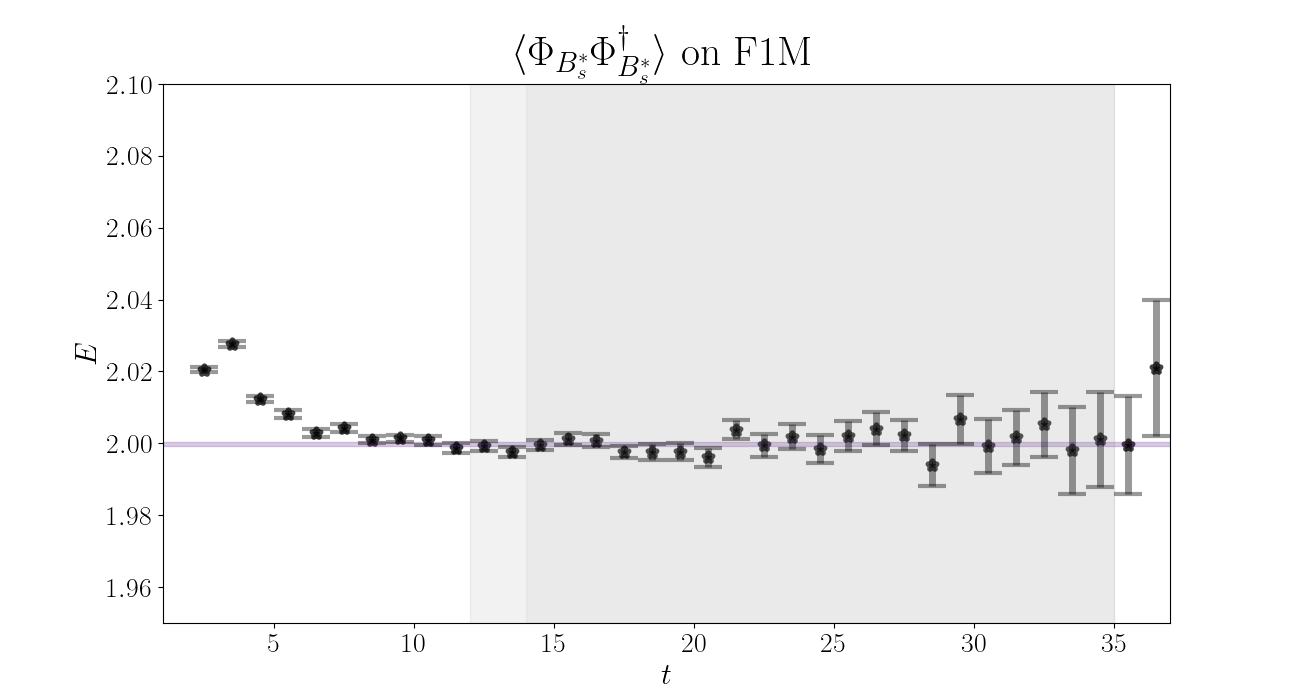}
    \caption{Like Fig.~\protect\ref{fig:negpardecayconstfitplots-C00078}, but for the F1M ensemble.\label{fig:negpardecayconstfitplots-F1M}}
\end{figure}

\begin{figure}[H]
    \centering
    \centering
    \includegraphics[width=0.49\linewidth]{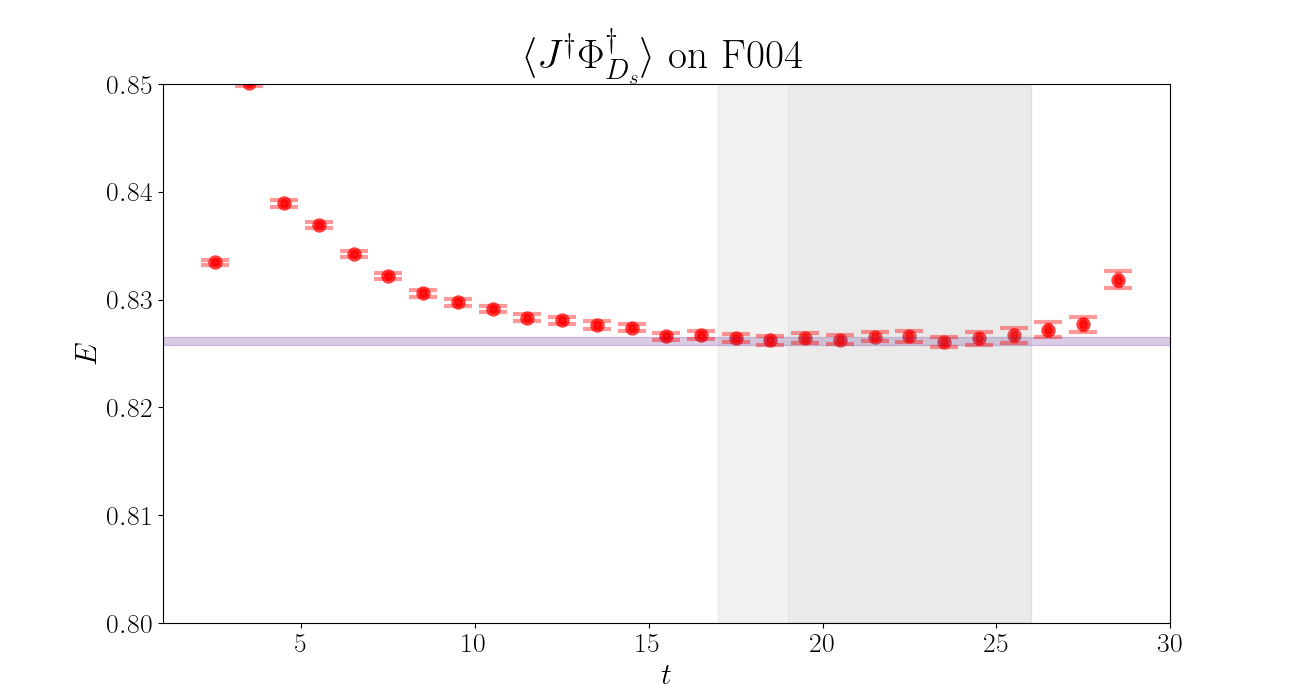}
    \hfill
    \includegraphics[width=0.49\linewidth]{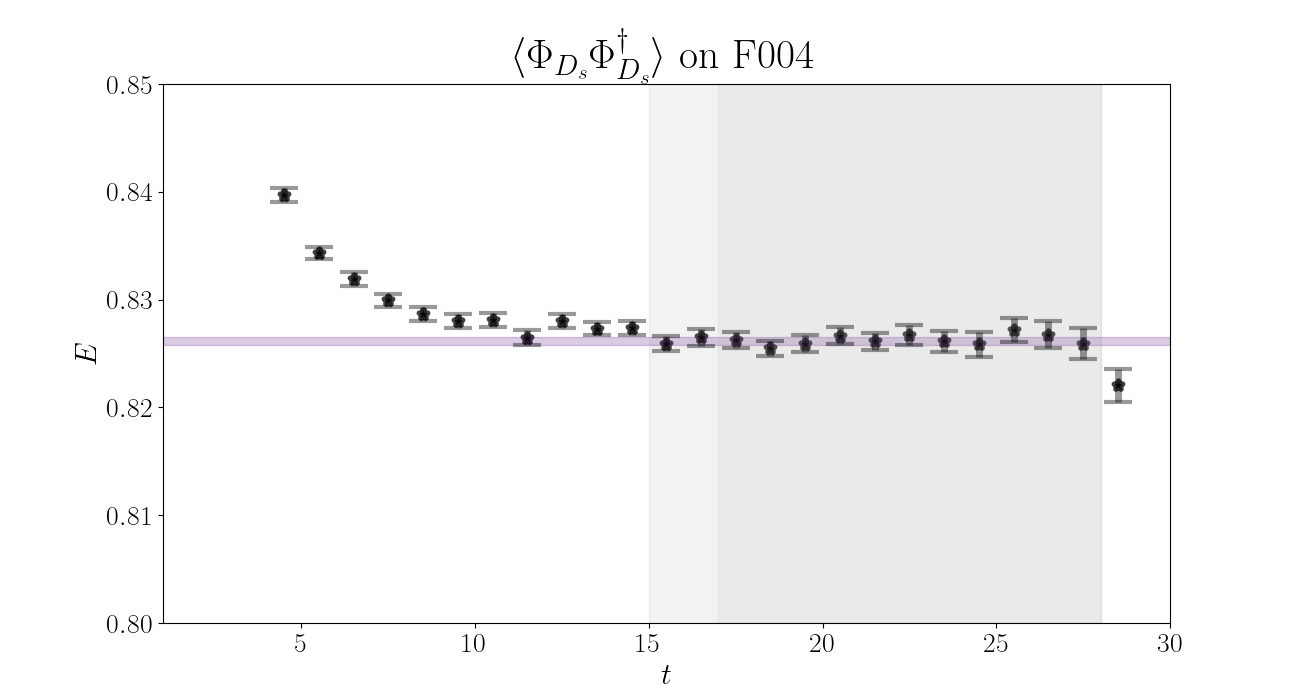}
    
    \includegraphics[width=0.49\linewidth]{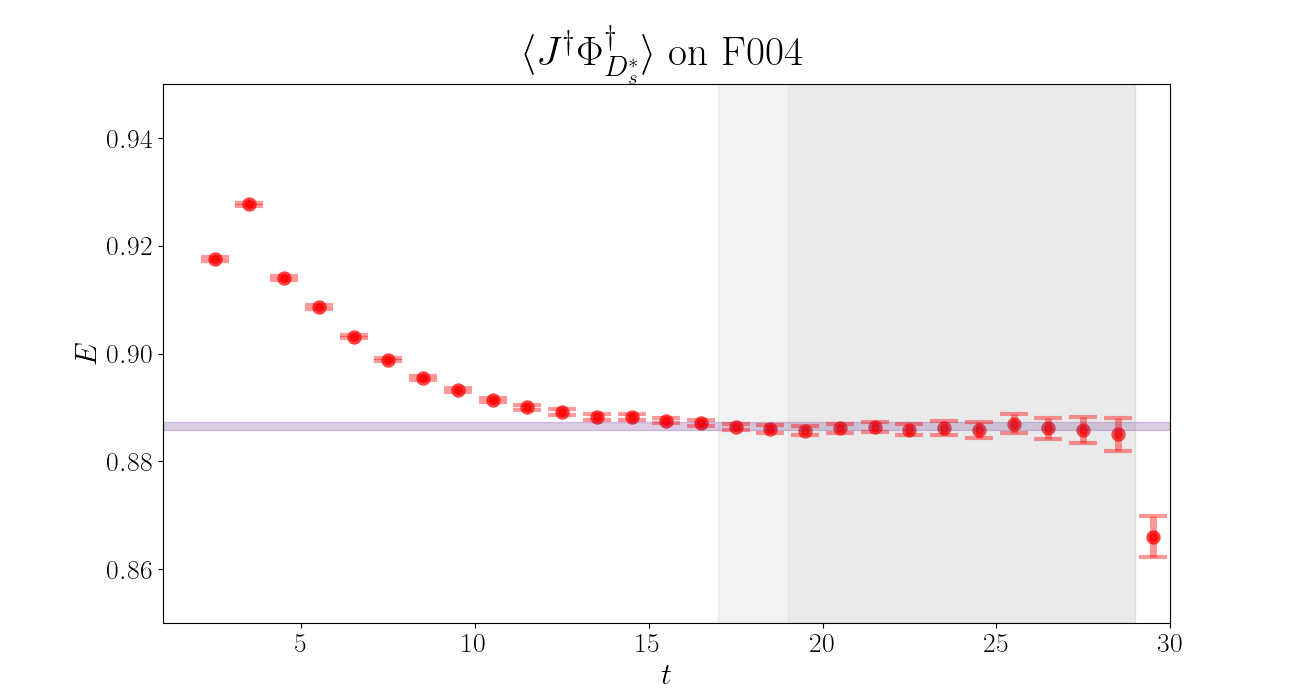}
    \hfill
    \includegraphics[width=0.49\linewidth]{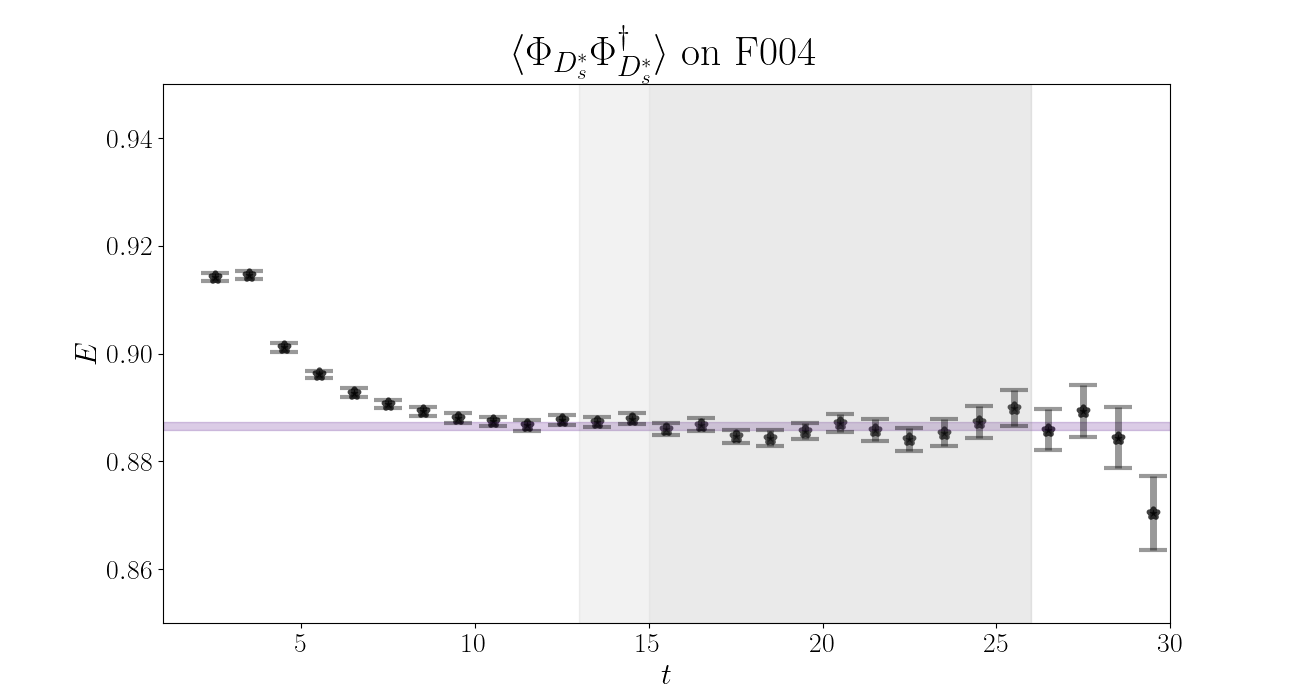}

    \includegraphics[width=0.49\linewidth]{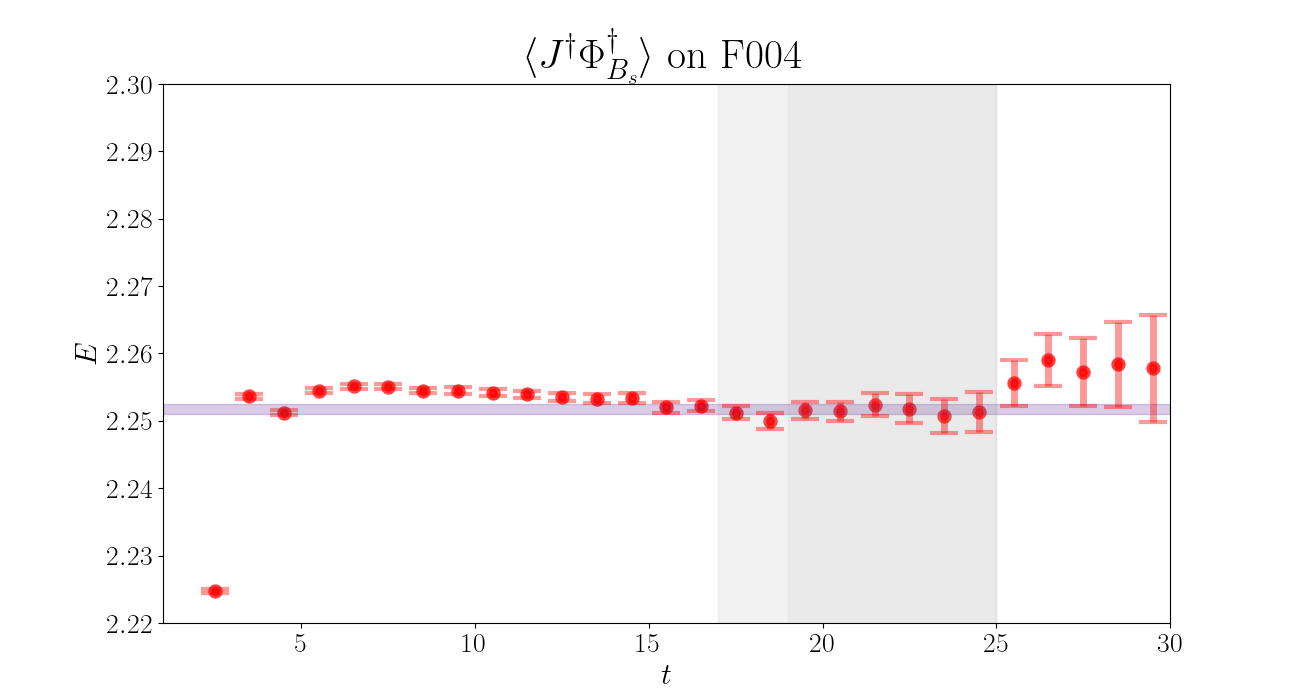}
    \hfill
    \includegraphics[width=0.49\linewidth]{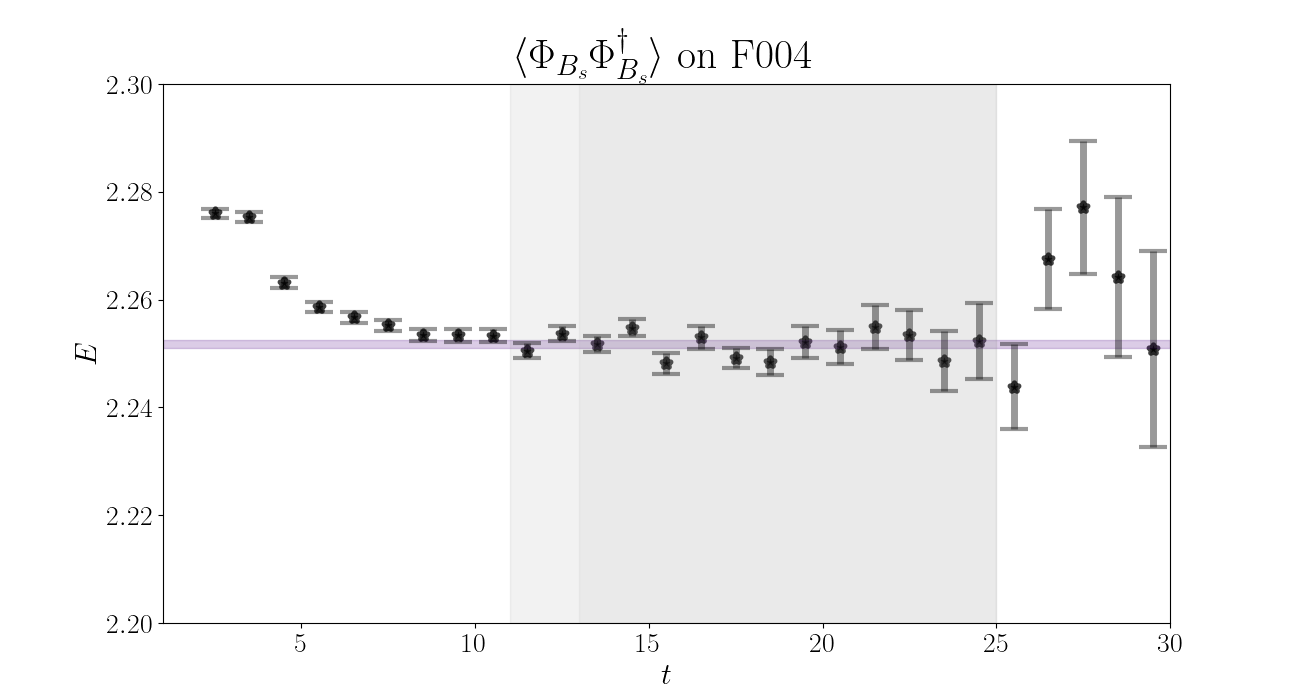}
    
    \includegraphics[width=0.49\linewidth]{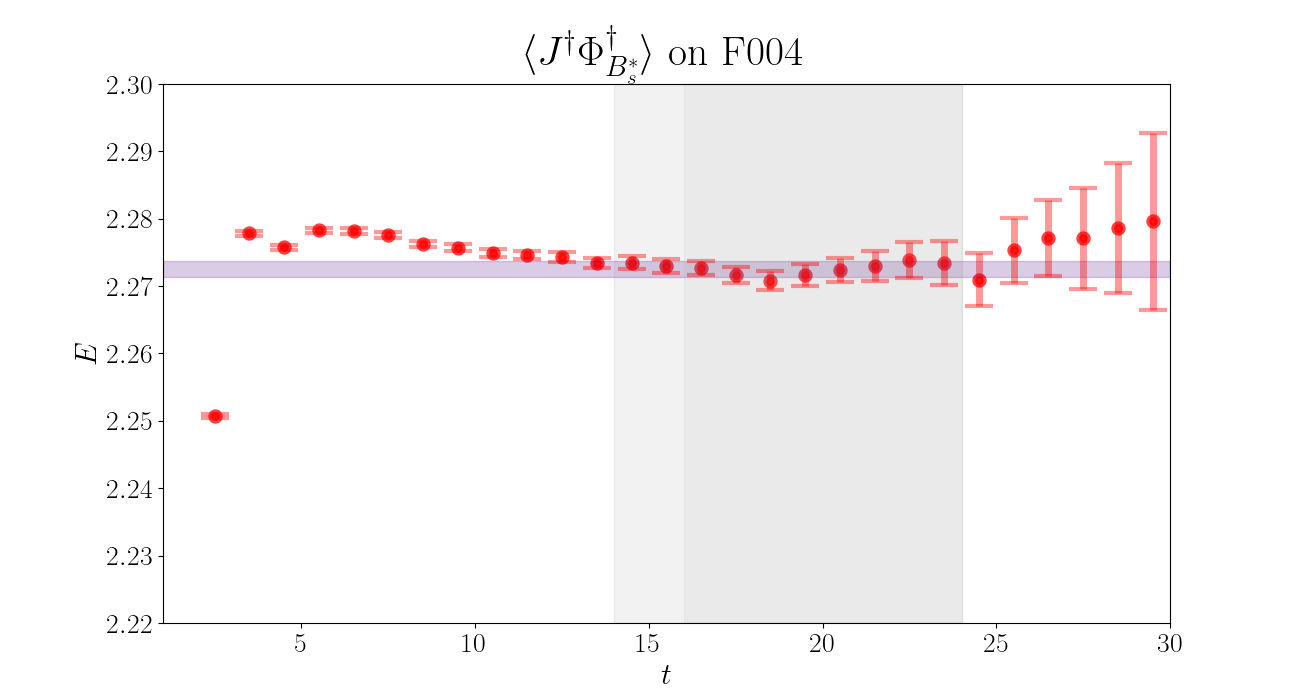}
    \hfill
    \includegraphics[width=0.49\linewidth]{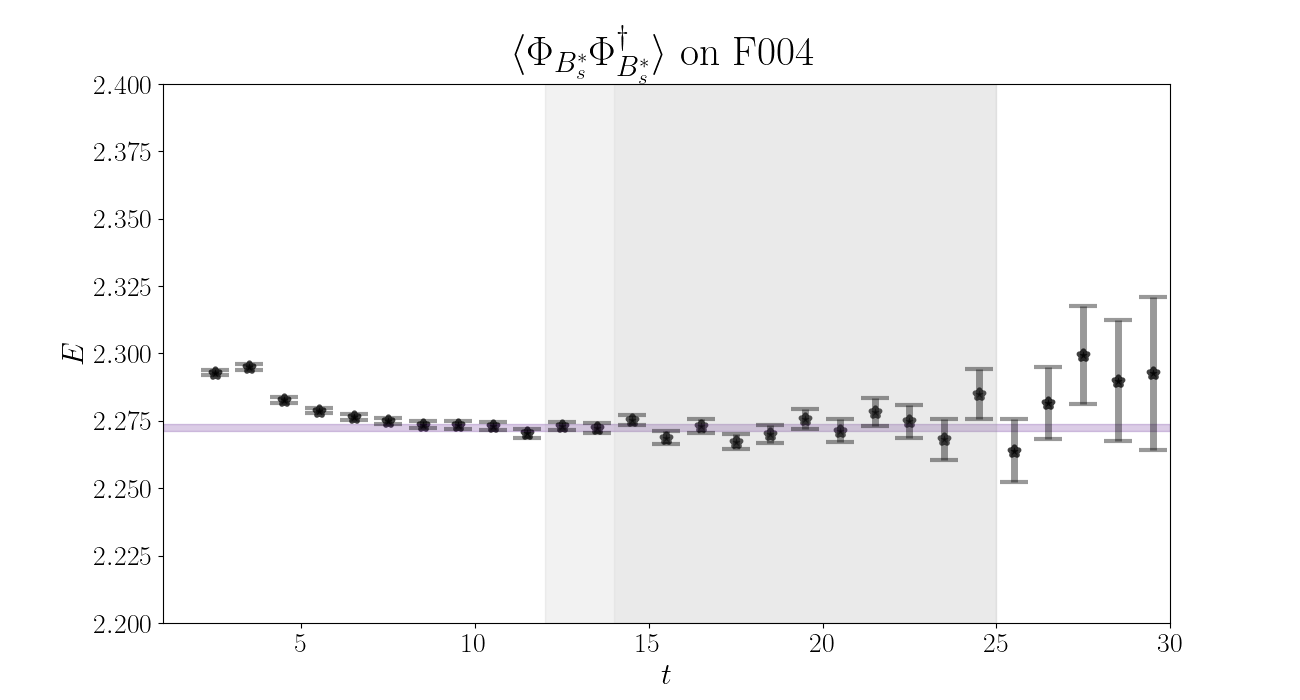}
    \caption{Like Fig.~\protect\ref{fig:negpardecayconstfitplots-C00078}, but for the F004 ensemble.\label{fig:negpardecayconstfitplots-F004}}
\end{figure}

\begin{figure}[H]
    \centering
    \centering
    \includegraphics[width=0.49\linewidth]{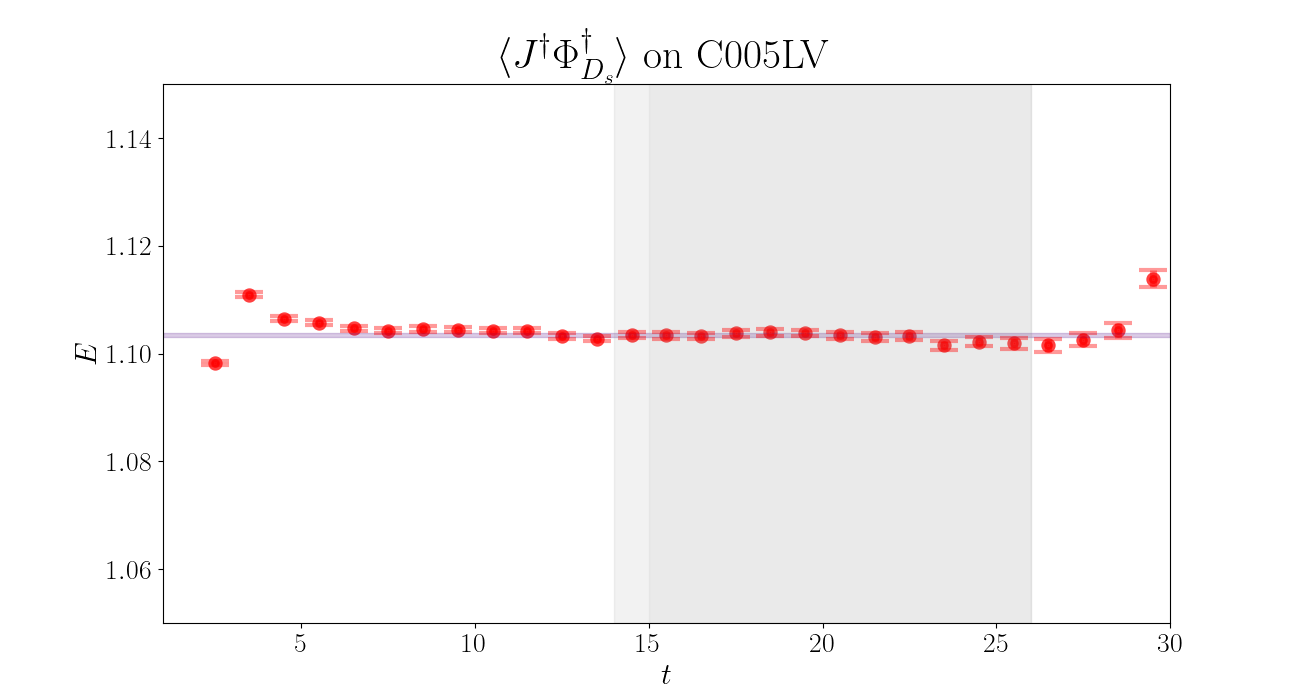}
    \hfill
    \includegraphics[width=0.49\linewidth]{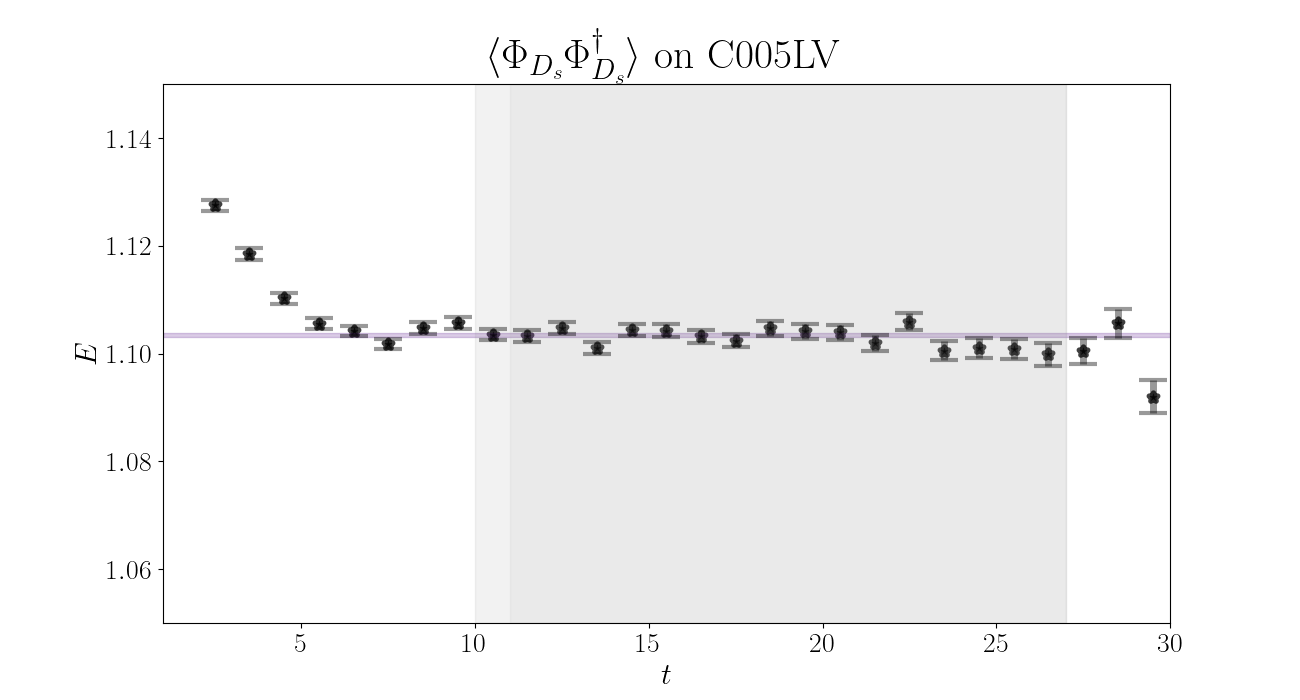}
    
    \includegraphics[width=0.49\linewidth]{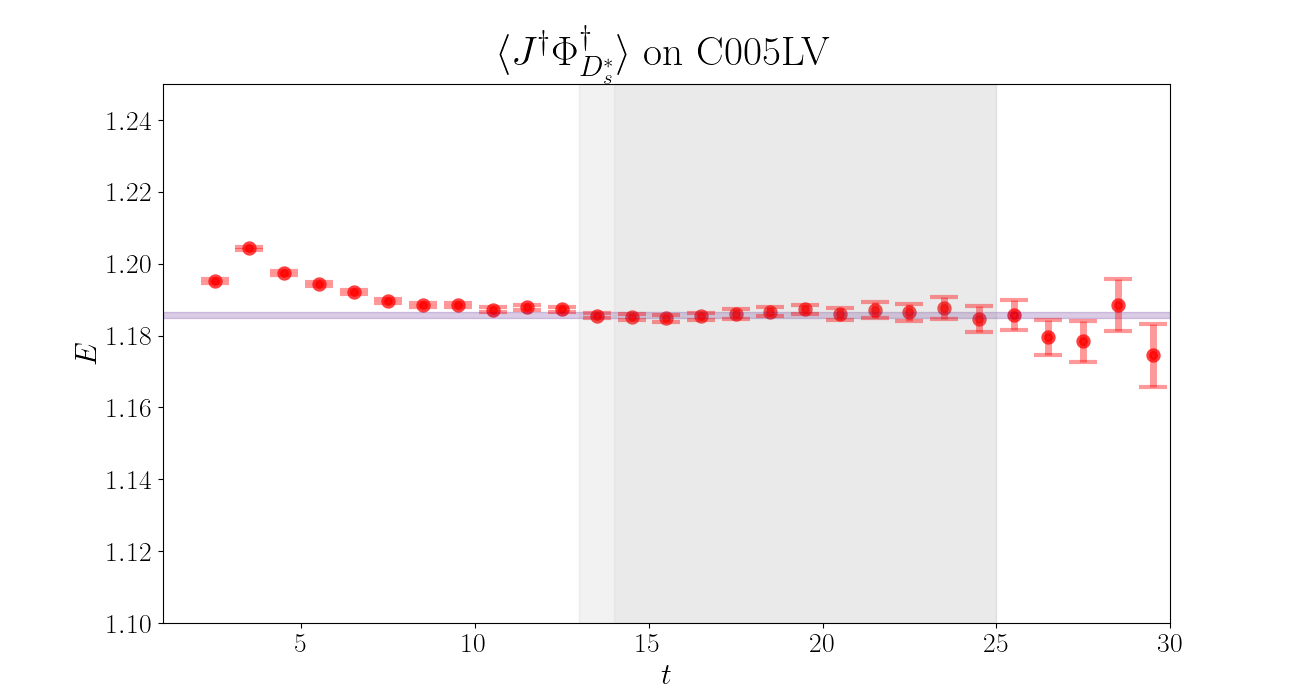}
    \hfill
    \includegraphics[width=0.49\linewidth]{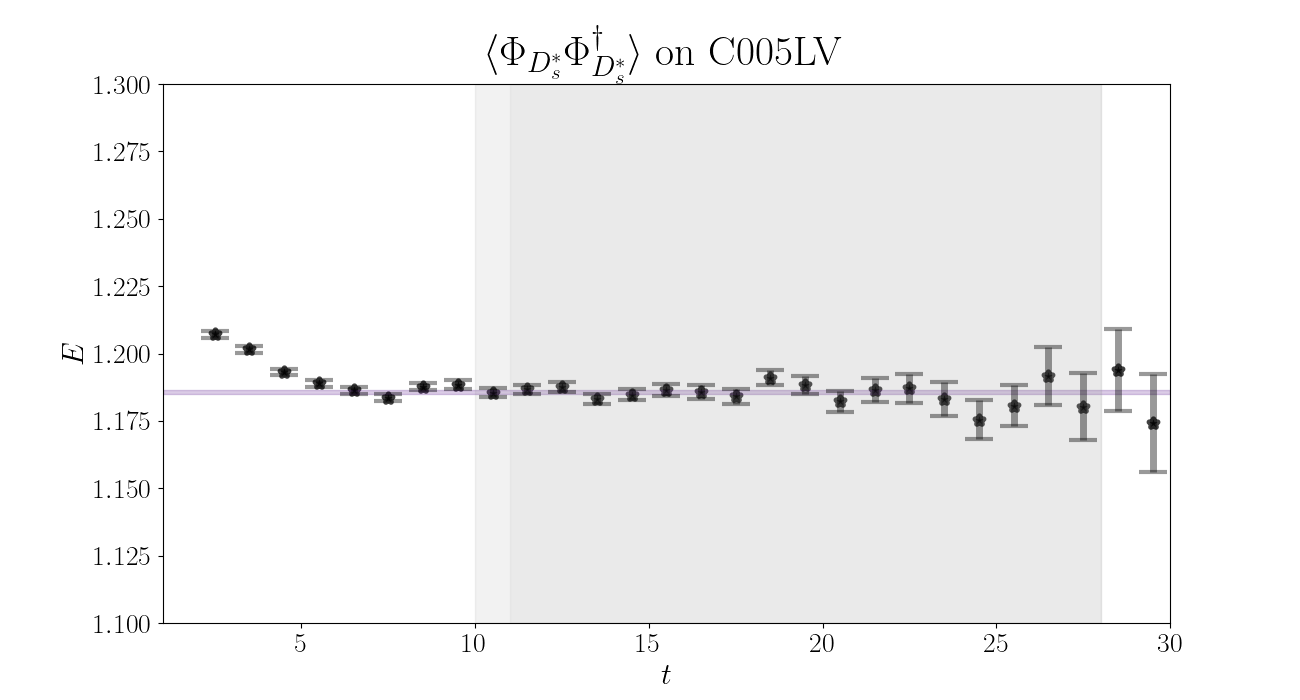}

    \includegraphics[width=0.49\linewidth]{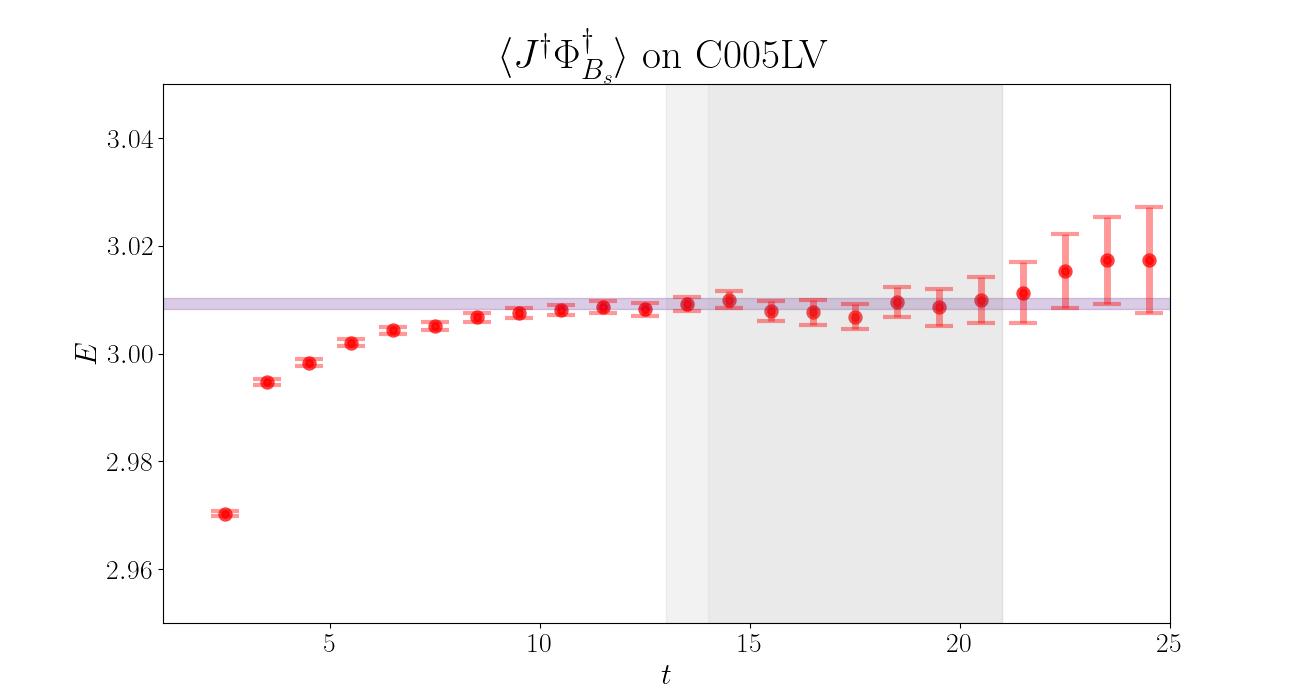}
    \hfill
    \includegraphics[width=0.49\linewidth]{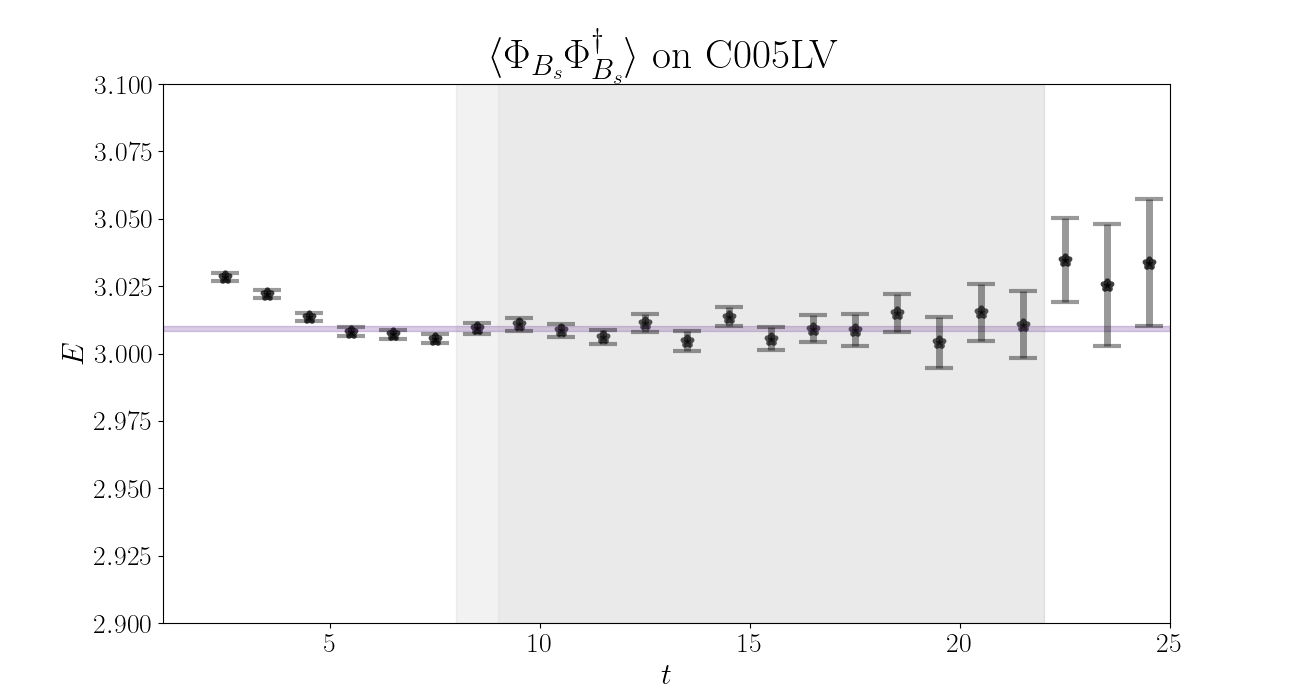}
    
    \includegraphics[width=0.49\linewidth]{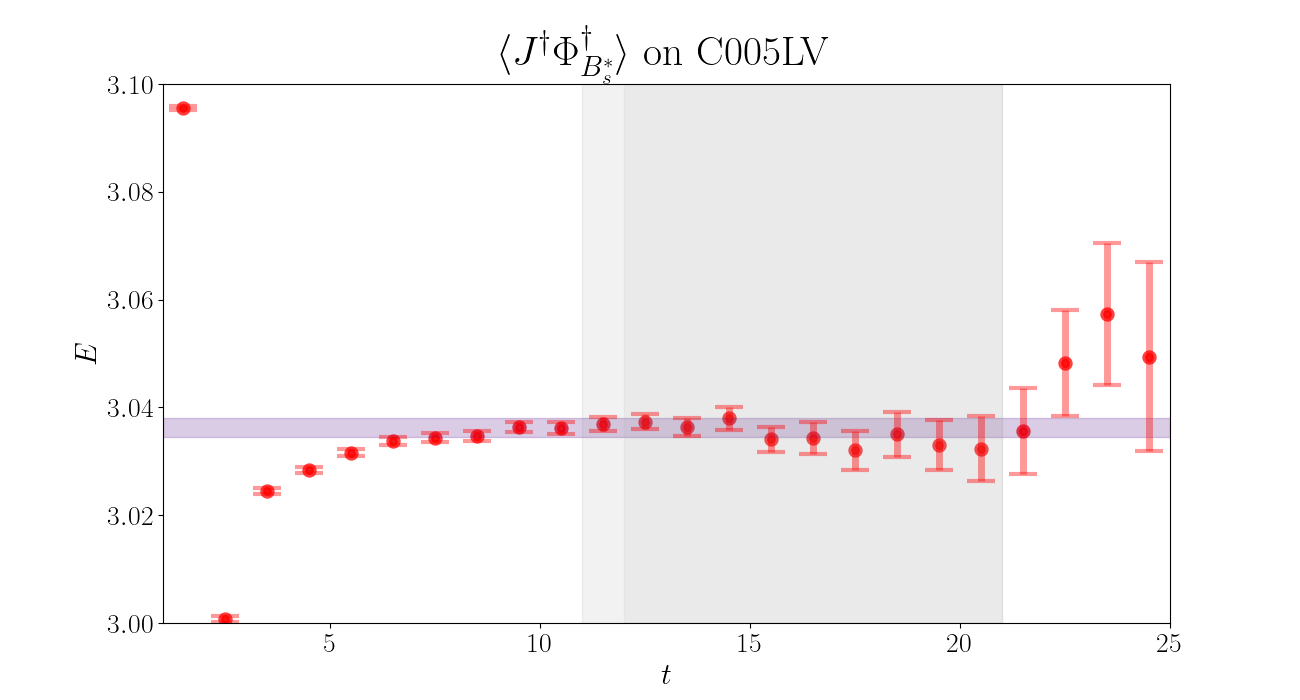}
    \hfill
    \includegraphics[width=0.49\linewidth]{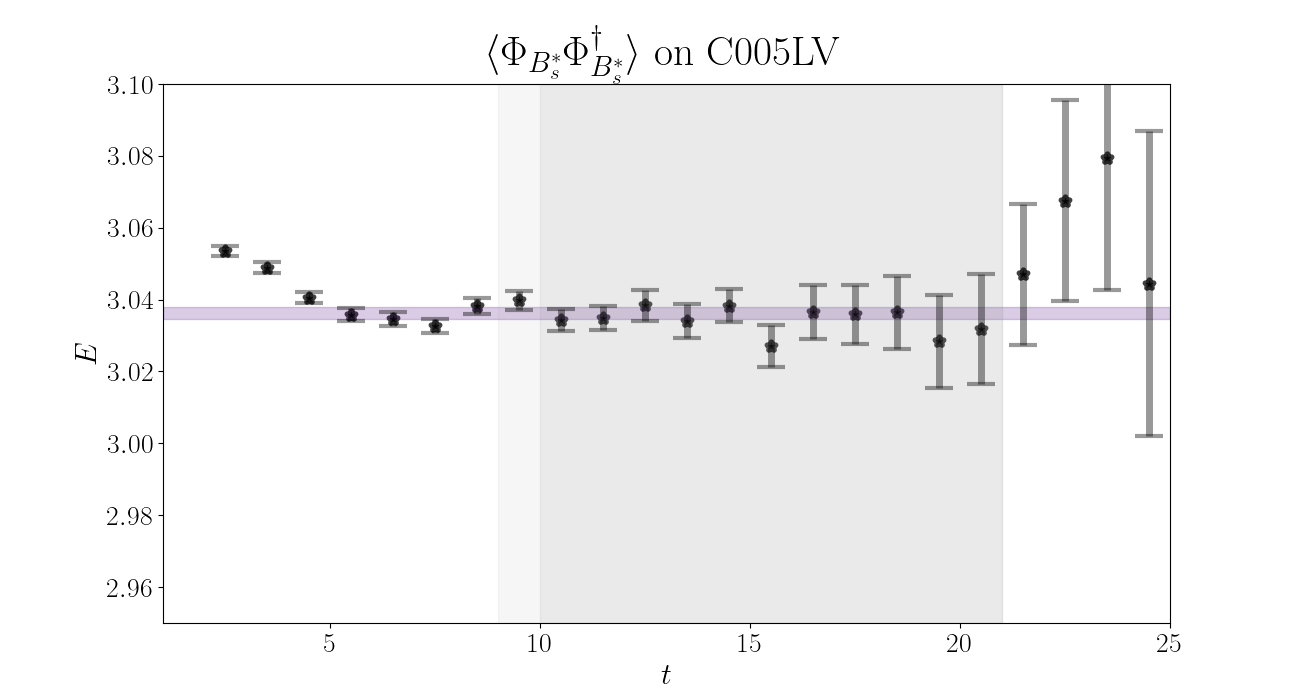}
    \caption{Like Fig.~\protect\ref{fig:negpardecayconstfitplots-C00078}, but for the C005LV ensemble.\label{fig:negpardecayconstfitplots-C005LV}}
\end{figure}

\begin{figure}[H]
    \centering
    \centering
    \includegraphics[width=0.49\linewidth]{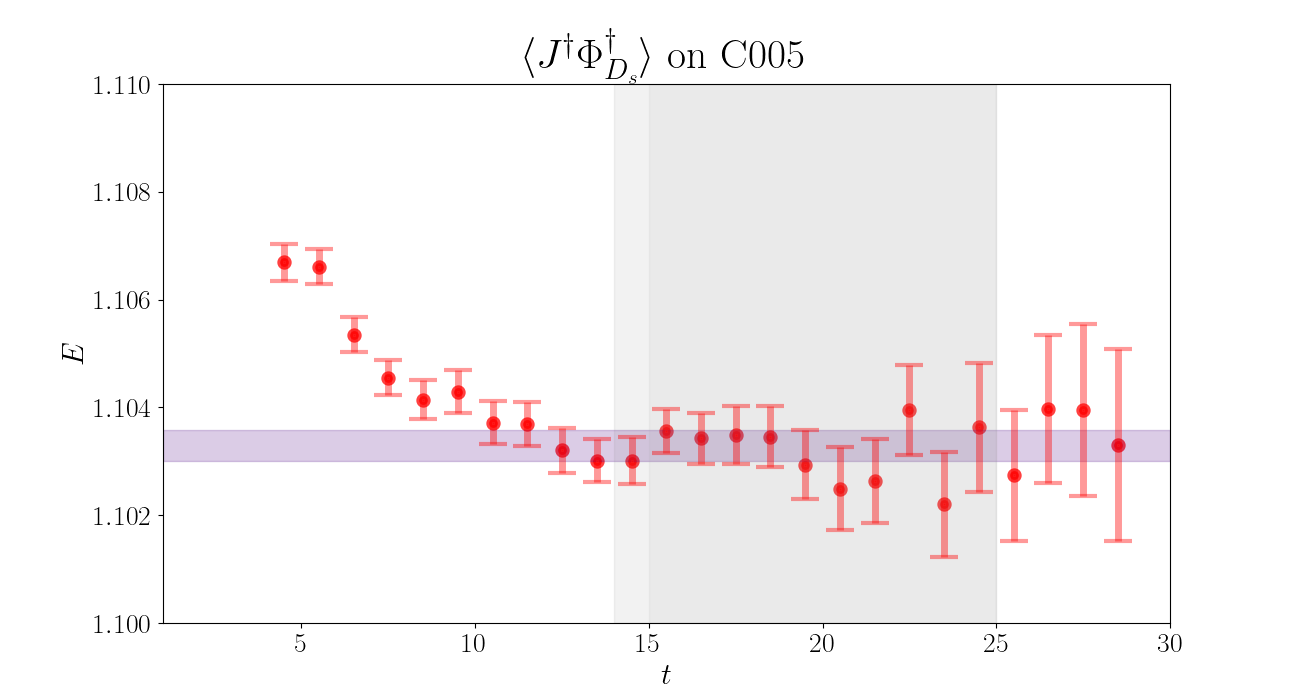}
    \hfill
    \includegraphics[width=0.49\linewidth]{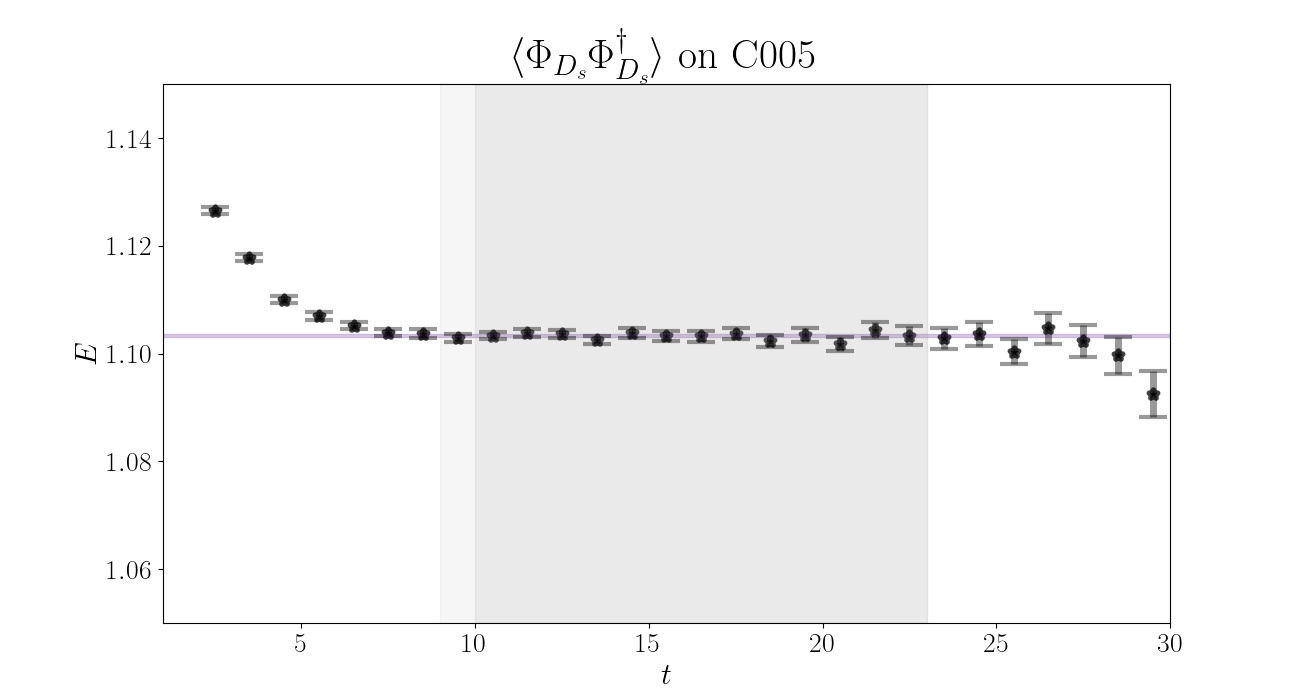}
    
    \includegraphics[width=0.49\linewidth]{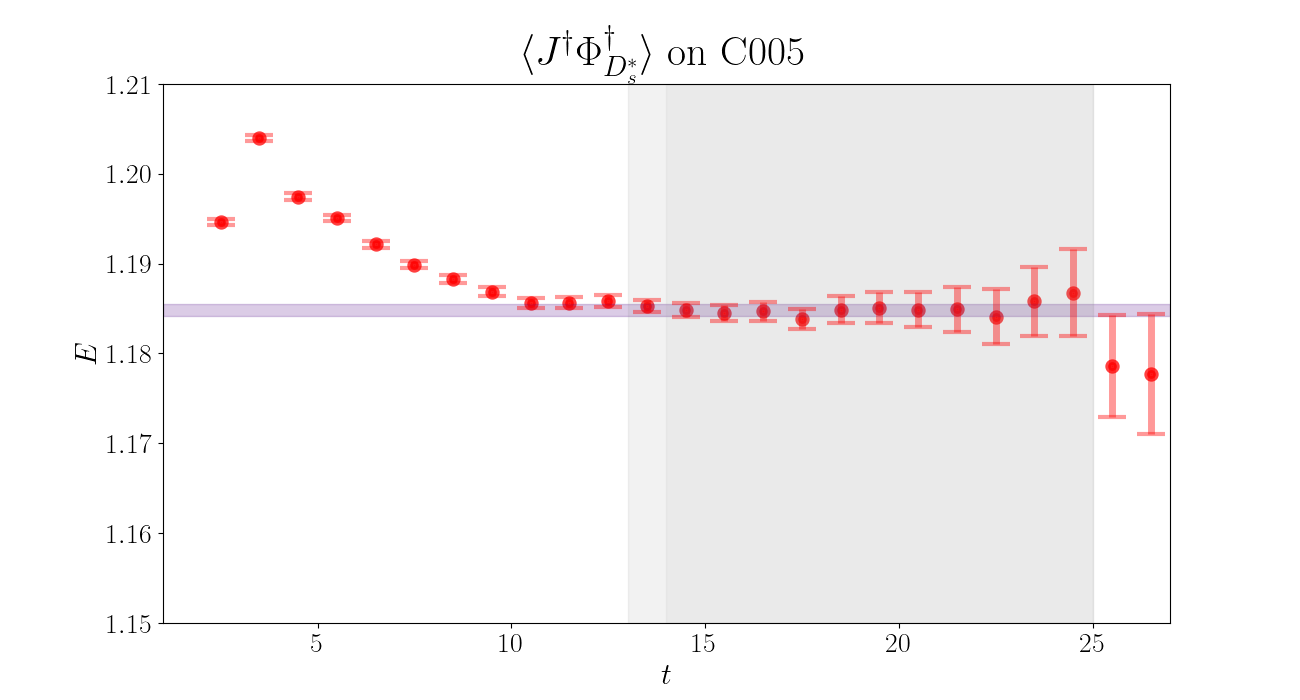}
    \hfill
    \includegraphics[width=0.49\linewidth]{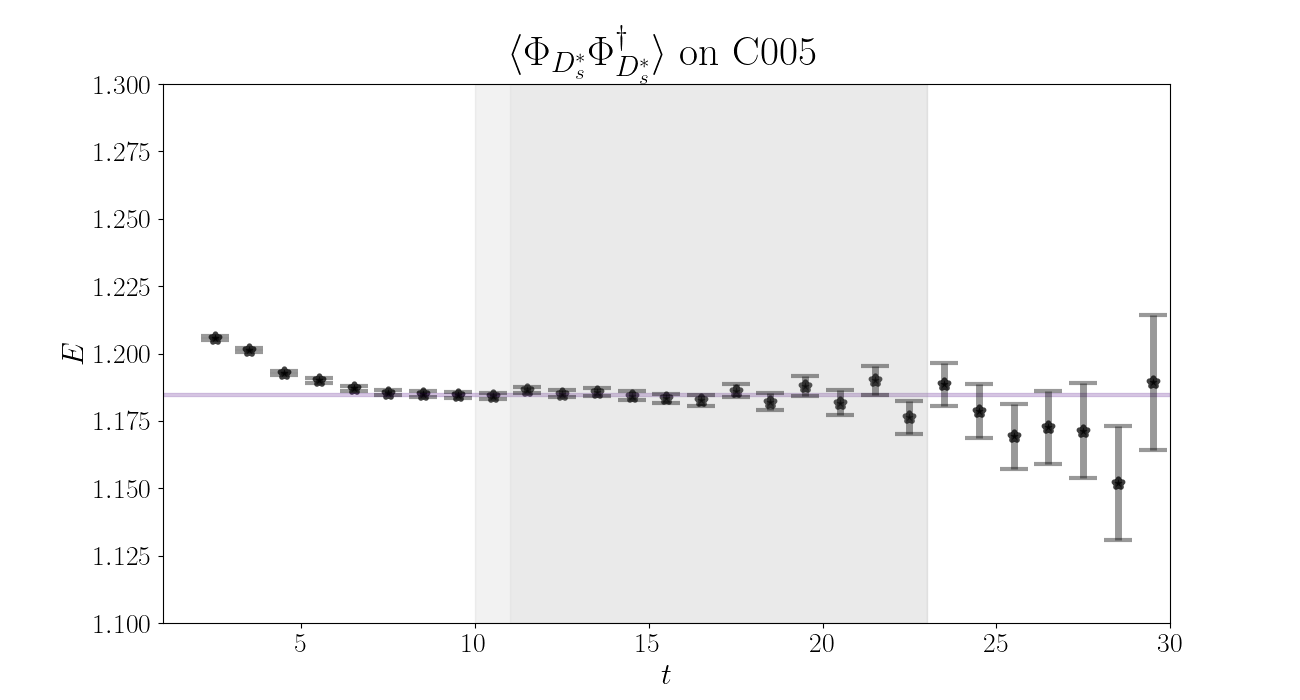}

    \includegraphics[width=0.49\linewidth]{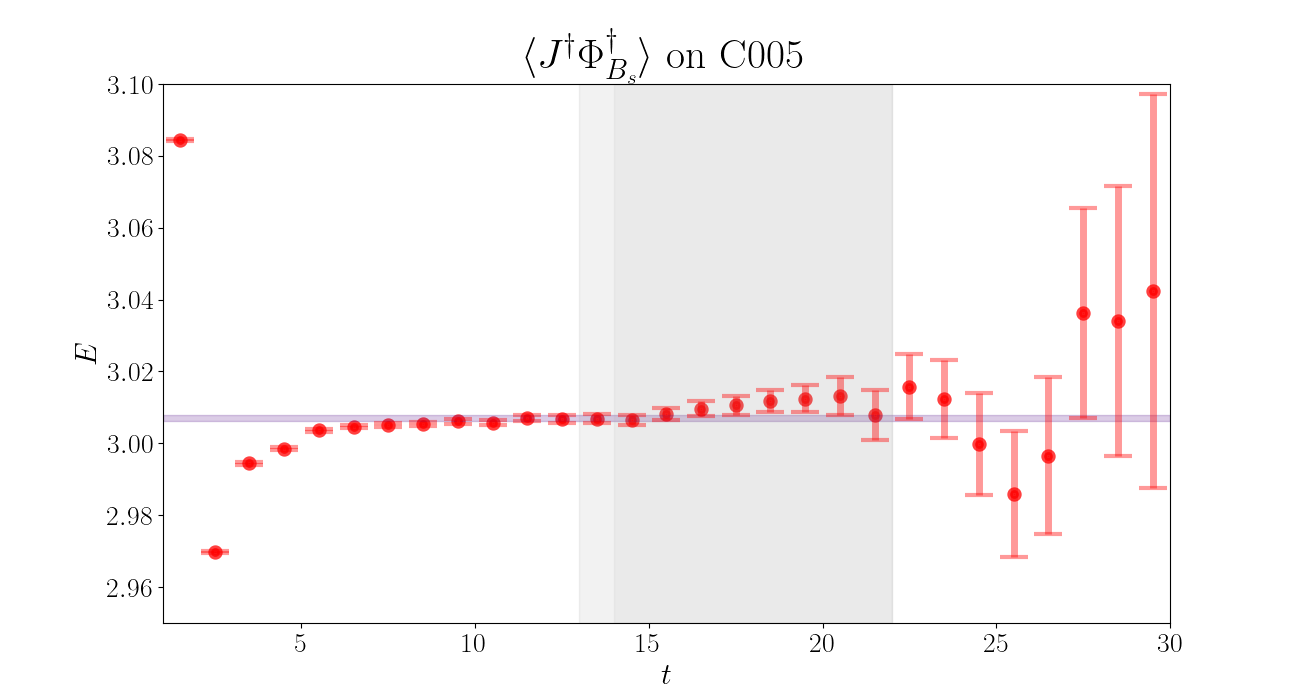}
    \hfill
    \includegraphics[width=0.49\linewidth]{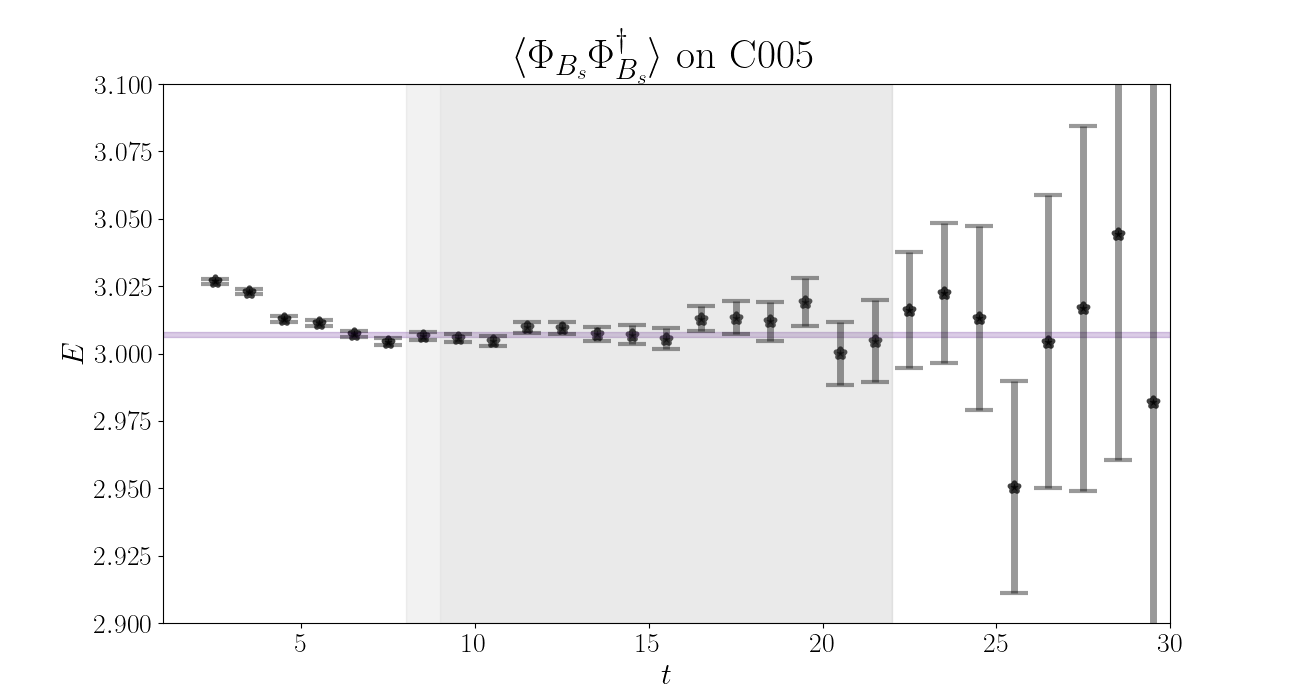}
    
    \includegraphics[width=0.49\linewidth]{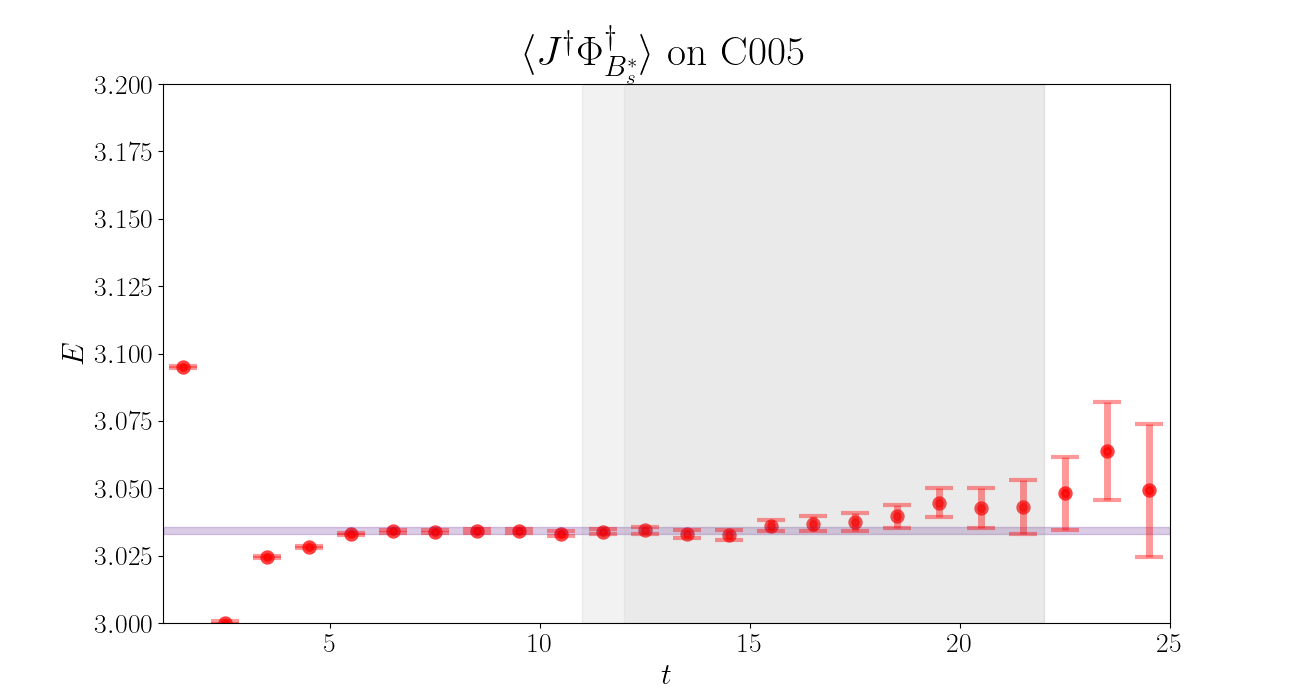}
    \hfill
    \includegraphics[width=0.49\linewidth]{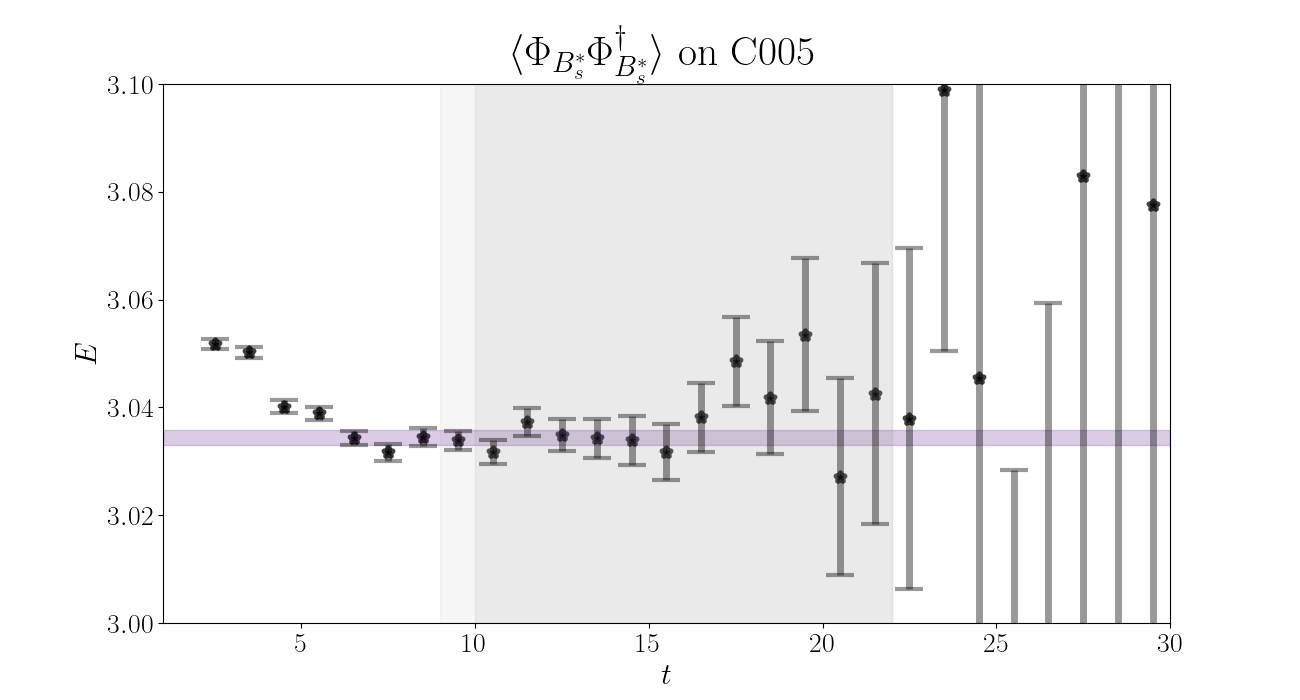}
    \caption{Like Fig.~\protect\ref{fig:negpardecayconstfitplots-C00078}, but for the C005 ensemble.\label{fig:negpardecayconstfitplots-C005}}
\end{figure}

\begin{figure}[H]
    \centering
    \centering
    \includegraphics[width=0.49\linewidth]{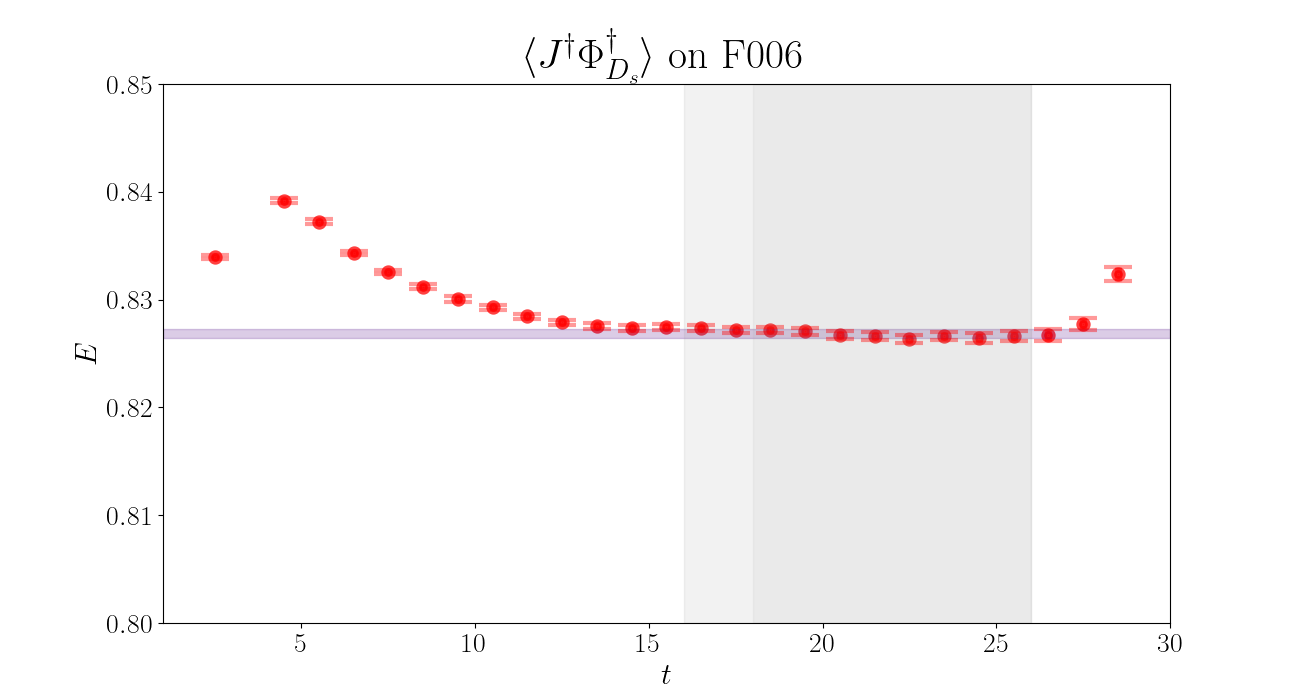}
    \hfill
    \includegraphics[width=0.49\linewidth]{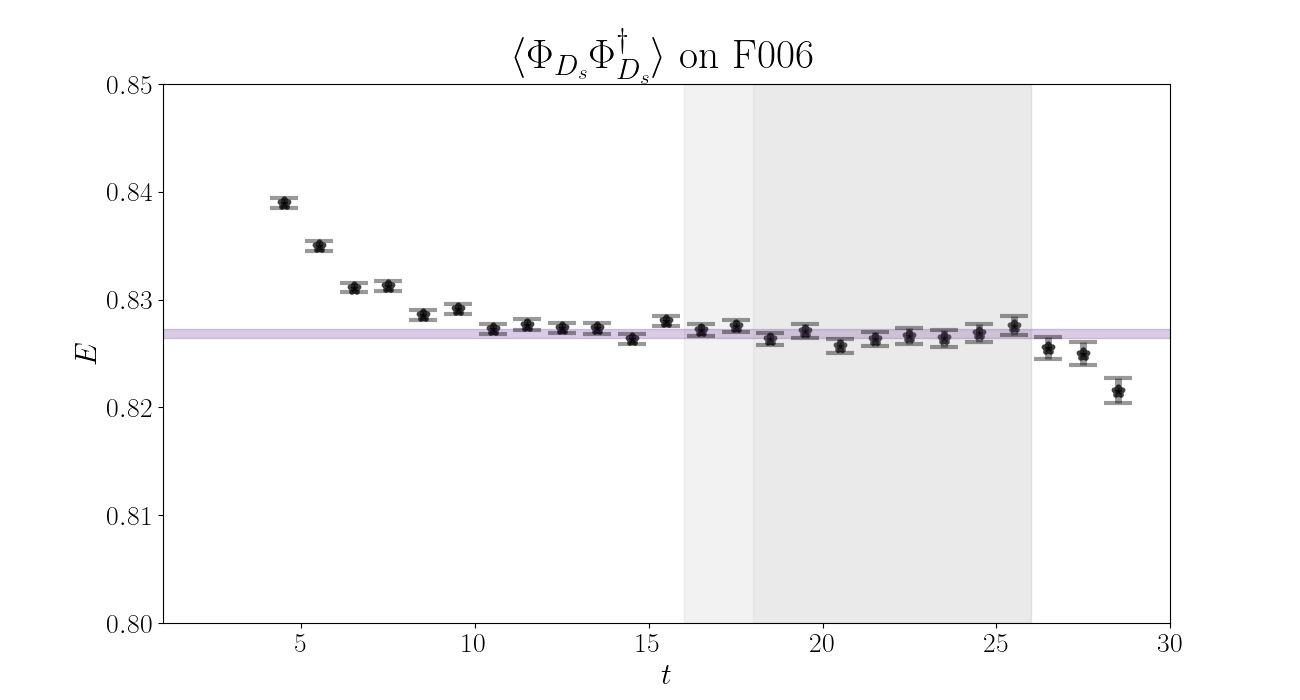}
    
    \includegraphics[width=0.49\linewidth]{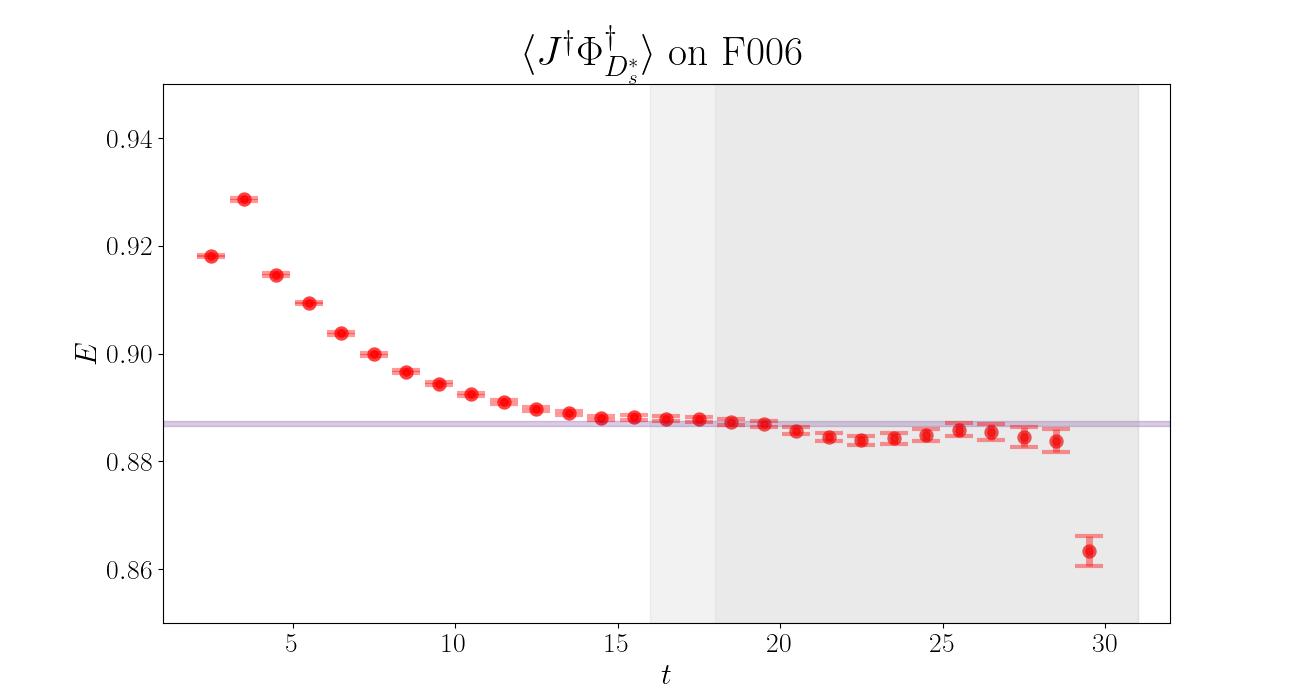}
    \hfill
    \includegraphics[width=0.49\linewidth]{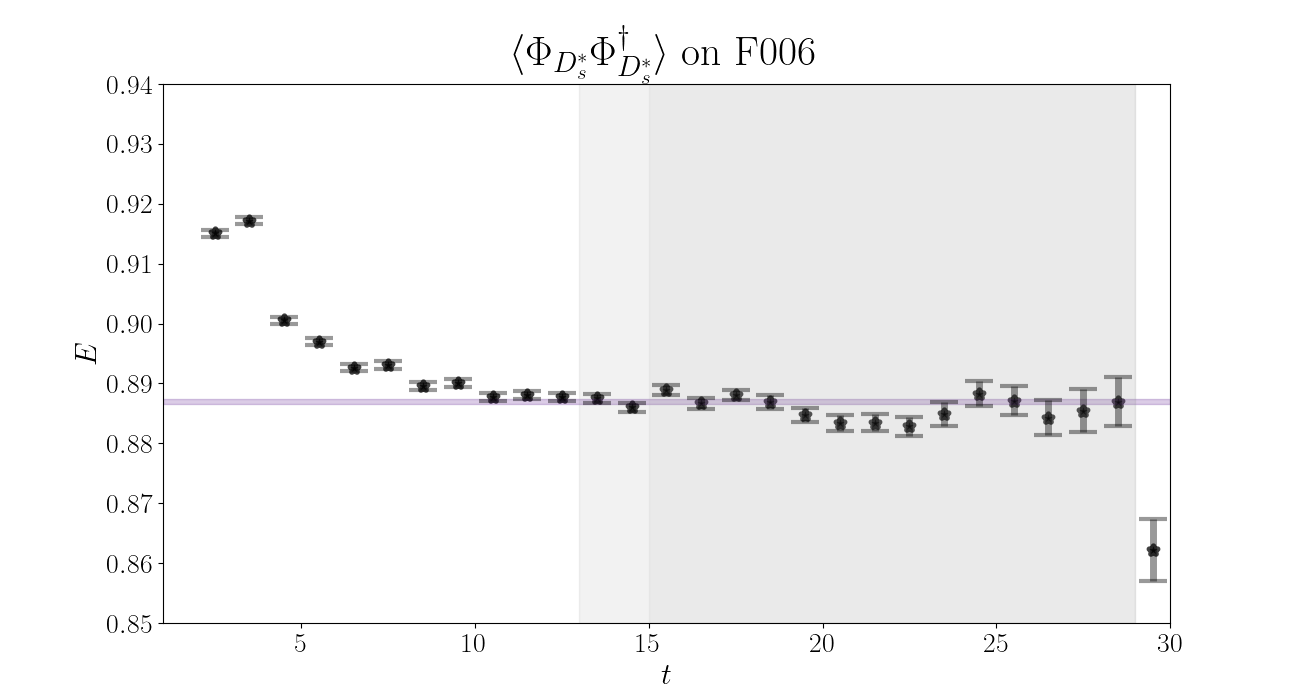}

    \includegraphics[width=0.49\linewidth]{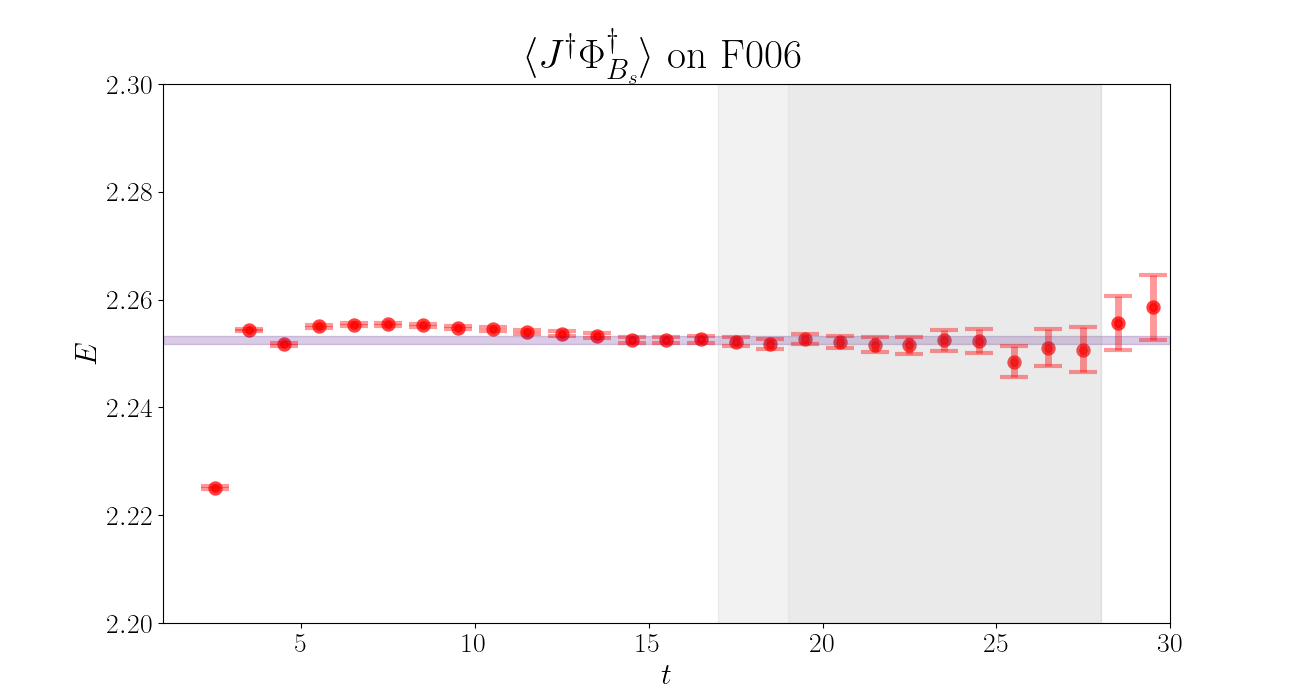}
    \hfill
    \includegraphics[width=0.49\linewidth]{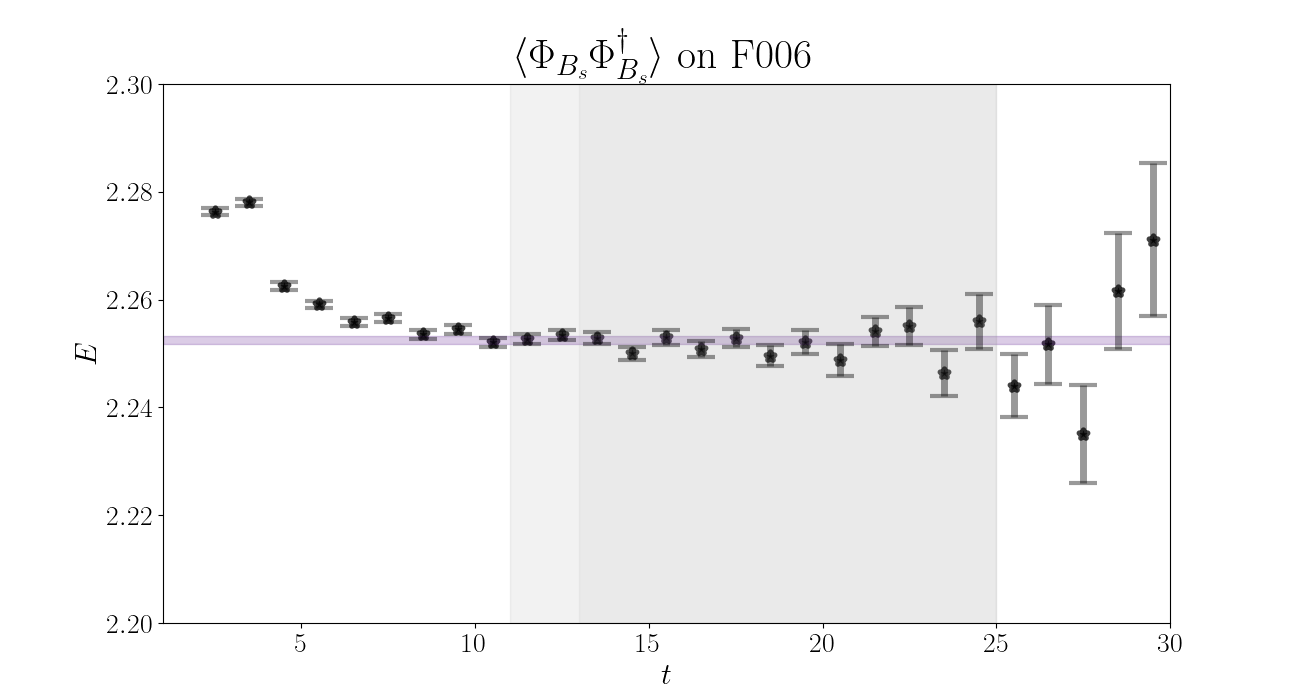}
    
    \includegraphics[width=0.49\linewidth]{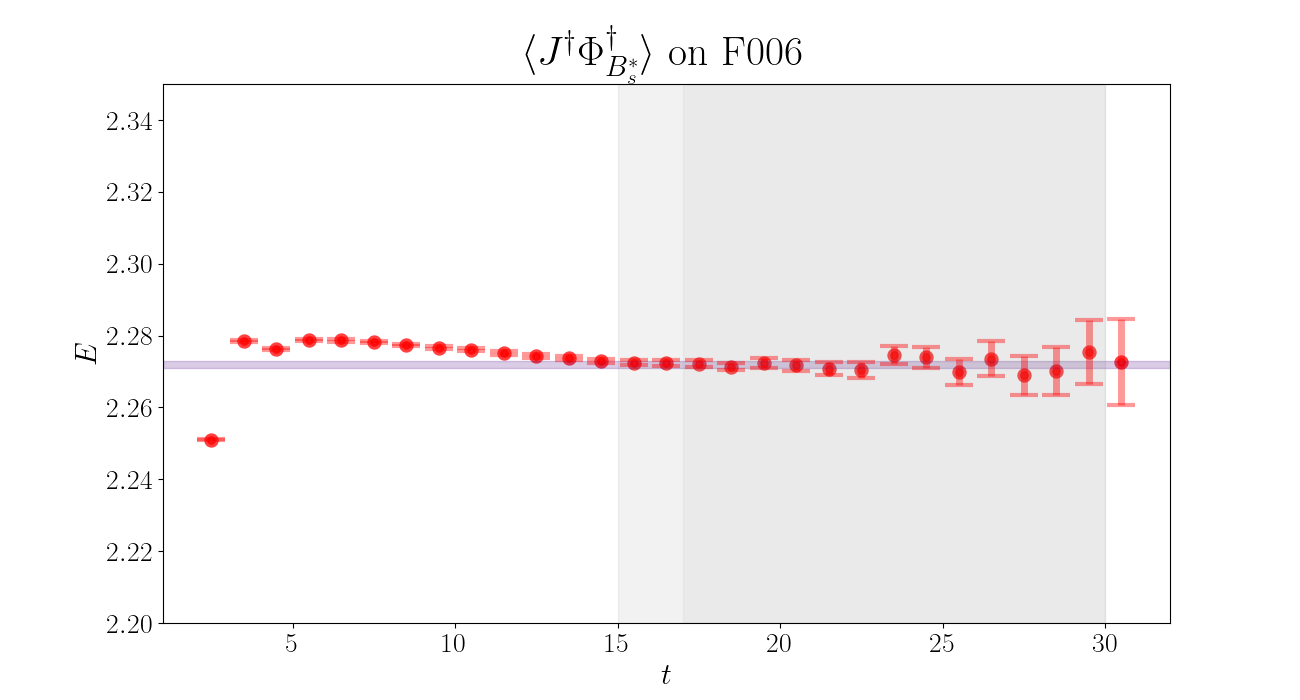}
    \hfill
    \includegraphics[width=0.49\linewidth]{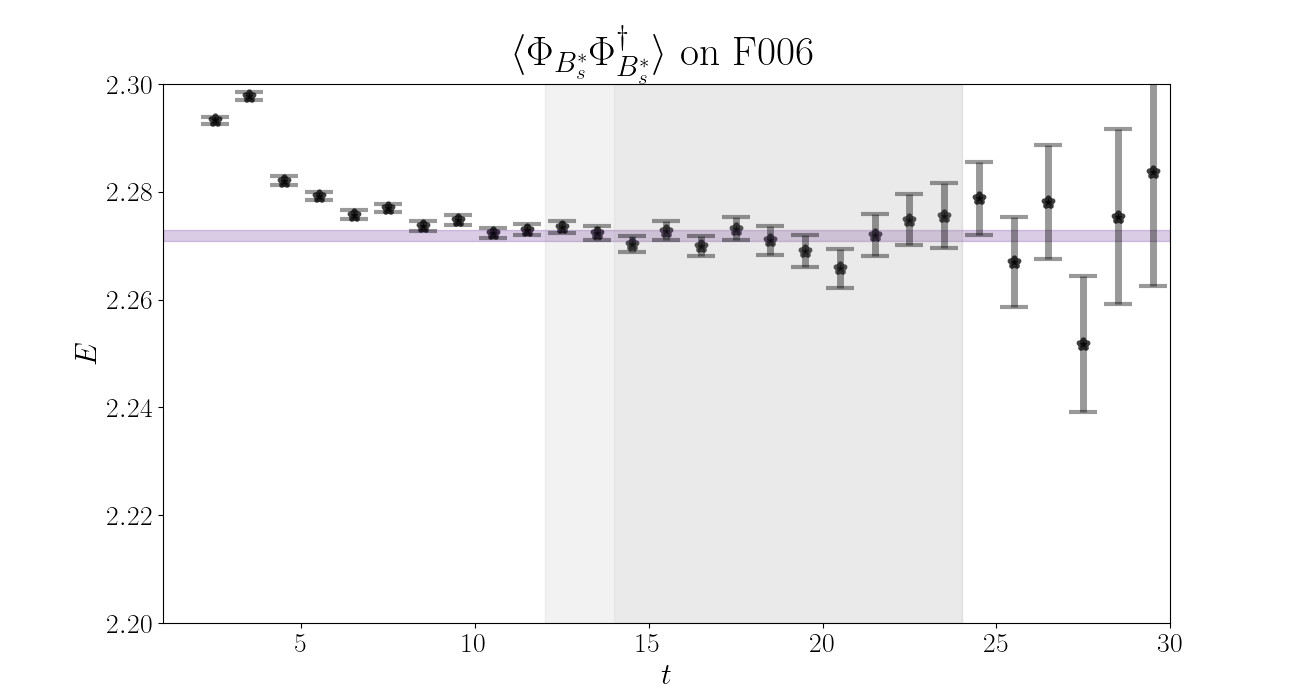}
    \caption{Like Fig.~\protect\ref{fig:negpardecayconstfitplots-C00078}, but for the F006 ensemble.\label{fig:negpardecayconstfitplots-F006}}
\end{figure}

\begin{figure}[H]
    \centering
    \centering
    \includegraphics[width=0.49\linewidth]{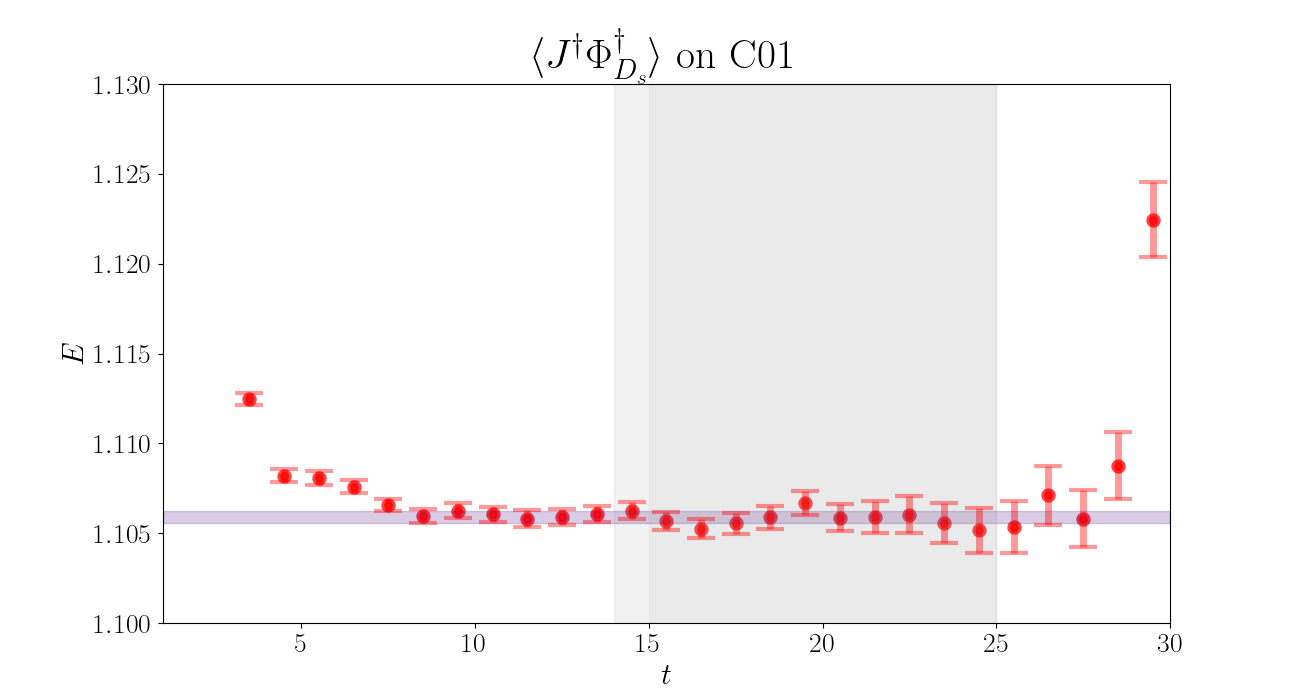}
    \hfill
    \includegraphics[width=0.49\linewidth]{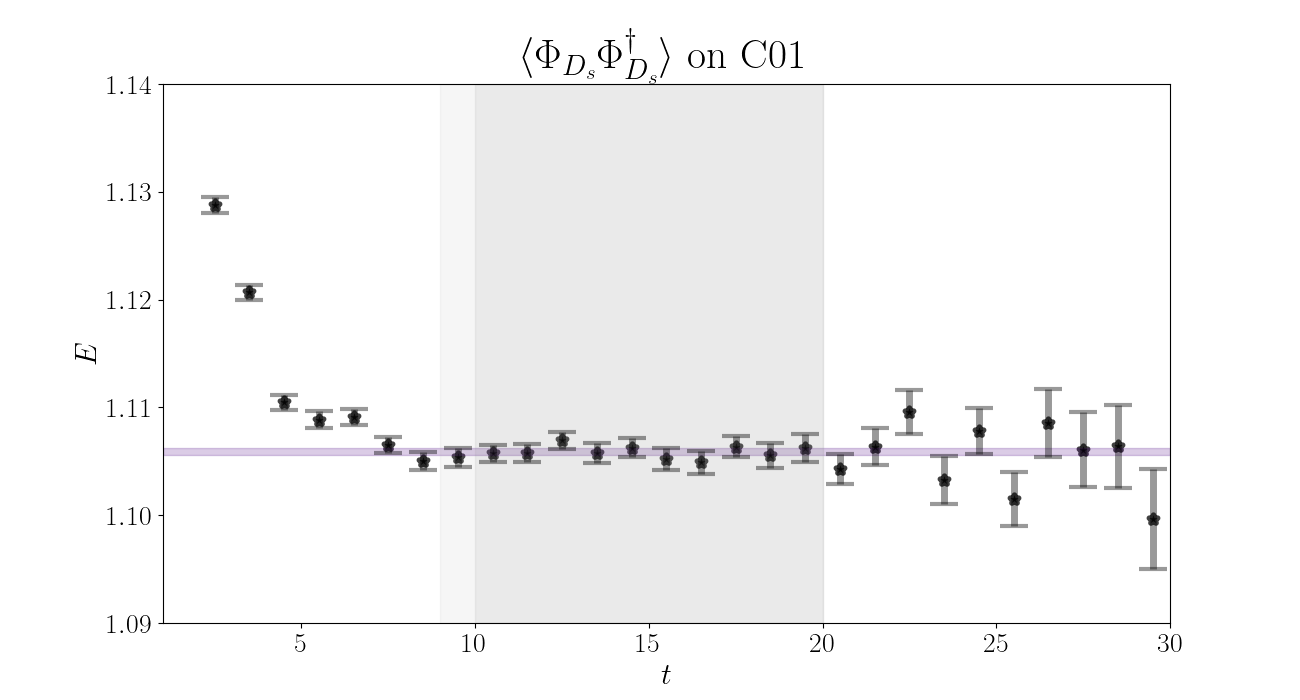}
    
    \includegraphics[width=0.49\linewidth]{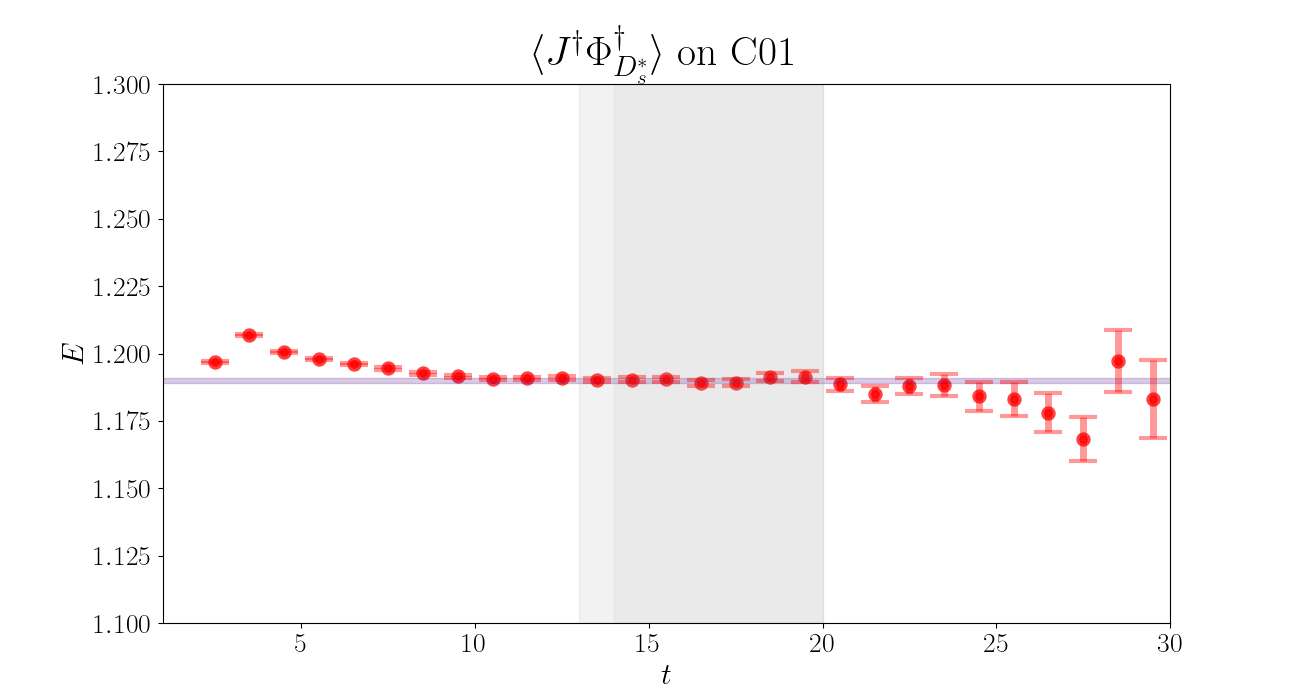}
    \hfill
    \includegraphics[width=0.49\linewidth]{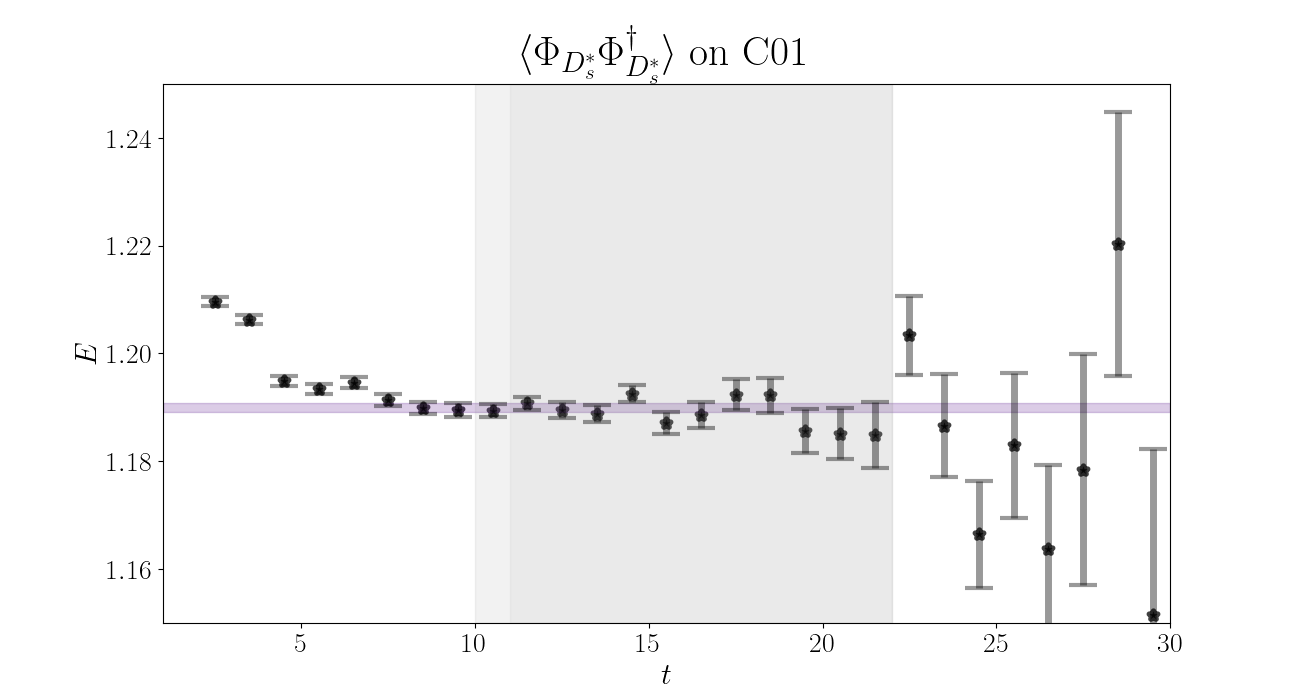}

    \includegraphics[width=0.49\linewidth]{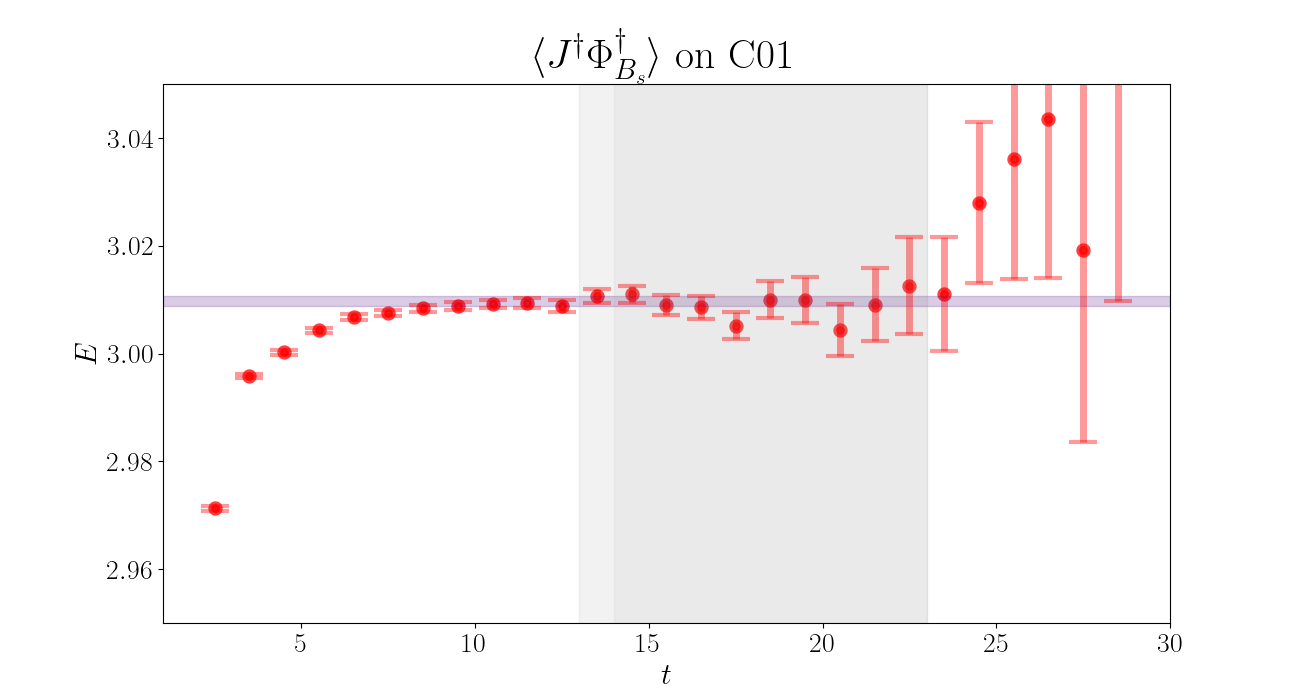}
    \hfill
    \includegraphics[width=0.49\linewidth]{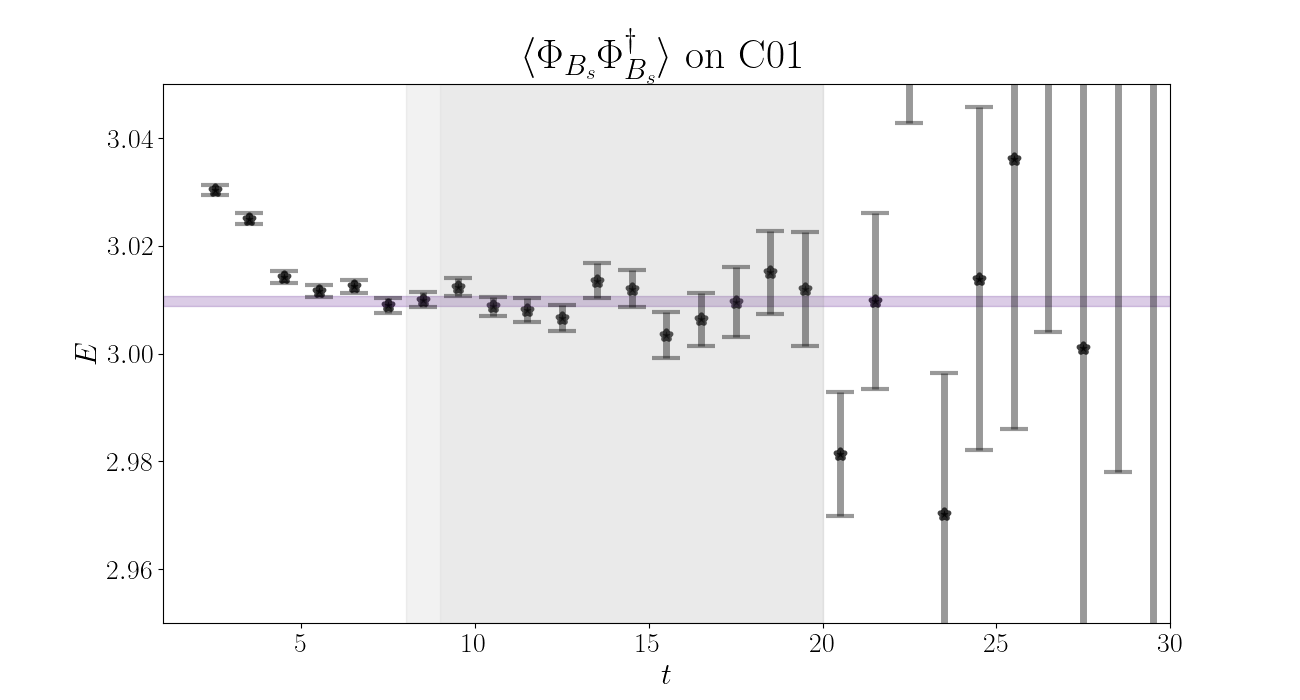}
    
    \includegraphics[width=0.49\linewidth]{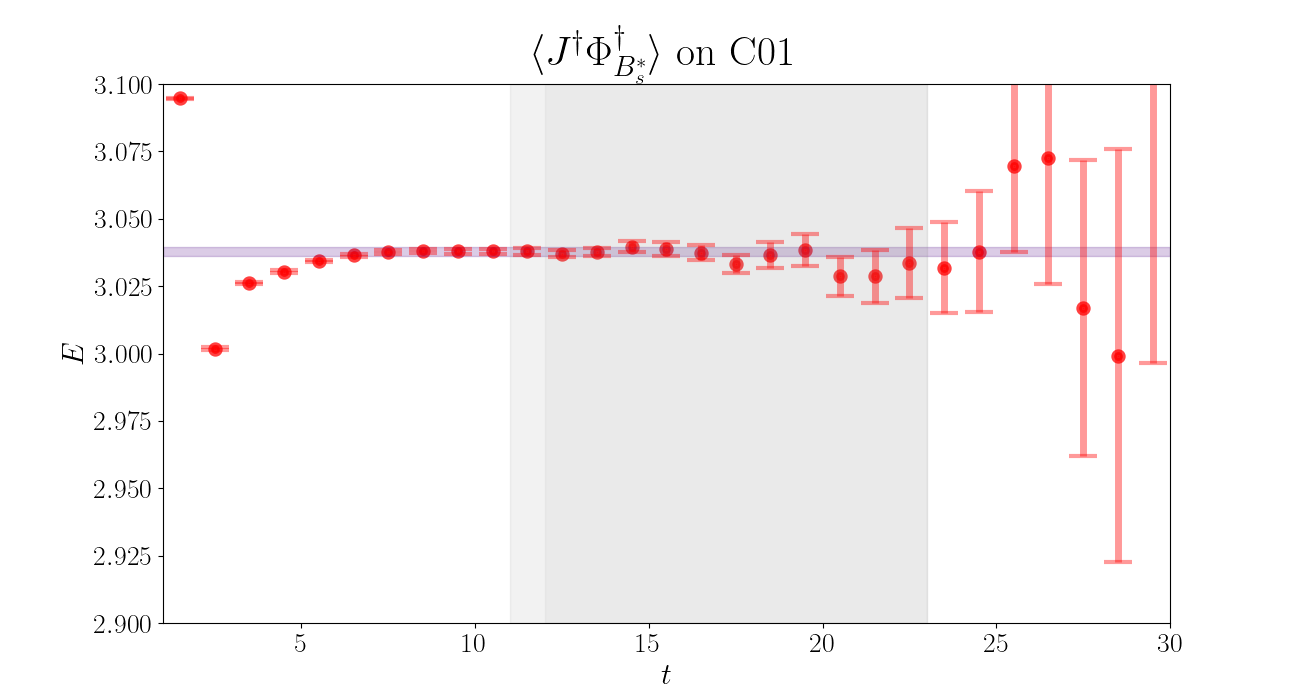}
    \hfill
    \includegraphics[width=0.49\linewidth]{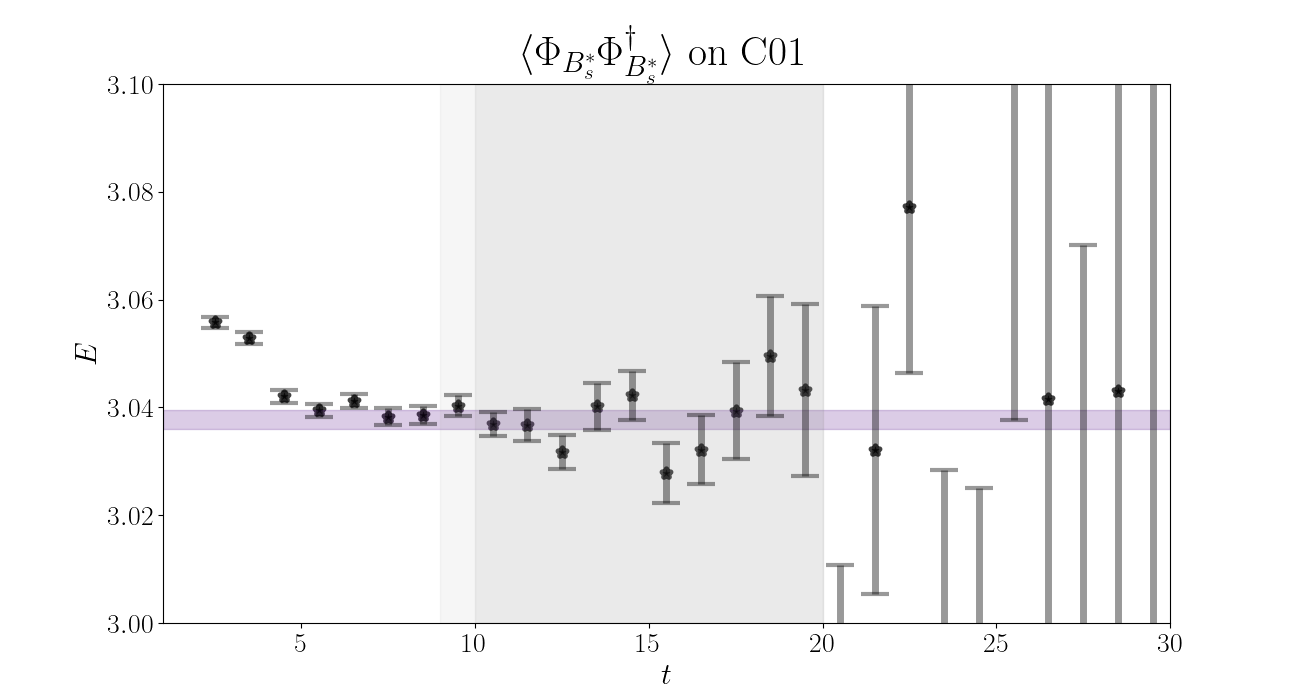}
    \caption{Like Fig.~\protect\ref{fig:negpardecayconstfitplots-C00078}, but for the C01 ensemble.\label{fig:negpardecayconstfitplots-C01}}
\end{figure}

Plots of the negative-parity hadron-current and hadron-hadron correlator effective energies are provided here for all ensembles in Figs.~\ref{fig:negpardecayconstfitplots-C00078}-\ref{fig:negpardecayconstfitplots-C01}.

\subsection{Positive-Parity Correlation Matrices}
\label{sec:corrplots}

\begin{figure}

    \includegraphics[width=0.49\linewidth]{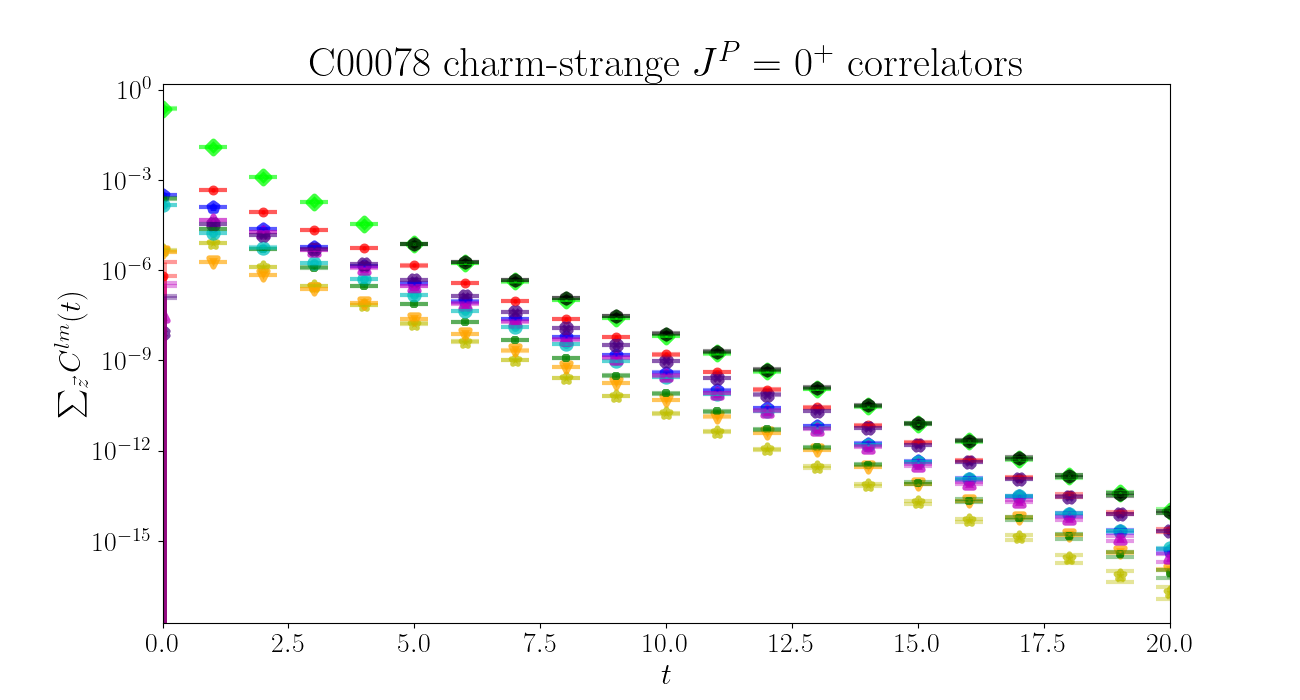}
    \hfill
    \includegraphics[width=0.49\linewidth]{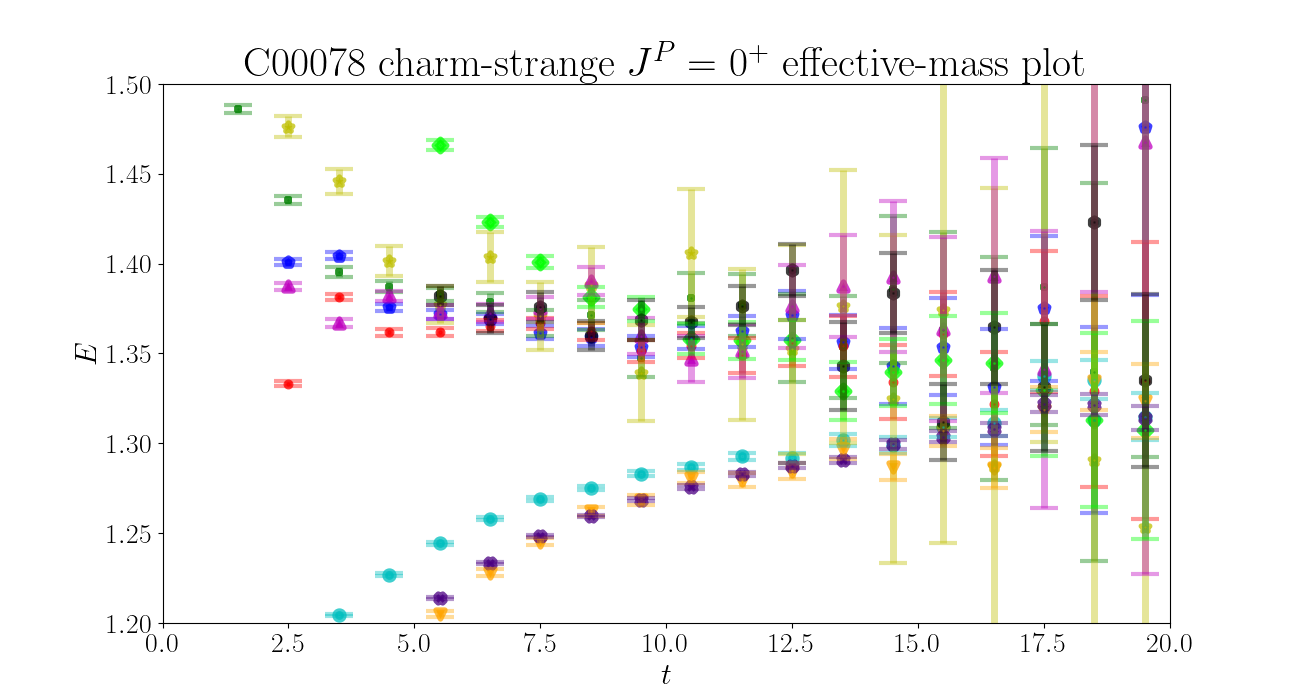}
    
    \hfill
        
    \includegraphics[width=0.49\linewidth]{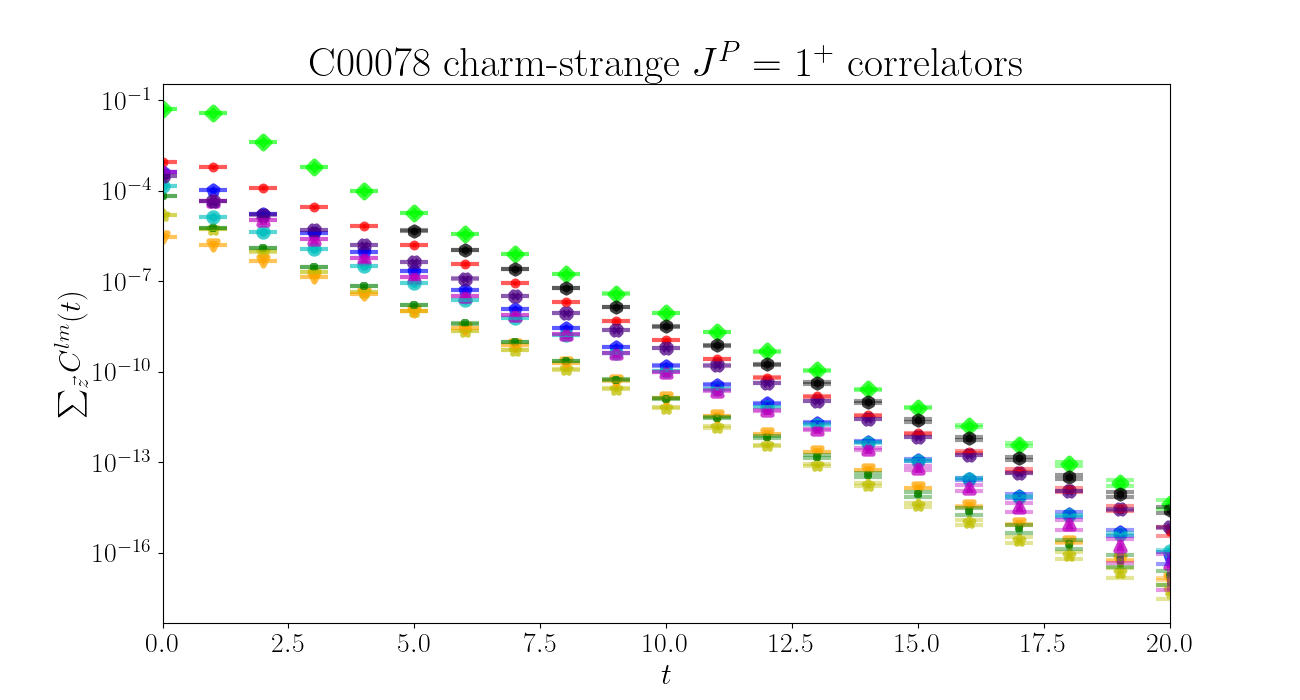}
    \hfill
    \includegraphics[width=0.49\linewidth]{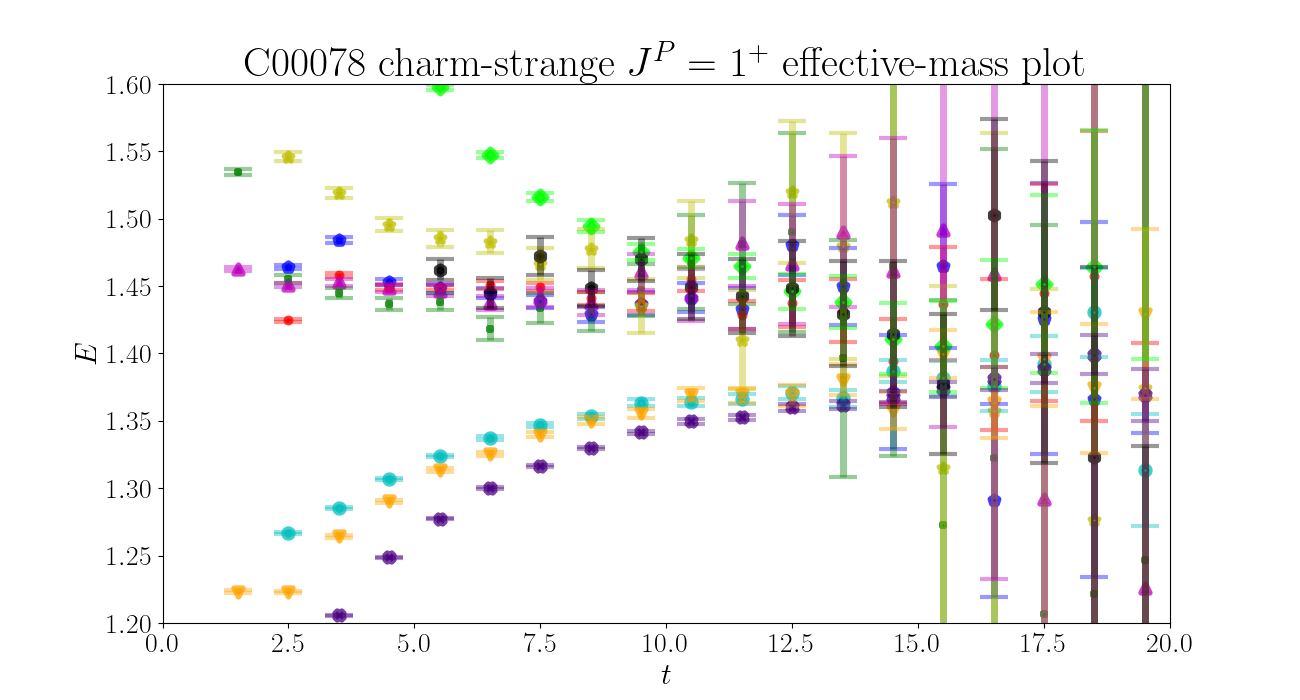}

    \includegraphics[width=0.49\linewidth]{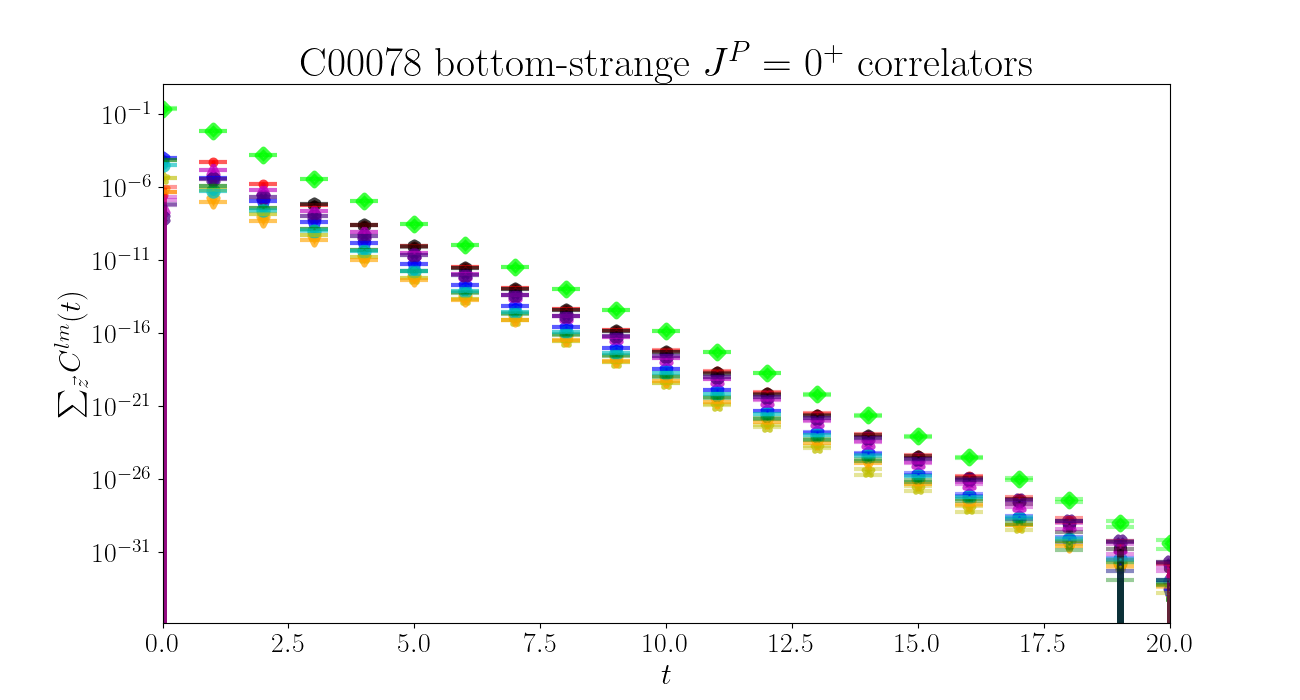}
    \hfill
    \includegraphics[width=0.49\linewidth]{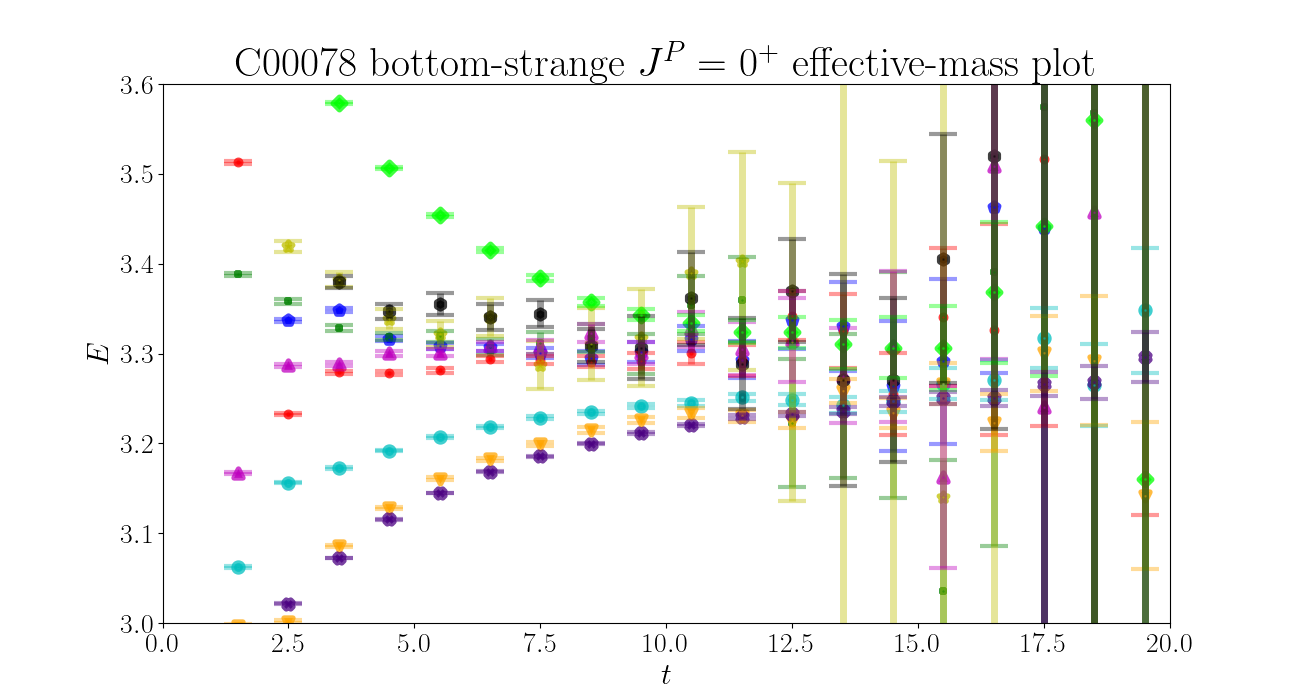}
    
    \hfill
        
    \includegraphics[width=0.49\linewidth]{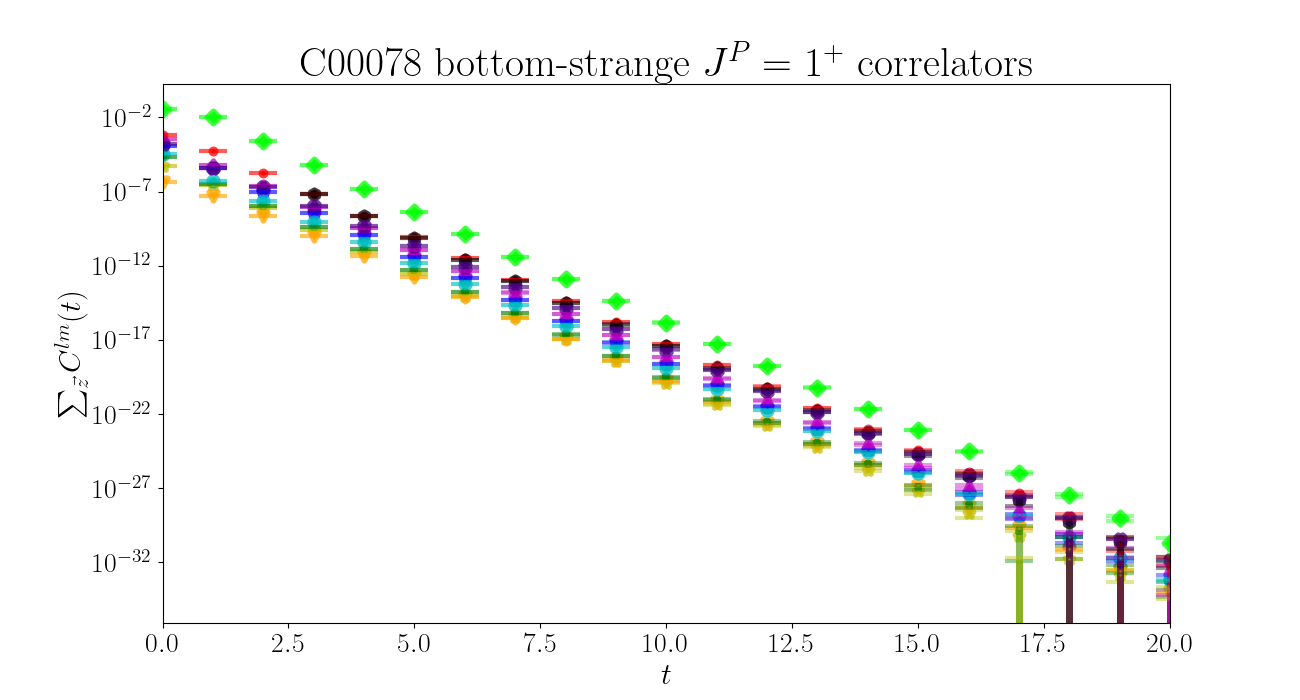}
    \hfill
    \includegraphics[width=0.49\linewidth]{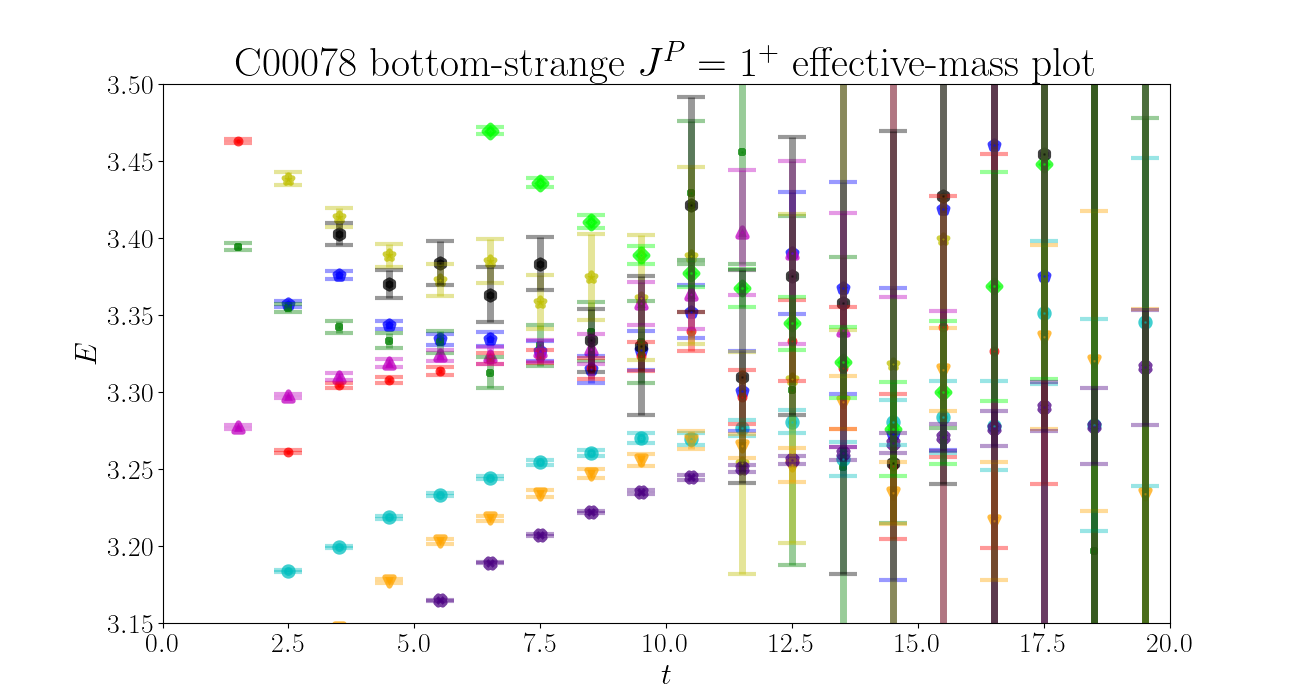}
    
    \centering
    \includegraphics[width=0.6\linewidth]{positive_parity/legend_new.pdf}
    \caption{Plots of the positive-parity correlation matrices $C^{lm}(t)$ (left) and associated effective energies (right) on the C00078 ensemble, in lattice unites. For the correlator plots, all of the negative off-diagonal elements have their sign flipped. A correlation function involving two unimproved currents $J_\text{unimp.}$ is additionally shown.$J_\text{unimp.}$ is given by dropping $\mathcal{O}(a)$-terms from Eq.~(\ref{eq:currents}). \label{fig:Clmplots-C00078}}
\end{figure}

\begin{figure}[H]

    \includegraphics[width=0.49\linewidth]{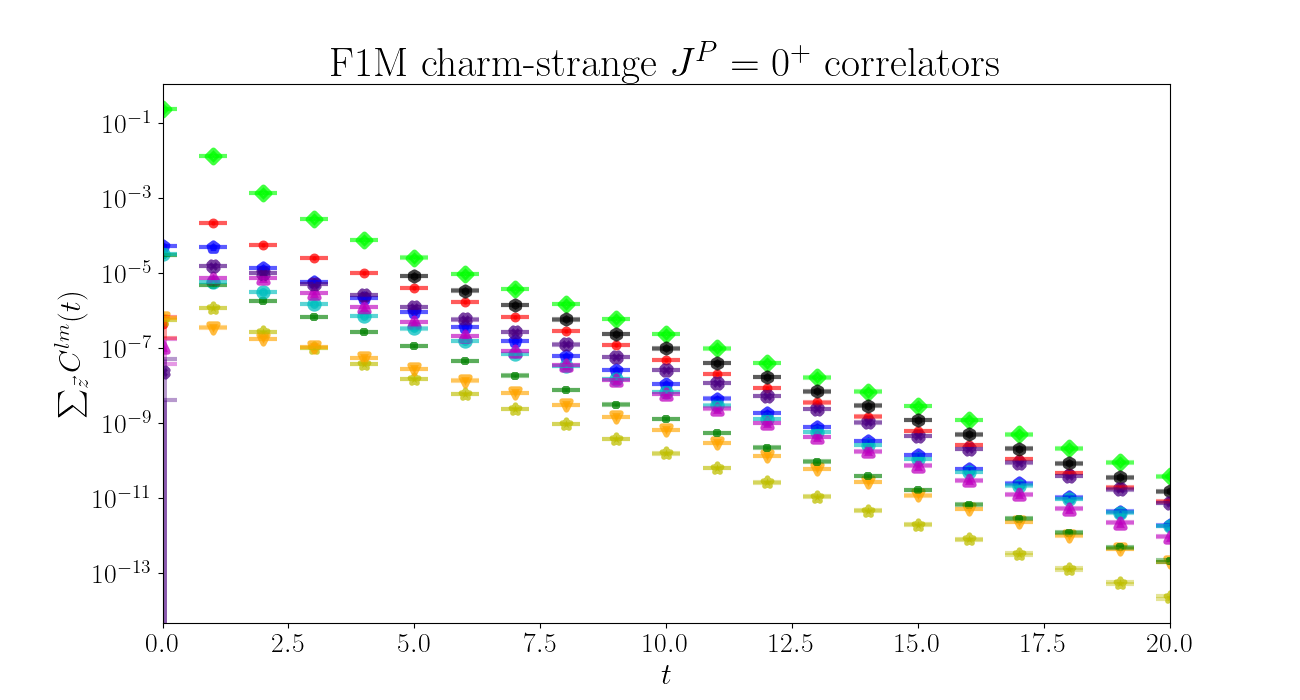}
    \hfill
    \includegraphics[width=0.49\linewidth]{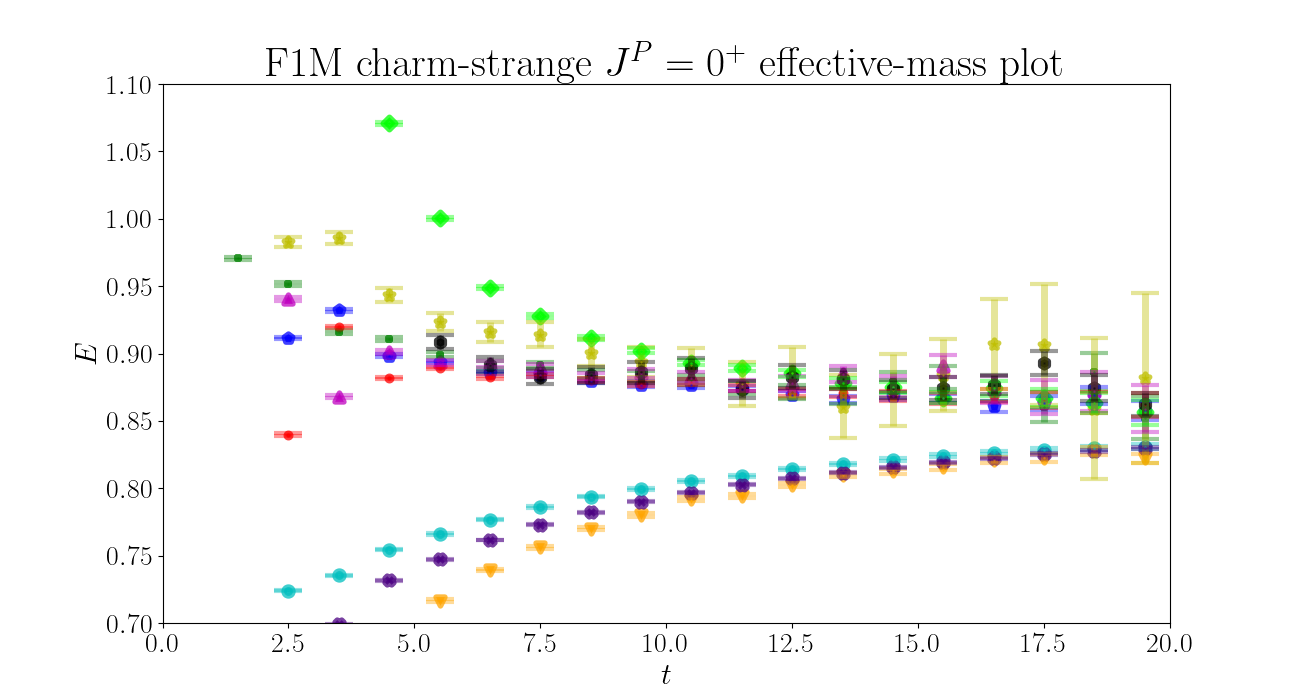}
    
    \hfill
        
    \includegraphics[width=0.49\linewidth]{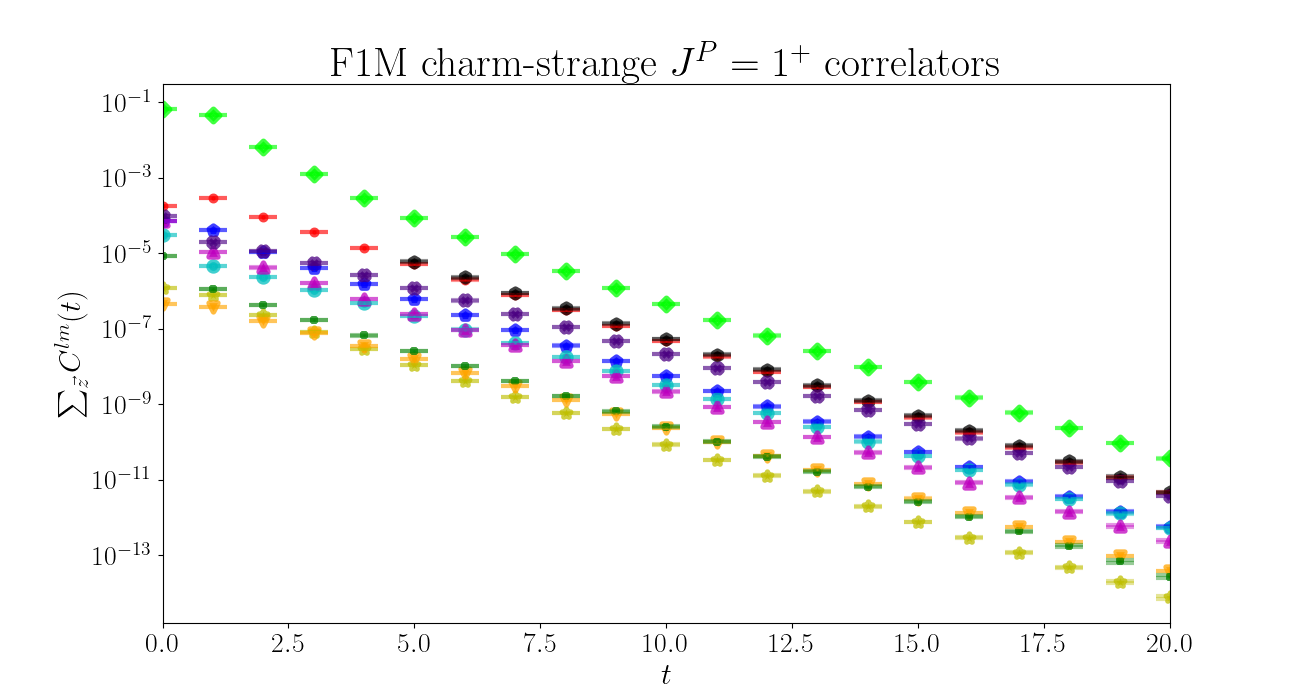}
    \hfill
    \includegraphics[width=0.49\linewidth]{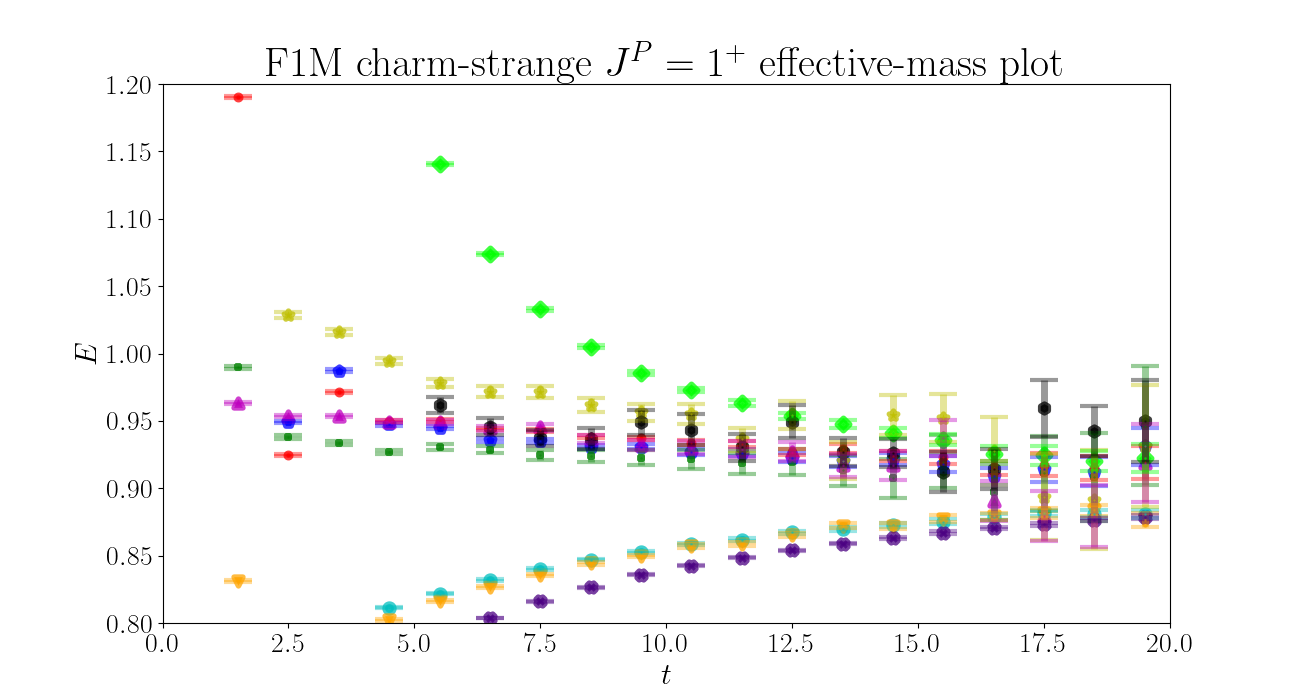}

    \includegraphics[width=0.49\linewidth]{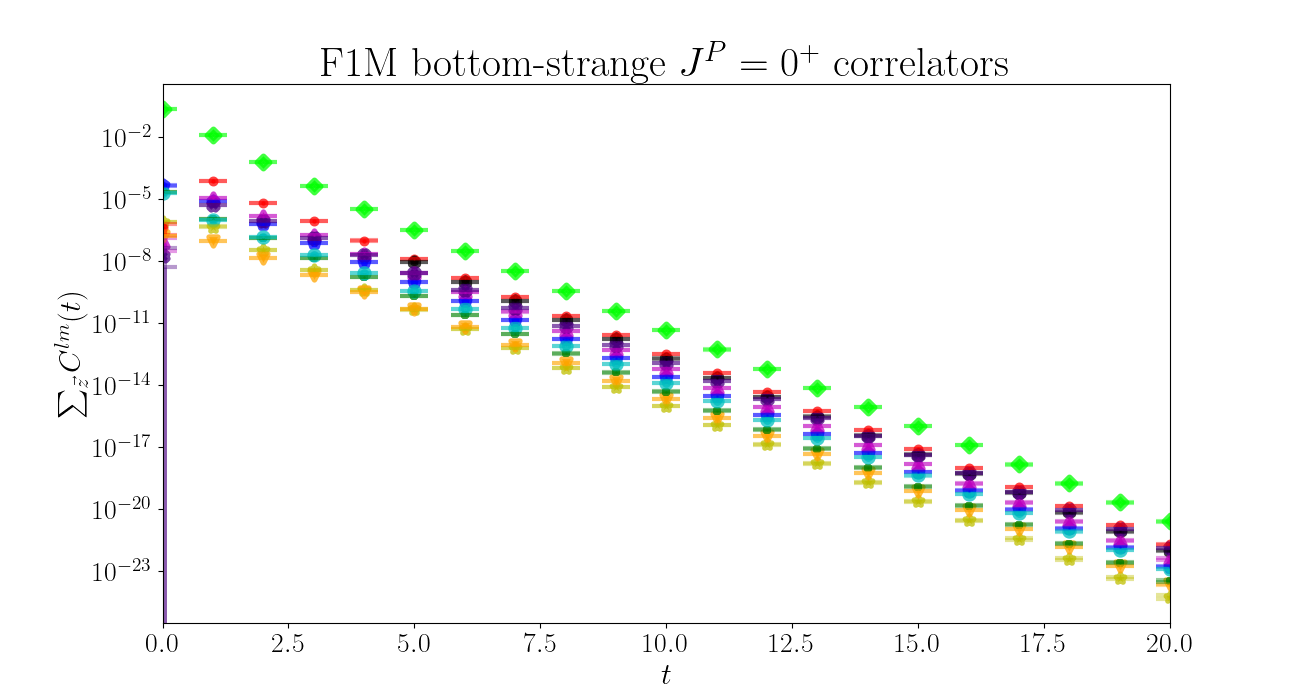}
    \hfill
    \includegraphics[width=0.49\linewidth]{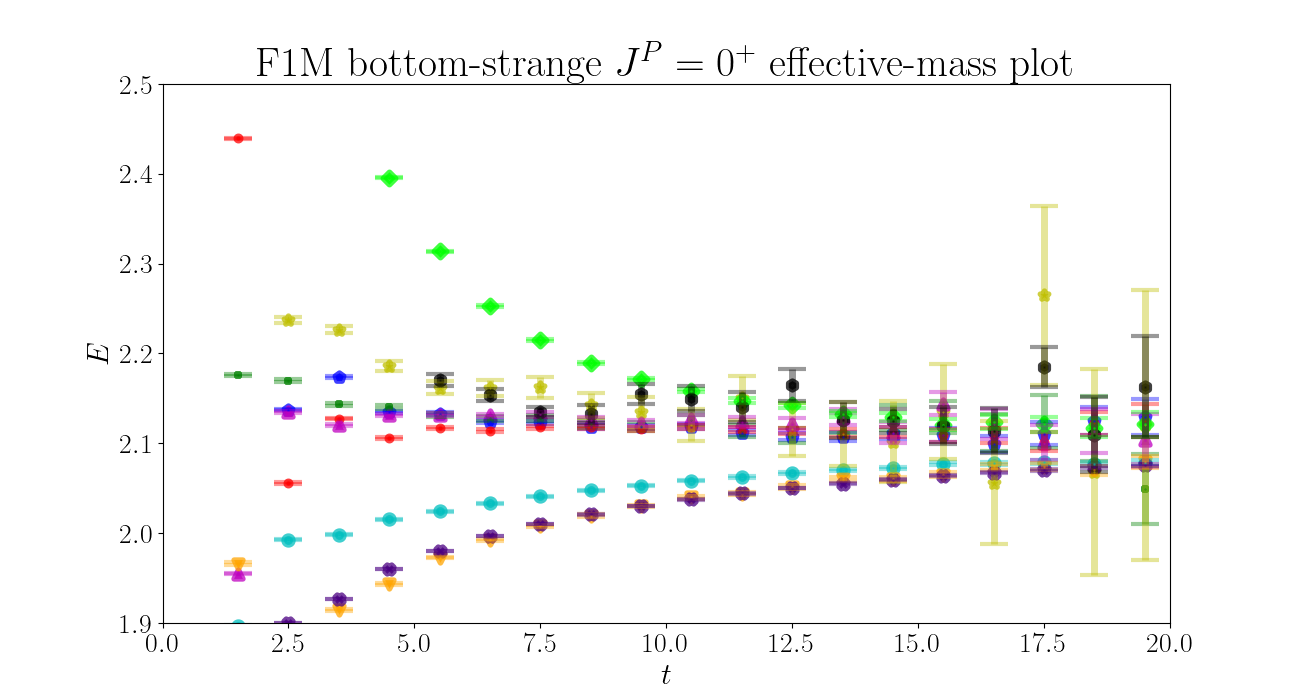}
    
    \hfill
        
    \includegraphics[width=0.49\linewidth]{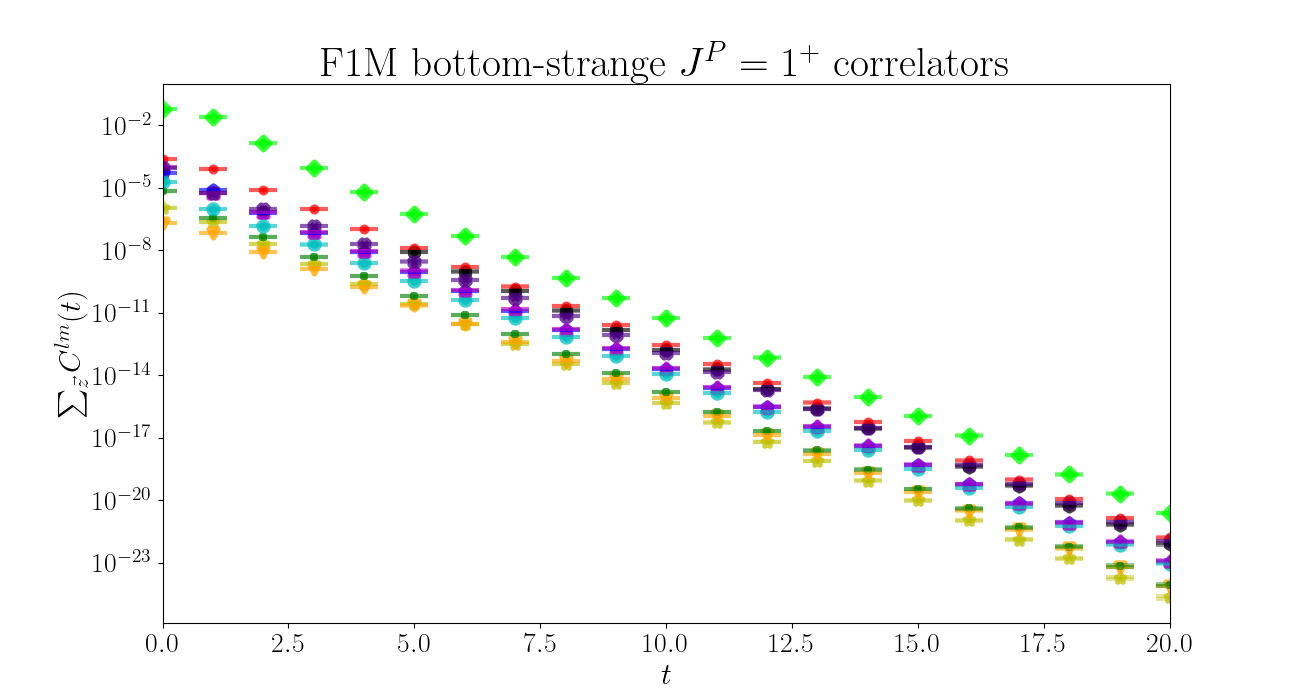}
    \hfill
    \includegraphics[width=0.49\linewidth]{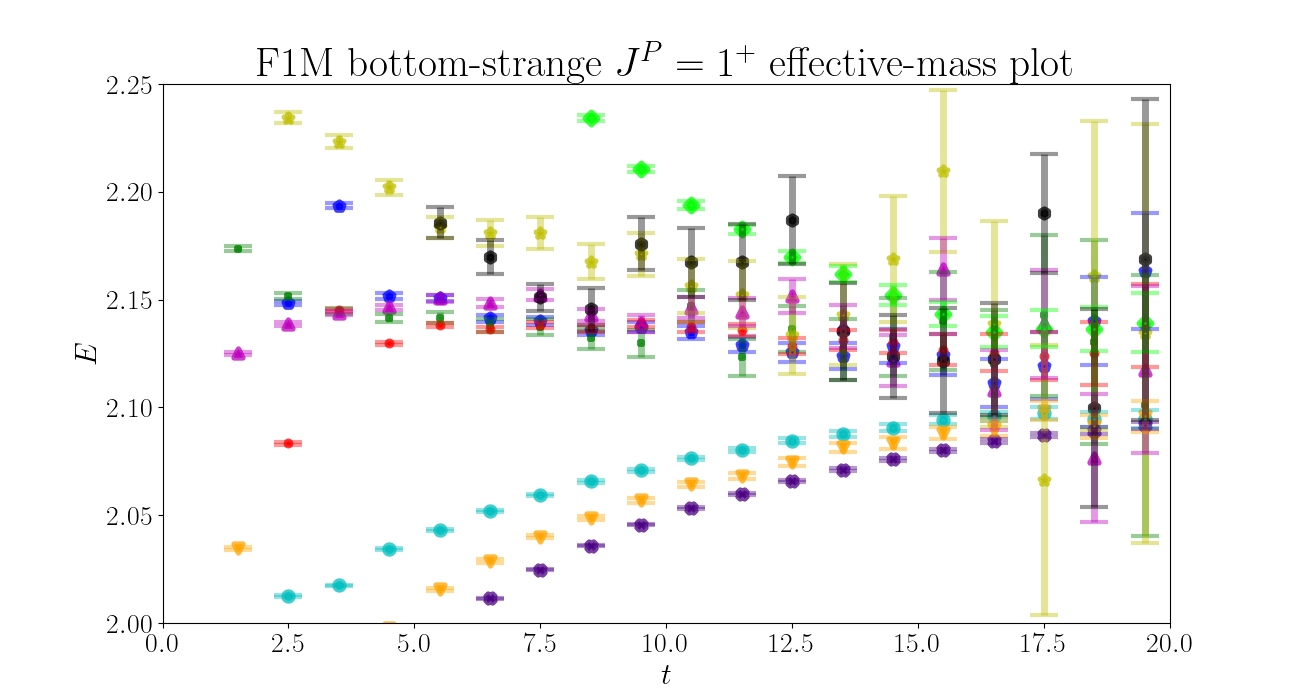}
    
    \centering
    \includegraphics[width=0.6\linewidth]{positive_parity/legend_new.pdf}
    \caption{Like Fig.~\protect\ref{fig:Clmplots-C00078}, but for the F1M ensemble. \label{fig:Clmplots-F1M}}
\end{figure}

\begin{figure}[H]

    \includegraphics[width=0.49\linewidth]{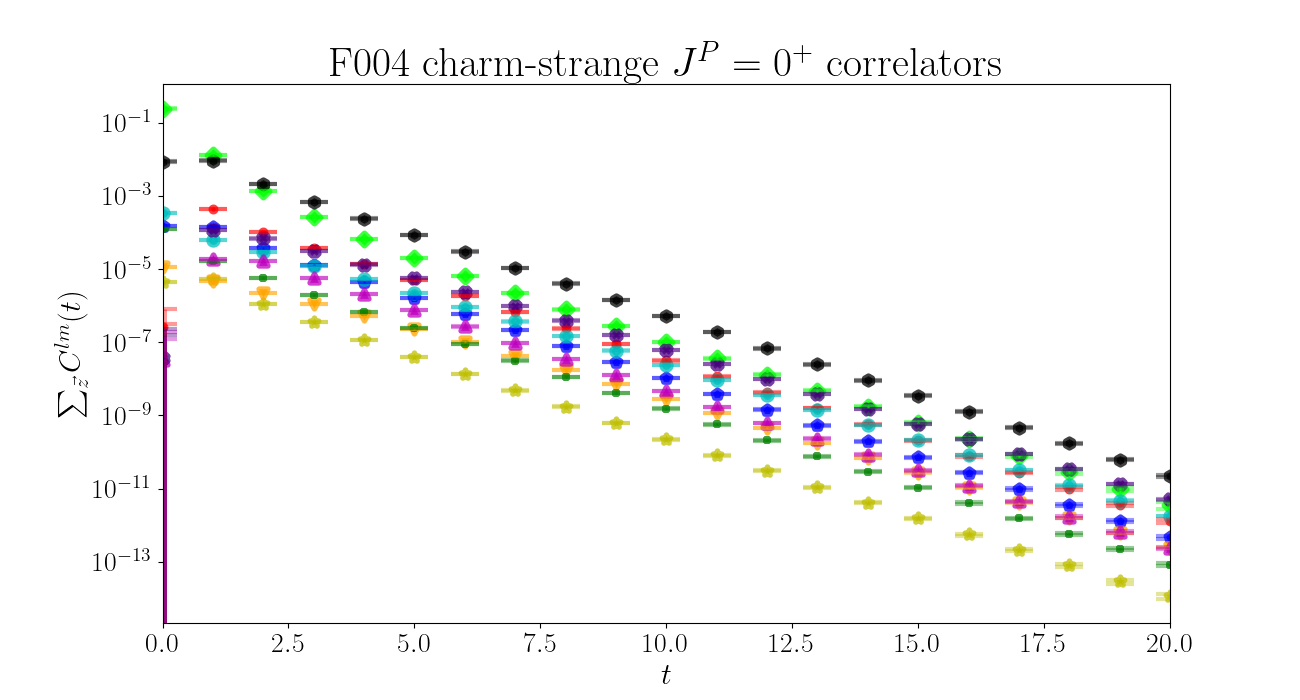}
    \hfill
    \includegraphics[width=0.49\linewidth]{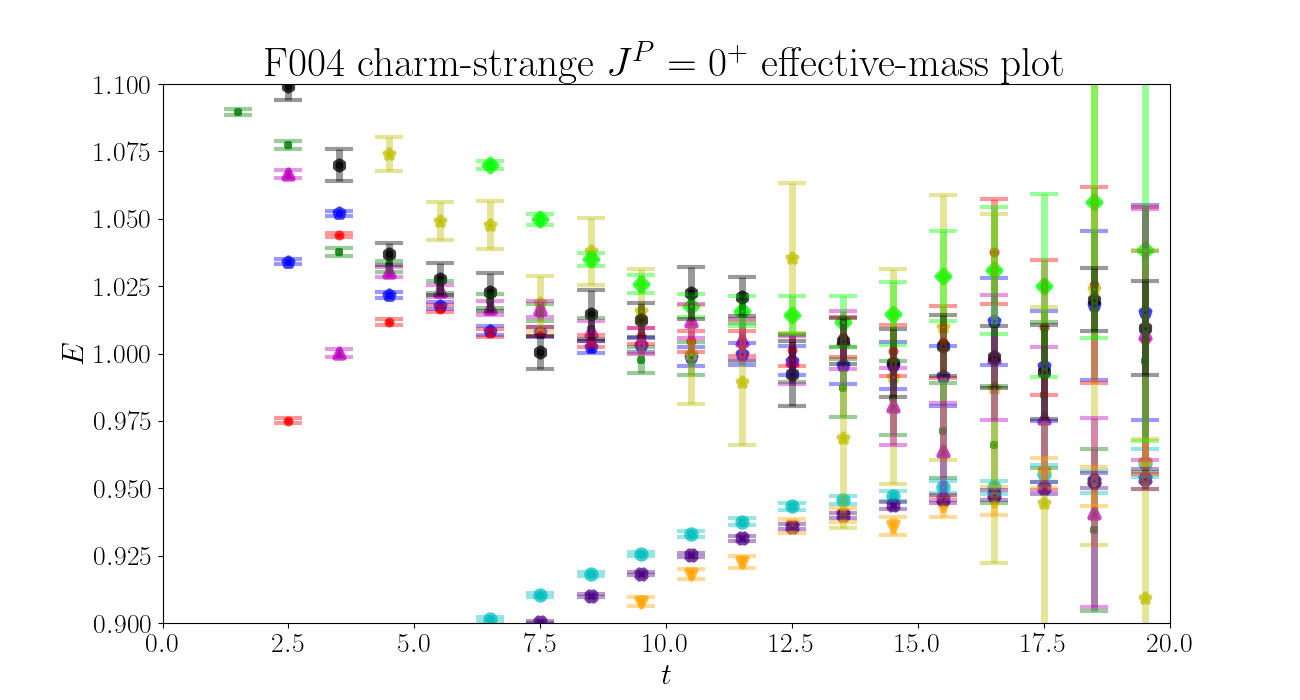}
    
    \hfill
        
    \includegraphics[width=0.49\linewidth]{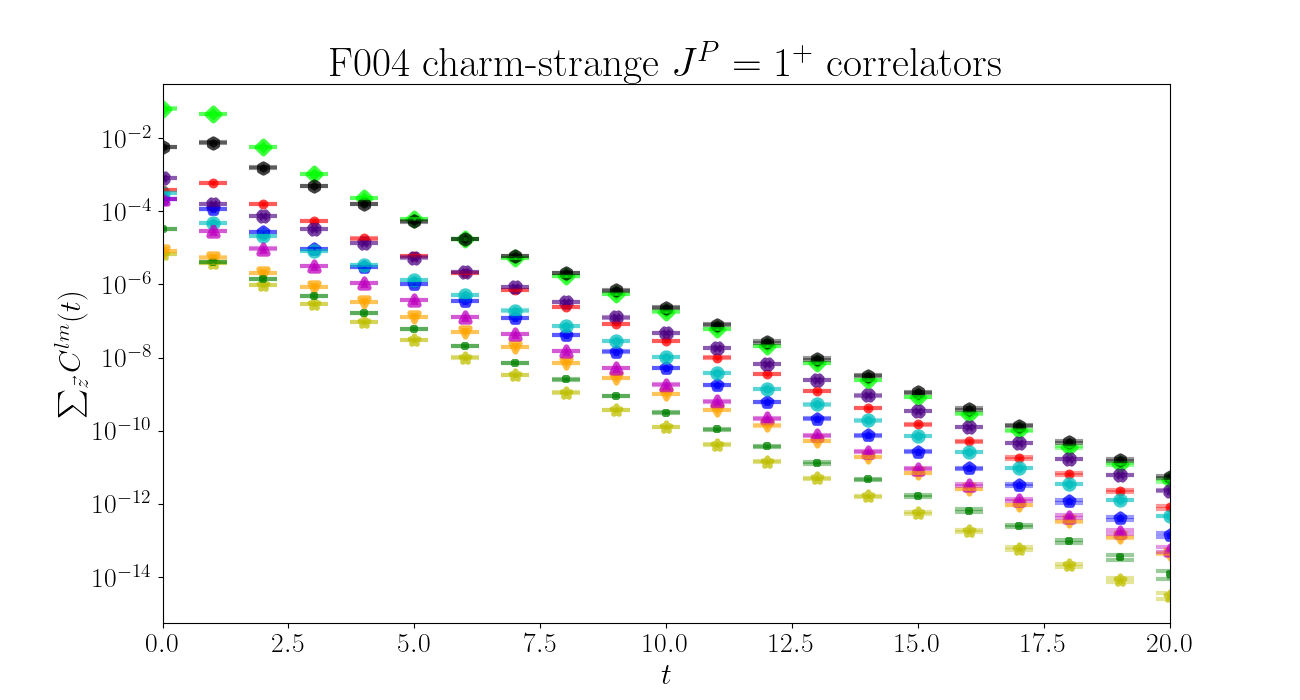}
    \hfill
    \includegraphics[width=0.49\linewidth]{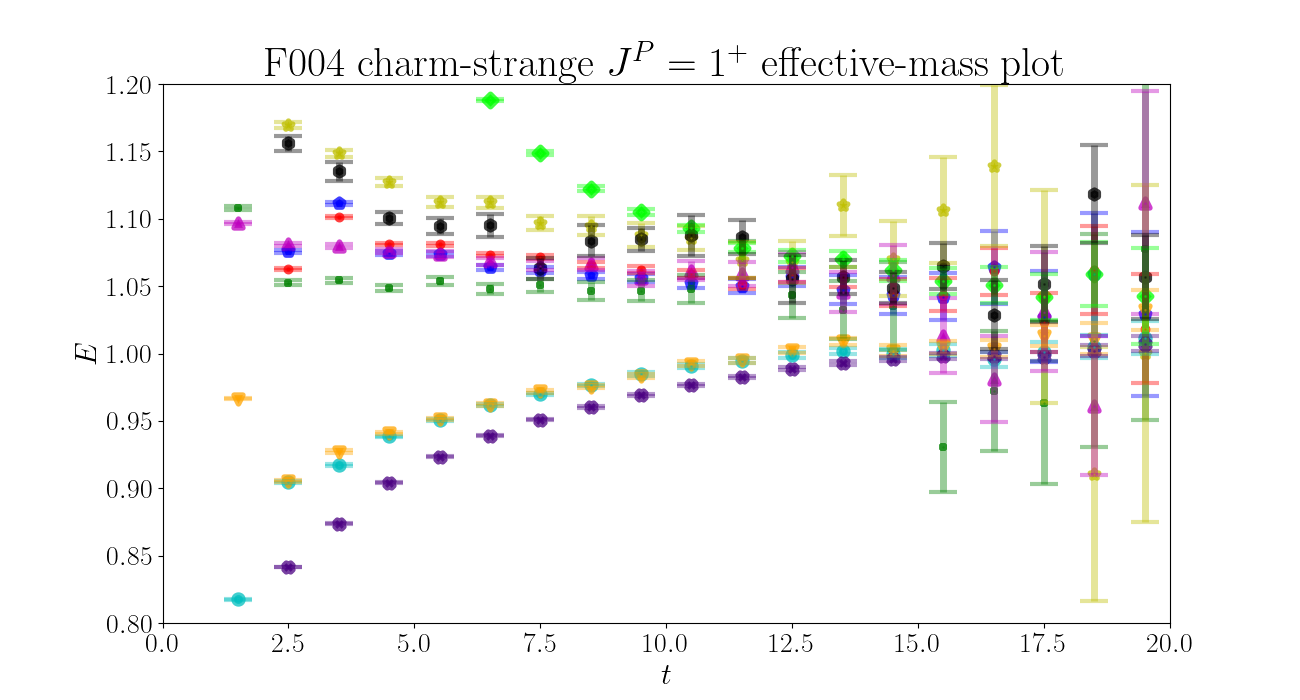}

    \includegraphics[width=0.49\linewidth]{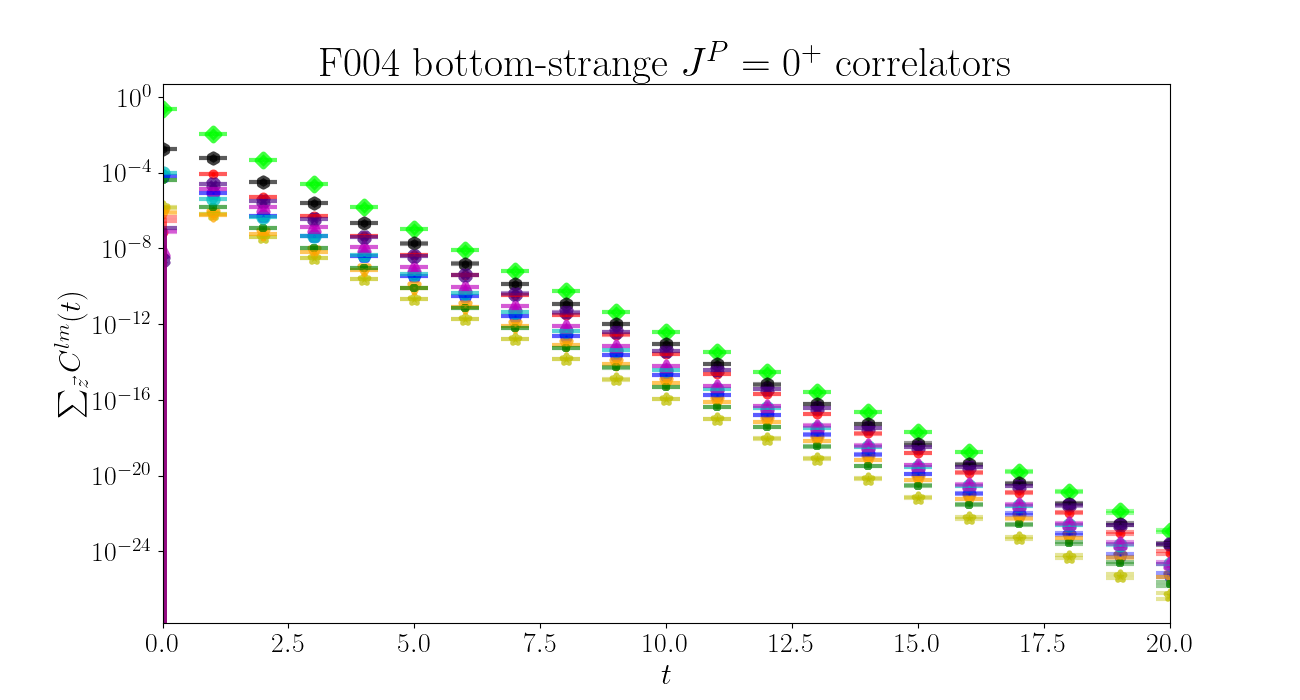}
    \hfill
    \includegraphics[width=0.49\linewidth]{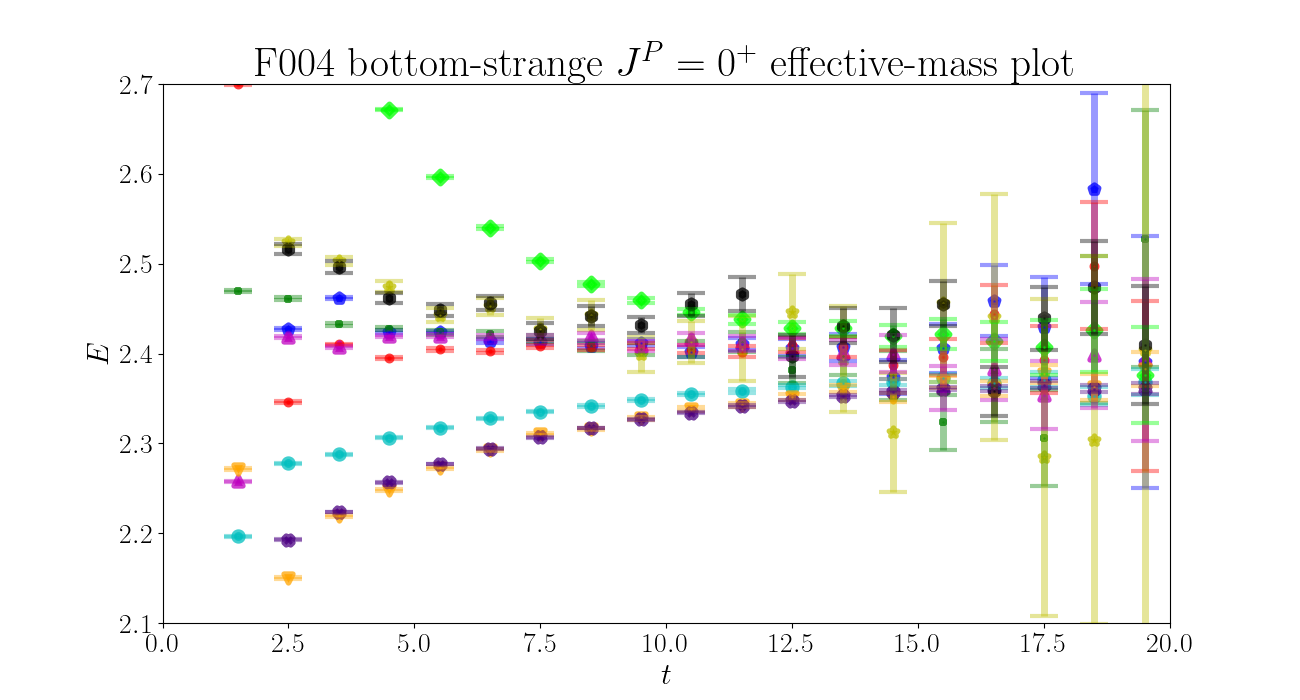}
    
    \hfill
        
    \includegraphics[width=0.49\linewidth]{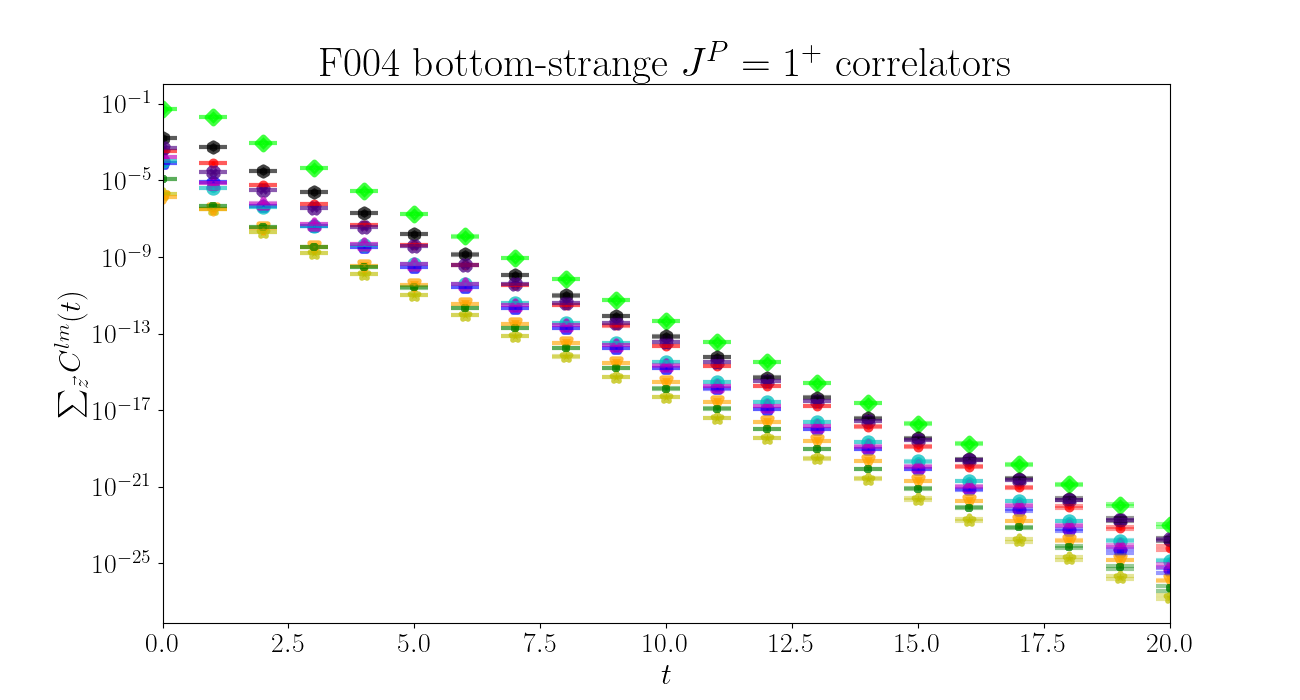}
    \hfill
    \includegraphics[width=0.49\linewidth]{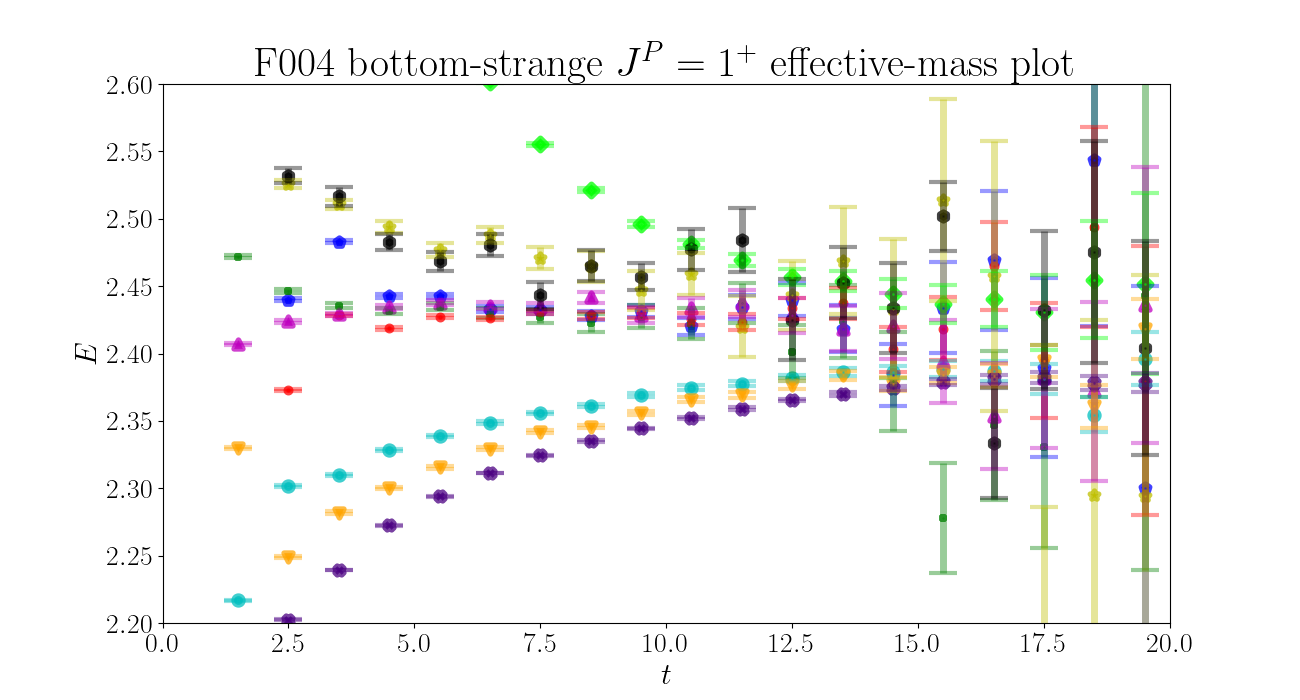}
    
    \centering
    \includegraphics[width=0.6\linewidth]{positive_parity/legend_new.pdf}
    \caption{Like Fig.~\protect\ref{fig:Clmplots-C00078}, but for the F004 ensemble. \label{fig:Clmplots-F004}}
\end{figure}

\begin{figure}[H]

    \includegraphics[width=0.49\linewidth]{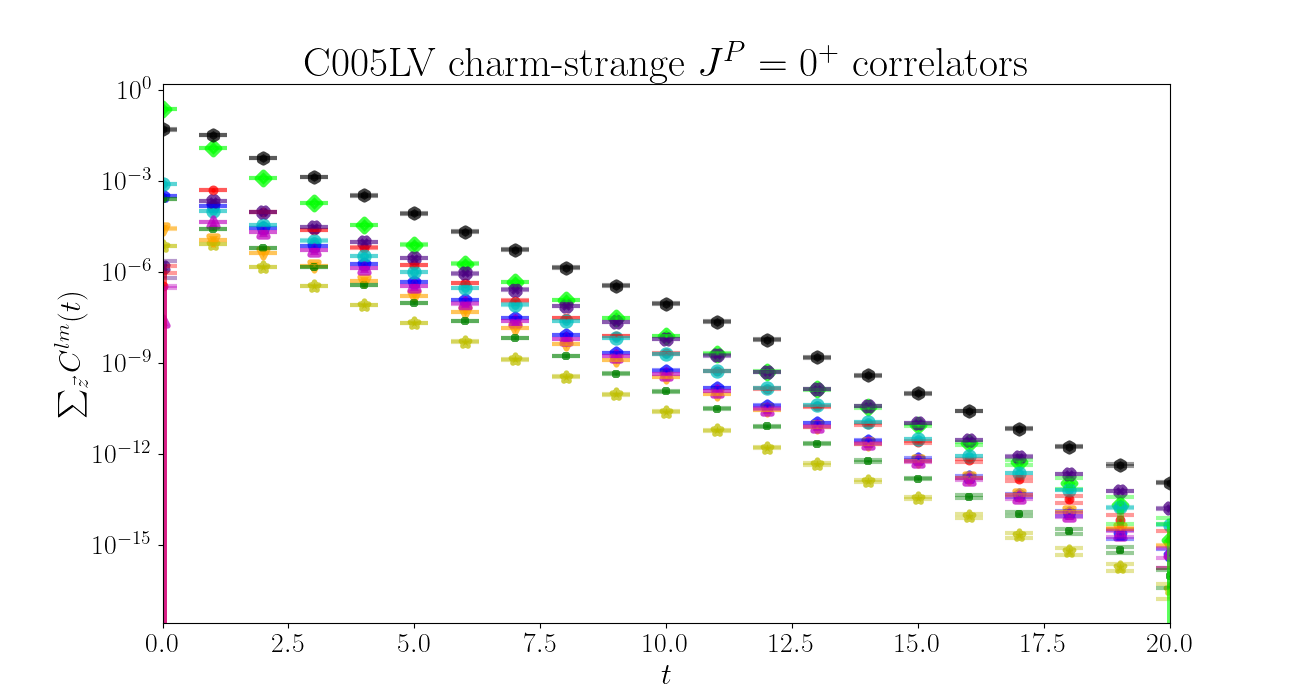}
    \hfill
    \includegraphics[width=0.49\linewidth]{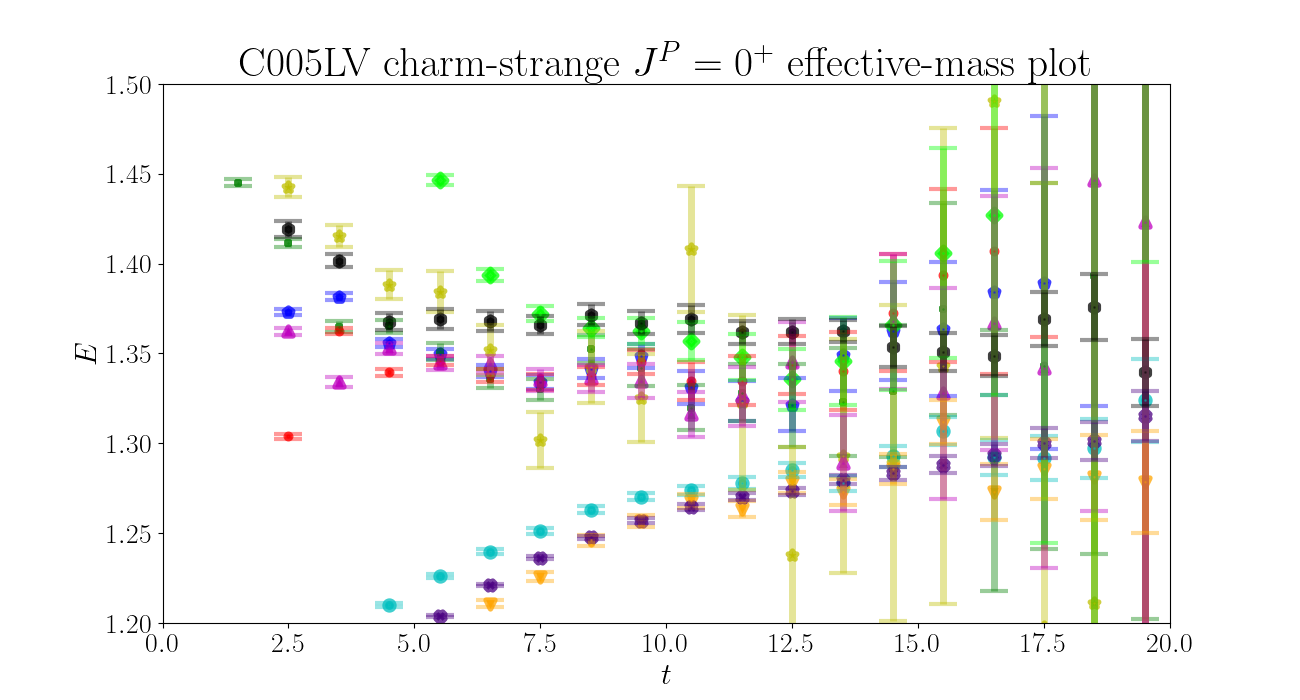}
    
    \hfill
        
    \includegraphics[width=0.49\linewidth]{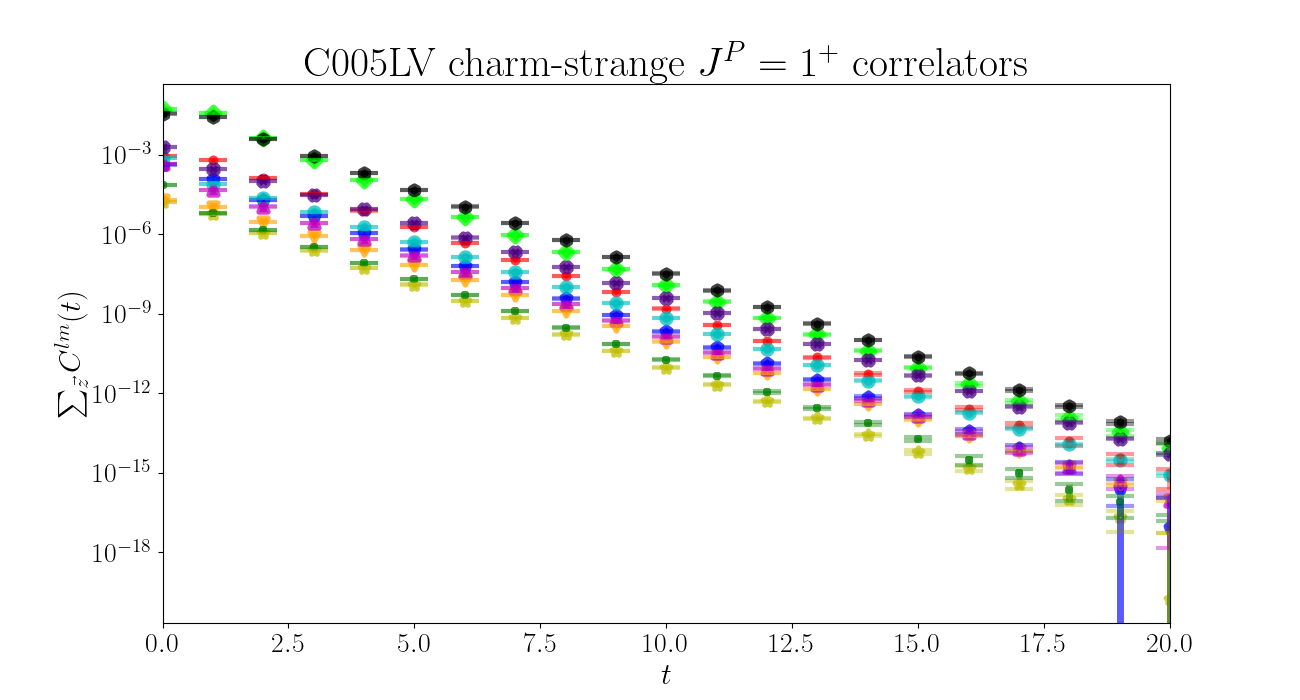}
    \hfill
    \includegraphics[width=0.49\linewidth]{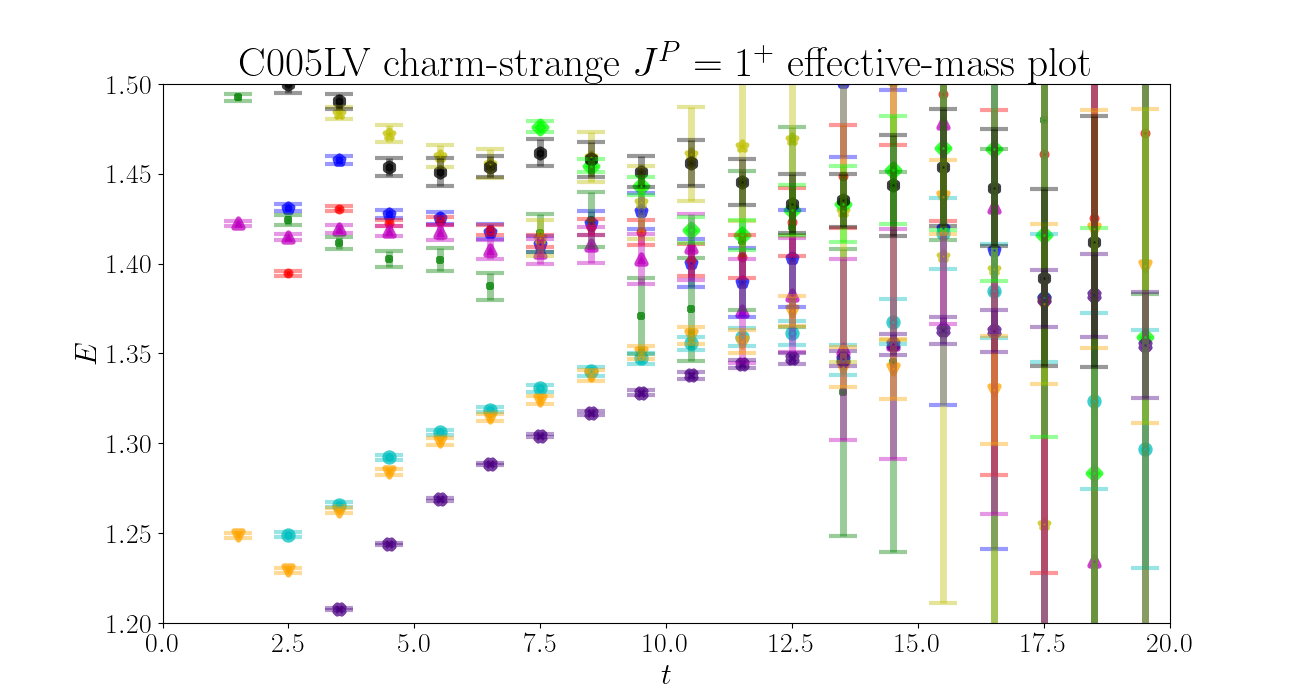}

    \includegraphics[width=0.49\linewidth]{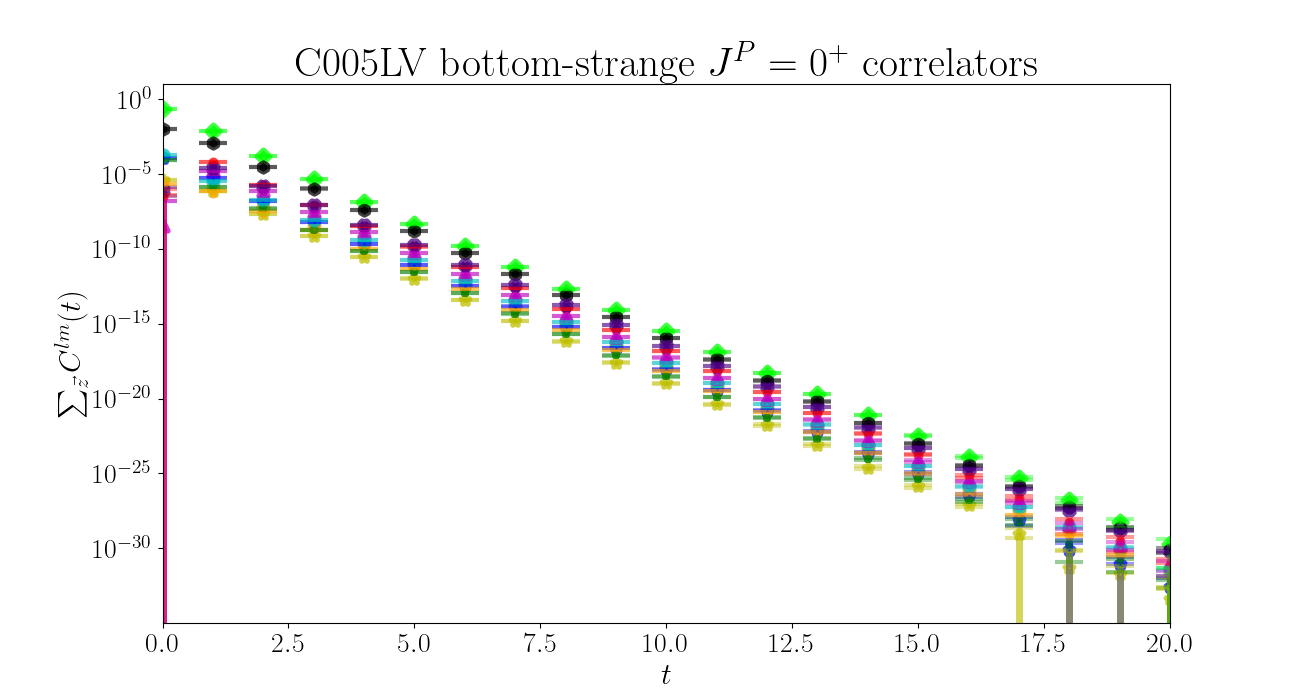}
    \hfill
    \includegraphics[width=0.49\linewidth]{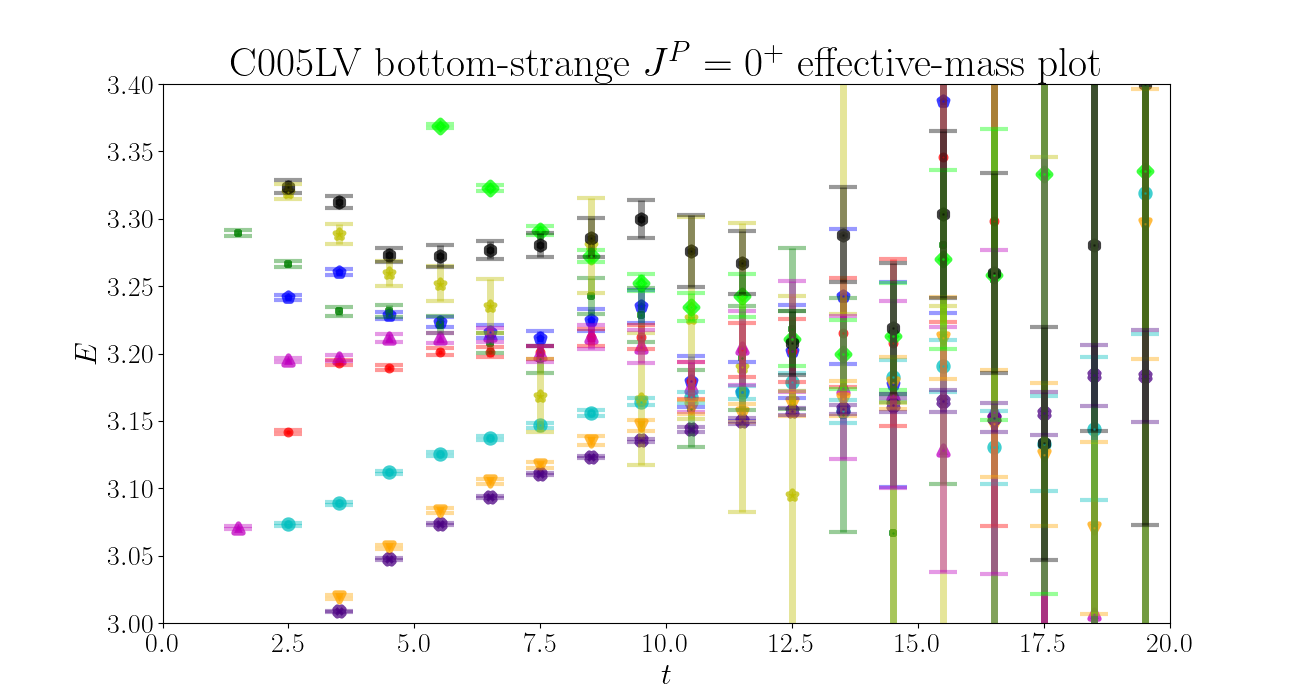}
    
    \hfill
        
    \includegraphics[width=0.49\linewidth]{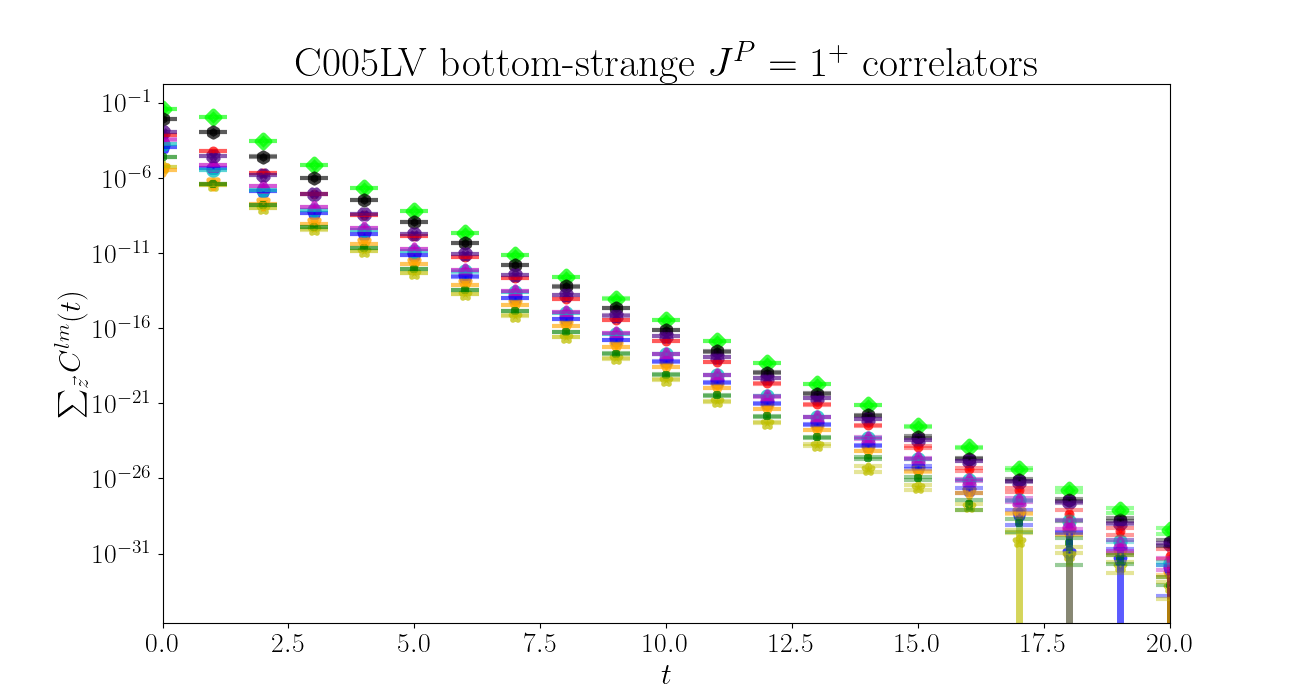}
    \hfill
    \includegraphics[width=0.49\linewidth]{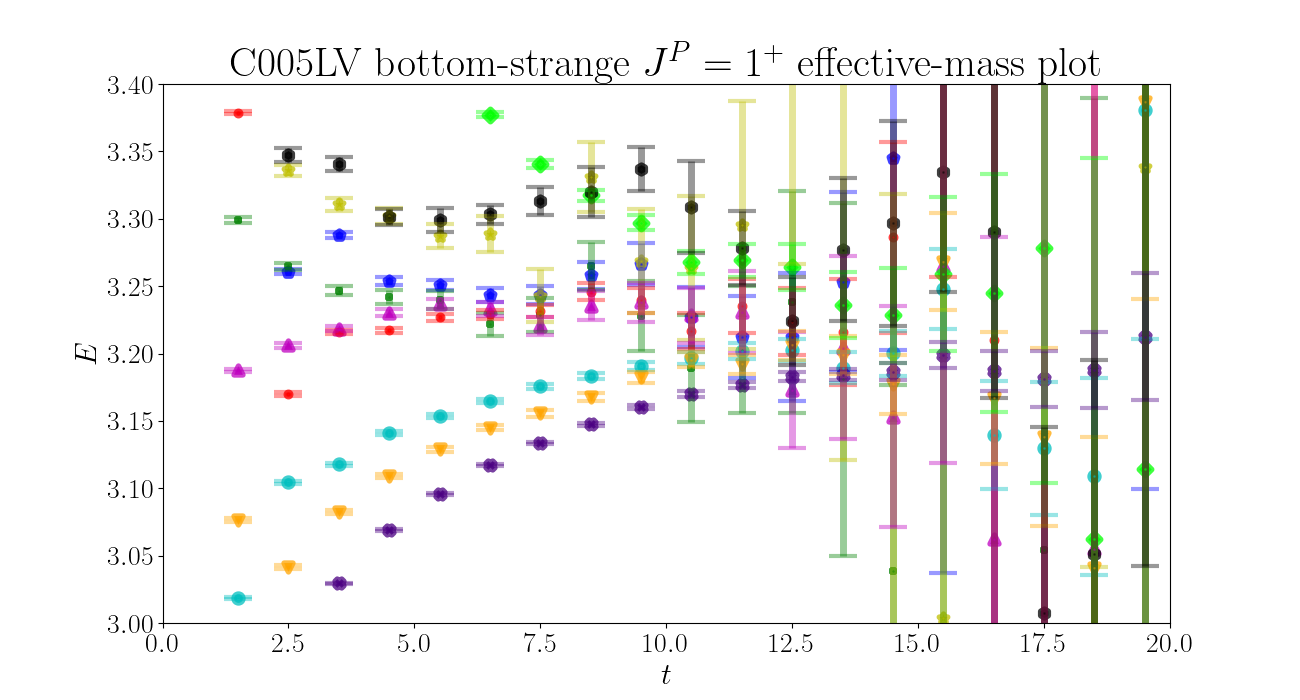}
    
    \centering
    \includegraphics[width=0.6\linewidth]{positive_parity/legend_new.pdf}
    \caption{Like Fig.~\protect\ref{fig:Clmplots-C00078}, but for the C005LV ensemble. \label{fig:Clmplots-C005LV}}
\end{figure}

\begin{figure}[H]

    \includegraphics[width=0.49\linewidth]{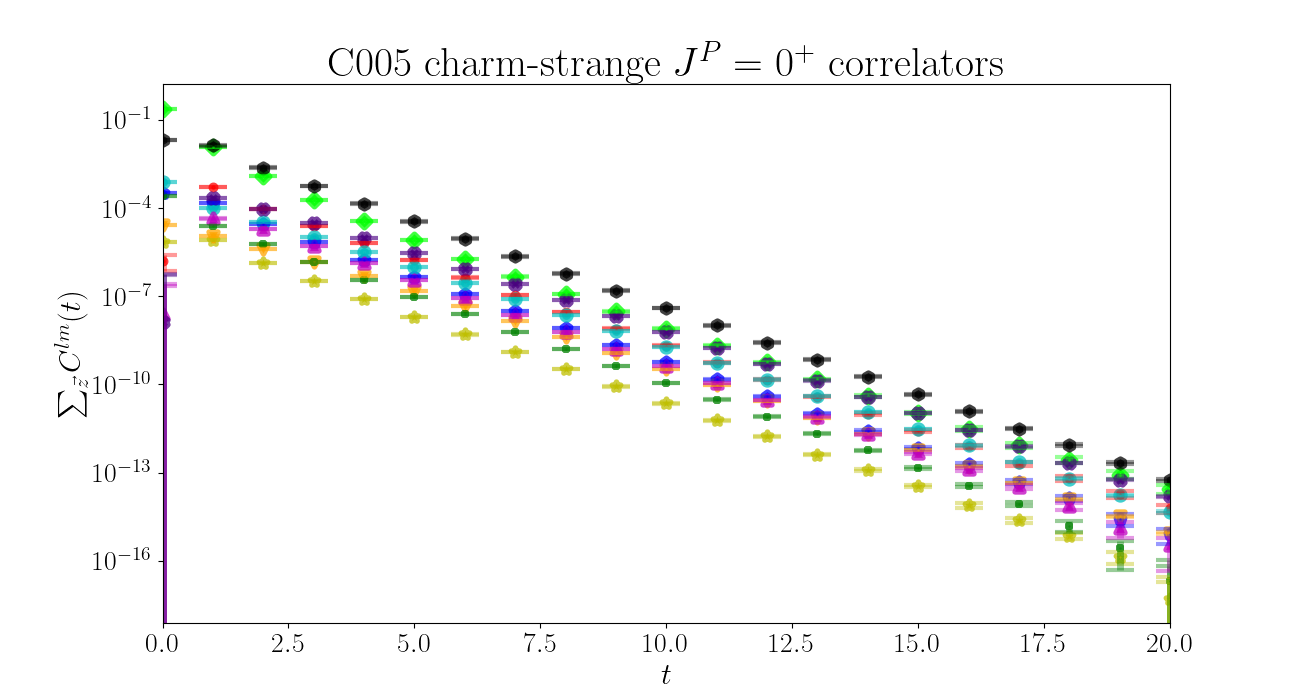}
    \hfill
    \includegraphics[width=0.49\linewidth]{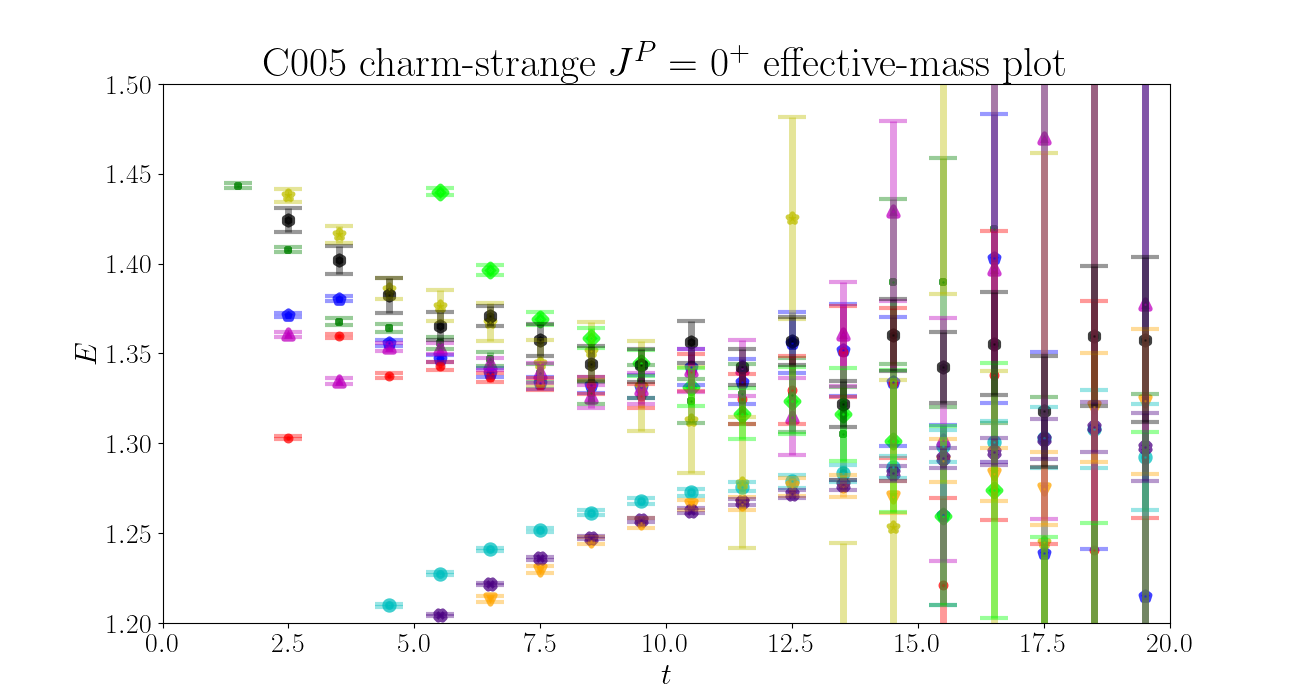}
    
    \hfill
        
    \includegraphics[width=0.49\linewidth]{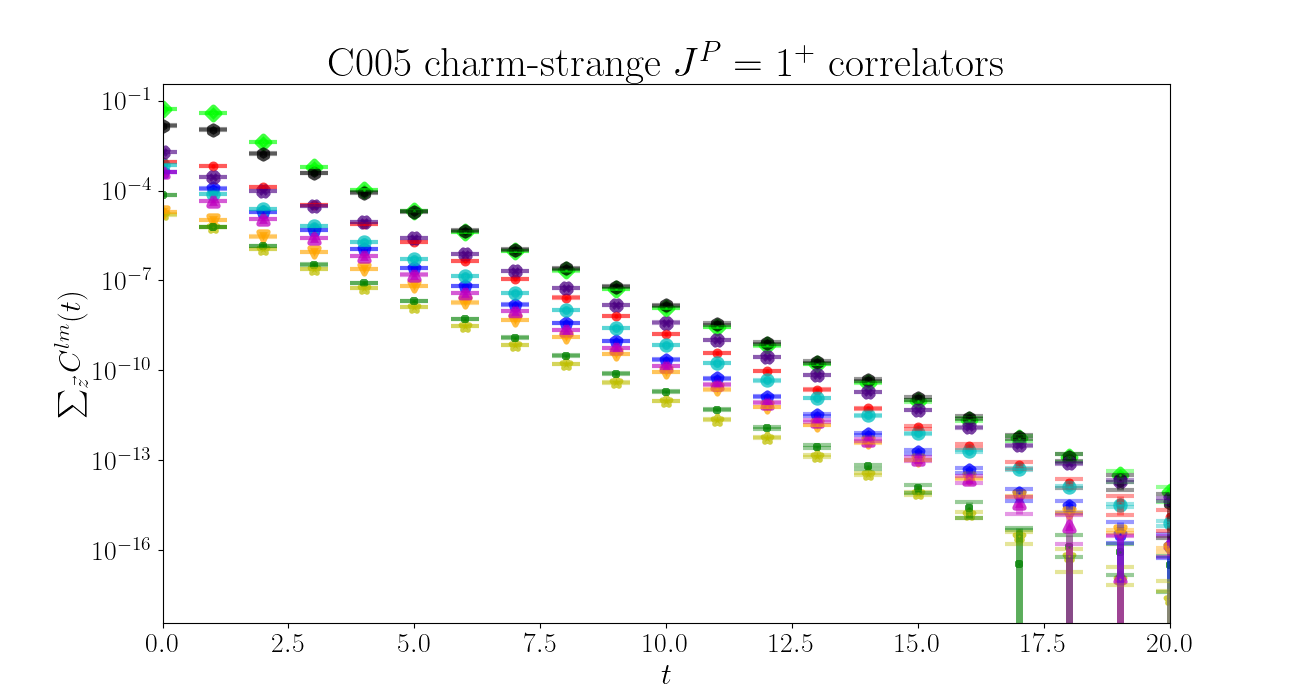}
    \hfill
    \includegraphics[width=0.49\linewidth]{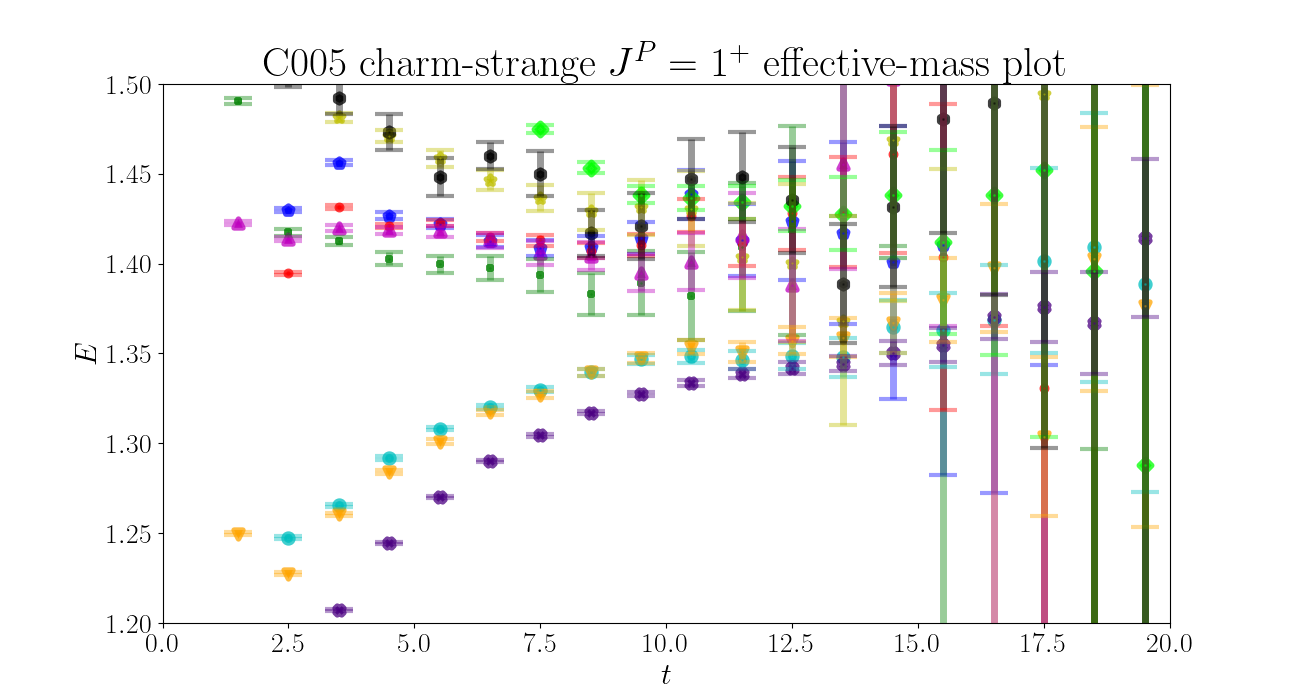}

    \includegraphics[width=0.49\linewidth]{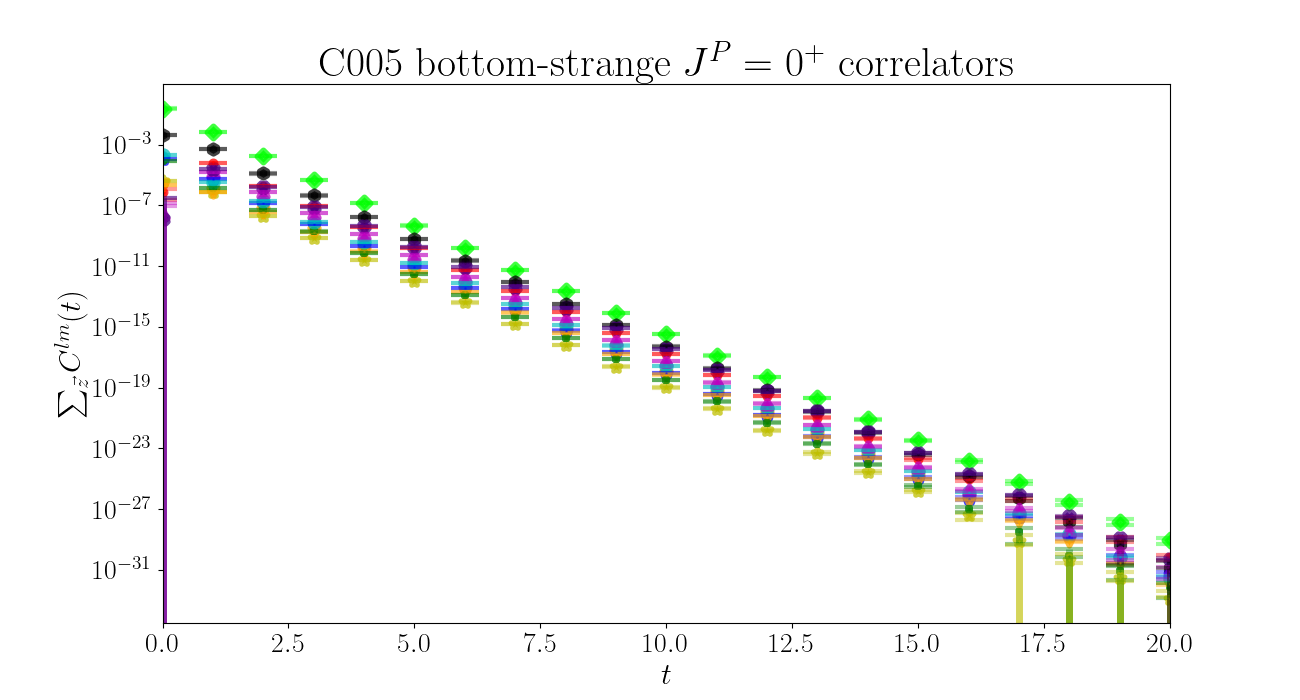}
    \hfill
    \includegraphics[width=0.49\linewidth]{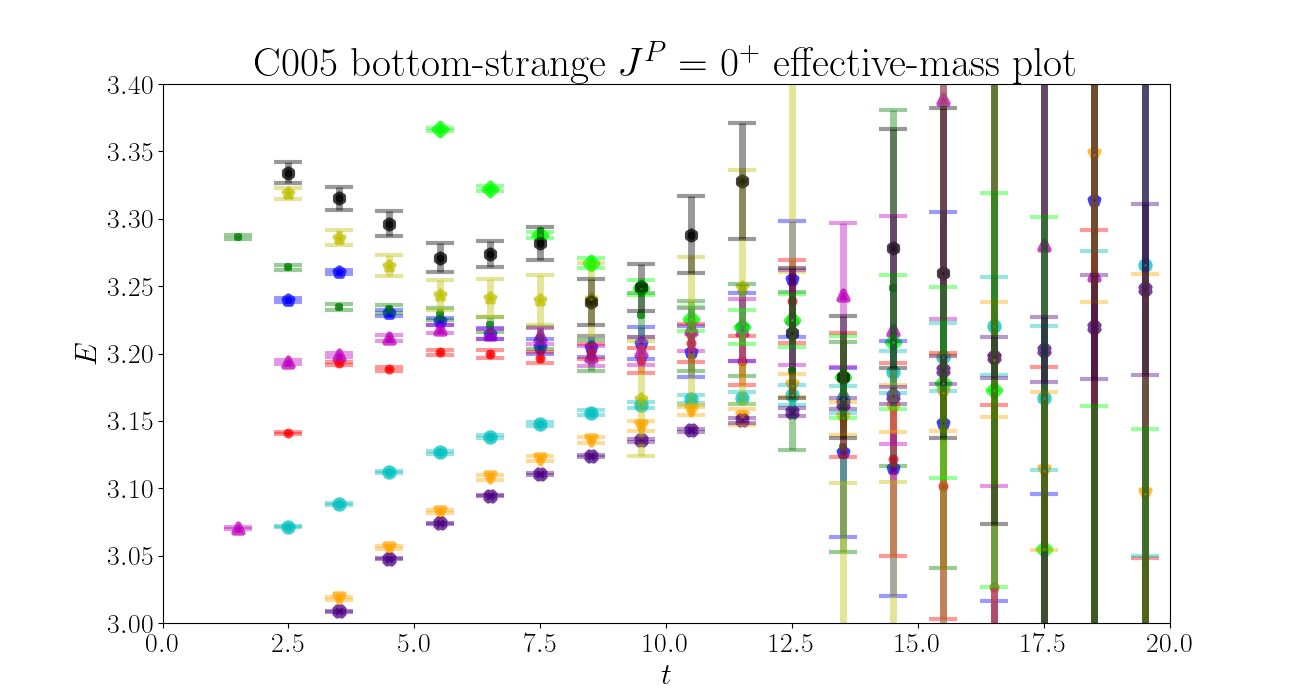}
    
    \hfill
        
    \includegraphics[width=0.49\linewidth]{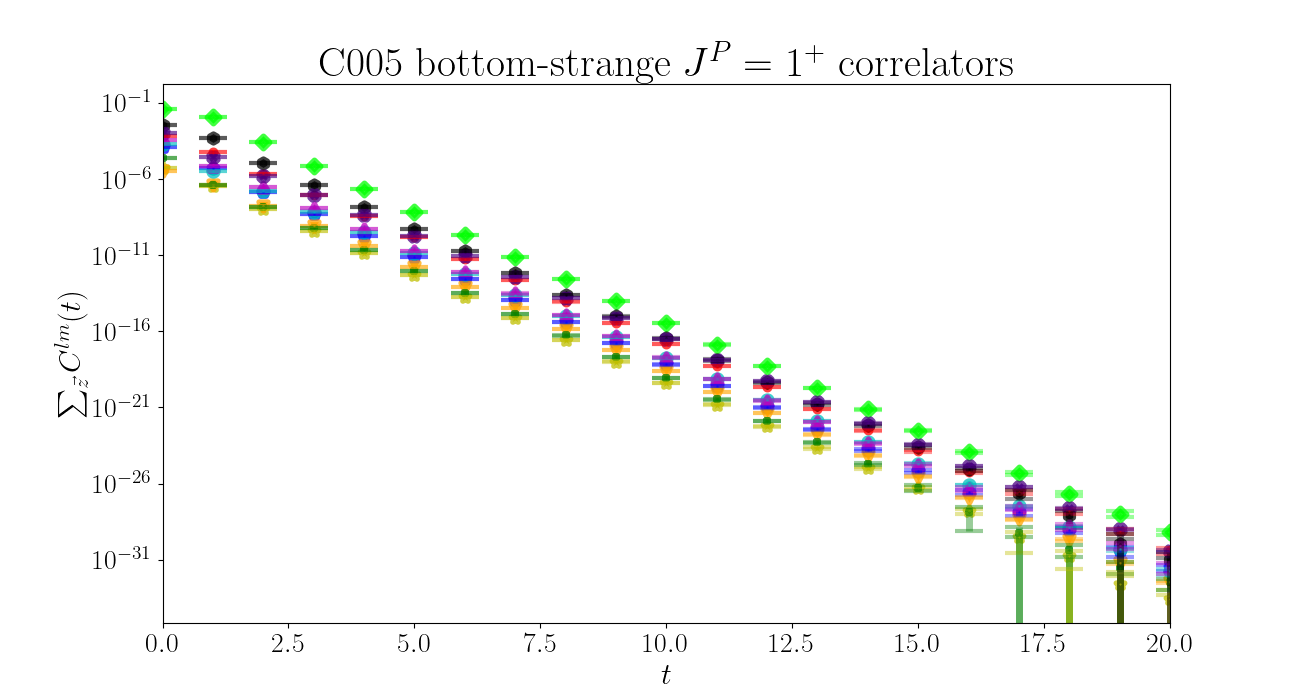}
    \hfill
    \includegraphics[width=0.49\linewidth]{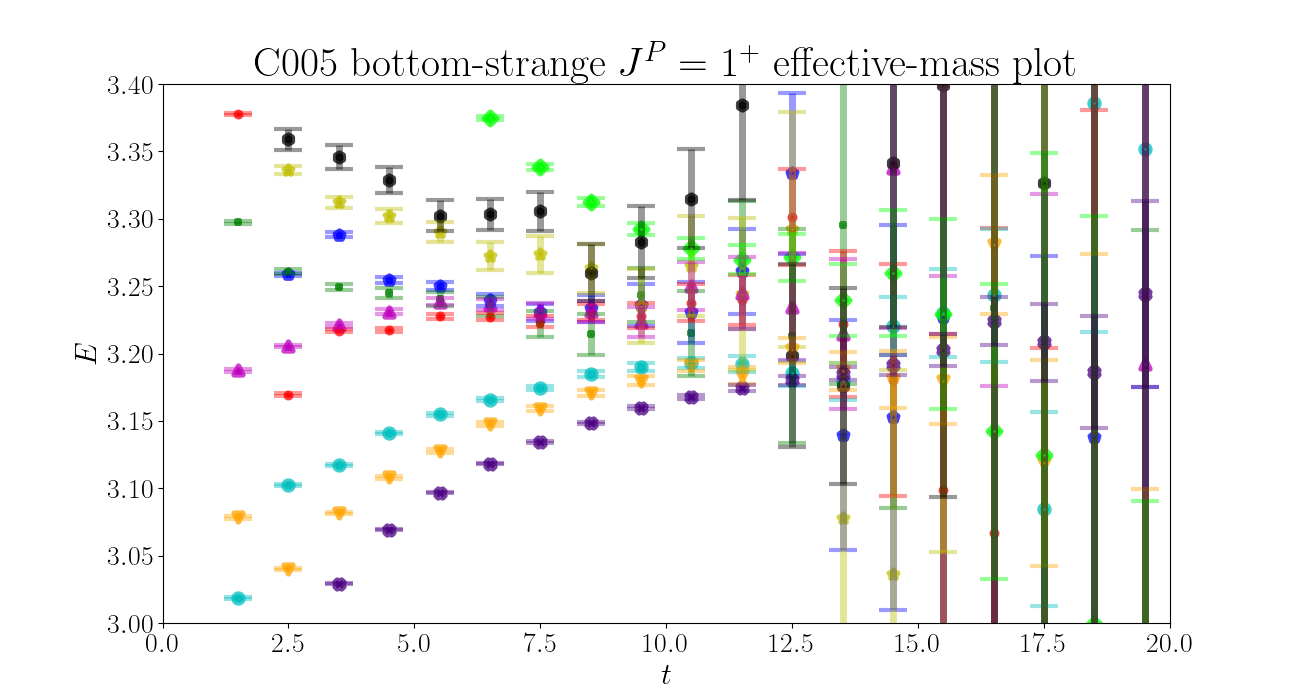}
    
    \centering
    \includegraphics[width=0.6\linewidth]{positive_parity/legend_new.pdf}
    \caption{Like Fig.~\protect\ref{fig:Clmplots-C00078}, but for the C005 ensemble. \label{fig:Clmplots-C005}}
\end{figure}

\begin{figure}[H]

    \includegraphics[width=0.49\linewidth]{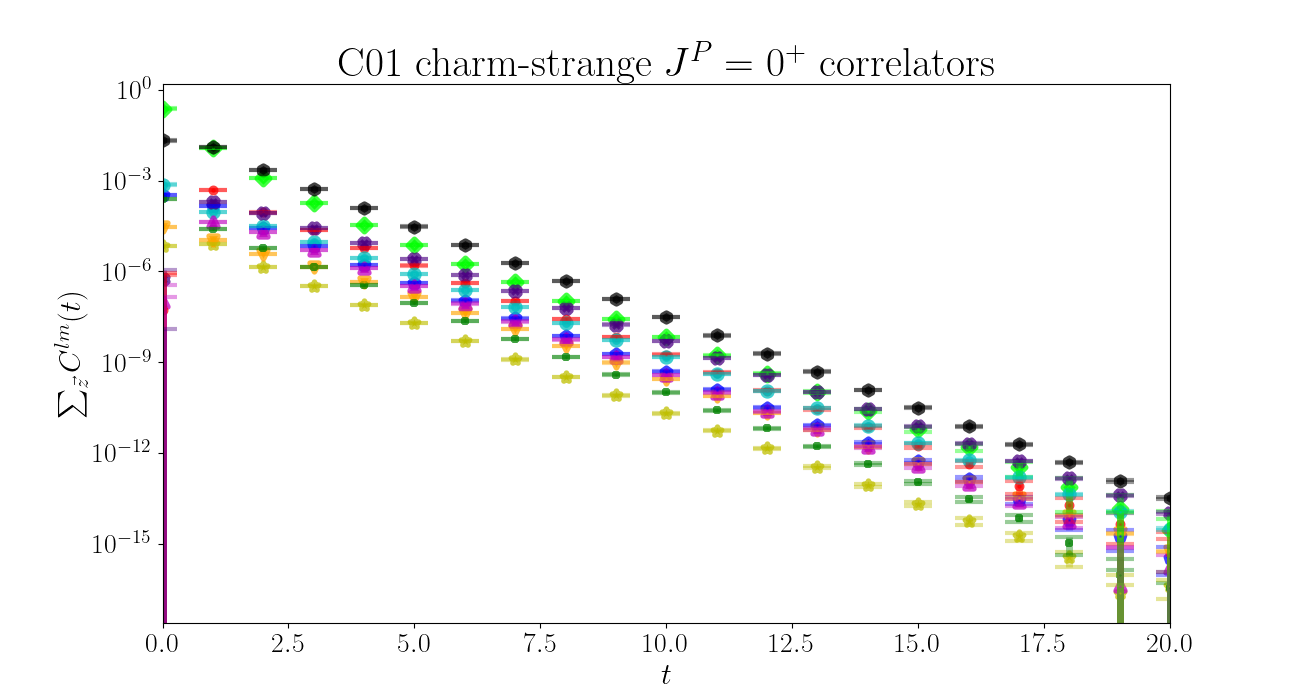}
    \hfill
    \includegraphics[width=0.49\linewidth]{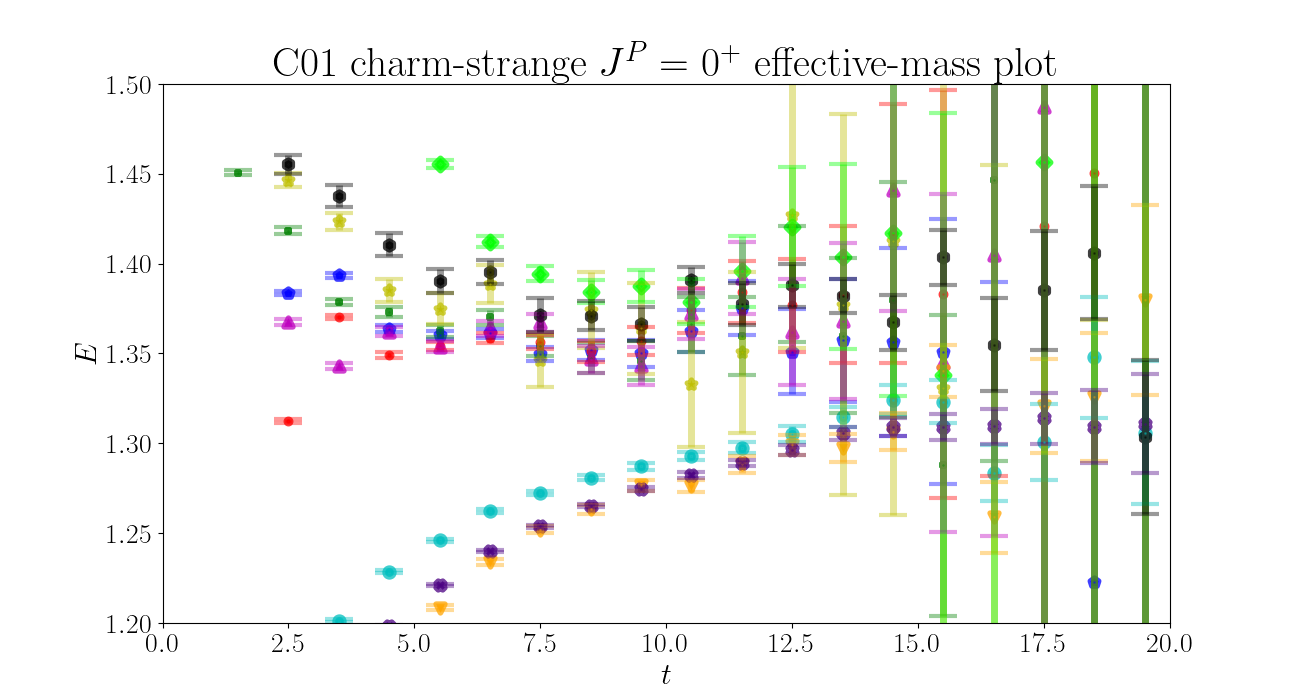}
    
    \hfill
        
    \includegraphics[width=0.49\linewidth]{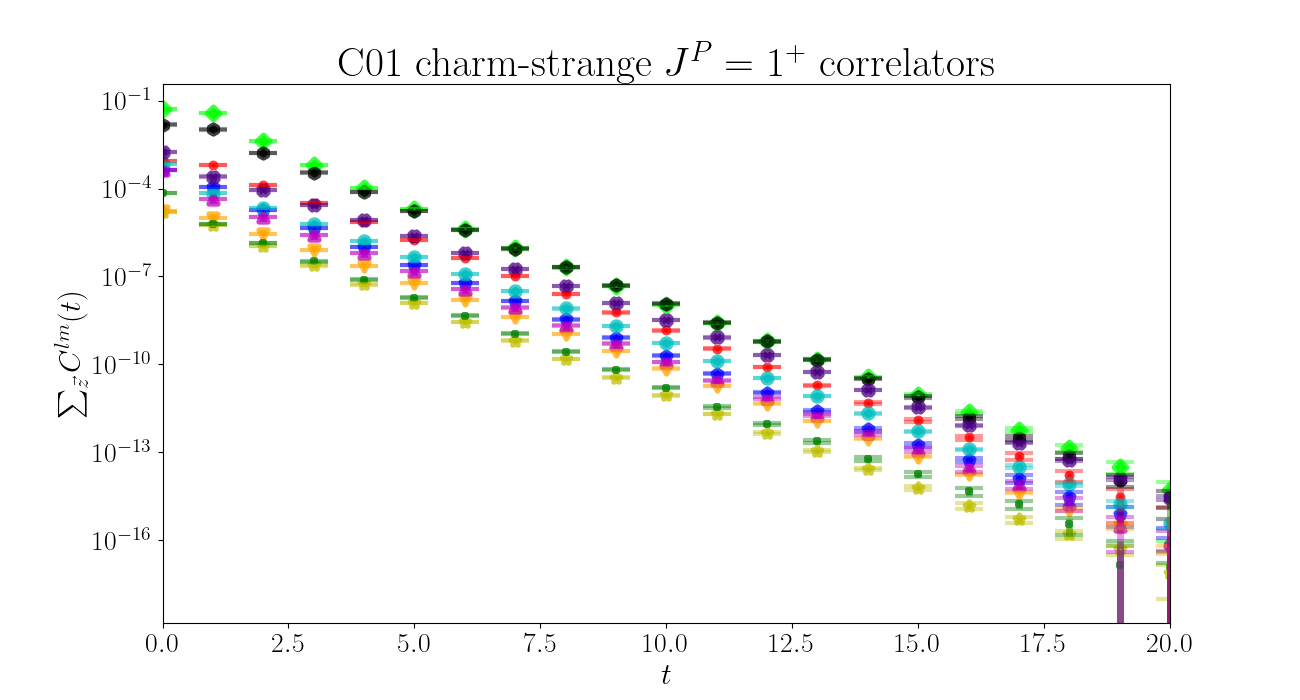}
    \hfill
    \includegraphics[width=0.49\linewidth]{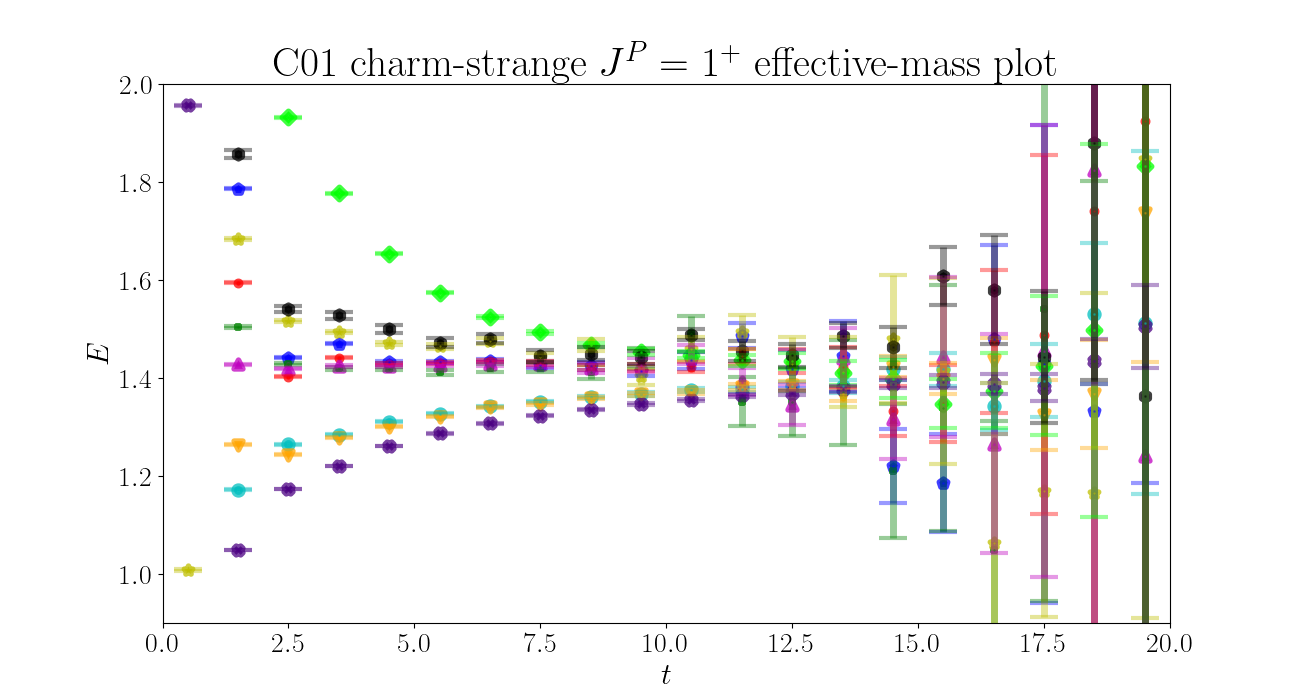}

    \includegraphics[width=0.49\linewidth]{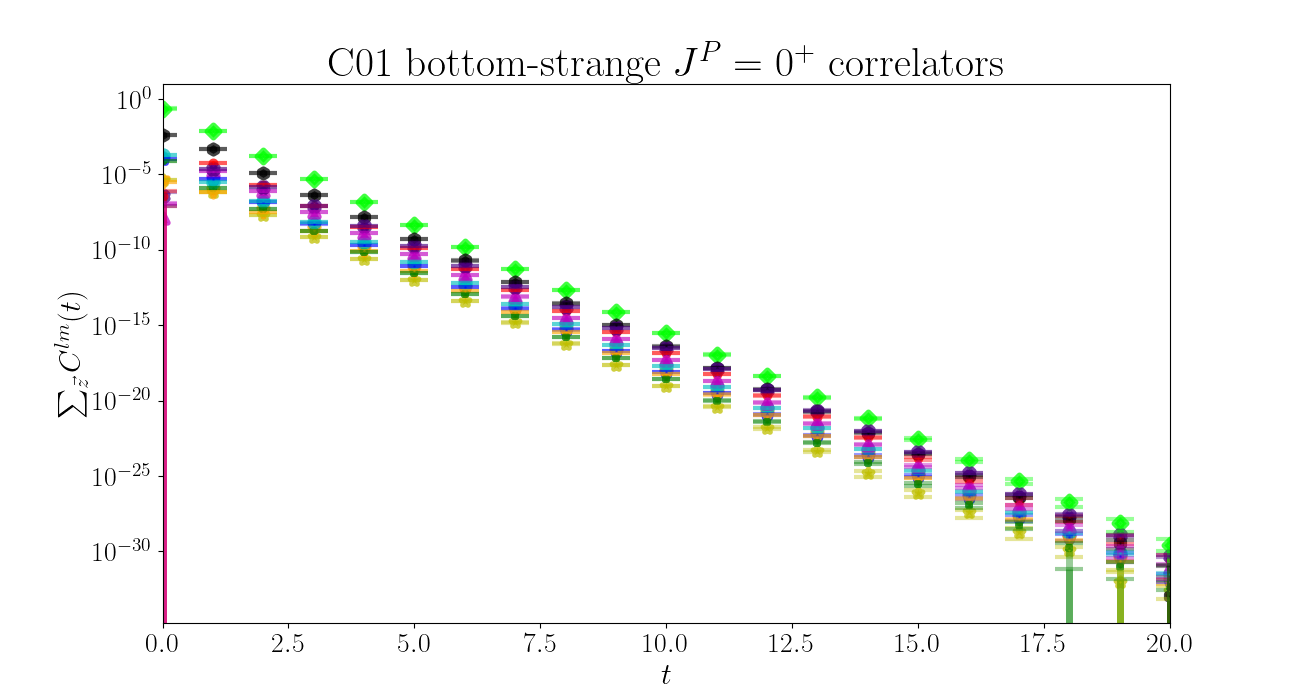}
    \hfill
    \includegraphics[width=0.49\linewidth]{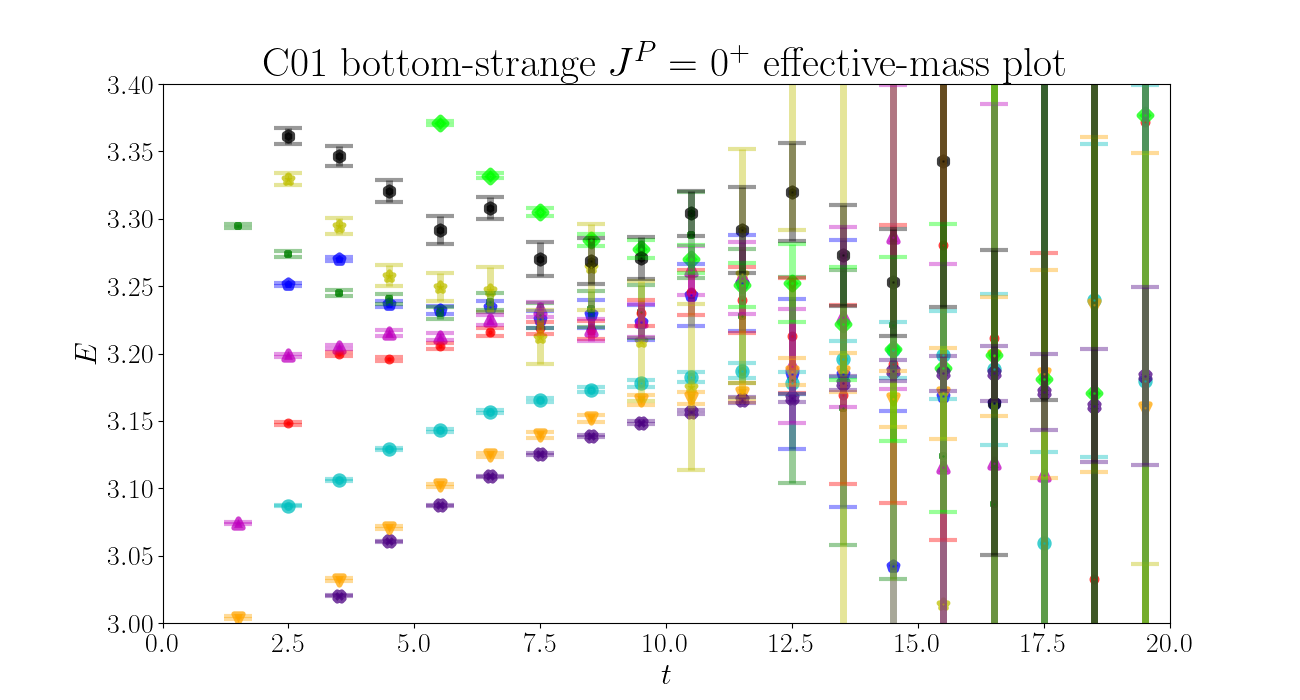}
    
    \hfill
        
    \includegraphics[width=0.49\linewidth]{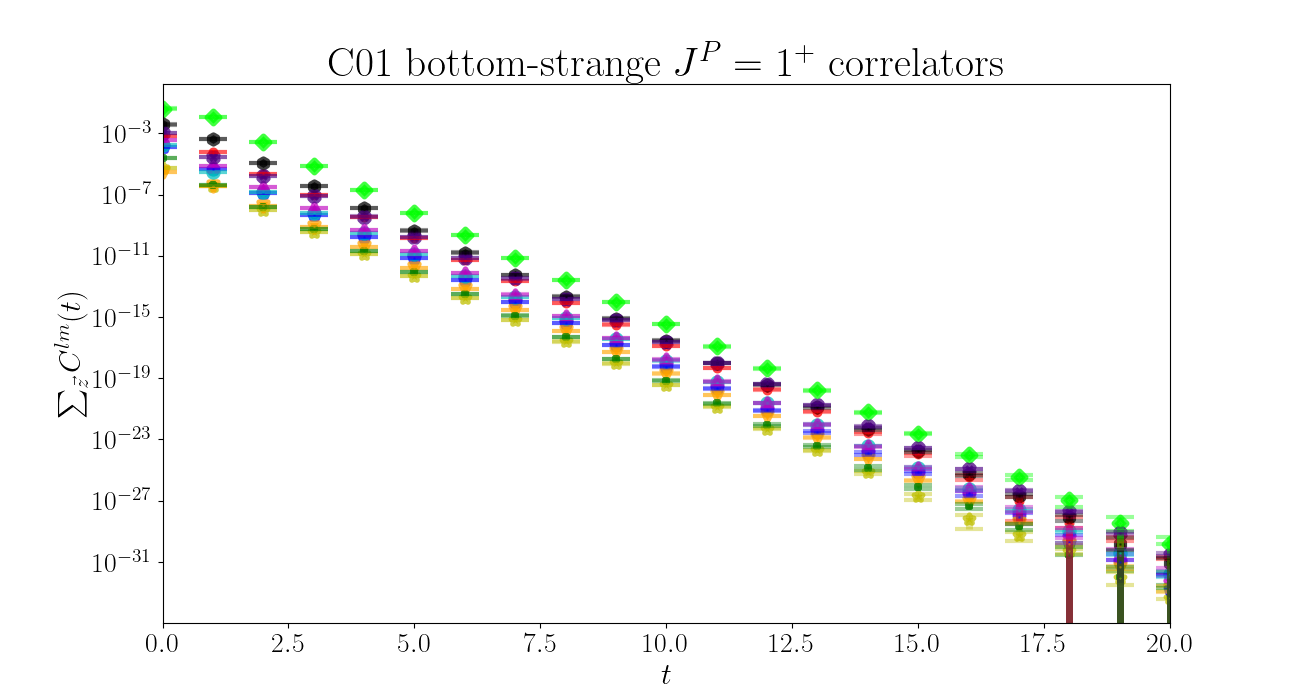}
    \hfill
    \includegraphics[width=0.49\linewidth]{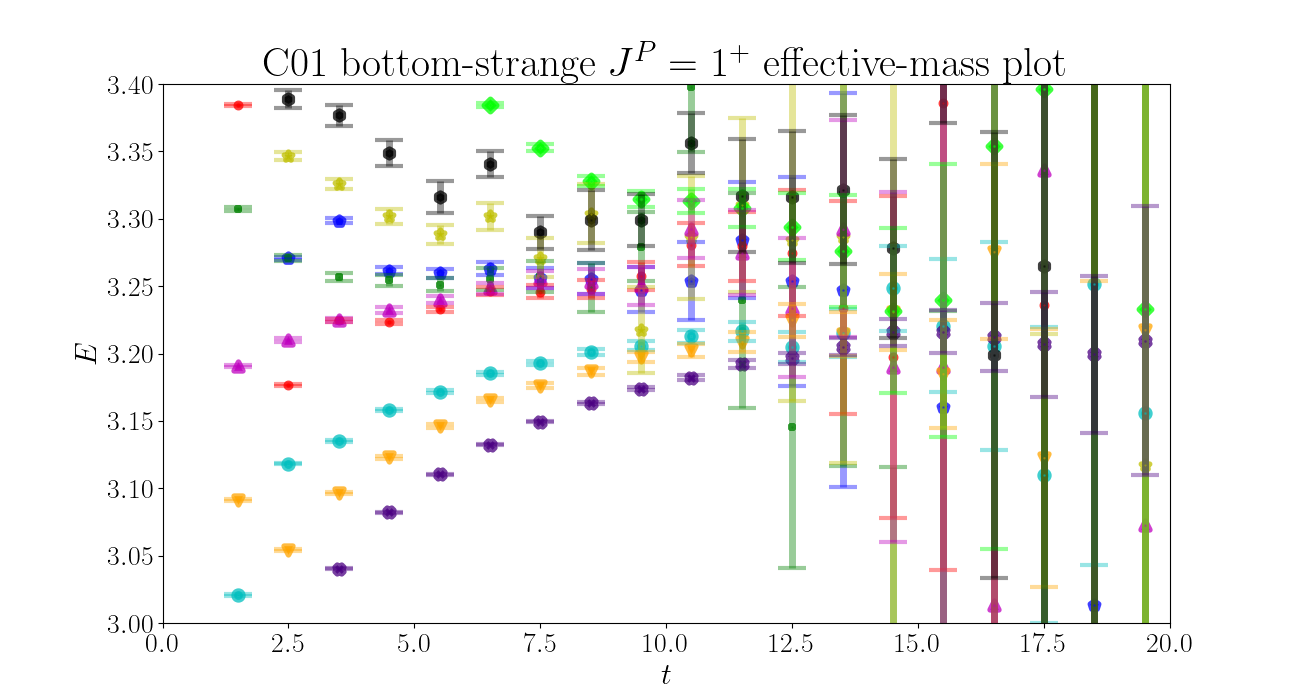}
    
    \centering
    \includegraphics[width=0.6\linewidth]{positive_parity/legend_new.pdf}
    \caption{Like Fig.~\protect\ref{fig:Clmplots-C00078}, but for the C01 ensemble. \label{fig:Clmplots-C01}}
\end{figure}

Plots of the positive-parity correlation matrices $C^{lm}(t)$ and associated effective-energy plots are provided here in Figs.~\ref{fig:Clmplots-C00078}-\ref{fig:Clmplots-C01} for all ensembles except F006, which can be found in the main text in Fig.~\ref{fig:Clmplots}.

\subsection{Positive-Parity Principal Correlators}

\begin{figure}[H]
    \centering
    \includegraphics[width=0.3\linewidth]{positive_parity/principalcorrlegend.pdf}
    
    \includegraphics[width=0.49\linewidth]{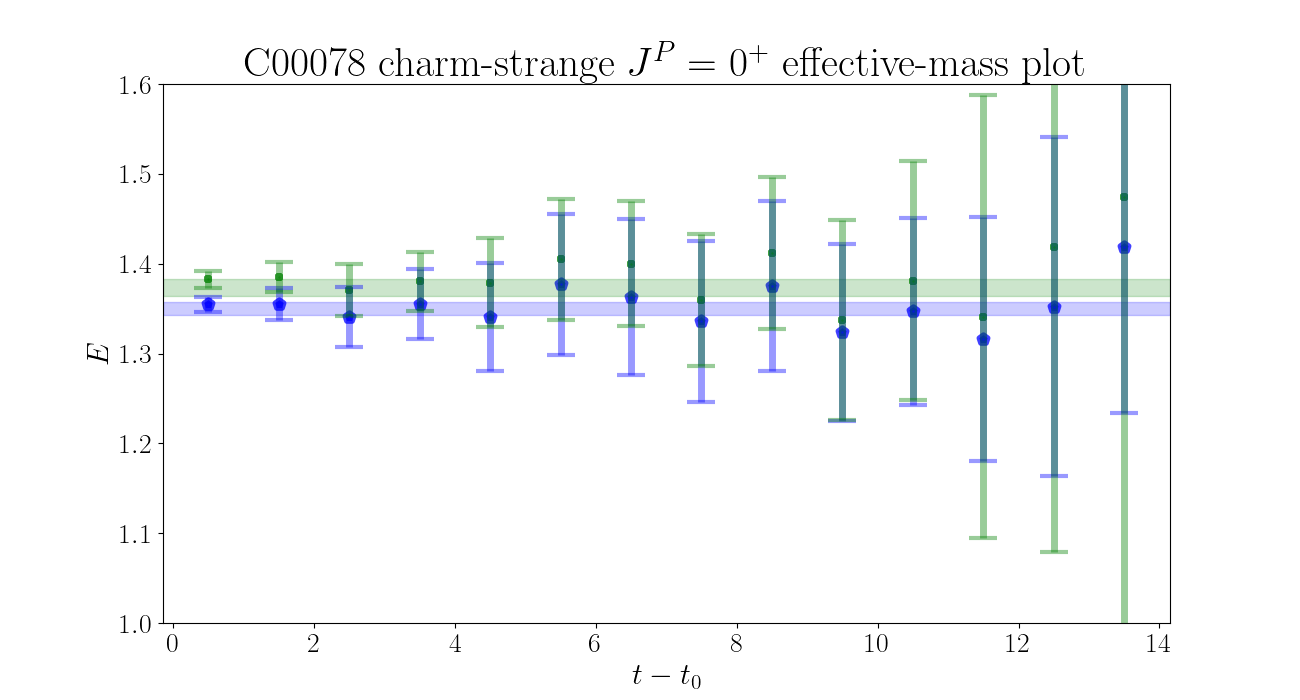}
    \hfill
    \includegraphics[width=0.49\linewidth]{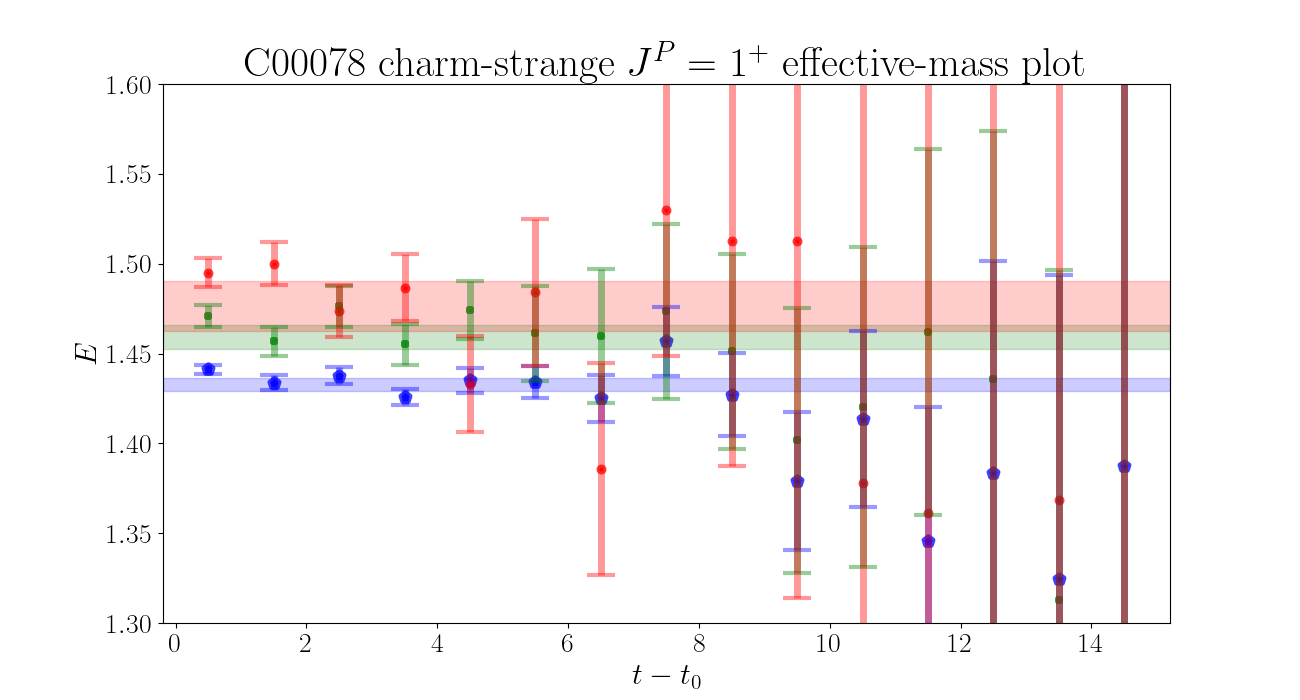}

    \includegraphics[width=0.49\linewidth]{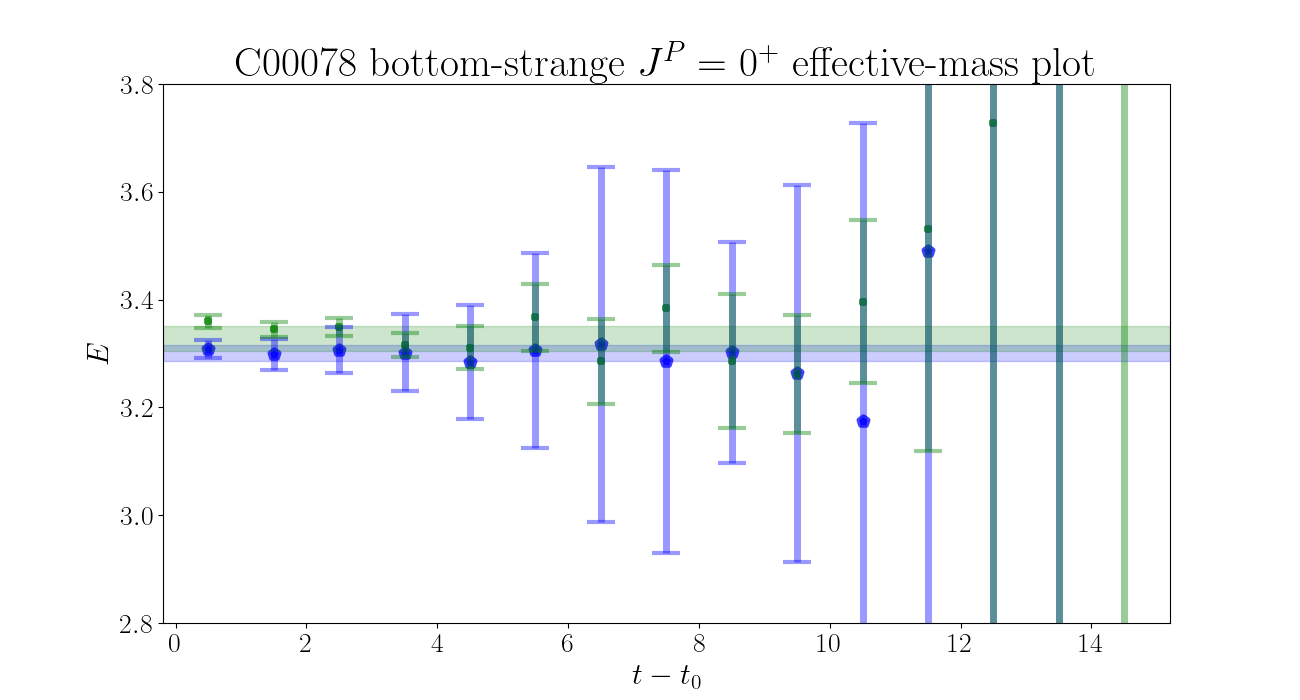}
    \hfill
    \includegraphics[width=0.49\linewidth]{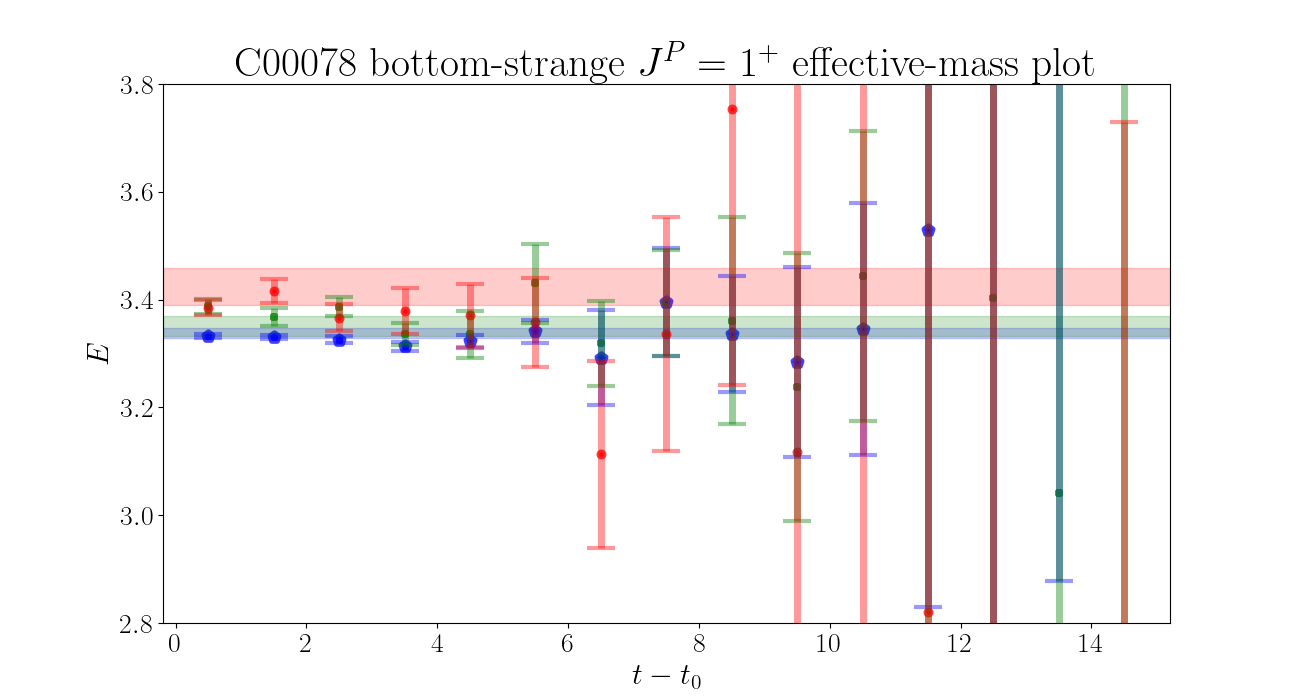}
    \caption{Like Fig.~\protect\ref{fig:principalcorr}, but for the C00078 ensemble. \label{fig:princcorr-C00078}}
\end{figure}

\begin{figure}[H]
    \centering
    \includegraphics[width=0.3\linewidth]{positive_parity/principalcorrlegend.pdf}
    
    \includegraphics[width=0.49\linewidth]{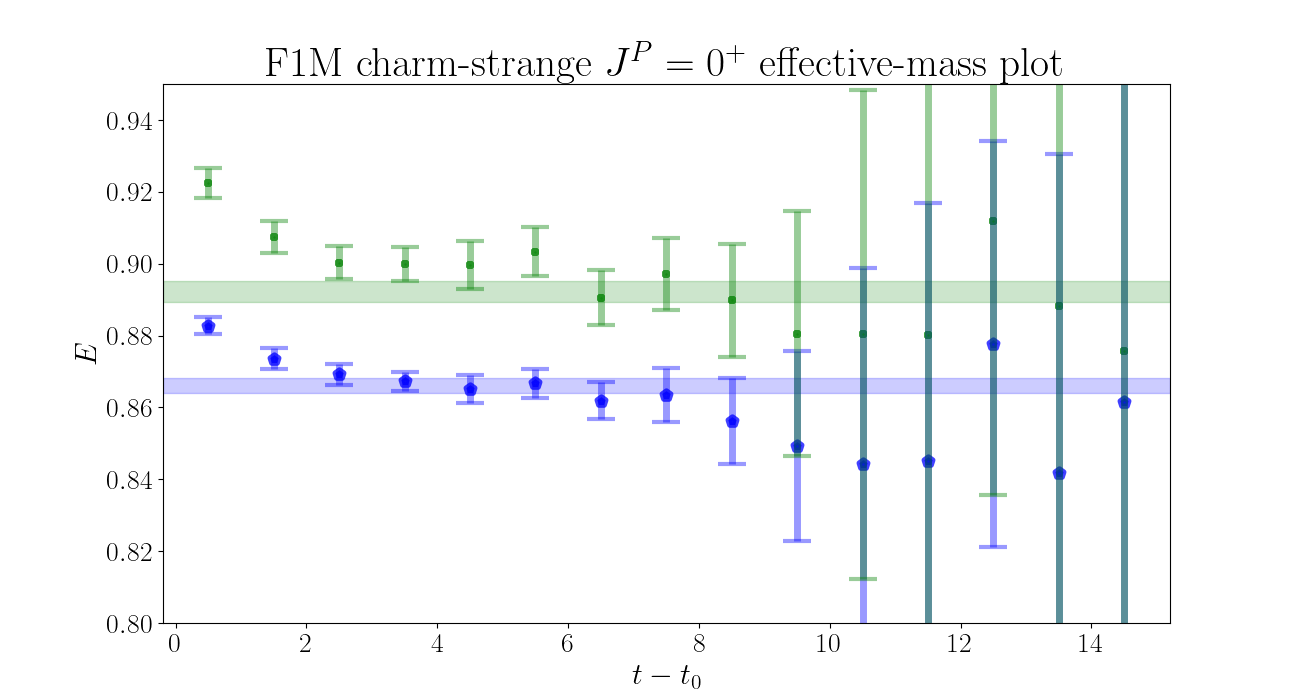}
    \hfill
    \includegraphics[width=0.49\linewidth]{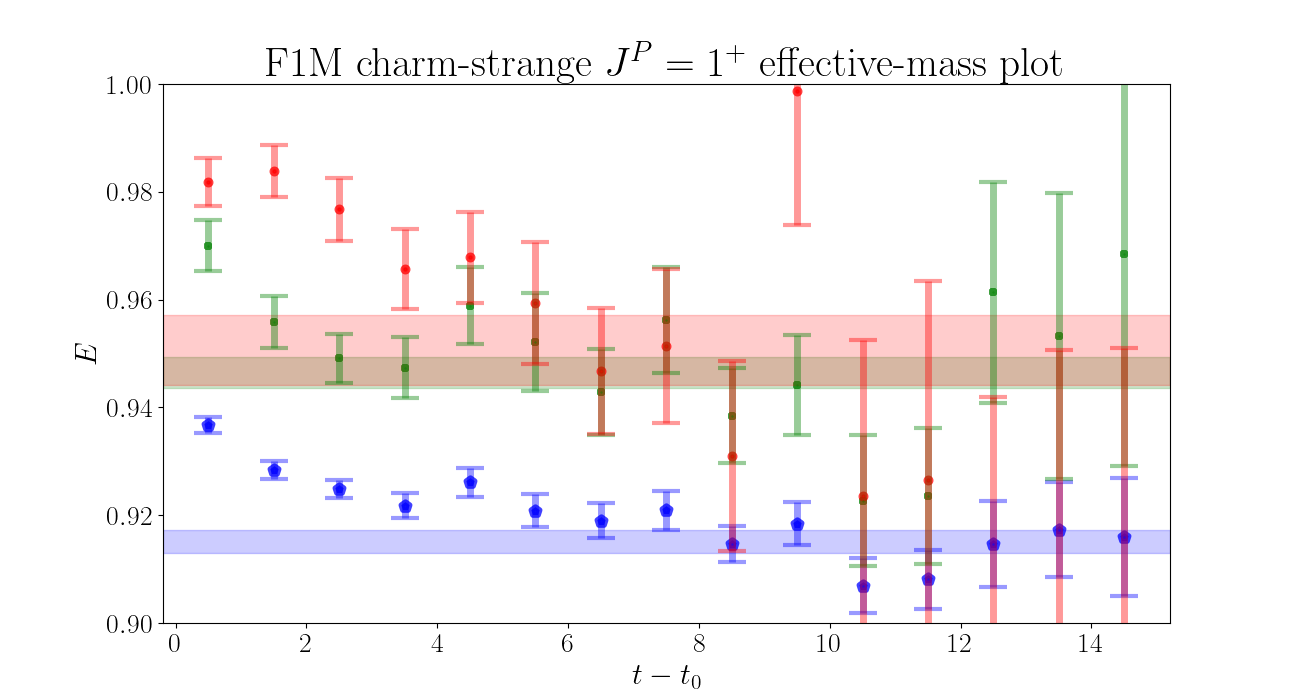}

    \includegraphics[width=0.49\linewidth]{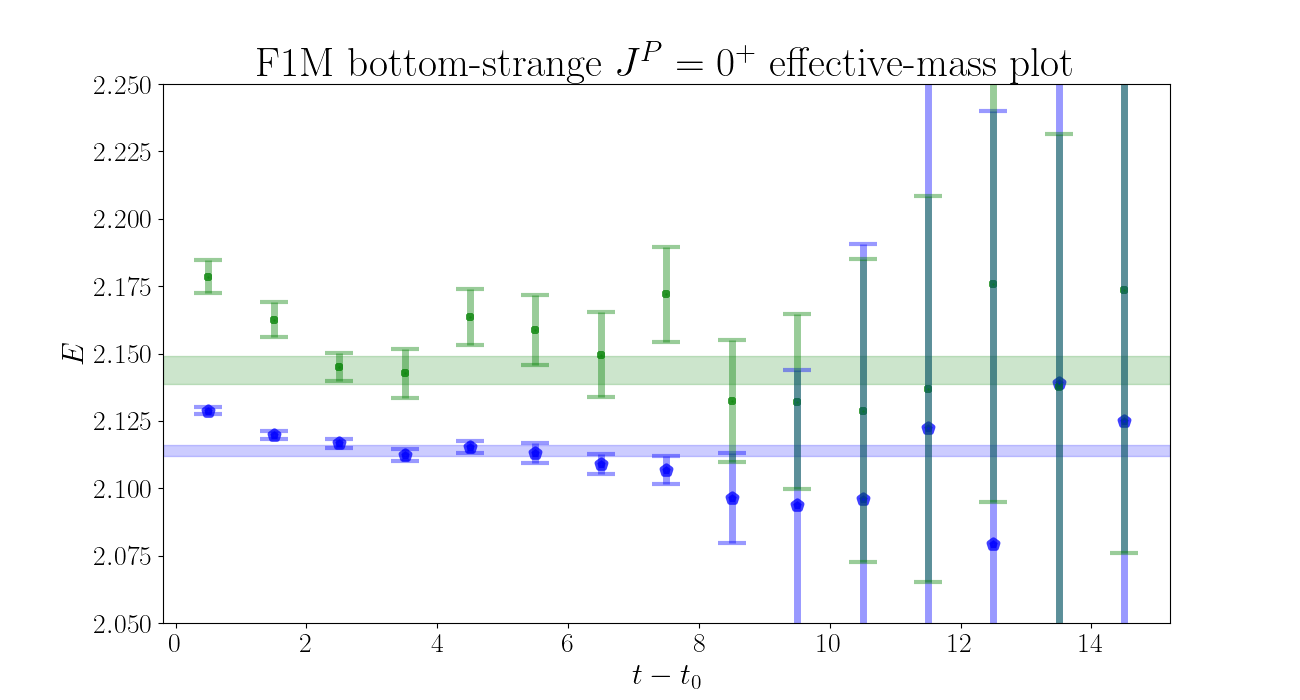}
    \hfill
    \includegraphics[width=0.49\linewidth]{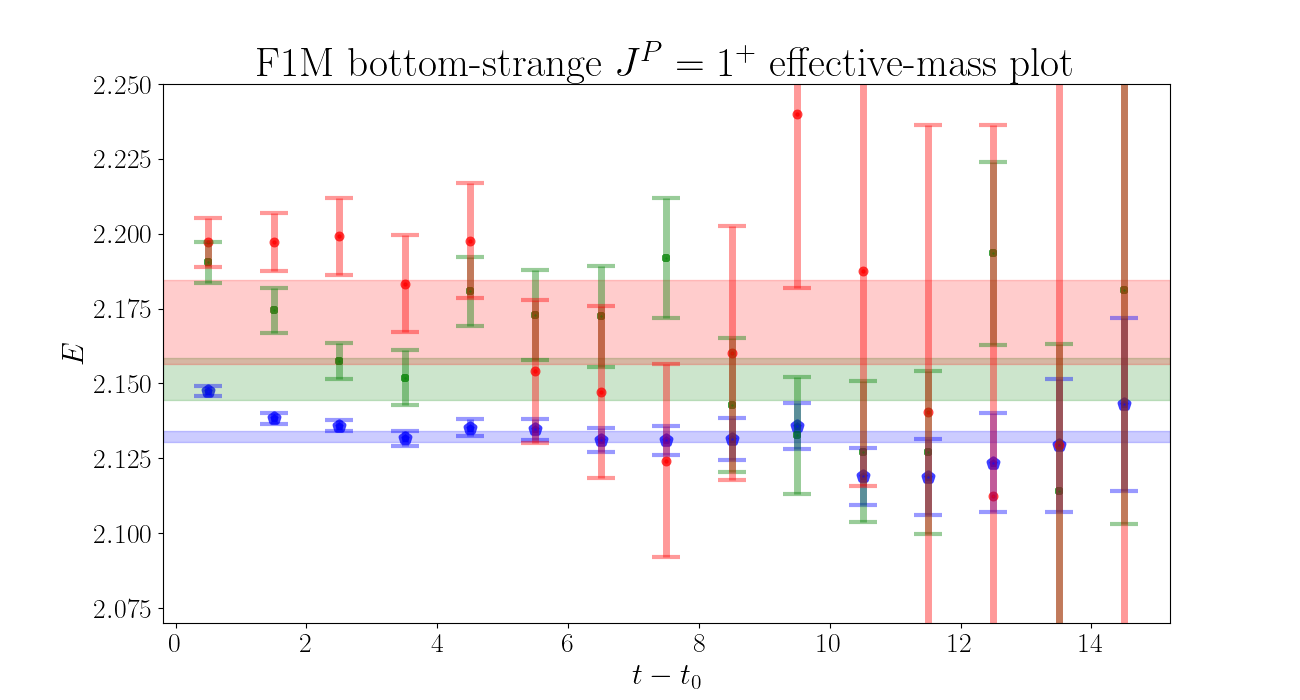}
    \caption{Like Fig.~\protect\ref{fig:principalcorr}, but for the F1M ensemble. \label{fig:princcorr-F1M}}
\end{figure}

\begin{figure}[H]
    \centering
    \includegraphics[width=0.3\linewidth]{positive_parity/principalcorrlegend.pdf}
    
    \includegraphics[width=0.49\linewidth]{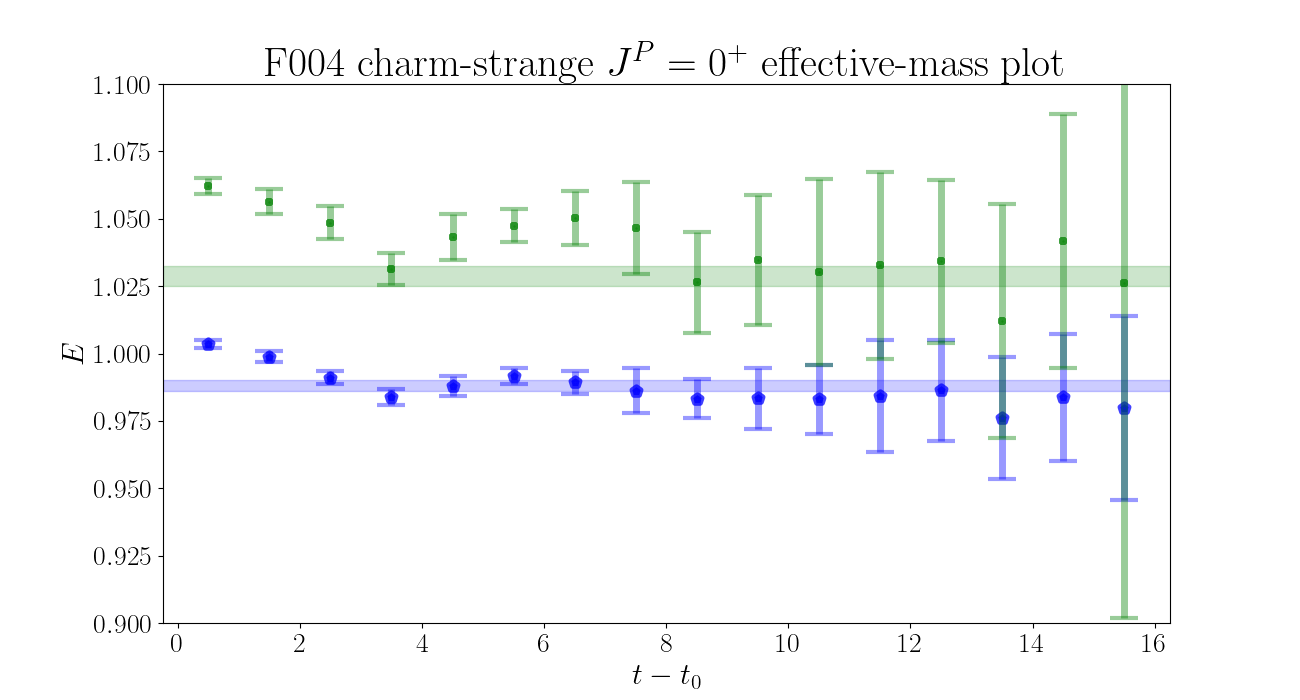}
    \hfill
    \includegraphics[width=0.49\linewidth]{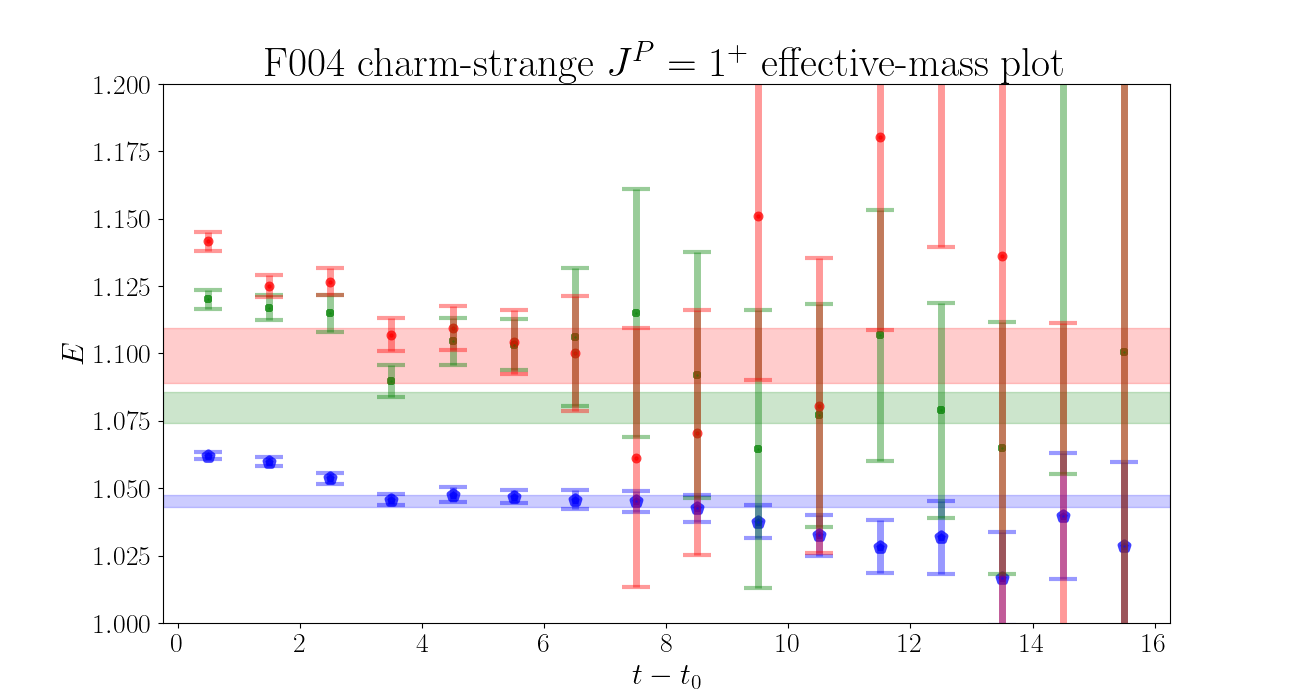}

    \includegraphics[width=0.49\linewidth]{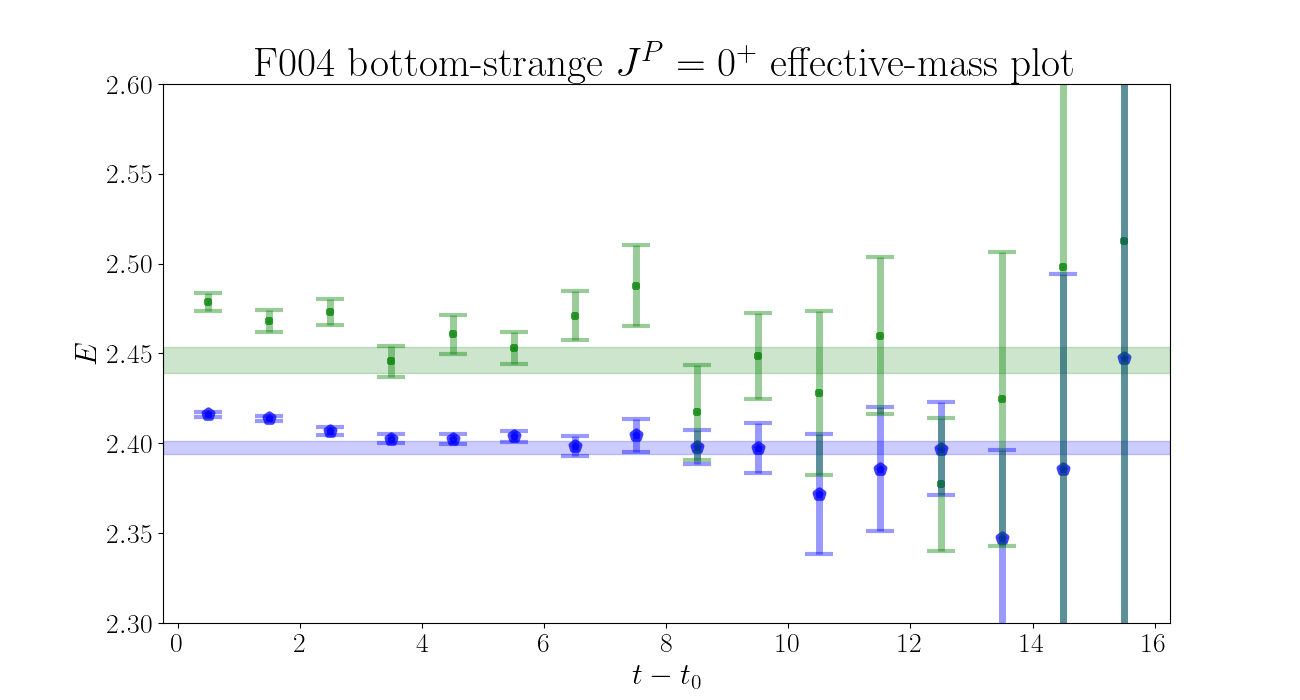}
    \hfill
    \includegraphics[width=0.49\linewidth]{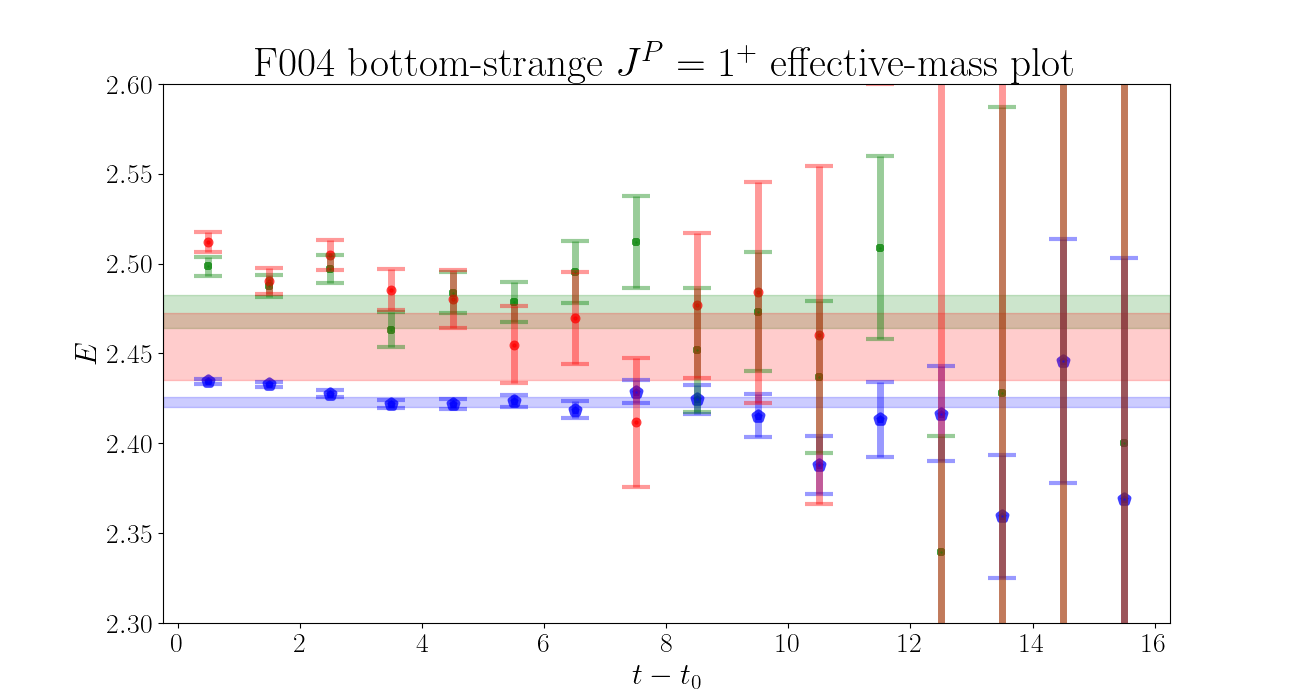}
    \caption{Like Fig.~\protect\ref{fig:principalcorr}, but for the F004 ensemble. \label{fig:princcorr-F004}}
\end{figure}

\begin{figure}[H]
    \centering
    \includegraphics[width=0.3\linewidth]{positive_parity/principalcorrlegend.pdf}
    
    \includegraphics[width=0.49\linewidth]{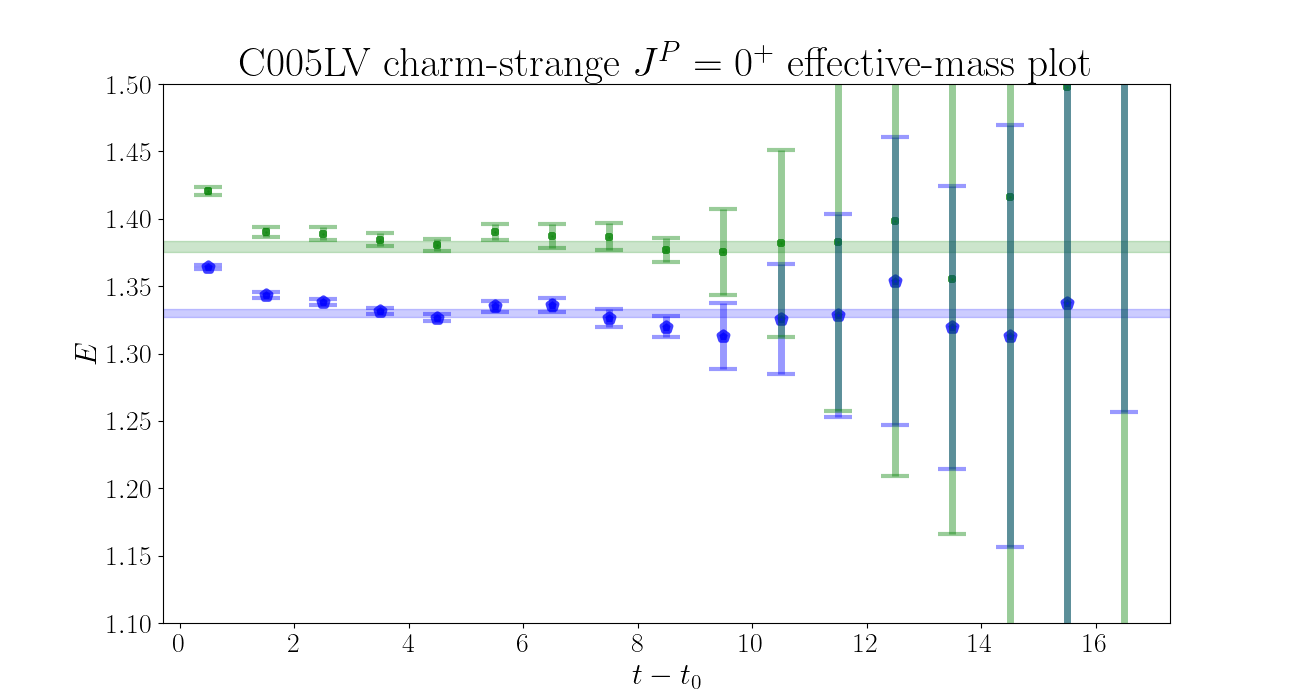}
    \hfill
    \includegraphics[width=0.49\linewidth]{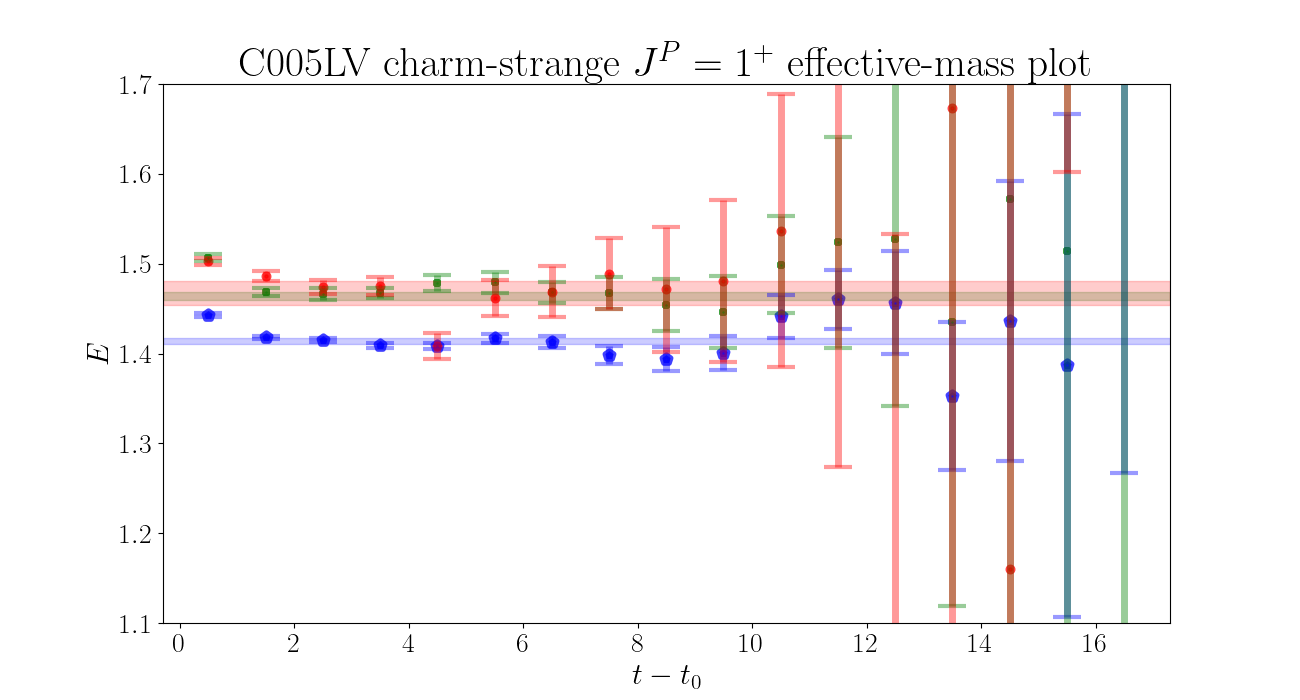}
    
    \includegraphics[width=0.49\linewidth]{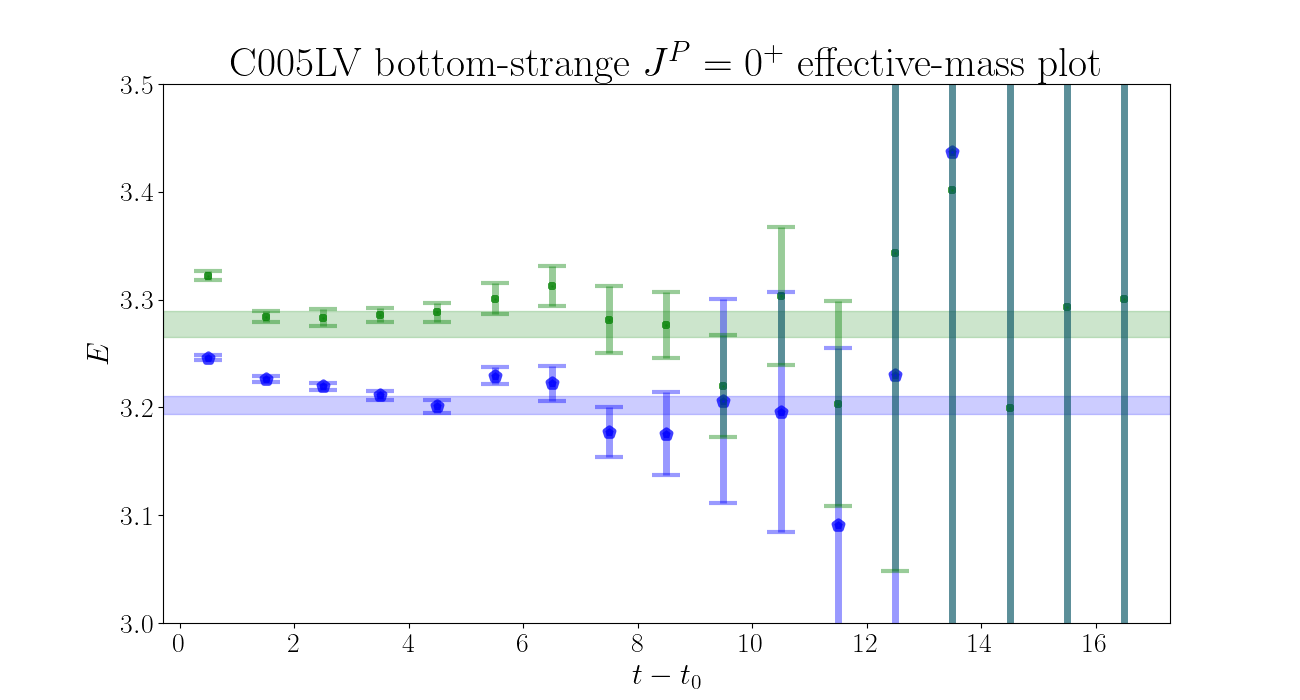}
    \hfill
    \includegraphics[width=0.49\linewidth]{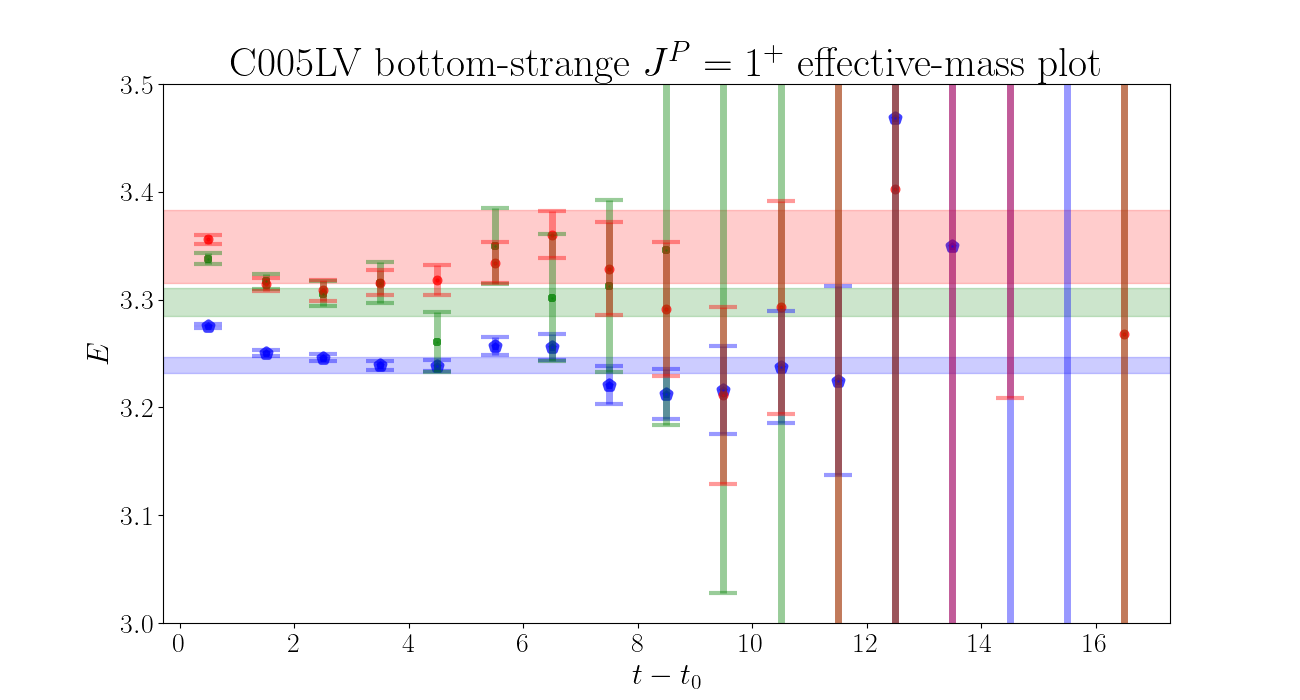}
    \caption{Like Fig.~\protect\ref{fig:principalcorr}, but for the C005LV ensemble. \label{fig:princcorr-C005LV}}
\end{figure}

\begin{figure}[H]
    \centering
    \includegraphics[width=0.3\linewidth]{positive_parity/principalcorrlegend.pdf}
    
    \includegraphics[width=0.49\linewidth]{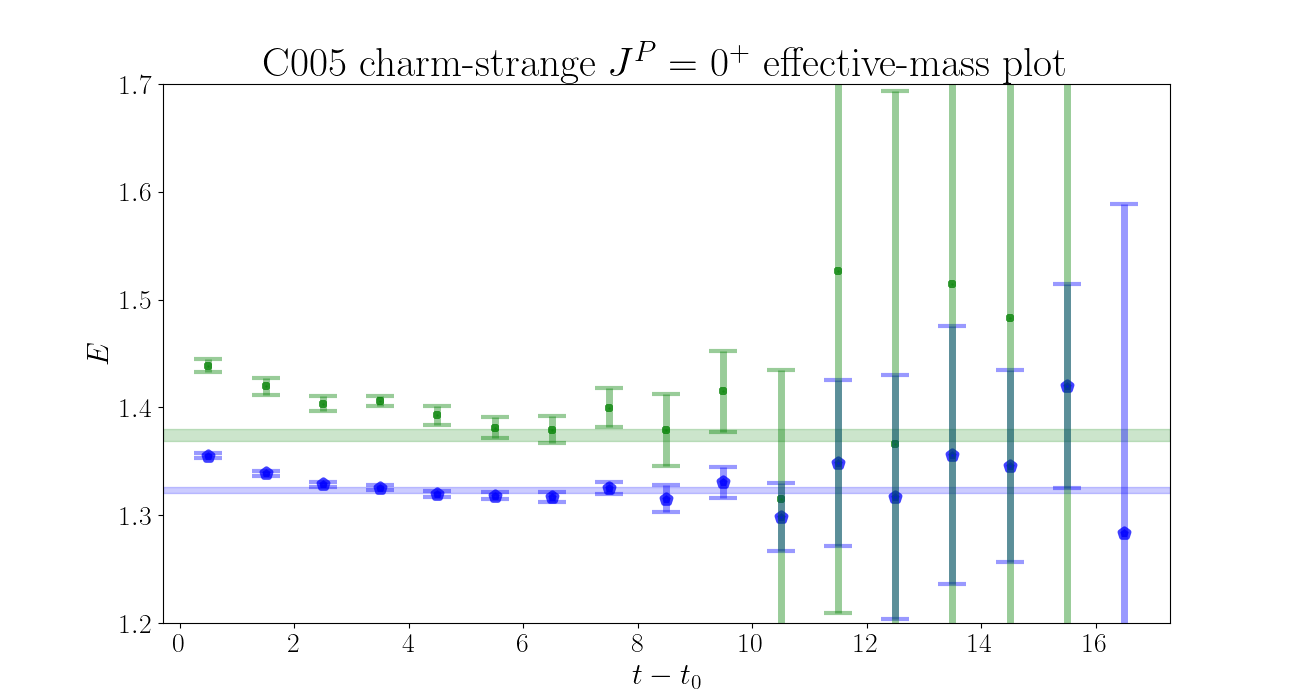}
    \hfill
    \includegraphics[width=0.49\linewidth]{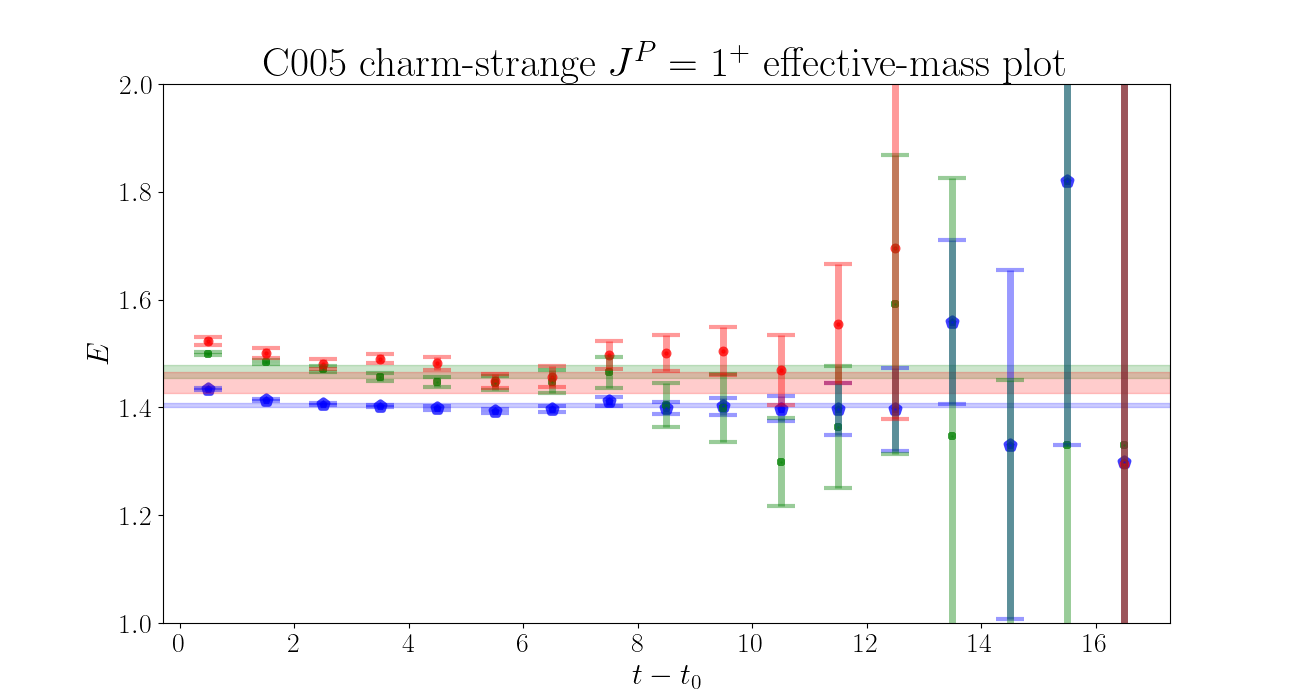}

    \includegraphics[width=0.49\linewidth]{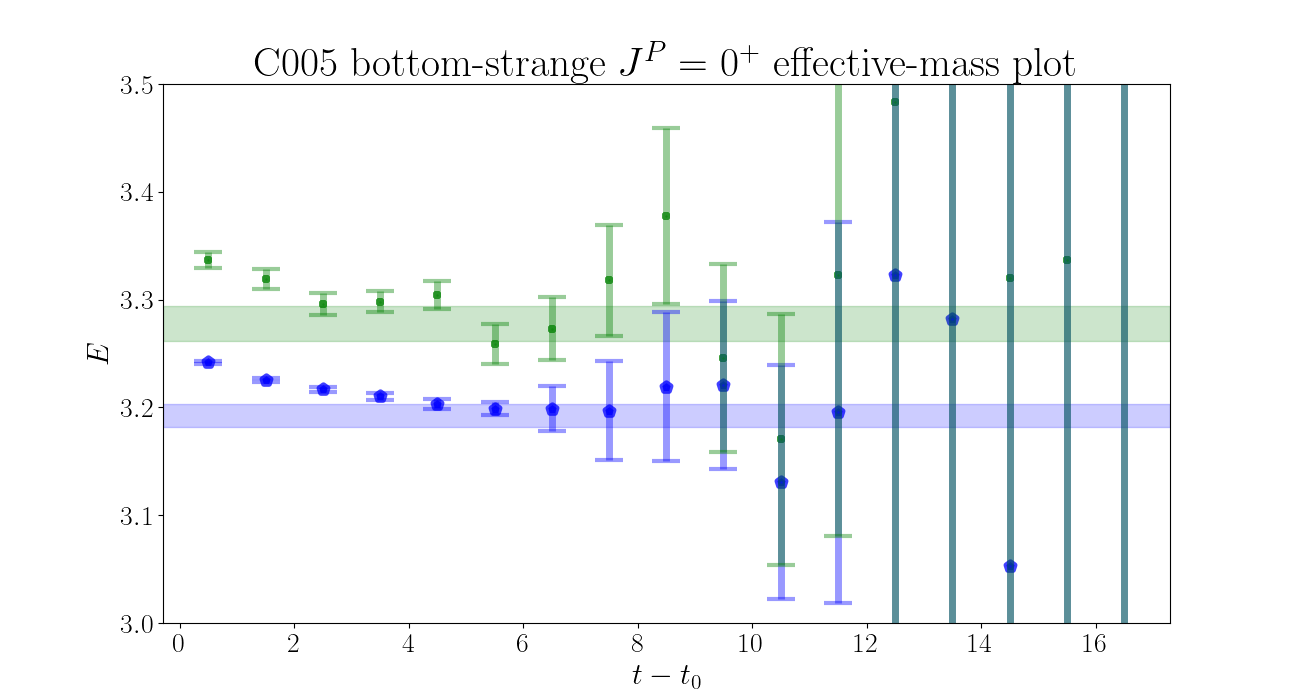}
    \hfill
    \includegraphics[width=0.49\linewidth]{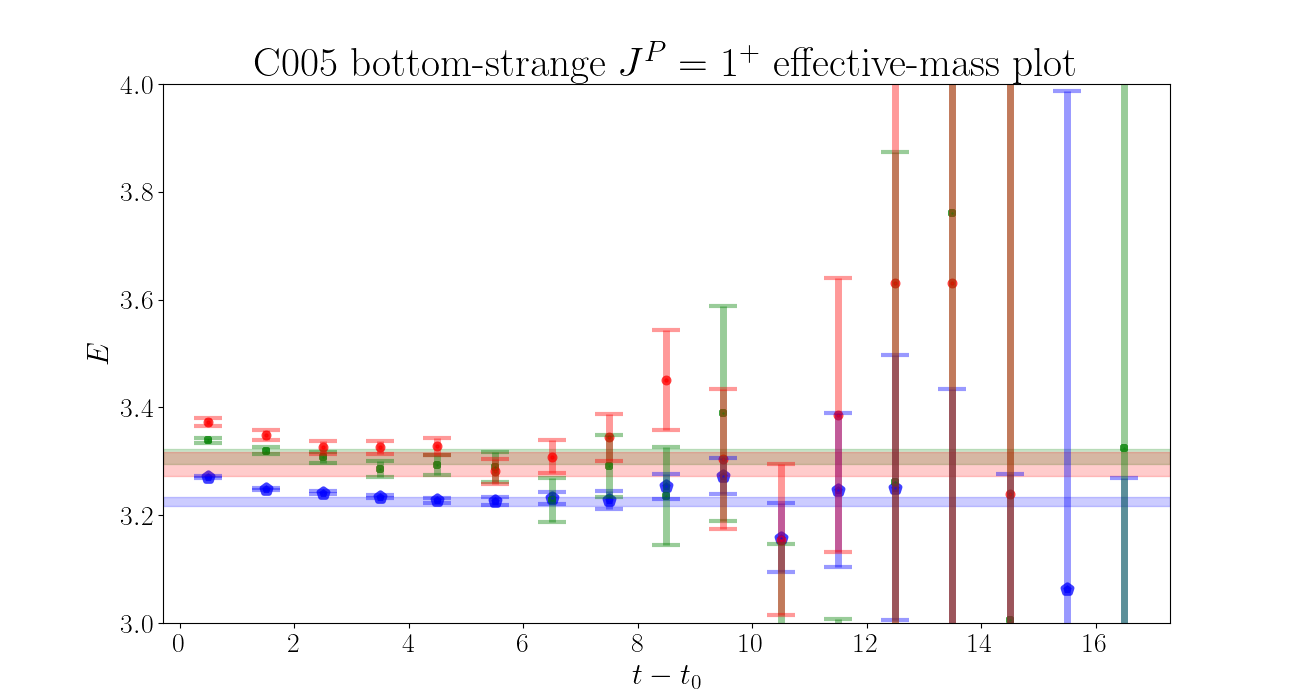}
    \caption{Like Fig.~\protect\ref{fig:principalcorr}, but for the C005 ensemble. \label{fig:princcorr-C005}}
\end{figure}

\begin{figure}[H]
    \centering
    \includegraphics[width=0.3\linewidth]{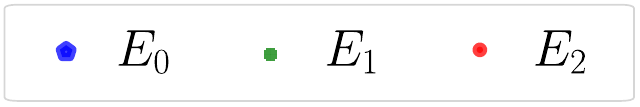}
    
    \includegraphics[width=0.49\linewidth]{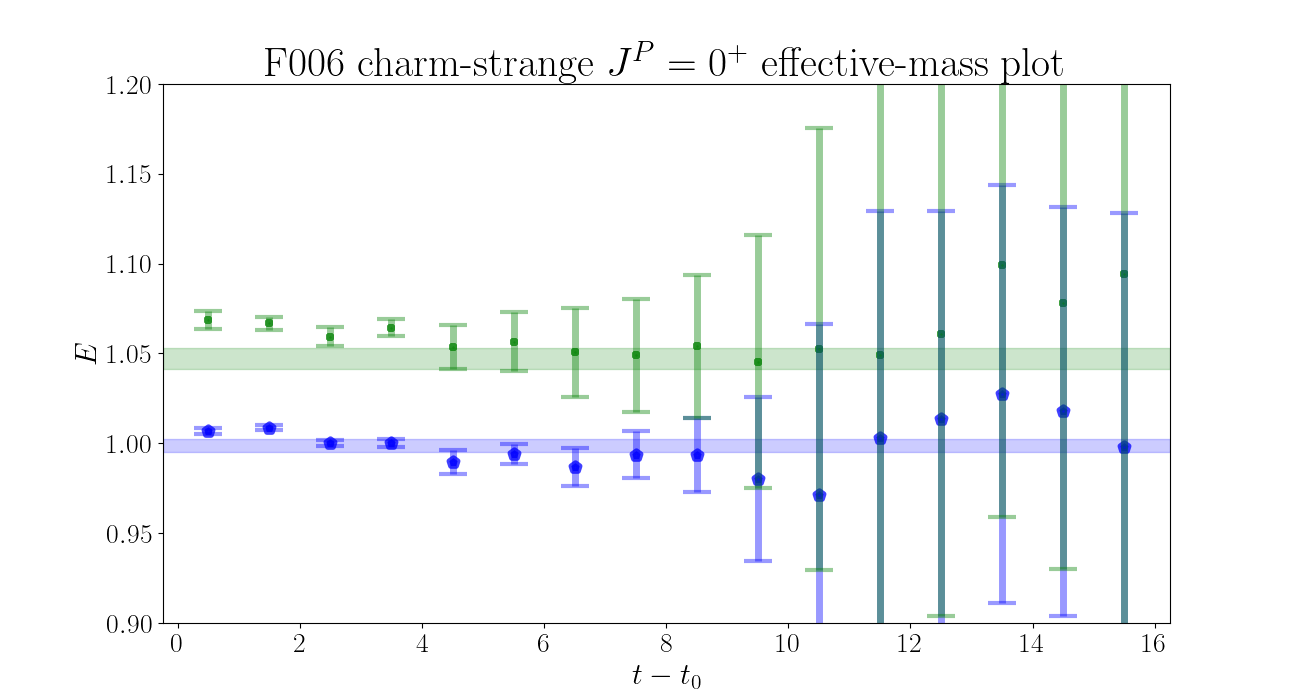}
    \hfill
    \includegraphics[width=0.49\linewidth]{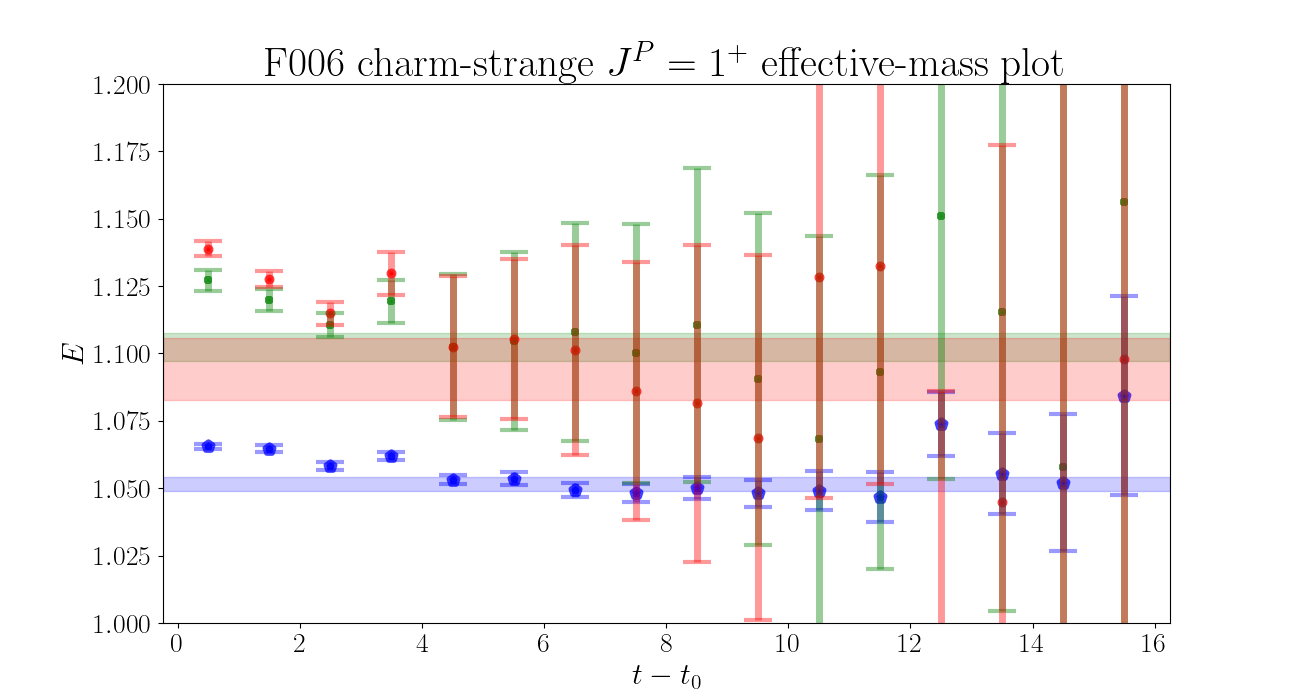}

    \includegraphics[width=0.49\linewidth]{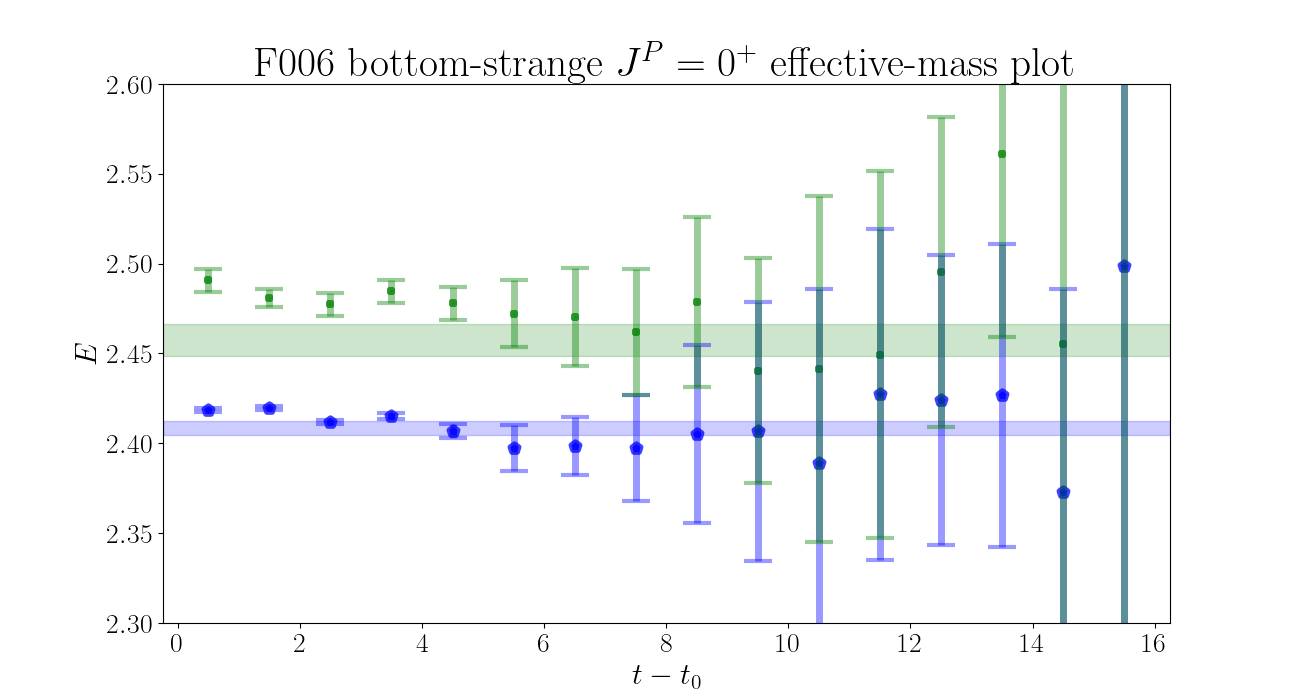}
    \hfill
    \includegraphics[width=0.49\linewidth]{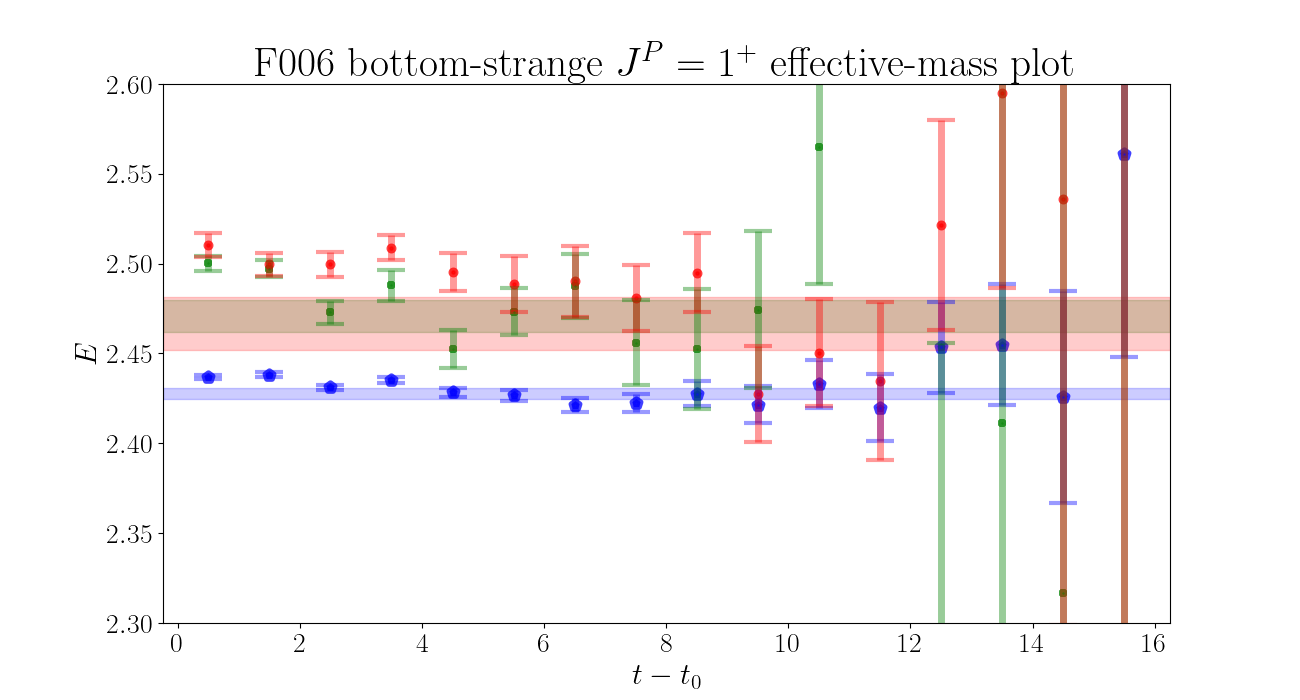}
    \caption{Like Fig.~\protect\ref{fig:principalcorr}, but for the F006 ensemble. \label{fig:princcorr-F006}}
\end{figure}

\begin{figure}[H]
    \centering
    \includegraphics[width=0.3\linewidth]{positive_parity/principalcorrlegend.pdf}
    
    \includegraphics[width=0.49\linewidth]{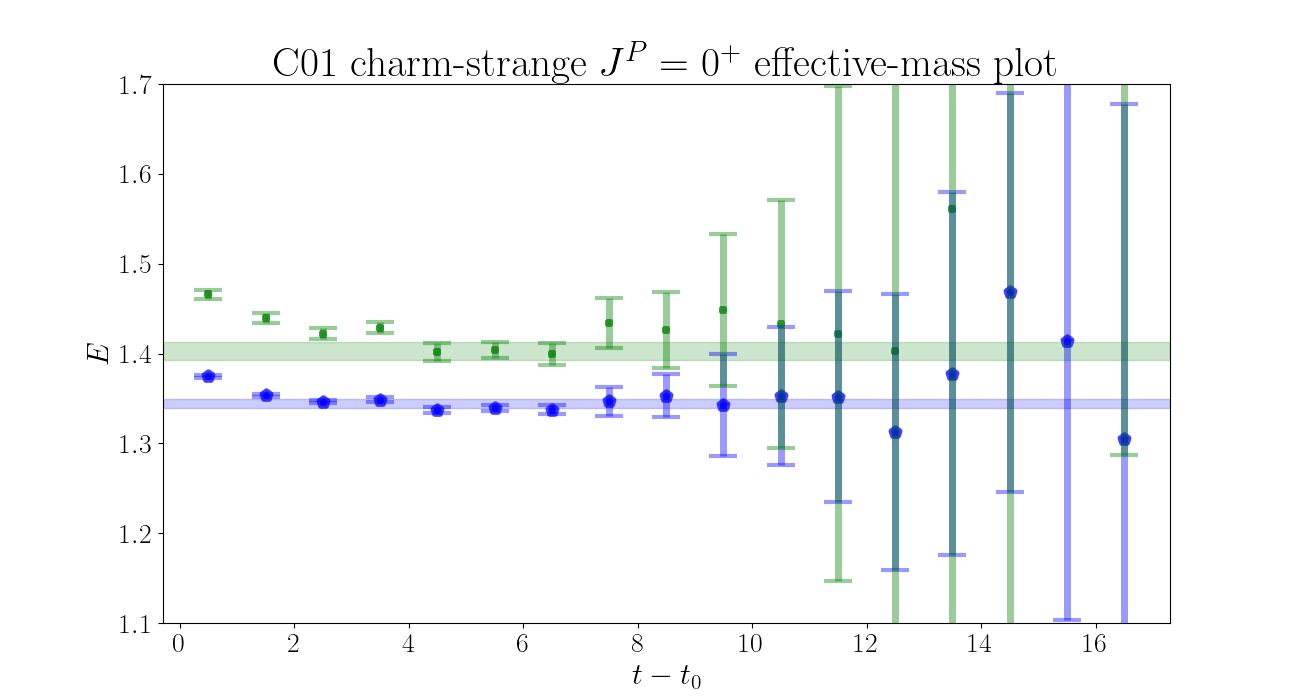}
    \hfill
    \includegraphics[width=0.49\linewidth]{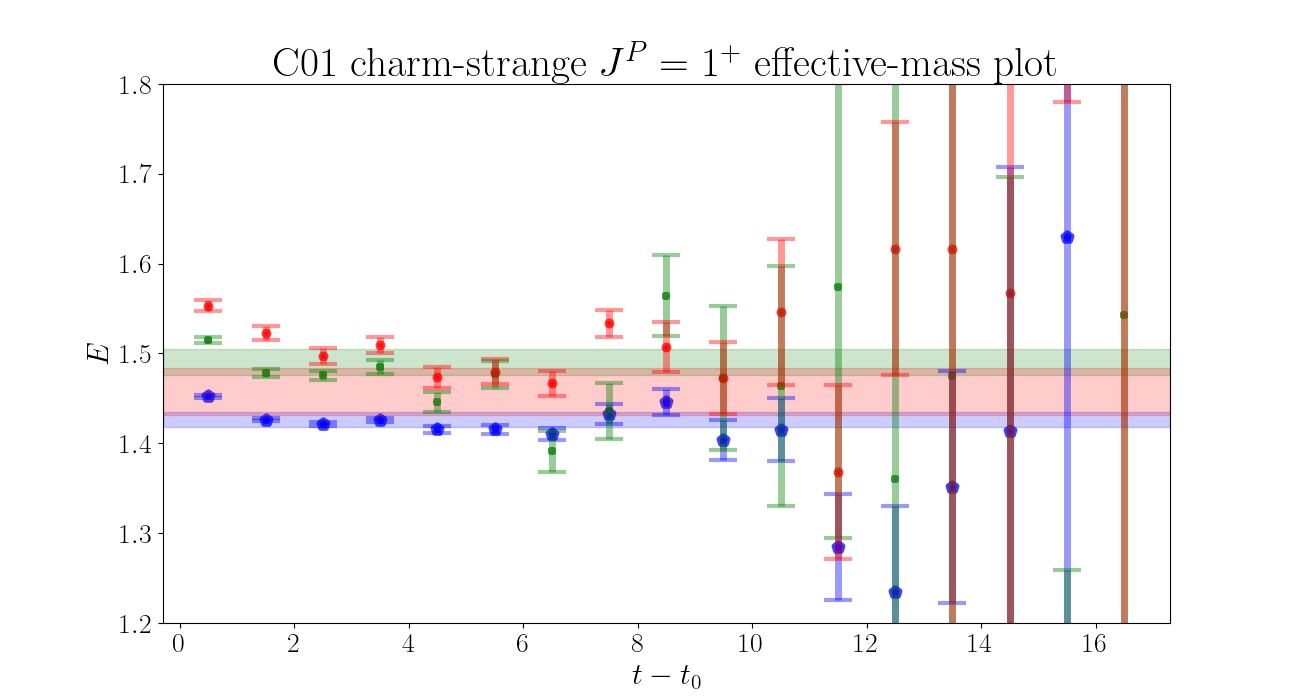}

    \includegraphics[width=0.49\linewidth]{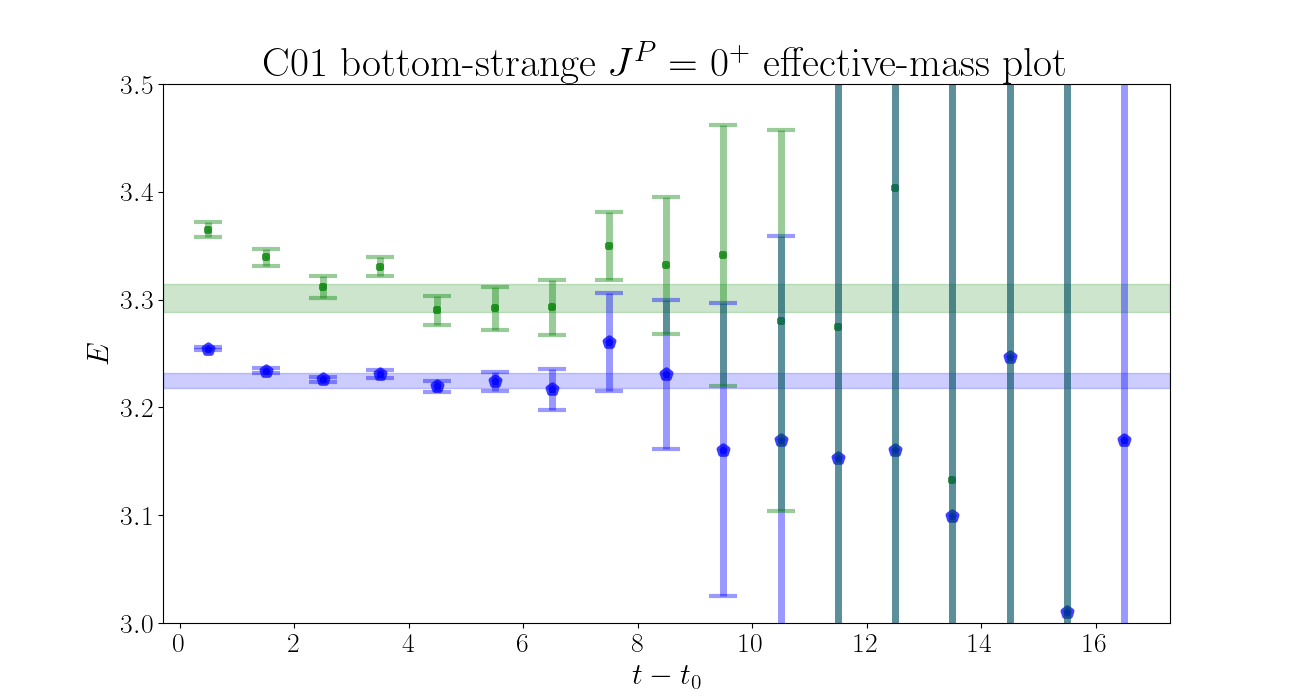}
    \hfill
    \includegraphics[width=0.49\linewidth]{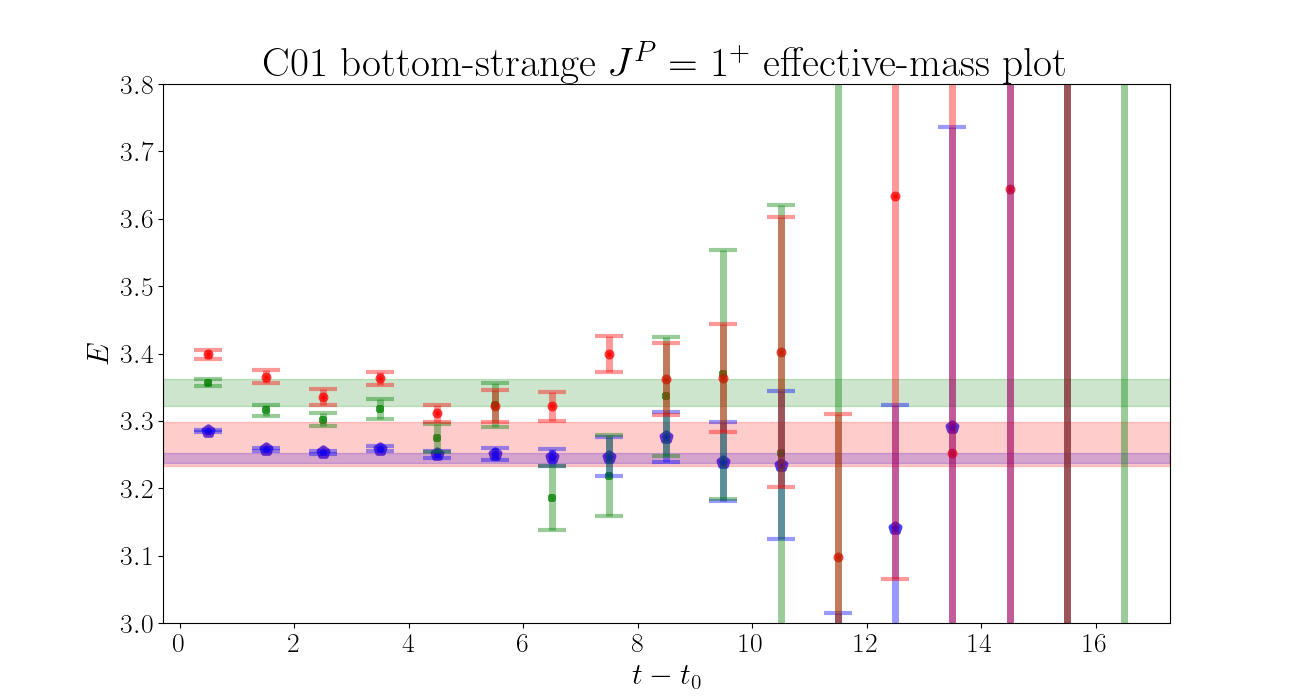}
    \caption{Like Fig.~\protect\ref{fig:principalcorr}, but for the C01 ensemble. \label{fig:princcorr-C01}}
\end{figure}

This section contains plots of the positive-parity principal-correlator effective energies in Figs.~\ref{fig:princcorr-C00078}-\ref{fig:princcorr-C01} for all ensembles except F006. The F006 plots were given in the main text in Fig.~\ref{fig:principalcorr}.

\label{sec:principalcorrplots}

\subsection{Positive-Parity $\langle J^\dag W^\dagger\rangle$ and $\langle W W^\dagger\rangle$}
\label{sec:wplots}

\begin{figure}[H]
    \centering
    \centering
    \includegraphics[width=0.49\linewidth]{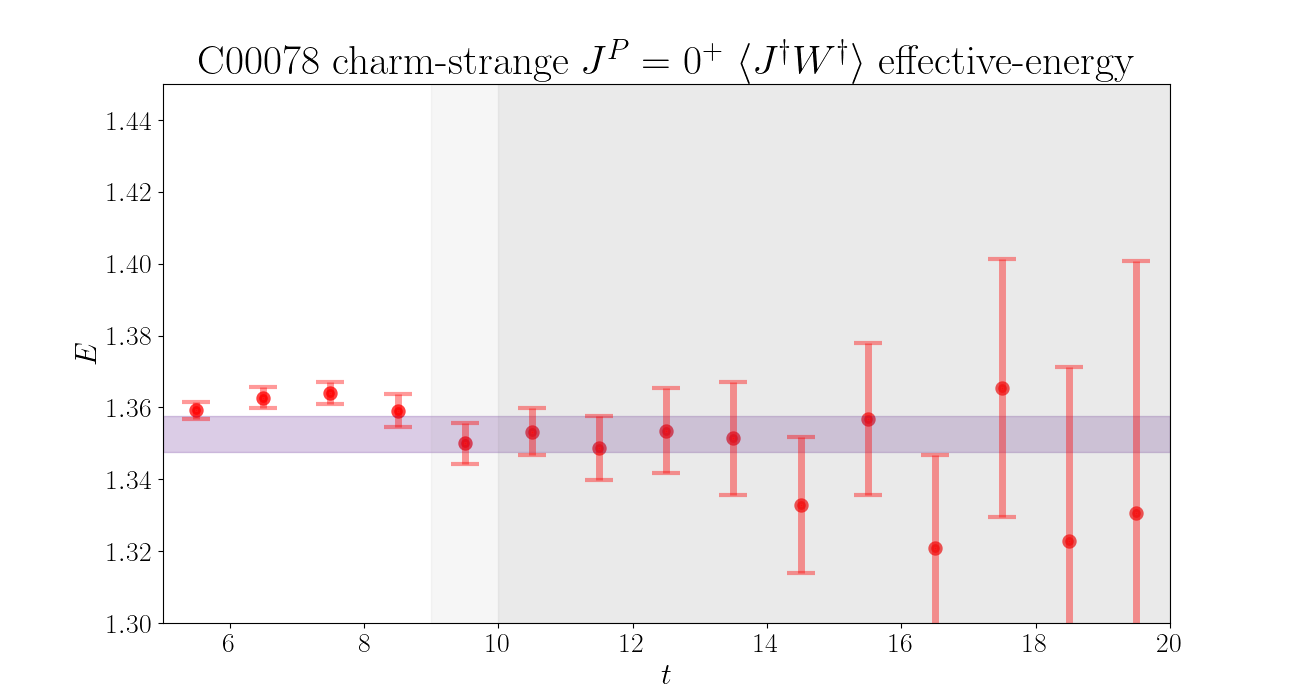}
    \hfill
    \includegraphics[width=0.49\linewidth]{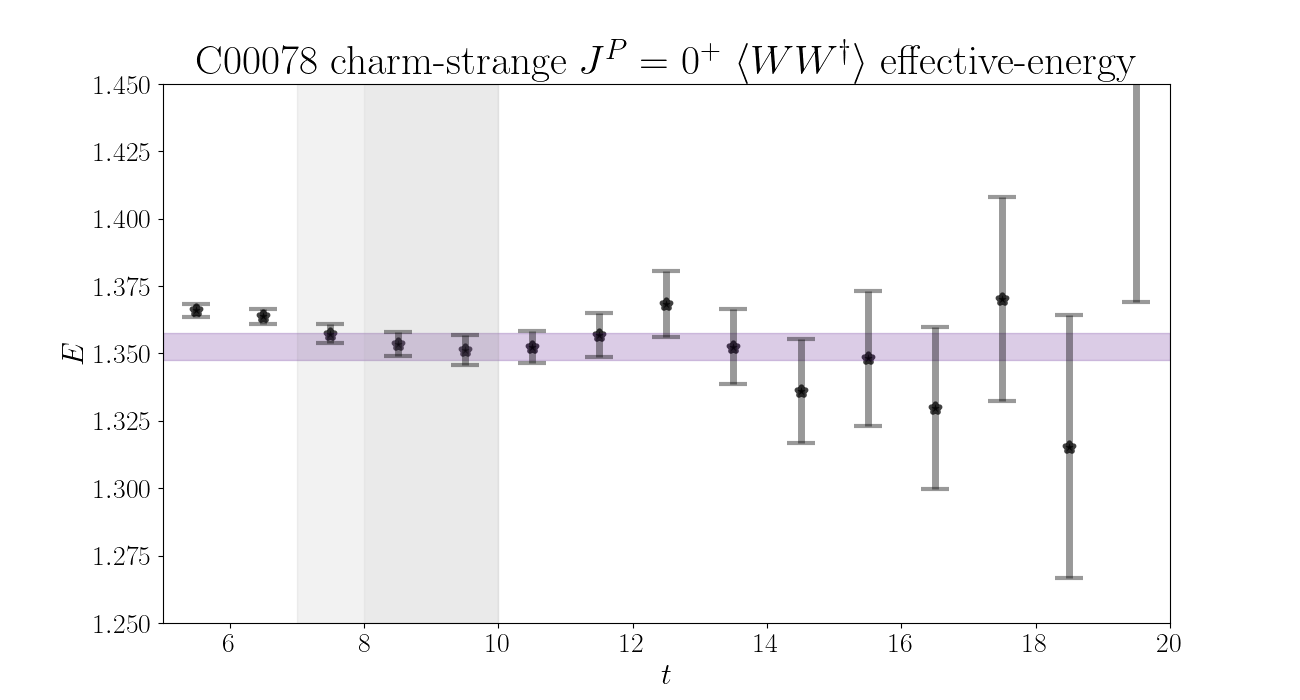}
    
    \includegraphics[width=0.49\linewidth]{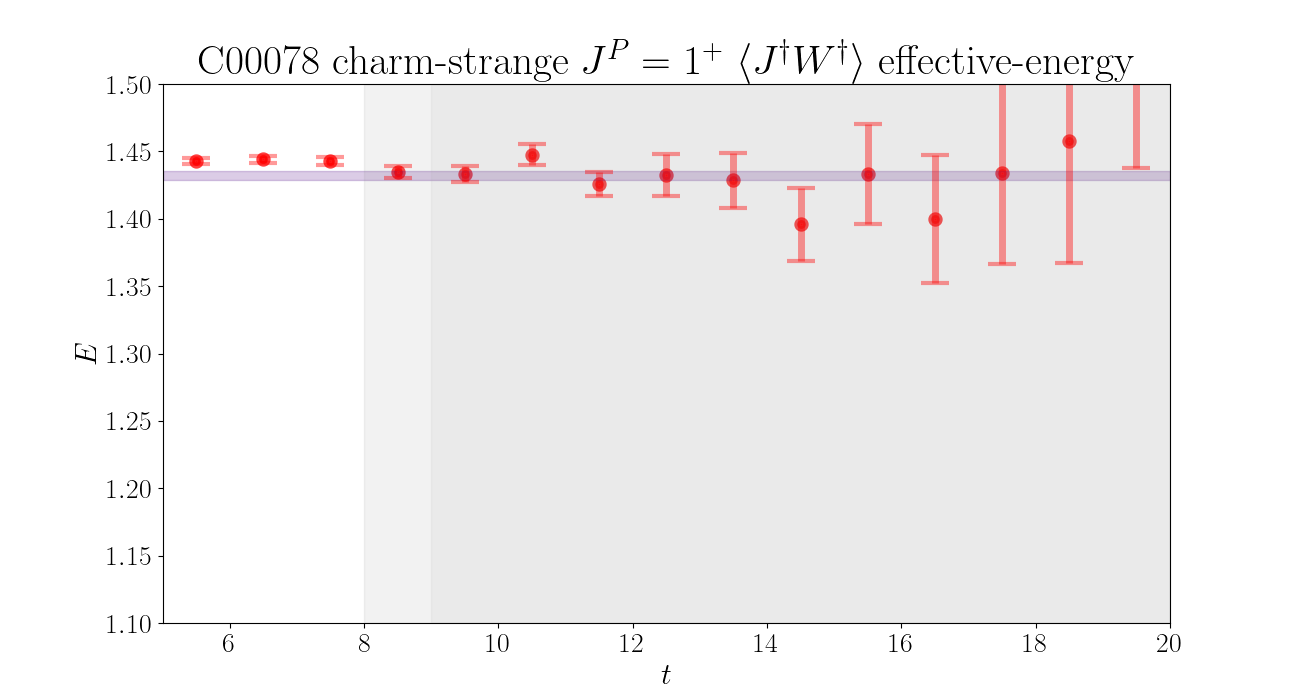}
    \hfill
    \includegraphics[width=0.49\linewidth]{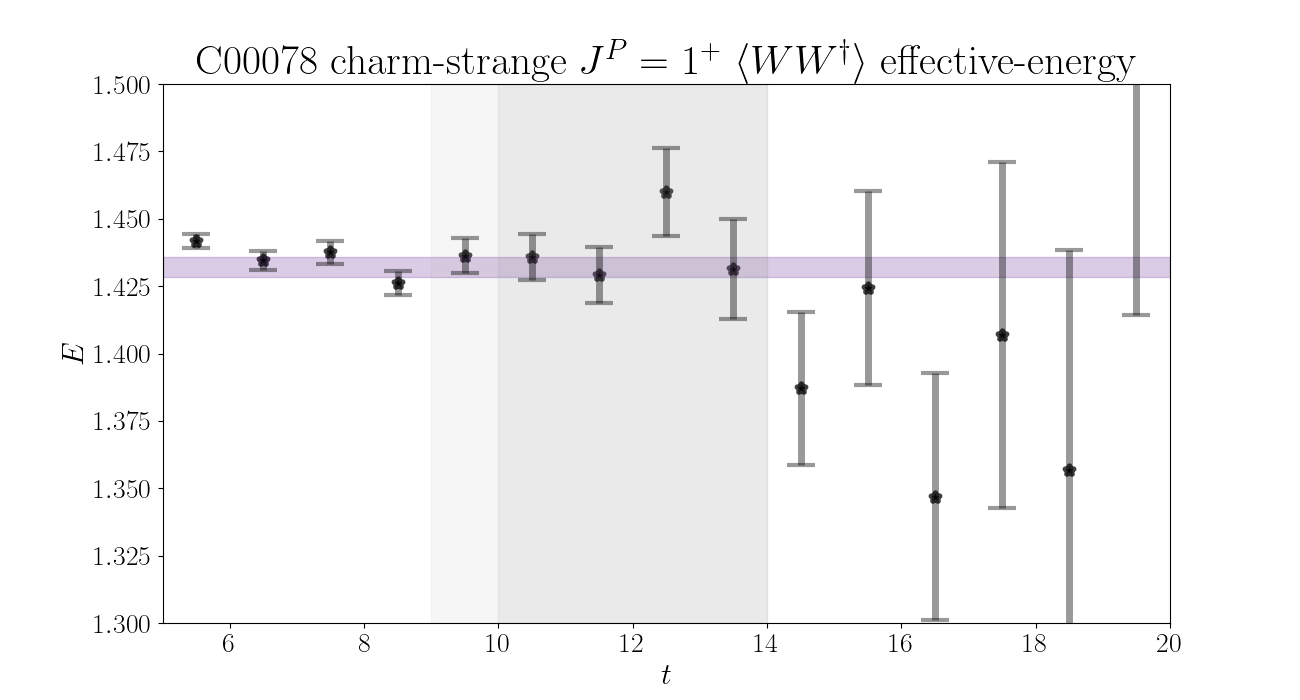}

    \includegraphics[width=0.49\linewidth]{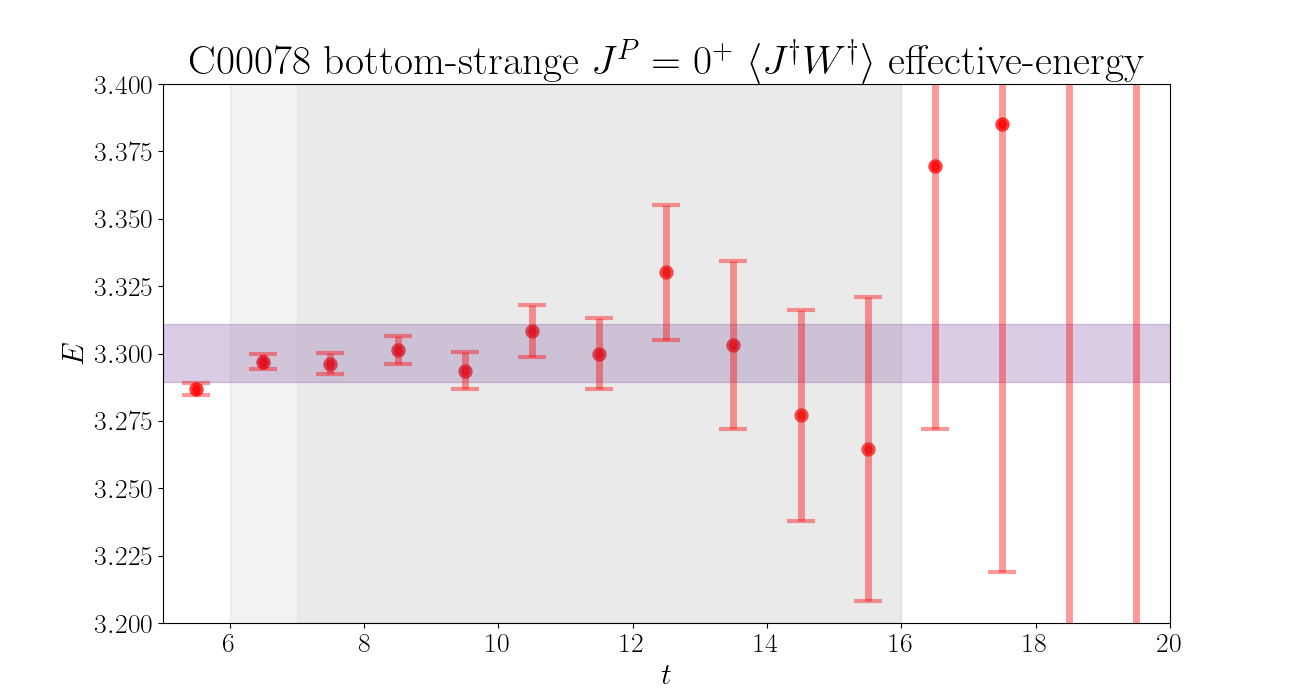}
    \hfill
    \includegraphics[width=0.49\linewidth]{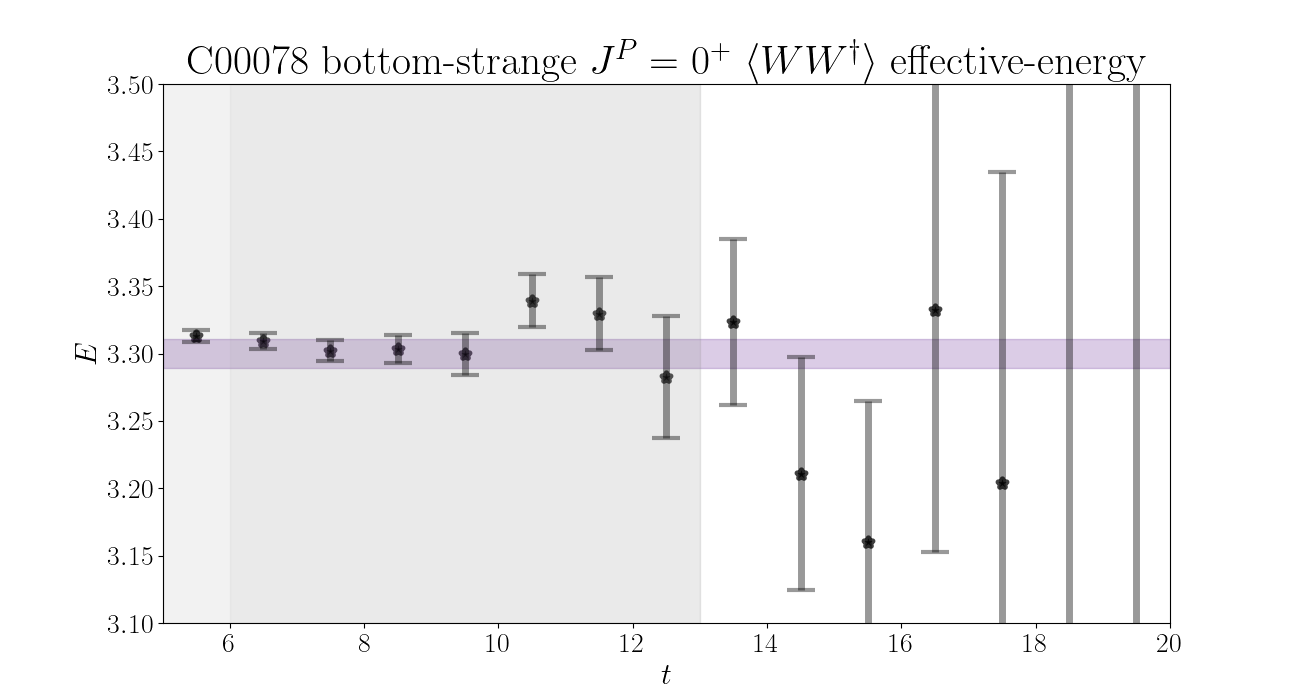}
    
    \includegraphics[width=0.49\linewidth]{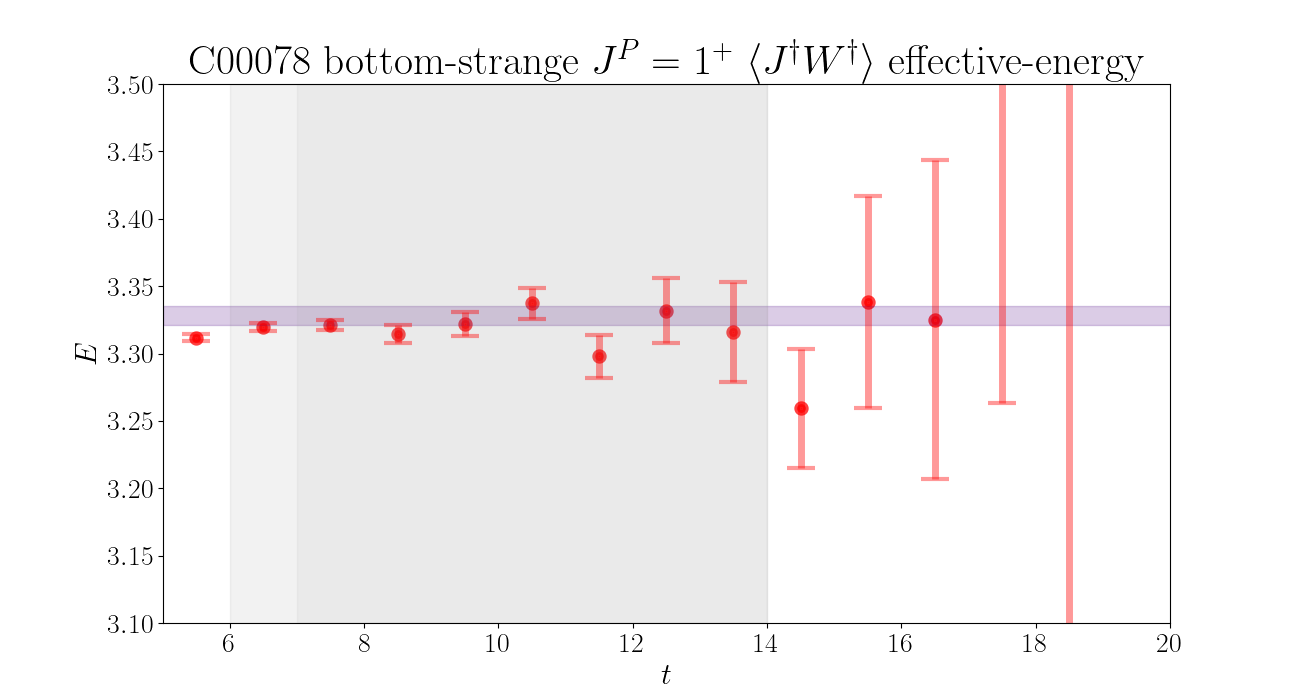}
    \hfill
    \includegraphics[width=0.49\linewidth]{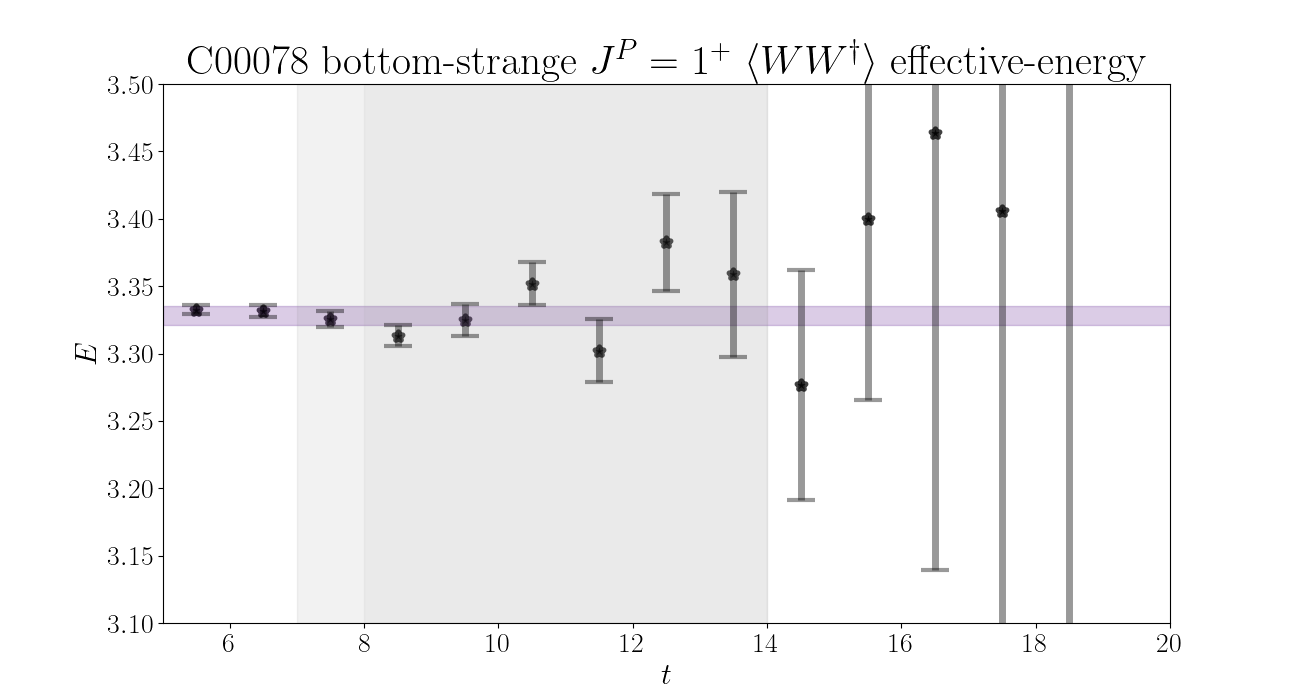}
    \caption{Like Fig.~\protect\ref{fig:F006wplots}, but for the C00078 ensemble. On this ensemble, the correlation-matrix element $C^{44}$, and hence $\langle J^\dagger W^\dagger \rangle$ and $\langle WW^\dagger \rangle$, were not calculated for $t<5$, and we thus cut the plots off there. \label{fig:JWplots-C00078}}
\end{figure}

\begin{figure}[H]
    \centering
    \centering
    \includegraphics[width=0.49\linewidth]{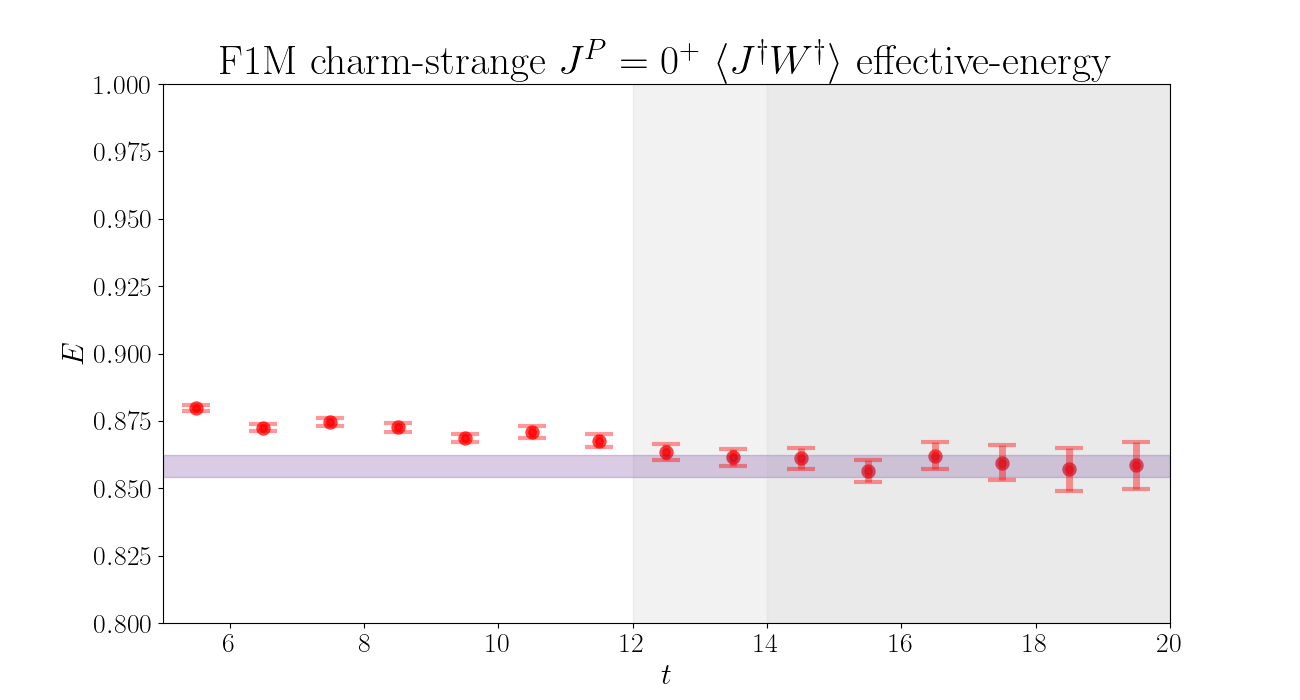}
    \hfill
    \includegraphics[width=0.49\linewidth]{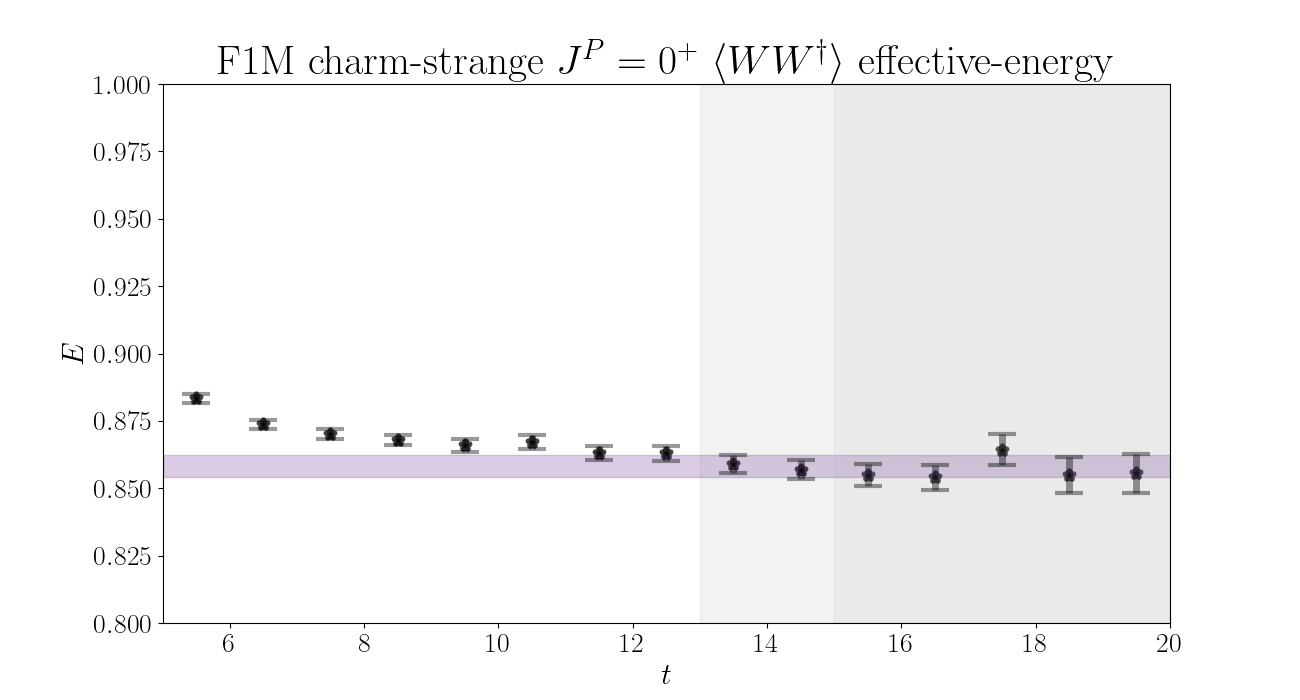}
    
    \includegraphics[width=0.49\linewidth]{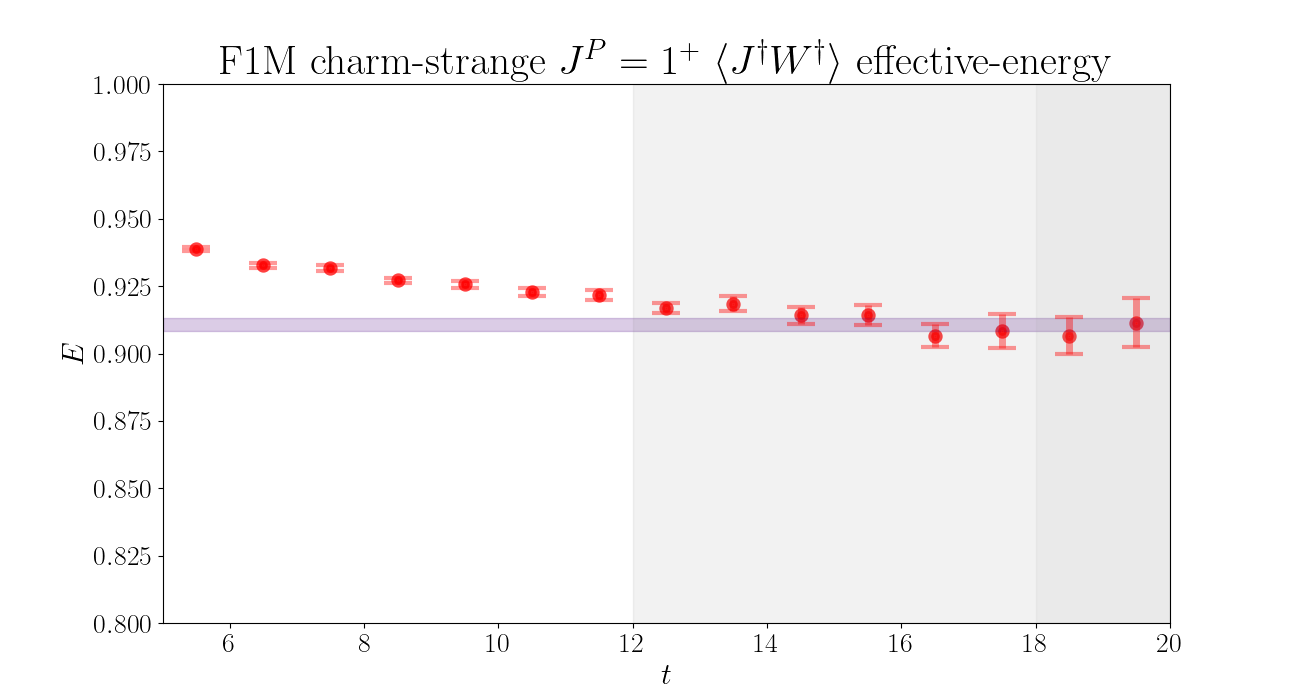}
    \hfill
    \includegraphics[width=0.49\linewidth]{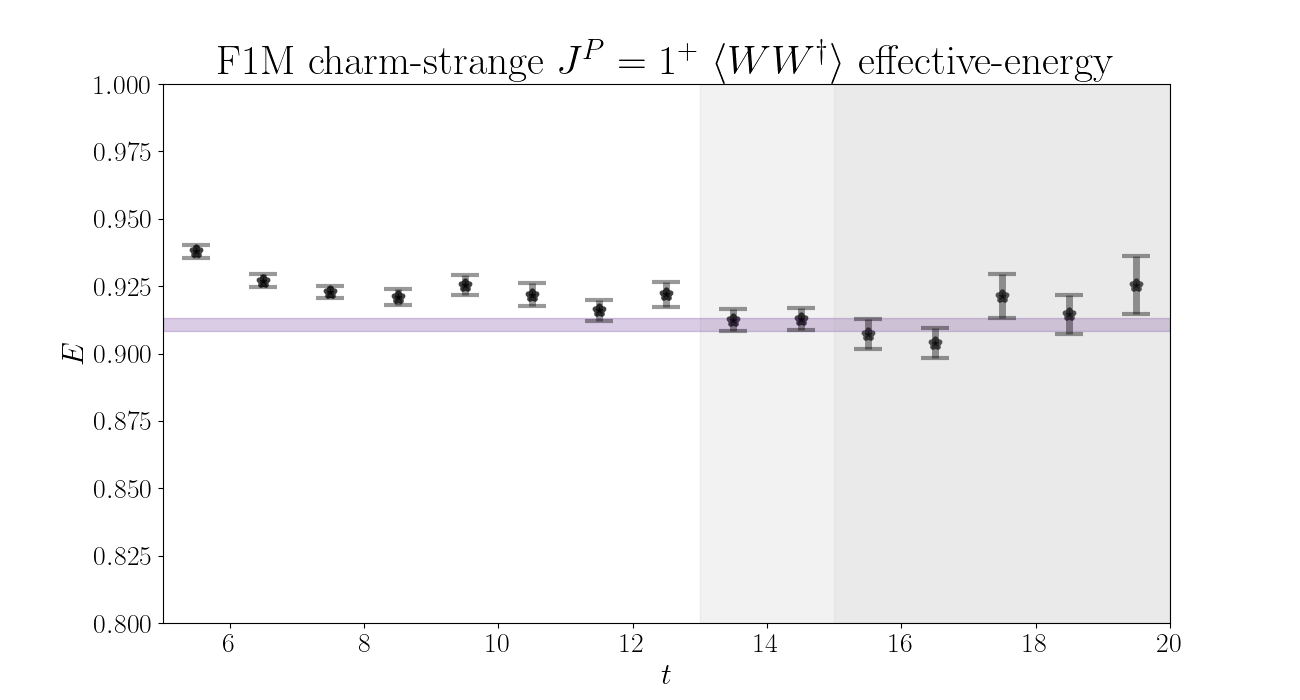}

    \includegraphics[width=0.49\linewidth]{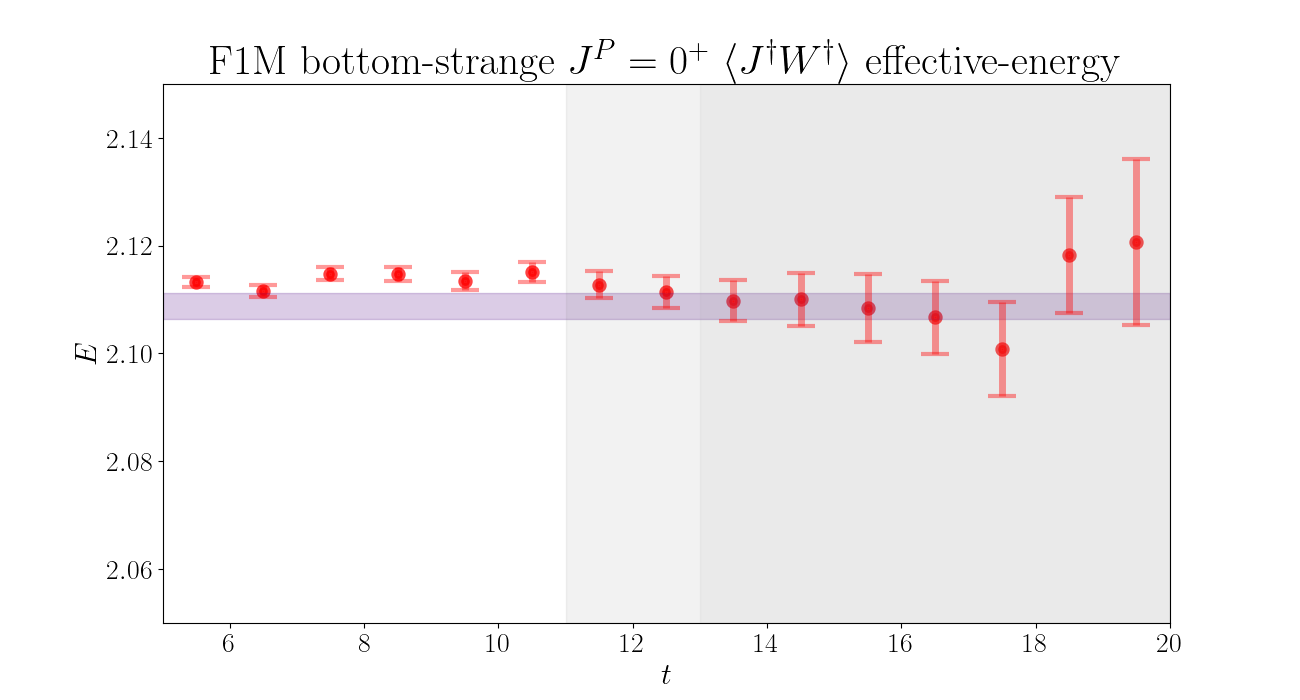}
    \hfill
    \includegraphics[width=0.49\linewidth]{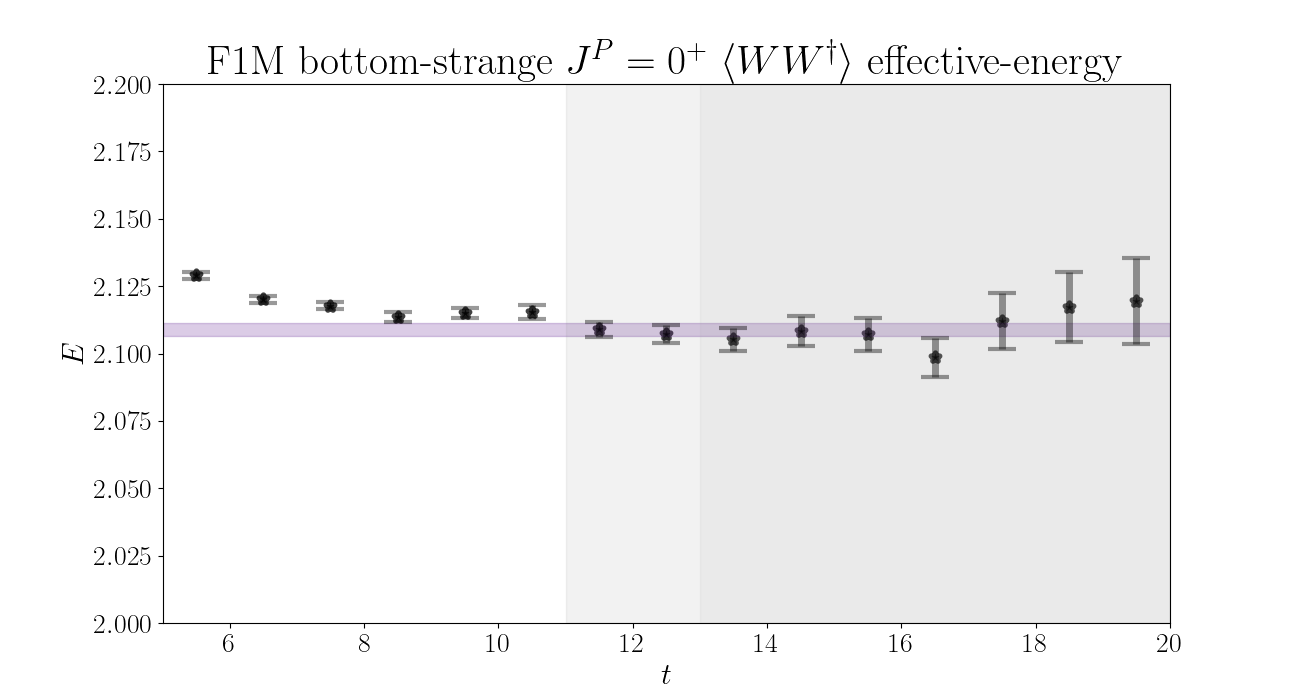}
    
    \includegraphics[width=0.49\linewidth]{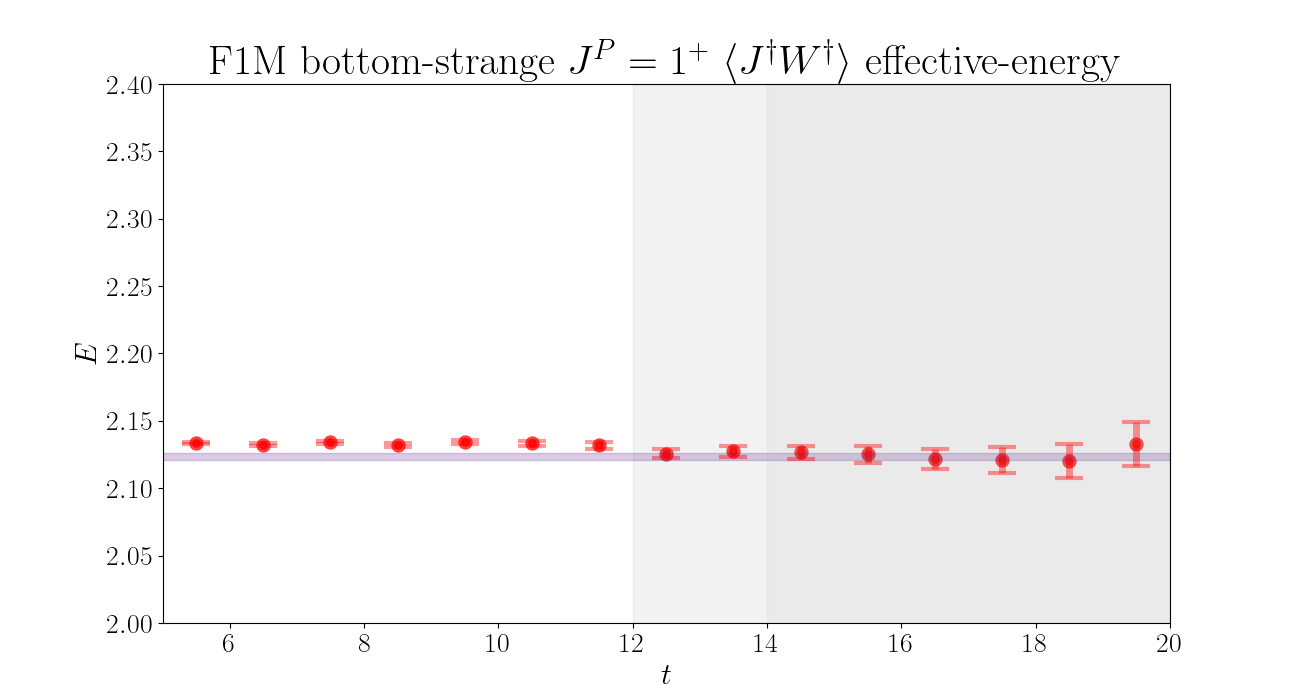}
    \hfill
    \includegraphics[width=0.49\linewidth]{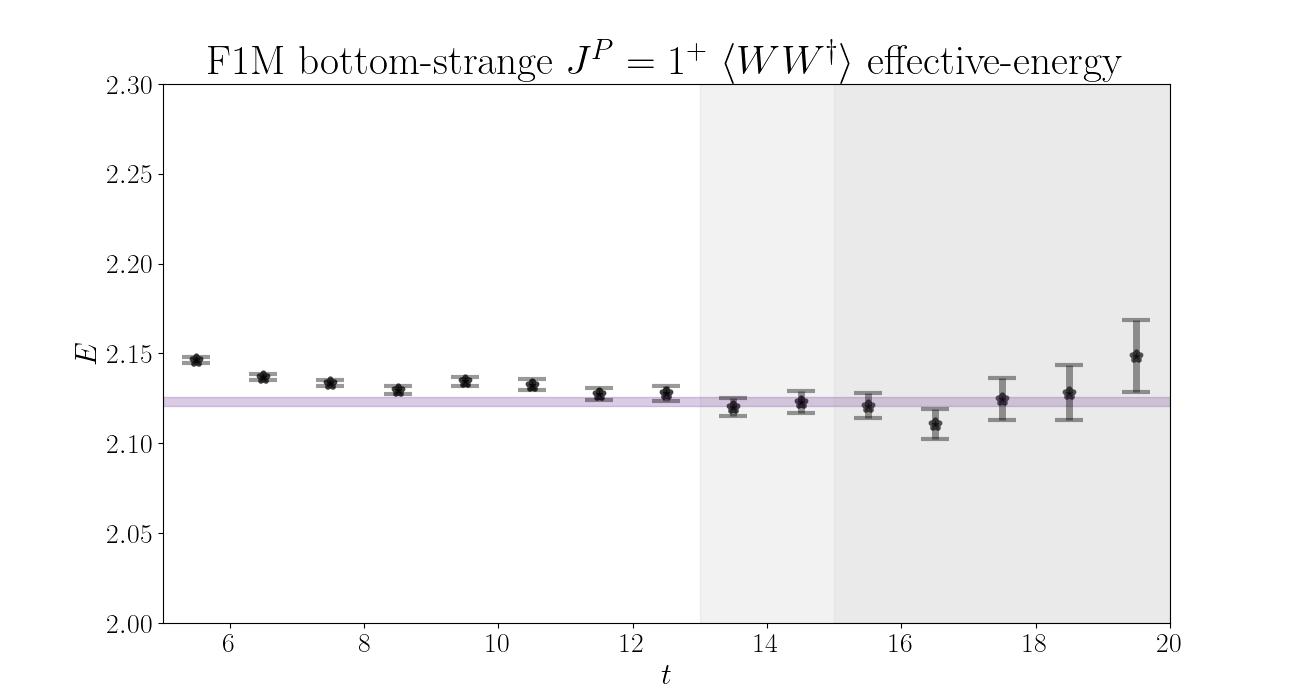}
    \caption{Like Fig.~\protect\ref{fig:F006wplots}, but for the F1M ensemble. On this ensemble, the correlation-matrix element $C^{44}$, and hence $\langle J^\dagger W^\dagger \rangle$ and $\langle WW^\dagger \rangle$, were not calculated for $t<5$, and we thus cut the plots off there. \label{fig:JWplots-F1M}}
\end{figure}

\begin{figure}[H]
    \centering
    \centering
    \includegraphics[width=0.49\linewidth]{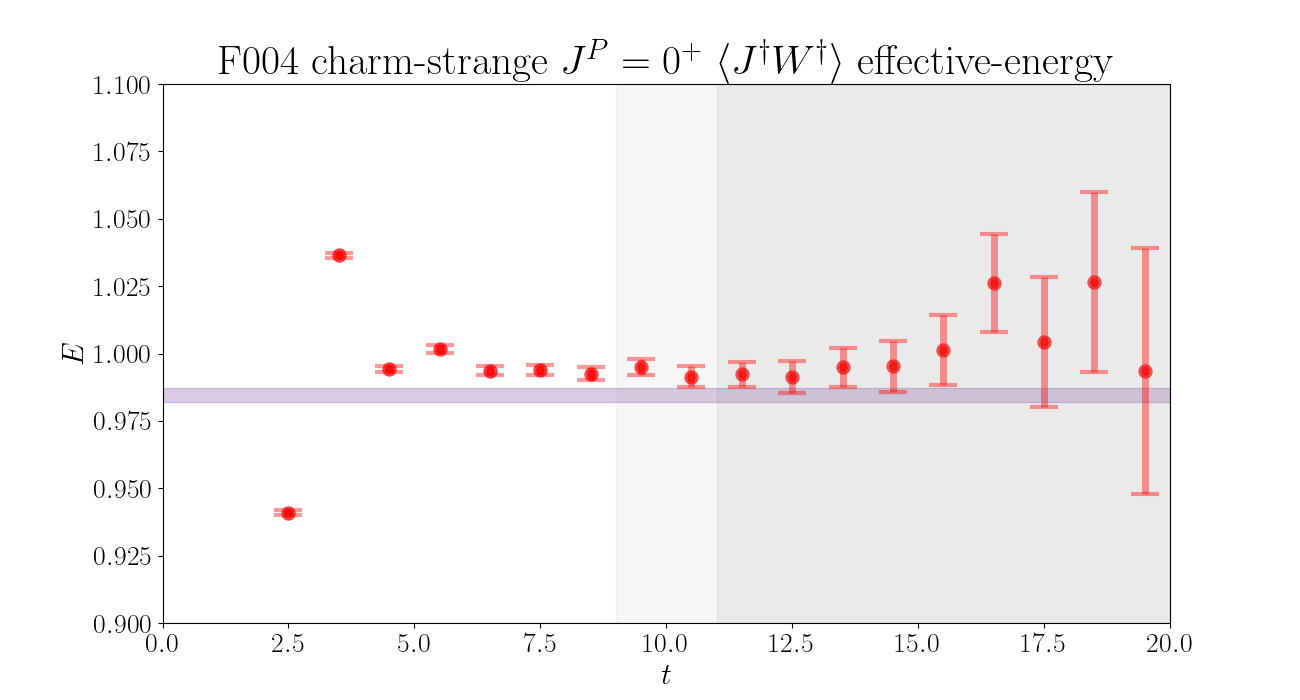}
    \hfill
    \includegraphics[width=0.49\linewidth]{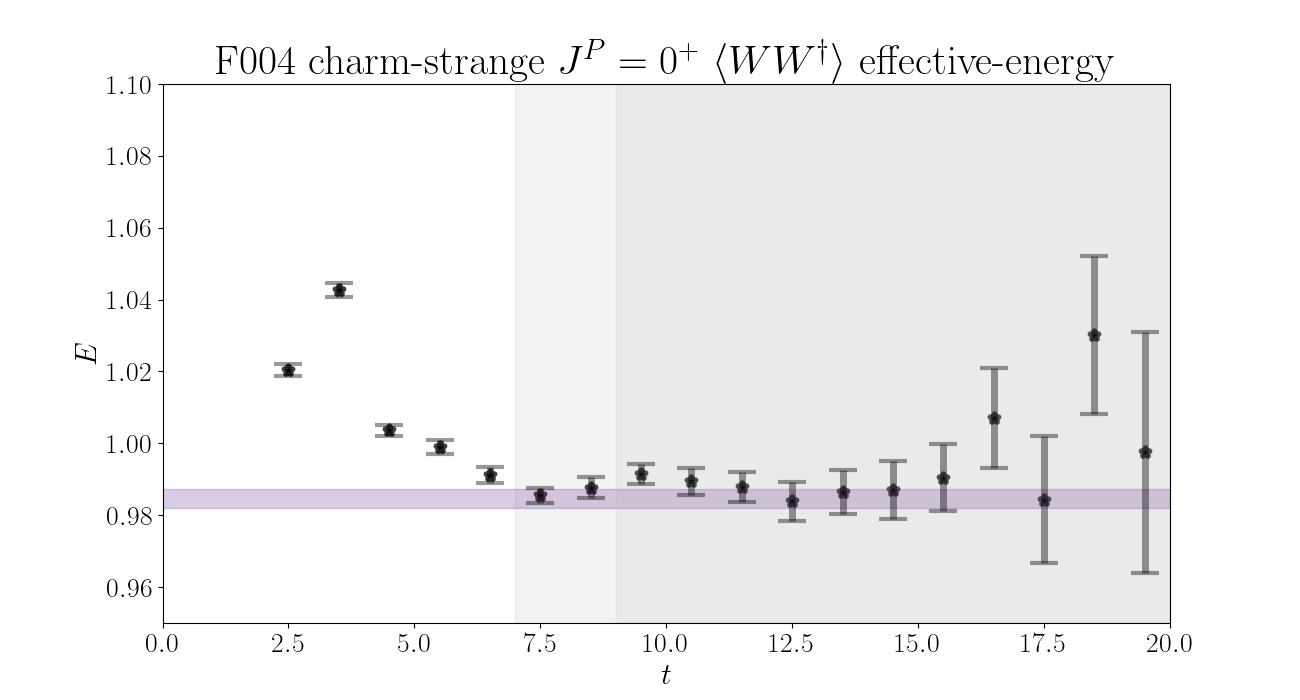}
    
    \includegraphics[width=0.49\linewidth]{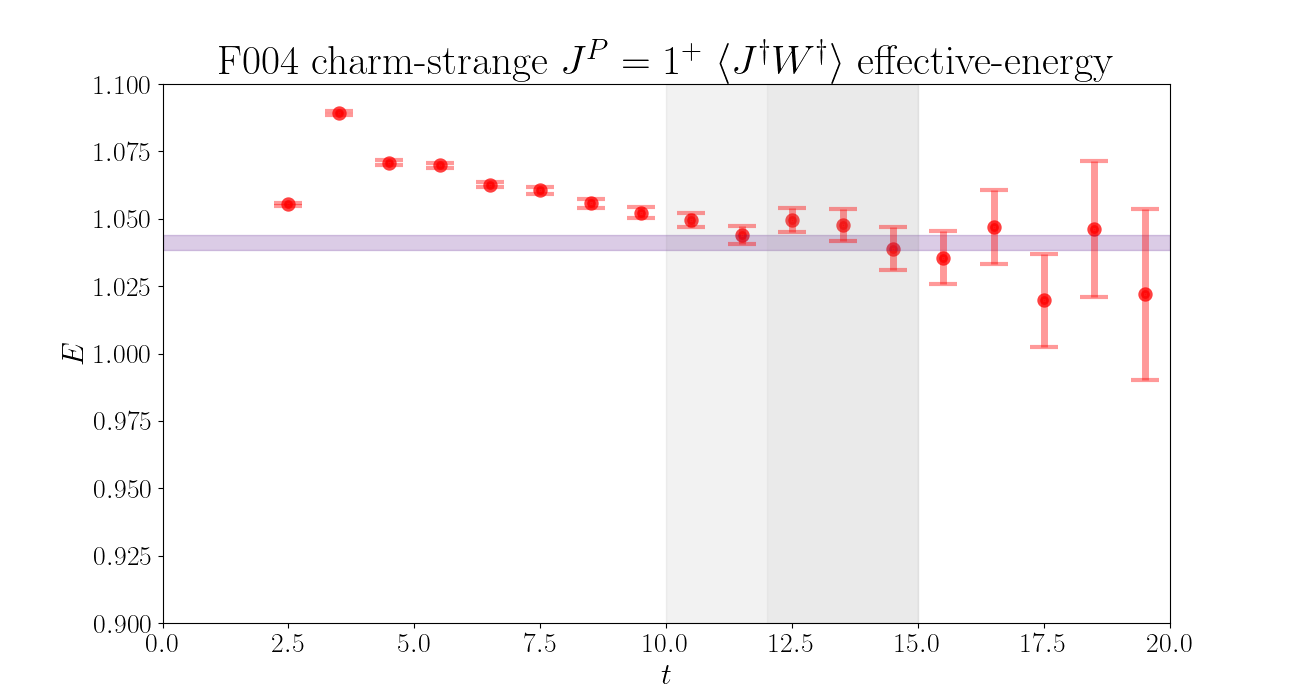}
    \hfill
    \includegraphics[width=0.49\linewidth]{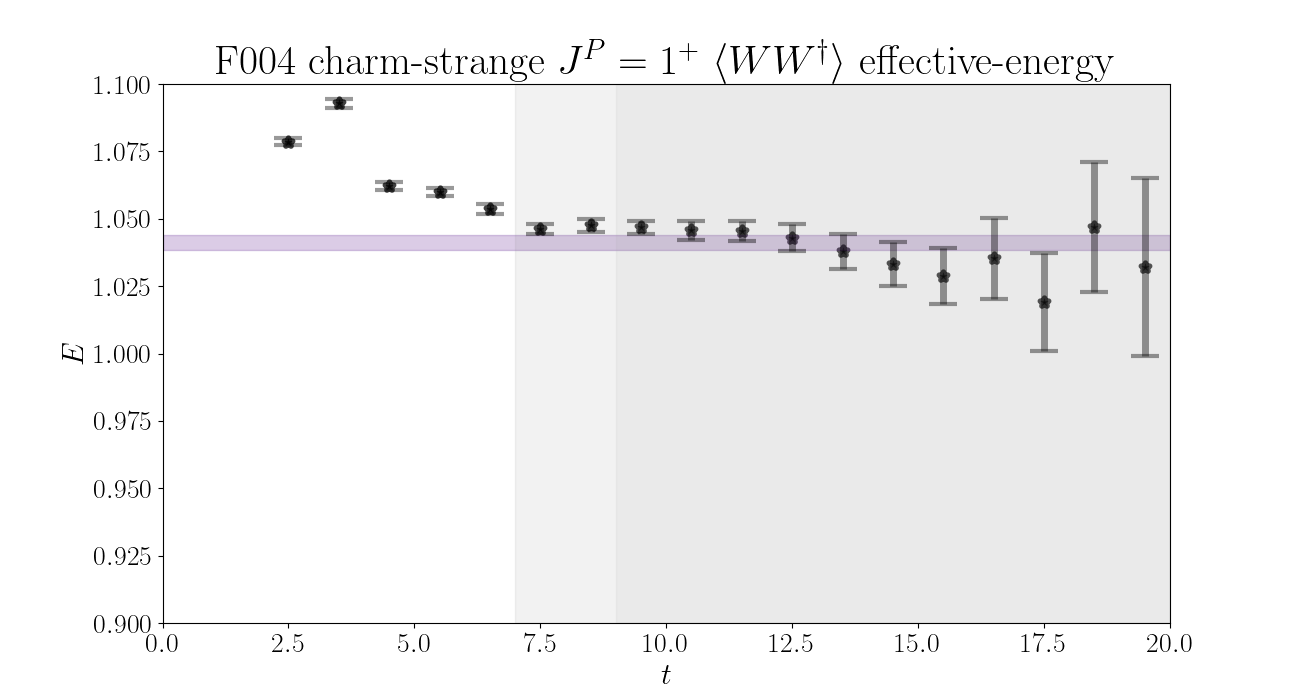}

    \includegraphics[width=0.49\linewidth]{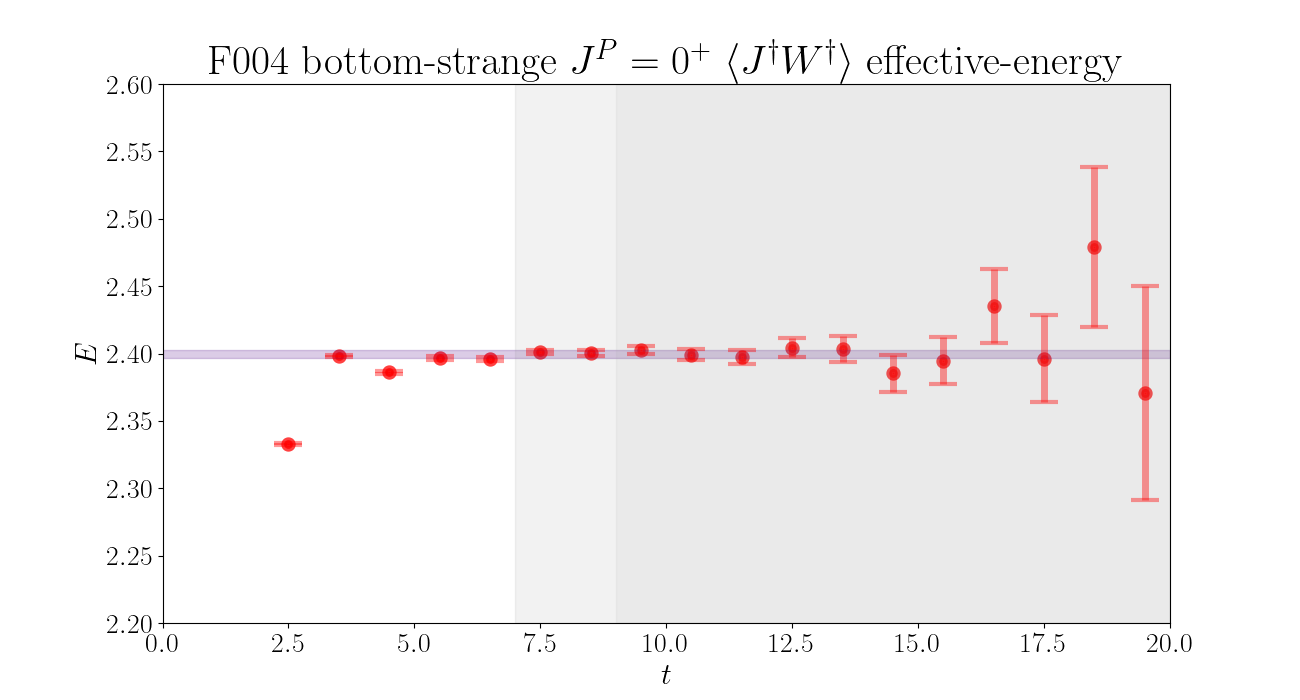}
    \hfill
    \includegraphics[width=0.49\linewidth]{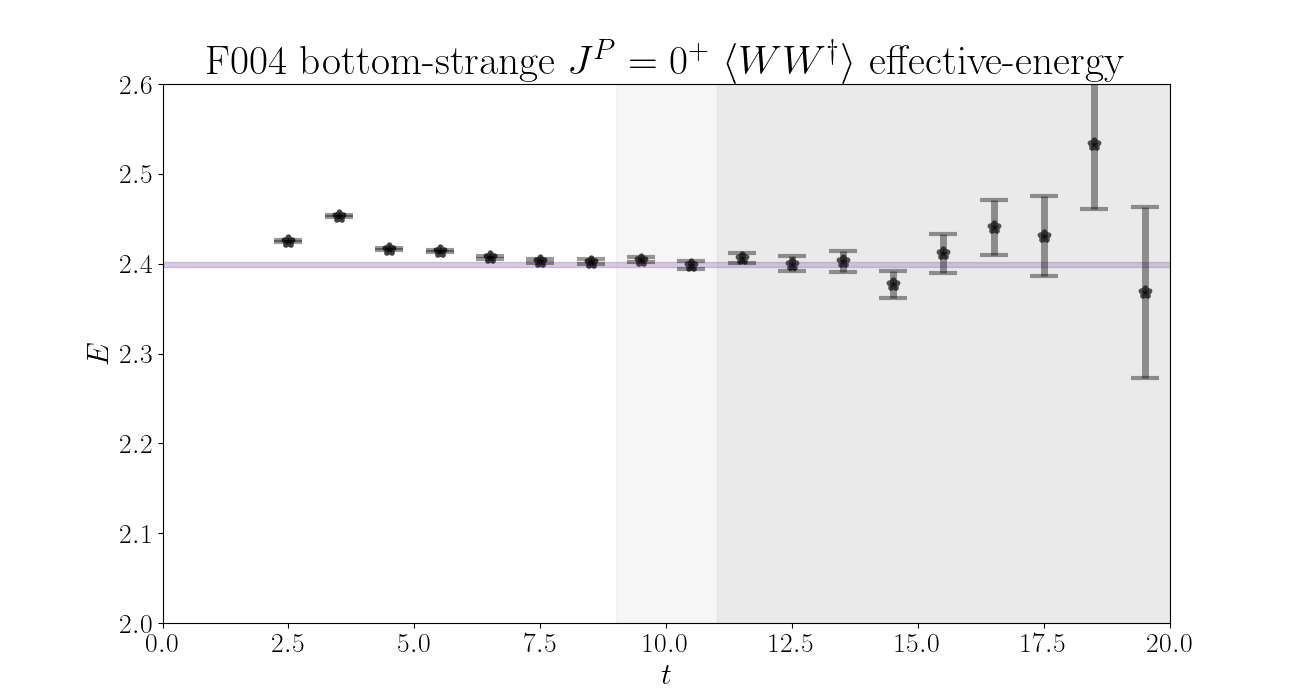}
    
    \includegraphics[width=0.49\linewidth]{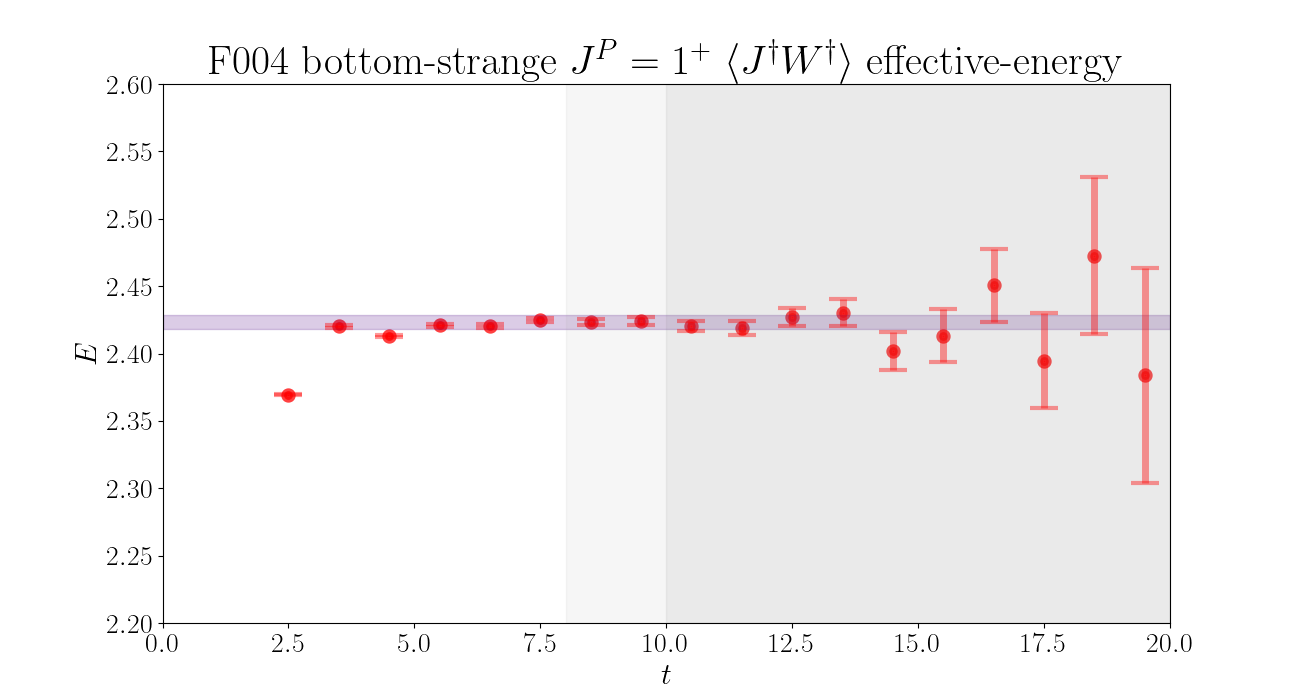}
    \hfill
    \includegraphics[width=0.49\linewidth]{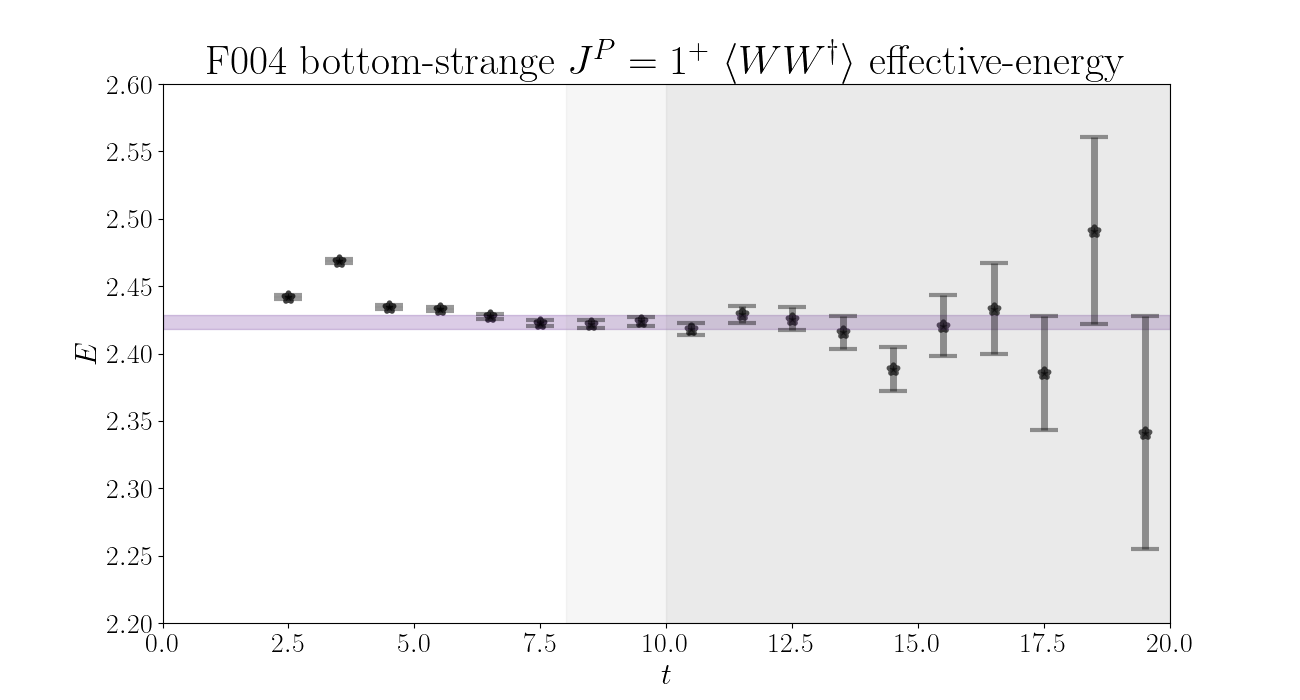}
    \caption{Like Fig.~\protect\ref{fig:F006wplots}, but for the F004 ensemble. \label{fig:JWplots-F004}}
\end{figure}

\begin{figure}[H]
    \centering
    \centering
    \includegraphics[width=0.49\linewidth]{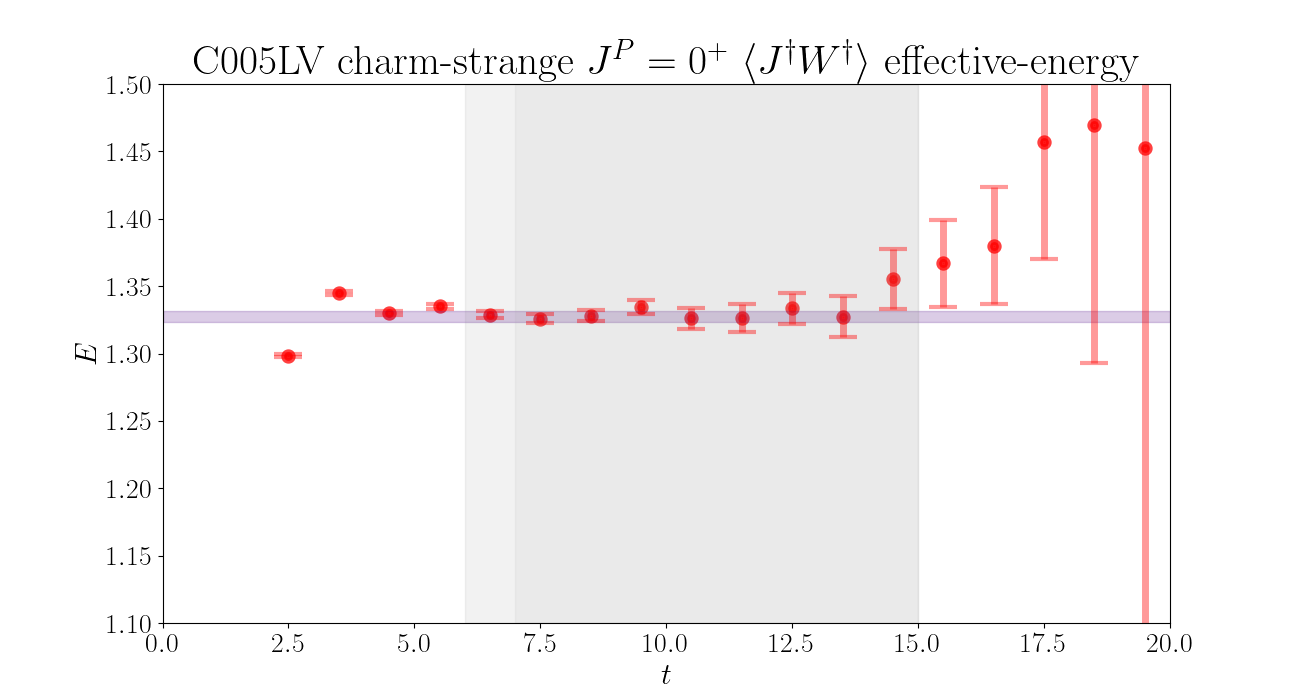}
    \hfill
    \includegraphics[width=0.49\linewidth]{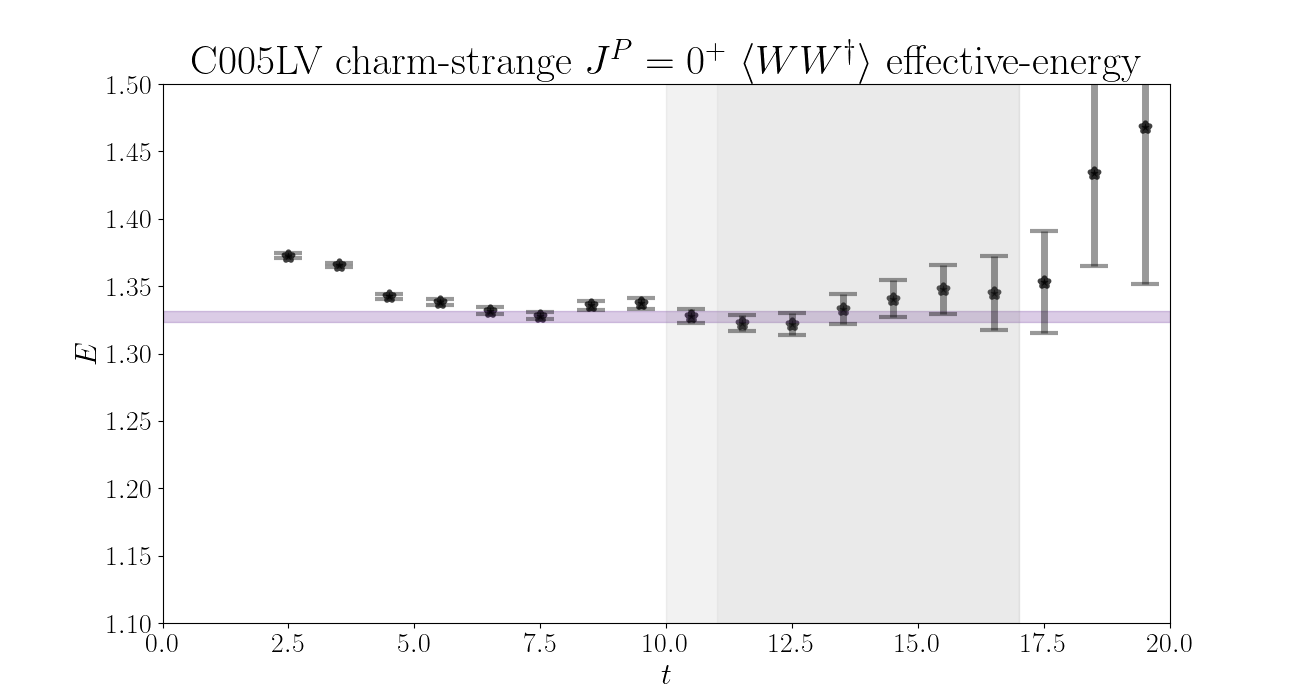}
    
    \includegraphics[width=0.49\linewidth]{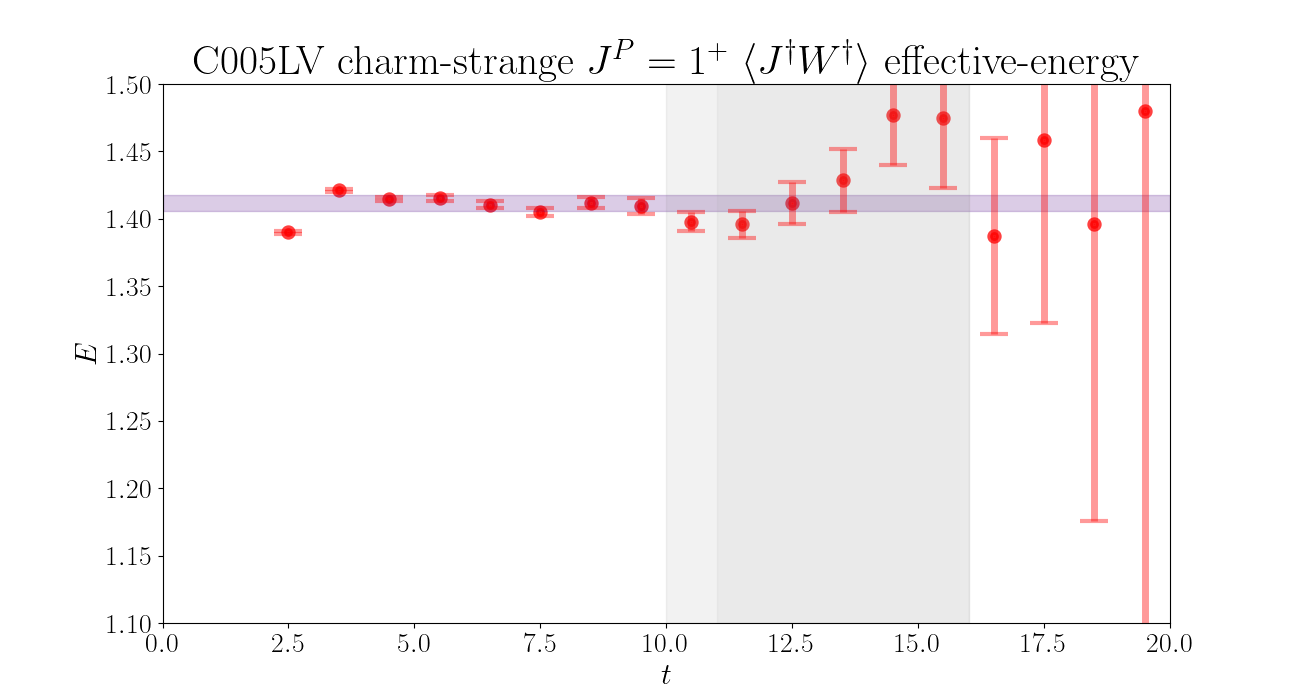}
    \hfill
    \includegraphics[width=0.49\linewidth]{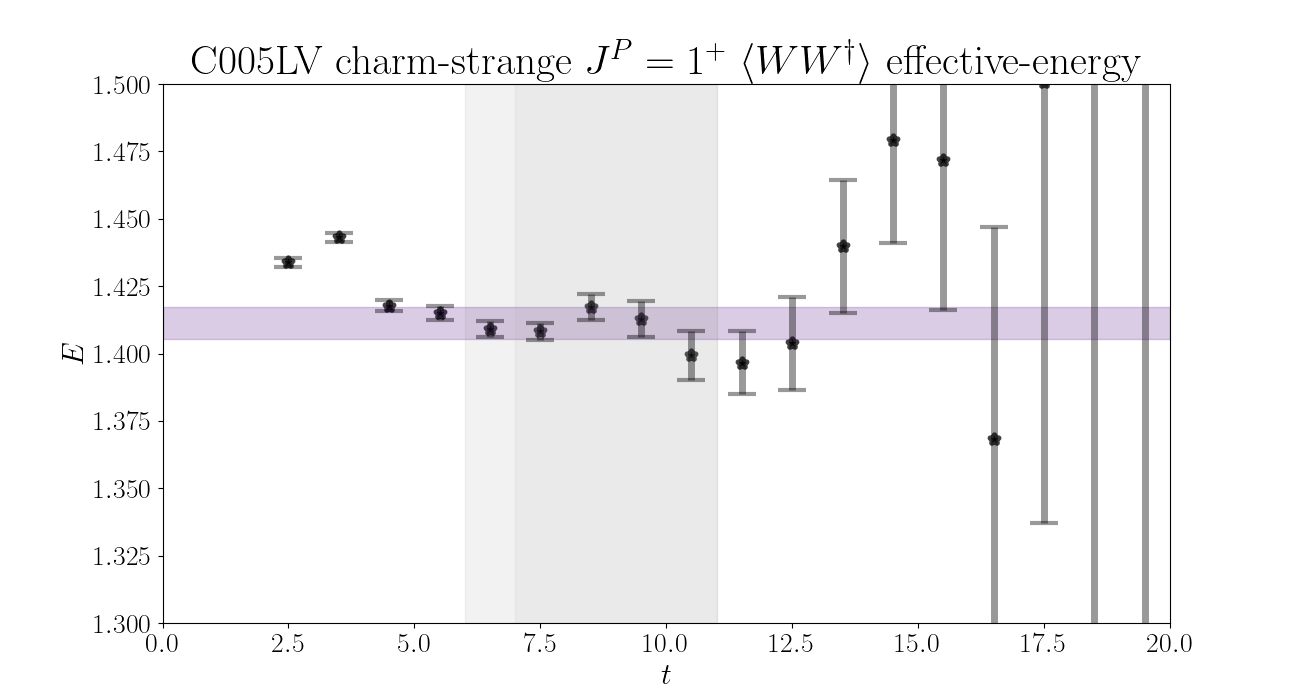}

    \includegraphics[width=0.49\linewidth]{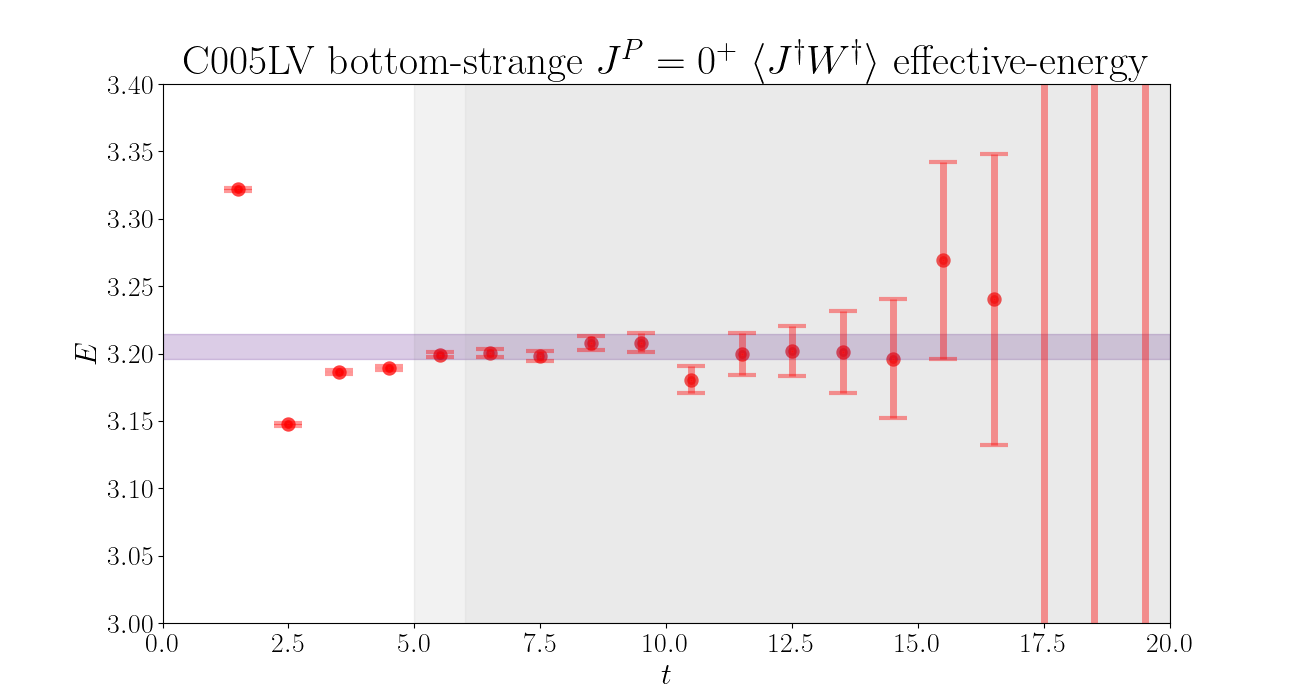}
    \hfill
    \includegraphics[width=0.49\linewidth]{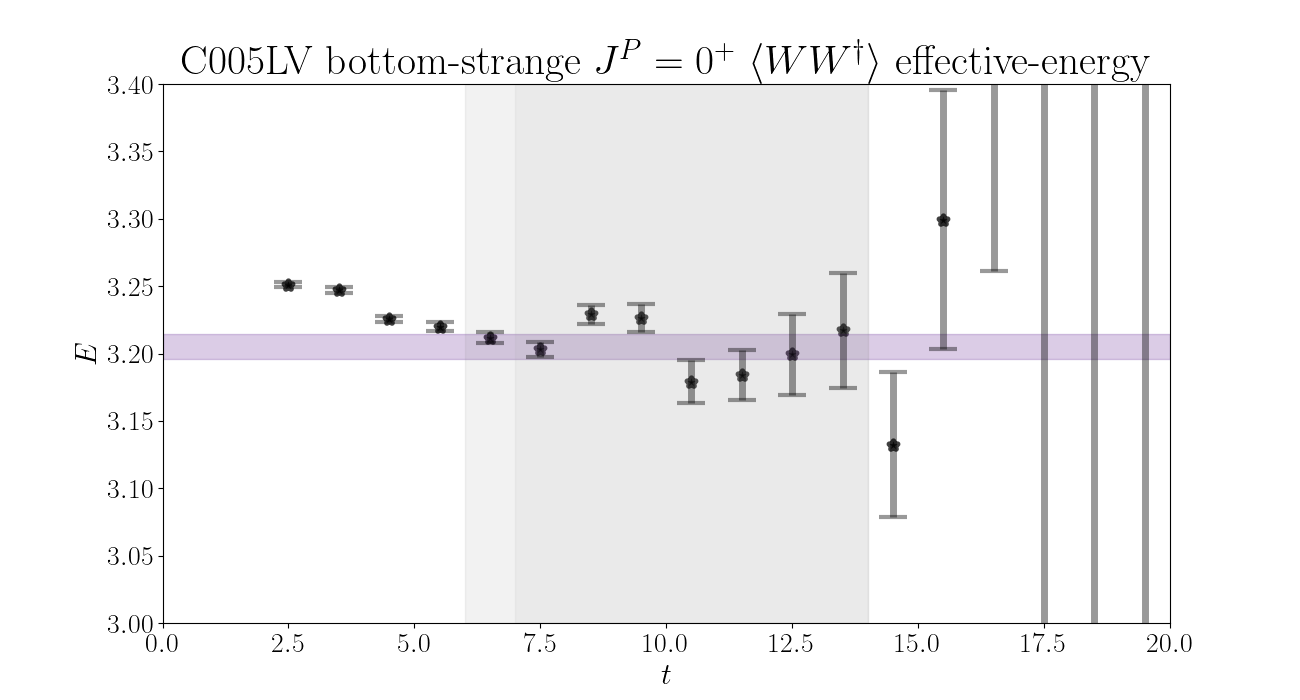}
    
    \includegraphics[width=0.49\linewidth]{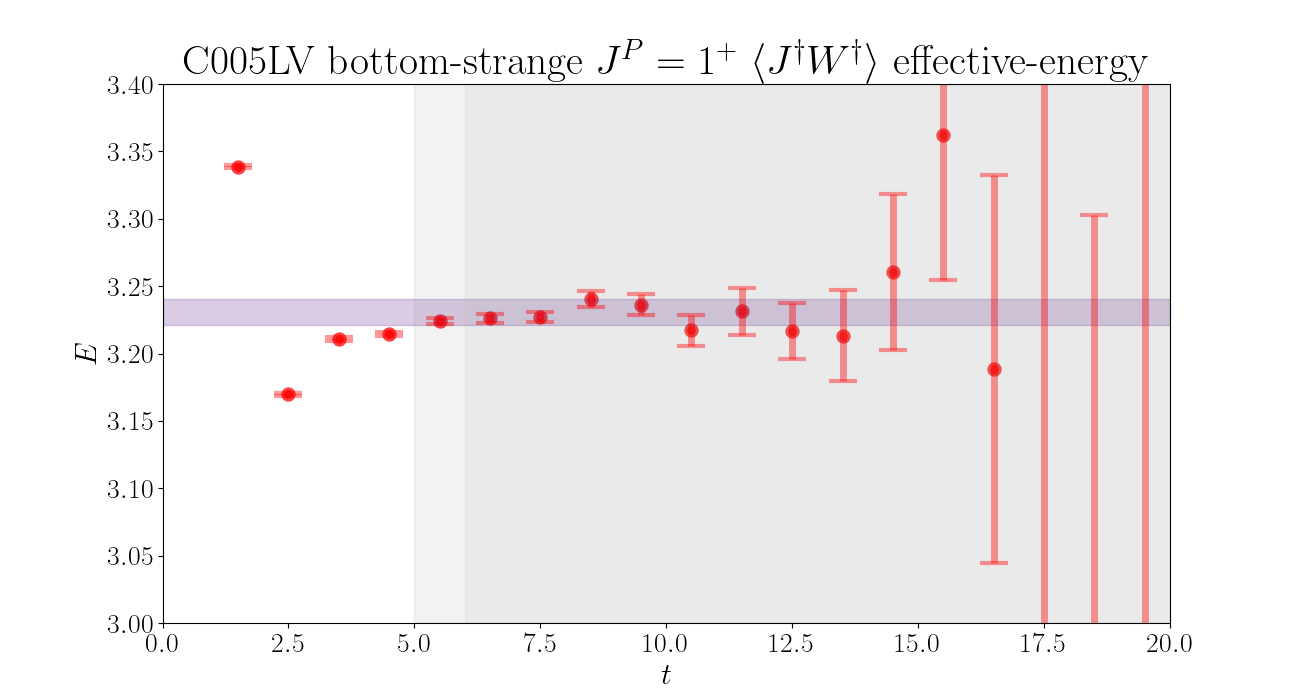}
    \hfill
    \includegraphics[width=0.49\linewidth]{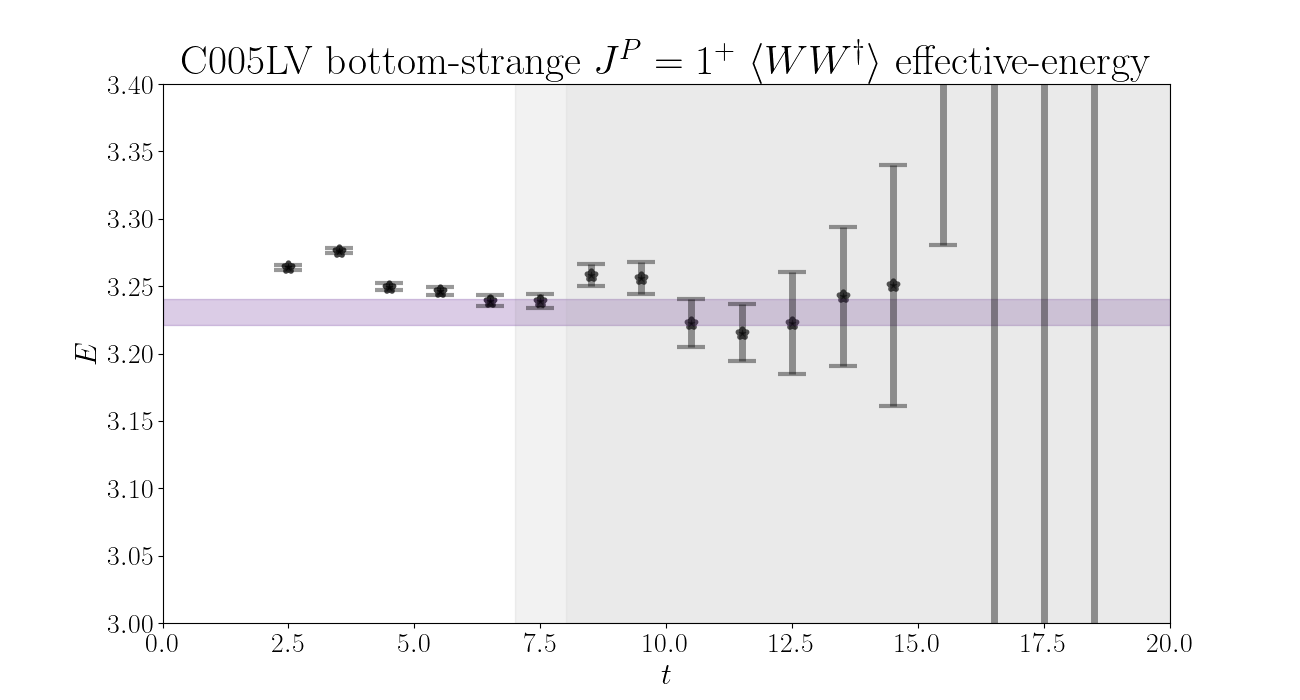}
    \caption{Like Fig.~\protect\ref{fig:F006wplots}, but for the C005LV ensemble. \label{fig:JWplots-C005LV}}
\end{figure}

\begin{figure}[H]
    \centering
    \centering
    \includegraphics[width=0.49\linewidth]{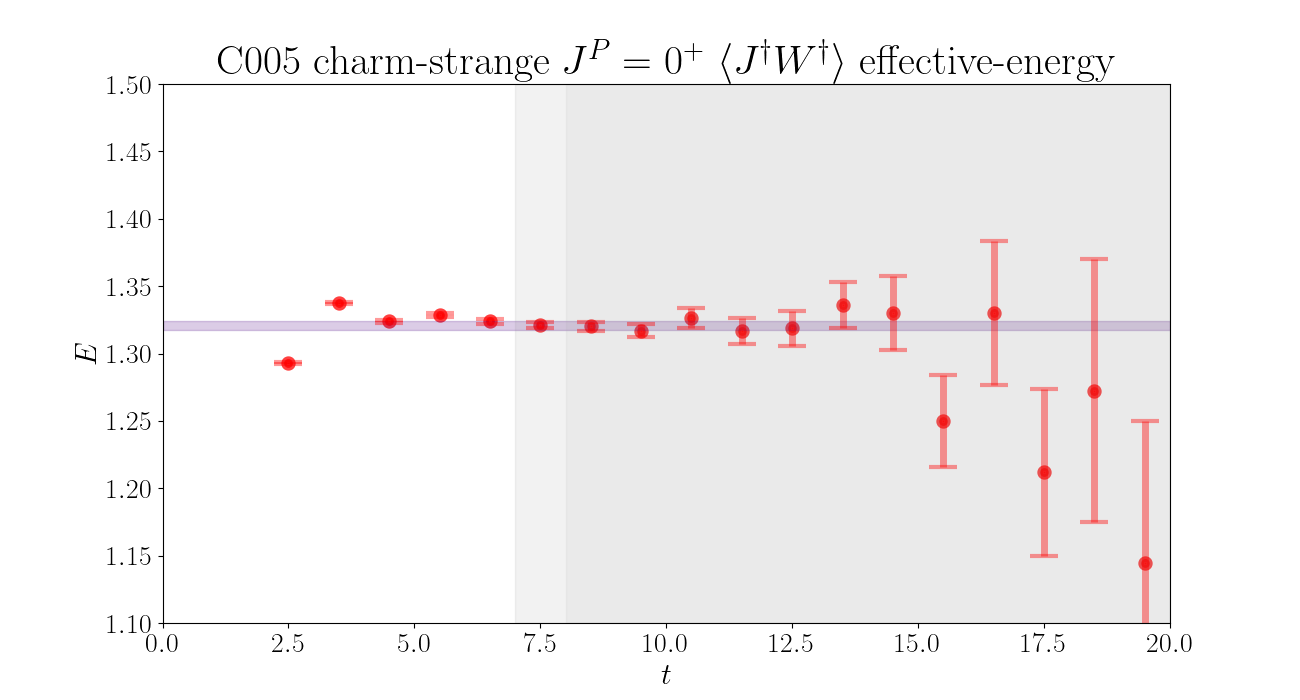}
    \hfill
    \includegraphics[width=0.49\linewidth]{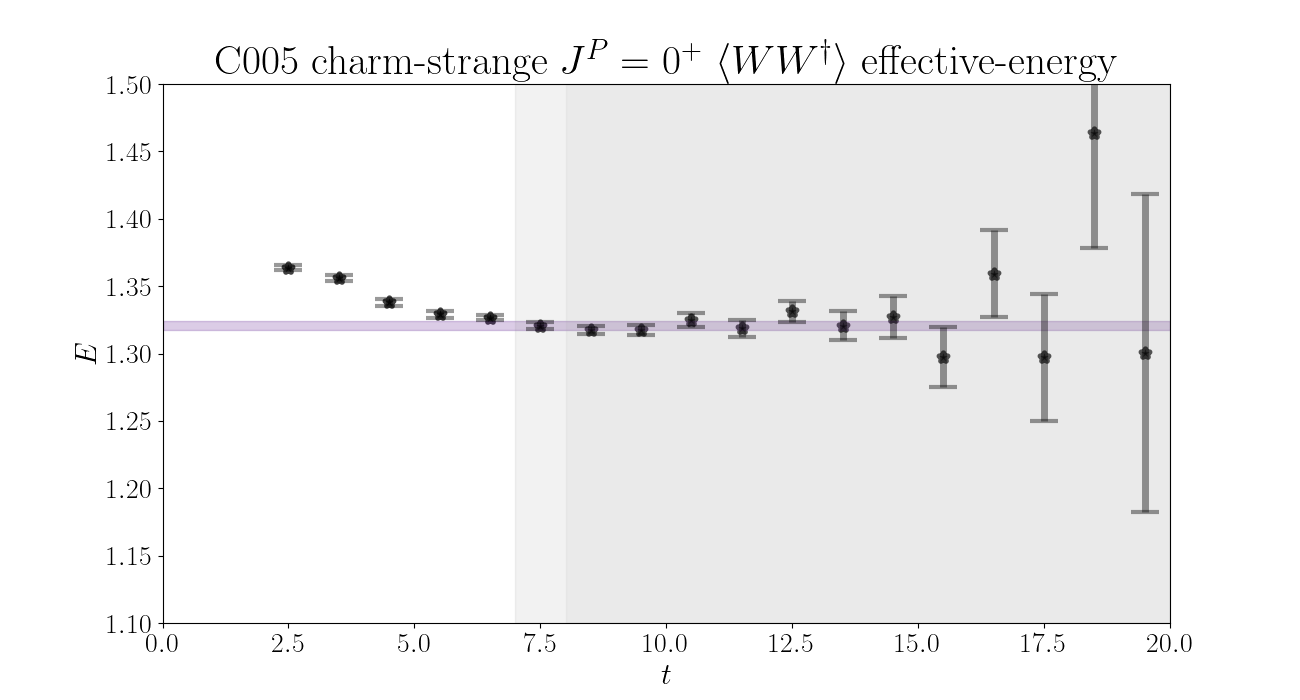}
    
    \includegraphics[width=0.49\linewidth]{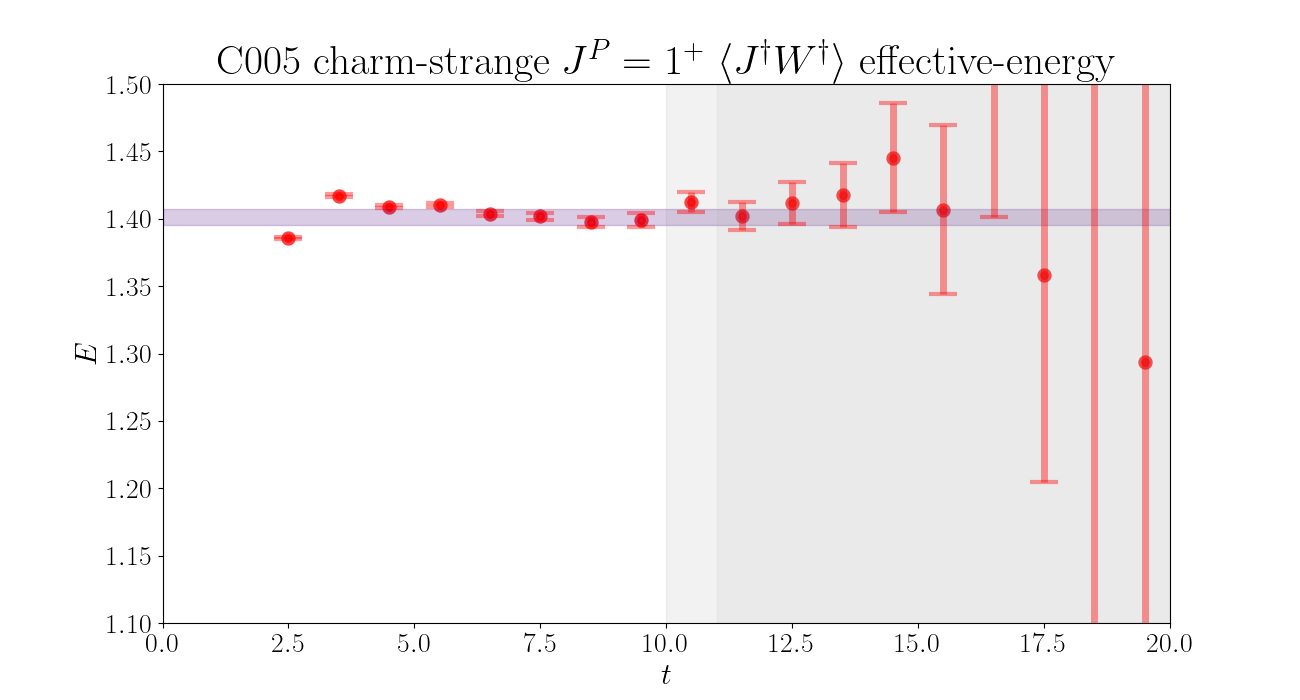}
    \hfill
    \includegraphics[width=0.49\linewidth]{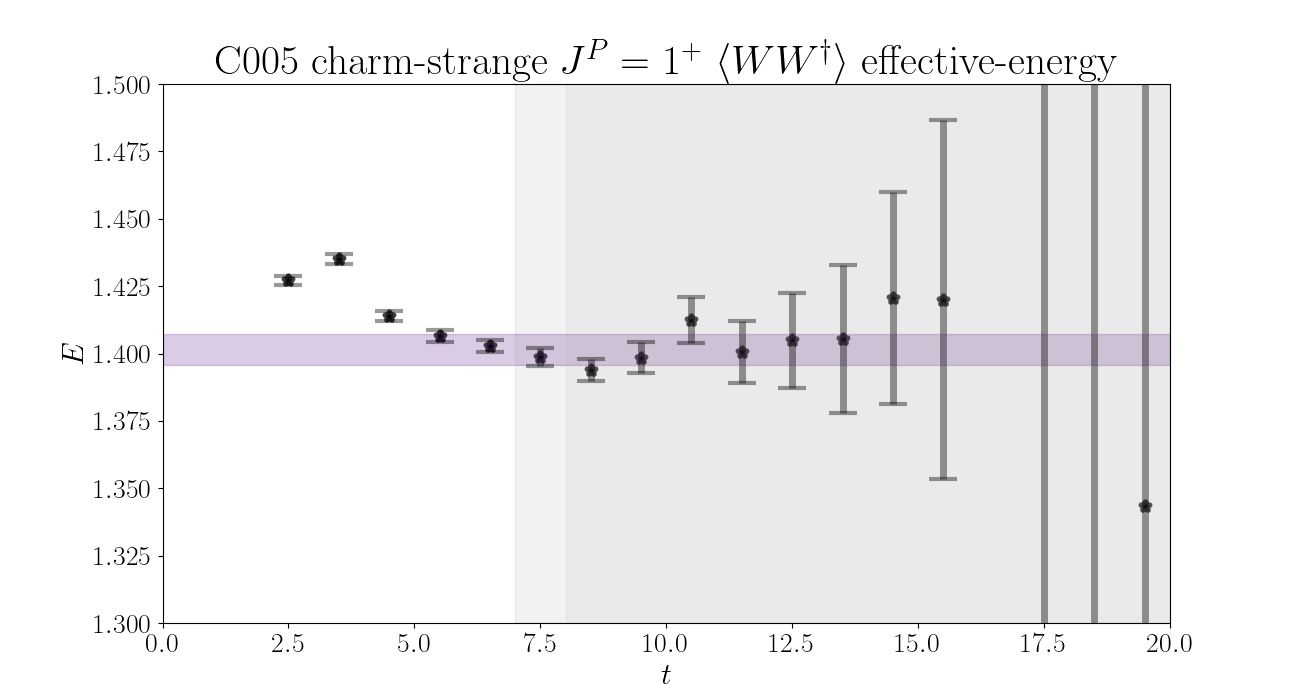}

    \includegraphics[width=0.49\linewidth]{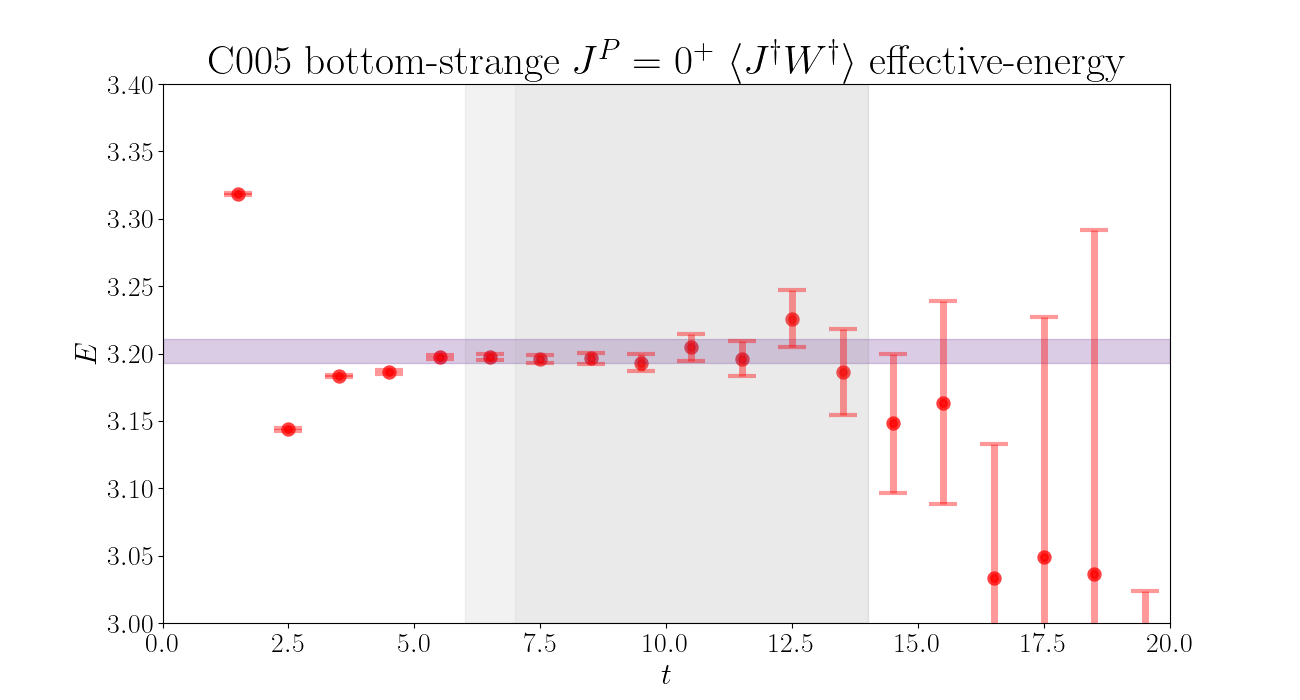}
    \hfill
    \includegraphics[width=0.49\linewidth]{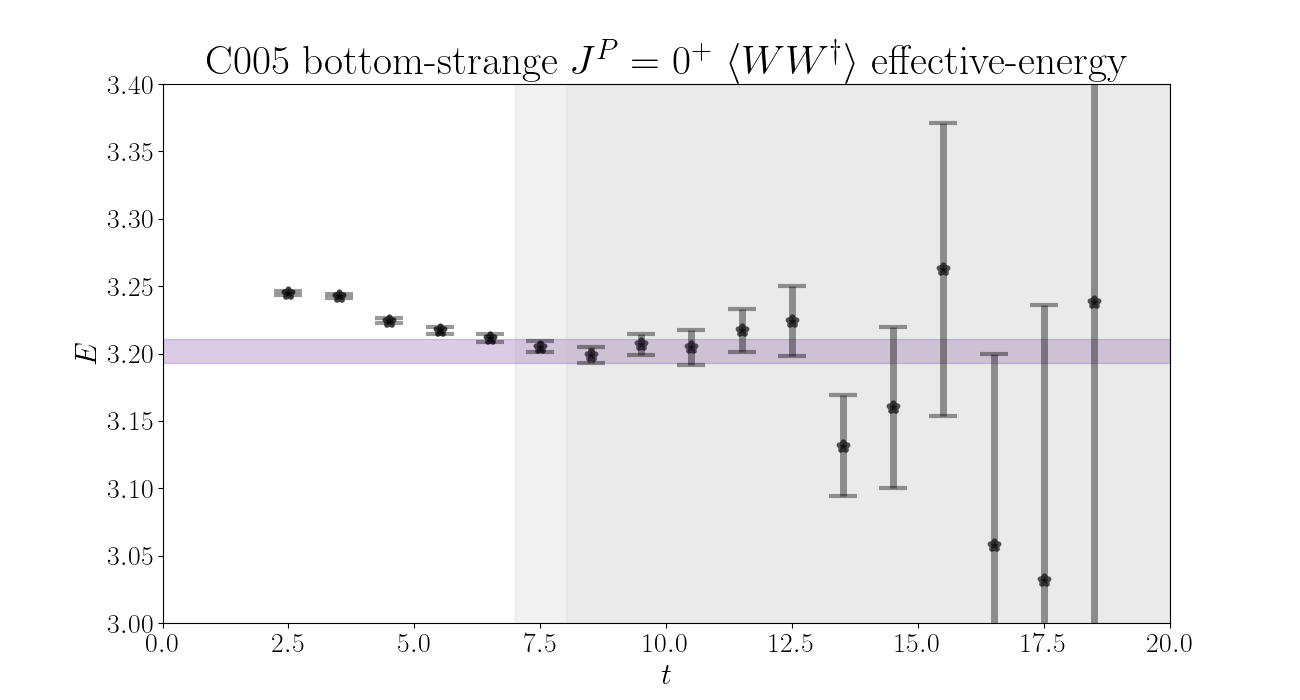}
    
    \includegraphics[width=0.49\linewidth]{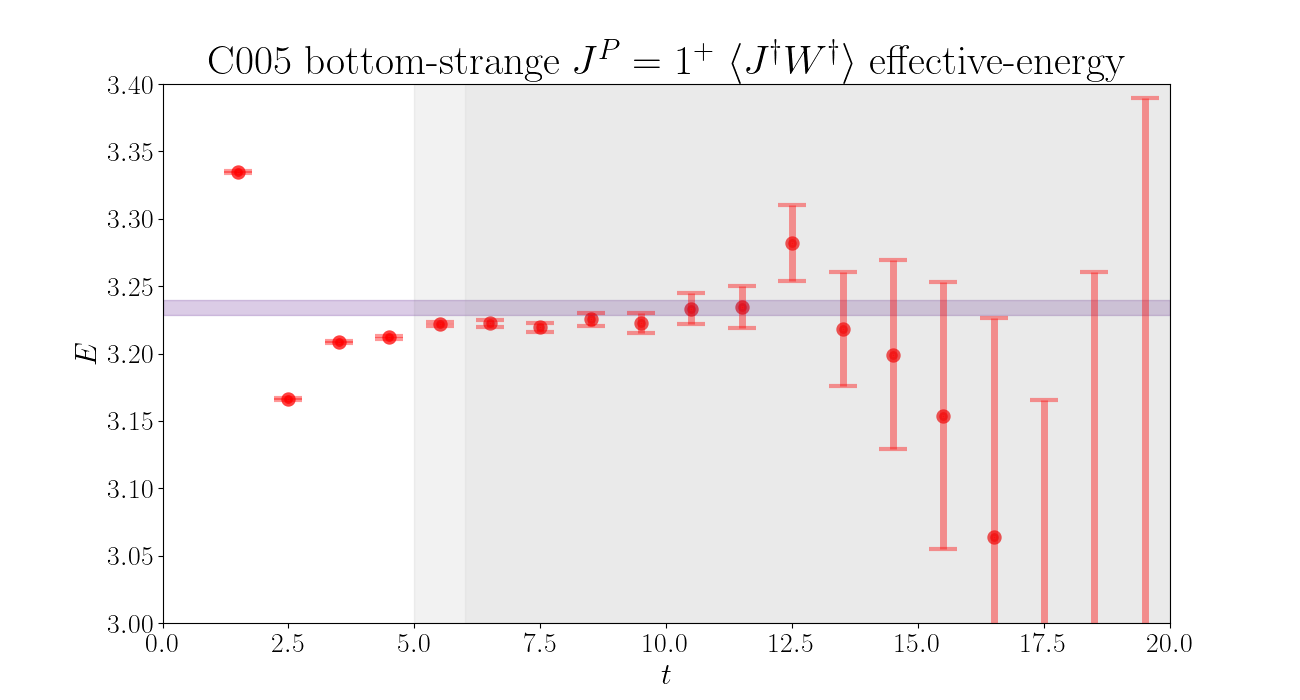}
    \hfill
    \includegraphics[width=0.49\linewidth]{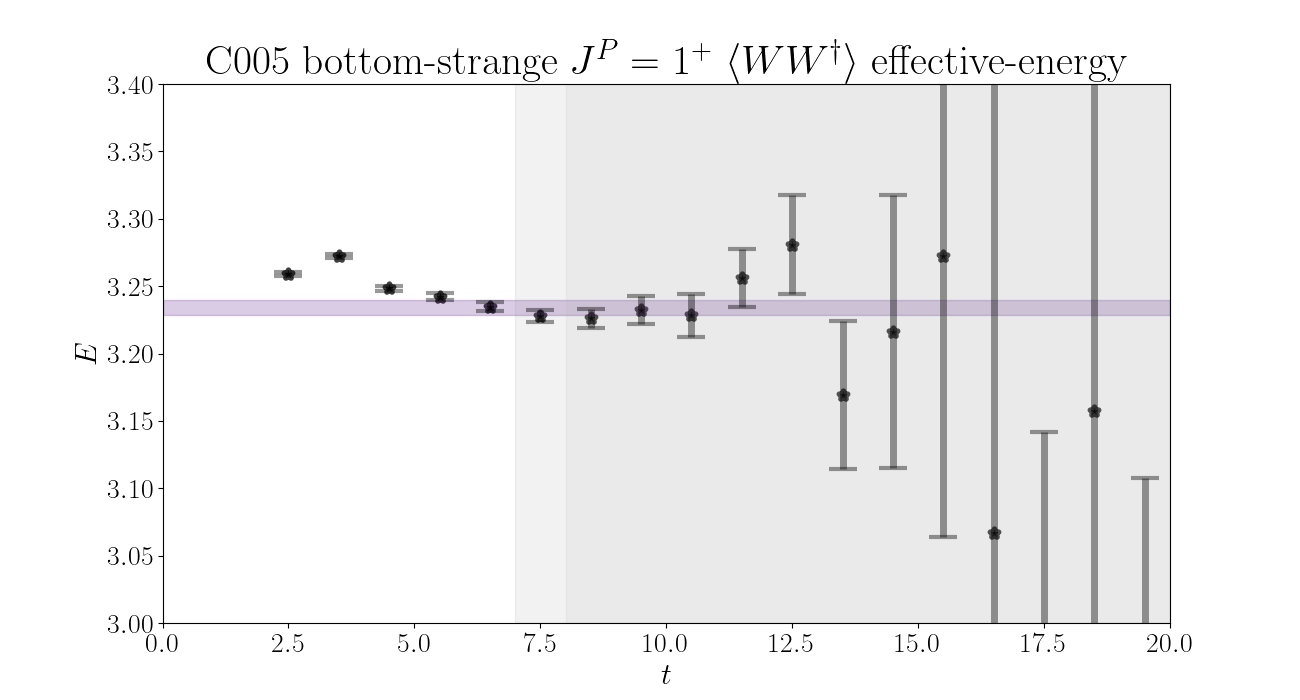}
    \caption{Like Fig.~\protect\ref{fig:F006wplots}, but for the C005 ensemble. \label{fig:JWplots-C005}}
\end{figure}

\begin{figure}[H]
    \centering
    \centering
    \includegraphics[width=0.49\linewidth]{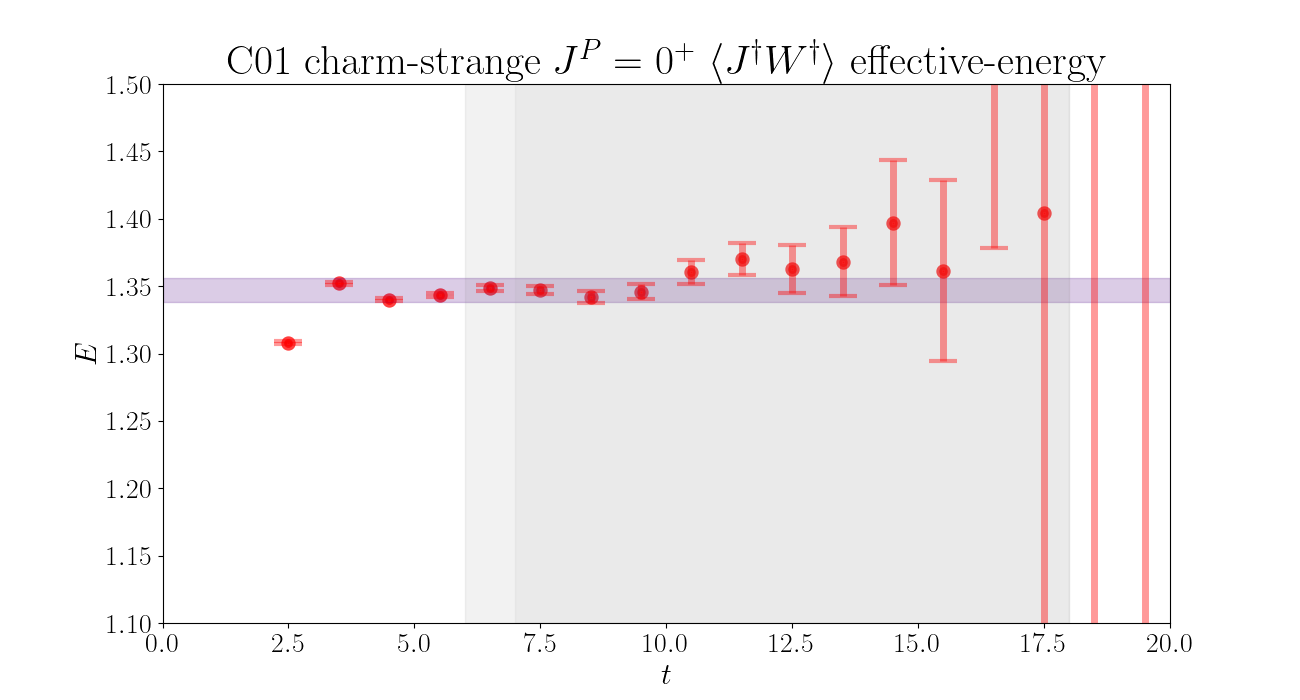}
    \hfill
    \includegraphics[width=0.49\linewidth]{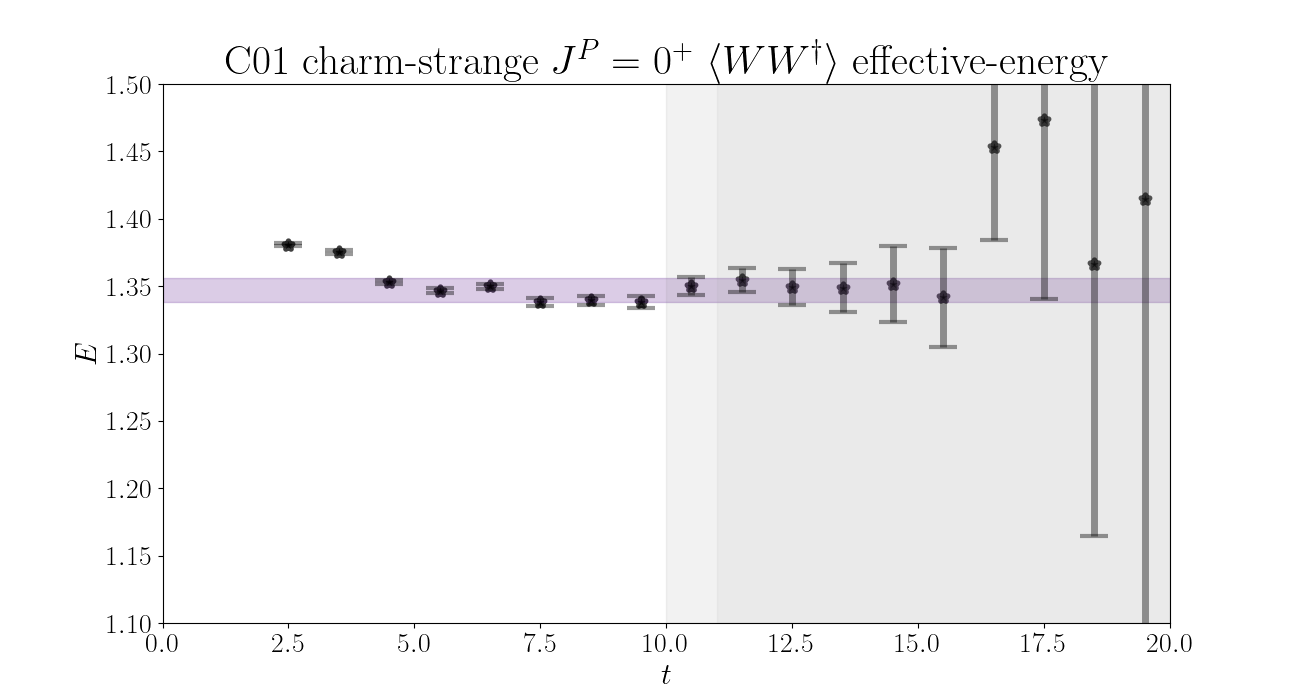}
    
    \includegraphics[width=0.49\linewidth]{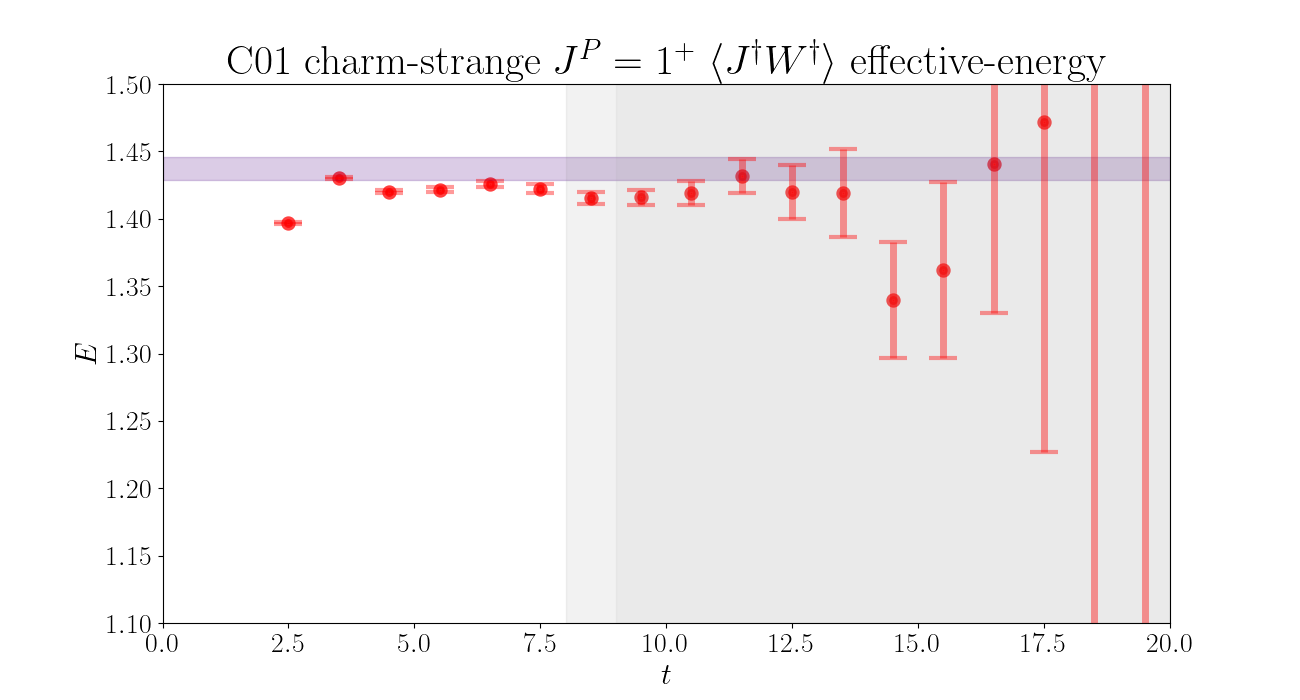}
    \hfill
    \includegraphics[width=0.49\linewidth]{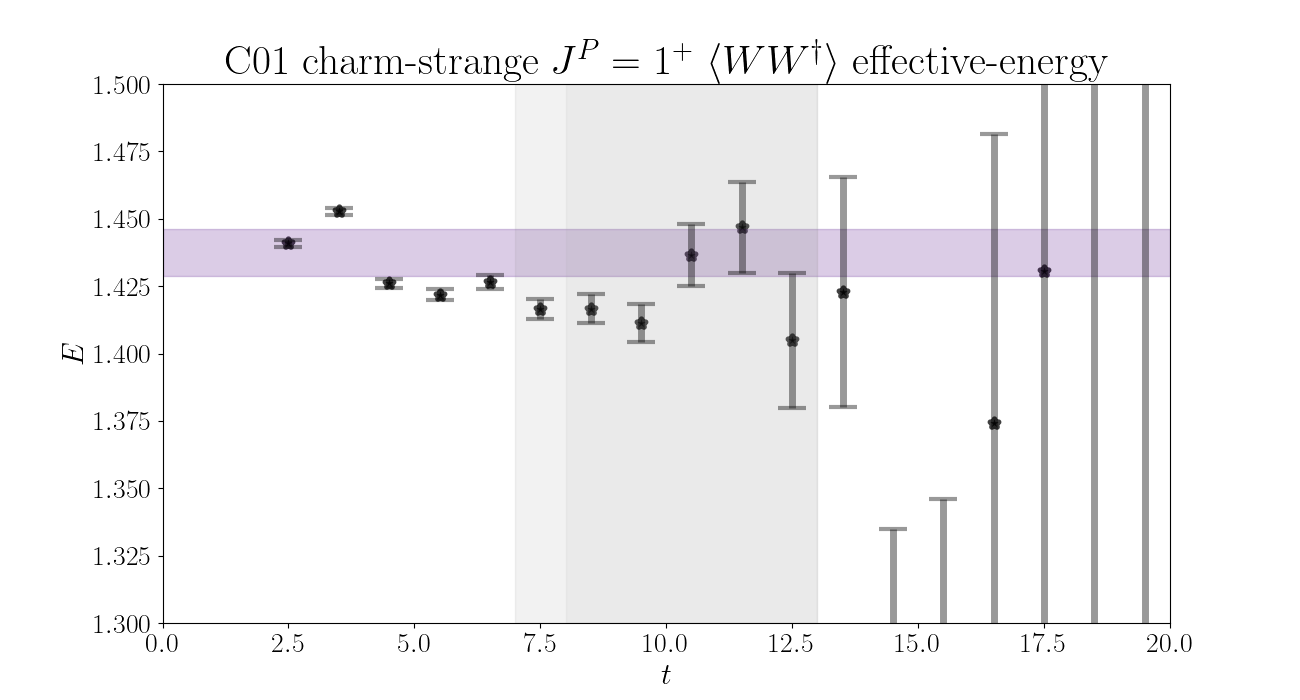}

    \includegraphics[width=0.49\linewidth]{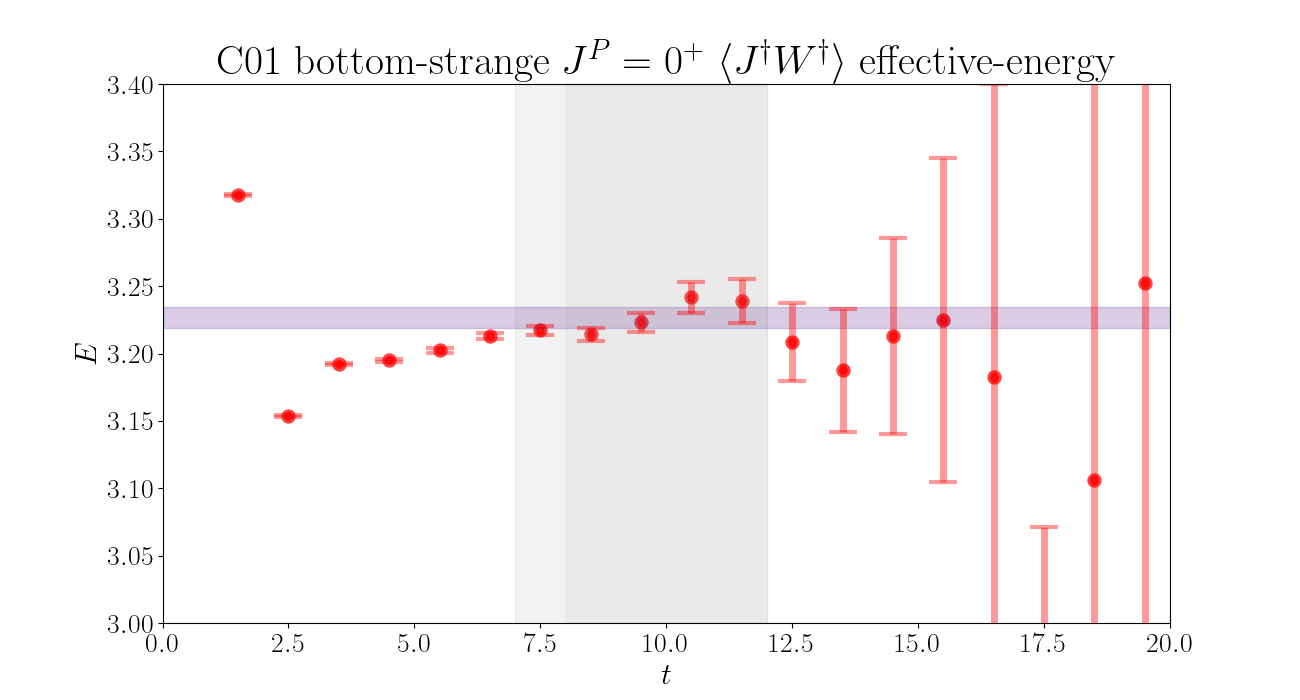}
    \hfill
    \includegraphics[width=0.49\linewidth]{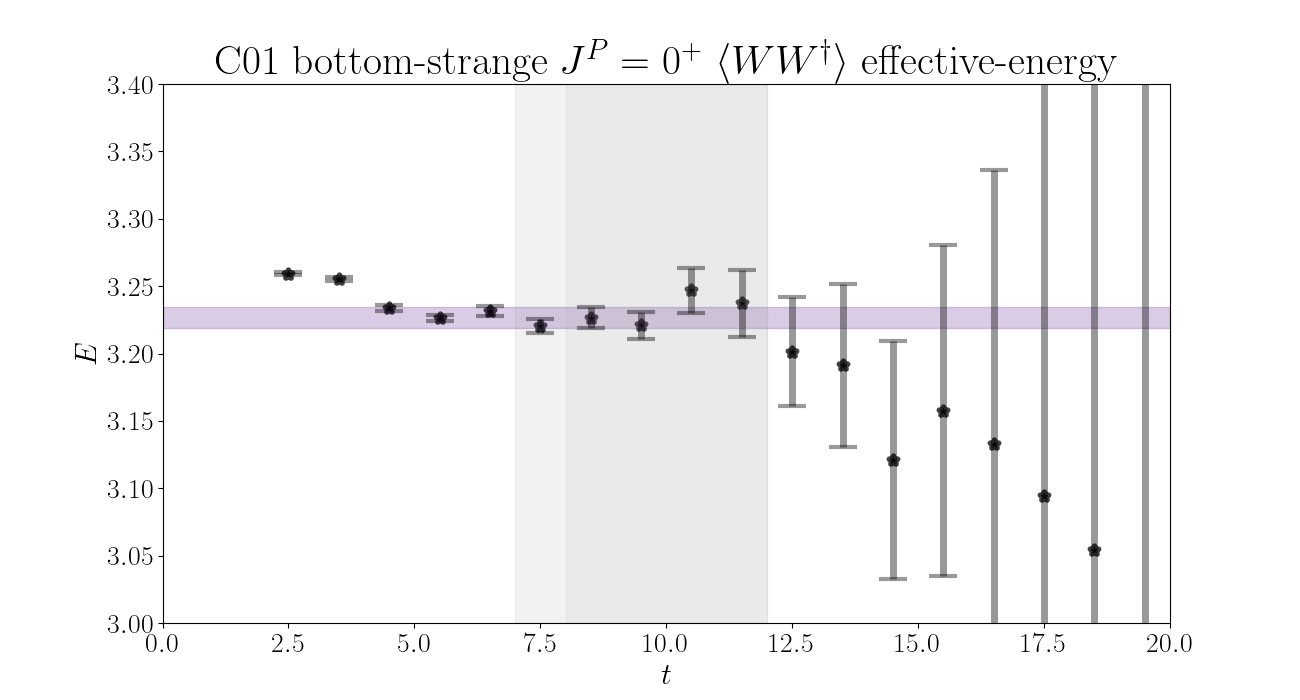}
    
    \includegraphics[width=0.49\linewidth]{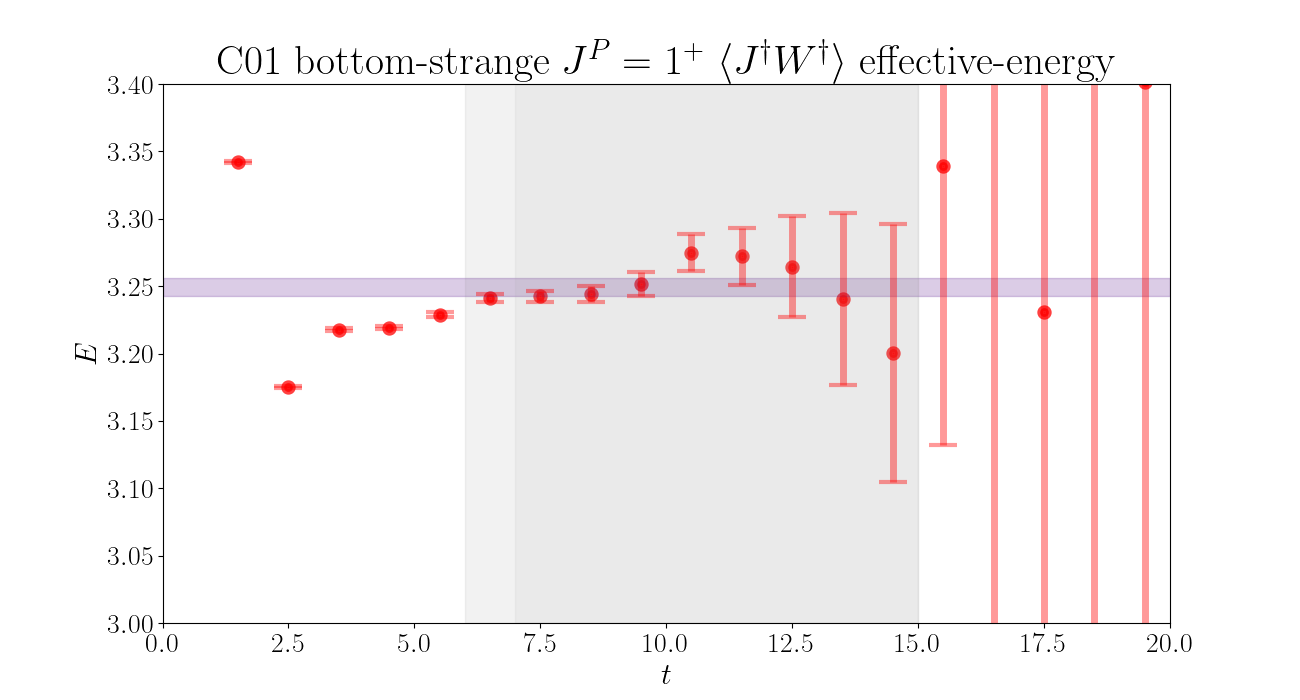}
    \hfill
    \includegraphics[width=0.49\linewidth]{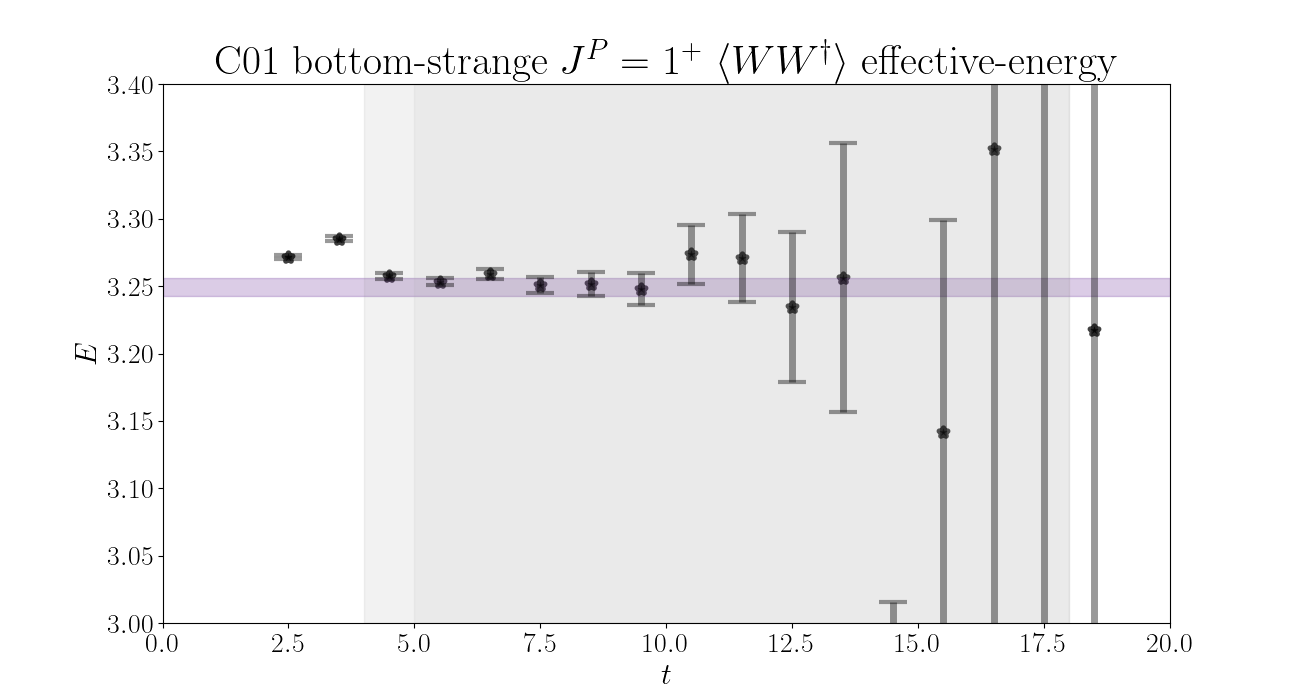}
    \caption{Like Fig.~\protect\ref{fig:F006wplots}, but for the C01 ensemble. \label{fig:JWplots-C01}}
\end{figure}

Plots of the positive-parity $\langle J^\dagger W^\dagger\rangle$ and $\langle W W^\dagger\rangle$ effective energies are provided here in Figs.~\ref{fig:JWplots-C00078}-\ref{fig:JWplots-C01} for all ensembles except F006, which can be found in main text in Fig.~\ref{fig:F006wplots}.

\subsection{Positive-Parity Chiral-Continuum Extrapolations using First-Order ERE}
\label{sec:chiralplots}

The plots of the chiral-continuum extrapolations of the ground-state binding energies obtained via the first-order ERE, discussed in Sec.~\ref{sec:chiralcontinuum}, are provided below in Figs.~\ref{fig:chiralcontEnextorderERE} and \ref{fig:chiralcontEnextorderEREnextorder}.

\begin{figure}[H]
    \centering
    \includegraphics[width=0.9\linewidth]{positive_parity/chiralcontiumlegend.pdf}
    \includegraphics[width=0.49\linewidth]{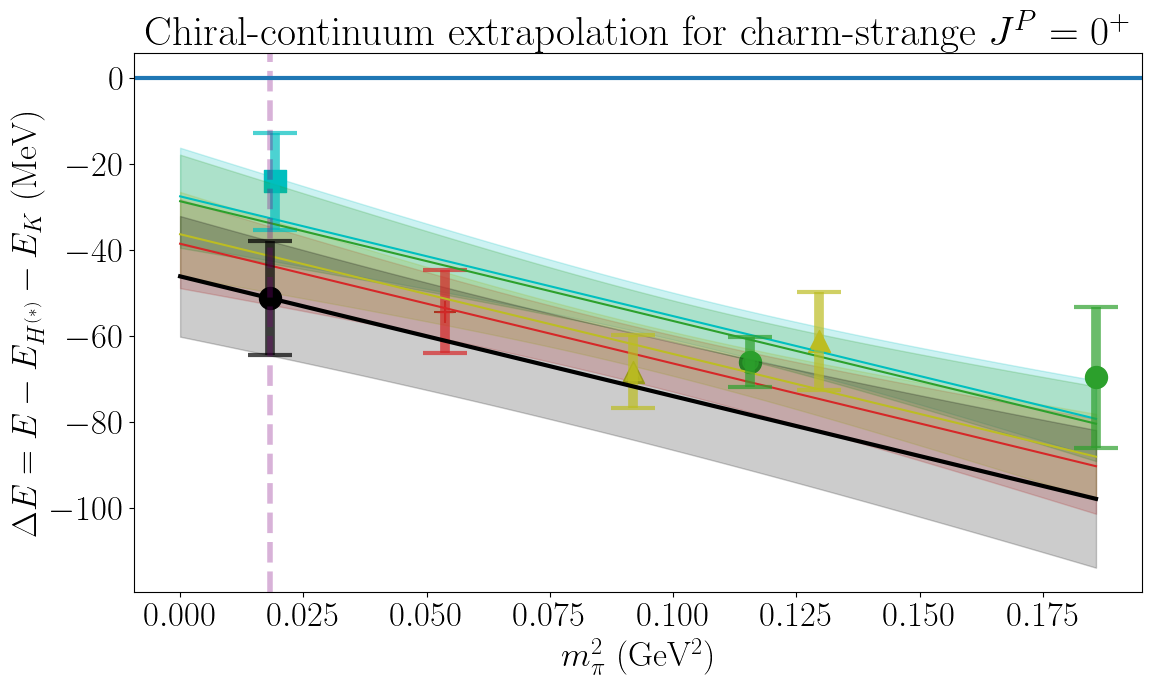}
    \hfill
    \includegraphics[width=0.49\linewidth]{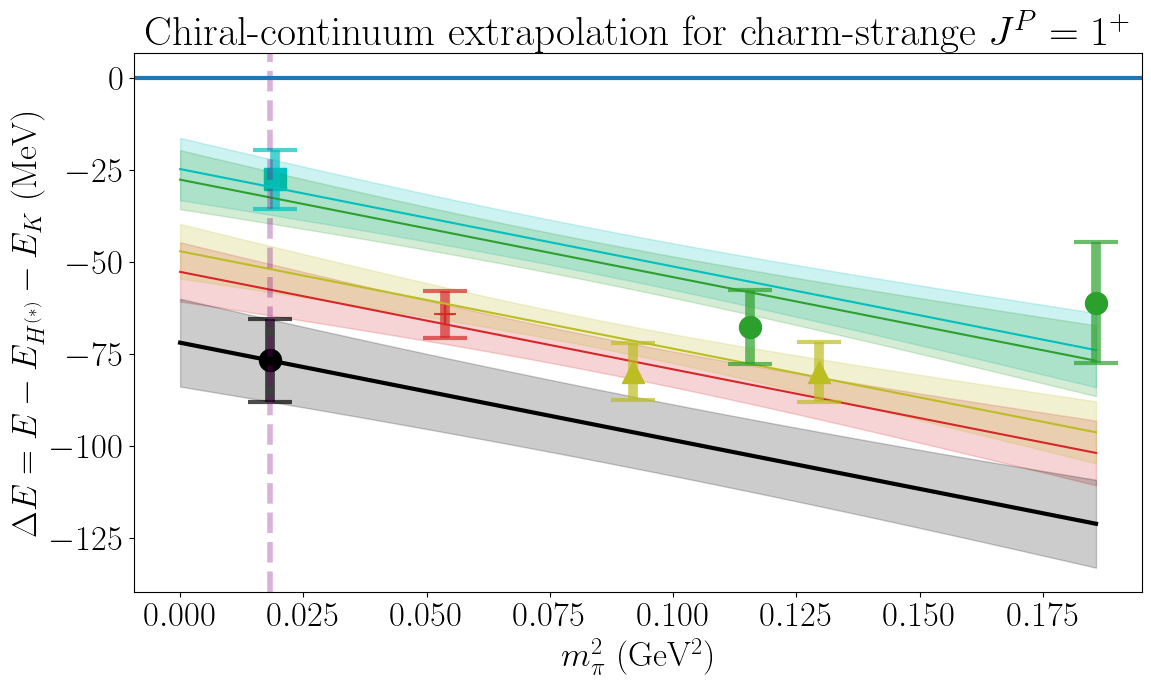}
    
    \includegraphics[width=0.49\linewidth]{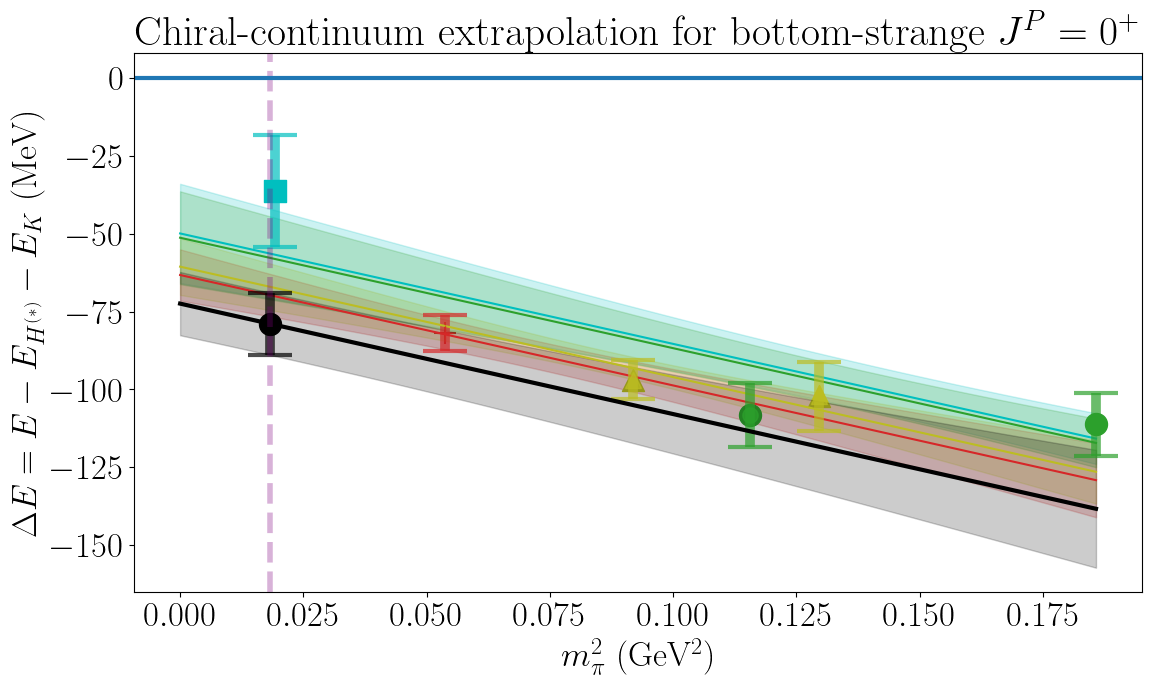}
    \hfill
    \includegraphics[width=0.49\linewidth]{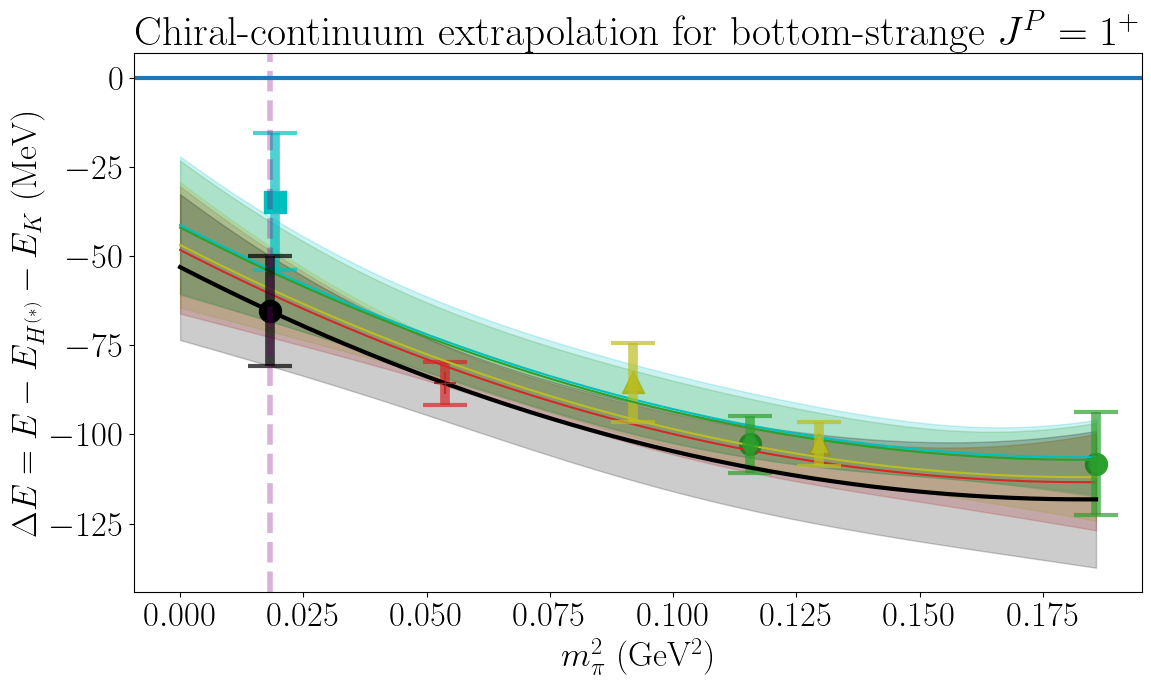}
    \caption{Linear-in-$m_\pi^2$ chiral-continuum extrapolations of the ground-state binding energies extracted from first-order ERE fits. The bands represent chiral extrapolations at fixed lattice spacing. The physical pion mass is indicated by the dashed line. The black band is the chiral extrapolation at $a=0$ and the black data point is $\Delta E_\text{phys}$.}
    \label{fig:chiralcontEnextorderERE}
\end{figure}

\begin{figure}[H]
    \centering
    \includegraphics[width=0.9\linewidth]{positive_parity/chiralcontiumlegend.pdf}
    \includegraphics[width=0.49\linewidth]{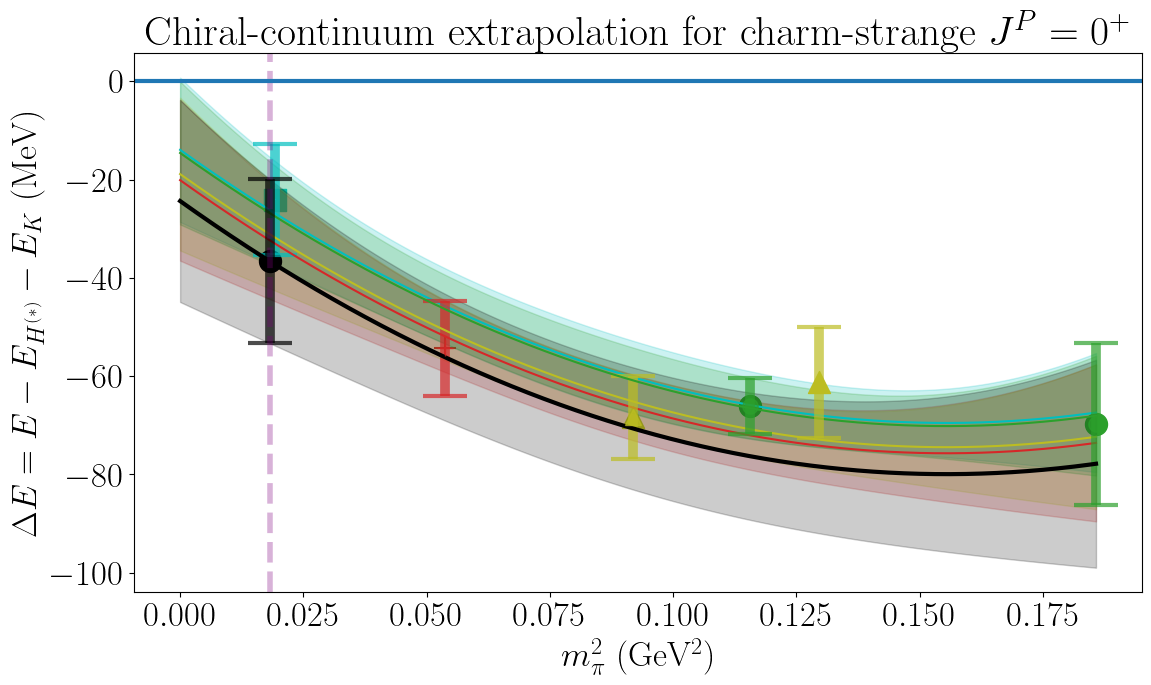}
    \hfill
    \includegraphics[width=0.49\linewidth]{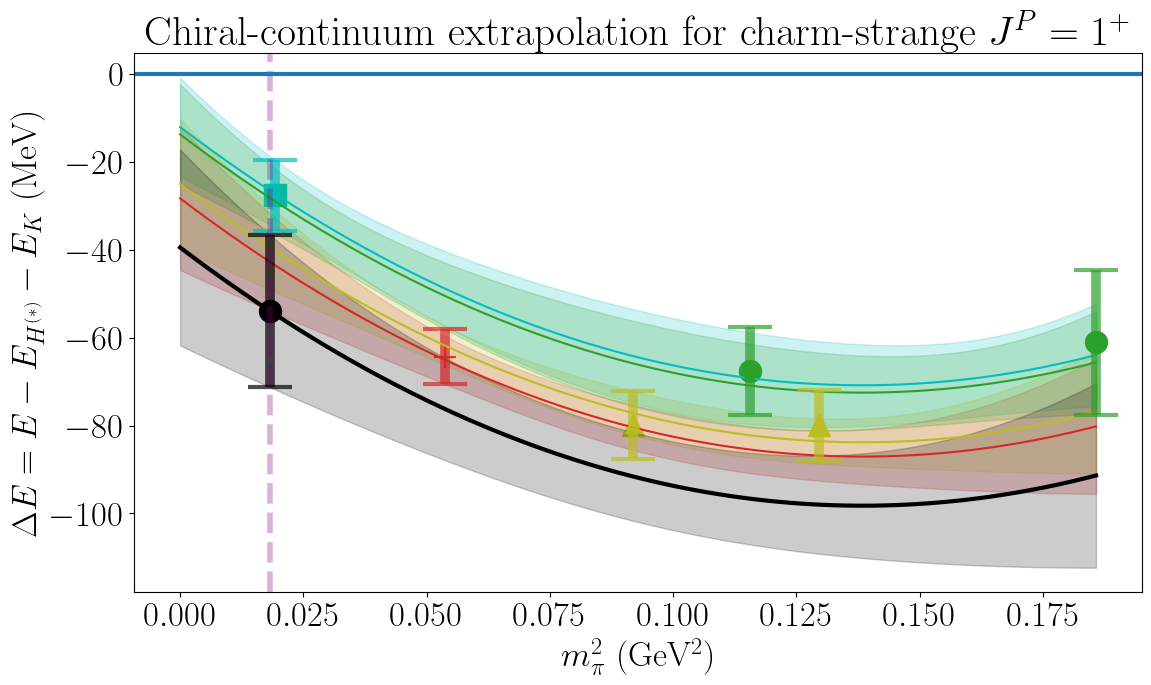}
    
    \includegraphics[width=0.49\linewidth]{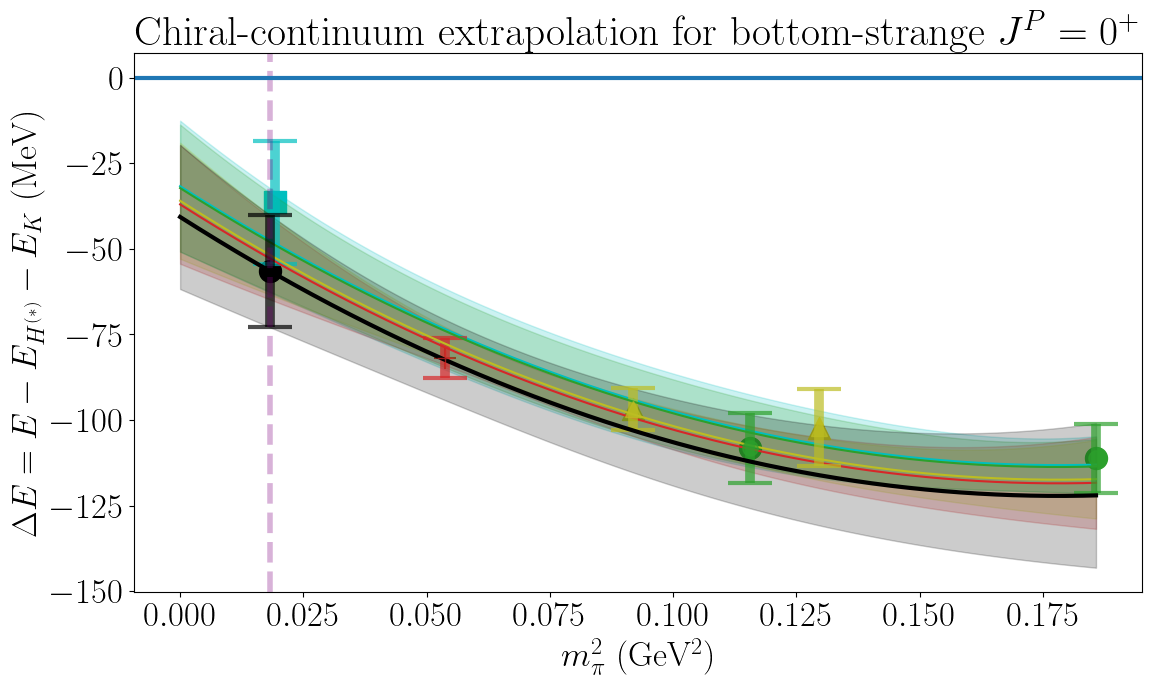}
    \hfill
    \includegraphics[width=0.49\linewidth]{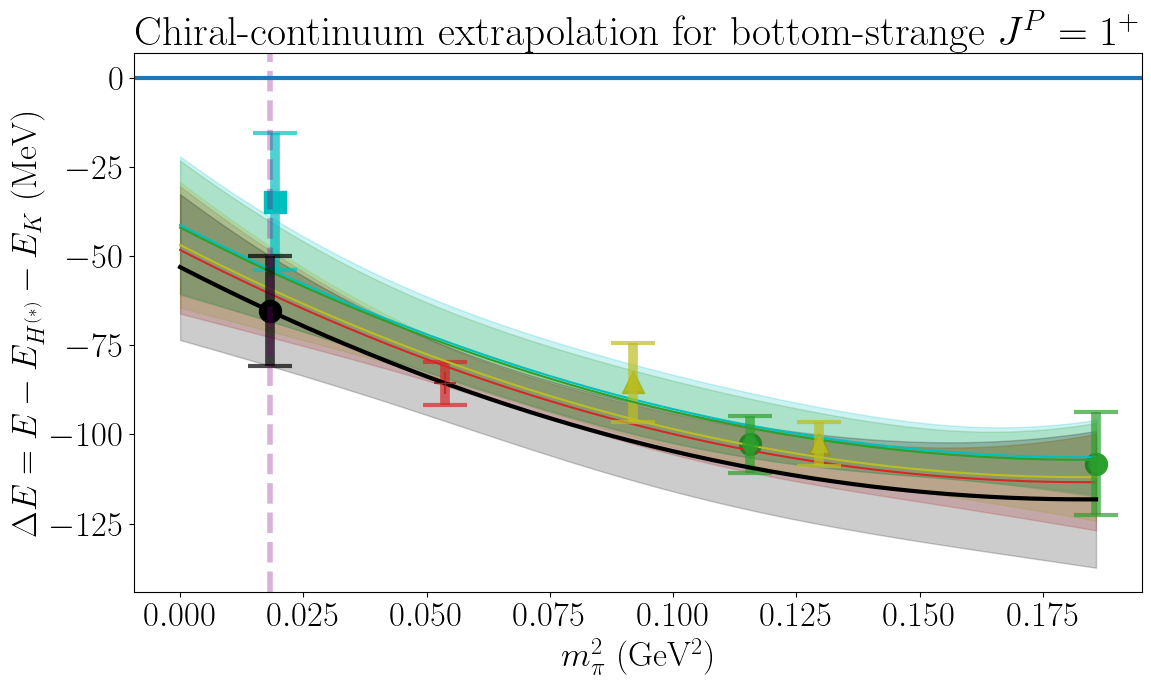}
    \caption{Quadratic-in-$m_\pi^2$ chiral-continuum extrapolations of the ground-state binding energies extracted from first-order ERE fits. The bands represent chiral extrapolations at fixed lattice spacing. The physical pion mass is indicated by the dashed line. The black band is the chiral extrapolation at $a=0$ and the black data point is $\Delta E_\text{phys}$.}
    \label{fig:chiralcontEnextorderEREnextorder}
\end{figure}

\bibliography{new}{}

\end{document}